\documentclass{cernyrep} 
\pdfoutput=1
\usepackage{texnames}
\usepackage{subfig}
\usepackage{graphicx}
\usepackage[T1]{fontenc}
\usepackage[bookmarks, colorlinks=true, linktoc=page, pdftex, linkcolor=blue, citecolor=blue, urlcolor=blue]{hyperref}
\usepackage{import}
\usepackage{longtable}
\usepackage{multirow}

\usepackage{amsmath}
\usepackage{amssymb}
\usepackage{hhline}

\begin{document}
\begin{flushright}
CERN-TH-2016-112
\end{flushright}
\title{Physics at a 100 TeV $pp$ collider: Standard Model processes}
 
\IfFileExists{../authors.tex}{\author{
M.L.~Mangano$^{1}$,\, 
G.~Zanderighi$^{1}$ (conveners),\,
J.A.~Aguilar Saavedra$^{2}$,\,
S. Alekhin$^{3,4}$,\,
S.~Badger$^{5}$,\,
C.W.~Bauer$^{6}$,\,
T.~Becher$^{7}$,\,
V.~Bertone$^{8}$,\,
M.~Bonvini$^{8}$,\,
S.~Boselli$^{9}$,\,
E.~Bothmann$^{10}$,\,
R.~Boughezal$^{11}$,\,
M.~Cacciari$^{12,13}$,\,
C.M.~Carloni Calame$^{14}$,\,
F.~Caola$^{1}$,\,
J.~M.~Campbell$^{15}$,\,
S.~Carrazza$^{1}$,\,
M.~Chiesa$^{14}$,\,
L.~Cieri$^{16}$,\,
F.~Cimaglia$^{17}$,\,
F.~Febres~Cordero$^{18}$,\,
P.~Ferrarese$^{10}$,\,
D.~D'Enterria$^{19}$,\,
G.~Ferrera$^{17}$,\,
X.~Garcia i Tormo$^{7}$,\,
M.~V.~Garzelli$^{3}$,\,
E.~Germann$^{20}$,\,
V.~Hirschi$^{21}$,\,
T.~Han$^{22}$,
H.~Ita$^{18}$,\,
B.~J\"ager$^{23}$,
S.~Kallweit$^{24}$,\,
A.~Karlberg$^{8}$,\, 
S.~Kuttimalai$^{25}$,\,
F.~Krauss$^{25}$,\,
A.~J.~Larkoski$^{26}$,\,
J.~Lindert$^{16}$,\,
G.~Luisoni$^{1}$,\,
P.~Maierh\"ofer$^{27}$,\,
O.~Mattelaer$^{25}$,\,
H.~Martinez$^{9}$,\,
S.~Moch$^{3}$,\,
G.~Montagna$^{9}$,\,
M.~Moretti$^{28}$,\,
P.~Nason$^{29}$,\,
O.~Nicrosini$^{14}$,\,
C.~Oleari$^{29}$,\,
D.~Pagani$^{30}$,\,
A.~Papaefstathiou$^{1}$,\,
F.~Petriello$^{31}$,\,
F.~Piccinini$^{14}$,\,
M.~Pierini$^{19}$,\,
T.~Pierog$^{32}$,\,
S.~Pozzorini$^{16}$,\,
E.~Re$^{33}$,\,
T.~Robens$^{34}$,\,
J.~Rojo$^{8}$,\,
R.~Ruiz$^{25}$,\,
K.~Sakurai$^{25}$,\,
G.~P.~Salam$^{1}$,\,
L.~Salfelder$^{23}$,\,
M.~Sch{\"o}nherr$^{28}$,\,
M.~Schulze$^{1}$,\,
S.~Schumann$^{10}$,\,
M.~Selvaggi$^{30}$,\,
A.~Shivaji$^{14}$,\,
A.~Siodmok$^{1,35}$,\,
P.~Skands$^{20}$,\,
P.~Torrielli$^{36}$,\,
F.~Tramontano$^{37}$,\,
I.~Tsinikos$^{30}$,\,
B.~Tweedie$^{22}$,\
A.~Vicini$^{17}$,\
S.~Westhoff$^{38}$,\
M.~Zaro$^{13}$,\,
D.~Zeppenfeld$^{32}$\,
\vspace*{2cm}
}

\institute{
$^{1}$ CERN, TH Department, CH-1211 Geneva, Switzerland. \\
$^{2}$ Departamento de F\'isica Te\'orica y del Cosmos,
Universidad de Granada, E-18071 Granada, Spain.\\
$^{3}$ 
II. Institute for Theoretical Physics, University of Hamburg,
Germany\\
$^{4}$  Institute for High Energy Physics, 142281 Protvino, Moscow
region, Russia. \\
$^{5}$ Higgs Centre for Theoretical Physics, School of Physics and Astronomy, The University of Edinburgh, Edinburgh EH9 3JZ, Scotland, UK.\\
$^{6}$ Ernest Orlando Lawrence Berkeley National Laboratory (LBNL), Physics
Division and University of California, Berkeley, CA 94720, USA\\
$^{7}$ Albert Einstein Center for Fundamental Physics, Institut f\"ur
Theoretische Physik, Universit\"at Bern, 
Sidlerstrasse 5, CH-3012 Bern, Switzerland \\
$^{8}$ Rudolf Peierls Centre for Theoretical Physics,
University of Oxford,
1 Keble Road, Oxford OX1 3NP, United Kingdom \\
$^{9}$ Dipartimento di Fisica, Universit\`a di Pavia and INFN Sezione di
Pavia, via A. Bassi 6, I-27100 Pavia, Italy \\
$^{10}$ II. Physikalisches Institut, Georg-August-Universit{\"a}t G{\"o}ttingen, 37077 G{\"o}ttingen, Germany.\\
$^{11}$ 
Argonne National Laboratory, 
High Energy Physics Division, Bldg 362, 
9700 South Cass Avenue,
Argonne, IL 60439, USA\\
$^{12}$ Universit\'e Paris Diderot, F-75013 Paris, France \\
$^{13}$ Sorbonne Universit\'es, UPMC Univ. Paris 06, UMR 7589, 
LPTHE, F-75005, Paris, France; CNRS, UMR 7589, LPTHE, F-75005, Paris, France\\
$^{14}$ INFN Sezione di Pavia, via A. Bassi 6, I-27100 Pavia, Italy\\
$^{15}$ Theory Group, Fermilab, PO Box 500, Batavia, IL, USA \\
$^{16}$ 
Universit\"at Z\"urich, Physik-Institut, Winterthurerstrasse 190, CH-8057
Z\"urich, Switzerland\\
$^{17}$ Dipartimento di Fisica, Universit\`a degli Studi di Milano 
and INFN, Sezione di Milano, Via Celoria 16, I-20133 Milano, Italy\\
$^{18}$ Fakult\"at f\"ur Mathematik und Physik, Physikalisches Institut,
Hermann-Herder-Straße 3, 79104 Freiburg, Germany \\
$^{19}$ CERN, EP Department, CH-1211 Geneva, Switzerland. \\
$^{20}$ School of Physics and Astronomy, Monash University, VIC-3800, Australia \\
$^{21}$ SLAC, National Accelerator Laboratory, 2575 Sand Hill Road, Menlo Park, CA 94025-7090, USA\\
$^{22}$ Department of Physics and Astronomy, Univ. of Pittsburgh, Pittsburgh, PA 15260, USA \\
$^{23}$ Institute for Theoretical Physics, University of T\"ubingen,  Auf der Morgenstelle 14, 72076 T\"ubingen, Germany.\\
$^{24}$
Institut f\"ur Physik \& PRISMA Cluster of Excellence,
Johannes Gutenberg Universit\"at, 55099 Mainz, Germany\\
$^{25}$ Institute for Particle Physics Phenomenology, Durham
University, Durham DH1 3LE, United Kingdom \\
$^{26}$ Center for the Fundamental Laws of Nature, Harvard University, Cambridge, MA 02138 USA\\
$^{27}$
Physikalisches Institut, Albert-Ludwigs-Universit\"at Freiburg,
79104 Freiburg, Germany\\
$^{28}$ Dipartimento di Fisica e Scienze della Terra, Universit\`a di Ferrara 
and INFN, Sezione di Ferrara, v. Saragat 1, I-44100 Ferrara, Italy\\
$^{29}$ Universit\`a di Milano-Bicocca and INFN, Sezione di Milano-Bicocca, Piazza della Scienza 3, 20126 Milano, Italy\\
$^{30}$ Center for Cosmology, Particle Physics and Phenomenology
(CP3), Universit\'e Catholique de Louvain, Chemin du Cyclotron 2, 
B-1348 Louvain-la-Neuve, Belgium\\
$^{31}$ 
Northwestern University, Department of Physics and Astronomy, 2145
Sheridan Road, 
Evanston, Illinois 60208-3112, USA\\
$^{32}$ Institute for Theoretical Physics, Karlsruhe Institute of
Technology,  76128 Karlsruhe, Germany.\\
$^{33}$ LAPTH, Univ. de Savoie, CNRS, B.P.110, Annecy-le-Vieux F-74941, France
\\
$^{34}$ IKTP, TU Dresden, Zellescher Weg 19, 01069 Dresden, Germany\\
$^{35}$ Institute of Nuclear Physics, Polish Academy of Sciences, ul. Radzikowskiego 152, 31-342 Krakow, Poland\\
$^{36}$ Dipartimento di Fisica, Universit\`a di Torino, and INFN, Sezione di Torino,
Via P.~Giuria~1, I-10125, Turin, Italy \\
$^{37}$ Universit\`a di Napoli ``Federico II'' and INFN, Sezione di Napoli, 80126 Napoli, Italy\\
$^{38}$ Institut f\"ur Theoretische Physik, Universit\"at Heidelberg, 69120 Heidelberg, Germany.\\
}
}{}

\maketitle 

\begin{abstract}
This report summarises the properties of Standard Model processes at
the 100~TeV $pp$ collider. We document the production rates
and typical distributions for a number of benchmark Standard Model
processes, and discuss new dynamical phenomena arising at the highest
energies available at this collider. 
We discuss the intrinsic physics interest in the measurement of these
Standard Model processes, as well as their role as backgrounds for New
Physics searches.
\end{abstract}
 
\tableofcontents

\newcommand{\met}{E\!\!\!\!/_T} 
\def\iab{ab$^{-1}$}
\def\ifb{fb$^{-1}$}
\def\ipb{pb$^{-1}$}
\def \gsim{\mathrel{\vcenter
     {\hbox{$>$}\nointerlineskip\hbox{$\sim$}}}}
\def \lsim{\mathrel{\vcenter
     {\hbox{$<$}\nointerlineskip\hbox{$\sim$}}}}

%
\clearpage 
\section{Foreword}
A 100~TeV $pp$ collider is under consideration, by the high-energy
physics community~\cite{benedikt,cepc_website}, as an important target
for the future development of our field, following the completion of
the LHC and High-luminosity LHC physics programmes. The physics
opportunities and motivations for such an ambitious project were
recently reviewed in~\cite{Arkani-Hamed:2015vfh}. The general
considerations on the strengths and reach of very high energy hadron
colliders have been introduced long ago in the classic pre-SSC EHLQ
review~\cite{Eichten:1984eu}, and a possible framework to establish
the luminosity goals of such accelerator was presented recently
in~\cite{Hinchliffe:2015qma}.

The present document is the result of an extensive study, carried out
as part of the Future Circular Collider (FCC) study towards a
Conceptual Design Report, which includes separate Chapters dedicated
to Standard Model physics (this paper), physics of the Higgs boson and
electroweak (EW) symmetry breaking~\cite{Higgs-report}, physics beyond
the Standard Model~\cite{BSM-report}, physics of heavy ion
collisions~\cite{Dainese:2016gch} and physics with the FCC injector
complex~\cite{inj-report}. Studies on the physics programme of an
$e^+e^-$ collider (FCC-ee) and $ep$ collider (FCC-eh) at the FCC
facility are proceeding in parallel, and preliminary results are
documented in~\cite{Gomez-Ceballos:2013zzn} (for FCC-ee) and
in~\cite{AbelleiraFernandez:2012cc} (for the LHeC precursor of
FCC-eh).

\section{Standard Model at 100~TeV: Introduction}
\label{sec:intro}
Standard Model particles play multiple roles in the 100~TeV collider
environment. In the context of BSM phenomena, and for most scenarios,
new BSM particles eventually decay to the lighter SM states, which
therefore provide the signatures for their production. BSM
interactions, furthermore, can influence the production properties of
SM particles, and the observation of SM final states can probe the
existence of an underlying BSM dynamics. SM processes therefore
provide both signatures and potential backgrounds for any exploration
of BSM phenomena. SM backgrounds have an impact on BSM studies in
different ways: on one side they dilute, and can hide, potential BSM
signals; on the other, SM processes influence the trigger strategies,
since they determine the irreducible contributions to trigger rates
and may affect the ability to record data samples of interest to the
BSM searches.

The observation of SM processes has also an interest per se. The huge
rates available at 100~TeV allow, in principle, to push to new limits
the exploration of rare phenomena (e.g. rare decays of top quarks or
Higgs bosons), the precision in the determination of SM parameters,
and the test of possible deviations from SM dynamics. The extremely
high energy kinematical configurations probe the shortest distances,
and provide an independent sensitivity to such deviations.

Finally, SM processes provide a necessary reference to benchmark the
performance of the detectors, whether in the context of SM
measurements, or in the context of background mitigation for the BSM
searches.

In this Chapter we review the key properties of SM processes at
100~TeV, having in mind the above considerations. This will serve as a
reference for future studies, and to stimulate new ideas on how to
best exploit the immense potential of this collider. We shall focus on
the production of key SM objects, such as jets, heavy quarks, gauge
bosons. The SM Higgs boson will be discussed in the Higgs Chapter of
this report~\cite{Higgs-report}.  We shall not address issues like the
current or expected precision relative to given processes. On one
side, and with some well understood exceptions notwithstanding,
leading-order calculations are typically sufficient to give a reliable
estimate of the production rates, and assess possible implications for
trigger rates, background contributions, and detector
specifications. On the other, any statement about the precision of
theoretical calculations made today will be totally obsolete by the
time this collider will operate, and assumptions about the accuracy
reach cannot but be overly conservative.

\clearpage 
\section{Parton distribution functions\footnote{Editor: J.~Rojo}}

\label{sec:pdf}
\def\be{\begin{equation}}
\def\ee{\end{equation}}
\def\bea{\begin{eqnarray}}
\def\eea{\end{eqnarray}}
\def\bi{\begin{itemize}}
\def\ei{\end{itemize}}
\def\ben{\begin{enumerate}}
\def\een{\end{enumerate}}
\def\la{\left\langle}
\def\ra{\right\rangle}
\def\lc{\left[}
\def\rc{\right]}
\def\lp{\left(}
\def\rp{\right)}
\def\as{\alpha_S}

\def\gev{\,\textrm{GeV}}
\def\tev{\,\textrm{TeV}}
\def\ord{\mathcal{O}}
\def\aNLO{{\sc\small MadGraph5\_aMC@NLO}}

\subsection{Introduction}
\label{sec:pdf_intro}

The accurate determination of the parton distribution functions (PDFs)
of the proton is an essential ingredient of the LHC physics
program~\cite{Forte:2013wc,Ball:2012wy,Rojo:2015acz,Butterworth:2015oua,Accardi:2016ndt},
and will be even more so at any future higher-energy hadron collider.
In particular, a new hadron collider with a center-of-mass energy of
$\sqrt{s}=100$ TeV will probe PDFs in several currently unexplored
kinematical regions, such as the ultra low-$x$ region, $x\lsim
10^{-5}$, or the region of very large momentum transfers, $Q^2 \ge
(10~{\rm TeV})^2$.
In addition, concerning the phenomenological implications of PDFs, the
situation is much more complex (and interesting) than simply assuming
that the FCC can be treated as a rescaled version of the LHC.
Indeed, understanding PDFs at 100 TeV involves addressing a number of
qualitatively new phenomena that have received limited attention up to
now.

It is extremely difficult to forecast what the status of our knowledge
about the proton structure will be in 20 or 25 years from now.
Progress in PDF
determinations~\cite{Ball:2014uwa,Dulat:2015mca,Harland-Lang:2014zoa,Alekhin:2013nda,Abramowicz:2015mha,Accardi:2016qay,Jimenez-Delgado:2014twa}
will strongly depend, on the one hand, on the full exploitation of the
information on PDF-sensitive measurements contained by LHC Run I and
Run II data~\cite{Rojo:2015acz}, as well as by the corresponding
HL-LHC measurements, and on the other hand, on the progress in
higher-order calculational techniques allowing to include many LHC
differential distributions in the PDF analysis at NNLO (and beyond),
see~\cite{Currie:2013dwa,Czakon:2015owf,Boughezal:2015dra} for some
recent examples.

Moreover, progress in global PDF analysis can also be driven by
methodological improvements, for instance in more efficient methods to
parametrize PDFs, or better techniques to estimate experimental,
model, and theoretical PDF uncertainties.
Another important factor to take into account is the fact that our
understanding of the proton structure would be substantially improved
in the case a new electron-nucleon collider would be operative before
the start-up of the FCC operations, such as the Large Hadron Electron
Collider (LHeC) at CERN~\cite{AbelleiraFernandez:2012cc} or the
Electron Ion Collider (EIC) in the U.S.A.~\cite{Boer:2011fh}.
In addition, in the long term, progress in non-perturbative lattice
calculations might also shed further light on the proton structure and
provide a useful complement to global PDF fits.

For these reasons, in this section we will concentrate on qualitative
aspects of PDFs that are important for a exploratory evaluation of the
physics potential of the FCC, which is the main goal of this report.
In particular we will focus on:
\begin{itemize}
\item What are the most relevant generic differences for PDFs when
  moving from the LHC energies, $\sqrt{s}=$14 TeV, to the FCC
  energies, $\sqrt{s}=$100 TeV.
  
  This includes the kinematical coverage in the $(x,Q^2)$ plane of a
  100 TeV collider, the ratios of PDF luminosities and their
  uncertainties between $\sqrt{s}=$100 TeV and $\sqrt{s}=$14 TeV, and
  the assessment of how available PDF sets extrapolate into the new
  kinematical regions covered by the FCC.
\item Qualitatively new phenomena about PDFs and DGLAP
  evolution that, while not
  essential for the exploitation of the LHC data, might become relevant
 at the extreme energies at which the FCC would operate.
 
 These include QED and weak effects in the PDF evolution, high-energy
 resummation effects, and the possibility of treating the top quark as
 a massless parton.
 In addition, we also study the role of photon-initiated contributions
 for electroweak processes at 100 TeV.
\end{itemize}

The outline of this section is the following.
In Sect.~\ref{sec:pdf_kinematics} we quantify the coverage of PDFs at
the FCC in the $(x,Q^2)$ plane, and study the behavior of PDFs in the
extreme large-$x$, large-$Q^2$ and small-$x$ regions accessible at the
FCC.
In Sect.~\ref{sec:pdf_lumi} we present a comparison of PDF
luminosities at 100 TeV for the most updated global PDF sets, and
compute various ratios of parton luminosities between 100 TeV and 14
TeV.
In Sect.~\ref{sec:pdf_top} we study the validity of the massless
approximation for the top quark at a 100 TeV collider.
In Sect.~\ref{sec:pdf_photon} we quantify the role of photon and
lepton-initiated contributions at 100 TeV, relevant when electroweak
corrections are accounted for.
In Sect.~\ref{sec:pdf_ew} we explore the possibility of treating
electroweak gauge bosons as massless and their inclusion into the
DGLAP evolution equations.
Finally in Sect.~\ref{sec:pdf_resum} we discuss the possible relevance
of high-energy (small-$x$) resummation effects for a 100 TeV collider.


\subsection{PDFs and their kinematical coverage at 100 TeV}
\label{sec:pdf_kinematics}

We begin by quantifying the kinematical coverage in the $(x,M_X)$
plane that PDFs probe in a 100 TeV hadron collider, with $M_X$ being
the invariant mass of the produced final states.
In Fig.~\ref{fig:kinplot} we represent the kinematical coverage in the
$(x,M_X)$ plane of a $\sqrt{s}=100$ TeV hadron collider compared with
the corresponding coverage of the LHC at $\sqrt{s}=14$ TeV.
  The dotted lines indicate regions of constant rapidity $y$ at the FCC.
  In this plot, 
  we also indicate the relevant $M_X$ regions for phenomenologically
  important processes, from low masses (such as Drell-Yan or low $p_T$ jets),
  electroweak scale processes (such as Higgs, $W,Z$, or top production),
  and possible new
  high-mass particles (such as a 2 TeV squark or a 20 TeV $Z'$).
  
\begin{figure}[t]
\centering
\includegraphics[width=0.93\textwidth]{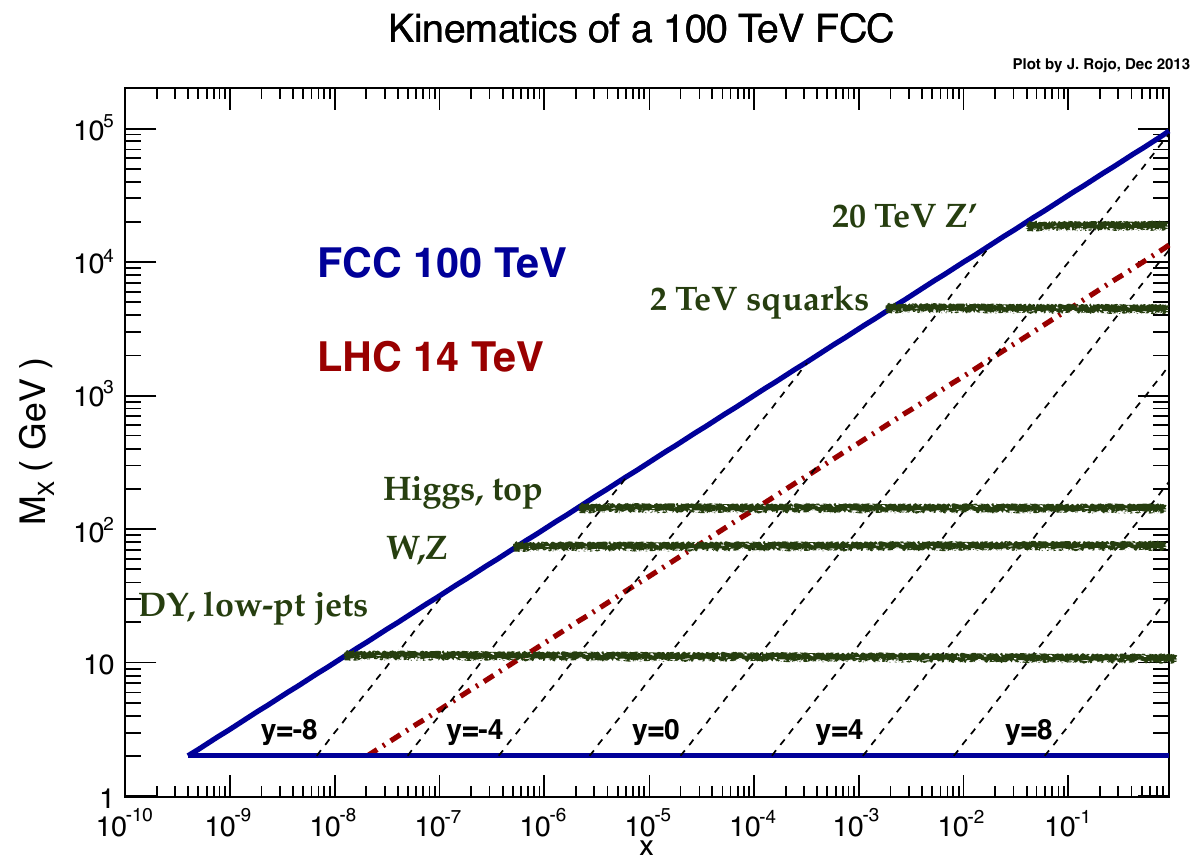}
\caption{\small Kinematical coverage in the $(x,M_X)$ plane of a
  $\sqrt{s}=100$ TeV
  hadron collider (solid blue line),
  compared with the corresponding coverage of the LHC at
  $\sqrt{s}=14$ TeV (dot-dashed red line).
  The dotted lines indicate regions of constant rapidity $y$ at the FCC.
  We also indicate the relevant $M_X$ regions for phenomenologically
  important processes, from low masses (Drell-Yan, low $p_T$ jets),
  electroweak scale processes (Higgs, $W,Z$, top), and possible new
  high-mass particles (squarks, $Z'$).  }
\label{fig:kinplot}
\end{figure}

In the low-mass region, for $M_X\le 10$ GeV, PDFs would be probed down
to $x\simeq 5\cdot 10^{-5}$ in the central region, $y\simeq 0$, and
down to $x\simeq 5 \cdot 10^{-7}$ at forward rapidities, $y\simeq 5$.
At even forward rapidities, for example those that can be probed by
using dedicated detectors down the beam pipe, PDFs could be probed
down to $x\simeq 10^{-8}$.
While these extreme regions of very low $x$ are not relevant for
neither electroweak scale physics nor for high-mass New Physics
searches, they are crucial for the tuning of soft and semi-hard
physics in Monte Carlo event generators~\cite{Skands:2014pea} and
therefore it is important to ensure that the PDFs exhibit a sensible
behaviour in this region.
Moreover, forward instrumentation would also be relevant for the
measurement of the total $pp$ cross-section at 100 TeV as well as to
provide input for the modelling of ultra-high energy cosmic ray
collisions~\cite{d'Enterria:2011kw}.
The prospects for soft physics at the FCC is studied in detail in
Sect.~\ref{sec:mb} of this report.

Concerning the production of electroweak particles such as weak gauge
bosons, the Higgs boson and top quarks, PDFs are probed down to
$x\simeq 5\cdot 10^{-4}$ in the central region, $y\simeq 0$, and down
to $x\simeq 2 \cdot 10^{-6}$ at forward rapidities, $y\simeq 5$.
This indicates that a good coverage of the forward region is also
instrumental for electroweak scale physics, whose production is much
less central than at the LHC.
In the case of Higgs production, if the Higgs can be reconstructed up
to rapidities of $y\simeq 4$, then this process would probe PDFs down
to $x\simeq 10^{-5}$.
Therefore, at a 100 TeV hadron collider a good knowledge of small-$x$
PDFs becomes crucial not only for soft and semi-hard physics, or for
low scale processes such as low-mass Drell-Yan or charm production,
but also for electroweak scale processes.

In the high-invariant mass region, $M_X\ge 5$ TeV, only medium and
large-$x$ PDFs would be probed, and these are currently known with
reasonable accuracy, except for very high $M_X$ values.
For instance, for the pair-production of 2 TeV squarks, only the
knowledge of PDFs for $x\gsim 10^{-3}$ is required.
The production of multi-TeV heavy particles is of course very central,
requiring instrumentation only down to $|y|\simeq 3$ at most.
For the heavier particles that can be probed at the FCC, such as a 20
or 30 TeV $Z'$, PDFs have large uncertainties since the very large-$x$
region is being probed, and this region is affected by the lack of
direct constraints, as we discuss below.

In Table~\ref{tab:kincov} we summarize the kinematical coverage in the
$(x,M_X)$ plane for various phenomenologically important processes at
the FCC, both for central, intermediate and forward rapidities.
For each value of the invariant mass $M_X$ and the absolute rapidity
$|y|$, the smallest value of Bjorken-$x$ required corresponds to
$x_{\rm min}=(M_X/\sqrt{s})\exp(-|y|)$.
This table conveys a similar message to that of
Fig.~\ref{fig:kinplot}: at a 100 TeV hadron collider, accurate
knowledge of PDFs is required in a very wide kinematical region,
ranging from ultra low-$x$ to very large-$x$, and from momentum
transfers close to $\Lambda_{\rm QCD}$ up to the highest values where
the FCC has sensitivity for new heavy particles, $M_X\simeq 50$ TeV.
That is, a huge range spanning 8 orders of magnitude in $x$ and 10 in
$Q^2$.

\begin{table}[h]
  \centering
  \small
  \begin{tabular}{|c|c|c|c|c|}
    \hline
    Process  &   $M_X$  & \multicolumn{3}{c|}{$x_{\rm min}$} \\
      &    &  $y=0$ & $|y|=2$ & $|y|=4$  \\
    \hline
    \hline
    Soft QCD  &  \multirow{3}{*}{1~(10) GeV}  & \multirow{3}{*}{$10^{-5}~(10^{-4})$}  &
    \multirow{3}{*}{$ 1.4 \cdot 10^{-6}~(1.4 \cdot 10^{-5})$}   &  \multirow{3}{*}{
    $ 1.8 \cdot 10^{-7}~(1.8 \cdot 10^{-6})$} \\
    Charm pair production  & & & & \\
    Low-mass Drell-Yan  & & & & \\
    \hline
    $W$ and $Z$ production &  \multirow{3}{*}{80~(400) GeV}  & \multirow{3}{*}{$8\cdot 10^{-4}~(4\cdot 10^{-3})$}  &
    \multirow{3}{*}{$1.1\cdot 10^{-4}~(5.4\cdot 10^{-4})$}   &  \multirow{3}{*}{$1.5\cdot 10^{-5}~(7.3\cdot 10^{-5})$} \\
    Top pair production  & & & & \\
    Inclusive Higgs  & & & & \\
    \hline
    Heavy New Physics  & 5~(25) TeV  &  0.05~(0.25)  & 0.007~(-)  & --  \\
     \hline
  \end{tabular}
  \caption{\small Kinematical coverage in the $(x,M_X)$ plane for representative
    processes at a 100 TeV hadron collider.
    For each type of process (low mass, electroweak scale processes, and heavy new physics)
    we indicate the relevant range for the final-state invariant mass $M_X$
    and the approximate minimum value of $x$ probed
    in the PDFs, $x_{\rm min}=(M_X/\sqrt{s})\exp(-|y|)$, for central ($y=0$), intermediate ($|y|=2$)
    and forward ($|y|=4$) rapidities.
\label{tab:kincov}
  }
\end{table}

Given this, it is important to verify that available PDF sets have a
sensible behaviour in all the relevant kinematical regions, specially
in the extrapolation regions at very small-$x$ and very large $Q^2$
which are not relevant for most LHC applications.
The goal here is not to understand similarities or differences between
PDF sets, but to ensure that PDF sets that will be used for FCC
simulations have a physical behaviour in the entire range of $x$ and
$Q$ required.

In the following, PDFs are accessed through the {\tt LHAPDF6}
interface~\cite{Buckley:2014ana}, version {\tt 6.1.5}, with the most
updated grid data files.
It should be emphasized the importance of using this specific version,
since previous versions had different options for the default PDF
extrapolations.
In addition, both the interpolation accuracy and the treatment of the
extrapolation regions, as well as the overall computational
performance, have been substantially improved in {\tt LHAPDF6} as
compared to its Fortran counterpart {\tt LHAPDF5}, and therefore the
use of the latter for FCC studies should be discouraged.\footnote{In
  {\tt LHAPDF5} the default extrapolation was simply to freeze the PDF
  below some value of $x_{\rm min}$, which could be as high as
  $10^{-5}$ for some widely used PDF sets, which can potentially lead
  to incorrect results if used for FCC studies.}

We begin by discussing the PDF behavior in the small-$x$ extrapolation
region.
As shown in Fig.~\ref{fig:kinplot}, for low scales and forward
rapidities, as those required for the description of soft QCD physics
and for Monte Carlo tuning, knowledge of PDFs would be required down
to $x\gsim 10^{-9}$.
In Fig.~\ref{fig:xpdf-smallx} we show the central values of the gluon
(upper) and the up quark PDFs (lower plots), comparing ABM12, CT14,
MMHT14 and NNPDF3.0 for $Q^2=4$ GeV$^2$.
All PDF sets shown are NLO except for ABM12 where the NNLO set is
used.
The comparison is performed down to $x=10^{-9}$, to ensure that the
entire region relevant for FCC studies is covered.
In all cases we observe a sensible extrapolation into the very
small-$x$ region.
Here we use the default extrapolating settings of {\tt LHAPDF6.1.5},
and we verified that the behaviour was instead unphysical if older
versions were used, where PDFs were frozen for some $x\le x_{\rm min}$
threshold.

\begin{figure}[t]
\centering
\includegraphics[width=0.49\textwidth]{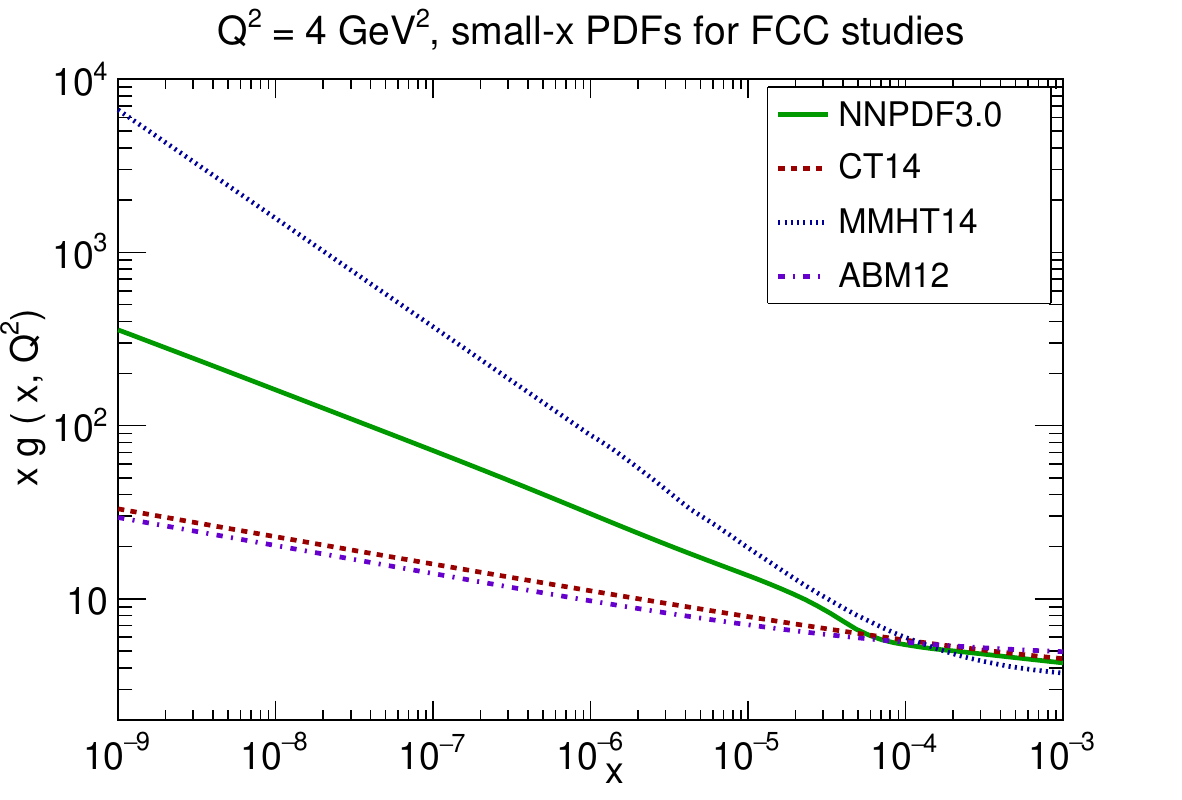}
\includegraphics[width=0.49\textwidth]{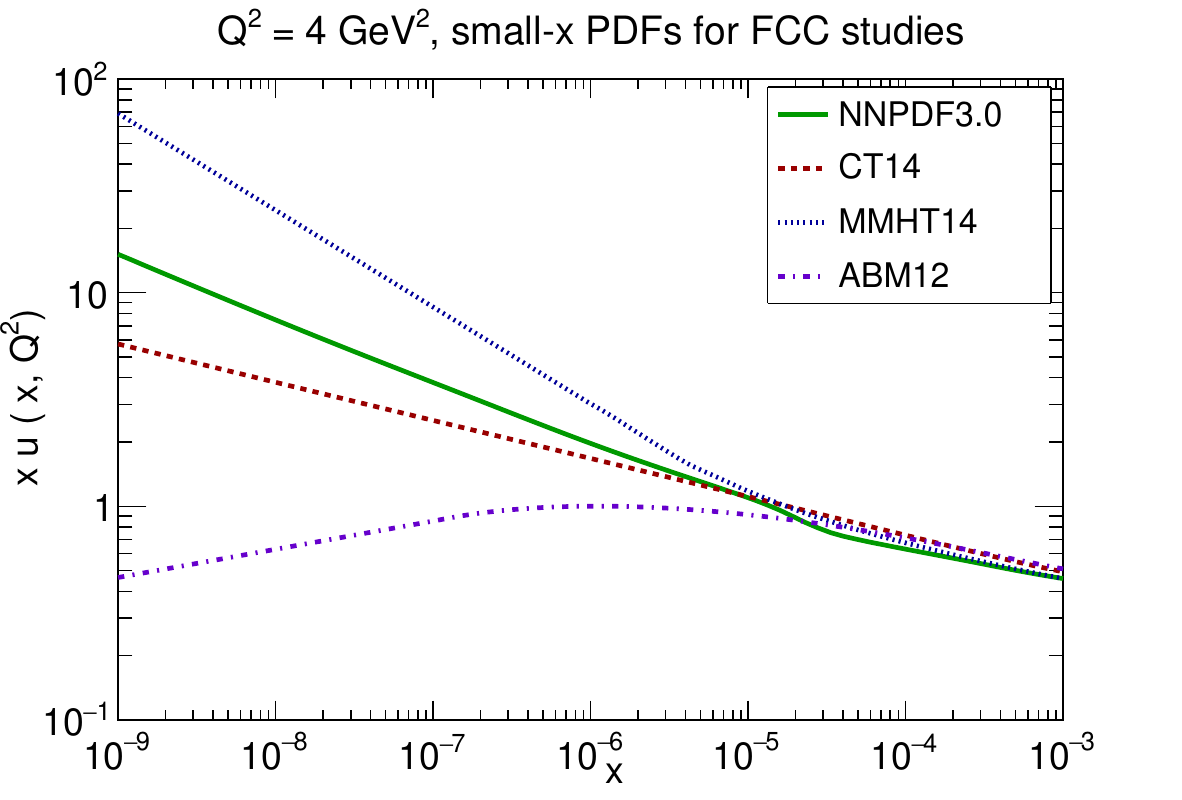}
\caption{\small Central values of the gluon (left)
  and the up quark PDFs (right) at NLO, comparing
  the ABM12,  CT14 and MMHT14 and NNPDF3.0 sets for $Q^2=4$ GeV$^2$.
  All PDF sets shown are NLO except for ABM12 where the NNLO set is used.
  In this small-$x$ region, PDF uncertainties (not shown here)
  can be large, see Fig.~\ref{fig:xpdf-smallx-err}.
}
\label{fig:xpdf-smallx}
\end{figure}

While in Fig.~\ref{fig:xpdf-smallx} we only show the central values of the three PDF sets,
in the
small-$x$ region these are affected by substantial
uncertainties~\cite{Gauld:2015yia} due to the lack of
direct experimental constraints, for instance, the HERA
structure functions data stops at $x_{\rm min}\simeq 5\cdot 10^{-5}$,
see for instance the measurement of the longitudinal structure
function $F_L(x,Q^2)$~\cite{Collaboration:2010ry,Abramowicz:2015mha}.
To illustrate this point, in Fig.~\ref{fig:xpdf-smallx-err}
we show the relative 68\% CL PDF uncertainties at $Q^2=100$ GeV$^2$
in the small-$x$ region for the ABM12,
CT14, MMHT14 and NNPDF3.0 NNLO sets.
Depending on the model chosen to parametrize PDF uncertainties towards
the region without experimental constraints, we observe a rapid increase in PDF
uncertainties for some sets (CT14, MMHT, NNPDF3.0),
where for $x < 10^{-5}$ uncertainties are already larger than 50\%, while
other sets (ABM12, but also CJ15, JR14 not shown here)
display small PDF uncertainties down to $x=10^{-7}$.
Recently, a number LHC measurements to constrain PDFs at small-$x$ has
been proposed.
The use of charmed meson forward production from LHCb has been recently
shown~\cite{Gauld:2015yia,Cacciari:2015fta,Zenaiev:2015rfa}
to provide useful constraints on the small-$x$ gluon PDF.\footnote{The PDF dependence of heavy quark production at a 100 TeV collider
is discussed in more detail in Sect.~\ref{sec:hvq} of this report.}
Another possibility is the use of forward quarkonium production, such
as $J/\Psi$, which has a similar sensitivity in $x$~\cite{Jones:2015nna}.
Taking this into account, one expects that before
the FCC start-up our knowledge of the small-$x$ PDFs would be substantially
improved.
The corresponding measurements at the FCC have the potential to extend
the constrains on the small-$x$ PDF by almost two orders of magnitude,
though here the instrumentation of the forward region will be crucial.
Measurements of very-small-$x$ PDFs are also of direct importance for
particle astrophysics, such as the ultra-high-energy neutrino
cross-sections~\cite{CooperSarkar:2011pa} and the prompt lepton
fluxes~\cite{Gauld:2015kvh,Bhattacharya:2015jpa,Garzelli:2015psa} that
are required for the interpretation of the IceCube astrophysical
neutrinos~\cite{Aartsen:2014gkd}.
%

\begin{figure}[t]
\centering
\includegraphics[width=0.49\textwidth]{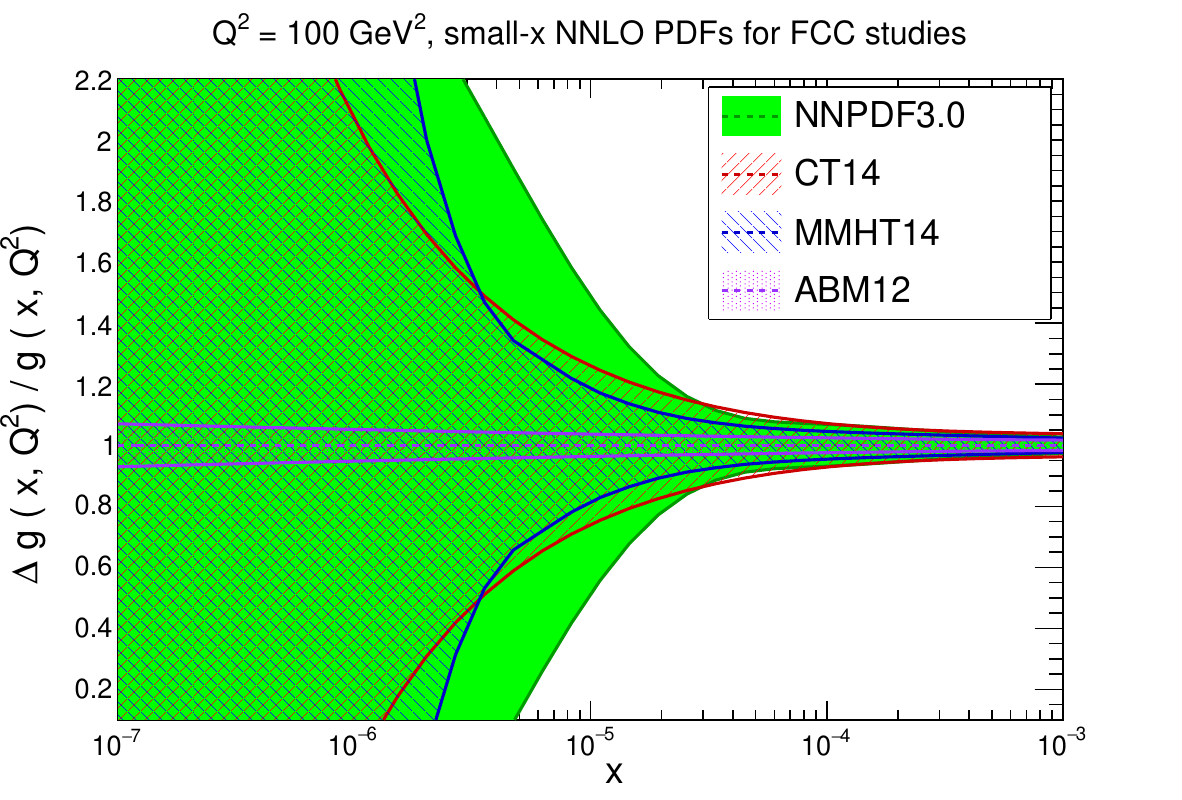}
\includegraphics[width=0.49\textwidth]{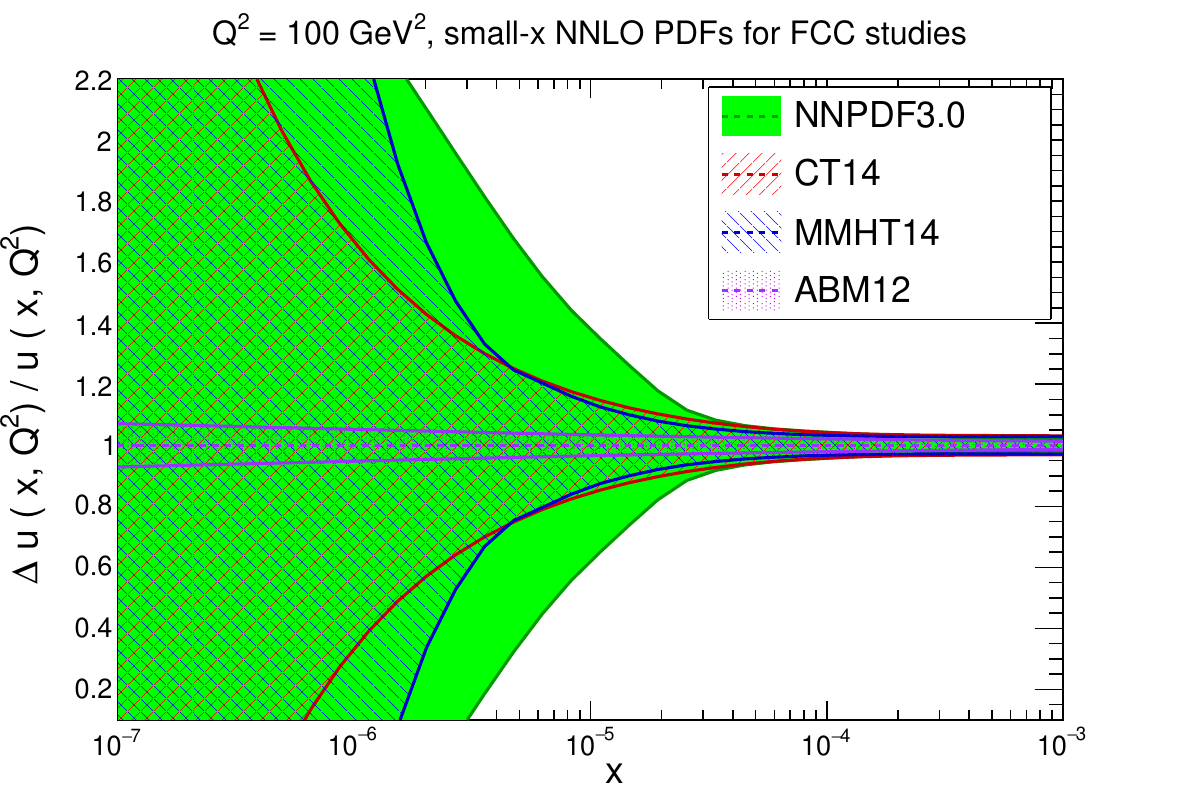}
\caption{\small The relative 68\% CL PDF uncertainties at $Q^2=100$ GeV$^2$
  in the small-$x$ region computed with the
  ABM12, CT14, MMHT14 and NNPDF3.0 NNLO sets.
  With the exception of ABM12, one finds a rapid increase in PDF
  uncertainties as we move towards the small-$x$ region $x\lsim 5\cdot
  10^{-5}$, where current experimental information is limited.  }
\label{fig:xpdf-smallx-err}
\end{figure}

\begin{figure}[t]
\centering
\includegraphics[width=0.46\textwidth]{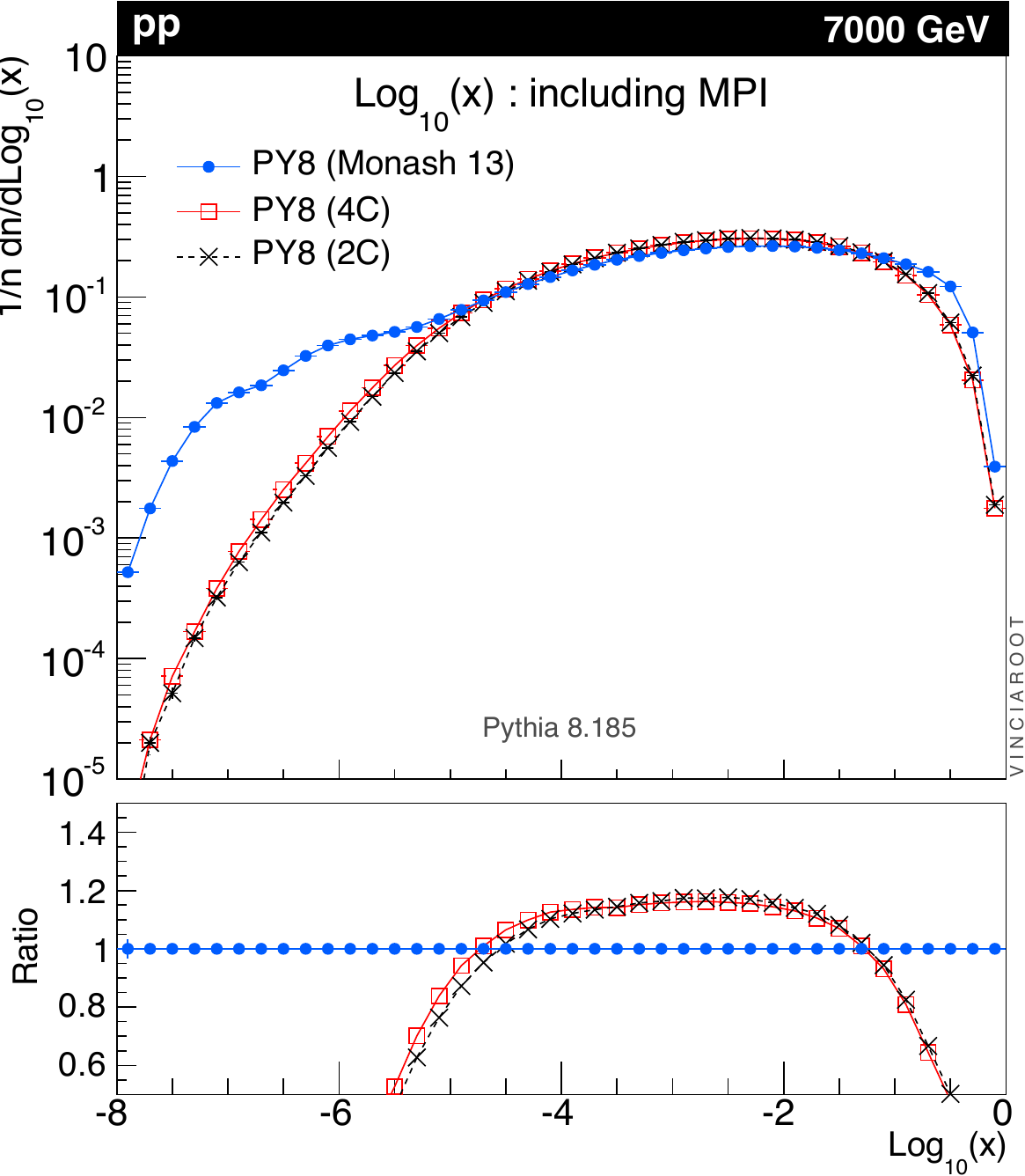}
\includegraphics[width=0.49\textwidth]{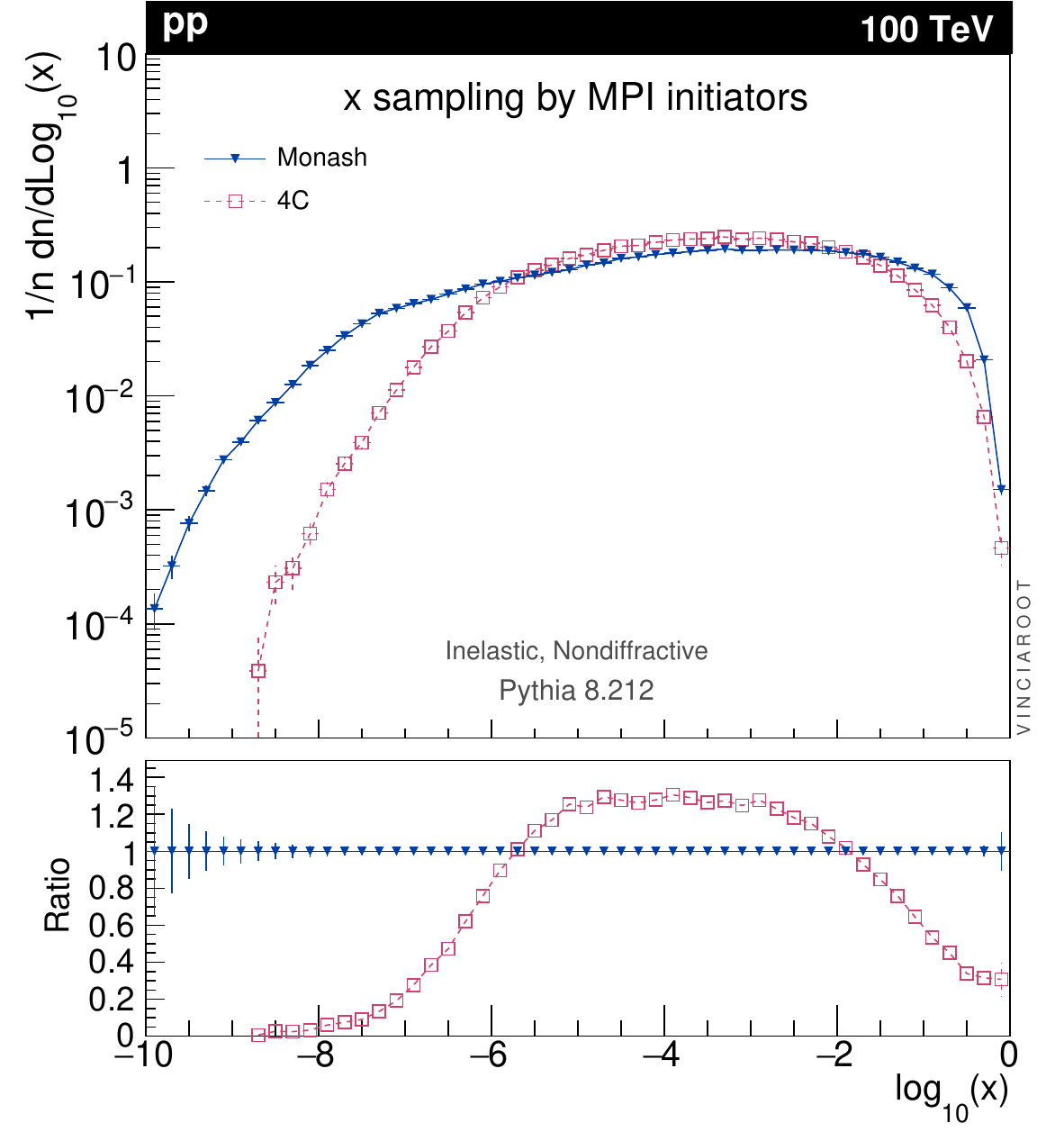}
\caption{\small Sampling of the values of Bjorken-$x$
probed in Multiple Parton Interactions (MPI) at the LHC 7 TeV (left) and
at the FCC 100 TeV (right plot) in {\tt Pythia8} with the Monash 2013 tune, compared
to older tunes, 4C and 2C (at 100 TeV we show only the comparison with the 4C tune).
The lower panel shows the ratio between the Monash 2013 tune and the
older {\tt Pythia8} tunes.  }
\label{fig:monash}
\end{figure}

Another strategy to quantify the relevant range of Bjorken-$x$ for
which PDFs are required in the modeling of soft and semi-hard physics
at the FCC is by sampling of the values of $x$ of the PDFs required in
the calculation of Multiple Parton Interactions (MPI) for different
values of the collider center-of-mass energy $\sqrt{s}$.
In Fig.~\ref{fig:monash} we compare the MPI sampling of $x$ between
the LHC 7 TeV and the FCC 100 TeV using {\tt
  Pythia8.2}~\cite{Sjostrand:2014zea}.
The results of the most update tune, Monash 2013~\cite{Skands:2014pea}
are compared with the older tunes 2C and 4C~\cite{Skands:2010ak}.
From this comparison we observe that, with the Monash 2013 tune, at
LHC7, PDFs with $x\gsim 10^{-6}$ lead to a sizable contribution,
$\gsim 5\%$, to the MPI distribution.
With the same settings, the FCC100 samples values of $x$ down to
$x\gsim 10^{-8}$, a region far from any direct experimental
constraint.
This illustrates the relevance of ultra-low $x$ PDFs for the modelling
of soft QCD at a 100 TeV collider.

Now we turn to discuss the region of large values of Bjorken-$x$.
This region is also affected by substantial PDF uncertainties due to
the limited direct experimental constraints.
To estimate the coverage in the large-$x$ region, it is useful to use
the result that for the production of a final state with invariant
mass $M_X$ and rapidity $y$ at a hadron collider with center-of-mass
energy $\sqrt{s}$, the LO values of the PDF momentum fractions $x_1$
and $x_2$ are $x_{1,2}=(M_X/\sqrt{s})\exp(\pm y)$.
Therefore, for a centrally produced final-state ($y=0$) of invariant
mass $M_X\simeq 7$ TeV (50 TeV) at $\sqrt{s}=$14 TeV (100 TeV) we will
have $\la x_{1,2}\ra\simeq 0.5$, while already for slightly
non-central production, $y\simeq 0.5$, PDFs are being probed up to
$x_1\simeq 0.8$ for both colliders.

In Fig.~\ref{fig:largexPDFs} we illustrate
the large-$x$ behaviour of the up, down, anti-up quark
and gluon PDFs,  evaluated at $Q=100$ GeV.
We compare the results of the ABM12, CT14, NNPDF3.0 and MMHT14 NNLO
PDF sets, with the corresponding 68\% CL PDF uncertainty in each case,
normalising to the central value of CT14.\footnote{In these plots, the
  ABM12 curves have been obtained using the internal interpolation
  routine provided by the authors, since the {\tt LHAPDF6} results
  were found to exhibit poor numerical stability at large $x$.}
As discussed above, the central production of a heavy system with
$M_X=10~(30~{\rm or}~50)$ TeV would probe the large-$x$ PDFs for
$x\gsim 0.1$~(0.3 or 0.5) at a 100 TeV collider.
As we can see, while for valence quarks (up and down) PDF
uncertainties in the region relevant for heavy particle production at
the FCC are moderate, for the gluon and anti-quarks PDF uncertainties
are large, thus degrading the accuracy of any theory prediction that
requires knowledge of PDFs in this region.
In addition, there is a significant spread between the central values
of the four sets.

\begin{figure}[t]
\centering
\includegraphics[width=0.49\textwidth]{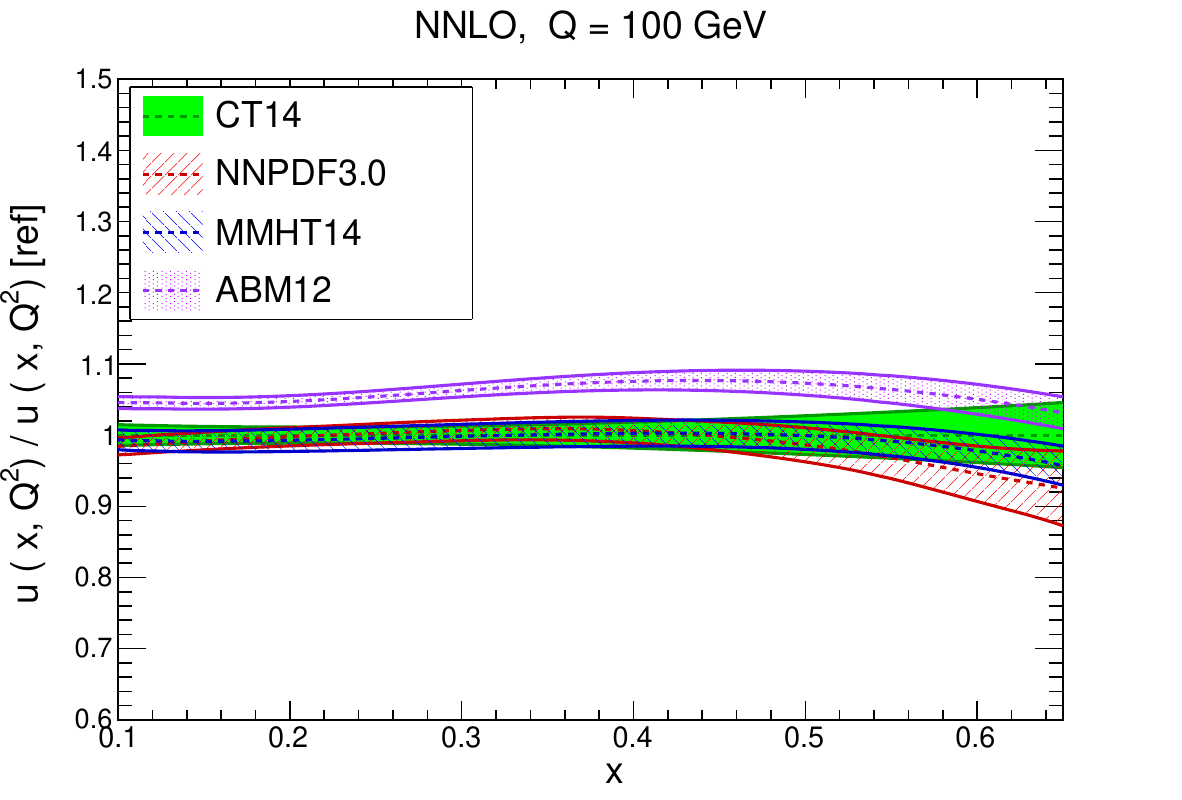}
\includegraphics[width=0.49\textwidth]{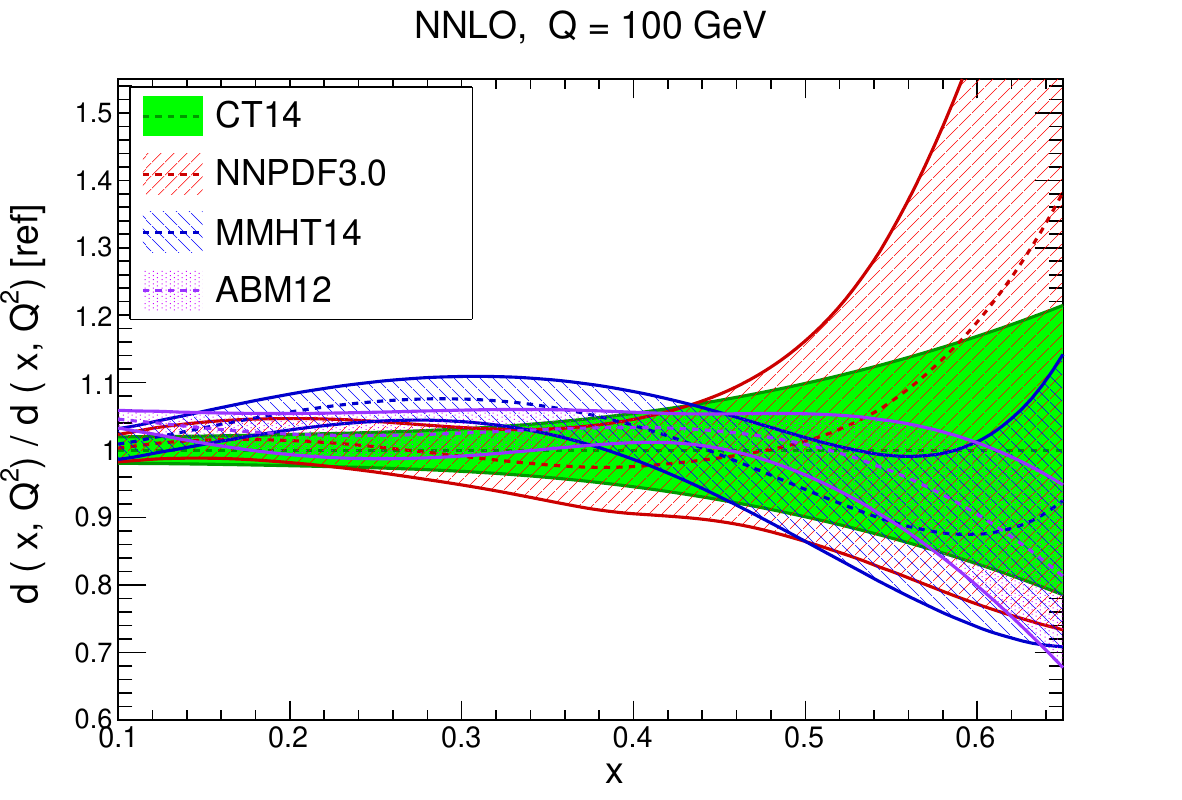}
\includegraphics[width=0.49\textwidth]{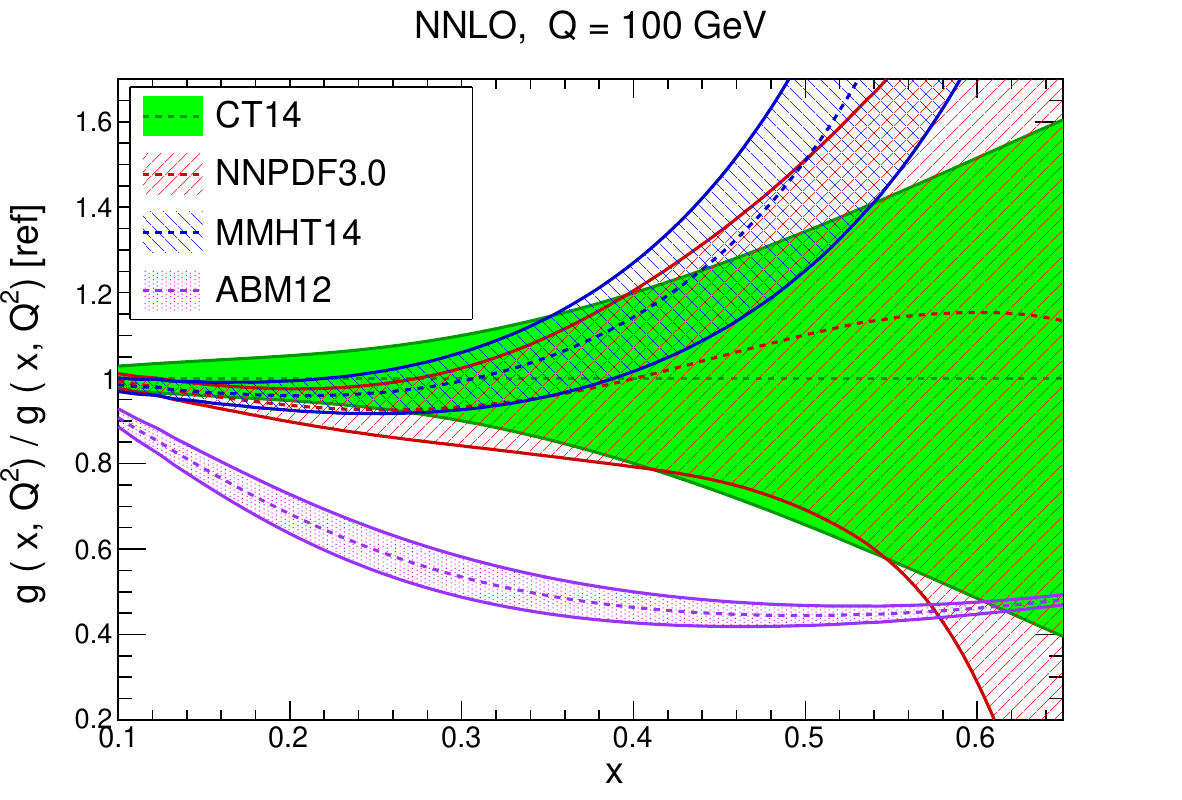}
\includegraphics[width=0.49\textwidth]{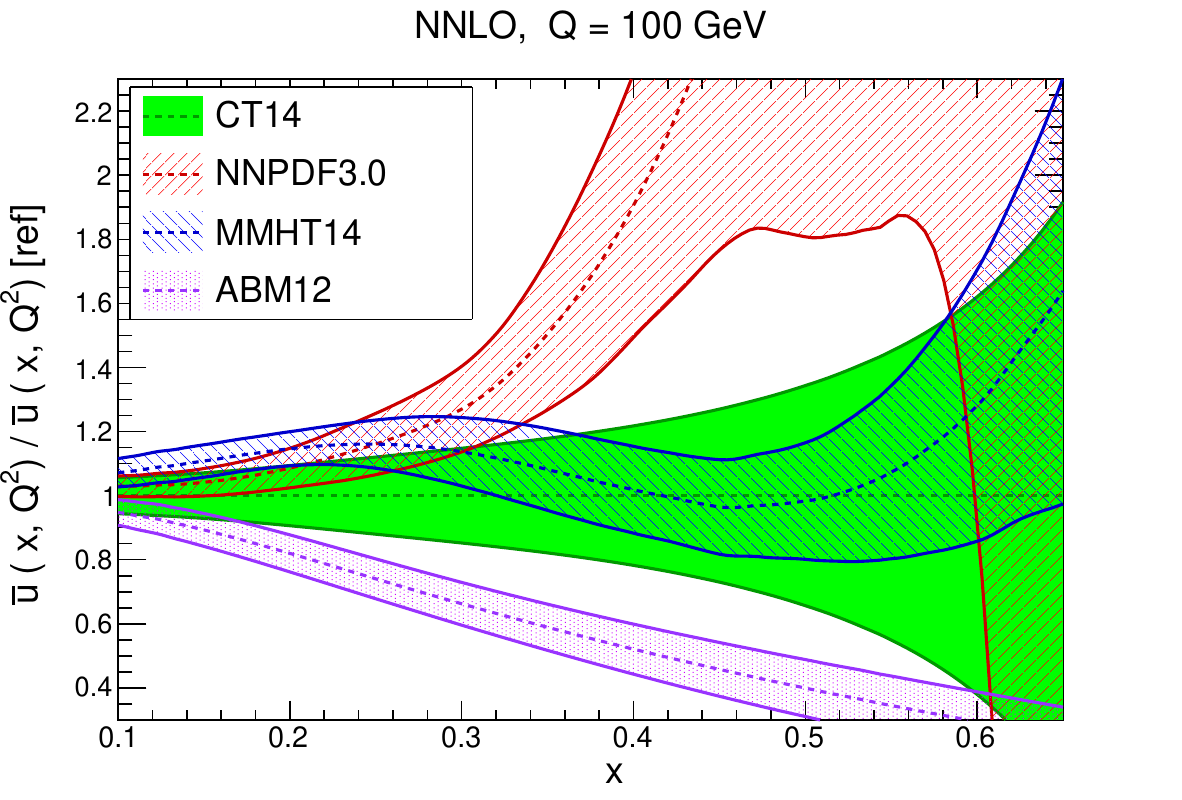}
\caption{\small The large-$x$ behaviour of the up, down, anti-up quark
  and gluon PDFs evaluated at $Q=100$ GeV.
  We compare the results of ABM12, CT14, MMHT14 and NNPDF3.0 NNLO, with the corresponding
  68\% CL PDF uncertainty in each case.
  The comparison is presented normalising to the central value of
  CT14.  }
\label{fig:largexPDFs}
\end{figure}

As in the case of small-$x$, new measurements from the LHC and other experiments
should allow to substantially reduce these PDF uncertainties before the
start of FCC operations.
For instance, the large-$x$ gluon can be constrained with data
on inclusive and differential top quark pair
production~\cite{Czakon:2013tha,Guzzi:2014wia}.
Moreover, since at large-$x$ and large-$Q$ the gluon and sea quark
distributions receive large contributions from radiation off valence
quarks, measurements aiming to constrain these will also lead to
improved gluons and sea quarks in the kinematic region relevant for
the FCC.
This effect is illustrated in Fig.~\ref{fig:largexPDFsPsup}, where we
show the ratio of parton distributions, $p(x, \mu)$ with respect to
the initial parametrization $p(x,\mu_0)$ for the gluon and sea
distributions at large $x$.
The solid line denotes the initial PDF $p(x,\mu_0)$ suppressed by a
factor of $(1−x)$ for the gluon (left panel) and up quarks (right
panel) and the dotted and dashed-dotted lines the respective results
of the evolution up to $\mu$ = 20 TeV.
One should also note that, as in the case of small-$x$,
the behaviour of PDFs in the large-$x$ extrapolation region is sensitive
to the underlying assumptions concerning the PDF parametrization~\cite{Ball:2016spl}.

PDFs at large-$x$ are also affected by a number of theoretical
uncertainties, from potential higher twists, enhanced higher-order
threshold logarithmic corrections, or nuclear effects from the
inclusion in the PDF fit of deuteron and heavy nuclear data.
A version of NNPDF3.0 including large-$x$ threshold resummation was
presented in~\cite{Bonvini:2015ira}, and then
used~\cite{Beenakker:2015rna} to produce threshold-improved NLO+NLL
predictions for high-mass squark and gluino production cross-sections
at the LHC.
This study showed that threshold logarithms in PDF fits are much smaller than PDF uncertainties,
provided NNLO calculations are used.
Therefore, PDFs with threshold resummation do not appear to be
required for FCC studies, since NNLO and N3LO calculations will be the
standard by then.
Likewise, other theory uncertainties like higher twists and nuclear
effects are subleading as compared to PDF uncertainties (see the
discussion in~\cite{Ball:2013gsa,Butterworth:2015oua} and references
therein), and moreover by the time the FCC starts operation, reliable
collider-only PDF sets, free of these ambiguities, will be available.

\begin{figure}[t]
\centering
\includegraphics[width=0.89\textwidth]{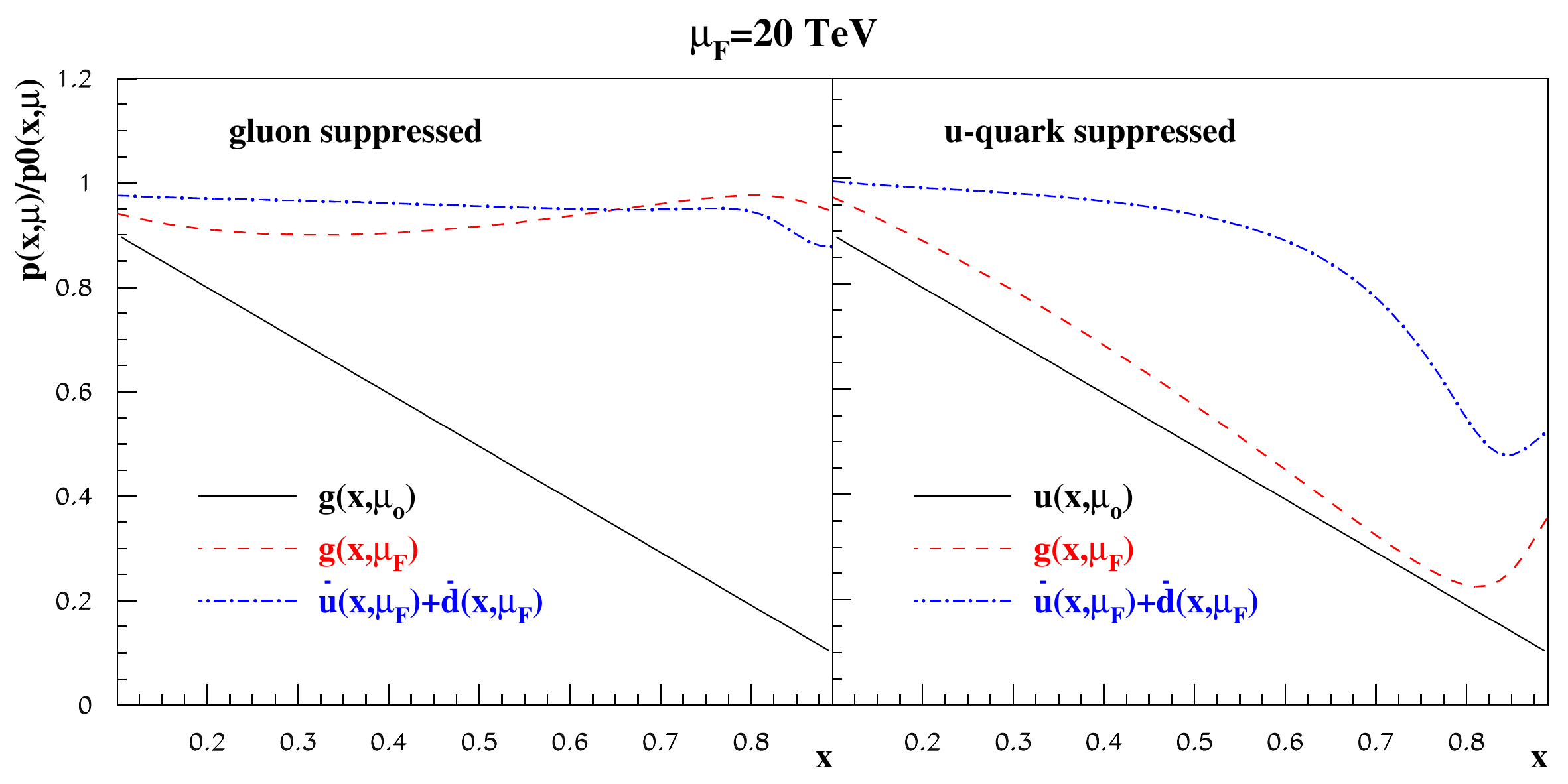}
\caption{\small The ratio of parton distributions, $p(x, \mu)$
  with respect to the initial parametrization $p(x,\mu_0)$
  for the gluon and sea distributions at large $x$.
  The solid line denotes the initial PDF $p(x,\mu_0)$  suppressed by a factor of $(1 - x)$
  for the gluon (left panel) and up quarks (right panel) and the
  dotted and dashed-dotted lines the respective results of the evolution up to $\mu$ = 20 TeV. 
}
\label{fig:largexPDFsPsup}
\end{figure}

The other kinematic region for which knowledge of PDFs will be
required in a previously unexplored region is that of very large
momentum transfers, for values of $Q$ between 5 TeV and 50 TeV.
This region is relevant for the production of possible massive BSM
particles.
As opposed to the small- and large-$x$ regions, the extrapolation into
very high $Q^2$ values is determined purely by perturbative DGLAP
evolution, and therefore the only requirement is that that available
PDF tabulations of current sets extend up to 100 TeV.
We have verified that this is the case for the modern PDF sets
discussed in this chapter.
However, the argument above however holds only for QCD evolution.
It should be taken into account that differences in the upwards
evolution in $Q^2$ can arise if the evolution equations are modified,
for instance in the case of electroweak corrections to DGLAP
evolution, Sect.~\ref{sec:pdf_ew}, or in the presence of high-energy
resummation effects, Sect.~\ref{sec:pdf_resum}.

In Fig.~\ref{fig:xpdf-largeq} we compare, for $x=0.001$, the evolution
of the central values of the gluon and up quark PDFs for the NNPDF3.0,
CT14 and MMHT14 NNLO sets, from a very low scale, $Q=2$ GeV, up to the
highest possible scales that the FCC can reach, $Q=100$ TeV.
It can be verified, by comparing with public PDF evolution packages
such as {\tt HOPPET}~\cite{Salam:2008qg} or {\tt
  APFEL}~\cite{Bertone:2013vaa}, that the tabulated extrapolation up
to very high $Q^2$ of modern PDF sets is consistent with DGLAP
evolution as expected.\footnote{Again, this is not necessarily true
  for older PDF sets. In some cases the coverage in $Q^2$ was
  restricted to 10 TeV, and from there upwards an unphysical
  (non-DGLAP) extrapolation was used.
  As in the case of large and small-$x$, use of these older sets can
  lead to incorrect results in the context of FCC simulations.  } 
We
conclude that, provided modern PDF sets are used, the extrapolation of
the DGLAP evolution in $Q^2$ to the region relevant at the FCC is
reliable.

\begin{figure}[t]
\centering
\includegraphics[width=0.49\textwidth]{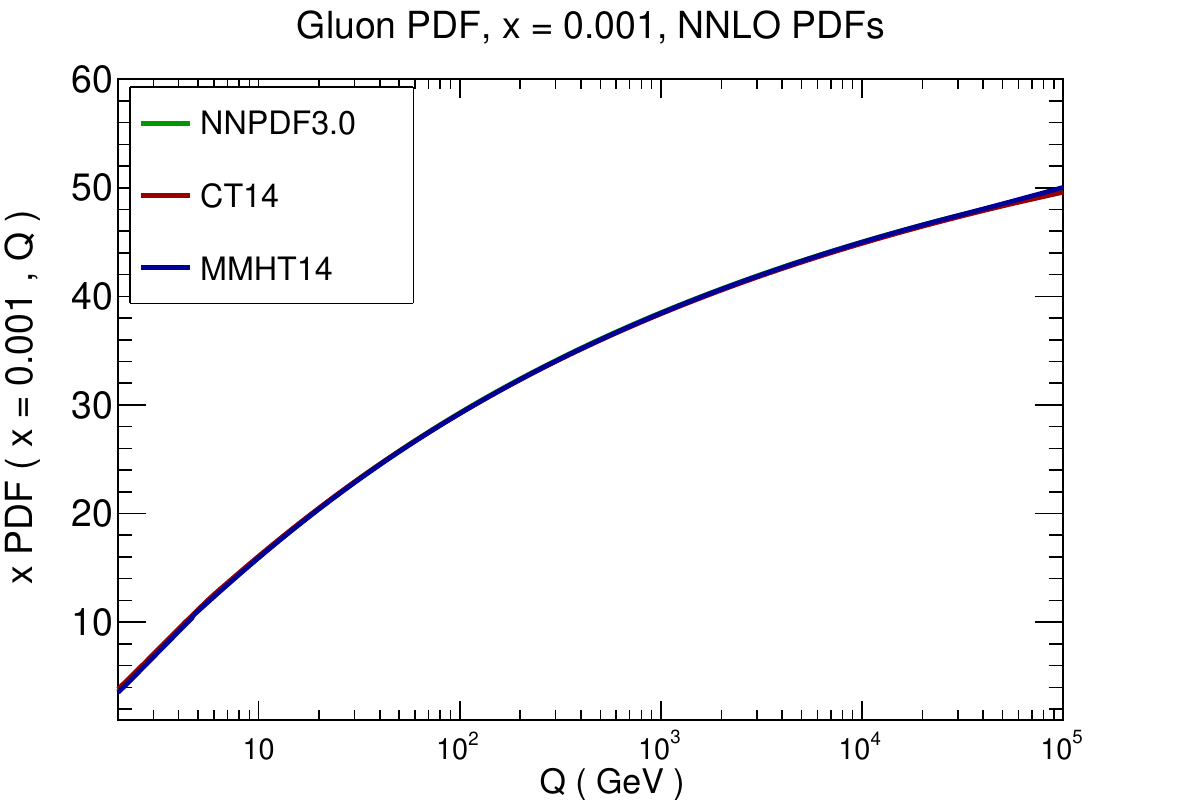}
\includegraphics[width=0.49\textwidth]{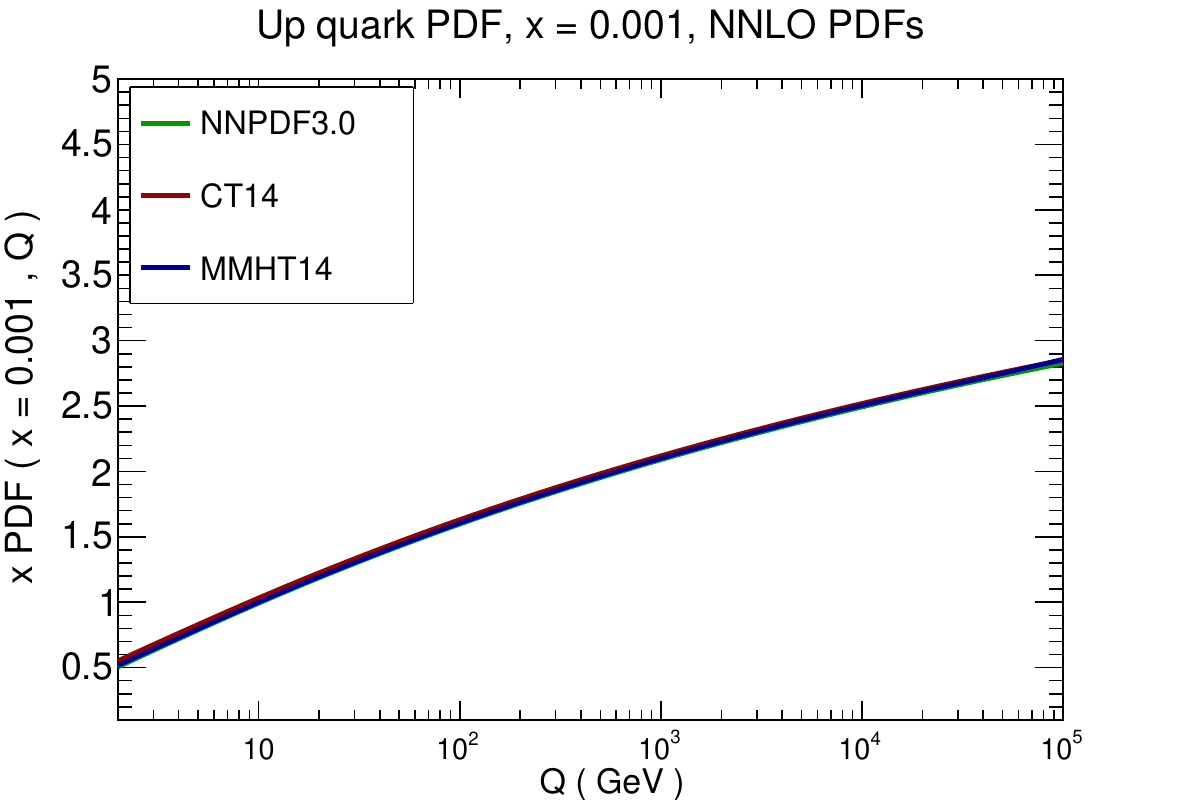}
\caption{\small The effects of DGLAP evolution on the central value of the
  up quark (left) and gluon PDFs (right plot) at $x=0.001$ when evolved
  from $Q=2$ GeV up to $Q=100$ TeV.
  We show the results from the NNPDF3.0, CT14 and MMHT14 NNLO sets,
  the three exhibiting a very similar behaviour.
  PDF uncertainties are not included in this comparison.
}
\label{fig:xpdf-largeq}
\end{figure}


\subsection{PDF luminosities at 100 TeV}
\label{sec:pdf_lumi}

Parton luminosities are useful to estimate the PDF dependence of
hadron collider cross-sections, by taking into account the most
relevant initial-state production channels.
While several definitions of the PDF luminosity can be adopted, in the
following we will use the luminosities as a function of the invariant
mass of the produced final state, $M_X$, defined as
\begin{equation}
  \label{eq:lumdef_sec3}
  \mathcal{L}_{ij}\left( M_X,\sqrt{s}\right)\equiv\frac{1}{s}\int_{\tau}^1
  \frac{dx}{x} f_i\left(x,M_X\right) f_j\left( \tau/x,M_X\right)\,,
\end{equation}
where $i$ and $j$ are PDF flavour indices, $\tau=M_X^2/s$, and
$\sqrt{s}$ is the collider center-of-mass energy.
Another useful way of representing PDF luminosities is as
two-dimensional functions of rapidity $y$ and invariant mass $M_X$ of
the final state,
\begin{equation}
  \label{eq:lumi-def}
  \widetilde{{\cal L}}_{ij}(M_X, y,\sqrt{s}) = \frac{1}{s} 
  f_{i}\left(\frac{M_Xe^{y}}{\sqrt{s}}, M_X\right)
  f_{j}\left(\frac{M_Xe^{-y}}{\sqrt{s}}, M_X\right)\, ,
\end{equation}
which leads to Eq.~(\ref{eq:lumdef_sec3}) upon integration
over the kinematically allowed range for the rapidity $y$, that is,
\be
\mathcal{L}_{ij}(M_X,\sqrt{s}) = \int_{-\ln \sqrt{s}/M_X}^{\ln \sqrt{s}/M_X} dy
~\widetilde{\mathcal{L}}_{ij}(M_X,y,\sqrt{s}) \, .
\ee
Typically PDF luminosities are presented summing over
quark flavor indices, and here we adopt the following convention:
\bea
\label{eq:qg}
{\cal L}_{qg}(M_X) &=& \nonumber
\sum_{i=-5}^5 ({\cal L}_{i0}(M_X) + {\cal L}_{0i}(M_X))\, , \qquad  i \ne 0\,,
\\ 
{\cal L}_{qq}(M_X) &=& 
\sum_{i=-5}^5\; \sum_{j=-5}^5 {\cal L}_{ij}(M_X) \, , \qquad i \ne 0\,,\,\, j
\ne 0\,,
\\ \nonumber
{\cal L}_{q\bar q}(M_X) &=& 
\sum_{i=-5}^5 {\cal L}_{i,-i}(M_X) \, , \qquad i \ne 0\,,
\eea
for the luminosities integrated in
rapidity Eq.~(\ref{eq:lumdef_sec3}), and similar definitions
for the double differential luminosities Eq.~(\ref{eq:lumi-def}).
Eq.~(\ref{eq:qg}) can be trivially generalized to the case
in which the top quark is treated as a massless parton.

In Fig.~\ref{fig:lumicomp} we show the
rapidity-integrated PDF luminosities Eq.~(\ref{eq:lumdef_sec3}),
as a function of the invariant mass of the system $M_X$,
for
the {\tt PDF4LHC15\_nnlo\_mc} PDF
set~\cite{Butterworth:2015oua,Carrazza:2015hva}, with
PDF uncertainties computed at 68\% confidence levels.

We show the gluon-gluon, quark-gluon, quark-quark and
quark-antiquark luminosities, normalized to the corresponding
central value, for the case of a $\sqrt{s}=100$ TeV collider.
Similar comparisons in the case of the LHC 14 TeV can be found
in~\cite{Butterworth:2015oua}.
We find PDF uncertainties are at the 5\% level for $200~{\rm GeV}
\lsim M_X \lsim 5$ TeV for all four PDF luminosities.
They become more important at larger values of $M_X$, relevant for
heavy particle searches, and for smaller values of $x$, relevant for
electroweak physics and semi-hard QCD.
For instance, at $M_X\simeq 20$ TeV the gluon-gluon PDF luminosity has
an associated uncertainty of around 20\%.
For the production of electroweak scale particles PDF uncertainties
are increased when going from the LHC to the FCC, due to the smaller
values of $x$ probed in the latter case.
For $M_X\simeq 100$ GeV, relevant for inclusive Higgs and weak gauge
boson production, PDF uncertainties are around the 10\% level.
It can also be instructive to plot the absolute PDF luminosities in each
channel together with the corresponding PDF uncertainties, this is done
in Sect.~\ref{sec:pdf_photon} later in this chapter.

\begin{figure}[t]
\centering
\includegraphics[width=0.49\textwidth]{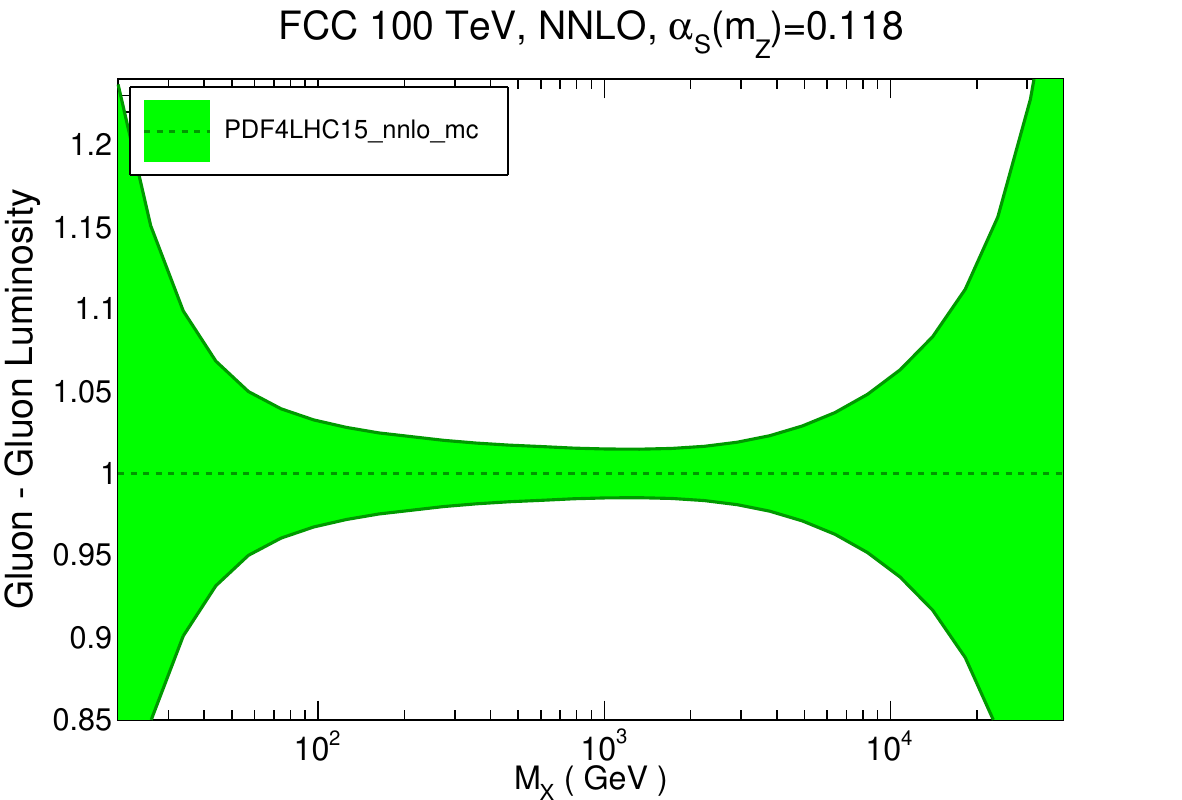}
\includegraphics[width=0.49\textwidth]{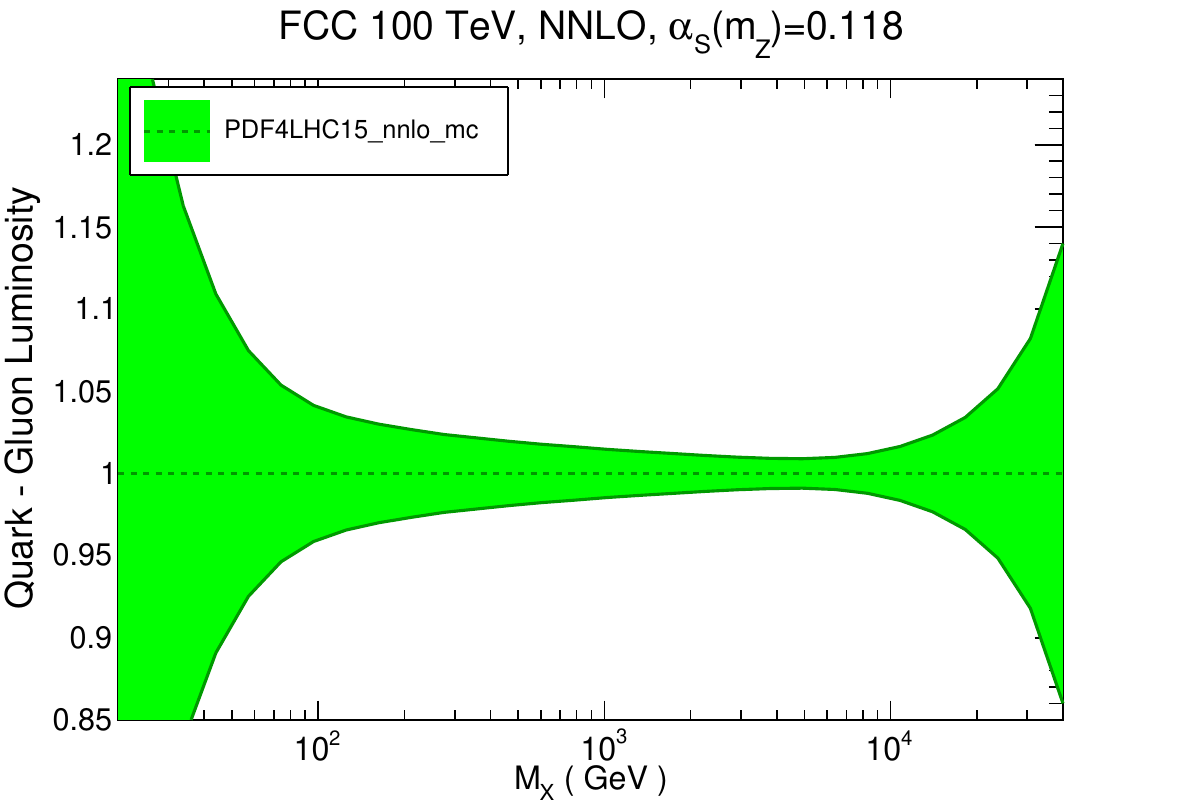}
\includegraphics[width=0.49\textwidth]{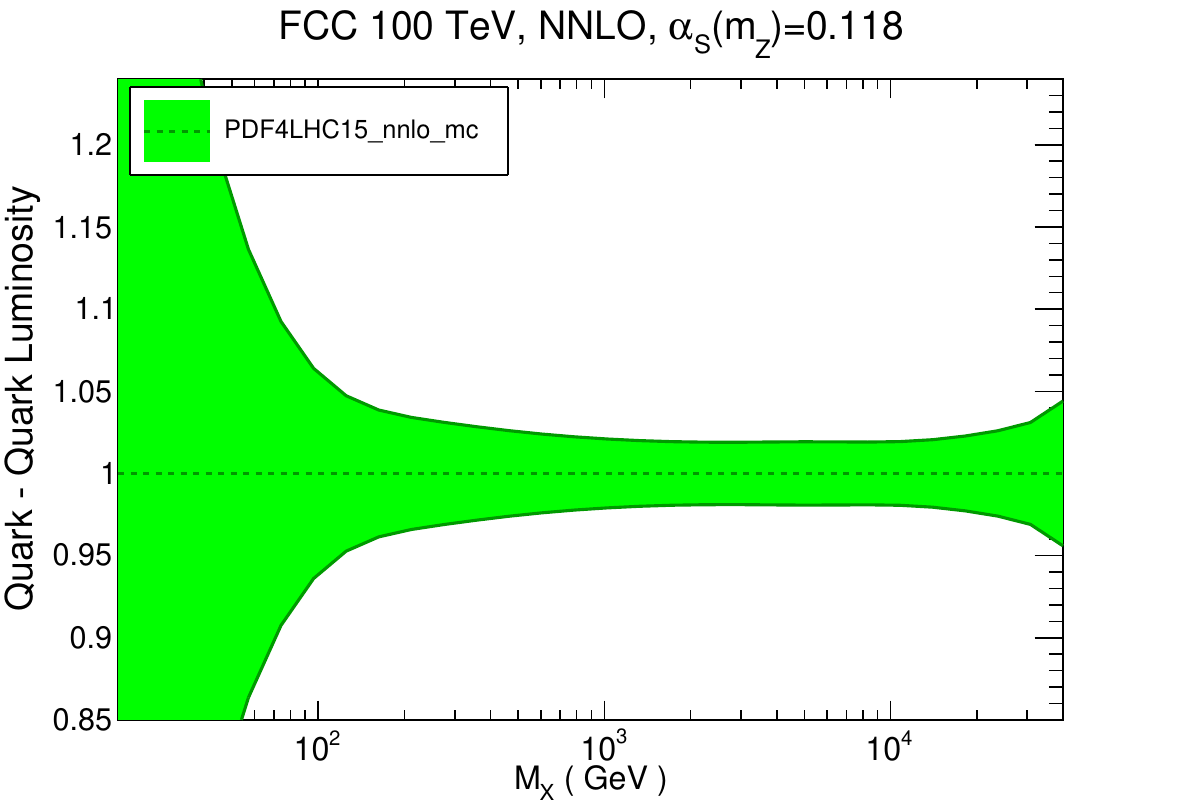}
\includegraphics[width=0.49\textwidth]{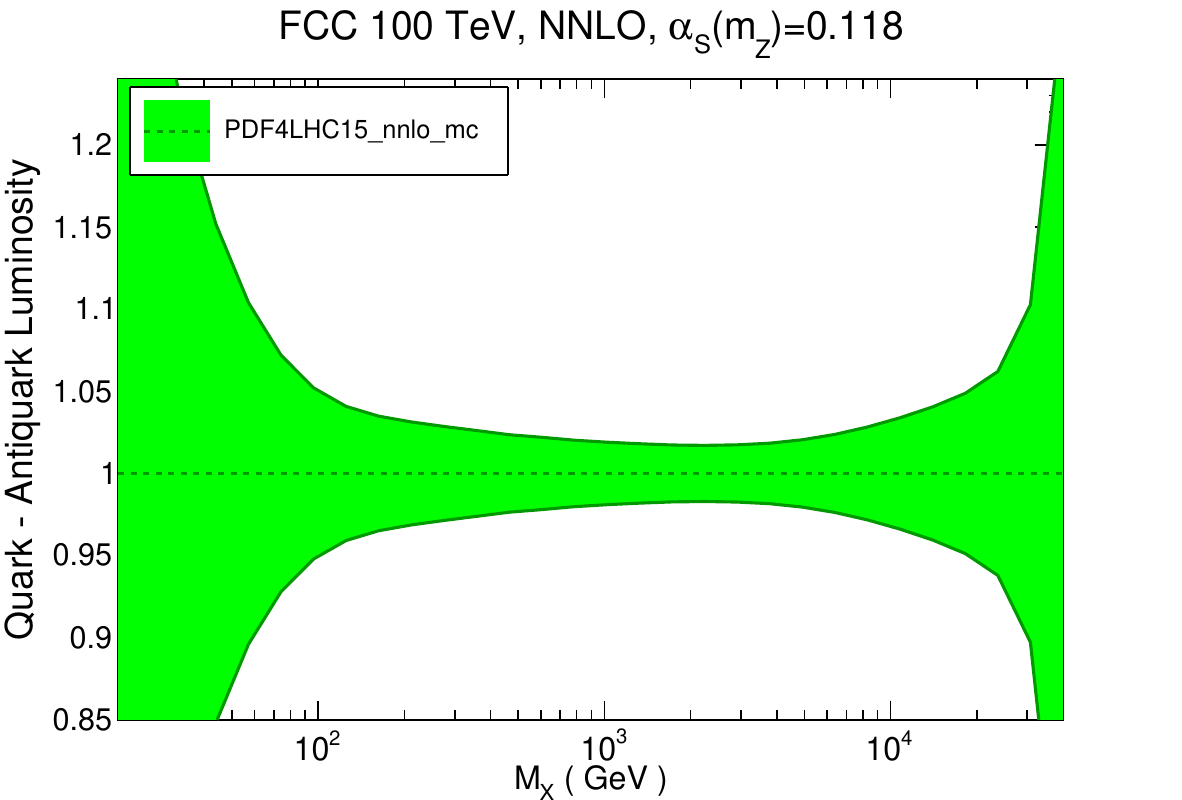}
\caption{\small  The relative uncertainties
  in the rapidity-integrated PDF luminosity at the FCC with $\sqrt{s}=100$ TeV
  computed with the {\tt PDF4LHC15\_nnlo\_mc} set, as a function
  of the final state invariant mass $M_X$.
  From left to right and from top to bottom we show the gluon-gluon,
  quark-gluon, quark-quark and quark-antiquark luminosities.  }
\label{fig:lumicomp}
\end{figure}

We now turn to discuss the double-differential PDF luminosities,
Eq.~(\ref{eq:lumi-def}), evaluated for a center of mass energy
$\sqrt{s}=100$ TeV.  In Fig.~\ref{fig:pdf-lumi-scan} shows the PDF
uncertainties, evaluated as 68\% CL, on the luminosities as a function
of $M_X$ and of the rapidity $y$.
As above, the {\tt PDF4LHC15\_nnlo\_mc} PDF set is used as input.
Fig.~\ref{fig:pdf-lumi-scan} represents the contours of constant PDF
uncertainties in the different flavour combinations.
One sees that for all flavour combinations, the uncertainties are
smallest, of the order of $1{-}2\%$, for pair invariant masses of the
order of a TeV.
They also all have a characteristic dip at rapidities of about
$|y|=1{-}2$.
One may speculate that this is a consequence of an anti-correlation
between moderately large and small-$x$ parton distributions caused by
momentum conservation.
For partonic-pair masses at the electroweak scale and in the region
above a few TeV, uncertainties grow larger.
In all cases PDF uncertainties grow large near the kinematic
boundaries, since these are sensitive to to PDFs at small and
large-$x$ that currently are constrained by few experimental
measurements.

\begin{figure}[t]
  \centering
  \includegraphics[width=0.49\textwidth,page=1]{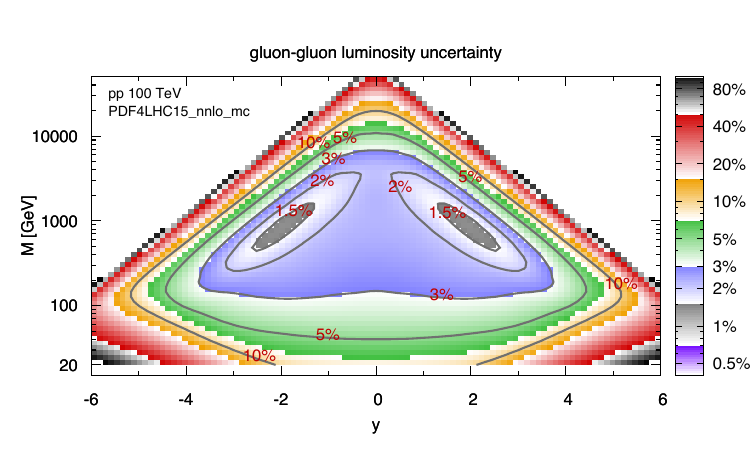}\hfill
  \includegraphics[width=0.49\textwidth,page=2]{figs/lumis-nosig-100TeV-PDF4LHC15_nnlo_mc.pdf}\\
  \includegraphics[width=0.49\textwidth,page=3]{figs/lumis-nosig-100TeV-PDF4LHC15_nnlo_mc.pdf}\hfill
  \includegraphics[width=0.49\textwidth,page=4]{figs/lumis-nosig-100TeV-PDF4LHC15_nnlo_mc.pdf}
  \caption{\small The contours of constant PDF uncertainty for the double-differential
    PDF luminosities Eq.~(\ref{eq:lumi-def}) evaluated for
a center of mass energy $\sqrt{s}=100$ TeV, with {\tt PDF4LHC15\_nnlo\_mc} PDF set used as input.
  }
  \label{fig:pdf-lumi-scan}
\end{figure}

Next we compute the ratio of the
rapidity-integrated PDF luminosities between 100 TeV and 14 TeV, for
different initial-state partonic channels, defined as:

\be
\label{eq:lumirat}
R_{ij}(M_X,\sqrt{s_1},\sqrt{s_2})\equiv\frac{\mathcal{L}_{ij}\left( M_X,\sqrt{s_1}\right)}{\mathcal{L}_{ij}\left( M_X,\sqrt{s_2}\right)} \, ,
\ee
with $\sqrt{s_1}=100$ TeV and $\sqrt{s_2}=14$ TeV.
Such ratios provide a convenient rule of thumb to rescale production
cross-sections between 14 and 100 TeV, for processes dominated by a
single initial-state luminosity.
Eq.~(\ref{eq:lumirat}) can thus be used to estimate ratios of
cross-sections between the different center-of-mass energies.
These cross-section ratios, in addition to providing stringent SM
tests and potential PDF-constraining information, could also be used
as an alternative method to search for new physics at the
FCC~\cite{Mangano:2012mh}.

In Fig.~\ref{fig:lumirat} we show the ratio of PDF luminosities,
Eq.~(\ref{eq:lumirat}), between the FCC $\sqrt{s_1}=100$ TeV and the
LHC $\sqrt{s_2}=14$ TeV, for the four different initial-state
channels.
These ratios have been computed with both the NNPDF3.0 and the {\tt
  PDF4LHC15\_nnlo\_mc} NNLO sets, to illustrate that the generically
$R_{ij}(M_X,\sqrt{s_1},\sqrt{s_2})$ depends only very mildly of the
specific input PDF set used.
In Fig.~\ref{fig:lumirat} we also include the 68\% CL PDF
uncertainties in the luminosity ratio, accounting for the correlations
between the results at the two values of the center-of-mass energy.
The ratio is computed between $M_X=10$ GeV and $M_X=6$ TeV, the highest
invariant masses that the LHC can reach.
From this comparison, we observe that for low invariant masses, $M_x
\lsim 100$ GeV, the increase in parton luminosities when going from
the LHC to the FCC is moderate, a factor 10 at most.
In this region the luminosity ratio is affected by large PDF
uncertainties, arising from the production of a small $M_X$ final
state at the FCC, which probes small-$x$ PDF.

\begin{figure}[t]
\centering
\includegraphics[width=0.49\textwidth]{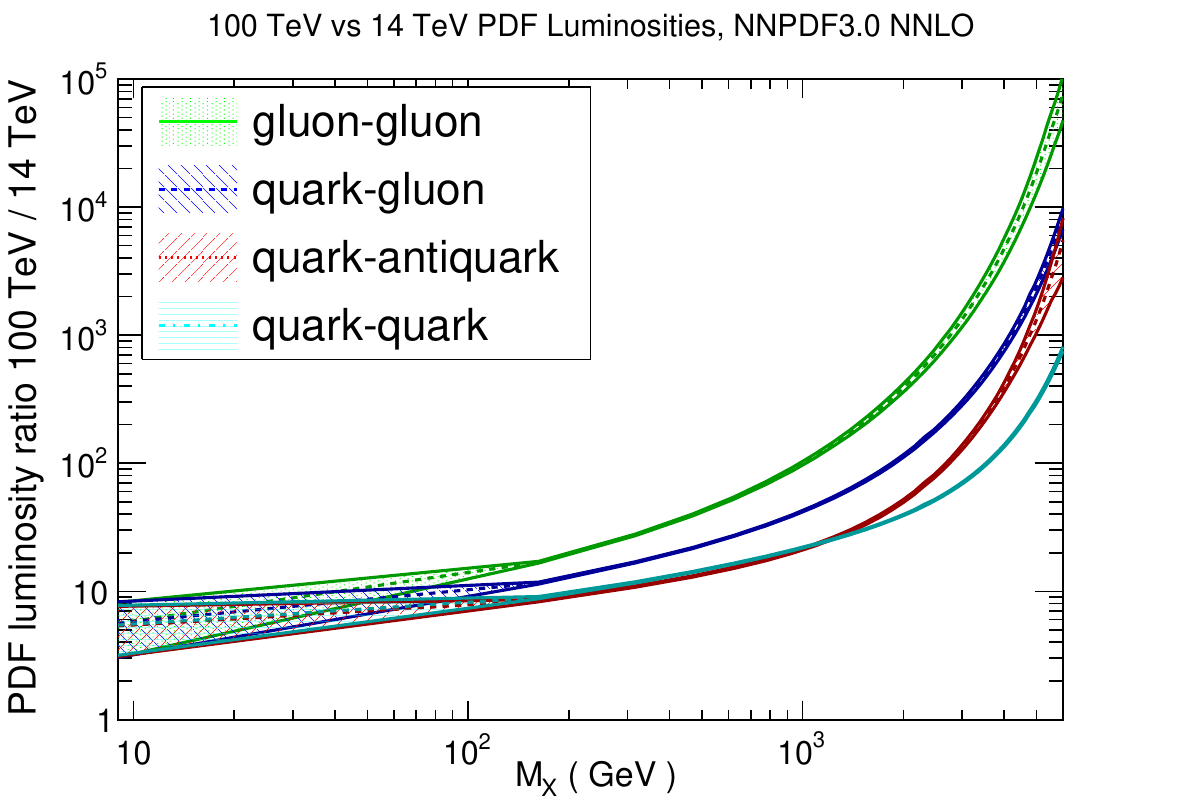}
\includegraphics[width=0.49\textwidth]{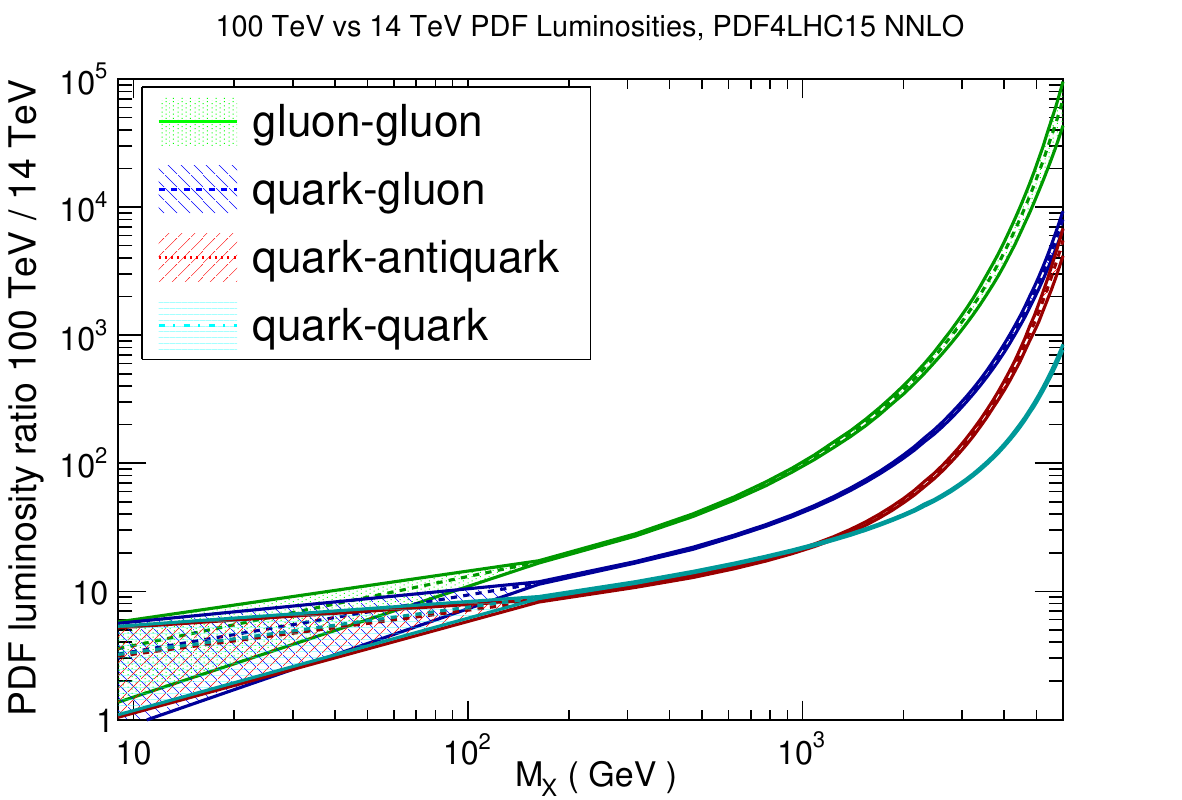}
\caption{\small The ratio of PDF luminosities,
  Eq.~(\ref{eq:lumirat}), between the FCC $\sqrt{s_1}=100$ TeV
  and the LHC $\sqrt{s_2}=14$ TeV center-of-mass energies,
  for different initial-state channels,
  together with the corresponding
  68\% CL PDF uncertainties.
  These ratios have been computed with the NNPDF3.0 (left plot)
  and PDF4LHC15 (right plot) NNLO PDFs.
}
\label{fig:lumirat}
\end{figure}

On the other hand, the luminosity ratio increases rapidly as we move
away from the electroweak scale, since these the increase in energy of
the FCC dramatically dominates over the large-$x$ fall-off of the PDFs
at the LHC.
For invariant masses around $M_X\simeq 1$ TeV, for instance, the $gg$,
$qg$, $q\bar{q}$ and $qq$ luminosity ratios are $\simeq 100, 50, 20$
and 10, respectively.
Gluon-initiated processes are those that benefit more from the
increase in center-of-mass energy due to the rapid rise of the gluon
PDF at medium- and small-$x$ from DGLAP evolution.
For the highest invariant masses that can be probed at the LHC,
$M_X\simeq 7$ TeV, the values of the ratios (in the same order) are
$10^5$, $10^4$, $5\cdot 10^3$ and 200.
The hierarchy $R_{gg} > R_{qg}>R_{q\bar{q}}> R_{qq}$ is maintained for
all invariant masses above $M_X\ge 200$ GeV.

The results in Fig.~\ref{fig:lumirat} can be used to compare with the
various ratios of cross-sections between 100 TeV and 14 TeV collected
elsewhere in this report.
In order to facilitate the comparison with ratios of cross-sections
between different center-of-mass energies presented elsewhere in this
report, in Table~\ref{tab:pdflumiValues} we provide the corresponding
numerical values of the PDF luminosity ratios show in
Fig.~\ref{fig:lumirat} for the case of the {\tt PDF4LHC15\_nnlo\_mc}
PDF set.

\begin{table}[th]
  \small
\begin{center}
\begin{tabular}{|c|c|c|c|c|}
\hline
$M_X$ (GeV)  &   $\mathcal{L}^{(100)}_{gg}/\mathcal{L}^{(14)}_{gg}$   & $\mathcal{L}^{(100)}_{qg}/\mathcal{L}^{(14)}_{qg}$& $\mathcal{L}^{(100)}_{q\bar{q}}/\mathcal{L}^{(14)}_{q\bar{q}}$& $\mathcal{L}^{(100)}_{qq}/\mathcal{L}^{(14)}_{qq}$ \\
\hline
\hline
50    &    8.8  $\pm$  0.3   &   6.9  $\pm$  0.3   &   5.7 $\pm$  0.4   &   5.9  $\pm$  0.4   \\    
58    &    9.5  $\pm$  0.3   &   7.4  $\pm$  0.3   &   5.9  $\pm$  0.4   &   6.3  $\pm$  0.4   \\    
68    &    10.3  $\pm$  0.3   &   7.9  $\pm$  0.3   &   6.3  $\pm$  0.3   &   6.6  $\pm$  0.3   \\    
80    &    11.2  $\pm$  0.3   &   8.5  $\pm$  0.3   &   6.6  $\pm$  0.3   &   7.0  $\pm$  0.3   \\    
94    &    12.2  $\pm$  0.3   &   9.1  $\pm$  0.2   &   7.0  $\pm$  0.2   &   7.3  $\pm$  0.2   \\    
111    &    13.4  $\pm$  0.3  &   9.7  $\pm$  0.2   &   7.4  $\pm$  0.2   &   7.8  $\pm$  0.2   \\    
130    &    14.7  $\pm$  0.3  &   10.5  $\pm$  0.2   &   7.8  $\pm$  0.2   &   8.2  $\pm$  0.2   \\    
152    &    16.2  $\pm$  0.4   &   11.3  $\pm$  0.2   &   8.2  $\pm$  0.2   &   8.7  $\pm$  0.2   \\    
178    &    18.0  $\pm$  0.4   &   12.2  $\pm$  0.2   &   8.7  $\pm$  0.2   &   9.2  $\pm$  0.2   \\    
209    &    20.0  $\pm$  0.4   &   13.3  $\pm$  0.2   &   9.2  $\pm$  0.2   &   9.8  $\pm$  0.2   \\    
245    &    22.5  $\pm$  0.4   &   14.5  $\pm$  0.2   &   9.8  $\pm$  0.2   &   10.5  $\pm$  0.2   \\    
287    &    25.4  $\pm$  0.5   &   15.8  $\pm$  0.2   &   10.5  $\pm$  0.2   &   11.3  $\pm$  0.2   \\    
336    &    28.9  $\pm$  0.6   &   17.4  $\pm$  0.2   &   11.1  $\pm$  0.2   &   11.9  $\pm$  0.3   \\    
394    &    33.2  $\pm$  0.8   &   19.3  $\pm$  0.3   &   12.0  $\pm$  0.2   &   13.0  $\pm$  0.2   \\    
462    &    38.6  $\pm$  0.9   &   21.5  $\pm$  0.3   &   13.0  $\pm$  0.2   &   14.0  $\pm$  0.2   \\    
541    &    45.1  $\pm$  1.2   &   24.2  $\pm$  0.3   &   14.1  $\pm$  0.2   &   15.2  $\pm$  0.2   \\    
634    &    54.0  $\pm$  1.6   &   27.4  $\pm$  0.4   &   15.4  $\pm$  0.3   &   16.5  $\pm$  0.2   \\    
744    &    65.3  $\pm$  2.2   &   31.4  $\pm$  0.5   &   17.0  $\pm$  0.3   &   18.1  $\pm$  0.2   \\    
872    &    80.8  $\pm$  2.8   &   36.4  $\pm$  0.6   &   19.0  $\pm$  0.4   &   19.9  $\pm$  0.3   \\    
1022    &    101  $\pm$  4   &   42.9  $\pm$  0.7   &   21.6  $\pm$  0.5   &   22.0  $\pm$  0.3   \\    
1198    &    131  $\pm$  6   &   51.6  $\pm$  1.0   &   25.1  $\pm$  0.6   &   24.7  $\pm$  0.3   \\    
1403    &    173  $\pm$  9   &   63.5  $\pm$  1.4   &   29.9  $\pm$  0.8   &   27.9  $\pm$  0.4   \\    
1646    &    238  $\pm$  14   &   80.3  $\pm$  1.8   &   37.0  $\pm$  1.1   &   32.5  $\pm$  0.7   \\    
1928    &    341  $\pm$  25   &   105  $\pm$  3   &   47.6  $\pm$  1.8   &   37.7  $\pm$  0.6   \\    
2260    &    517  $\pm$  45   &   143  $\pm$  5   &   65.0  $\pm$  2.9   &   45.4  $\pm$  0.7   \\    
2649    &    837  $\pm$  90   &   207  $\pm$  9   &   94.7  $\pm$  4.9   &   56.7  $\pm$  1.0   \\    
3105    &    1454  $\pm$  200   &   322  $\pm$  15   &   151  $\pm$  9   &   74.8  $\pm$  1.4   \\    
3639    &    2815  $\pm$  512   &   546  $\pm$  33   &   269  $\pm$  18   &   106  $\pm$  2   \\    
4265    &    6233  $\pm$  1395   &   1047  $\pm$  84   &   549  $\pm$  50   &   168  $\pm$  5   \\    
5000    &    16646  $\pm$  4557   &   2356  $\pm$  249   &   1366  $\pm$  207   &   308  $\pm$  10   \\  
\hline
\end{tabular}
\end{center}
\caption{\small \label{tab:pdflumiValues}
Numerical values of the ratios of PDF luminosities, Eq.~(\ref{eq:lumirat}), between
$\sqrt{s_1}=100$ TeV and $\sqrt{s_2}=14$ TeV computed with 
the {\tt PDF4LHC14\_nnlo\_mc} set.
The graphical representation of these ratios is presented
in Fig.~\ref{fig:lumirat} (right).
}
\end{table}

\subsection{The top quark as a massless parton}
\label{sec:pdf_top}

At a 100~TeV hadron collider, particles with masses around the
electroweak scale appear as comparably light as the bottom quark at
the Tevatron collision energy of $\sqrt{s}\sim2\,\text{TeV}$.
When a very heavy scale is involved in the process, the gluon
splitting into a top-antitop pair may present a large logarithmic
enhancement.
For $Q\sim 10\,\text{TeV}$, for instance,
$\alpha_s(Q)\log(Q^2/m_t^2)\sim 0.6$, which makes a perturbative
expansion of the hard process questionable.
Therefore, one might wonder if the concept of top quark PDF is
relevant at the FCC, just as charm and bottom PDFs are commonly used
in LHC calculations.
The question is then what is more suitable and advantageous, from a
calculational point of view, to use at the FCC: a fixed-flavor number
(FFN) scheme, where the top is a massive quark, or a variable-flavor
number (VFN) scheme, where the top is a massless parton?
The discussion is thus completely analogous to the case of bottom
quarks at the LHC~\cite{Maltoni:2012pa}.

As with the charm and bottom quarks, introducing a PDF for the top
quark inside the proton allows us to resum potentially large collinear
logarithms of the form $\alpha_s^n(Q)\log^n(Q^2/m_t^2)$ to all orders
in perturbation theory.
The generalization of the DGLAP evolution equations to include a top
PDF up to NNLO is straightforward, and indeed most modern PDF sets
provide variants where the maximum number of light quarks in the PDF
evolution is set to $n_f=6$.
Indeed, the majority PDF fits are performed in a VFN scheme with a
maximum of $n_f=5$ light partons, since in the fitted hard
cross-sections top is always treated as a massive quark, and the
resulting PDFs at $\mu_F=m_t$ can then be used as boundary condition
to construct the $n_f=6$ PDFs including a top quark.

In Fig.~\ref{fig:topPDF} we show the top quark PDF, evaluated at
$Q=10$ TeV, compared with the other light partons, in the case of the
NNPDF2.3NNLO $n_f=6$ PDF set~\cite{Ball:2012cx}.
We observe that the top quark PDF can be of a similar size as the
light quark PDFs, in particular at medium and small-$x$, the region
where the effects of DGLAP evolution are dominant.
We also see that the charm and bottom PDFs are essentially
indistinguishable from the light quark PDFs for $x\lsim 10^{-3}$.
In Fig.~\ref{fig:topPDF} we also show the ratio between the gluon PDF
between the $n_f=5$ and $n_f=6$ schemes, as a function of the
factorization scale $Q$.
We observe that the differences between the two schemes can be up to
several percent for $Q\ge 1$ TeV, a region well covered by the FCC
kinematics.
Therefore, the use of the $n_f=6$ scheme would also have implications
for precision calculations involving gluons and light quarks, and not
only those with initial state top quarks.

\begin{figure}[t]
\begin{center}
\includegraphics[width=16cm]{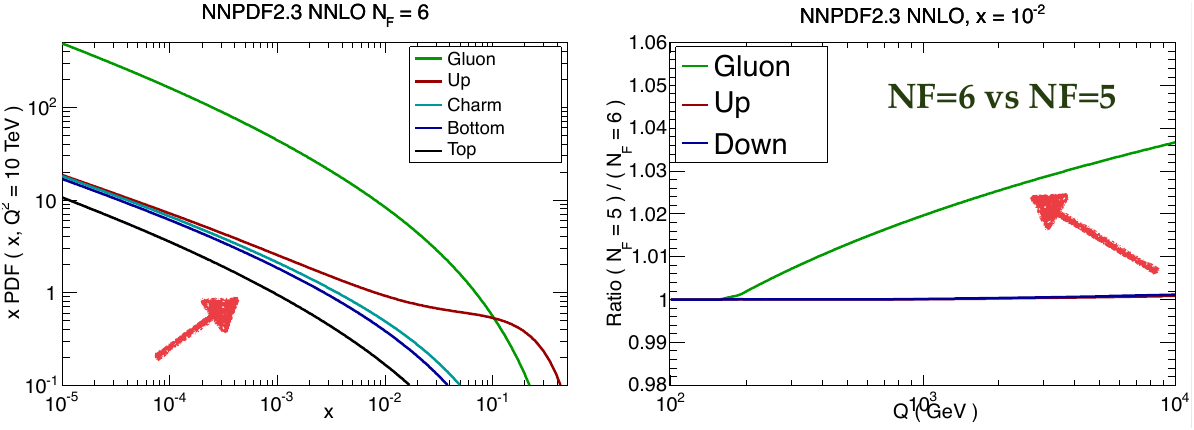}
\end{center}
\caption{Left plot: the top quark PDF compared with the other
  light partons, in the NNPDF2.3NNLO $n_f=6$ PDF set
  evaluated at $Q=10$ TeV. 
  Right plot: Ratio between the gluon PDF in the $n_f=5$ and $n_f=6$
  factorization schemes, as a function of the factorization scale $Q$.
}
\label{fig:topPDF}
\end{figure}

So while technically generating a top quark PDF is straightforward, it
still needs to be determined if it provides any calculational
advantage over using the standard FFN scheme approach, where the top
quark is always treated as massive, even for the extreme energies of a
100 TeV collider.
This issue has been recently studied
in~\cite{Han:2014nja,Dawson:2014pea}, both reaching similar
conclusions:
a purely massless treatment of top quarks is unreliable even at the
FCC, but the concept of a top quark PDFs is certainly relevant in the
context of matched calculations.
To illustrate this point, in the left plot of Fig.~\ref{fig:schemes1},
taken from~\cite{Han:2014nja}, we show a comparison of calculations in
the 5-flavor, massless 6-flavor, and ACOT
matched\cite{Aivazis:1993pi,Collins:1998rz} schemes for the inclusive
production of a hypothetic heavy scalar, labeled $H^0$, at a $100$ TeV
proton-proton collider.
This calculation uses as input the NNPDF2.3NLO $n_f=6$
set~\cite{Ball:2012cx}.
The ACOT scheme shows the desired behavior of interpolating between
the region near the top threshold and the very high energy limit
(where collinear logarithms in the top quark mass become large).
It should be stressed that the simplest LO $n_f=6$ calculation is
unreliable even for masses as large as $10$~TeV, indicating that the
minimum scale above which a parton interpretation for the top quark
becomes justified is much larger than the top mass itself.

The fact that the massless approximation for top production works
rather worse than for charm and bottom quarks can be traced back, at
least partially, to the fact that the resummed collinear logarithms
are substantially smaller as compared to the other heavy quarks.
This is illustrated by the right plot in Fig.~\ref{fig:schemes1},
taken from~\cite{Dawson:2014pea}, which compares the size of the
collinear logarithm $\alpha_s(\mu)\ln \mu^2/m_q^2$ as a function of
the ratio $\mu/m_q$ for the three heavy quarks: charm, bottom and top.
Even for very large ratios $\mu/m_q\sim 100$, the need for resummation
of collinear logarithms in the top quark mass is not evident, since
$\alpha_s(\mu)\ln \mu^2/m_q^2$, while being large, is perturbative in
the relevant kinematical range.
This is opposed to charm, and to a lesser extend bottom, whose
corresponding logarithms eventually become non-perturbative and
require collinear resummation.
One reason that partially explains this difference is the fact that
$\alpha_s(m_t)\ll \alpha_s(m_c)$, which allows a much larger lever arm
in $Q$ before resummation is required.

\begin{figure}[t]
\begin{center}
  \includegraphics[width=8.2cm]{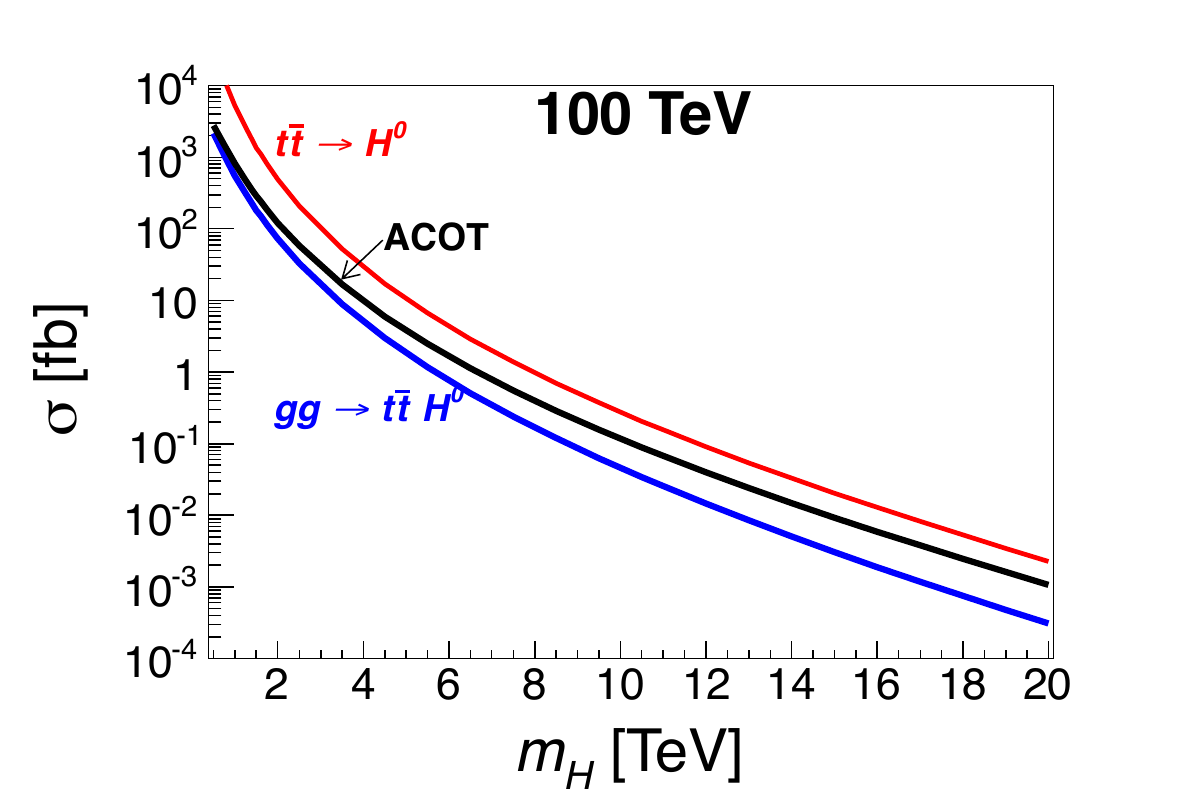}
  \includegraphics[width=7.7cm]{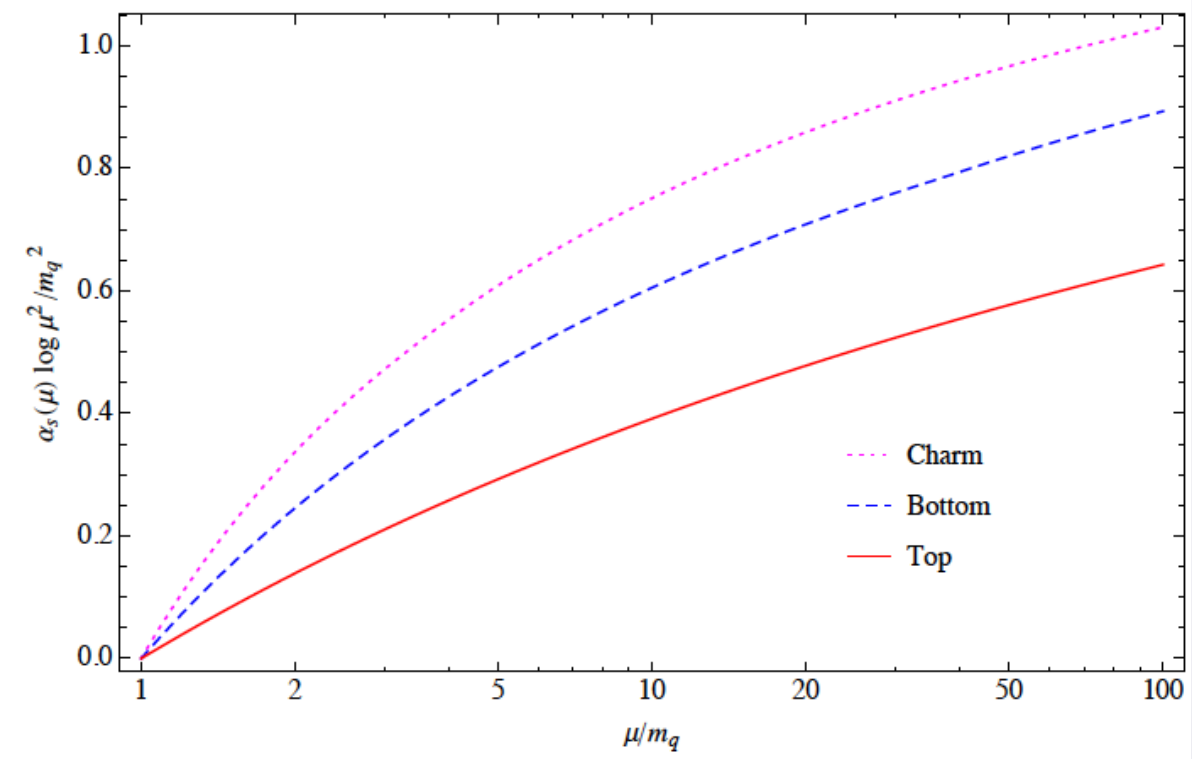}
\end{center}
\vspace{-0.6cm}
\caption{Left plot:
  Inclusive cross section for $H$ production with Yukawa
  coupling $y=1$ at $100$ TeV versus its mass $m_{H}$, in the 5-flavor scheme (bottom blue), the
  6-flavor scheme (upper red), and the ACOT scheme (middle black),
  from~\cite{Han:2014nja}.
  Right plot: the size of the collinear logarithm $\alpha_s(\mu)\ln
  \mu^2/m_q^2$ as a function of $\mu/m_q$ for charm, bottom and top,
  from~\cite{Dawson:2014pea}.  }
\label{fig:schemes1}
\end{figure}

So in conclusion, current studies indicate, while the purely massless
approximation for top quarks is unreliable even at the extreme FCC
energies, the concept of top quark PDF is certainly useful in order to
construct matched calculations.
This way, one can supplement and improve massive fixed-order
calculations with all-order resummations of collinear logarithms in
matched schemes such as ACOT~\cite{Aivazis:1993pi,Collins:1998rz} or
FONLL~\cite{Cacciari:1998it,Forte:2010ta}.
For example, as shown in the heavy quark chapter of this report,
Sect.~\ref{sec:hvq}, it is possible to generalize the FONLL
calculation for the $p^h_T$ distribution in for heavy flavor
differential distributions~\cite{Cacciari:1998it} to the case of top
quark production at the FCC.
The matched calculation is found to provide a more precise estimate in
the region of transverse momenta up to 10 TeV.
Eventually, this matching can be performed up to NNLO order, using the
corresponding calculations for jet production~\cite{Currie:2013dwa}
and for top quark production~\cite{Czakon:2015owf}.
%

\subsection{Photon- and lepton-initiated processes at 100 TeV}
\label{sec:pdf_photon}

A 100 TeV Future Circular Collider is bound to probe the interactions
of elementary particles at extreme energies with high accuracy.
In order to correctly identify possible BSM effects, the theoretical
predictions for the SM processes have to match the precision of the
corresponding experimental measurements.
In other words, the impact of higher-order corrections on
phenomenological predictions has to be under control.
To this purpose, the computation of NLO QCD corrections is necessary,
but often not sufficient.
In fact, at fixed order the inclusion of the NNLO QCD corrections in
QCD as well as of the EW corrections is in general desirable and in
particular cases even essential.
The implications of higher-order EW corrections to matrix elements for
FCC processes is discussed elsewhere in this report.

In order to formally achieve the desired level of accuracy, not only
the matrix elements of the hard processes, but also the parton
distribution functions (PDFs) of the proton, have to be known at the
same level of precision.
While most PDF groups provide since some time PDF sets accurate up to
NNLO in QCD~\cite{Ball:2014uwa,Harland-Lang:2014zoa,Dulat:2015mca,
  Abramowicz:2015mha,Alekhin:2013nda}, for EW corrections the
situation is less satisfactory.
Indeed, EW corrections require the calculation of photon-induced
processes, and thus PDFs both with QED effects in the evolution and
with a determination of the photon PDF $\gamma(x,Q)$ are necessary for
consistent calculations.
In this respect, a number of PDF sets with a photon PDF and QED
effects are available in LHAPDF: MRST2004QED~\cite{Martin:2004dh},
NNPDF2.3QED~\cite{Ball:2013hta} and the recent
CT14QED~\cite{Schmidt:2015zda}. 
In addition, PDF evolution with QED effects has been implemented in
the {\tt APFEL} PDF evolution
program~\cite{Carrazza:2014gfa,Bertone:2013vaa} at LL, and inclusion
of NLL QED splitting functions~\cite{deFlorian:2015ujt} is underway.

The determination of the photon PDF obtained by the three
groups differ in a number of
important aspects.
First, different data sets are used in the fits.
Second, the form of the photon distribution at the initial scale $Q_0$
is different. Finally, the DGLAP evolution from $Q_0$ to the final
scale $Q$ is not the same in all cases.
As far as the functional form of the PDF at the initial scale
$\gamma(x,Q_0)$ is concerned, NNPDF2.3QED only assumes that the
photon PDF is positive-definite. In a first step, PDF
replicas for all partons are fitted to deep-inelastic
structure functions, which only provide very loose constraints
on the photon PDF.
In a second step, the photon PDF is constrained from LHC Drell-Yan
data. This constraint enters at LO, however, because the
photon-initiated component of Drell-Yan production is small, even the
relatively precise LHC data constrain the photon PDF only weakly. In
particular, since no data is available at large $x$ and no functional
form is assumed, in this region PDF uncertainties on $\gamma(x,Q)$
turn out to be quite large.\footnote{High-statistics Drell-Yan
  measurements at the LHC such as the recent ATLAS 8 TeV high-mass
  analysis~\cite{Aad:2016zzw} should provide additional information in
  this region.}

In contrast to the NNPDF2.3QED determination, the CT14QED and
MRST2004QED sets are based on the assumption that the functional form
of the photon PDF at the initial scale can be determined by the
valence-quark distributions.
In essence, they are given by a convolution of valence-quark
distributions with the $P_{\gamma q}$ splitting functions, with a
normalization for the up- and down-type distributions that differ in
the two approaches.  Determining the photon PDF reduces then to fixing
one or two parameters in the CT14QED and MRST2004QED approaches,
respectively.  For the CT14QED, set the constraints are obtained by
fitting ZEUS data for the production of isolated photons in DIS, while
for MRST2004QED an assumption is made for the normalization
coefficients and no data are used to constraint the photon PDF.

Recently the CT collaboration also released a photon PDF that includes
the elastic component of the photon PDF (CT14QEDinc) obtained in the
so-called photon equivalent approximation, which involves an
integration over the proton electromagnetic form factors.
In fact, the photon PDF, unlike the quark and gluon PDFs, has a large
elastic component in which the proton remains intact
(see~\cite{Martin:2014nqa,Harland-Lang:2016apc,Harland-Lang:2016lhw}
and references therein).
This component has not been discussed in the NNPDF2.3QED and
MRST2004QED fits, but is included in the photon PDF determination
of~\cite{Harland-Lang:2016apc}.
Another important difference is connected to the DGLAP evolution: in
the evolution of the CT14QED and MRST2004QED the scale is evolved
simultaneously for the QCD and for the QED evolution, while in the
NNPDF2.3QED approach the two scales run independently.
Very recently, in Ref.~\cite{Bertone:2016ume}, the NNPDF3.0QED set has
been derived, which combines the NNPDF3.0 quark and gluon PDFs with
the NNPDF2.3 photon PDF using the same solution of the DGLAP equations
as CT14QED and MRST2004QED.

All these differences result in predictions for the photon PDF from
different sets that are not always compatible. In particular, as
compared to NNPDF2.3QED, the CT14QED and MRST2004QED photon
distribution functions are softer at large $x$, and exhibit smaller
PDF uncertainties due to their more restrictive parametrizations.
It will be important to understand and resolve the sources of these
differences between QED PDF sets.
In the following we will present results based on the NNPDF2.3QED set,
with the caveat that conclusions could be rather different if other
QED sets were used as input to the calculations.

At very high energies, even PDFs for electroweak massive gauge bosons
might be required, and this possibility is discussed in
Sect.~\ref{sec:pdf_ew} below.
On top of the photon-induced processes, higher-order EW corrections
also induce lepton-initiated channels whose computation formally
requires the knowledge of the leptonic content of the
proton~\cite{Bertone:2015lqa}.
To determine the lepton PDFs, the first step is to include them in the
DGLAP evolution equations with QED corrections~\cite{Bertone:2015lqa},
which mixes the evolution of the lepton and photon PDFs with that of
quarks and gluons.
Next one needs to adopt suitable boundary conditions $i.e.$, the
initial scale lepton PDFs.
Since a determination of lepton PDFs from data is hardly achievable
because of their smallness, here we assume that the light lepton PDFs,
$i.e.$, electrons and muons, are purely generated by photon splitting
at the respective mass scales.

Under this assumption, one can approximate their distributions at the
initial scale $Q_0\simeq1$ GeV as:
\begin{equation}\label{eq:ansatz}
  \ell^{\pm }(x,Q_0) =
  \frac{\alpha(Q_0)}{4\pi} \ln\left(\frac{Q_0^2}{m_\ell^2}\right)
  \int_x^1\frac{dy}{y} P_{\ell\gamma}^{(0)}\left(\frac{x}{y}\right)
  \gamma(y,Q_0)\,, \quad l^{\pm}=e^{\pm},\mu^{\pm}\, ,
\end{equation}
with $\alpha$ the QED running coupling constant.
The $\tau^{\pm}$ lepton PDFs are then dynamically generated from
threshold using the standard variable-flavour-number
scheme~\cite{Forte:2010ta,Aivazis:1993pi,Thorne:1997ga}.
Here we will use use the \texttt{apfel\_nn23qednlo0118\_lept} set of
PDFs~\cite{Bertone:2015lqa} generated starting from the NNPDF2.3QED
NLO set using the Ansatz in Eq.~(\ref{eq:ansatz}) for the light lepton
PDFs.

Before studying the size of photon- and lepton-initiated processes at
a 100 TeV collider, it is useful to study the behaviour of the parton
luminosities of the different initial states, by including also photon
and leptons initiated processes.
Parton luminosities can either be defined as a function of $M_X$, the
invariant mass of the final state, as done
in Eqns.~(\ref{eq:lumdef_sec3})--(\ref{eq:qg}),
or in terms of  $y$, the rapidity of the final state,
integrating over the invariant mass,
\begin{equation}\label{eq:lumdefy}
  \Psi_{ij}(y) \equiv 
  2e^{-2y}\int_{\sqrt{\tau_{\rm cut}}e^{y}}^{e^{-y}} dx\,x
  f_{i}(x,\sqrt{s}xe^{-y}) f_{j}(xe^{-2y},\sqrt{s}xe^{-y})\,,
\end{equation}
with $\tau_{\rm cut} \equiv M_{X,\rm cut}^2/s$.
In Eq.~(\ref{eq:lumdefy}) the lower bound of the integral,
proportional to $\sqrt{\tau_{\rm cut}}$, implies that $M_{X}\geq
M_{X,\rm cut}$.

\begin{figure}[t]
  \centering
  \includegraphics[scale=0.49]{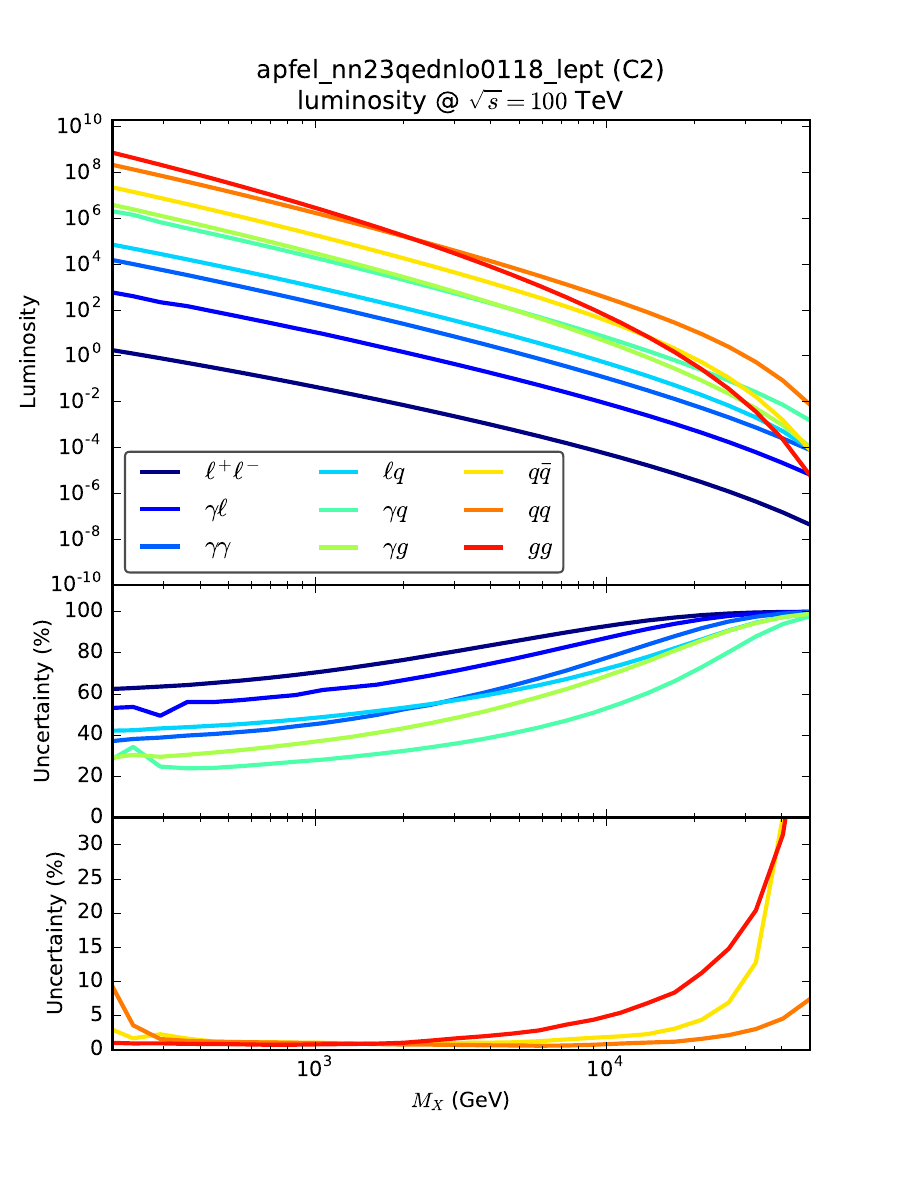}
  \includegraphics[scale=0.49]{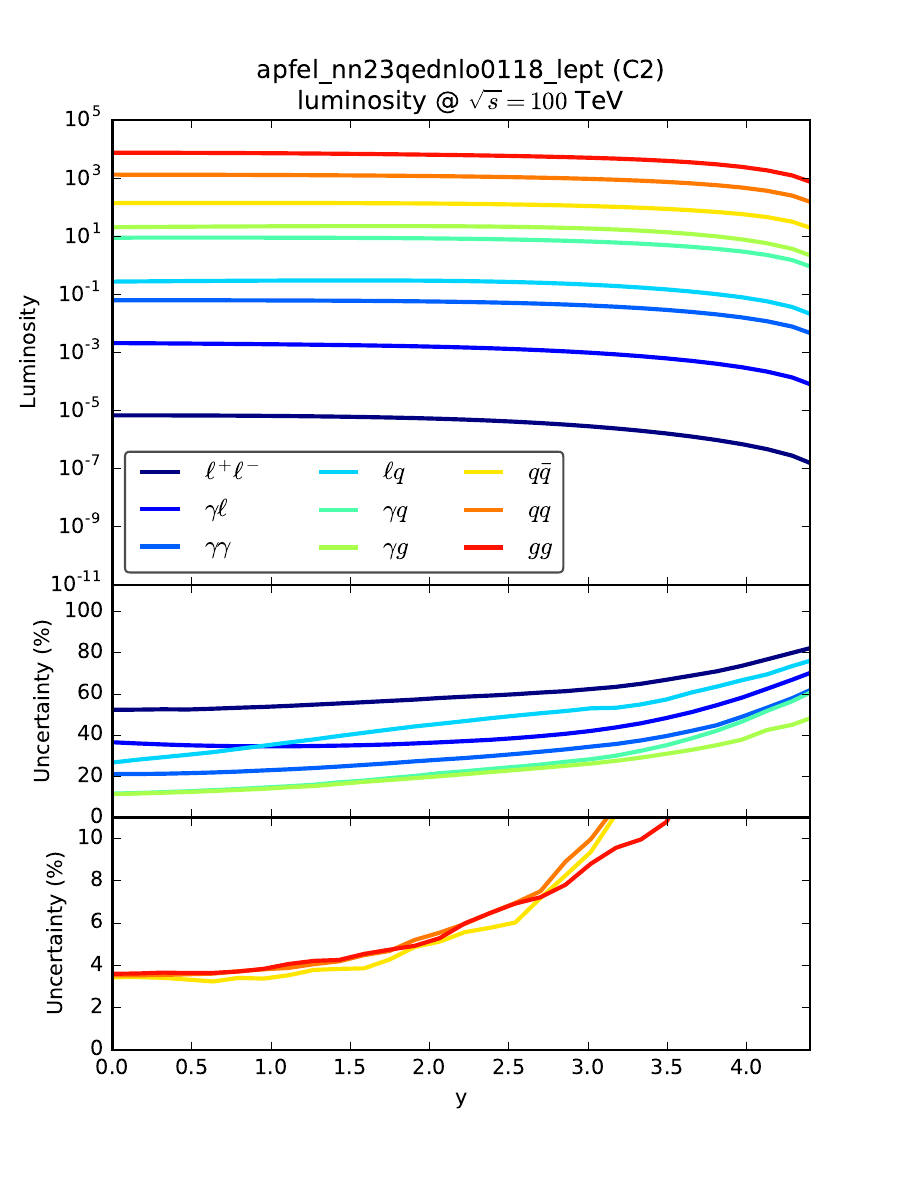}
  \vspace{-0.5cm}
  \caption{PDF luminosities for the quark-quark ($qq$),
    quark-antiquark ($q\bar{q}$) and gluon-gluon ($gg$) initial states
    compared with
    the different photon- and
    lepton-initiated channels, as a function of the invariant mass
    $M_X$ of the final-state system (left) and of its rapidity $y$
    (right).
    The central and lower panels show the corresponding 68\% confidence level
    PDF uncertainties in the various cases.
    Note that in the right plot the rapidity $y$ is that of the final-state
    system, not the rapidity of the final-state particles.
    In the rapidity-dependent luminosity, the minimum value of the
    final-state invariant mass is set to $M_{X,\rm cut} =10$ GeV.
  }
  \label{fig:lumileptM}
\end{figure}

In Fig.~\ref{fig:lumileptM} we compare the size and the shape of the
different parton luminosities for an hadron collider with a center of
mass energy $\sqrt{S}=100$ TeV, both as a function of the invariant
mass $M_X$ (left plot) and of the final-state system rapidity $y$
(right plot).
For the rapidity-dependent luminosities, we impose a cut of $M_{X,\rm
  cut} = 10$ GeV.
In Figs.~\ref{fig:lumileptM} we also plot the corresponding 68\%
confidence level PDF uncertainties for each luminosity type,
separating the luminosities involving photon or lepton PDFs (central
panel) and those involving only quarks or gluons (lower panel).
The central value of the luminosities is assigned to be the midpoint
of the 68\% confidence level interval, and thus by construction
PDF uncertainties will not exceed 100\%, as can be seen 
in the central panel of Fig.~\ref{fig:lumileptM} (left).

The relative size of the plotted luminosities follows the expected
pattern. In general, the photon PDF suppresses the luminosity by a
factor of $\alpha\simeq 10^{-2}$ with respect to the (anti)quark PDFs
and, analogously, the lepton PDFs suppress the luminosity by an
additional factor of $\alpha$ with respect to the photon PDF.
This can be easily seen in Fig.~\ref{fig:lumileptM}, {\it e.g.} by
comparing $\Phi_{\gamma\ell}$($\Psi_{\gamma\ell}$) with
$\Phi_{\gamma\gamma}$($\Psi_{\gamma\gamma}$) and $\Phi_{\ell ^+ \ell
  ^-}$($\Psi_{\ell ^+ \ell ^-}$), the three lowest curves.
However, from Fig.~\ref{fig:lumileptM} we also notice that this
hierarchy is not satisfied at large invariant masses.
In this kinematic region, large $M_X$, one is probing PDFs at rather
large values of $x$, and here the pure-QCD luminosity combinations,
$\Phi_{q\bar{q}}$, $\Phi_{gq}$ and $\Phi_{gg}$, become closer to the
luminosities involving photons and leptons, with important
phenomenological implications: as opposed to the naive expectation,
photon- and lepton-initiated contributions can become as large as the
standard quark-initiated contributions. However, it is important to
keep in mind that the uncertainty in the NNPDF2.3QED luminosity
determinations involving photons (shown in the middle panel of
Fig.~\ref{fig:lumileptM}) is very large, and that the NNPDF2.3QED
results are not compatible with other determinations that instead
predict a lower photon PDF effects at large $M_X$.
In the NNPDF approach, it can be shown that this effect is partially
caused by the relative behaviour of the strong coupling $\alpha_s$
with respect to the QED coupling $\alpha$ as functions of the scale
$M_X$, together with the fact that PDF uncertainties for the photon
(and thus for the lepton) PDFs at large-$x$ are huge, $\ge 50\%$ for
$M_X\ge 10$ TeV, see the central panel of Fig.~\ref{fig:lumileptM}
(left).

From Fig.~\ref{fig:lumileptM} we also see that, contrary to the
$\Phi_{ij}(M_X)$ luminosities, the rapidity-dependent luminosities
$\Psi_{ij}(y)$ maintain the same hierarchy all over the range in $y$.
 The reason for this is that the value of
 the final state system rapidity $y$ is not directly related to the value of
$M_X$, which also in this case is used as factorisation scale.
Thus, the previous argument justifying the suppression of the QCD
luminosities with respect to the QED ones does not apply for the case
of $\Psi_{ij}$.
Note that for the rapidity-dependent luminosity in
Fig.~\ref{fig:lumileptM}, the rapidity $y$ is that of the final-state
system (say a $Z'$ boson in inclusive $Z'$ production), not the
rapidity of the final-state particles (in this case the leptons from
the $Z'$ decay).

Following this discussion of the PDF luminosities including photon-
and lepton-initiated channels, now we present predictions for
electroweak production processes at a 100 TeV hadron collider.
We concentrating on the differential distributions as a function of
the final state invariant mass $M_X$, allowing a direct mapping with
the corresponding PDF luminosities collected in
Fig.~\ref{fig:lumileptM}.
Our results have been obtained with {\aNLO}~\cite{Alwall:2014hca}
using the \texttt{apfel\_nn23qednlo0118\_lept} PDF set.
The relevant SM input parameters have been set to the following
values:
\begin{eqnarray}
&\alpha_s(m_{Z})=0.118\, ,\quad &G_{F}=1.16639\times 10^{-5}\,,
\nonumber\\ & m_{Z}=91.1876~\gev\, , \quad & m_{W}=80.385~\gev\, , \nonumber\\
& m_{H}=125~\gev\,, \quad &\Gamma_{Z}=2.4952~\gev\, , \quad \Gamma_{W}=2.085~\gev\, . \label{eq:inputs}
\end{eqnarray}
The masses of all quarks (except the top quark) and leptons are
neglected.
We set the renormalisation and factorisation scales to
$\mu_{F}=\mu_{R}=H_{T}/2$, where $H_{T}$ is the scalar sum of the
transverse masses of the final-state particles.
We restrict ourselves to LO results at the parton level, since NLO
corrections and parton shower effects would not modify qualitatively
the results.
We separately identify the contributions from initial states with only
(anti)quarks and gluons, initial states with at least one photon and
no leptons, and initial states with at least one lepton.

In this report, we aim only to disentangle the contributions of the
photon and lepton-initiated channels compared to the quark and gluon
initiated channels for 100 TeV processes.
A more refined phenomenological study of these processes would require
to include the NLO EW corrections, which in general cannot be
neglected.
The interplay between photon-initiated processes and NLO EW
corrections have been studied, among others,
in~\cite{Boughezal:2013cwa} for neutral current Drell-Yan
and~\cite{Baglio:2013toa} for $WW$ production, as well as
in~\cite{Hollik:2015lha} for squark-antisquark production.

We start by considering the case of the production of an
electron-positron pair at $\sqrt{s}=100$ TeV.
At leading order we have the usual quark-antiquark annihilation
diagram (neutral current Drell-Yan), and in the presence of EW
corrections we also need to account for the photon-photon
electron-positron initial states.
Similarly, also $\mu^+ \mu^-$ and $\tau^+ \tau^-$ initial states can
contribute to the corresponding final states.
Each initial state leads to a different contribution to the $M_X$
invariant mass distributions: $q\bar{q}$ has a $s$-channel diagram,
$\gamma\gamma$ has $t$-and $u$-channel diagrams, while the $e^+e^-$
initial state has $s$- and $t$-channel diagrams.
These three partonic processes yield LO cross sections of
$\ord(\alpha^2)$, thus they all contribute to the same order in the
perturbative expansion.

In Fig.~\ref{fig:epem-fcc1} we show the invariant mass distribution of
the lepton pair in neutral Drell-Yan production at a 100 TeV hadron
collider for $m_{e^+e^-} \ge 5\tev$.
We also investigate how the results are modified in the presence of
realistic acceptance cuts.
In the left plot of Fig.~\ref{fig:epem-fcc1}, the transverse momentum
of the leptons must satisfy a $p_T^{e^\pm} \ge 10~\gev$ cut, while in
the right plot the $p_T$ selection requirement is $p_T^{e^\pm} \ge
100~\gev$ and in addition there is a rapidity acceptance requirement
of $|\eta_{e^\pm}| \le 4$.
The center panels shows the relative contribution of each initial
state, while the lower panel shows the corresponding PDF uncertainty
in each case.

\begin{figure}
  \centering
  \includegraphics[clip=true, trim=0.cm 3.5cm 0.7cm 1cm, width=0.49\textwidth]{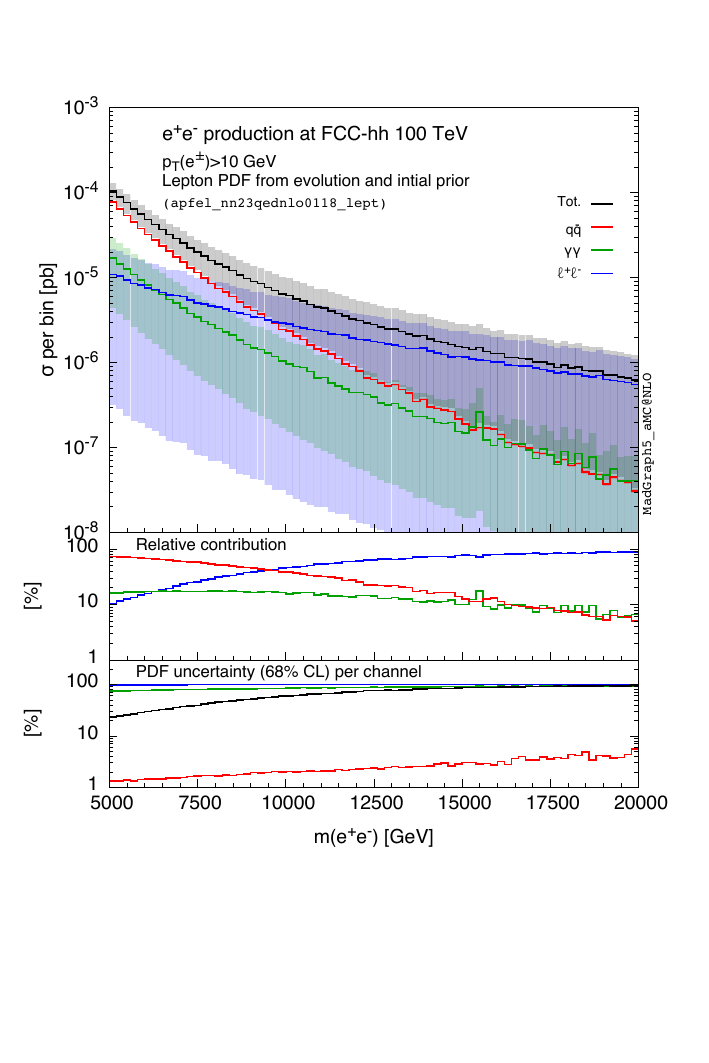}
  \includegraphics[clip=true, trim=0.cm 3.5cm 0.7cm 1cm, width=0.49\textwidth]{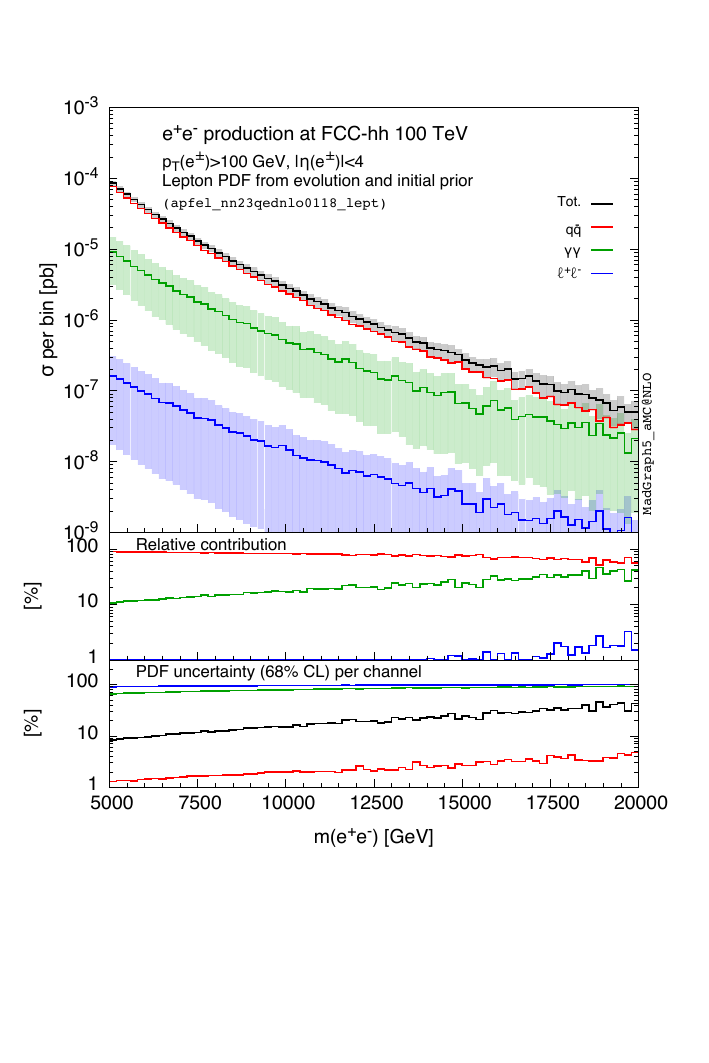}
  \caption{\label{fig:epem-fcc1}
    Upper panel: the invariant mass distribution in $e^+e^-$ production
    at a 100 TeV hadron collider.
    In the left plot, the transverse momentum of the leptons must
    satisfy a
    $p_T^{e^\pm} \ge 10~\gev$ cut, while in the right plot the $p_T$ selection
    requirement is
    $p_T^{e^\pm} \ge 100~\gev$ and in addition there is a rapidity
    acceptance requirement of $|\eta_{e^\pm}| \le 4$.
    The center panels shows the relative contribution of each initial
    state, while the lower panel shows the corresponding
    PDF uncertainty in each case.
  }
\end{figure}

We see that in the case of loose (and unrealistic) acceptance cuts,
left plot of Fig.~\ref{fig:epem-fcc1}, the contribution of the
$\ell^+\ell^-$ channel is not negligible and is even dominant for
$m_{e^+e^-} \ge 5\tev$.
This behaviour is due to the fact that the partonic cross-section for
the $e^+e^- \rightarrow e^+e^-$ process with massless electrons has a
collinear divergence for electrons collinear to the beam pipe.
However, once a reasonable acceptance cuts in the lepton transverse
momentum $p_T^{e^\pm} \ge 100\gev$ and in their rapidity
$|\eta^{e^\pm}|\le 4$, the contribution of the $\ell^+\ell^-$ initial
state is strongly suppressed (right plot).

Note also that, even for realistic acceptance cuts, the photon-photon
initiated contribution is $\ge 10\%$ for all the range in invariant
mass, although with very large associated uncertainties, and thus is
mandatory to include it in any precision calculation.
Part of this effect is the consequence of the relative behaviour of
the $M_X$-differential luminosities shown in the left panel of
Fig.~\ref{fig:lumileptM} where the $\Phi_{\gamma \gamma}$ luminosity
is relatively less suppressed as compared to $\Phi_{q\bar{q}}$ at
large invariant masses.
Moreover, the $q\bar{q}$-channel receives an additional kinematic
suppression due to $s$-channel diagrams that are instead absent in the
$\gamma\gamma$-channel.
We also note that the $\gamma\gamma$ contribution is affected by very
large PDF uncertainties, but these will have been greatly reduced
before the start of the operations of the FCC thanks to the full
exploitation of the constrains from the LHC data~\cite{Rojo:2015acz}.

In Fig.~\ref{fig:epem-fcc1-integrated} we show the total integrated
cross-section for the production of a dilepton pair at $\sqrt{s}=100$
TeV with invariant mass above a given threshold $m_{\rm min}$.
The final-state leptons are required to have a transverse momentum
$p_T^{e^\pm}\ge 100$ GeV and to lie in the rapidity range
$|\eta^{e^\pm}|\le 2.5~(4.0)$ in the left (right) plot.
Given the integrated luminosities expected at the FCC, we see that one
can expect sizable rates of dilepton events with invariant masses
above 20 TeV.
As in Fig.~\ref{fig:epem-fcc1}, the contribution of the lepton PDFs is
negligible once the calculation is restricted to the experimentally
accessible region.
At the highest possible invariant masses, the contribution from the
$\gamma\gamma$ initial state could be as large as that from the
$q\bar{q}$ initial state, although current uncertainties on the photon
PDF are still too large to draw any definitive conclusion.

\begin{figure}
  \centering
  \includegraphics[clip=true, trim=0.cm 3.5cm 0.7cm 1cm, width=0.49\textwidth]{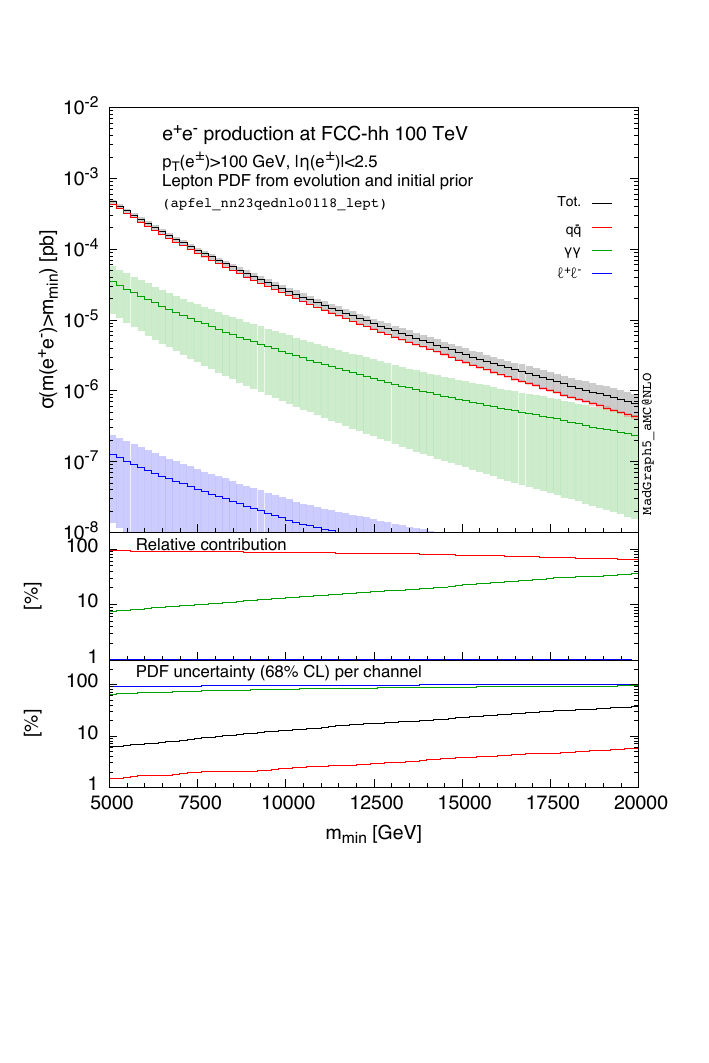}
  \includegraphics[clip=true, trim=0.cm 3.5cm 0.7cm 1cm, width=0.49\textwidth]{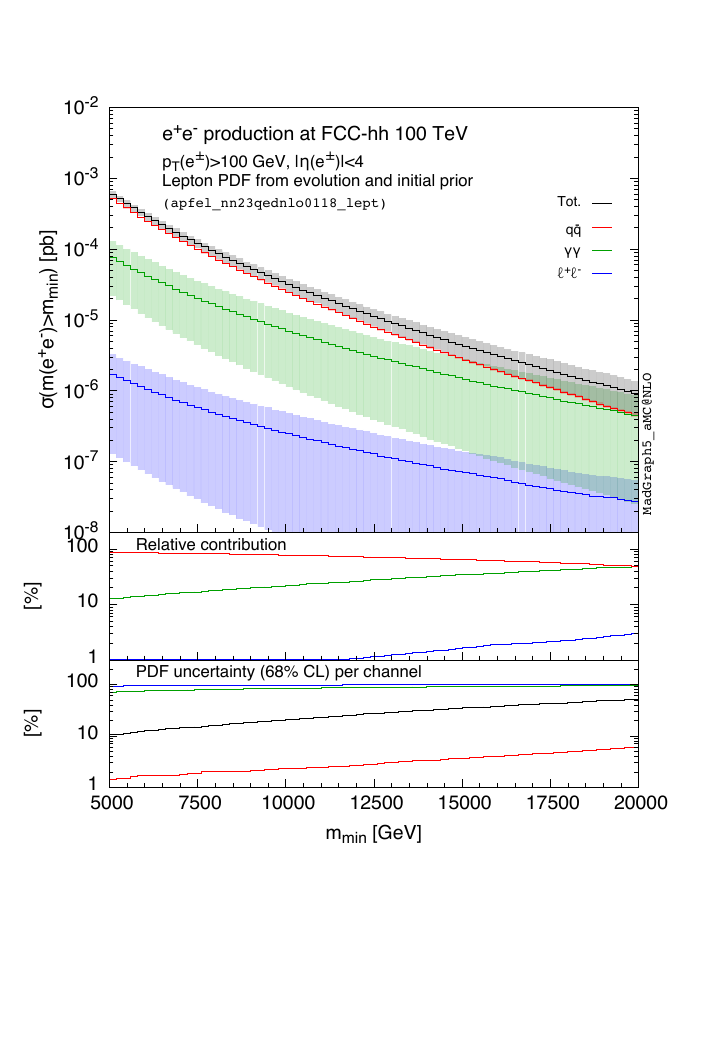}
  \caption{\label{fig:epem-fcc1-integrated}
    Same as Fig.~\ref{fig:epem-fcc1}, now
    we showing the total integrated cross-section above a minimum value
    of the invariant mass of the dilepton pair $m_{\rm min}$.
    The leptons are required to have a transverse momentum
    $p_T^{e^\pm}\ge 100$ GeV and to lie in the rapidity range
    $|\eta^{e^\pm}|\le 2.5~(4.0)$ in the left (right) plot.  }
\end{figure}

Next we turn to the production of electroweak gauge boson pairs at 100
TeV, in particular, we consider at $W^+W^-$ production with undecayed
$W$ bosons.
A more detailed study of di-boson production at 100 TeV can be found
in Sect.~\ref{sec:VV} of this report.
In the calculation, we keep the $W$ boson stable so that we can
estimate the effects due only to the $\ell^{+}\ell^{-}$ luminosity, as
opposed to also the matrix-element enhancements.
In Fig.~\ref{fig:ww} we show the differential distributions for the
invariant mass of the di-boson pair $m_{W^+ W^-}$ using the same
format as for di-lepton production in Fig.~\ref{fig:epem-fcc1}.
In the left plot we have not imposed any acceptance cut, while in the
right plot the rapidity of the electroweak gauge bosons is required to
satisfy $|\eta_{W^{\pm}}|\le 4$.

First of all, we observe that also for $W^+W^-$ production the
contribution from the lepton PDFs can be safely neglected, as was the
case in di-lepton production.
On the other hand, the photon-initiated contribution dominates over
the quark-antiquark annihilation for $m_{W^+ W^-} \ge 7.5$ TeV in the
case of realistic selection cuts.
One should however take into account that this $\gamma\gamma$
contribution is affected by very substantial PDF uncertainties for all
the relevant range of $m_{W^+ W^-}$ values.

As in the case of di-lepton production, the increase of the relative
importance of the $\gamma\gamma$ channel for large $m_{W^+ W^-}$ is
consistent with the behaviour of the $\Phi_{\gamma \gamma}$ and
$\Phi_{q\bar{q}}$ luminosities shown in Fig.~\ref{fig:lumileptM}.
Again, no suppression from $s$-channel
diagrams is present in $\gamma \gamma \rightarrow W^+ W^-$ production,
leading to a further relative enhancement with respect to the $q\bar{q}$
channel at high  $m_{W^+ W^-}$.
On the other hand, in the $\gamma\gamma$-channel the $W$ bosons are
produced more peripherally than in the $q\bar q$-channel.
Therefore, the cut in pseudorapidity reduces the relative impact of
the $\gamma\gamma$ channel, but it does not modify the qualitative
conclusions.

\begin{figure}
  \centering
  \includegraphics[clip=true, trim=0.cm 3.5cm 0.7cm 1cm, width=0.49\textwidth]{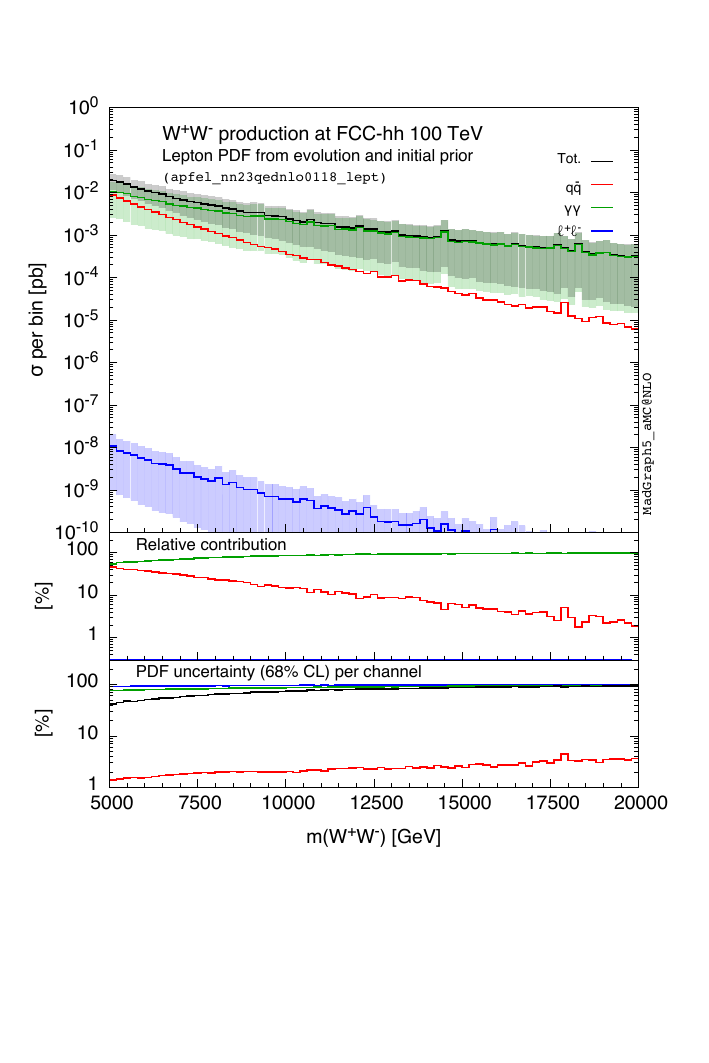}
  \includegraphics[clip=true, trim=0.cm 3.5cm 0.7cm 1cm, width=0.49\textwidth]{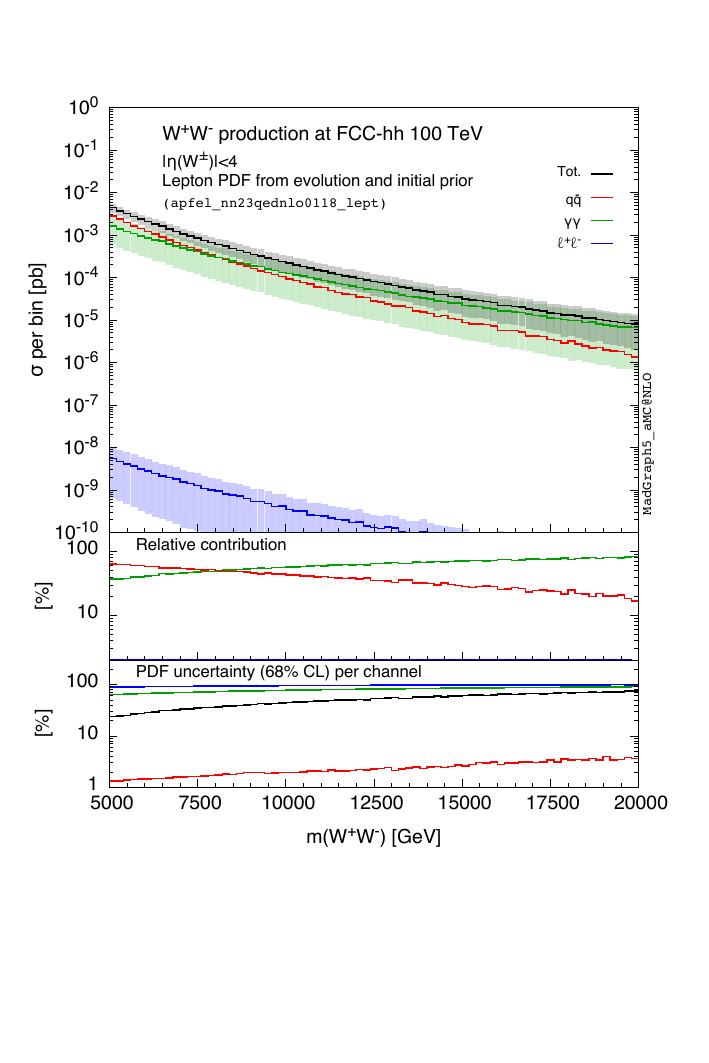}
  \caption{\label{fig:ww} Same as Fig.~\ref{fig:epem-fcc1}
    for the production of $W^+W^-$ pairs at a 100 TeV hadron collider.
    In the left plot we have not imposed any acceptance cut, while
    in the right plot the rapidity of the electroweak gauge bosons
    is required to satisfy $|\eta_{W^{\pm}}|\le 4$.
  }
\end{figure}

In Fig.~\ref{fig:ww-integrated} we show a similar comparison as that
in Fig.~\ref{fig:ww}, but now plotting the total integrated
cross-section above a minimum value of the invariant mass of the
$W^+W^-$ pair $m_{\rm min}$, rather than the cross-section per bin.
The rapidity of the $W$ bosons is restricted to lie in the
$|\eta^{W^\pm}|\le 2.5~(4.0)$ region in the left (right) plot.
Therefore, the rates for di-boson production will be substantial even
for invariant masses as large as $m_{\rm min}\simeq 20$ TeV, specially
if also hadronic decay channels can be reconstructed.

\begin{figure}
  \centering
  \includegraphics[clip=true, trim=0.cm 3.5cm 0.7cm 1cm, width=0.49\textwidth]{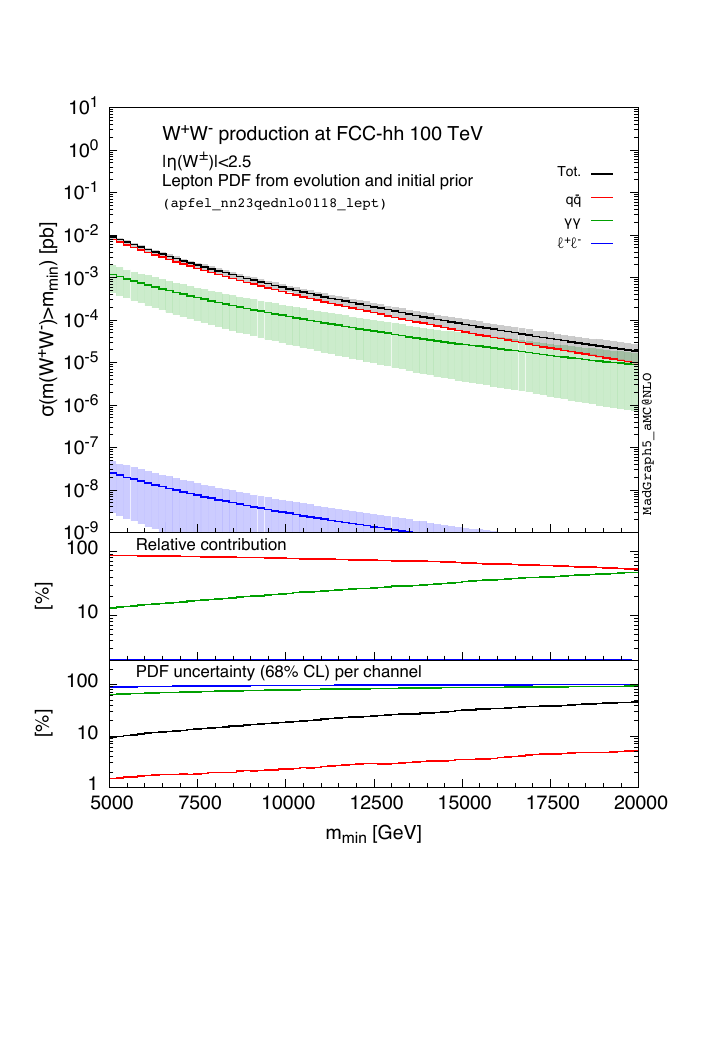}
  \includegraphics[clip=true, trim=0.cm 3.5cm 0.7cm 1cm, width=0.49\textwidth]{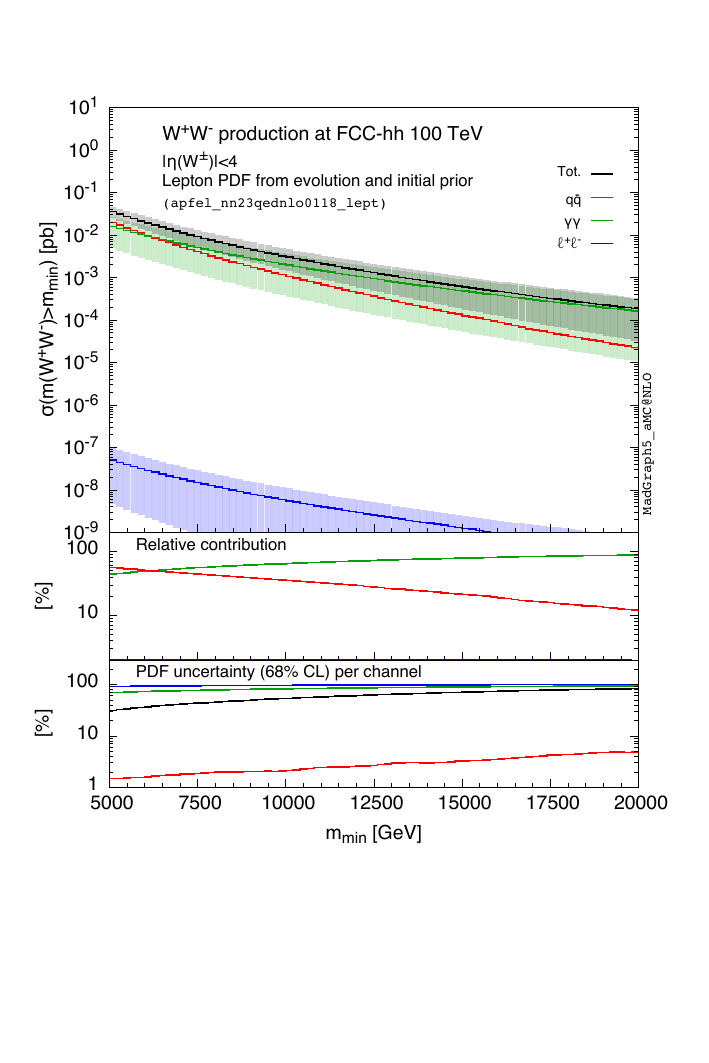}
  \caption{\label{fig:ww-integrated}
    Same as Fig.~\ref{fig:ww},
now showing the total integrated cross-section above a minimum value
    of the invariant mass of the $W^+W^-$ pair $m_{\rm min}$.
    The rapidity of the $W$ bosons is restricted to lie in the
    $|\eta^{W^\pm}|\le 2.5~(4.0)$ region in the left (right) plot.  }
\end{figure}

To summarize, in this contribution we have explored the impact of
photon- and lepton-initiated contributions to electroweak processes at
a 100 TeV hadron collider.
We find that both for Drell-Yan and for $WW$ production, the
contribution from the $\gamma\gamma$ initial state is comparable to
that from $q\bar{q}$ annihilation within the large uncertainties of
the former.
While the photon-initiated contribution currently is affected by large
PDF uncertainties, this should not be a major issue at the FCC since
these uncertainties can be substantially reduced using the information
from available and future LHC measurements.
We also find that, provided realistic acceptance cuts are imposed, the
contribution from lepton-initiated processes is as expected completely
negligible.

\subsection{Electroweak gauge bosons as massless partons}
\label{sec:pdf_ew}

For processes that involve energies much greater than the electroweak
scale, it might be more adequate to treat massive electroweak gauge
bosons as massless partons, in a way similar to what can be done with
heavy quarks; see Sect.~\ref{sec:pdf_top}.
The justification to consider EW bosons as initial-state partons at
very high energies is discussed in more detail in
Sect.~\ref{sec:dyan_highenergy}, where relevant technical issues are
addressed.
In this section, we present instead some preliminary results for the
effects of including weak gauge bosons as massless partons into the
DGLAP evolution equations for parton distributions.

Electroweak evolution equations are substantially more involved than
their QED and QCD counterparts; see~\cite{Ciafaloni:2005fm} and
references therein.
However, one can obtain a first approximation of their effects by
studying the fixed-order splitting rates of quarks into $W$ and $Z$
bosons.
This approach, which generalizes the usual Weizs\"acker-Williams
calculation for collinear photon radiation off a relativistic charge,
is known as the effective $W$
approximation\cite{Dawson:1984gx,Kane:1984bb}.
Note that this approximation formally breaks down when the
interference between transverse and longitudinal polarizations is
important~\cite{Dawson:1984gx}.
Sub-dominant contributions to this approximation include power
corrections of $\mathcal{O}(M_{W/Z}^2/Q^2)$~\cite{Kunszt:1987tk} as
well as higher-order perturbative
QCD~\cite{Kunszt:1987tk,Dawson:1988ai} and EW
corrections~\cite{Kunszt:1987tk}.

One major novelty is the appearance of longitudinal polarization
modes.
For radiation of a $W$ boson off an unpolarized light quark $q$ in the
initial state, carrying an energy fraction $z \equiv E_W / E_q$, in
the limit where $E_W \gg m_W$ the leading-order transverse and
longitudinal $W$ content of the proton~\cite{Dawson:1984gx} is then
given by:
\begin{eqnarray}
&& f_{W_T \in q}(z,Q^2) = \frac{C_V^2 + C_A^2}{8\pi^2}\left(\frac{1 + (1-z)^2}{z}\right) \log\left(\frac{Q^2}{M_W^2}\right)\, , \label{effwapprox} \\
&& f_{W_0 \in q}(z) = \frac{C_V^2 + C_A^2}{4\pi^2} \frac{(1-z)}{z}\, ,
 \quad C_V = -C_A = \frac{g}{2\sqrt{2}} \nonumber \, .
\end{eqnarray}
Up to the different gauge couplings $C_V$ and $C_A$, the expressions
for the $Z$ boson radiation off quarks are
identical~\cite{Dawson:1984gx,Kane:1984bb}.
It should be mentioned though that in some cases, interference with
photon emissions might become sizable, requiring a coherent
mixed-state treatment~\cite{Ciafaloni:2005fm}.

The scale $Q$ appearing in Eq.~(\ref{effwapprox}) in the logarithm for
transverse emission is a maximum (space-like) virtuality cutoff or
transverse momentum cutoff, typically set by the scale of the hard
process in which the $W$ is participating.
For $Q \gg M_{W}$, the logarithm asymptotically diverges,
necessitating collinear resummation, in close analogy with massless
gauge theories.
Numerically, the impact of this resummation at FCC energies has not
yet been assessed.
In this respect, the interplay with QCD evolution might be
particularly important, as the quark PDFs that source the heavy vector
PDFs does evolve appreciably between $O$(100~GeV) and $O$(10~TeV).
The integrated longitudinal structure function in
Eq.~(\ref{effwapprox}), by contrast, does not contain a
logarithm. This is because longitudinal emission off of massless
fermions is only possible at transverse momentum scales of order
$M_{W}$, and does not receive further contributions as we integrate
out to higher momentum scales. This behavior is a manifestation of the
Goldstone Boson equivalence
theorem~\cite{Lee:1977eg,Chanowitz:1985hj}: when the transverse
momentum becomes much larger than the weak scale, longitudinal gauge
bosons act like Goldstone bosons, and thus decouple from light
fermions.

Fixed-order (unresummed) weak boson PDFs for the proton can be
obtained by a convolution of the above distributions with the standard
quark PDFs,
\begin{equation}
  f_{W \in P}(\xi,Q_W^2,Q_q^2) = \sum_{q}\int^1_\xi \frac{dz}{z} ~ f_{W\in q}(z,Q_W^2) ~ f_{q\in P}\left(\frac{\xi}{z},Q_q^2\right) \, .
\label{eq:pdf}
\end{equation}
Note that in performing this procedure, the energy fractions of the
electroweak gauge bosons are implicitly bounded from below by $M_W /
E_q$, else the effective $W$ approximation is not valid.
In Eq.~(\ref{eq:pdf}) we have also allowed for independent
factorization scales for the quarks and for the vector boson.
Due to the strong-ordering effect, we must have $Q_W \ge Q_q$.

For transverse vector bosons, $Q_W$ should be evaluated near the hard
process scale. To the extent that the fixed-order approach is
adequate, also choosing $Q_q$ near the hard process scale is naively
appropriate.
However, since quarks of a given virtuality can only source vectors at
{\it larger} virtualities, there is intrinsically some error implicit
in this choice.
Similarly, a choice $Q_q \sim M_{W}$ would miss potentially
$O(10\%$--$100\%)$ corrections from QCD evolution.
The best scale choice of $Q_q$ for a fixed-order treatment of the
electroweak PDFs likely lies somewhere in between.
For longitudinal vectors there is less ambiguity.
Since they are only resolved out of the quarks at $Q \sim M_{W}$,
quark PDFs evaluated near $M_{W}$ are likely adequate.
As explained above, the longitudinal structure functions $f_{W_0\in
  q}$ do not contain explicit scale dependence.
In the following we will for simplicity set $Q_q = Q_W$ for transverse
bosons, and $Q_q = M_W$ for longitudinal.

\begin{figure}[!t]
\centering
\includegraphics[width=0.60\textwidth]{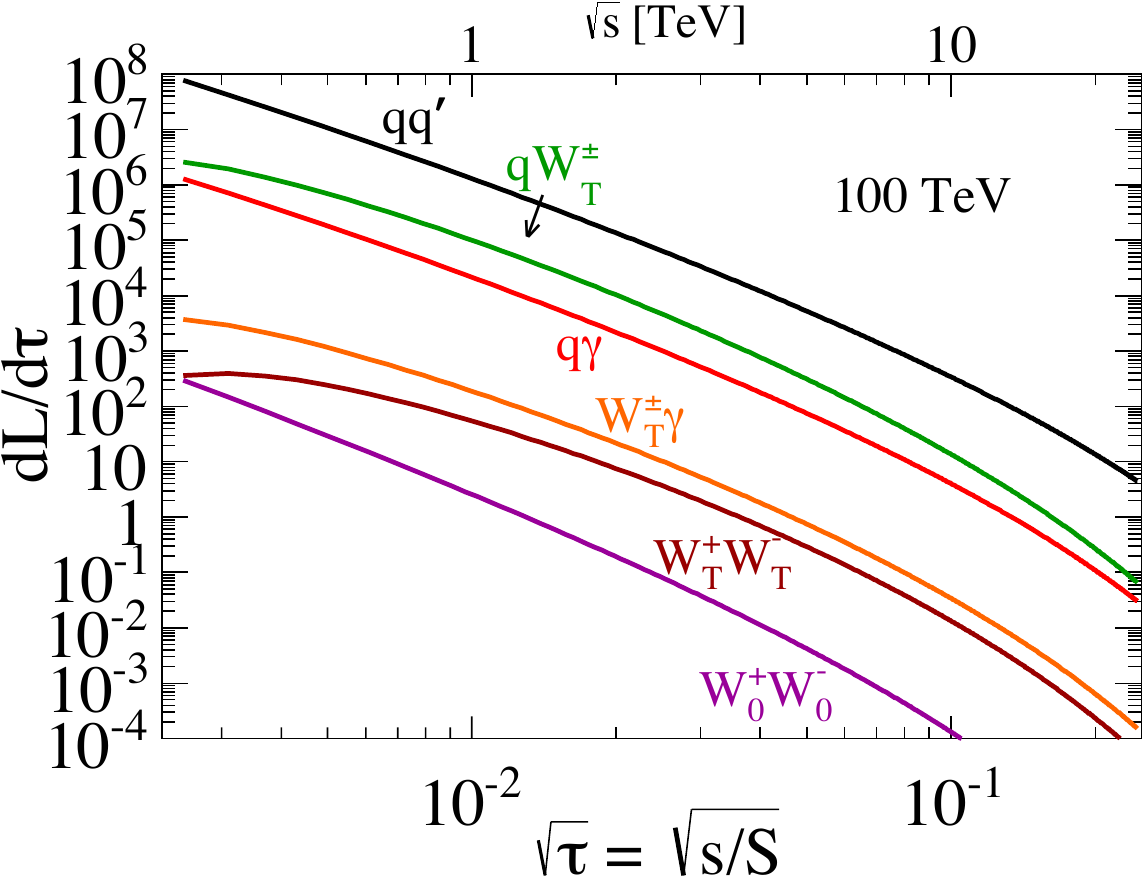}		\label{fig:sppcLumi}
\caption{Partonic luminosities, $dL/d\tau$
  Eq.~\ref{eq:lumi}, for a hadronic center-of-mass energy of $\sqrt{S}=100$ TeV, as a function of
  $\sqrt{\tau}=\sqrt{s/S}$, the ratio of the partonic $\sqrt{s}$ and hadronic $\sqrt{S}$
  center-of-mass energies.
  We compare the standard $qq'$ luminosity with the luminosities
  involving photons and electroweak gauge bosons.  }
\label{fig:ewLumi}
\end{figure}

At a hadron collider, we define partonic luminosities by the general
formula
\begin{equation}
\frac{d\mathcal{L}_{ij}}{d\tau}(Q_i^2,Q_j^2) = \frac{1}{(\delta_{ij}+1)} \int^1_\tau \frac{d\xi}{\xi}
 \left[ 
 f_{i\in P}\left(\xi,Q_i^2\right)f_{j \in P}\left(\frac{\tau}{\xi},Q_j^2\right) + 
 f_{i\in P}\left(\frac{\tau}{\xi},Q_i^2\right)f_{j\in P}\left(\xi,Q_j^2\right) 
 \right], \label{eq:lumi}
\end{equation}
where $\tau \equiv s/S$ is the ratio between the partonic $\sqrt{s}$
and hadronic $\sqrt{S}$ center-of-mass energies squared.
Luminosities involving massive vector bosons can be derived by
plugging Eq.~(\ref{eq:pdf}) into Eq.~(\ref{eq:lumi}).
Leaving for simplicity the factorization scales implicit, one finds
the following result for the $qW$ luminosity
\begin{equation}
 \frac{d\mathcal{L}_{qW}}{d\tau} = \int^1_\tau\frac{d\xi}{\xi} ~\int^1_{\tau/\xi}\frac{dz}{z} ~ \sum_{q'}
 \left[
 f_{q\in P}(\xi)f_{W\in q'}(z)f_{q'\in P}\left(\frac{\tau}{\xi z}\right) + f_{q\in P}\left(\frac{\tau}{\xi z}\right)f_{W\in q'}(z)f_{q'\in P}(\xi)
 \right]
\end{equation}
while for the different $WW$ processes the corresponding luminosities are instead
\begin{eqnarray}
 \frac{d\mathcal{L}_{WW'}}{d\tau} =
 \frac{1}{(\delta_{WW'}+1)}
 \int^1_\tau\frac{d\xi}{\xi}\int^1_{\tau/\xi}\frac{dz_1}{z_1}~\int^1_{\tau/\xi/ z_1}\frac{dz_2}{z_2} \cdot \qquad
 \qquad \qquad
 \qquad
 \qquad
 \qquad
 \nonumber
 \nonumber
 \\
 \sum_{q,q'}
 \Bigg[
   f_{W\in q}(z_2)f_{W'\in q'}(z_1)~f_{q\in P}(\xi)f_{q'\in P}\left(\frac{\tau}{\xi z_1 z_2}\right) +
   f_{W\in q}(z_2)f_{W'\in q'}(z_1)~f_{q\in P}\left(\frac{\tau}{\xi z_1 z_2}\right)f_{q'\in P}(\xi) 
   \Bigg].
\end{eqnarray}

In Fig.~\ref{fig:ewLumi}, we represent the parton luminosities
Eq.~(\ref{eq:lumi}) for various initial states at a 100 TeV
hadron-hadron collider.
We include as well the photon PDF, derived analogously to the
transverse $W$ PDF, using an effective virtuality cutoff at
$\Lambda_\gamma =\sqrt{1.5~\text{GeV}^2} \approx 1.22$ GeV, and again
ignoring possible coherence effects with $Z$ emission within the
region $Q \gsim M_Z$.
Note that below the cut-off $\Lambda_\gamma$, the PDF should be
matched with the non-perturbative intrinsic photon
PDF~\cite{Drees:1994zx,Alva:2014gxa,Ball:2013hta}, see
Sect.~\ref{sec:pdf_photon} for a discussion of recent determination
non-perturbative photon PDF.
For most of the luminosities, a common factorization scale of $Q^2 =
s/4$ is used, with $\sqrt{s}$ the partonic CoM energy.
For the longitudinal $W$, we choose to use instead $Q^2 = M_W^2$.
In Fig.~\ref{fig:ewLumiRatio} we also show the ratio the various partonic luminosities
shown in Fig.~\ref{fig:ewLumi} between center-of-mass
energies of 100 TeV and 14 TeV.
Note also that the photon-initiated luminosities can be substantially
enhanced once the non-perturbative photon PDF $\gamma(x,Q_0^2)$ is
taken into account.

\begin{figure}[!t]
\centering
\includegraphics[width=0.60\textwidth]{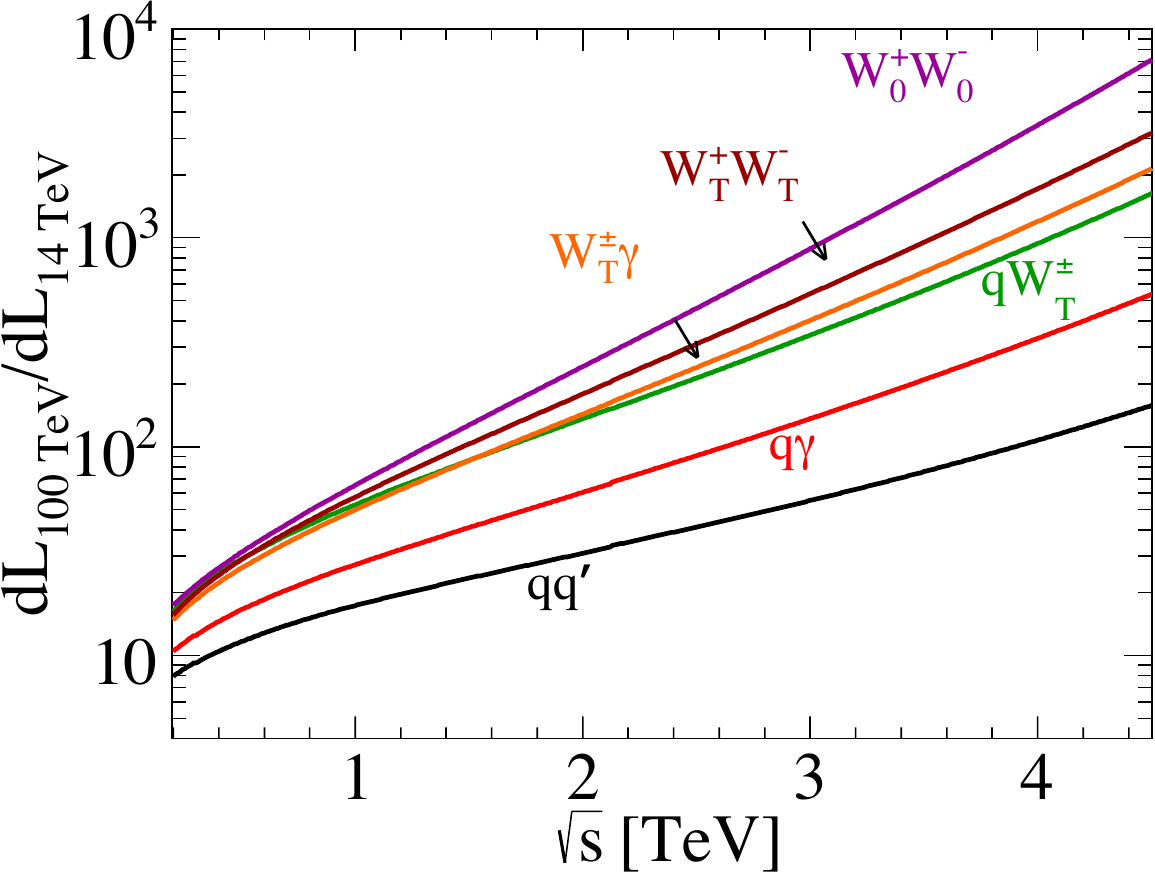}	\label{fig:sppcLumiRatio}
\caption{Ratio of the various partonic luminosities shown in Fig.~\ref{fig:ewLumi} between center-of-mass
energies of 100 TeV and 14 TeV.}
\label{fig:ewLumiRatio}
\end{figure}

One immediate observation from comparing the $W_T\gamma$ and $W_T W_T$
luminosities is their similarity.  That transverse weak bosons begin
to appear on the same footing as photons is a manifestation of the
restoration of EW symmetry.
The longitudinal bosons are sourced from the quarks as described above
without a logarithmic enhancement, and hence with individual splitting
rates that are $\log(s/4m_W^2) \sim \mathcal{O}(3-10)$ times smaller
than their transverse counterparts at multi-TeV energies.  This leads
to $\mathcal{O}(10-100)$ times smaller luminosities.

For the vector-boson-fusion (VBF) process initiated by the
longitudinal bosons, the electroweak PDF approach effectively
integrates out the usual forward tagging jets, treating them as part
of the ``beam''.
This of course becomes progressively more justifiable at higher
partonic CM energies, as the tagging jets with $p_T\sim M_W$ will
appear at extremely high rapidities, and may anyway become a less
distinctive feature to discriminate against backgrounds in the
presence of copious QCD initial-state radiation at similar values of
the transverse momentum.
From a practical perspective, the ability to treat VBF as a
$2\rightarrow2$ process rather than $2\rightarrow4$ would
significantly reduce the computational burden for event simulation.
The tagging jets can then be resolved using the usual initial-state
radiation machinery, appropriately adapted for this unique electroweak
splitting process.
In particular, merging with a matrix element description for higher
$p_T$ may remain important for obtaining a detailed understanding of
central jet vetoes.
Nonetheless, this might still be simplified to a $2\to 3$ scattering
question by exploiting the electroweak PDFs.

Similar considerations apply to other processes involving longitudinal
weak bosons in the initial state, such as the production of heavy top-
or bottom-partners through $W_0 b$ or $Z_0 b$ fusion.
The electroweak PDF approach may also be useful for new physics
processes involving initial-state transverse bosons, such as resonant
production of a heavy graviton or enhanced continuum scattering from
higher-dimensional operators.
In particular, due to the large $SU(2)_L$ non-abelian self-coupling,
collinear-enhanced secondary radiation of weak bosons from the initial
state may become relevant at the level of 10's of percent.
Subtly, emissions of this type will affect not only the energy
spectrum of the initial weak bosons, but also their isospin
composition.
These effects can only be efficiently captured in the fully
interleaved QCD+EW DGLAP evolution.

\subsection{High-energy resummation of PDF evolution}
\label{sec:pdf_resum}

When Bjorken-$x$ is small enough, logarithms of the form $\ln^k 1/x$
in the DGLAP splitting functions and in partonic matrix elements
become numerically large, and might hamper the standard perturbative
expansion.
In principle these logarithms should thus be resummed to all orders in
the strong coupling $\as(Q^2)$ for those processes that probe the
small-$x$ region.
On the other hand, so far there is no conclusive evidence for the
onset of high-energy resummation in HERA or LHC data, though the
recently reported instability of QCD fits to the legacy HERA
combination in the small-$x$ and $Q^2$ region is certainly
tantalizing~\cite{Caola:2010cy,Caola:2009iy,Abramowicz:2015mha}.
As summarized in Fig.~\ref{fig:kinplot}, the FCC will probe small
values of $x$ for many relevant processes, and thus it is important to
assess the importance of such logarithms and of their resummation in
the context of FCC phenomenology.  It is the purpose of this section
to provide a qualitative estimate of the size and impact of
high-energy resummation for a $100$~TeV hadron collider.

These small-$x$ logarithms arise from radiation of highly energetic
gluons, and appear as single logarithms of the form $\as^n\ln^kx$ with
$k\leq n$ to all orders $n$ in $\as$.  In the $\overline{\text{MS}}$
scheme, or a variant of this scheme often considered in small-$x$
resummation called $Q_0\overline{\text{MS}}$, both the PDF evolution
(in the singlet sector) and the partonic coefficient functions are
affected by small-$x$ logarithmic enhancement.
Therefore, to properly account for small-$x$ resummation effects,
refitting PDFs with resummed splitting functions and coefficient
functions is mandatory.
This is very important, because for many processes most of the
resummation effect is expected to come from the resummation in PDF
evolution, which is always leading in the singlet sector, while
resummation of coefficient functions starts at NLLx for processes
which are quark initiated at tree level.

Small-$x$ resummation is based on the fundamental $k_t$ factorization
theorem~\cite{Catani:1990xk,Catani:1990eg,Catani:1993ww,Catani:1993rn,Catani:1994sq},
valid in the high-energy limit $s\gg Q^2$.
It generalises the standard collinear factorization to the case of
off-shell initial-state partons, and reduces to it in the on-shell
limit.
Resummation of small-$x$ logarithms in the evolution of parton
distribution can be achieved using the duality between the
complementary BFKL and DGLAP evolutions, which describe the evolution
of the PDFs in $x$ and $Q^2$ respectively, both derivable from the
high-energy factorization.
This duality can be exploited to resum to all orders in $\as$ singular
small-$x$ contributions to the DGLAP gluon anomalous dimensions up to
NLLx~\cite{Altarelli:1999vw}.  Obtaining perturbatively stable and
reliable resummed anomalous dimensions requires the addition of some
extra ingredients, namely the resummation of anti-collinear
contributions~\cite{Salam:1998tj,Altarelli:2005ni} and resummation of
subleading running coupling
contributions~\cite{Camici:1997ta,Altarelli:2001ji,Ball:2007ra}.
Finally, resummation of quark anomalous dimensions and coefficient
functions can be performed (to the lowest non-trivial logarithmic
order) from high-energy
factorization~\cite{Catani:1994sq,Altarelli:2008aj}.

Despite the fact that the formalism for consistently resumming DGLAP
anomalous dimensions as well as the coefficient functions for the main
processes entering a PDF fit has been available for quite some time,
no global PDF analysis has been performed including the effects of
small-$x$ resummation.\footnote{See also Ref.~\cite{White:2006yh} for
  a study of the impact of small-$x$ resummation in the MRST fits.}
Therefore, unfortunately no consistent application of NLLx small-$x$
resummation to hadron collider phenomenology has been
performed.\footnote{On the other hand, NNPDF fits with large-$x$
  threshold resummation have recently become
  available~\cite{Bonvini:2015ira}.}
Part of the reason for this resides in the complexity of the small-$x$
resummation formalism which makes a reliable numerical implementation
challenging.

\begin{figure}[t]
  \centering
  \includegraphics[width=0.495\textwidth,page=1]{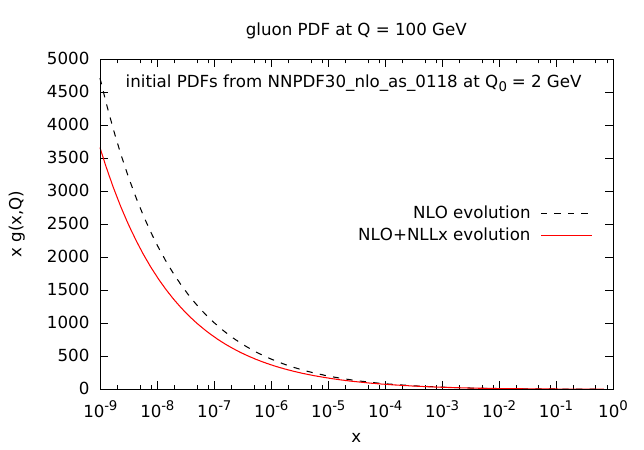}
  \includegraphics[width=0.495\textwidth,page=2]{figs/gluon_PDF_small-x_comparison_new.pdf}
  \caption{Gluon (left) and the total quark singlet (right) PDFs obtained evolving the \texttt{NNPDF30\_nlo\_as\_0118} PDFs
    from an initial scale $Q_0=2$~GeV up to the EW scale $Q=100$~GeV with fixed NLO (black dashed) and resummed NLO+NLLx (red solid) DGLAP evolution.}
  \label{fig:gluonsmallxres}
\end{figure}

Ongoing work~\cite{BMPinpreparation} aims at providing resummed
anomalous dimensions and coefficient functions through a fast
\texttt{C++} code named \texttt{HELL} (standing for High Energy Large
Logarithms).
This code has been interfaced to the
\texttt{APFEL}~\cite{Bertone:2013vaa} PDF evolution package, which is
then able to perform DGLAP evolution with LLx and NLLx small-$x$
resummation matched to the fixed order LO and NLO.
Ongoing developments aim at including also the small-$x$ resummation
of deep-inelastic coefficient functions in \texttt{HELL} and thus also
in \texttt{APFEL}.
Once the implementation has been finalized, it will be possible to
perform for the first time fully consistent PDF fits with small-$x$
resummation; preliminary results obtained in the context of the NNPDF
methodology~\cite{NNPDFsmallxRes-inpreparation} are reported below.

It is possible to estimate the effect of small-$x$ resummation at a
100 TeV collider as follows.
We show in Fig.~\ref{fig:gluonsmallxres} how the gluon (left plot) and
the quark singlet (right plot) PDFs are modified when performing fixed
NLO DGLAP evolution as compared to resummed NLO+NLLx evolution.
The initial condition for the evolution is the
\texttt{NNPDF30\_nlo\_as\_0118} set at $Q_0=2$~GeV, which is evolved
upwards in $Q^2$ using \texttt{APFEL+HELL} up to a typical electroweak
scale $Q=100$~GeV.
Recall from Fig.~\ref{fig:kinplot} that at the FCC, for $Q=100$~GeV,
the kinematic region down to $x\simeq 10^{-5}$ will be probed
(assuming a rapidity coverage of $|y|\lesssim 4$).

In Fig.~\ref{fig:gluonsmallxresRatio} (left) we show the corresponding
ratio of the gluon and quark singlet PDFs evolved with resummed
NLO+NLLx evolution to the same PDF evolved with fixed-order NLO
evolution at $Q=100$~GeV.
In this comparison, we also include the 68\% CL uncertainties, to compare them with the shift
induced by the small-$x$ resummation effects.
We observe a sizable effect of reducing the gluon and quark singlet
PDFs, by approximately $-20\%$ for $x\lesssim10^{-6}$, but also by as
much as $-5\%$ at intermediate values of $x\simeq 10^{-3}$.
%

\begin{figure}[t]
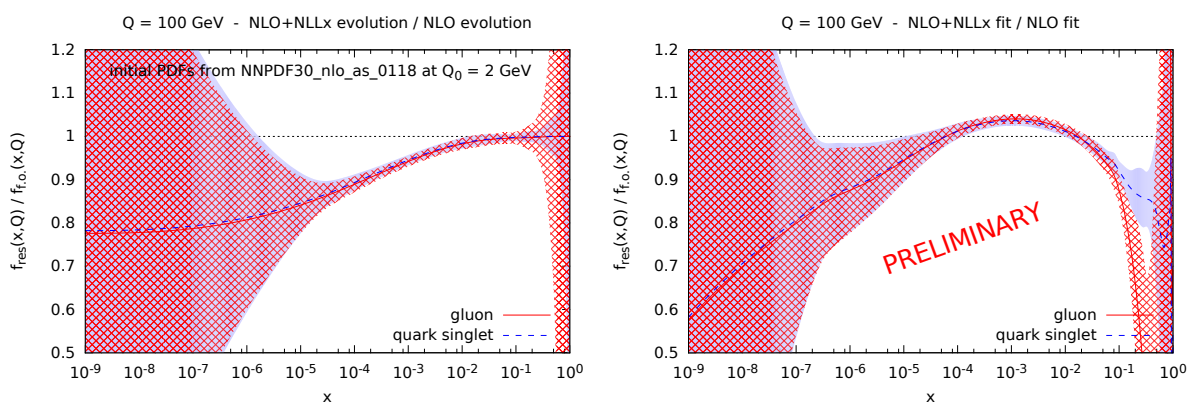

  \centering
  \includegraphics[width=0.49\textwidth,page=3]{figs/gluon_PDF_small-x_comparison_new.pdf}
  \includegraphics[width=0.49\textwidth,page=4]{figs/gluon_PDF_small-x_comparison_new.pdf}
  \caption{Left plot: ratio of the gluon (solid red) and quark singlet (dashed blue) PDFs evolved with resummed NLO+NLLx evolution
    to the same input PDF evolved with fixed-order NLO evolution at $Q=100$~GeV.
    In this case the input PDF set at $Q_0=2$ GeV is NNPDF3.0 NLO.
    We also show the 68\% CL PDF uncertainty band for the numerator of this ratio.
    Right plot: same comparison, now using as input a preliminary DIS-only NNPDF fit performed using the resummed NLO+NLLx splitting
    functions in the DGLAP evolution, resulting in different input PDFs at the initial
    parametrization scale $Q_0$.
  }
  \label{fig:gluonsmallxresRatio}
\end{figure}

However, we note that in general refitting the PDFs with resummed
evolution and coefficient functions will modify also the PDFs at the
input parametrization scale, partially compensating some of the
observed differences.
Therefore, the actual effects of small-$x$ resummation will be
different as compared to what Fig.~\ref{fig:gluonsmallxresRatio}
indicates; in fact, fitted resummed PDFs could even be larger than
their fixed-order counterparts for some values of $x$.
To illustrate this point, in Fig.~\ref{fig:gluonsmallxresRatio}
(right) we show a similar comparison as in
Fig.~\ref{fig:gluonsmallxresRatio} (left), but now using as input
boundary condition for the evolution a preliminary NNPDF DIS-only
small-$x$ resummed fit.
In this preliminary small-$x$ fit, DGLAP evolution has been performed
with NLO+NLLx DGLAP splitting functions rather than the standard NLO
ones used in fixed-order fits (but coefficients functions are still
fixed-order NLO).
As compared to using NLO PDFs as input, we observe that, when using a
(partially) consistent resummed PDF set as input, the suppression at
small-$x$ appears to be reduced, and a moderate enhancement of the
PDFs at intermediate $x$ is found, followed by a further suppression
at large-$x$.
While all these results are very preliminary,
Figs.~\ref{fig:gluonsmallxres}--\ref{fig:gluonsmallxresRatio} strongly
suggest that the small-$x$ resummation effects will be relevant for
precision physics at a 100 TeV collider.

It is also useful to estimate the potential impact of small-$x$
resummation effects for physical observables at the FCC.
To do so, we consider the effect of resummed PDFs on a process which
is directly sensitive to the medium and small-$x$ gluons, namely Higgs
production in gluon fusion.
We define
\begin{equation}\label{eq:Higgsratio}
  R_h \equiv \frac{\sigma_{\rm NLO}(\text{NLO+NLLx PDFs})}{\sigma_{\rm NLO}(\text{NLO PDFs})}
\end{equation}
to be the ratio of the NLO cross section obtained with resummed
NLO+NLLx PDFs to the NLO cross section obtained with NLO PDFs.  In
absence of fully consistent fitted resummed PDFs, we use the same
approximate strategy used above of evolving with resummed NLO+NLLx
anomalous dimensions the NLO PDFs from $Q_0=2$~GeV up to the Higgs
mass ($m_h=125$~GeV).

We find that $R_h\simeq 0.96$ for the LHC at $13$~TeV while $R_h\simeq
0.89$ for the FCC at $100$~TeV.
Consistently with Fig.~\ref{fig:gluonsmallxresRatio}, the cross
section is reduced by a sizable amount, $-4\%$ at LHC and $-11\%$ at
FCC, where the larger effect at the FCC arises because the gluons
fusing into the Higgs are on average at smaller $x$.
Using refitted resummed PDFs will of course modify these estimates,
most likely reducing the effect of the resummation, or even giving an
enhancement of the cross section.
Indeed, if one repeats the exercise using the preliminary fitted PDFs,
the effect turns out to be an enhancement of $+0.5\%$ at LHC and of
$+7\%$ at FCC.
None of these estimates is fully reliable, however they clearly show
that small-$x$ resummation will have a sizable impact at FCC.
Note that for the specific case of Higgs pair production in gluon
fusion, one should also consistently resum the small-$x$ logarithms in
the partonic cross section: the effect of this resummation has not
been studied yet, but the small-$x$ contributions are known to be
non-negligible for high collider energies~\cite{Ball:2013bra}, and
will be another fundamental ingredient for precision phenomenology at
FCC.

\clearpage
\def\ptmin{\ensuremath{\rm p_{\perp}^{\rm min}}\xspace} 
\def\ptminnought{\ensuremath{\rm p_{\perp,0}^{\text{min}}}\xspace}
\def\preco{\ensuremath{\rm p_{\rm reco}}\xspace}
\def\ptch{\ensuremath{\rm p_{\perp\rm ch}}\xspace}
\def\pto{\ensuremath{\rm p_{\perp0}}\xspace}
\def\figRef#1{fig.~\ref{#1}\xspace}
\def\eqRef#1{eq.~\eqref{#1}\xspace}
\def\ttt#1{\texttt{#1}}

\def\sqrts{\sqrt{\rm s}}
\def\pp{p-p}
\def\ppbar{p-${\rm \bar p}$}
\def\epem{e^+e^-}
\def\pTrm{\ensuremath{\rm p_{_{\rm T}}}}
\def\dNdeta{\rm dN_{\rm ch}/d\eta}
\def\dNdetaZero{\rm dN_{\rm ch}/d\eta|_{\eta=0}}
\def\dNdetaZeroInel{\rm dN_{\rm ch}^{^{\rm inel}}/d\eta|_{\eta=0}}
\def\dNdetaZeroNSD{\rm dN_{\rm ch}^{^{\rm NSD}}/d\eta|_{\eta=0}}
\def\dEdeta{\rm dE/d\eta}
\def\dNdpT{\rm dN_{\rm ch}/dp_{_{\rm T}}}
\def\PNch{\rm P(N_{\rm ch})}
\def\meanpt{\rm \left< p_{\rm T} \right>}
\def\Lambdaqcd{\Lambda_{_{\rm QCD}}}
\def\sigmainel{\sigma_{\rm inel}}
\def\sigmahard{\sigma_{_{\rm pQCD}}}

\def\Pom{\mathbb{P}}

\def\phojet{\textsc{phojet}\xspace}
\def\epos{\textsc{epos}\xspace}
\def\eposlhc{\textsc{epos-lhc}\xspace}
\def\qgsjet{\textsc{qgsjet}\xspace}
\def\qgsjetII{\textsc{qgsjet-ii}} 
\def\sibyll{\textsc{sibyll}\xspace}

\def\pythia{\textsc{pythia}\xspace}
\def\herwig{\textsc{herwig}\xspace}
\def\lhapdf{\textsc{lhapdf}\xspace}

\def\cO#1{{{\cal{O}}}\left(\rm #1\right)}

\def\cmDDE{c.m.\@\xspace}
\def\ieDDE{i.e.\@\xspace}
\def\egDDE{e.g.\@\xspace}

\section{Global event properties\footnote{Editors: D.~d'Enterria, P.~Skands}}
\label{sec:mb}

Unlike hard SM and BSM probes, which target small fractions of the
total $pp$ cross section, the aim with minimum-bias (MB) physics
studies is to examine highly inclusive event samples, subject only to
detector-acceptance limits and minimal trigger conditions (hence the
name ``minimum bias''\footnote{A ``minimum-bias'' trigger typically
  relies on hits in a set of forward detectors to ensure that at least
  a minimal amount of observable activity was produced. If hits are
  required on both sides of the event, the term
  ``non-single-diffractive'' (NSD) is also sometimes used. Triggers
  with zero bias are also possible, typically provided by a simple
  synchronisation with the bunch-crossing clock --- hence a zero-bias
  sample can include some empty events where nothing actually
  happened.}).  The absence of any explicit requirement of hard
activity implies that the particle production in such events is
dominated by soft and semihard QCD processes.
On the soft side, given the extended composite nature of hadrons, even
at asymptotically large energies, a non-negligible fraction of
inelastic \pp\ interactions involve ``peripheral'' scatterings with
small transverse momentum transfer, described in terms of a Pomeron
($\Pom$) contribution, identified perturbatively with a colour-singlet
multi-gluon exchange, responsible for diffractive
dissociation. Elastic and diffractive scatterings account for a
noticeable fraction, about a third, of the total \pp\ cross section at
high energies. In the semihard domain, at increasingly larger \cmDDE\
energies the inelastic cross section receives major contributions from
the region of low parton fractional momenta ($x=p_{_{\rm
    parton}}/p_{_{\rm hadron}}$), where the gluon distribution rises
very fast. As a matter of fact, at $\sqrts$~=~100~TeV the partonic
cross section saturates the total inelastic cross section (\ieDDE\
$\sigmahard\approx\sigmainel\approx$~100~mb) at momenta much larger
than $\Lambda_{QCD}$, $\pTrm\approx$~10~GeV/c~(see
\egDDE~\cite{Skands:2014jea}).  Such a ``divergent'' behaviour (taking
place {\it well} above the infrared regime around
$\Lambdaqcd\approx$~0.2~GeV) is solved by reinterpreting this
observation as a consequence of the increasing number of multiparton
interactions (MPI) occurring in a single \pp\ collision.
The energy evolution of such MPI and low-$x$ effects is implemented
phenomenologically in all MCs through a transverse momentum cutoff
$\pto$ of a few GeV that tames the fastly-rising $1/\pT^4$ minijet
cross section (\eg\ in \pythia\ the cutoff is introduced through a
multiplicative $1/(\pT^2+\pto^2)^2$ factor). This $\pto$ regulator is
commonly defined so as to run with \cmDDE\ energy following a slow
power-law (or logarithmic) dependence, closely mimicking the
``saturation scale'' $\rm Q_{\rm sat}$ that controls the onset of
non-linear (gluon fusion) effects saturating the growth of the PDFs as
$x\to$~0~\cite{Gribov:1984tu}.  Last but not least, all MC generators,
both based on pQCD or Reggeon Field Theory (RFT) alike, use
parton-to-hadron fragmentation approaches fitted to the experimental
data -- such as the Lund string~\cite{Andersson:1983ia}, area
law~\cite{Artru:1974hr} or cluster
hadronisation~\cite{Marchesini:1991ch} models -- to hadronise the
coloured degrees of freedom once their virtuality evolves below
$\cO{1~GeV}$.

Closely connected to multiparton interactions is the ``underlying
event'' (UE) activity denoting the global enhancement of softer
particle production that accompanies the hardest partonic interactions
in the event, contributing a ``pedestal'' term to jet energies and
reducing particle isolation. Finally, for high-luminosity colliders,
the additional ``pileup'' events that are recorded in the same bunch
crossing as a primary triggered event are essentially
unbiased\footnote{Note however, that a trigger event accompanied by an
  upwards fluctuation in pileup activity, is more likely to pass a
  given jet $p_\perp$ trigger threshold than the same event
  accompanied by a low pileup level, hence the bias is not completely
  zero.}, hence the determination of pileup characteristics also falls
under the minimum-bias physics program.

Notwithstanding the challenges posed by understanding and modelling
semihard and non-perturbative dynamics, very large minimum-bias event
samples can typically be accumulated in a matter of days, allowing for
excellent high-statistics studies of a large range of physical
observables which in turn furnish important constraints on
phenomenological QCD models, hypotheses, and fits.  The questions
asked are often rather simple, such as: \emph{what does the average
  collision look like?} and \emph{how sizable are the event-to-event
  fluctuations?} Indeed, the charged-particle multiplicity
distribution is typically the first physics measurement that a new
collider experiment publishes. But importantly, the tails of
distributions are also coming under increasing scrutiny, in particular
towards large multiplicities and by using rare particles (such as ones
containing multiple strange quarks, or $c$ and $b$ quarks) as tracers
of the underlying physics mechanisms. The term ``minimum bias'' is
perhaps then slightly misleading.  Nonetheless, since these studies
still do not rely on any conventional ``hard trigger'', we retain the
term MB as a catch-all phrase, including also diffractive and elastic
scattering as well as more exclusive (biased) subsets of the MB data
sample.

\subsection{Minimum bias collisions}

The general-purpose Monte Carlo (MC) models used in high-energy
collider physics, such as \pythia~6~\cite{Sjostrand:2006za},
\pythia~8~\cite{Sjostrand:2014zea}, \herwig++~\cite{Bahr:2008pv}, and
\textsc{sherpa}~\cite{Gleisberg:2008ta}, are fully based on a pQCD
framework which then incorporates soft diffractive scatterings in a
more or less ad hoc manner. In contrast, MC models commonly used in
cosmic-ray physics~\cite{d'Enterria:2011kw} such as
\epos~\cite{Werner:2005jf,Pierog:2009zt,Pierog:2013ria},
\qgsjet~01~\cite{Kalmykov:1997te,Kalmykov:1993qe},
\qgsjetII~\cite{Ostapchenko:2005nj,Ostapchenko:2004ss,Ostapchenko:2007qb,Ostapchenko:2010vb}
and \sibyll~\cite{Ahn:2009wx}, as well as
\phojet~\cite{Engel:1994vs,Engel:1995yda,Engel:1995sb} and {\sc
  dpmjet}~\cite{Ranft:1994fd,Roesler:2000he} mostly used for collider
environments, are based on simple quantum field-theory principles
--such as unitarity and analyticity of scattering amplitudes as
implemented in the RFT model~\cite{Gribov:1968fc}.  The latter MCs
start off from a construction of the hadron-hadron elastic scattering
amplitude to determine the total, elastic and inelastic (including
diffractive) cross sections, extended to include hard processes via
``cut (hard) Pomerons'' (also known as ``parton ladder'') diagrams.
In this section, we compare the basic properties of the MB observables
characterising the final states produced in proton-proton collisions
at $\sqrts$~=~100~TeV, predicted by pQCD- and RFT-based hadronic
interaction models~\cite{d'Enterria:2016aur}.

The basic ingredients of the \pythia~6 and 8 event generators are
leading-order (LO) pQCD $2\to 2$ matrix elements, complemented with
initial- and final-state parton radiation (ISR and FSR), folded with
PDFs (interfaced here via the \lhapdf\ v6.1.6
package~\cite{Bourilkov:2006cj}), and the Lund string
model~\cite{Andersson:1983ia} for parton hadronisation. The
decomposition of the inelastic cross section into non-diffractive and
diffractive components is based on a Regge
model~\cite{Schuler:1993wr}.  For the minimum-bias studies we use the
\pythia\ event generator in two flavours: the Fortran version
6.428~\cite{Sjostrand:2006za}, as well as the C++ version \pythia\
8.170~\cite{Sjostrand:2007gs}.  We consider two different ``tunes'' of
the parameters governing the non-perturbative and semi-hard dynamics
(ISR and FSR showering, MPI, beam-remnants, final-state
colour-reconnection (CR), and hadronisation). For \pythia\ 6.4 we use
the Perugia 2011 tune (\ttt{MSTP(5)=350})~\cite{Skands:2010ak}, while
for \pythia\ 8 we use the Monash 2013 (\ttt{Tune:ee=7;
  Tune:pp=14})~\cite{Skands:2014pea}. Both sets of parameters
(Table~\ref{tab:pythiaTunes}) have been obtained from recent (2011 and
2013 respectively) analysis of MB, underlying-event (UE), and/or
Drell-Yan data in \pp\ collisions at $\sqrts$~=~7~TeV.

\begin{table}[htbp]
\centering
{\footnotesize
\begin{tabular}{lccccccccc}\hline
Version \hspace*{-5mm}& Tuning          & Diffraction & \multicolumn{2}{c}{Semihard dynamics} & \multicolumn{2}{c}{Initial state} & \multicolumn{2}{c}{Final state}  \\ 
        & &             & $\rm \pto(7~\mathrm{TeV})$ & \hspace*{-3mm}power $b$ &   PDF  & p-p overlap  & CR & hadr \\ 
\hline
6.428 & Perugia 2011& Regge~\cite{Schuler:1993wr} &  2.93 GeV & 0.265 & CTEQ5L       &
$\exp(-r^{1.7})$  & moderate & Lund model\\ 
8.170 & Monash 2013  & Improved~\cite{Navin:2010kk} & 2.28 GeV & 0.215 & \hspace*{-3mm}NNPDF2.3 LO\hspace*{-3mm}  &
$\exp(-r^{1.85})$  & moderate & Lund model \\ 
\hline
\end{tabular}
}
\caption{Comparison of the various ingredients controlling the non-perturbative and semihard (MPI, saturation)
dynamics in the two \pythia\ MCs used in this work. See text for details.}
\label{tab:pythiaTunes}
\end{table}

For the initial state, \pythia~6 (Perugia 2011) uses the CTEQ5L parton
densities~\cite{Lai:1999wy} and \pythia~8 (Monash) the more recent
NNPDF2.3 LO set~\cite{Ball:2013hta}. For the description of the
transverse parton density, both models use a proton-proton overlap
function proportional to $\exp(-r^{n})$, with slightly different
exponents ($n$~=~1.7 and 1.85 respectively). The Perugia-2011 choice
results in a slightly broader \pp\ overlap function which thereby
enhances the fluctuations in the number of MPI relative to the
Monash-2013 choice. The perturbative MPI cross sections are suppressed
below a regularisation scale, $\pto$, whose evolution with \cmDDE
energy is driven by a power law,
\begin{equation}
\rm \pto^2(s)=\pto^2(s_0)\cdot(s/s_0)^b~,
\label{eq:power}
\end{equation} 
with the parameters quoted in Table~\ref{tab:pythiaTunes} (with
$\sqrt{s_0}=7~\mathrm{TeV}$).  Given that the generation of additional
parton-parton interactions in the UE is suppressed below $\pto$, a
{\it higher} scaling power $b$ implies a {\it slower} increase of the
overall hadronic activity. Thus, the Monash tune results in a slower
evolution of $\pto$, yielding larger MPI activity at 100~TeV compared
to the Perugia tune.  The treatment of diffraction has improved in
\pythia~8 compared to 6, by viewing a diffractive system as a
Pomeron-proton collision which can include hard scatterings subject to
all the same ISR/FSR and MPI dynamics as for a ``normal''
parton-parton process~\cite{Navin:2010kk,Rasmussen:2015iid}.  For the
final-state, the two tunes have strong final-state colour
reconnections (implemented through different
models~\cite{Sandhoff:2005jh,Skands:2007zg,Sjostrand:2014zea}), which
act to \emph{reduce} the number of final-state particles (for a given
\pto value), or, equivalently, lower the \pto value that is required
to reach a given average final-state multiplicity.  The Lund
hadronisation parameters for light- and heavy-quarks have been updated
in \pythia~8
compared to \pythia~6 by refitting updated sets of LEP and SLD data~\cite{Skands:2014pea}.\\

The RFT-based models used in this work differ in various
approximations for the collision configurations (\egDDE\ the
distributions for the number of cut Pomerons, and for the
energy-momentum partition among them), the treatment of diffractive
and semihard dynamics, the details of particle production from string
fragmentation, and the incorporation or not of other final-state
effects (Table~\ref{tab:RFT_MCs}).  Whereas the RFT approach is
applied using only Pomerons and Reggeons in the case of \qgsjetII\ and
\phojet, \epos\ extends it to include partonic
constituents~\cite{Drescher:2000ha}. In the latter case, this is done
with an exact implementation of energy sharing between the different
constituents of a hadron at the amplitude level. The evolution of the
parton ladders from the projectile and the target side towards the
center (small $x$) is governed by the DGLAP
equations~\cite{Gribov:1972ri,Altarelli:1977zs,Dokshitzer:1977sg}.
For the minijet production cutoff, \phojet\ uses dependence of the
form $\rm \pto(s) \sim \pto + C \cdot \log(\sqrts)$, whereas \epos\
and \qgsjetII\ use a fixed value of $\rm \pto$. The latter resums
low-$x$ effects dynamically through enhanced diagrams corresponding to
multi-Pomeron
interactions~\cite{Ostapchenko:2005nj,Ostapchenko:2006vr,Ostapchenko:2006nh}.
In that framework, high mass diffraction and parton screening and
saturation are related to each other, being governed by the chosen
multi-Pomeron vertices, leading to impact-parameter and
density-dependent saturation at low momenta~\cite{Ostapchenko:2005yj}.
LHC data were used to tune the latest \qgsjetII-04
release~\cite{Ostapchenko:2010vb} shown here.  \epos\, on the other
hand, uses the wealth of RHIC proton-proton and nucleus-nucleus data
to parametrise the low-$x$ behaviour of the parton densities in a more
phenomenological way~\cite{Werner:2005jf} (correcting the $\Pom$
amplitude used for both cross-section and particle production).  The
\epos\ MC is run with the LHC tune~\cite{Pierog:2013ria} which
includes collective final-state string interactions which result in an
extra radial flow of the final hadrons produced in more central pp
collisions. Among all the MC models presented here, \phojet\ is the
only one which does not take into account any retuning using LHC data
(its last parameter update dates from year 2000).

\begin{table}[htbp]
\centering
\scriptsize\begin{tabular}{lcccc}\hline
Model (version)                        & Diffraction  & \multicolumn{2}{c}{Semihard dynamics}   & Final state \\
                                       &              &    $\rm \pto$      &  evolution        &       \\\hline
\eposlhc~\cite{Pierog:2013ria}         & effective diffractive $\Pom$  & 2.0 GeV  & power-law corr. of $\Pom$  & area law hadronisation + collective flow  \\
\qgsjetII-04~\cite{Ostapchenko:2005nj,Ostapchenko:2004ss,Ostapchenko:2007qb} & G.-W.~\cite{Good:1960ba} + $\Pom$ cut-enhanced & 1.6 GeV  & enhanced $\Pom$-graphs  & simplified string hadronisation\\
\phojet\ 1.12~\cite{Engel:1994vs,Engel:1995yda} & G.-W.~\cite{Good:1960ba} & 2.5 GeV & $\rm \pto(s)\propto\log(\sqrts)$ & hadronisation via \pythia\ 6.115 \\ \hline
\end{tabular}
\caption{Comparison of the main ingredients controlling the non-perturbative and semi-hard dynamics
present in the RFT-based event generators used in this work.}
\label{tab:RFT_MCs}
\end{table}

The results are presented, in the case of \pythia~6 and 8, for primary
charged particles, defined as all charged particles produced in the
collision including the products of strong and electromagnetic decays
but excluding products of weak decays, obtained by decaying all
unstable particles\footnote{\pythia\ 6.4:
  \ttt{MSTJ(22)=2,PARJ(71)=10}. \pythia\ 8:
  \ttt{ParticleDecays:limitTau0 = on}, \ttt{ParticleDecays:tau0Max =
    10}.}  for which $c\tau<$~10~mm. For the RFT MCs, unless stated
otherwise, the results correspond to the primary charged hadrons (with
the same $c\tau$ requirement) but without charged leptons which,
nonetheless, represent a very small correction (amounting to about
1.5\% of the total charged yield, mostly from the Dalitz $\pi^0$
decay). Unless explicitly stated, no requirement on the minimum $\pT$
of the particles is applied in any of the results presented.

\subsubsection{Inelastic pp cross section}

The most inclusive quantity measurable in \pp\ collisions is the total
hadronic cross section $\sigma_{\rm tot}$ and its separation into
elastic and inelastic (and, in particular, diffractive) components.
In both \pythia\ 6 and 8, the total hadronic cross section is
calculated using the Donnachie-Landshoff
parametrisation~\cite{Donnachie:1992ny}, including Pomeron and Reggeon
terms, whereas the elastic and diffractive cross sections are
calculated using the Schuler-Sj\"ostrand model~\cite{Schuler:1993wr}.
The predictions for the inelastic cross sections in \pp\ at
$\sqrts$~=~100~TeV, obtained simply from $\sigma_{\rm tot} -
\sigma_{\rm el}$, yield basically the same value, $\sigmainel \approx
107$~mb, for both \pythia\ 6 and 8. The RFT-based MCs, based on $\Pom$
amplitudes, predict slightly lower values $\sigmainel = 105.4, 104.8,
103.1$~mb for \eposlhc, \qgsjetII\ and \phojet\ respectively.  The
$\sqrts$ dependence of the inelastic cross section predicted by all
models is shown in Fig.~\ref{fig:sigma_pp_vs_sqrts} together with the
available data from \ppbar\ (UA5~\cite{Alner:1986iy},
E710~\cite{Amos:1991bp} and CDF~\cite{Abe:1993xy}) and \pp\
(ALICE~\cite{Abelev:2012sea}, ATLAS\cite{Aad:2011eu,atlas13TeV},
CMS~\cite{Chatrchyan:2012nj,cms13TeV}, TOTEM~\cite{Antchev:2011vs})
colliders, as well as the AUGER result at $\sqrts$~=~57~TeV derived
from cosmic-ray data\footnote{Note: AUGER measures p-Air cross
  sections and extrapolates to \pp\ via a Glauber
  model.}~\cite{Auger:2012wt}. Interestingly, all model curves cross
at about $\sqrts\approx$~60~TeV, and predict about the same inelastic
cross section at the nominal FCC-pp \cmDDE\ energy of 100~TeV.
\begin{figure}[htp]
\centering
\includegraphics[width=0.5\textwidth]{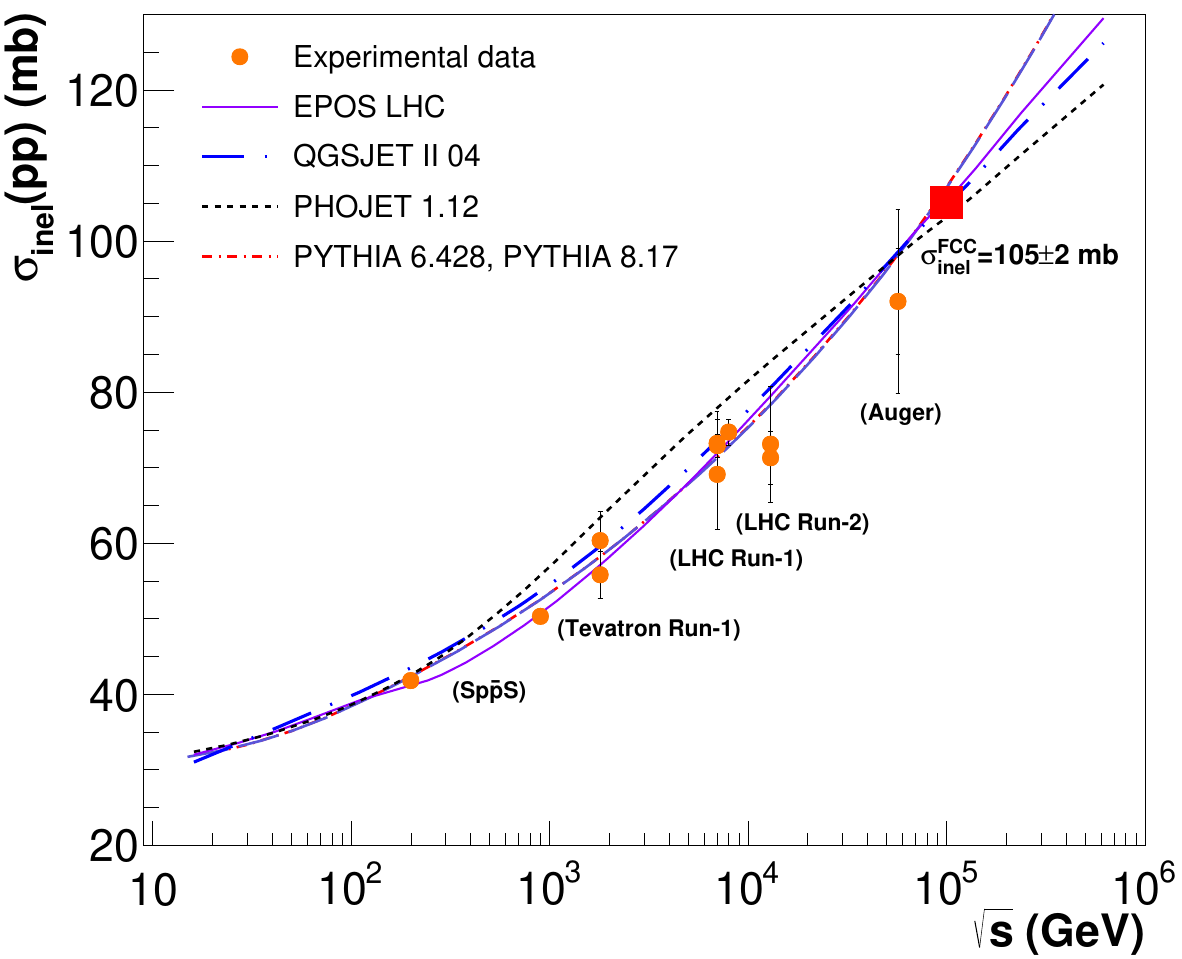}
\caption{Inelastic p-p cross section $\sigmainel$ as a function of \cmDDE\ energy in the range
  $\sqrts~\approx$~10~GeV--500~TeV. Experimental data points at various collider and cosmic-ray
  energies~\cite{Alner:1986iy,Amos:1991bp,Abe:1993xy,Abelev:2012sea,Aad:2011eu,Antchev:2011vs,atlas13TeV,Chatrchyan:2012nj,cms13TeV,Auger:2012wt} 
  are compared to the predictions of \eposlhc, \qgsjetII-04, \phojet~1.12, and \pythia~(both 6.428 and 8.17
  predict the same dependence). The red box indicates the average prediction of all models at 100~TeV.}
\label{fig:sigma_pp_vs_sqrts}
\end{figure}
A simple average among all predictions yields $\sigmainel(\rm
100\;TeV) = 105.1 \pm 2.0$~mb, whereas larger differences in the
energy evolution of $\sigmainel$ appear above the
$\sqrts\approx$~300~TeV, \ieDDE\ around and above the maximum energy
observed so far in high-energy cosmic rays impinging the Earth
atmosphere~\cite{d'Enterria:2011kw}.  The expected increase in the
inelastic \pp\ cross section at FCC(100~TeV) is about 45\% compared to
the LHC results at 13~TeV
($\sigmainel$~=~73.1~$\pm$~7.7~mb~\cite{atlas13TeV} from ATLAS, and
preliminarily 71.3~$\pm$~3.5~mb~\cite{cms13TeV} from CMS).

\subsubsection{Particle and energy pseudorapidity densities}

Figure~\ref{fig:dNdeta_100TeV} shows the distribution of charged
particles produced per unit of pseudorapidity, as a function of
pseudorapidity ($\dNdeta$) in \pp\ collisions at 100 TeV, as predicted
by the different models.
\begin{figure}[htpb]
\centering
\includegraphics[width=0.49\textwidth]{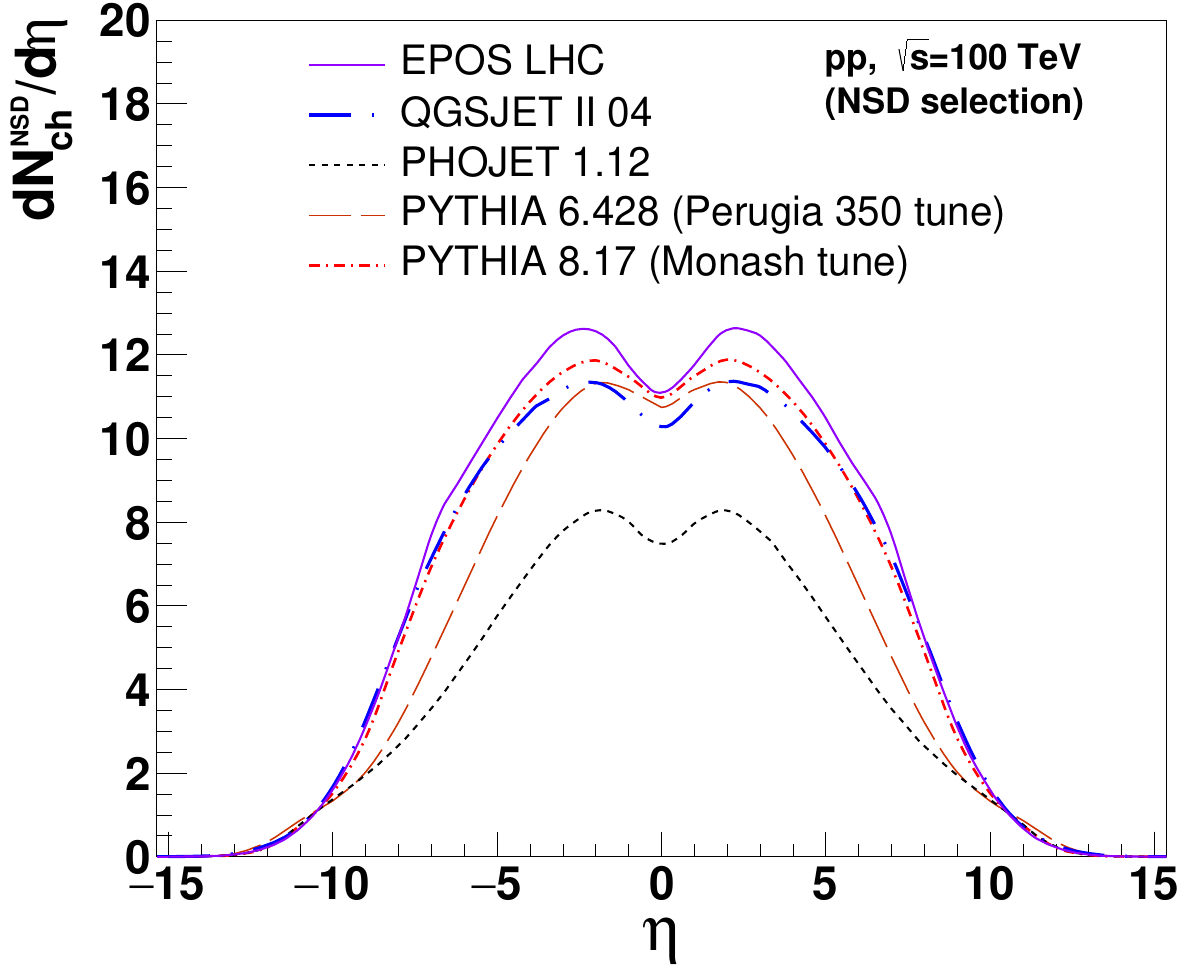}
\includegraphics[width=0.49\textwidth]{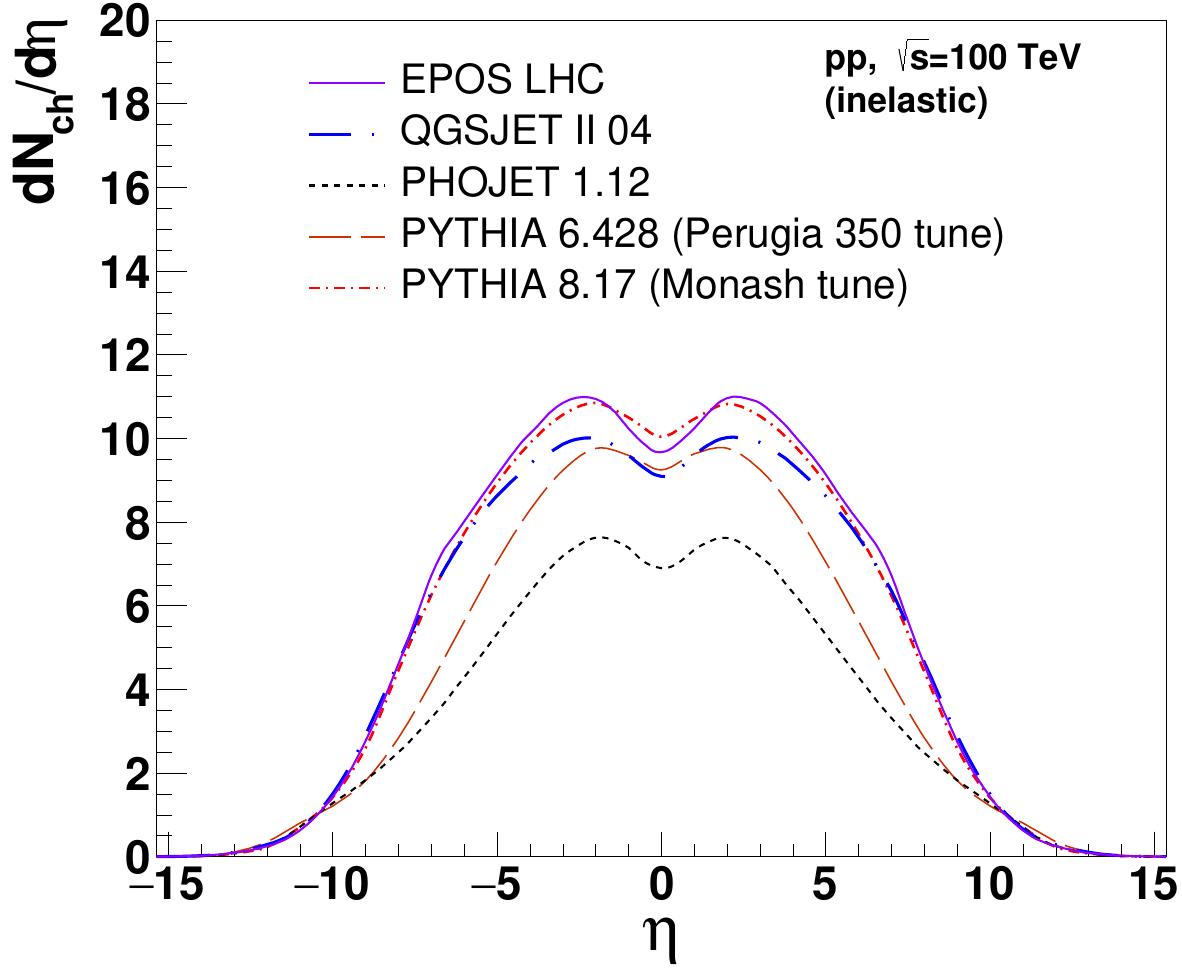}
\caption{Distributions of the pseudorapidity density of charged particles in non single-diffractive (left) and
  inelastic (right) \pp\ collisions at $\sqrts$~=~100~TeV, predicted by different hadronic MC generators.}
\label{fig:dNdeta_100TeV}
\end{figure}
About 10 charged particles are produced at midrapidity at FCC-pp.  The
left plot shows the so-called ``non single-diffractive'' (NSD)
distribution, mimicking the typical experimental requirement of a
two-arm trigger\footnote{In \pythia\ 6 and 8 this is achieved by
  directly switching off single-diffractive contributions via:
  \ttt{\scriptsize MSUB(92)=MSUB(93)=0}, and \ttt{\scriptsize
    SoftQCD:singleDiffraction=off}.} with particles in opposite
hemispheres to eliminate backgrounds from beam-gas collisions and
cosmic-rays.  Such NSD topology reduces significantly the detection
rate of (single) diffractive collisions characterised by the survival
of one of the colliding protons and particle production in just one
hemisphere. The right plot shows the inclusive inelastic distribution
which, including lower-multiplicity diffractive interactions, has a
smaller average number of particles produced. At midrapidity
($\eta=0$), all models (except \phojet) predict very similar number of
hadrons produced. Taking a (non-weighted) average of all the
predictions (except \phojet\ which is systematically lower by
$\sim$40\%), we obtain: $\dNdetaZeroNSD = 10.8 \pm 0.3$ and
$\dNdetaZero = 9.6 \pm 0.2$.  However, at forward rapidities
(equivalent to small $x\approx \pT/\sqrts\cdot e^{-\eta}$) \pythia~6
and \phojet\ predict noticeably ``thinner'' distributions than the
rest, due to lower underlying gluon densities at scales around $\rm
\pto$, than those from the NNPDF 2.3 LO set used in
\pythia~8~\cite{Skands:2014pea}.  A significant fraction of the
particles produced issue from the fragmentation of partons from
semihard MPI, the hardest partonic collision in the MB event producing
only a small fraction of them. The fact that \phojet\ misses about
$\sim$40\% of the particles yields is indicative of missing
multiparton contributions in this Monte Carlo generator.

The energy dependence of the charged hadron pseudorapidity density at
$\eta = 0$ predicted by the different models in the range
$\sqrts$~=~10~GeV--700~TeV is presented in
Fig.~\ref{fig:dNdeta0_vs_sqrts} compared to the existing NSD (left
panel) and inelastic (right panel) data measured at Sp$\bar{\mbox p}$S
(UA1~\cite{Albajar:1989an}, and UA5~\cite{Alner:1986xu}), Tevatron
(CDF~\cite{Abe:1989td,Abe:1988yu}) and LHC
(ALICE~\cite{Aamodt:2010ft,Aamodt:2010pp}, ATLAS~\cite{Aad:2010ac} and
CMS~\cite{Khachatryan:2010xs}) colliders.  As aforementioned, the NSD
selection has central densities which are about 15\% larger than those
obtained with the less-biased INEL trigger, which has less particles
produced on average as it includes (most of) diffractive
production. All models (except \phojet, whose results are not actually
trustable beyond $\sqrts$~=~75~TeV) more or less reproduce the
available experimental data up to LHC, and show a very similar trend
with $\sqrts$ up to FCC energies. Beyond 100 TeV, however, \eposlhc\
tends to produce higher yields than the rest of MCs.

\begin{figure}[tpb]
\centering
\includegraphics[width=0.49\textwidth,height=7.cm]{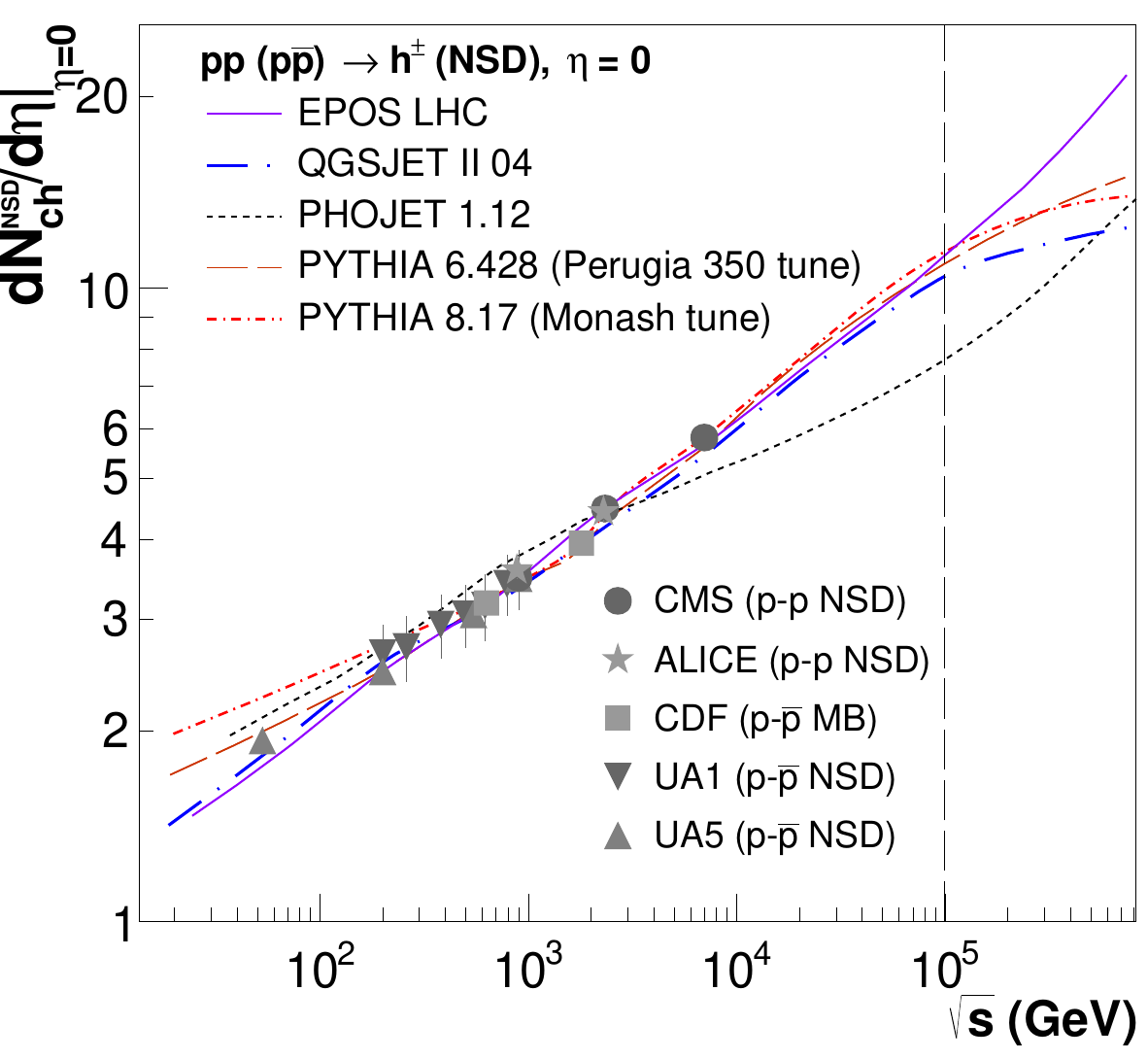}
\includegraphics[width=0.49\textwidth,height=7.cm]{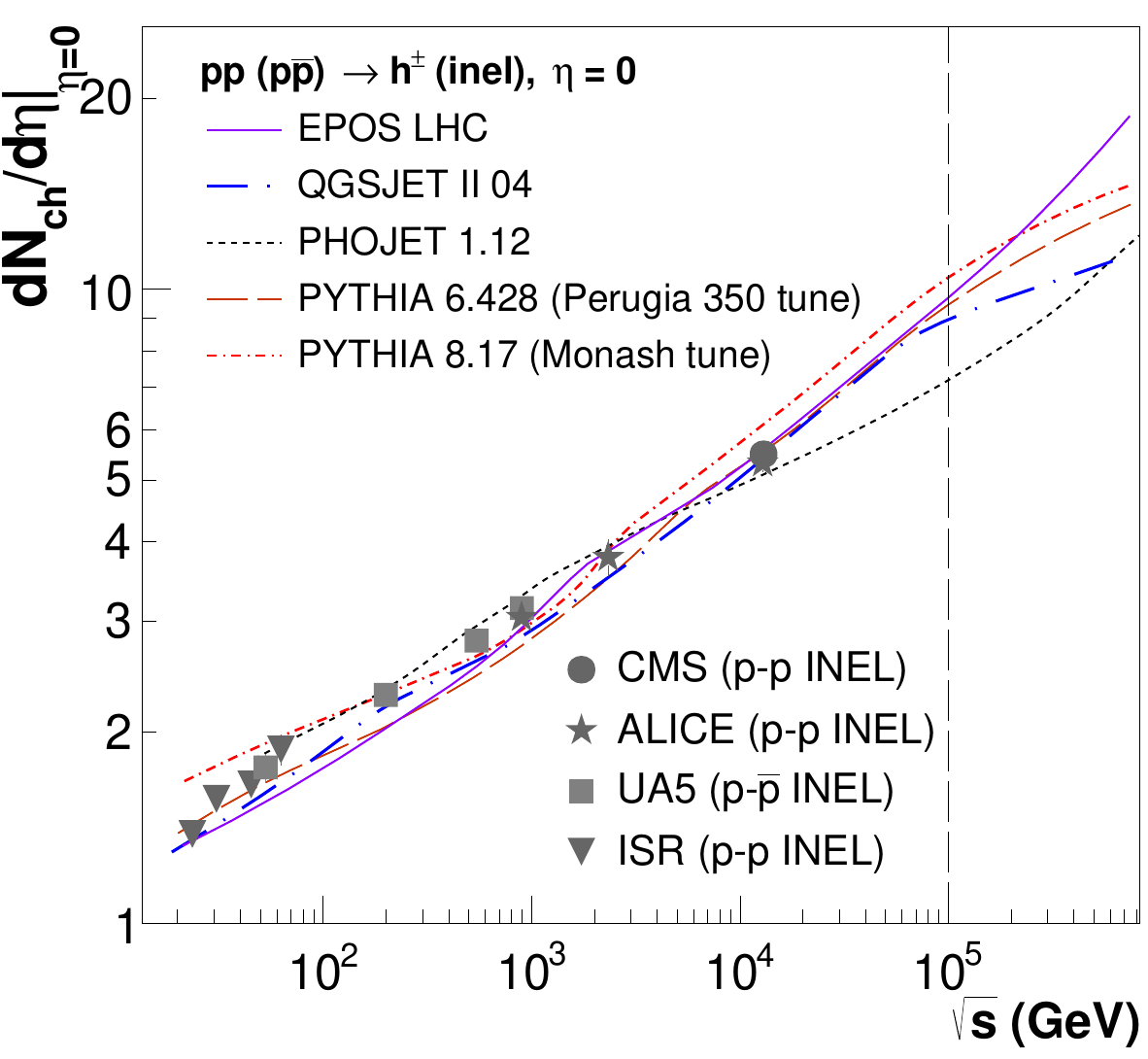}
\caption{Evolution of the charged particle pseudorapidity density at
  midrapidity, $\dNdetaZero$, as a function of collision energy,
  $\sqrts$, for non-single diffractive (left) and inelastic (right)
  \pp\ collisions. The data points show existing collider
  data~\cite{Albajar:1989an,Alner:1986xu,Aad:2010ac,Khachatryan:2010xs,Khachatryan:2015jna}.
  The vertical line indicates the FCC energy at 100~TeV.}
\label{fig:dNdeta0_vs_sqrts}
\end{figure}

The FCC experiments aim at fully tracking coverage in the central
$|\eta|<5$ region. The total number of charged particles expected in
the tracker system is obtained by integrating the $\dNdeta$
distributions over that interval, which yields an average of $\rm
N_{_{\rm ch}}(\Delta\eta$=10$)\approx$~100.  For the expected FCC
pileups, in the range $\cO{200-1000}$, this value implies that the
trackers would sustain on average a total number of 20--100 thousand
tracks per bunch crossing. Such a value is of the same order of
magnitude as a {\it single} central Pb-Pb collision at LHC
energies~\cite{Adam:2015ptt}, and thus perfectly manageable for the
high-granularity FCC tracker designs.  Further integrating the
$\dNdeta$ distributions over all pseudorapidities, one obtains the
total number of charged particles produced in an average \pp\
collision at 100~TeV. The \epos, \pythia~8 and \qgsjetII\ models
predict the largest total charged multiplicities, $\rm N_{_{\rm
    ch}}~(N_{_{\rm ch}}^{^{\rm NSD}})$~=~161~(184), 160~(170),
152~(172) respectively; followed by \pythia~6, $\rm N_{_{\rm
    ch}}~(N_{_{\rm ch}}^{^{\rm NSD}}) = 131~(150)$; and \phojet, $\rm
N_{_{\rm ch}}~(N_{_{\rm ch}}^{^{\rm NSD}}) = 103~(111)$.

The plots in figure~\ref{fig:dEdeta_100TeV} show the energy density as
a function of pseudorapidity. The left plot shows the distribution for
total energy, and the right one for the energy carried by charged
particles above a minimum $\pT = 100$~MeV/c. \phojet\ predicts the
lowest energy produced at all rapidities (consistent with the lower
particle yields produced by the model) whereas \pythia~8 predicts the
highest.
\begin{figure}[htpb]
\centering
\includegraphics[width=0.49\textwidth]{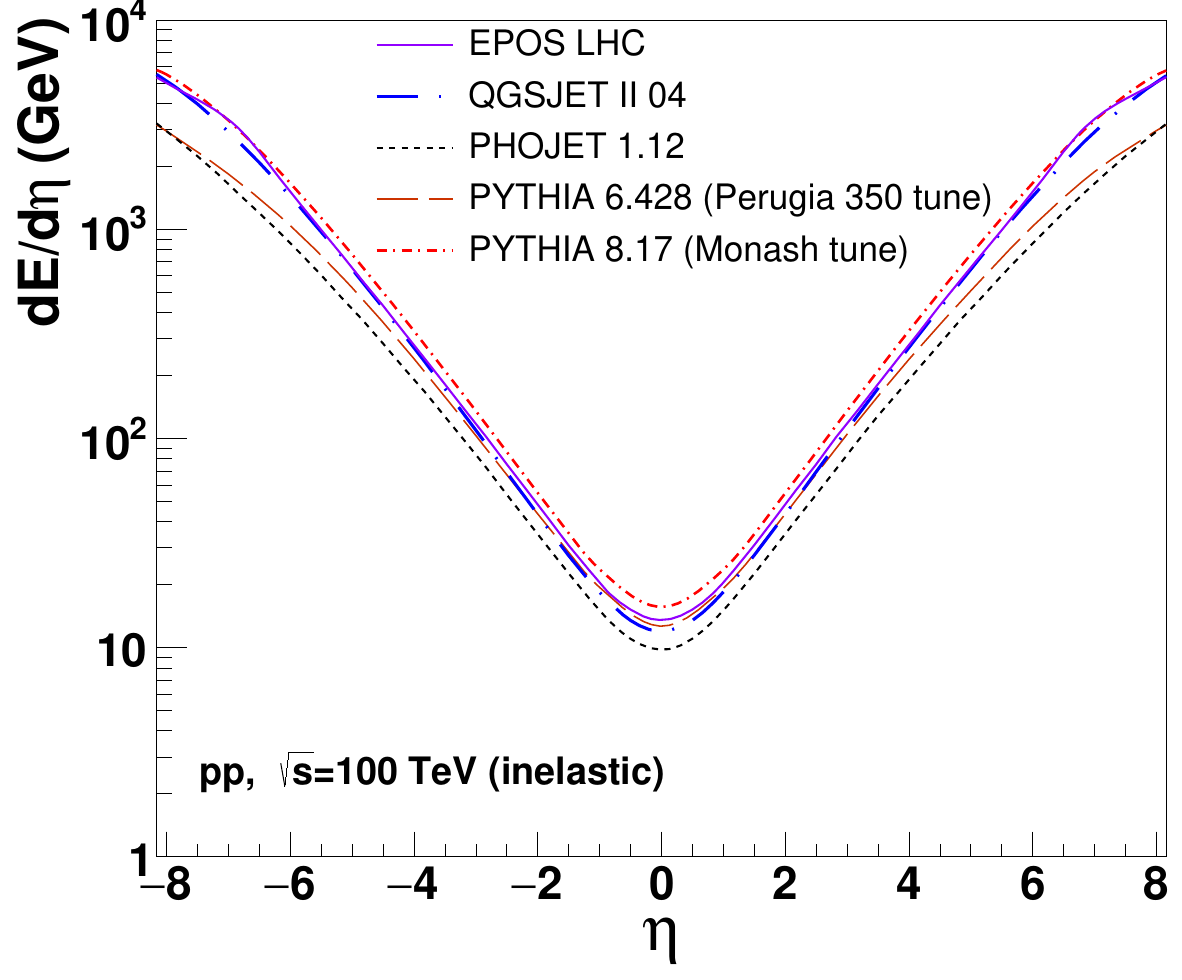}
\includegraphics[width=0.49\textwidth]{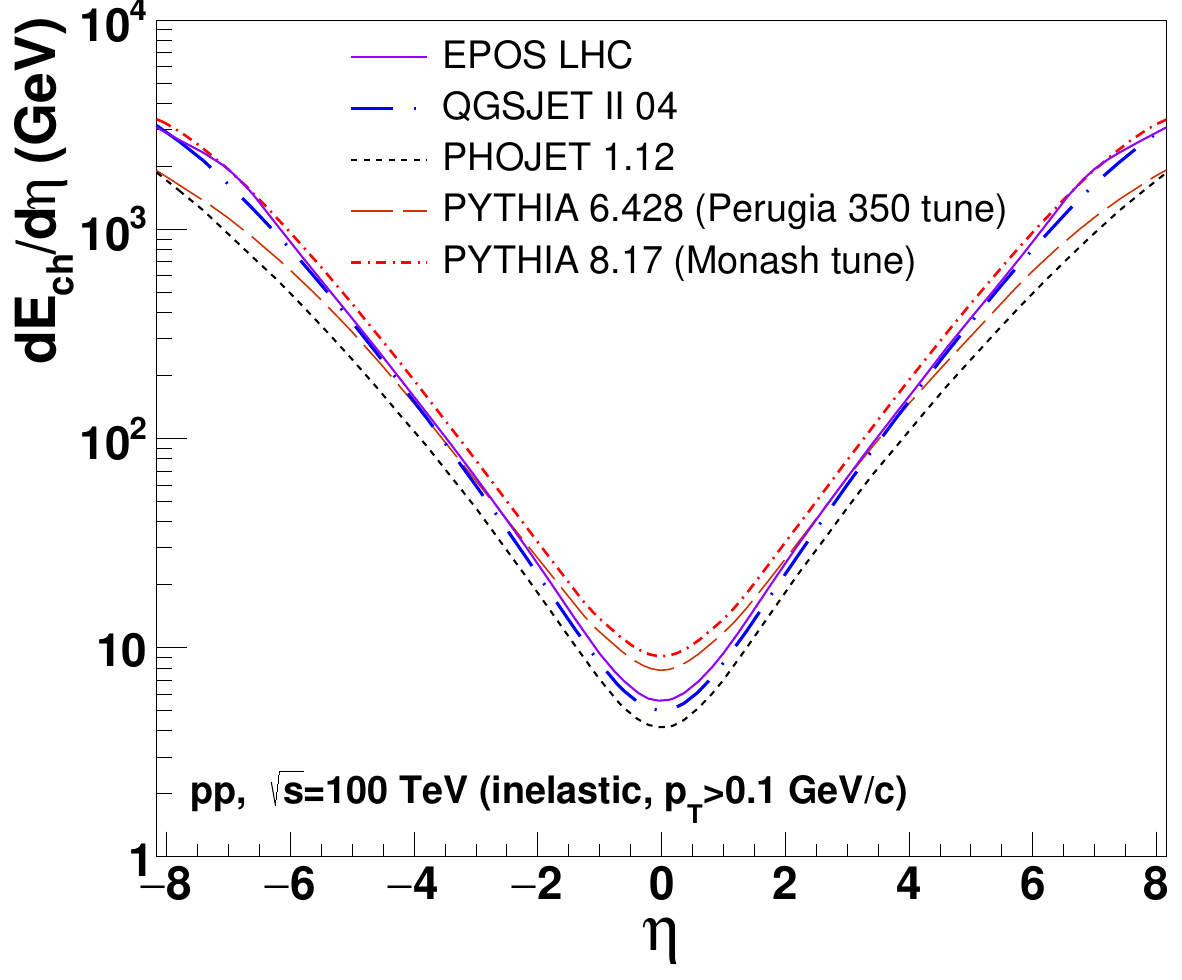}
\caption{Distribution of the energy pseudorapidity density of all
  particles (left) and of charged particles with $\pT>$~0.1~GeV/c
  (right) in inelastic \pp\ collisions at $\sqrts$~=~100~TeV,
  predicted by the different MCs considered in this work.}
\label{fig:dEdeta_100TeV}
\end{figure}
At $\eta = 0$, the total energy produced per unit rapidity is $\rm
dE/d\eta = $~9.9, 12.2, 12.6, 13.7 and 15.6~GeV for \phojet,
\qgsjetII, \pythia~6, \epos-LHC and \pythia~8 respectively. The same
values at the forward edges of typical detector coverages ($|\eta| =
5$) are $\rm dE/d\eta\approx$~410, 525, 670, 700 and 760~GeV for
\phojet, \pythia~6, \qgsjetII, \epos-LHC and \pythia~8
respectively. The trend for \pythia~6 is to predict a smaller relative
increase of energy density as a function of rapidity compared to the
rest of models due, again, to a more relatively depleted underlying
gluon density at the increasingly lower $x$ values probed at forward
$\eta$.

\subsubsection{Multiplicity distribution}

The multiplicity distribution $\rm P(N_{\rm ch})$, \ieDDE\ the
probability to produce $\rm N_{\rm ch}$ charged particles in a \pp\
event, provides important differential constraints on the internal
details of the hadronic interaction models.
Figure~\ref{fig:PNch_100TeV} shows the distribution for charged
particles produced at central rapidities ($|\eta|<1$) in inelastic
\pp\ collisions at the FCC. The tail of the $\rm P(N_{\rm ch})$
distribution (left) gives information on the relative contribution of
multiparton scatterings (multi-Pomeron exchanges), whereas the low
multiplicity part (right) is mostly sensitive to the contributions
from diffraction (single Pomeron exchanges).
\begin{figure}[tpb]
\centering
\includegraphics[width=0.49\textwidth,height=7.cm]{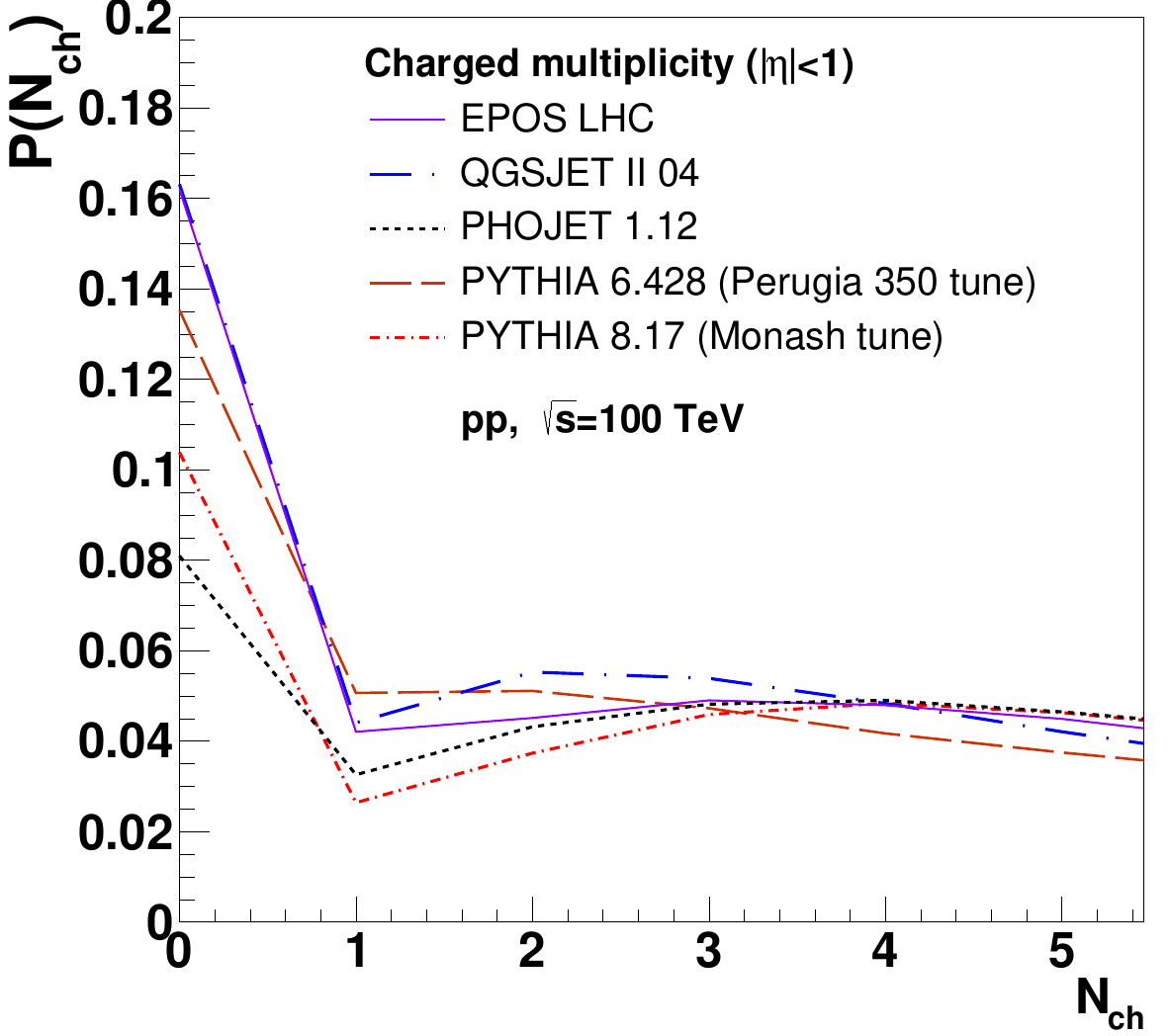}
\includegraphics[width=0.49\textwidth,height=7.cm]{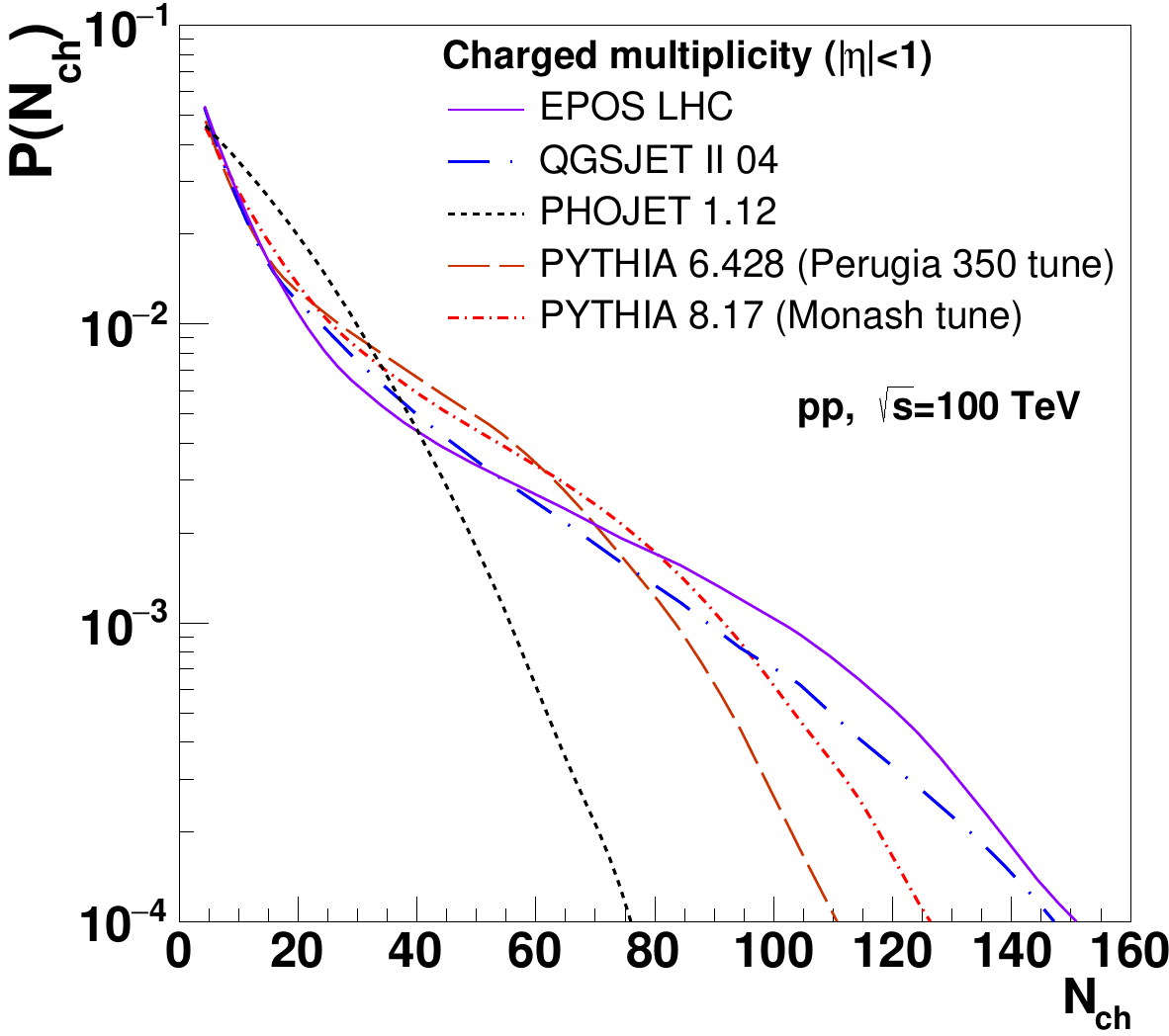}
\caption{Per-event charged particle probability (within $|\eta|<$~1)
  in inelastic \pp\ collisions at $\sqrts$~=~100~TeV: full
  distribution (right), zoom at low multiplicities $\PNch<5)$ (left).}
\label{fig:PNch_100TeV}
\end{figure}
The different MCs predict quite different distribution in both ends of
the spectrum. The RFT-based models \eposlhc\ and \qgsjetII\ both
predict higher yields at very low ($\rm N_{\rm ch}<3$) and very high
($\rm N_{\rm ch}>100$) particle multiplicities, whereas \pythia~6 and
8 feature higher yields in the intermediate region $\rm N_{\rm
  ch}\approx$~30--80, and \phojet\ has a very similar $\rm P(N_{\rm
  ch})$ distribution to \pythia\ but clearly produces much fewer
particles at intermediate and high multiplicities, compared to the
rest of models (which is, again, indicative of missing MPI
contributions in this MC).

\subsubsection{Transverse momentum distribution}

\begin{figure}[htpb!]
\centering
\includegraphics[width=0.46\textwidth,height=7.2cm]{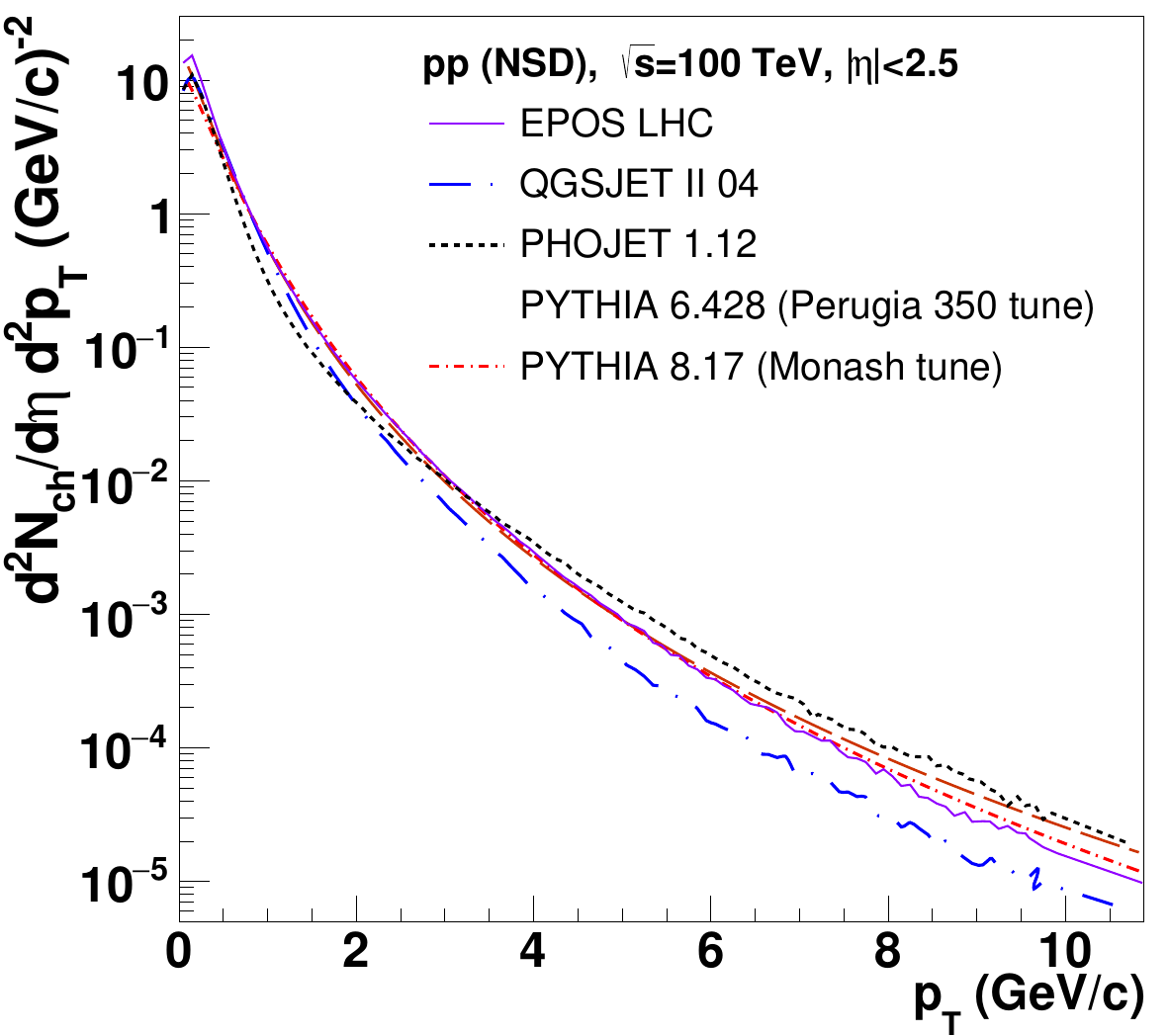} 
\includegraphics[width=0.53\textwidth,height=7.4cm]{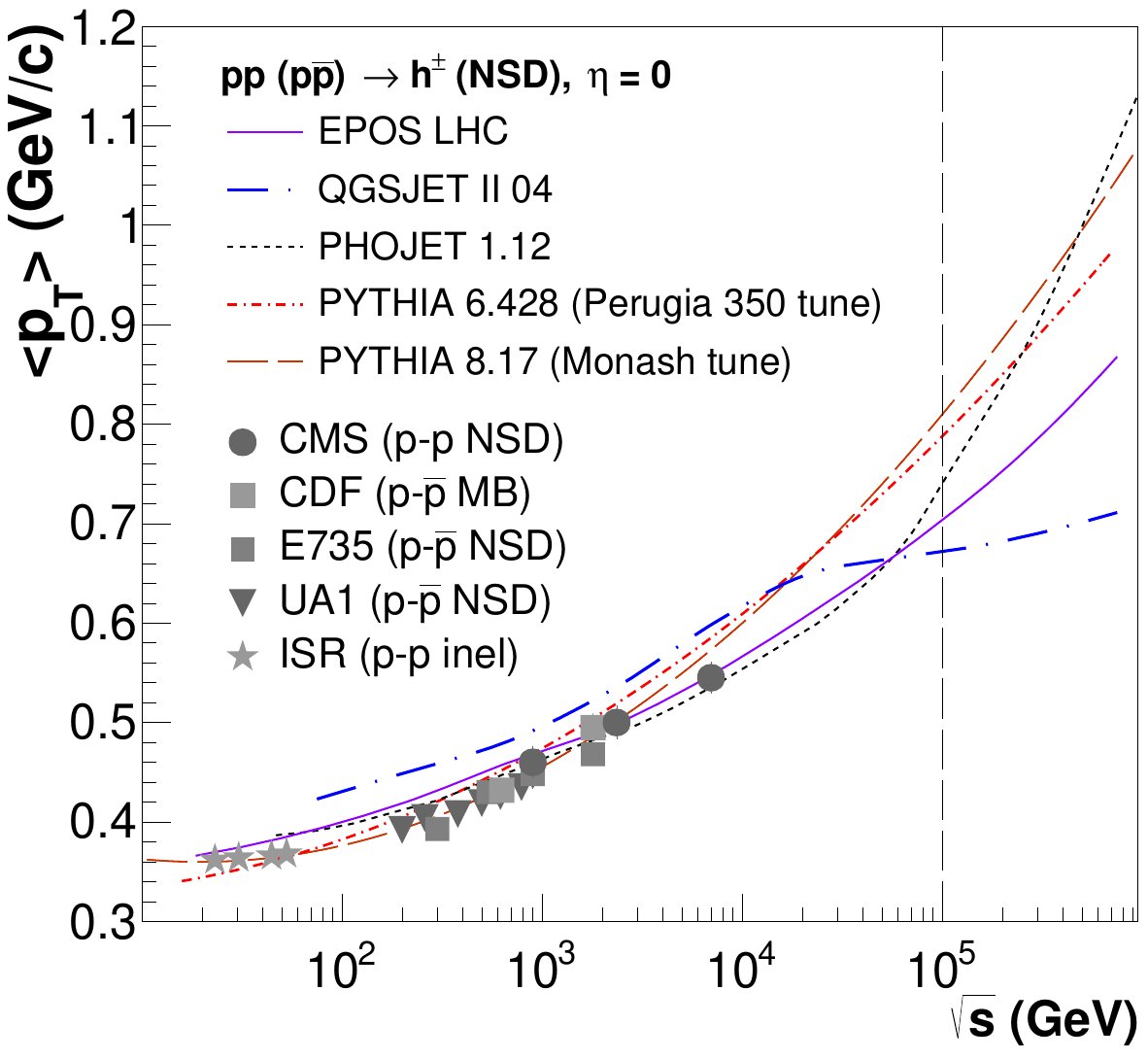}
\caption{Left: Transverse momentum spectrum in \pp\ collisions at
  $\sqrts$~=~100~TeV predicted by the different MCs considered in this
  work (absolutely normalised at a common value at $\pT \approx
  0.5$~GeV/c).  Right: Evolution of $\meanpt$ at midrapidity as a
  function of \cmDDE\ energy $\sqrts$. Data points show existing
  collider
  results~\cite{Rossi:1974if,Albajar:1989an,Abe:1988yu,Khachatryan:2010xs,Rossi:1974if,Alexopoulos:1988na},
  and the vertical line indicates the FCC(100~TeV) energy.}
\label{fig:dNdpT}
\end{figure}
Figure~\ref{fig:dNdpT} (left) shows the $\pTrm$-differential
distributions of charged particles at midrapidity ($|\eta|<2.5$) in
NSD \pp\ collisions at FCC(100 TeV) predicted by all models except
\phojet. All spectra have been absolutely normalised at their value at
$\pTrm \approx 0.5$~GeV to be able to easily compare their
shapes. Both \pythia~6 and 8 feature the largest yields at the
high-$\pT$ end of the distributions (not shown here), \qgsjetII\
features the ``softest'' spectrum, whereas \epos\ shows higher yields
in the region $\pT\approx$~1--5~GeV/c, due to collective partonic flow
boosting the semihard region of the spectra, but then progressively
falls below the pure-pQCD \pythia\ MC generators. The \phojet\
spectrum has a more convex shape, being comparatively depleted at
intermediate $\pT\approx$~1--3~GeV/c but rising at its tail.  Studying
the $\sqrts$-evolution of the average $\pTrm$ of the spectra provides
useful (integrated) information.  At high energies, the peak of the
perturbative cross section comes from interactions between partons
whose transverse momentum is around the saturation scale, $\rm \meanpt
\approx Q_{\rm sat}$, producing (mini)jets of a few GeV which fragment
into hadrons. As explained in the introduction, \pythia\ and \phojet\
MCs have an energy-dependent $\pTrm$ cutoff that mimics the power-law
evolution of $\rm Q_{\rm sat}$, while \epos\ and \qgsjetII\ have a
fixed $\pTrm$ cutoff and the low-$x$ saturation dynamics is
implemented through corrections to the multi-Pomeron dynamics. The
different behaviors are seen in the $\sqrts$-evolution of the average
$\pTrm$ shown in Fig.~\ref{fig:dNdpT} (right).  All MCs, except
\qgsjetII, predict a (slow) powerlaw-like increase of $\meanpt$ with
energy. Both \pythia~6 and 8 --whose dynamics is fully dominated by
(mini)jet production-- predict a higher $\meanpt$ than the rest of
models, yielding $\meanpt\approx 0.82$~GeV/c at FCC(100 TeV) to be
compared with 0.71 and 0.67 GeV/c from \eposlhc\ and \qgsjetII\
respectively. Above $\sqrts\approx$~20~TeV, \qgsjetII\ predicts a
flattening of $\meanpt$ whereas the \eposlhc\ evolution continues to
rise due to the final-state collective flow which increases the
$\meanpt$ with increasing multiplicity.

\subsubsection{Minimum bias summary}
\label{sec:summary}

In summary, the global properties of the final states produced in
hadronic interactions of protons at center-of-mass energies of the
Future Hadron Collider, have been studied with various Monte Carlo
event generators used in collider physics (\pythia~6, \pythia~8, and
\phojet) and in ultrahigh-energy cosmic-rays studies (\epos, and
\qgsjetII). Despite their different underlying modeling of hadronic
interactions, their predictions for proton-proton (\pp) collisions at
$\sqrts$~=~100 TeV are quite similar (excluding \phojet, whose
parameters have not been improved with the available collider data in
the last 15 years).
\begin{table}[htbp]
\centering
{\footnotesize
\begin{tabular}{l|c|c|c|c|c|c}\hline\hline
 &\pythia~6& \pythia~8 & \eposlhc & \qgsjetII\ & \phojet &  \hspace{0.0cm}Average$^\star$\hspace{0.0cm} \\ \hline
$\sigmainel$ (mb)  & $106.9$ & $107.1$ & $105.4$ & $104.8$ & $103.1$ & $105.1 \pm 2.0$\\ \hline
$\rm N_{_{\rm ch}} (N_{_{\rm ch}}^{^{\rm NSD}})$ &  131 (150) & 160~(170) & 161 (184) & 152 (172) &  101 (121) & 150 (170) $\pm$ 20\\
$\dNdetaZero$      & $ 9.20 \pm 0.01$ & $10.10 \pm 0.06$ & $ 9.70 \pm 0.16$ & $9.10 \pm 0.15$ & $6.90 \pm 0.13$ & $9.6 \pm 0.2$ \\
$\dNdetaZeroNSD$   & \hspace{0.0cm} $10.70 \pm 0.06$ \hspace{0.0cm} & \hspace{0.0cm} $10.90 \pm 0.06$
\hspace{0.0cm} & \hspace{0.0cm} $11.10 \pm 0.18$ \hspace{0.0cm} & \hspace{0.0cm} $10.30 \pm 0.17$ \hspace{0.0cm} & \hspace{0.0cm} $7.50 \pm 0.15$ \hspace{0.0cm} & \hspace{0.0cm} $10.8 \pm 0.3$ \hspace{0.0cm}\\ \hline
$\rm dE/d\eta|_{\eta=0}$ (GeV) & $12.65 \pm 0.07$ & $ 15.65 \pm 0.02$  & $ 13.70 \pm 0.02$ & $ 12.2 \pm 0.02$ & $9.9 \pm 0.01$ & $ 13.6 \pm 1.5$ \\
$\rm dE/d\eta|_{\eta=5}$ (GeV) & $525 \pm 4$ & $760 \pm 1$ & $ 700 \pm 1$ & $ 670 \pm 1$ & $410 \pm 1$ & $ 670 \pm 70$ \\ \hline
$\rm P(N_{ch}< 5)$  & $0.28$ & $0.22$ & $0.35$ & $0.36$ & $0.25$   & -- \\
$\rm P(N_{ch}>100)$ & $3.3\cdot10^{-3}$ & $0.011$ & $0.025$ & $0.018$ & $10^{-5}$ & -- \\ \hline
$\meanpt$ (GeV/c)   & $0.80 \pm 0.02$ & $0.84 \pm 0.02$ & $0.71 \pm 0.02$ & $0.67 \pm 0.02$ & $0.73 \pm 0.02$ & $0.76 \pm 0.07$ \\\hline\hline
\end{tabular}
}
\caption{Comparison of the basic properties of particle production in \pp\ collisions at $\sqrts$~=~100~TeV,
  predicted by \pythia~6 and 8, \eposlhc, \qgsjetII, and \phojet: Inelastic cross section $\sigmainel$;
  total charged multiplicities ($\rm N_{_{\rm ch}}$), and pseudorapidity charged particle densities at
  midrapidity ($\dNdetaZero$) for inelastic and NSD selections; energy densities at midrapidity ($\rm
  dE/d\eta|_{\eta=0}$), and at more forward rapidities ($\rm dE/d\eta|_{\eta=5}$); typical values of the charged
  multiplicity probabilities $\rm P(N_{ch})$ (over $|\eta|<1$) for low and high values of $\rm N_{ch}$; and
  mean charged particle transverse  momentum $\meanpt$ over $|\eta|<2.5$. The quoted uncertainties on the individual
  predictions are just the MC statistical ones. The last column indicates the average of all MCs
  (except \phojet)$^\star$\ for each observable, with uncertainties approximately covering the range of the predictions.} 
\label{tab:MB_summary}
\end{table}
Table~\ref{tab:MB_summary} lists the predictions of these basic
kinematical observables for all MCs considered.  The averages of all
MC predictions (except \phojet) for the different observables are: (i)
\pp\ inelastic cross sections $\sigmainel$~=~105~$\pm$~2~mb (to be
compared with $\sigmainel \approx$~72~mb at the LHC(13~TeV), \ie\ a
$\sim$45\% increase), (ii) total charged multiplicity $\rm N_{_{\rm
    ch}}~(N_{_{\rm ch}}^{^{\rm NSD}})$~=~150~(170)~$\pm$~20,
(iii) charged particle pseudorapidity density at midrapidity $\dNdetaZero = 9.6 \pm 0.2$ 
(to be compared with the LHC(13~TeV) result of $\dNdetaZero$~=~5.4~$\pm$~0.2, \ie\ an increase of
$\sim$80\%), and $\dNdetaZeroNSD = 10.8 \pm 0.3$ for the NSD selection, 
(iv) energy density at midrapidity $\rm dE/d\eta|_{\eta=0} =  13.6 \pm 1.5$~GeV,
and energy density at the edge of the central region $\rm dE/d\eta|_{\eta=5} = 670 \pm 70$~GeV, and 
(v) average transverse momenta at midrapidities $\meanpt = 0.76 \pm
0.07$~GeV/c (to be compared with =~0.55~$\pm$0.16 at the LHC(8~TeV),
\ie\ a $\sim$40\% increase).  The per-event multiplicity probabilities
$\PNch$, have been also compared: \eposlhc\ and \qgsjetII\ both
predict higher yields at very low ($\rm N_{\rm ch}<3$) and very high
($\rm N_{\rm ch}>100$) particle multiplicities, whereas \pythia~6 and
8 feature higher yields in the intermediate region $\rm N_{\rm
  ch}\approx$~30--80. These results are useful to estimate the
expected detector occupancies and energy
deposits from pileup collisions at high luminosities of relevance for planned FCC detector designs.\\

\subsection{Underlying event in high-\pTrm triggered events}
The fact that hard jets (or more generally, high-$p_\perp$ triggered
events of any kind) are accompanied by a global ``pedestal'' of
additional particle production, called the ``underlying event'' (UE),
has been known since the days of UA1~\cite{Arnison:1983gw}. It
originates from the same additional parton-parton interactions (or cut
Pomerons, depending on the language) as those that drive the tail
towards large multiplicities in MB events. However, the imposition of
a hard trigger biases the event selection towards events with many MPI
(each of which has a chance to be the trigger reaction).  Average
particle multiplicities and $E_T$ sums in the UE are therefore
typically several times larger than in MB events at the same \cmDDE
energy.

The average properties of the UE have been well established by
measurements at RHIC, the Tevatron, and LHC, and are generally well
reproduced by MC models that include hard (perturbative) QCD
interactions and MPI. We here consider extrapolations to 100 TeV of
several recent UE tunes of \herwig 7~\cite{Bahr:2008pv,Bellm:2015jjp}
(version 3.0) and \pythia 8~\cite{Sjostrand:2014zea} (version 8.215),
which incorporate slightly different MPI models, described below. To
facilitate comparisons between the MB and UE results presented in this
study, we choose one of the \pythia 8 tunes to be the same (Monash
2013) as in the plots in the previous subsection.

The amount of transverse energy associated with the UE is relevant to
a broad range of studies, since it enters as an additive term in jet
energy calibrations. Independently of the details of jet algorithms
and calibration techniques, the average $E_T$ density (per unit
$\Delta\eta\times\Delta\phi$) furnishes a salient basic
characterisation of the UE level, and we take this as our main
observable for this study. The relative improvement obtainable from
calibration techniques that take the in-situ (per-event) UE level into
account can be estimated from the event-by-event fluctuations, which
are sizable (larger than a naive Poissonian $\sqrt{\left<E_T\right>}$)
at the LHC~\cite{Aad:2010fh}. Thus we also include the standard
deviation of the $E_T$ density.  To put these results in a tracking
context, we also include results for the charged-track densities and
the average track $p_\perp$.  We do not attempt to include an estimate
of the inhomogeneities in the UE distribution within each event.

We consider a fictitious detector spanning $|\eta|<6$ (which can
roughly be considered the ``central'' rapidity plateau at 100 TeV
energies, spanning the seagull-shaped peak of the
$dN_\mathrm{ch}/d\eta$ distribution, cf.~the preceding subsection) and
use the conventional ``Transverse Region'' to define the UE phase
space, covering the azimuth range $60^\circ < |\Delta\phi| <
120^\circ$ with respect to the highest-$p_\perp$ track in the
event. Within the transverse region, we include all stable charged
final-state particles, <, subject to two different $p_\perp$ cuts, at
$100$ and $500$ MeV respectively. These cuts are carried over from the
ATLAS study this analysis is modelled on~\cite{Aad:2010fh}. The lower
one includes the peak of soft particle production with transverse
momenta $p_\perp\sim\Lambda_\mathrm{QCD}$ while the higher one focuses
on the tail with $p_\perp>\Lambda_\mathrm{QCD}$.

These observables are intended to give a first idea of what the UE may
look like at 100 TeV, for detector-design and physics-analysis /
jet-calibration estimation purposes. They do not address the more
detailed physics studies of the UE \emph{dynamics} that could be
possible at 100 TeV. It is nonetheless worth emphasising that an
increasing number of such studies are now being undertaken at the LHC,
driven by tantalising hints of non-trivial global hadronisation
effects in \pp\ collisions which go beyond the ability of most current
models to describe. Among the most intriguing observations are the
appearance in high-multiplicity \pp\ collisions of qualitative
features that are traditionally associated with collective / flow-like
effects and/or with an increased energy scale for particle
production. Examples are the CMS ``ridge''
effect~\cite{Khachatryan:2010gv}, the by now well-established gradual
increase of $\left<p_\perp\right>$ with multiplicity, and the
seemingly increased rates of strangeness and baryon production,
relative to models that correctly describe equivalent observables in
$e^+e^-$ environments (see, e.g., the plots available on
\texttt{mcplots.cern.ch}~\cite{Karneyeu:2013aha}). We expect that an
analogous fruitful programme of new measurements exploring the UE
dynamics in further detail can be carried out at 100 TeV.  From the
point of view of detector design, we note that hadron-flavour
dependence (and hence particle identification capability) has emerged
as a powerful
tool~\cite{Abe:1989hy,Acosta:2005pk,Abelev:2006cs,Aamodt:2011zza,Baranov:2011ch,Abelev:2012jp,Aaij:2012ut,Chatrchyan:2013qsa,Bierlich:2015rha,Martin:2016igp}
to disentangle the trends along axes of mass, strangeness, spin, and
baryon number.

\subsubsection{MC Models}

The current MPI model in \herwig 7 includes hard \cite{Bahr:2008dy}
(similar to the JIMMY \cite{Butterworth:1996zw} package) and soft
components \cite{Bahr:2009ek} of multiple partonic interactions as
well as improved colour reconnection models \cite{Gieseke:2012ft}.
The main parameters of the model are $\ptmin$ which sets a transition
scale between the hard and soft (non-perturbative) components, $\mu$
which can be interpreted as the inverse radius of the proton
(governing the difference in matter overlap between central and
peripheral \pp\ collisions), and $\preco$ which parametrises the
probability of colour reconnection.  The value of $\ptmin$ is allowed
to vary with \cmDDE energy according to the same power law as in
\pythia, \eqRef{eq:power}, and, in fact, it is $\ptminnought$ and $b$
that are fit to data, with $E_0=7$~TeV. (Note that \ptmin is the only
parameter in \herwig 7 which varies explicitly with the energy,
similarly to the case in \pythia.)  The detailed description of how
the MPI parameters were fitted to the experimental data can be found
in~\cite{Seymour:2013qka}. The most recent and default tune of \herwig
7.0 (H7-UE-MMHT) gives a good description of the underlying event data
from Tevatron's lowest energy point \cite{Aaltonen:2015aoa}, $\sqrt{s}
= 300$ GeV to the LHC's \cite{Aad:2010fh} highest $\sqrt{s} = 13$~TeV
(although the LHC's highest energy UE data \cite{atlas-ue13} was not
used for the tune). Therefore, we use H7-UE-MMHT as ``the best''
prediction of \herwig 7 for 100 TeV UE analysis. For comparison we
also show results of an older \herwig++ tune UE-EE-4.

In \pythia 8, there is no sharp distinction between soft and hard
MPI~\cite{Sjostrand:1987su}; instead there is a single eikonalised
$p_\perp$-ordered framework, with interleaved
evolution~\cite{Sjostrand:2004ef} of parton showers and MPI. The
baseline implementation in \pythia 8 is described in
\cite{Corke:2010yf}. Similarly to \herwig, the main model parameters
are: 1) an IR regularisation scale for the QCD $2\to 2$ cross section,
\pto; 2) a parameter governing the assumed transverse shape of the
proton mass distribution, and 3) a parameter controlling the strength
of final-state colour reconnections. In the original \pythia
modeling~\cite{Sjostrand:1987su}, the energy dependence of the total
cross section was taken as the guideline for the energy evolution of
the $\pto$ parameter, with a power $b=0.16$ in \eqRef{eq:power}
motivated by a cross section scaling like $s^{0.08}$. This \pto
scaling was ruled out by Tevatron measurements~\cite{Acosta:2004wqa}
as producing a too fast growth of the UE with \cmDDE energy, though it
was occasionally retained for variations. Modern tunes have $b$ values
in the range $0.21 - 0.26$. The Monash 2013 tune~\cite{Skands:2014pea}
uses a relatively low value, $b=0.215$ (see
table~\ref{tab:pythiaTunes}), and this was left unchanged in the ATLAS
A14 tune~\cite{ATL-PHYS-PUB-2014-021}. Preliminary comparisons at 13
TeV~\cite{atlas-ue13} indicate continued good agreement, though a
slightly higher scaling power around $b=0.23$ (resulting in a slower
$\sqrt{s}$ scaling of UE and MB quantities) may be preferred. In this
study, we include the baseline Monash 2013 and A14 tunes, as well as a
``Fast Scaling'' variant of the Monash tune that uses the old $b=0.16$
scaling power, for a conservative upper-limit estimate of the
extrapolated activity.

\subsubsection{Results: UE Extrapolations to 100 TeV}

\begin{figure}[t]
\centering
\includegraphics[width=0.495\textwidth]{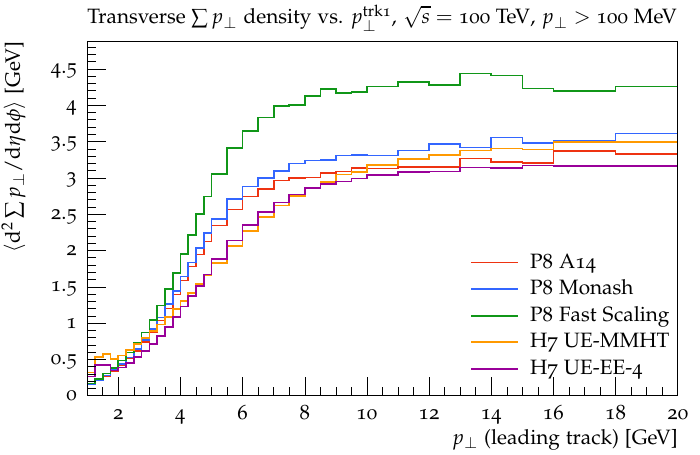}
\includegraphics[width=0.495\textwidth]{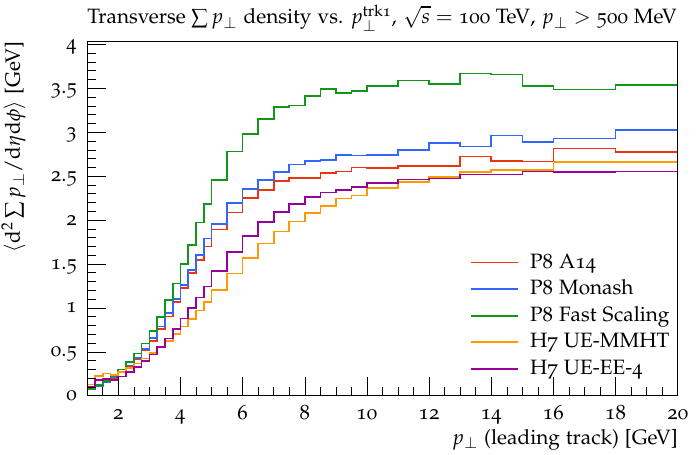}\\
\hspace*{0.05\textwidth}\begin{minipage}[b]{0.405\textwidth}
\caption{$pp$ collisions at 100 TeV.
Predictions for the transverse-region charged-particle
$\Sigma p_\perp$ density, with $\ptch >
100$~MeV (top left) and $\ptch>500$~MeV (top right) cuts. The bottom
right-hand plot shows the event-by-event fluctuations as measured by
the standard deviation for the $\ptch > 500$~MeV cut.
\label{fig:ue-sumpt}
}
\end{minipage}\hspace*{0.04\textwidth}
\includegraphics[width=0.495\textwidth]{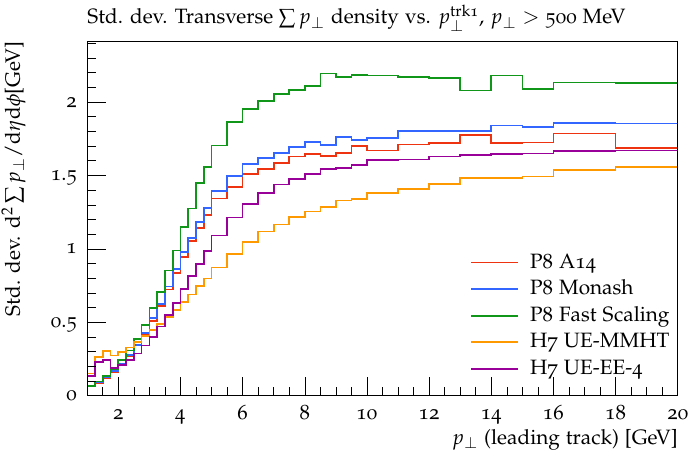}
\end{figure}
In \figRef{fig:ue-sumpt}, we show the \herwig and \pythia
extrapolations to 100 TeV for the summed charged-particle $p_\perp$
density in the transverse region, as defined above, focusing on the
region $p_{\perp\mathrm{lead}} < 20$ GeV in which the transition to
the UE plateau occurs. The top left- and right-hand plots show the two
different charged-particle $p_\perp$ cuts, while the bottom right-hand
one shows the standard-deviation fluctuations for the $\ptch > 500$
MeV cut. Given the order-of-magnitude extrapolation in \cmDDE energy,
there is a remarkable level of agreement between the central models
(i.e., excluding the extreme Fast Scaling one), with the
charged-particle UE plateau characterised by
\begin{eqnarray}
\left<\sum p_{\perp\mathrm{ch}}\right>_{\ptch > 100~\mathrm{MeV}}~\mbox{(per unit
  $\Delta\eta\Delta\phi$)} & = & 3.3 \pm
0.5~\mathrm{GeV} ~,\\
\left<\sum p_{\perp\mathrm{ch}}\right>_{\ptch > 500~\mathrm{MeV}} ~\mbox{(per unit
  $\Delta\eta\Delta\phi$)} & = &
2.7 \pm 0.4~\mathrm{GeV} ~,
\end{eqnarray}
within slightly inflated 15\% uncertainties, and the Fast Scaling
variant defining conservative upper-limit densities of 4.4 and 3.6
GeV, respectively. Note that the total summed $p_\perp$ in the
transverse region rises slowly with jet $p_\perp$, and that including
both charged and neutral particles would result in numbers
approximately a factor 1.6 higher.

We emphasise that there is some arbitrariness whether to use the lower
or higher cut to estimate UE contributions to jets. For the charged
component, particles with $p_\perp < 500$~MeV typically do not make it
to the calorimeter and hence do not contribute to calorimetric energy
measurements. On the other hand, low-$p_\perp$ neutral particles
(including photons) may or may not be absorbed in the inner
detector. A phenomenology calculation could therefore well use the
lower cut (assuming experimental results will be corrected for loss
effects) while a calorimeter study could use some combination of the
two.

For comparison, the Snowmass study in \cite{Skands:2013asa}, which
considered extrapolations to 100 TeV using the latest set of ``Perugia
2012'' tunes~\cite{Skands:2010ak} of the \pythia 6 event
generator~\cite{Sjostrand:2006za} (version 6.428), found, for a
reference sample of 100-GeV dijets, in the region $|\eta|<2.5$, a
neutral+charged $p_\perp$ density in the transverse region of
$4.4\pm0.45$ GeV. Translated to the phase-space region studied here,
this prediction is somewhat lower than the ones above, consistent with
the Perugia 2012 tune's higher \pto scaling power $b=0.24$.

Finally, we note that the small bumps on the HERWIG 7 curves at very
low $p_{\mathrm{lead}}$ are due to the colour structure of soft MPI
and will be addressed in the next release.

\begin{figure}[t]
\centering
\includegraphics[width=0.495\textwidth]{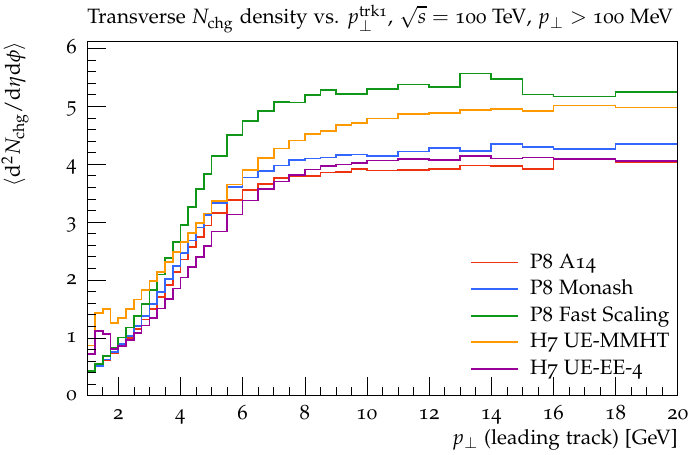}
\includegraphics[width=0.495\textwidth]{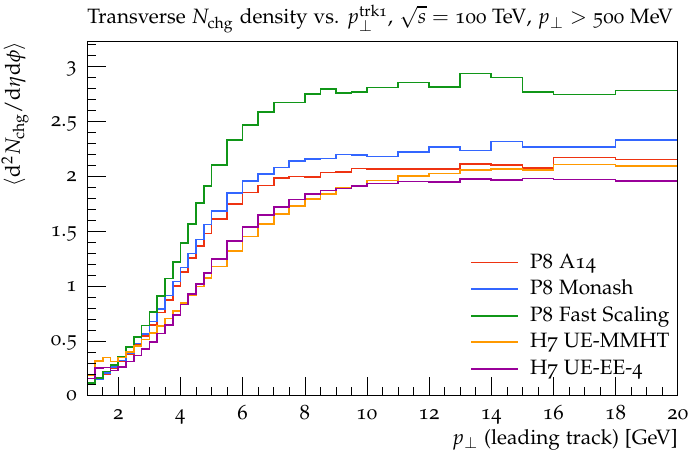}\\
\hspace*{0.05\textwidth}\begin{minipage}[b]{0.405\textwidth}
\caption{$pp$ collisions at 100 TeV.
Predictions for the transverse-region charged-particle density,
 with $p_\perp > 100$~MeV (top left) and $p_\perp>500$~MeV (top right)
 cuts. The bottom
right-hand plot shows the event-by-event fluctuations as measured by
the standard deviation for the $p_\perp > 500$~MeV cut.
\label{fig:ue-nch}}
\end{minipage}\hspace*{0.04\textwidth}
\includegraphics[width=0.495\textwidth]{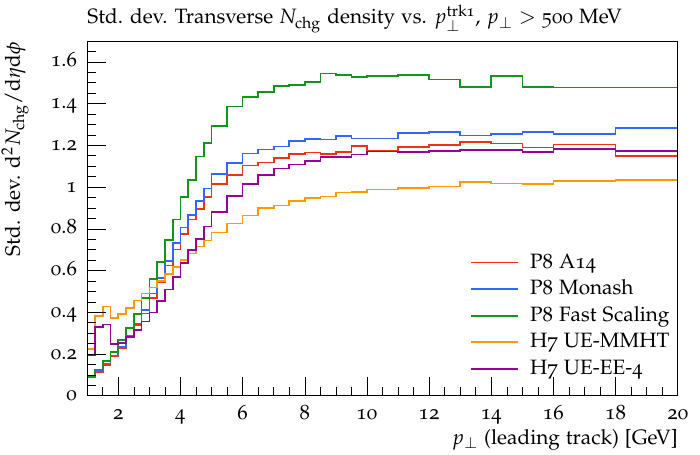}
\end{figure}
The charged-particle densities shown in \figRef{fig:ue-nch} exhibit a
larger spread between the models. In particular for the soft end of
the spectrum, highlighted by the top left-hand plot, the H7 UE-MMHT
model predicts the same density as the Fast Scaling \pythia tune,
30\% above the level of the other models. In the right-hand plot,
however, with the $p_\perp$ cut of $500$ MeV, the H7 UE-MMHT level
drops down to that of the other central tunes, while the Fast Scaling
\pythia tune remains above. Interestingly, the H7 UE-EE-4 level is
lower, but its fluctuations higher, than those of H7 UE-MMHT. We note
that the former has a smaller inverse proton size,
$\mu^2=1.11$ compared to  UE-MMHT $\mu^2=2.30$. 

\begin{figure}[t]
\centering
\includegraphics[width=0.495\textwidth]{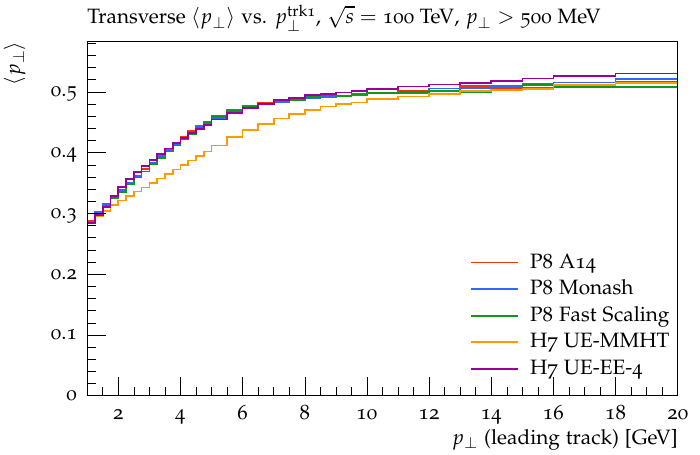}
\caption{$pp$ collisions at 100 TeV.
Predictions for the transverse-region average $p_\perp$ of charged
particles for the  $p_\perp>500$~MeV cut.
\label{fig:ue-avgpt}
}
\end{figure}
The final plot in \figRef{fig:ue-avgpt} displays a remarkable
agreement on the average $p_\perp$ of charged particles.  Despite the
underlying model differences, and the significant uncertainties
surrounding aspects such as colour reconnections, the predictions are
virtually indistinguishable, the only exception being the H7 UE-MMHT
model in the turn-on region below $p_{\perp\mathrm{lead}} = 10$ GeV.

\clearpage
\section{Inclusive vector boson production}
\label{ref:dyan}

The production of $W$ and $Z$ bosons is a valuable probe of both EW
and QCD dynamics. The total production rate of $W^\pm$ ($Z^0$) bosons
at 100~TeV is about 1.3 (0.4)~$\mu$b. This corresponds to samples of
$O(10^{11})$ leptonic ($e,\mu$) decays per \iab.  The production
properties are known today up to next-to-next-to-leading order (NNLO)
in QCD, leading to a precision of the order of the percent. A detailed
discussion of the implications of this precision, and of the possible
measurements possible with $W$ and $Z$ final states at 100~TeV, is
outside the scope of this review, also because the LHC has only
started exploiting the full potential of what can be done with them
(for a recent review, see Ref.~\cite{Mangano:2015ejw}).  We shall
therefore focus here on documenting some basic rates and
distributions, to show the extreme kinematical configurations that may
be accessed at 100~TeV, and to highlight some of the novel features of
EW interactions that will emerge at these energies.

\subsection{Inclusive $W/Z$ rates and distributions}
First of all, we compute the PDF uncertainties in the inclusive
cross-sections (and their ratios) for electroweak gauge boson
production at both 14 TeV and 100 TeV.
We use the NNLO inclusive calculation of Ref.~\cite{Anastasiou:2003ds}
as implemented in the {\tt VRAP} code.
We compare the results from four modern PDF sets:
ABM12~\cite{Alekhin:2013nda}, CT14~\cite{Dulat:2015mca},
MMHT14~\cite{Harland-Lang:2014zoa} and NNPDF3.0~\cite{Ball:2014uwa}.
These four NNLO sets have $\alpha_s(M_Z)=0.118$, except
ABM12 for which the native value is $\alpha_s(M_Z)=0.1132$.
The PDF sets are accessed via the {\tt LHAPDF6} interface.

In Table~\ref{tab:vrapNNLO14}
we show the total NNLO inclusive cross-sections (including the leptonic
branching fractions) and the corresponding
percentage PDF uncertainties
for weak gauge boson production at the LHC 14 TeV.
    We also indicate the shift in the central cross-section
    of the different PDFs as compared to a reference
    cross-section, which here is taken to be that
    of NNPDF3.0 NNLO.
    The corresponding results
    at 100 TeV are shown in
    Table~\ref{tab:vrapNNLO100}.
    We observe a substantial increase on the PDF systematics
    when going from 14 TeV to 100 TeV, specially
    for NNPDF3.0 but also for CT14 and MMHT14.

\begin{table}[h]
  \centering
  \small 
  \begin{tabular}{|c|c|c|c|c|c|c|c|c|}
     \hline
    &  \multicolumn{8}{c|}{14 TeV} 
     \\
    \hline
    &  \multicolumn{2}{c|}{NNPDF3.0} &
    \multicolumn{2}{c|}{CT14} &
    \multicolumn{2}{c|}{MMHT14} &
    \multicolumn{2}{c|}{ABM12} \\
    \hline
    &  $\sigma ({\rm nb}) \pm \delta_{\rm pdf}$  &  $\sigma/\sigma_{\rm ref}$
    &  $\sigma ({\rm nb}) \pm \delta_{\rm pdf}$  &  $\sigma/\sigma_{\rm ref}$
    &  $\sigma ({\rm nb}) \pm \delta_{\rm pdf}$  &  $\sigma/\sigma_{\rm ref}$
    &  $\sigma ({\rm nb}) \pm \delta_{\rm pdf}$  &  $\sigma/\sigma_{\rm ref}$ \\
    \hline
    \hline
    $ W^+$  &  12.2 $\pm $2.3\%  & 1
    &  12.4 $\pm$2.4\%  & 1.01
    &  12.5 $\pm$ 1.5\%   & 1.02
    &  12.7 $\pm$ 1.2\%  & 1.04\\
    \hline
     $ W^-$  & 9.1 $\pm$2.4\%  & 1
    & 9.2 $\pm$2.3\%   & 1.02
    & 9.3 $\pm$1.5\%   & 1.03
    & 9.3 $\pm$1.2\% & 1.03\\
    \hline
     $ Z $  & 2.0 $\pm$2.2\%   &1
    &  2.1 $\pm$2.2\%  & 1.01
    &  2.1 $\pm$1.6\%  & 1.02
    &  2.1 $\pm$1.2 \% & 1.00\\
    \hline
     $W^+/W^-$  & 1.4 $\pm$0.8\%   &1
    & 1.3 $\pm$2.4\%  & 1.00
    & 1.3 $\pm$1.5\%  & 1.00
    & 1.4 $\pm$1.2 \% & 1.01 \\
    \hline
     $W/Z$  &   10.5 $\pm$0.4\% &1
    & 10.5 $\pm$1.4\%  & 1.00
    & 10.5 $\pm$0.9\%  & 1.00
    & 10.5 $\pm$0.7\%  & 1.00 \\
    \hline
  \end{tabular}
  \caption{\small \label{tab:vrapNNLO14}
    The PDF uncertainties for the NNLO inclusive cross-sections
    for weak gauge boson production at the LHC 14 TeV.
    We also indicate the shift in the central cross-section
    of the different PDFs as compared to a reference
    cross-section, which here is taken to be that
    of NNPDF3.0.
    The calculation has been performed with the {\tt VRAP} code.
    The leptonic branching fractions have been included in the calculation.
  }
    \end{table}

\begin{table}[h]
  \centering
  \small
  \begin{tabular}{|c|c|c|c|c|c|c|c|c|}
     \hline
    &  \multicolumn{8}{c|}{100 TeV} 
     \\
    \hline
    &  \multicolumn{2}{c|}{NNPDF3.0} &
    \multicolumn{2}{c|}{ABM12} &
    \multicolumn{2}{c|}{CT14} &
    \multicolumn{2}{c|}{MMHT14} \\
      \hline
    &  $\sigma ({\rm nb}) \pm \delta_{\rm pdf}$  &  $\sigma/\sigma_{\rm ref}$
    &  $\sigma ({\rm nb}) \pm \delta_{\rm pdf}$  &  $\sigma/\sigma_{\rm ref}$
    &  $\sigma ({\rm nb}) \pm \delta_{\rm pdf}$  &  $\sigma/\sigma_{\rm ref}$
    &  $\sigma ({\rm nb}) \pm \delta_{\rm pdf}$  &  $\sigma/\sigma_{\rm ref}$ \\
    \hline
    \hline
    $ W^+$  & 77.0 $\pm$13.1\%   & 1
    & 74.9$\pm$7.2\%   & 0.97
    & 71.8 $\pm$4.8\%   & 0.93
    & 74.1 $\pm$2.0\%  & 0.96 \\
    \hline
     $ W^-$  & 63.4 $\pm$8.5\%   & 1
    & 62.9  $\pm$5.9\%  & 0.99
    & 61.3 $\pm$3.6\%  & 0.97
    & 62.2 $\pm$2.0\%  & 0.98 \\
    \hline
     $ Z$  &  14.1 $\pm$ 7.9\%  &1
    & 13.9 $\pm$5.7\%   & 0.99
    & 13.7 $\pm$3.7\%   & 0.97
    & 13.9 $\pm$2.0\%   & 0.98 \\
    \hline
     $ W^+/W^-$  & 1.2 $\pm$4.3\%    &1
    &  1.2 $\pm$7.1\%  & 0.98
    &  1.2 $\pm$4.8\%  & 0.97
    & 1.2 $\pm$2.0\%  & 0.98  \\
    \hline
     $ W/Z$  &  9.9 $\pm$2.9\%   &1
    &  9.9 $\pm$3.9\%   & 1.00
    &  9.7 $\pm$2.6\%  & 0.98
    &  9.8 $\pm$1.1\%  & 0.99\\
    \hline
  \end{tabular}
  \caption{\small \label{tab:vrapNNLO100}
    Same as Table~\ref{tab:vrapNNLO14} now for
    $\sqrt{s}=100$ TeV.
  }
    \end{table}

To investigate the impact of realistic acceptance cuts,
we have used {\tt MCFM} v7.0.1 to compute the NLO
cross-sections (using NNLO PDFs) including the decays
of the gauge bosons.
We have considered three different cases for the final-state cuts:
\begin{itemize}
\item No cuts
\item {\it LHC} cuts: $p_T^l\ge 20$ GeV, $|\eta_l|\le 2.5$
  \item {\it FCC} cuts: $p_T^l\ge 20$ GeV, $|\eta_l|\le 5$
\end{itemize}
In addition, jets are reconstructed with the anti-$k_t$ algorithm
with $R=0.4$, but no cuts are imposed on the kinematics of this jet.
The results are summarized in Table~\ref{tab:MCFM14}, where we show
the production cross-sections and the corresponding percentage PDF
uncertainties for weak gauge bosons at 14 TeV and 100 TeV with
different kinematical cuts on the final state particles.
The calculation has been performed at NLO with {\tt MCFM} v7.0.1,
using the NNPDF3.0 NNLO PDF set.
We observe that PDF uncertainties are reduced if the rapidity of the
final-state leptons is restricted to the central region, indicating
that the increase of PDF errors from 14 to 100 TeV arises from the
forward region, sensitive to the poorly-known small-$x$ PDFs.

\begin{table}[h]
  \centering
  \small
  \begin{tabular}{|c||c|c|c|c|c|}
     \hline
    \multicolumn{6}{|c|}{NNPDF3.0 NNLO} 
     \\
     \hline
   $\sigma(pp\to V\to l_1l_2)~{\rm [nb]}~(\pm\delta_{\rm pdf}\sigma$)  & \multicolumn{2}{c|}{14 TeV}
     & \multicolumn{3}{c|}{100 TeV}
     \\
    \hline
    &  No cuts &
        {\it LHC} cuts &
          No cuts &
        {\it LHC} cuts &
        {\it FCC} cuts \\
    \hline
    \hline
    $W^+$  & 12.2  (2.2\%)
    & 6.5  (2.2\%)
    & 77.3  (13.1\%)
    & 28.3  (3.3\%)
    & 54.3  (6.5\%)
    \\
    \hline
    $W^-$  & 9.2  (2.3\%)
    & 4.9  (2.3\%)
    & 64.3  (8.9\%)
    & 27.2  (3.3\%)
    & 45.5  (4.0\%)
    \\
    \hline
    $Z$  & 2.1  (2.1\%)
    & 1.5  (2.1\%)
    & 14.5  (7.7\%)
    & 8.3 (3.3\%)
    & 12.8  (5.0\%)
    \\
    \hline
  \end{tabular}
  \caption{\small \label{tab:MCFM14} The production cross-sections
    for weak gauge bosons at 14 TeV and 100 TeV, including
    the leptonic decays, with
    different kinematical cuts on the final state particles,
    see text for more details.
    We provide both the total cross-section and the corresponding
    percentage PDF uncertainty.
    The calculation has been performed at NLO with {\tt MCFM} v7.0.1,
    using the NNPDF3.0 NNLO PDF set.
  }
    \end{table}

\begin{figure}[h!]
\centering
\includegraphics[width=0.8\textwidth]{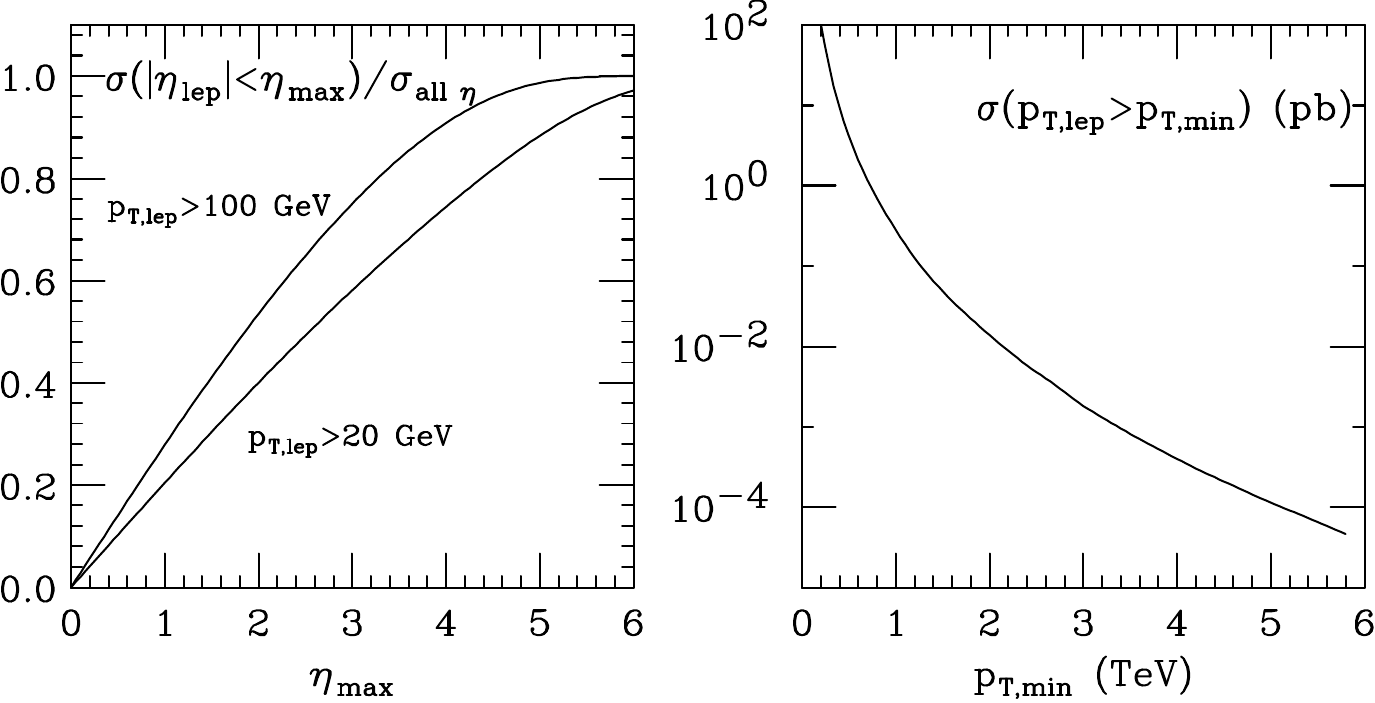}
\caption{Left: rapidity acceptance for leptons from inclusive $W$
  production and decay, for $p_T$ thresholds of 20 and 100~GeV. Right:
inclusive lepton $p_T$ spectrum.}
\label{fig:wleppt}
\end{figure}
At 100~TeV, gauge bosons will have a rather broad rapidity
distribution and, as shown in the left plot of Fig.~\ref{fig:wleppt},
more than 50\% of the leptons with $p_{T}>20$~GeV will be produced at
$\vert \eta \vert>2.5$ (w.r.t. $\sim 30\%$ at 14~TeV). Even leptons
with $p_{T}>100$~GeV will have a large forward rate, with about 40\%
of them at $\vert \eta \vert>2.5$ ($\sim 10\%$ at 14~TeV). Their $p_T$
spectrum will also extend to large values, as shown in the right plot
of Fig.~\ref{fig:wleppt}. The largest fraction of these high-$p_T$
leptons will arise from $W$'s produced at large $p_T$, in association
with jets.

We focus now on the charged lepton rapidity distributions.  In $pp$
collisions rapidity distributions are forward-backward symmetric and
only the positive (or negative) the rapidity range needs to be shown.
The $W^+$ bosons are produced at larger rapidities and with a larger
production rate than the $W^-$ bosons.  This is because the production
of $W^+$ ($W^-$) bosons is mainly controlled by $u \bar d$ ($\bar u
d$) quarks collisions.  The rapidity asymmetry is the result of
$u(x)/d(x)$ becoming larger at larger $x$. The total rate difference
is due to the $u$ quark density being larger than that of the $d$
density (the antiquark $\bar u$ and $\bar d$ densities are relatively
similar, especially at small values of parton momentum fractions).
Due to parity violation in the $W$ boson production and decay, the
charged (anti)lepton tends to be produced in the direction of the
initial-state (anti)quark. Therefore $\ell^-$ prefers the direction of
the $d$-type quark, and $\ell^+$ the direction of the $\bar{d}$-type
antiquark. The rapidity distribution of charged leptons is therefore
the result of opposite physical effects: the parton densities of the
colliding hadrons favour forward production of $W^+$ over $W^-$
bosons, but their decays favour forward emission of $\ell^-$ over
$\ell^+$ leptons. This leads to a peculiar structure of the leptonic
charge asymmetry, which changes sign at some $p_T$-dependent value of
rapidity.

In Fig.~\ref{fig:wzrap} we show the normalized rapidity distribution
of the $W^\pm$ and $Z$ bosons in NLO QCD computed with the DYNNLO
parton level Monte Carlo~\cite{Catani:2009sm} by using
NNPDF3.0~\cite{Ball:2014uwa} parton densities at NLO with
$\alpha_S(M_Z^2) = 0.118$. The leptonic charge asymmetry is shown in
Fig.~\ref{fig:asy-lep}, for various lepton $p_T$ thresholds. Notice
that, while at LHC energies the asymmetry changes sign at $\eta \sim 2.5$ for
$p_T\gsim 20$~GeV,  here the zero is shifted to much higher $\eta$
values, as a result of the much wider boson rapidity spectrum. The
asymmetry is also very small in the central $\eta$ region, since at
100 TeV, for the values of $x$ relevant to central $W$ production,
the valence component of quark densities is suppressed with respect
to the sea, and thus $u(x)\sim d(x)$.
\begin{figure}[h!]
\centering
\includegraphics[height=0.25\textheight]{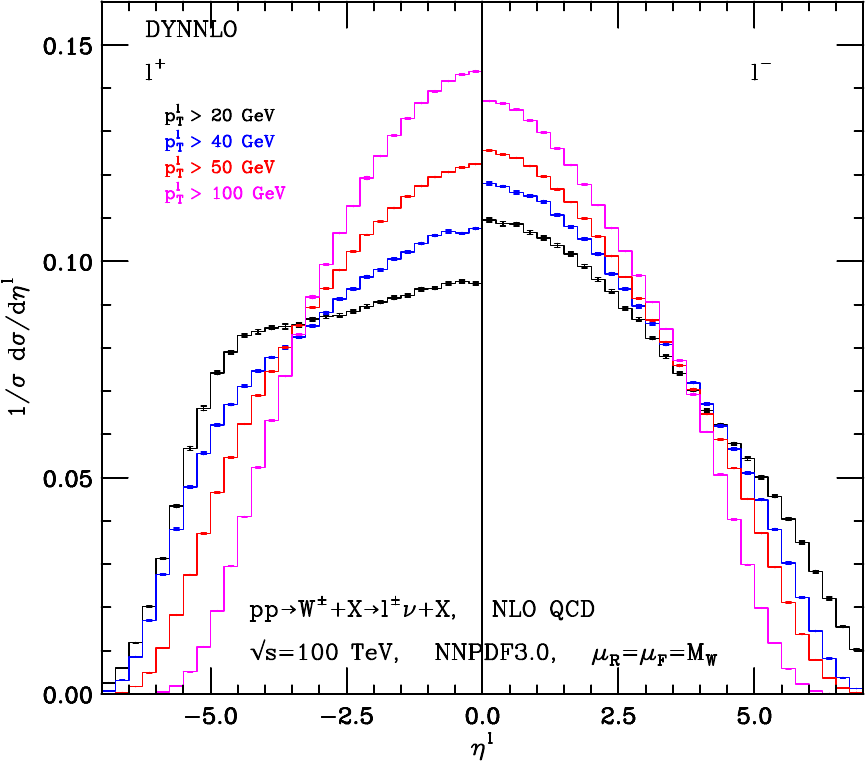}
\hfill
\includegraphics[height=0.25\textheight]{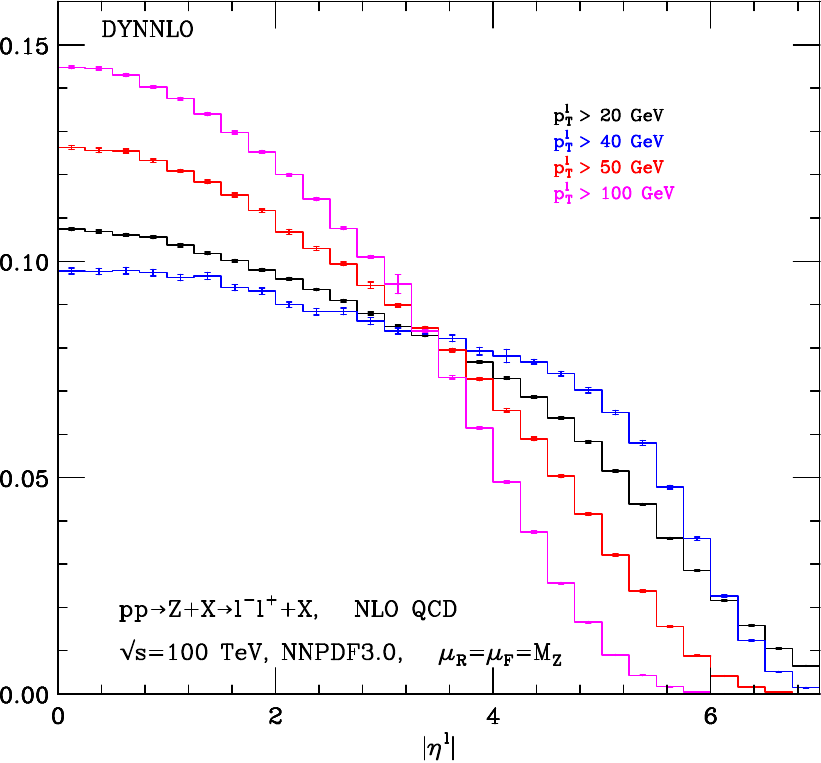}
\caption{
Normalized rapidity distribution of the charged leptons from
$W^\pm$ (left panel) and $Z$ (right panel)
boson decays, at  $\sqrt{s}=100$~TeV. 
The charged leptons are required to have a minimum $p_T$ of $20$, $40$, $50$ and $100$ GeV.
The error bars reported in the histograms refer to an estimate of the numerical error in the Monte Carlo
integration carried out by the DYNNLO code.
}
\label{fig:wzrap}
\end{figure}

\begin{figure}[h!]
\centering
\includegraphics[width=0.6\textwidth]{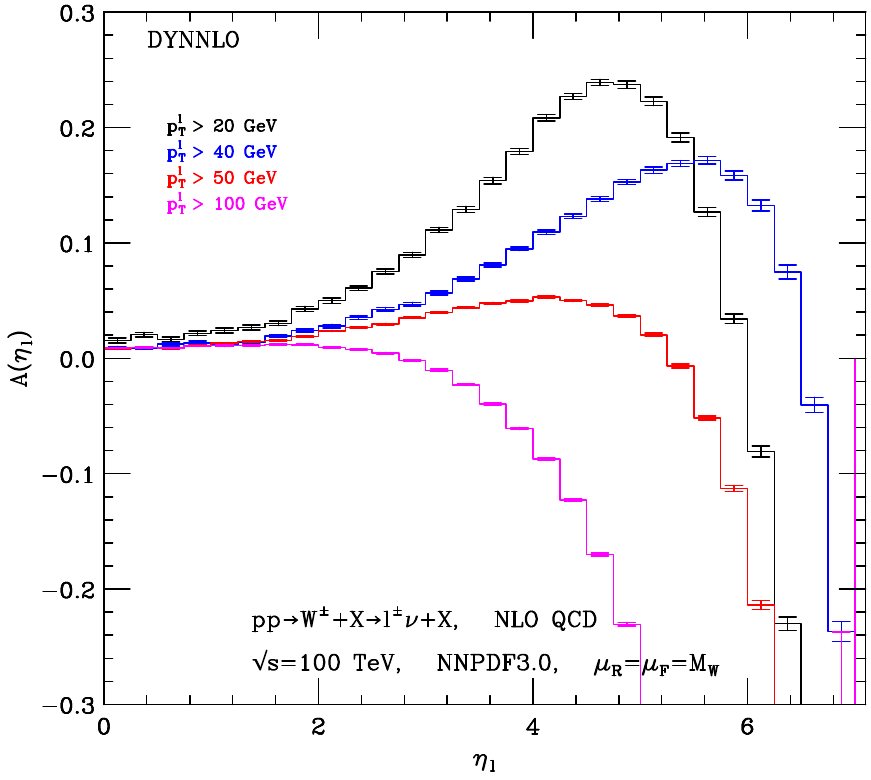}
\caption{
Leptonic charge asymmetry, for different lepton $p_T$ thresholds.
}
\label{fig:asy-lep}
\end{figure}

\subsection{$W/Z$ boson production at small $q_T$}
\label{sec:dyan_smallpt}

An observable particularly relevant in $W/Z$ boson production is the
transverse-momentum ($q_T$) distribution of the vector boson. In the
large $q_T$ region ($q_T \gtrsim M_V$) QCD corrections are known in
analytic form up to
$\mathcal{O}(\alpha_S^2)$~\cite{Ellis:1981hk,Arnold:1988dp,Gonsalves:1989ar}
and fully exclusive computations of $W/Z$ bosons in association with a
jet are available up to
$\mathcal{O}(\alpha_S^3)$~\cite{Boughezal:2015dva,Ridder:2015dxa}.

However the bulk of the $W/Z$ bosons are produced at small $q_T$ ($q_T
\ll M_V$) where the reliability of fixed-order calculations is spoiled
by large logarithmic corrections of the form $\alpha_S^n
(M_V^2/q_T^2)\ln^m(M_V^2/q_T^2)$ (with $0\leq m \leq 2n-1$) due to
soft and/or collinear parton emissions. At a centre--of--mass energy
of $100$~TeV about half of $W/Z$ bosons are produced in the region
where $q_T \lesssim 15$ GeV.  In order to restore the reliability of
perturbation theory in the small-$q_T$ region, these
logarithmically-enhanced terms have to be systematically resummed to
all perturbative orders. The resummed and fixed-order predictions can
be consistently matched at intermediate values of $q_T$ to obtain a
uniform theoretical accuracy in a wide range of transverse momenta.

We consider the processes $pp \to W^\pm \to l\nu_l$ and $pp \to Z \to
l^+ l^−$ at $\sqrt s = 100$~TeV centre--of--mass energy and we compute
the transverse-momentum distribution by using the resummation
formalism proposed in
Refs.~\cite{Catani:2000vq,Bozzi:2005wk,Bozzi:2007pn}.  The numerical
results are obtained by using the code \texttt{DYqT}, which is based
on the results presented in Refs.~\cite{Bozzi:2008bb,Bozzi:2010xn}.
An analogous but more general computation~\cite{Catani:2015vma}, which
includes the full dependence on the final-state lepton(s) kinematics,
is encoded in the numerical program \texttt{DYRes}.  We provide
predictions at NNLL+NNLO (NLL+NLO) accuracy by using
NNPDF3.0~\cite{Ball:2014uwa} parton densities at NNLO (NLO) with
$\alpha_S(M_Z^2) = 0.118$ and $\alpha_S$ evaluated at 3-loop (2-loop)
order.  As for the EW couplings, we use the values quoted in the PDG
2014~\cite{Agashe:2014kda} within the so called $G_\mu$ scheme, where
the input parameters are $G_F$, $M_Z$, $M_W$.

\begin{figure}[h!]
\centering
\includegraphics[height=0.25\textheight]{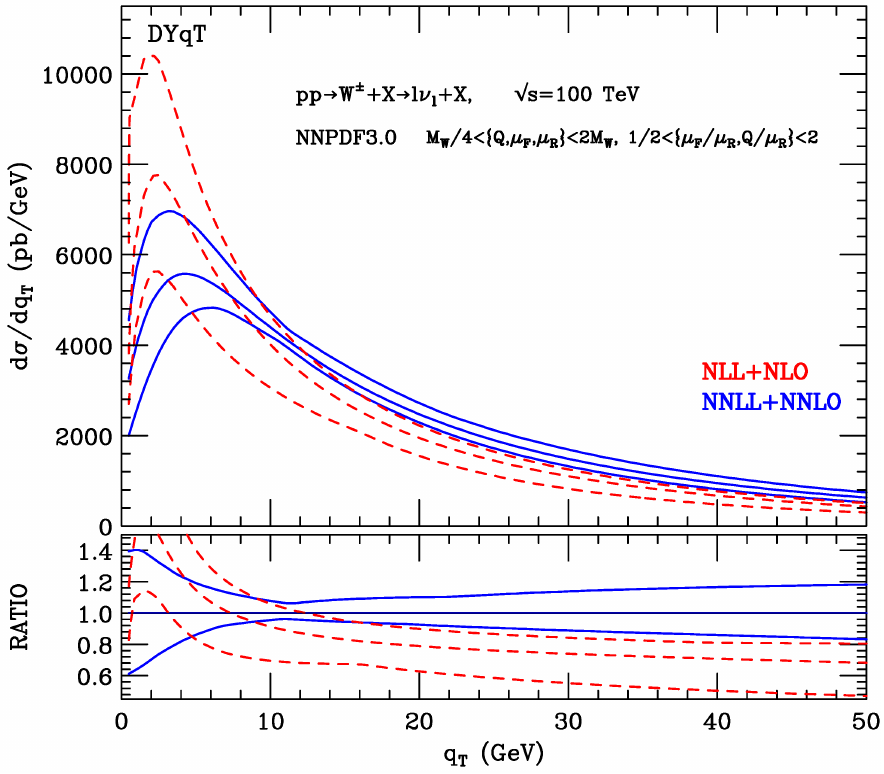}
\hfill
\includegraphics[height=0.25\textheight]{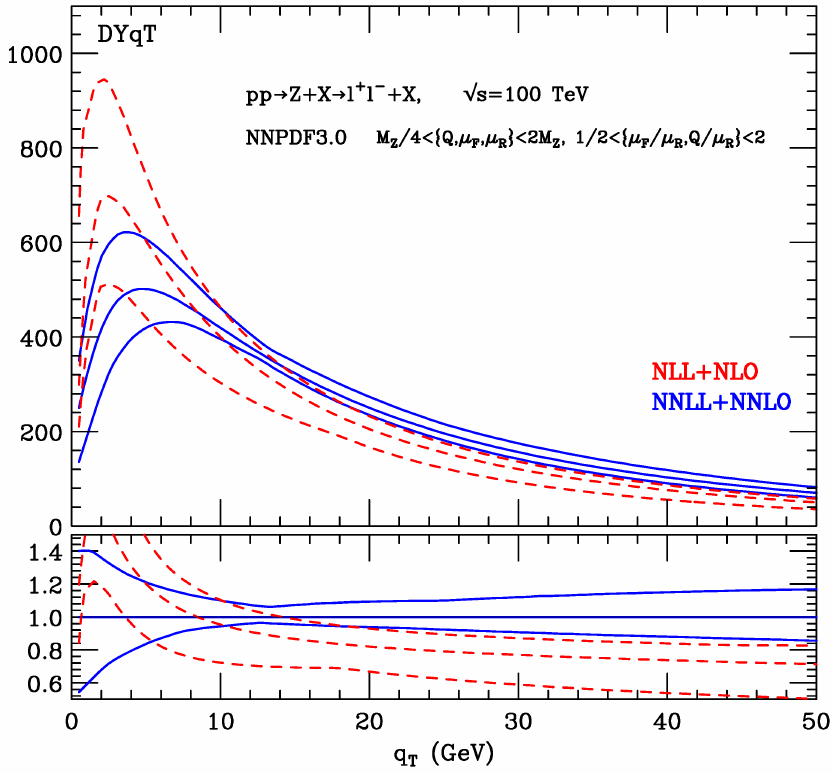}
\caption{The $q_T$ spectrum of $W^\pm$ (left panel) and $Z$ (right
  panel) bosons in $pp$ collisions at $\sqrt{s}=100$~TeV.  The bands
  are obtained by performing $\{\mu_F, \mu_R, Q \}$ variations (as
  described in the text) around the central value $M_W/2$.  The lower
  panel presents the ratio of the NLL+NLO and NNLL+NNLO bands with
  respect to the NNLL+NNLO result at the central value of the scales.
}
\label{fig:vqt}
\end{figure}

The NLL+NLO and NNLL+NNLO results for the $q_T$ spectrum of on-shell
$W$ and $Z$ bosons produced at $\sqrt s = 100$~TeV are presented in
Fig.~\ref{fig:vqt}.  The bands provide an estimate of the perturbative
uncertainties due to missing higher-order contributions.  The bands
are obtained through independent variations of factorization
($\mu_F$), renormalization ($\mu_R$) and resummation ($Q$) scales in
the range $M_V/4\leq \{\mu_F, \mu_R, Q\}\leq 2 M_V$ with the
constraints $0.5\leq \mu_F/\mu_R\leq 2$ and $0.5\leq Q/\mu_R\leq 2$.
The lower panels present the ratio of the scale-dependent NNLL+NNLO
and NLL+NLO results with respect to the NNLL+NNLO result at the
central value $\mu_F=\mu_R=Q=M_V/2$ of the scales.

The region of small and intermediate values of $q_T$ is shown in the
main panels of Fig.~\ref{fig:vqt}.  The shape of the $W$ and $Z$ $q_T$
spectra is qualitatively similar, with the $Z$ spectrum slightly
harder than the $W$ spectrum.  Both the $W/Z$ NNLL+NNLO $q_T$ spectra
are harder than the corresponding spectra at NLL+NLO accuracy with a
sensible reduction of the scale-variation band going from the NLL+NLO
to the NNLL+NNLO band.  The NLL+NLO and NNLL+NNLO bands overlap at
small transverse momenta and remain very close by increasing $q_T$.
The NNLL+NNLO uncertainty is about $\pm 20$\% at the peak, it
decreases to about $\pm 6$\% at $q_T \simeq 10-15$~GeV and increases
to about $\pm 15$\% at $q_T\sim 50$~GeV.

\subsection{DY production at large $p_T$ and at large mass}
\label{sec:dyan_largept}
The left plot in Fig.~\ref{fig:w_largept} shows the integrated $p_T$
spectrum of $W$ bosons, from a LO calculation.  With luminosities in excess of
1~\iab, data will extend beyond 15~TeV. The immense kinematical
reach of DY distributions at 100~TeV is also displayed by the right
plot in the same Figure, which shows the 
integrated dilepton invariant mass distribution, for one 
lepton family, with $\vert \eta_\ell \vert<2.5$. The DY statistics,
with the anticipated $O(20)$~\iab, will extend out to $M_{\ell\ell} \sim 20$~TeV.

\begin{figure}[h!]
\centering
\includegraphics[width=0.45\textwidth]{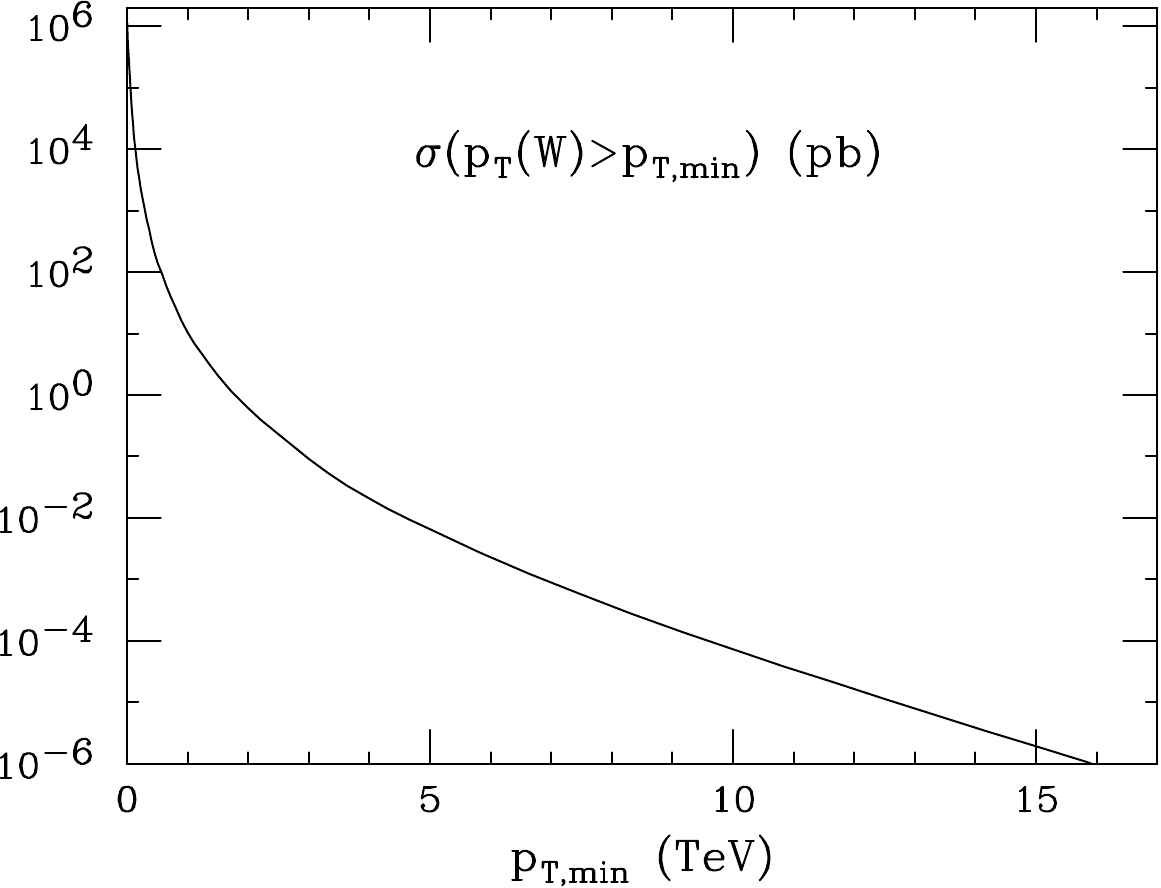}
\includegraphics[width=0.46\textwidth]{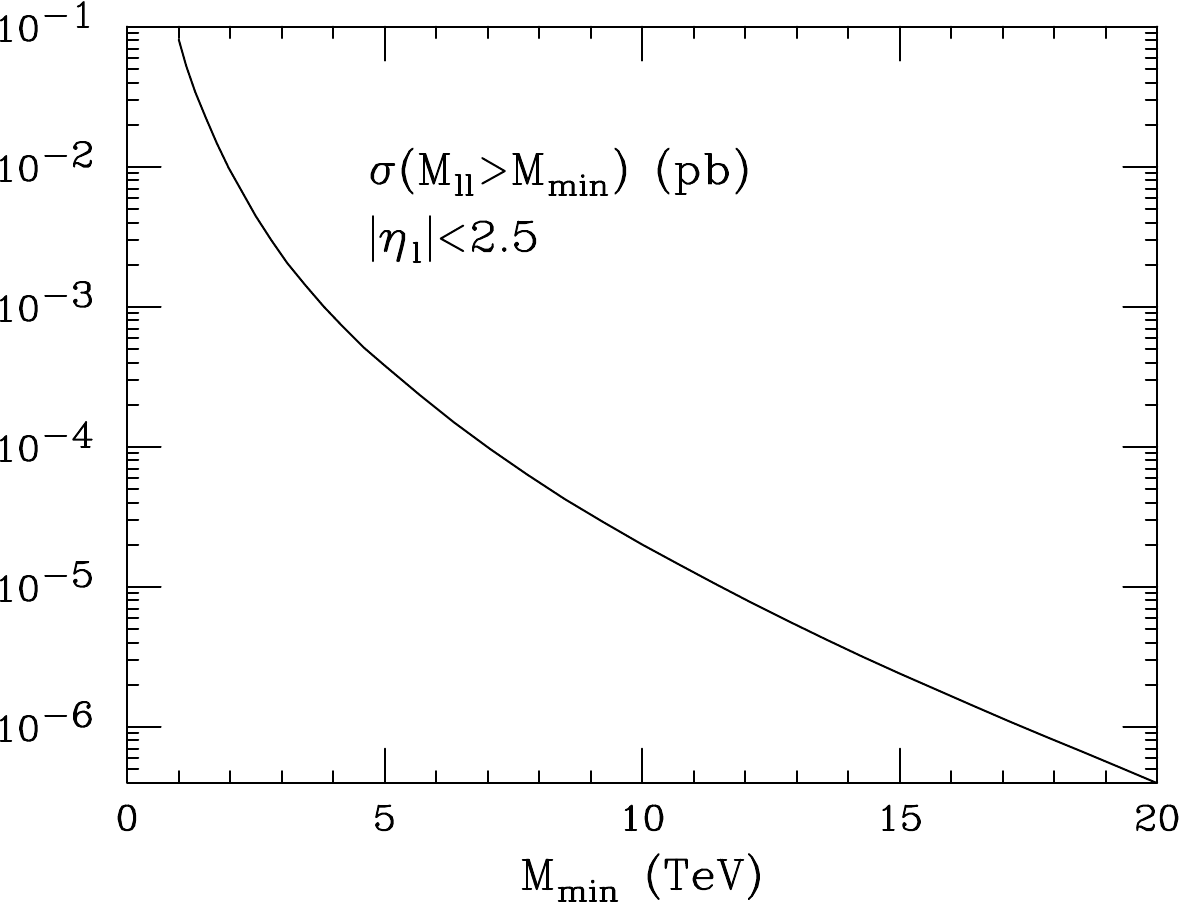}
\caption{Left: inclusive $p_T$ spectrum of $W$ bosons. Right:
  Integrated dilepton invariant mass distribution, for one 
  lepton family, with $\vert \eta_\ell \vert<2.5$.}
\label{fig:w_largept}
\end{figure}
NNLO results have recently become available for the $W/Z$+jet
transverse momentum
distributions~\cite{Boughezal:2015dva,Boughezal:2015ded,Ridder:2015dxa}.
For a gauge boson produced at large $p_T$, there is always at least
one jet recoiling against it, and therefore one can assume that this
calculation provides NNLO accuracy for the $W/Z$ inclusive $p_T$
spectrum.  The $p_T(W)$ differential distribution at 100~TeV is shown
in the left plot of Fig.~\ref{fig:w_largept_nnlo}, which shows also
the comparison with the NLO result. The
calculation~\cite{Boughezal:2015dva} was performed using anti-$k_t$
jets~\cite{Cacciari:2008gp} at R=0.4, $|\eta_J|<5$,
$\mu^2=M_W^2+\sum_{j} p_{T,j}^2$ and CT14 NNLO PDFs. A minimum
threshold $p_T>1$~TeV was applied to the leading jet: this biases the
$W$ $p_T$ spectrum in the region below $O(1.5)$~TeV, but has no impact
above that.  On the right of Fig.~\ref{fig:w_largept_nnlo} we show the
integrated spectrum of the leading jet in $W$+jet events. We notice
the huge increase from LO to NLO, due to the appearance at
$O(\alpha_s^2)$ of processes where two jets recoil against each other,
the $W$ being radiated from the initial state or from one of them
(this will be discussed more extensively in the section
subsection). The LO jet spectrum matches well the result of the $W$
spectrum in the left plot, corresponding to the LO configurations
where the $W$ recoils against a jet.  We point out that the NNLO/NLO
$K$ factors are very close to one, suggesting that after inclusion of
the new NLO topologies ones has reached a rather stable perturbative
expansion.  We also recall that this calculation only includes the QCD
effects. For $p_T$ beyond the TeV scale, the effects of virtual EW
corrections are known to lead to important
corrections~\cite{Denner:2011vu}, as will be discussed in
Section~\ref{sec:EW}.

\begin{figure}[h!]
\centering
\includegraphics[width=0.49\textwidth]{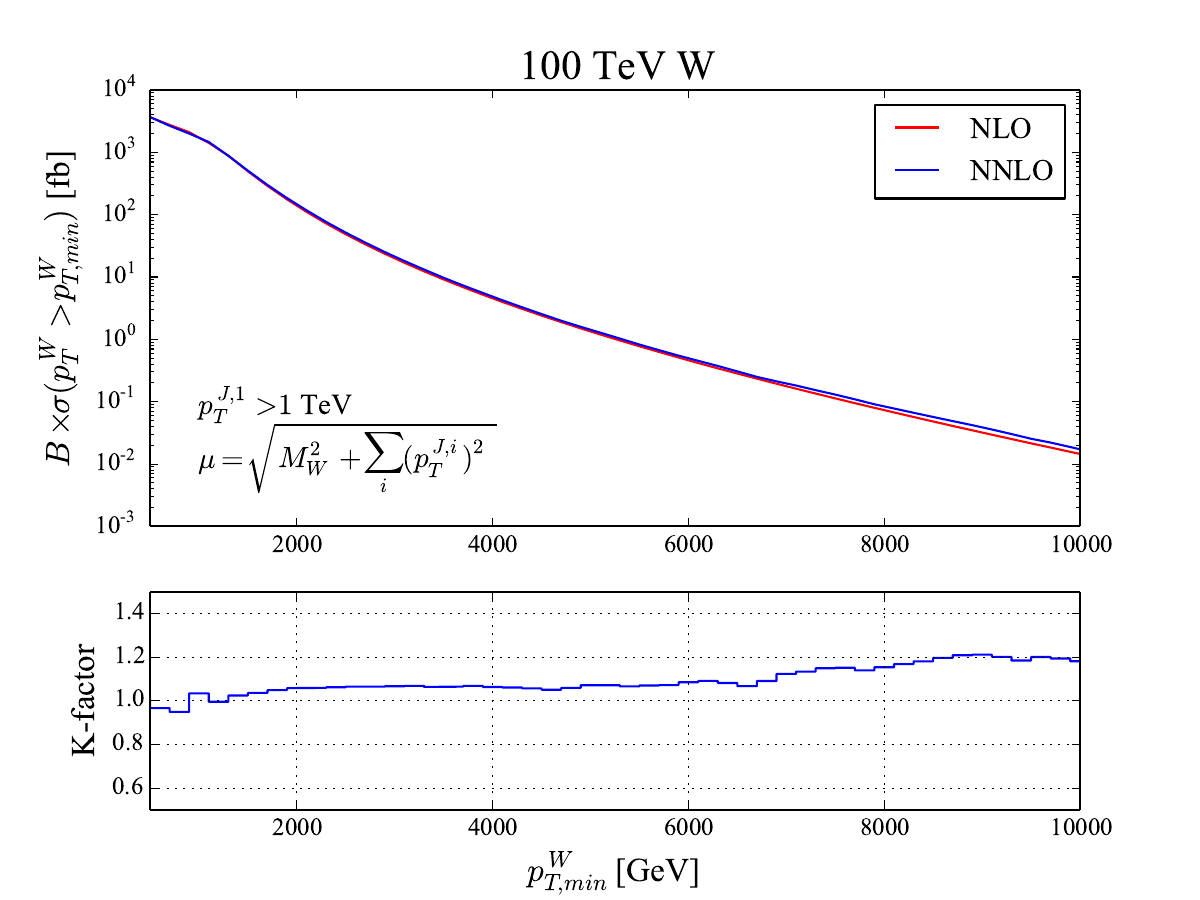}
\includegraphics[width=0.49\textwidth]{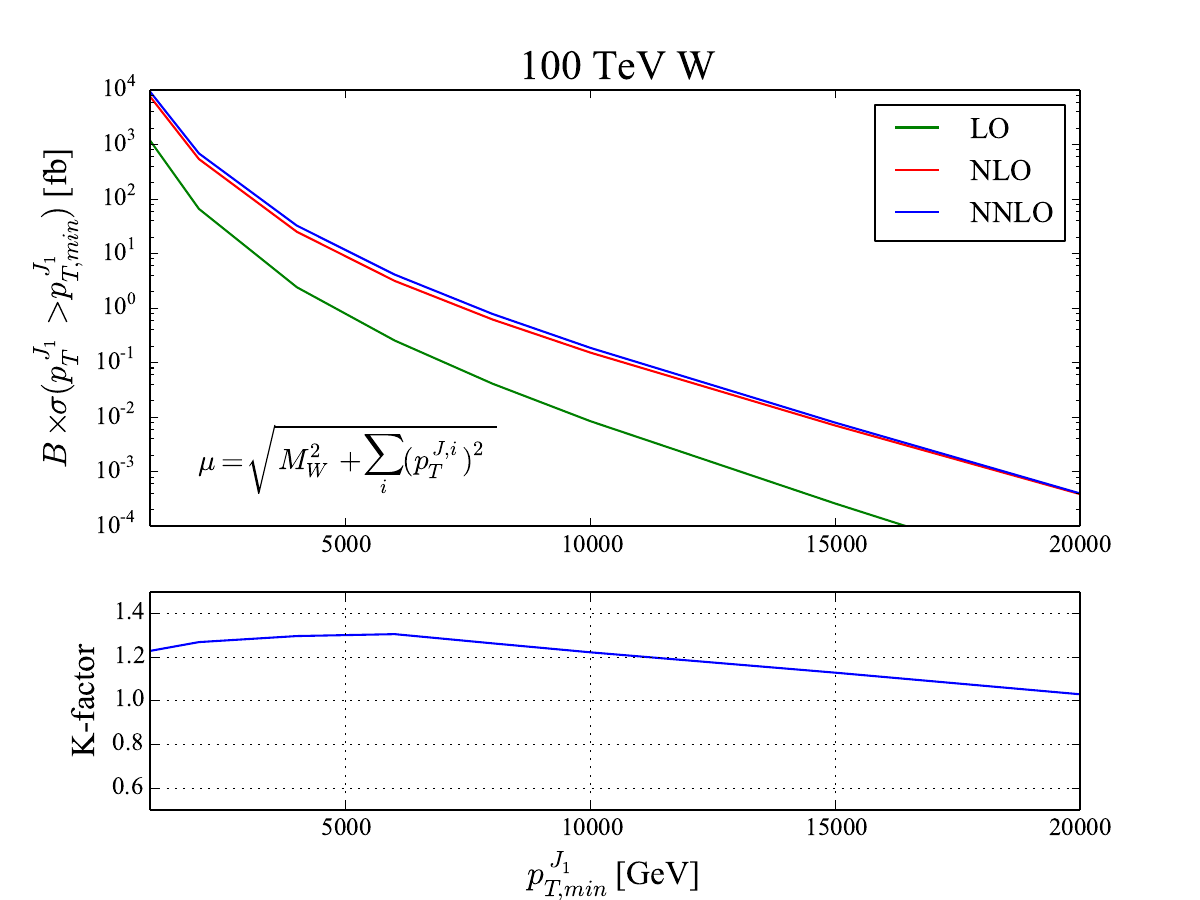}
\caption{$BR(W\to e\nu) \times \sigma(p_T^{X}>p_{T,min}^X)$ at NNLO
  and NLO, with $X=W$ (left) and $X=j_1$ (right)
  is the leading jet in $W+jet$ inclusive events. The lowest
  $p_{T,min}^{J_1}$ entry in the right plot corresponds to
  $p_{T,min}^{J_1}=1$~TeV. Lower
insets: the NNLO/NLO $K$ factors. }
\label{fig:w_largept_nnlo}
\end{figure}

\subsection{Production of gauge bosons at the highest energies}
\label{sec:dyan_highenergy}
For processes involving gauge bosons and jets at such large energies,
a very interesting new phenomenon emerges, namely the growth of the
gauge boson emission probability from high-$p_T$ jets. If we ask what
is the most likely mechanism to produce gauge bosons in final states
with at least one multi-TeV jet, it turns out that this is not the LO
QCD process where the gauge boson simply recoils against the jet, but
the higher-order process where it is a second jet that absorbs the
leading jet recoil, and the gauge boson is radiated off some of the
quarks~\cite{Rubin:2010xp}.  In other words, the parton-level
scattering $q q \to qq V$ dominates over $qg \to qV$ (for simplicity,
we do not show explicitly the possibly different quark flavour types
involved in the processes). The emission probability of gauge bosons
in this case is enhanced by large logarithms of $p_{T,jet}/m_V$, and
can reach values in the range of 10\% and more, as shown in
Fig.~\ref{fig:multiWrates}. This gives the emission probability for
one or more $W$ bosons in events in which there is at least one jet
above a given $p_T$ threshold. The kinematical properties of these
events are illustrated for various distributions in
Figs.~\ref{fig:WiniHighEtjet_1} (at LO) and ~\ref{fig:WiniHighEtjet_2}
(at (N)NLO). To highlight the kinematical evolution with jet $p_T$ we
show results for final states with a jet above 1~TeV, and above
10~TeV. In the case of largest $p_T$, we see the dominance of events
in which the two jets balance each other in transverse momentum, while
the $W$ carries a very small fraction of the leading jet momentum. One
third of the $W$'s are emitted within $\Delta R<1$ from the subleading
jet, with a large tail of emission at larger angles, due in part to
$W$ radiation from the initial state.

\begin{figure}[h!]
\centering
\includegraphics[width=0.7\textwidth]{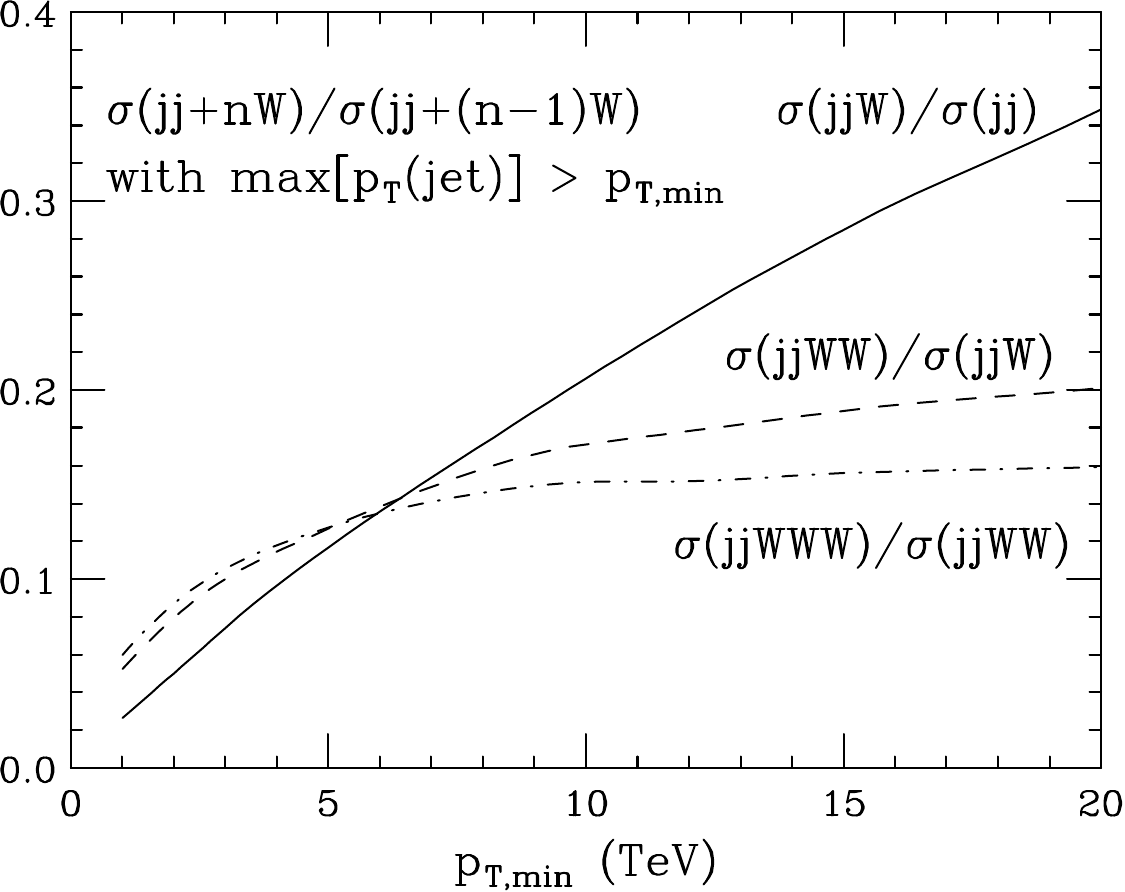}
\caption{Emission
probability for additional $W$ bosons in dijet events at large $p_T$. }
\label{fig:multiWrates}
\end{figure}

\begin{figure}[h!]
\centering
\includegraphics[width=0.9\textwidth]{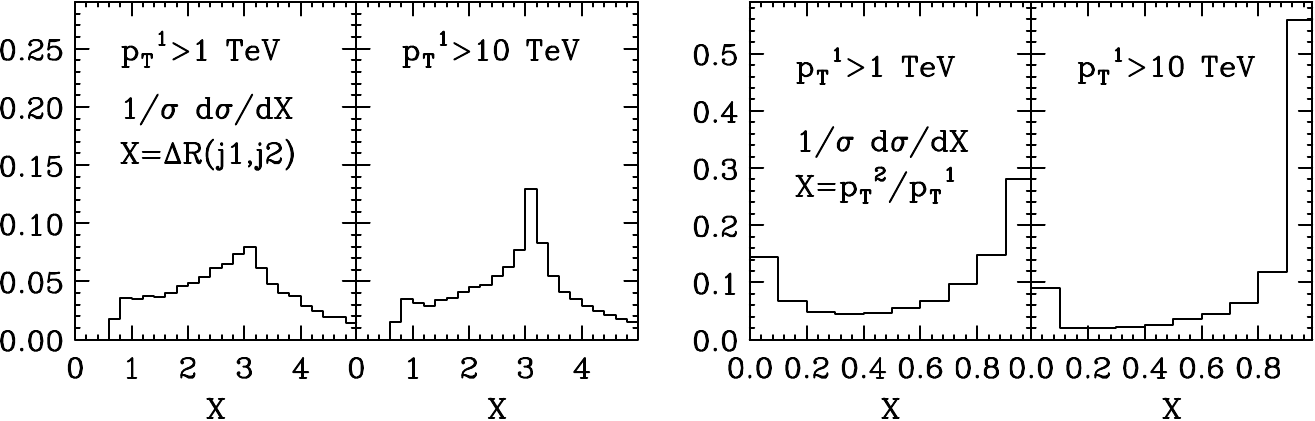}
\caption{Kinematical correlations, at LO, in high-$p_T$ jet events with $W$
  radiation, for values of the leading jet $p_T>1$ and 10~TeV.}
\label{fig:WiniHighEtjet_1}
\end{figure}

\begin{figure}[h!]
\centering
\includegraphics[width=0.49\textwidth]{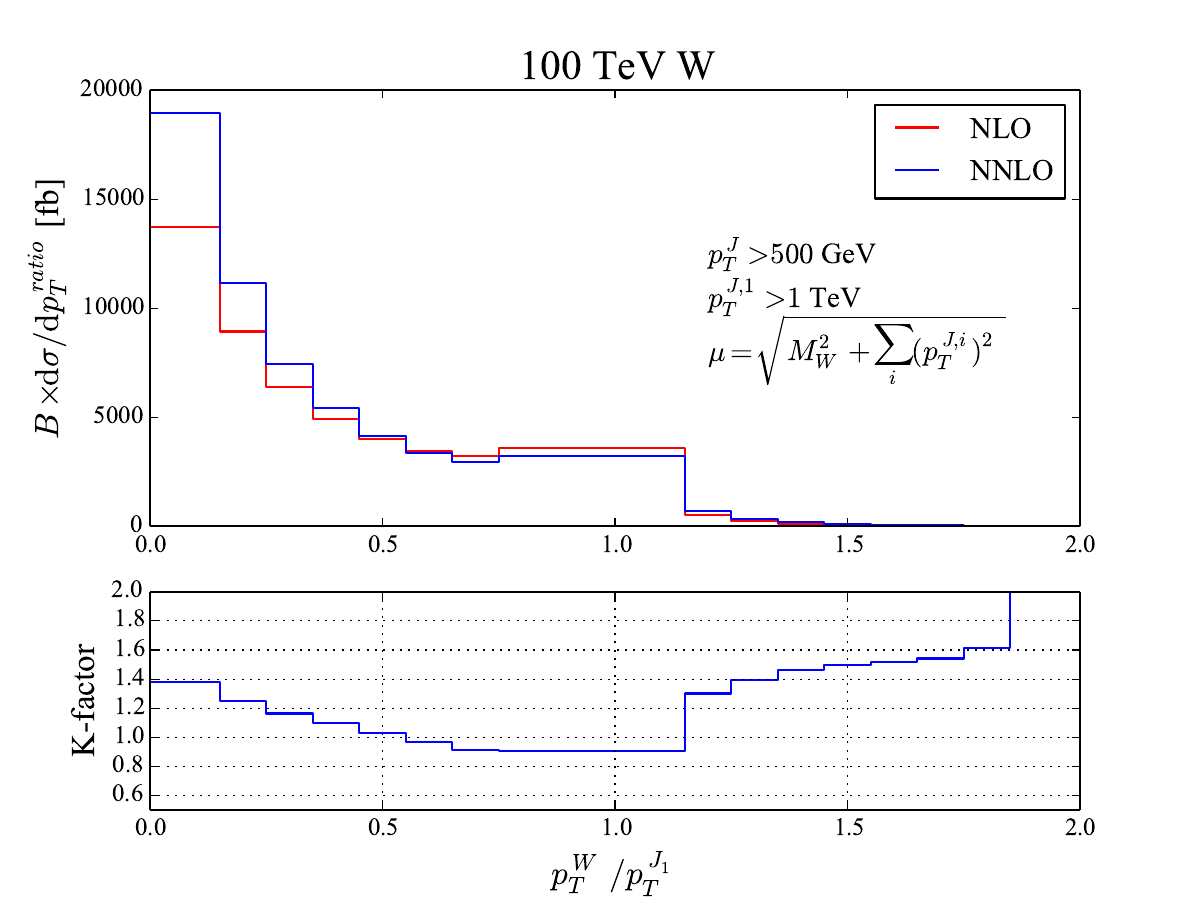}
\includegraphics[width=0.49\textwidth]{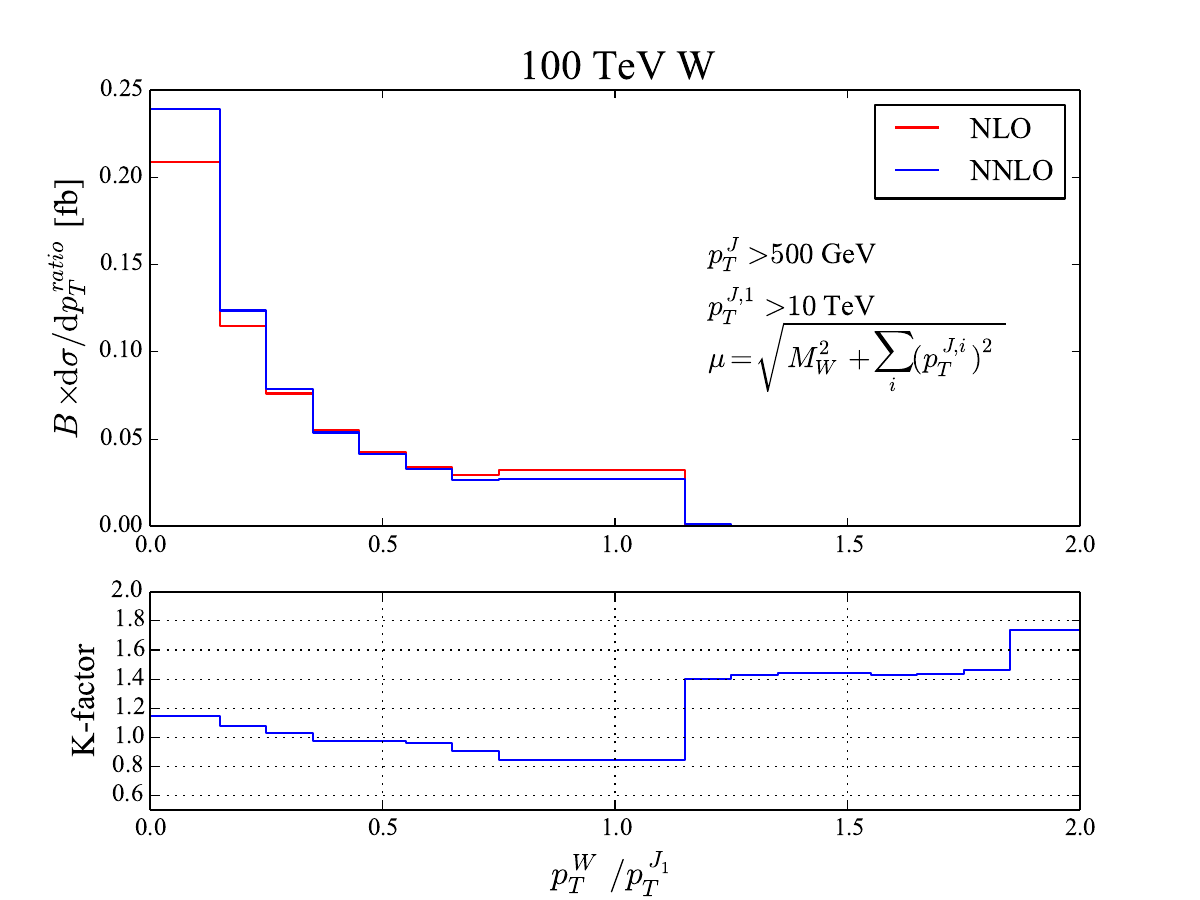}
\\
\includegraphics[width=0.49\textwidth]{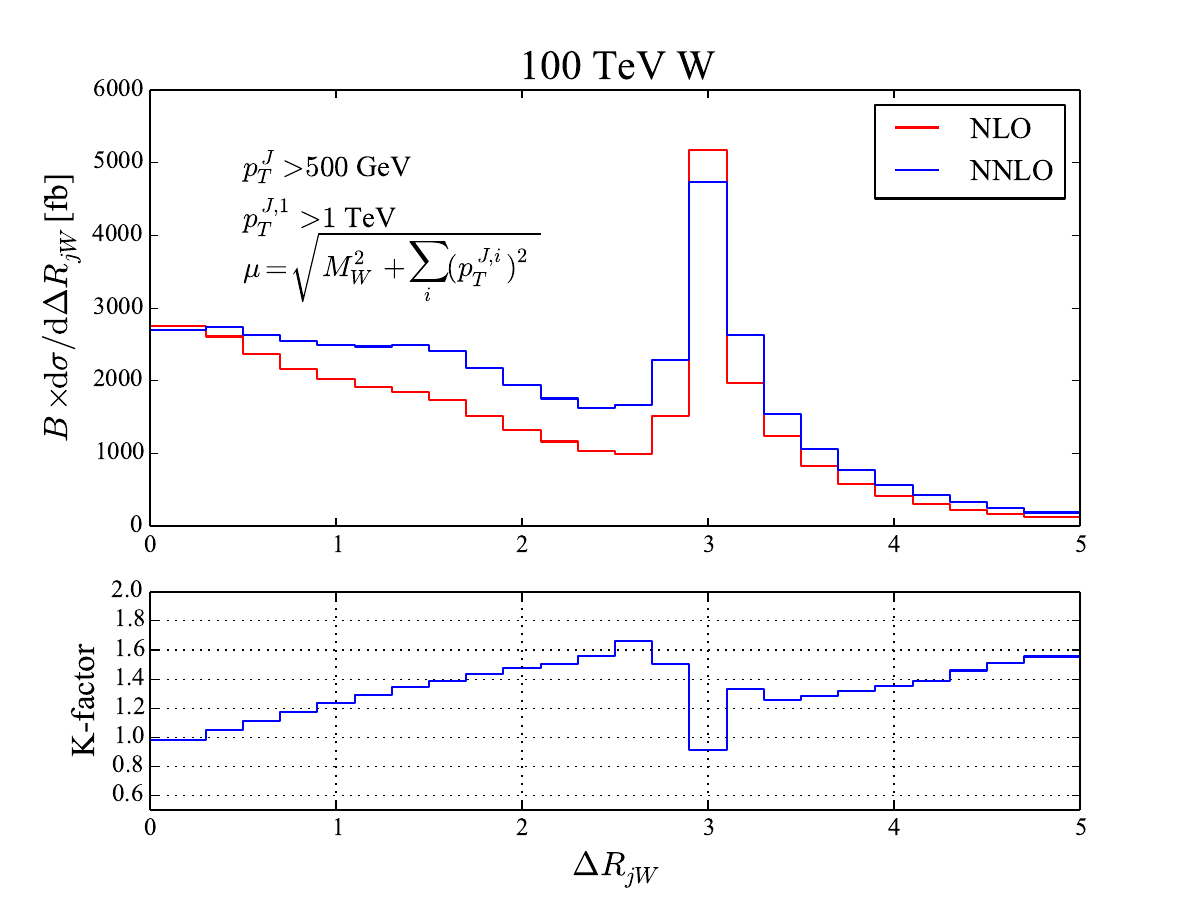}
\includegraphics[width=0.49\textwidth]{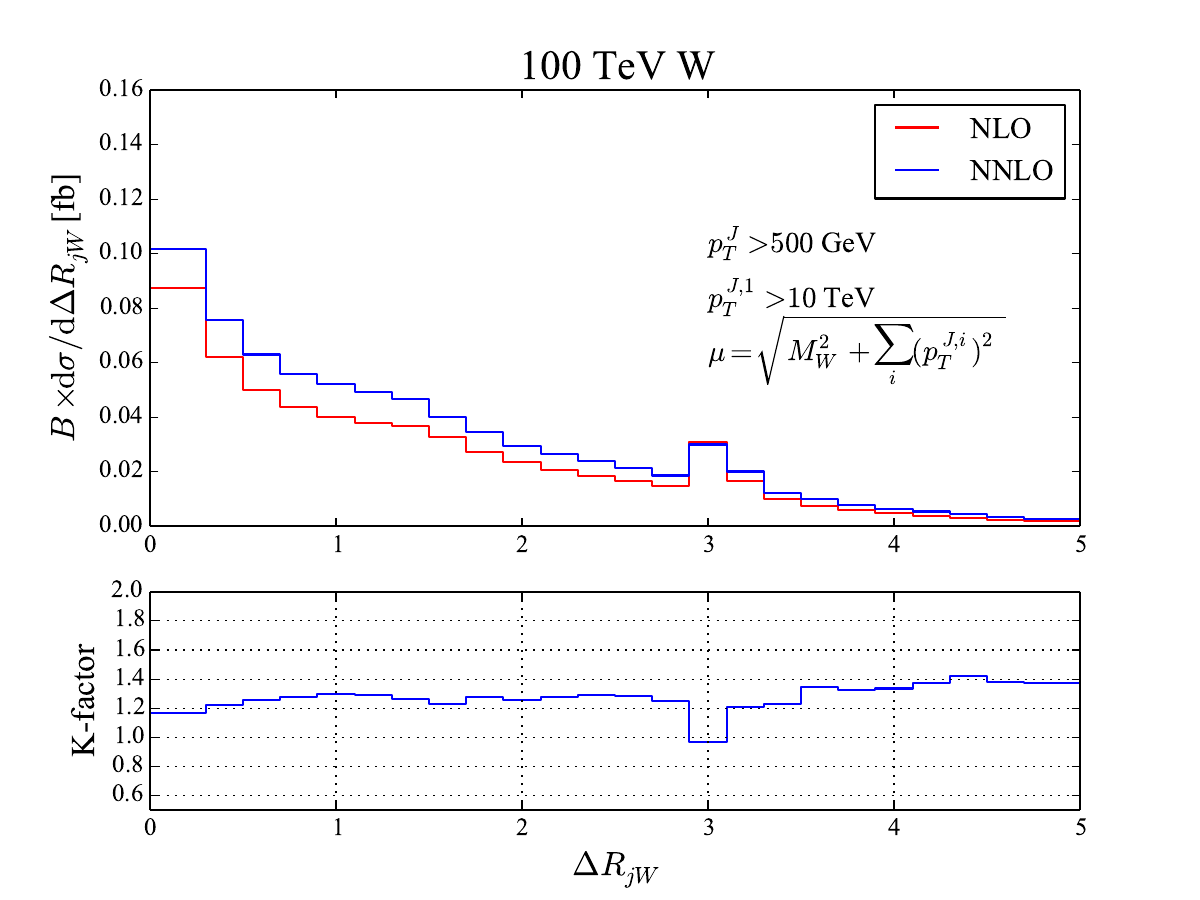}
\caption{Kinematical correlations at (N)NLO in $W+$jet(s) events,
  for values of the leading jet $p_T>1$ and 10~TeV.}
\label{fig:WiniHighEtjet_2}
\end{figure}

The process considered above is just one manifestation of the general
fact that, in hard electroweak interactions at multi-TeV energies, the
soft/collinear structure of almost {\it any} multi-TeV process can
become significantly altered, as the logarithmic enhancements familiar
from QED and QCD will become active for electroweak emissions (see,
e.g.,~\cite{Moretti:2006ea,Bell:2010gi,Dittmaier:2012kx,Christiansen:2014kba,Hook:2014rka}).
Obtaining correct descriptions of the complete event structure when
$\sqrt{\hat{S}} \gg m_W$ can be then greatly facilitated by
incorporating factorization and resummation, such as that provided by
parton showering and parton distribution functions.  In effect, we
will begin to see weak bosons (including the Higgs boson) behaving as
nearly-massless partons, in stark contrast to the conventional
perspective in which they are viewed as ``heavy'' particles.  Jets,
whether initiated by QCD processes, electroweak process, or new
physics processes, will be found to contain electroweak splittings
with probabilities at the $O(10\%)$ level.  Similarly, weak bosons can
usefully be thought of as collinear components of the protons, at the
same level as gluons and photons.

To develop some intuition of the collinear splitting behavior of
electroweak ``partons,'' it is useful to first consider a conceptual
limit with an unbroken $SU(2) \times U(1)$ gauge symmetry with
massless gauge bosons and fermions, supplemented by a massless scalar
doublet field $\phi$ without a VEV (the would-be Higgs doublet).  In
this limit, many processes are direct analogs of those in QED and QCD.
Fermions with appropriate quantum numbers may emit (transverse)
$SU(2)$ and $U(1)$ gauge bosons with both soft and collinear
enhancements.  The $SU(2)$ bosons couple to one another via their
non-abelian gauge interactions, and undergo soft/collinear splittings
of the schematic form $W\to WW$, similar to $g\to gg$.  All of the
electroweak gauge bosons may also undergo collinear-enhanced
splittings into fermion pairs, similar to $g\to q\bar q$ or $\gamma
\to f \bar f$.  Beyond these, the major novelty is the introduction of
the scalar degrees of freedom.  First, the scalars may themselves
radiate $SU(2)$ and $U(1)$ gauge bosons, with soft/collinear limits
identical to their counterparts with fermionic sources.  Second, the
electroweak gauge bosons can split into a pair of scalars, again in
close analog with splittings to fermion pairs.  Third, fermions with
appreciable Yukawa couplings to the scalar doublet can emit a scalar
and undergo a chirality flip.  Finally, the scalars can split into
collinear fermion pairs.

In the realistic case of spontaneously-broken symmetry, several
important changes take place.  Primarily, all of the soft and
collinear divergences associated with the above splittings become
physically regulated, effectively shutting off at $p_T \lsim m_W$ (or
$m_h$, $m_t$ where appropriate).  Roughly speaking, $m_W$ plays a role
similar to $\Lambda_{\rm QCD}$ in the QCD parton shower, albeit with
far less ambiguity of the detailed IR structure since this regulation
occurs at weak coupling.  Another major difference is the mixing of
the scalar doublet's Goldstone degrees of freedom into the $W$ and $Z$
gauge bosons, allowing for the appearance of longitudinal modes.  In
many cases, the longitudinal gauge bosons behave identically to the
original scalars, as dictated by the Goldstone equivalence
theorem~\cite{Lee:1977eg,Chanowitz:1985hj}.  For example the splitting
$W_T^+ \to W_{L}^+ Z_{L}$ is, up to finite mass effects, an exact
analog of $W_T^+ \to \phi^+ {\rm Im}(\phi^0)$ in the unbroken theory.
Similarly for longitudinal gauge boson emissions from heavy fermions,
such as the equivalence between $t_L \to Z_{L} t_R$ and $t_L \to {\rm
  Im}(\phi^0) t_R$.

But important exceptional cases now also occur for emissions near $p_T
\sim m_W$.  Most well known, even a massless fermion exhibits a kind
of soft/collinear-enhanced emission of $W_{L}$ and
$Z_{L}$~\cite{Kane:1984bb,Dawson:1984gx}.  These emissions have no
Goldstone equivalent analog, and are highly power-suppressed for $p_T
\gsim m_W$.  But the overall population of emissions at the boundary
between ``broken'' and ``unbroken'' behavior nonetheless grows
logarithmically with the fermion energy.  This is formally
sub-dominant to the double-logarithmic growth of transverse emissions,
but remains numerically important at multi-TeV energy scales.
Emissions from massless quarks also cause the energetic initial-state
protons to act as sources of longitudinal boson beams, allowing for
studies of the high-energy interactions of the effective Goldstone
bosons through weak boson scattering (discussed further below).
Similar types of emissions occur in the splittings of transverse
bosons, such as $W_T^+ \to Z_{L} W_T^+ \, / \, Z_T W_{L}^+$.

\begin{table}
\begin{center}
\begin{tabular}{ l | c | c | c }
Process \ &  ${\mathcal P}(p_T)$  & \ ${\mathcal P}(1~{\rm TeV})$ \  & \ ${\mathcal P}(10~{\rm TeV})$ \  \\   \hline
$f \to V_Tf$ \ &  $(3\times10^{-3})\left[\log\frac{p_T}{m_{\rm EW}} \right]^2$ &  1.7\% &  7\%    \\
$f \to V_{L}f$ \ &  $ (2\times10^{-3})\log\frac{p_T}{m_{\rm EW}}$ &  0.5\% &  1\%   \\
$V_T \to V_T V_T$ \ & $(0.01)\left[\log\frac{p_T}{m_{\rm EW}} \right]^2$  &  6\% &  22\%   \\
$V_T \to V_{L}V_T$ \ & $(0.01)\log\frac{p_T}{m_{\rm EW}}$  &  2\% &  5\%   \\
$V_T \to f\bar f$ \ & $(0.02)\log\frac{p_T}{m_{\rm EW}}$  &  5\% &  10\%   \\
$V_T \to V_{L} h$ \ & $(4\times10^{-4})\log\frac{p_T}{m_{\rm EW}}$  &  0.1\% &  0.2\%   \\
$V_{L} \to V_T h$ \ & $(2\times10^{-3})\left[\log\frac{p_T}{m_{\rm EW}} \right]^2$  &  1\% &  4\%   \\
\end{tabular}
\end{center}
\caption{An illustrative set of approximate total electroweak
  splitting rates in final-state showers \cite{EWshower}.}
\label{table:splittingRates}
\end{table}

Table~\ref{table:splittingRates} provides a few estimates for total
splitting rates of individual final-state particles, including
approximate numerical values for particles produced at $p_T = 1$~TeV
and 10~TeV.  The $SU(2)$ self-interactions amongst transverse gauge
bosons tend to give the largest rates, quickly exceeding 10\% as the
energy is raised above 1~TeV (these rates are slightly lower than
those extracted from Fig.~\ref{fig:multiWrates}, since there an
important contribution to $W$ emission came from initial state
radiation).  This has significant impact on processes with prompt
transverse boson production such as $W/Z/\gamma$+jets, and especially
on multiboson production including transverse boson scattering.
Generally, it is important to appreciate that {\it any} particle in an
event, whether initial-state or final-state, or even itself produced
inside of a parton shower, can act as a potential electroweak
radiator.  Consequently, the total rate for finding one or more
electroweak splittings within a given event must be compounded, and
can sometimes add up to $O(1)$.

\clearpage
\def\Wpj{$W^+\,\!+\,1$}
\def\Wmj{$W^-\,\!+\,1$}
\def\Wpjj{$W^+\,\!+\,2$}
\def\Wpjjj{$W^+\,\!+\,3$}
\def\Wmjjj{$W^-\,\!+\,3$}
\def\Wpjjjj{$W^+\,\!+\,4$}
\def\Wmjjjj{$W^-\,\!+\,4$}
\def\Vjj{$V\,\!+\,2$}
\def\Vjjjj{$V\,\!+\,4$}
\def\Wpjx{$W^+\,\!+\,1,2$}
\def\Wpjjx{$W^+\,\!+\,2,3$}
\def\Wpjjja{$W^+,\!+\,1,2,3$}
\def\Wpjjjx{$W^+\,\!+\,3,4$}
\def\Wpjn{$W^+\!+\,n$}
\def\Wmjn{$W^-\!+\,n$}
\def\Wmjnm{$W^-\!+\,(n-1)$}
\def\Zjn{$Z\!+\,n$}
\def\Vjn{$V\!+\,n$}
\def\draftnote#1{{\it #1}}
\def\SI#1#2{#1\,\rm{#2}}
\def\barn{\rm{b}}
\def\micro{\ensuremath{\mu}}

\def\tableHeaderVPhantom{$\vphantom{\displaystyle{\frac{2^n_A}{2^n_A}}}$}
\def\minptcellNoLine#1{\multicolumn{1}{c}{\SI{#1}{GeV}\tableHeaderVPhantom}}
\def\minptcell#1{\multicolumn{1}{c|}{\SI{#1}{GeV}\tableHeaderVPhantom}}
\def\minptcellDoubleline#1{\multicolumn{1}{c||}{\SI{#1}{GeV}}}
\def\processcell#1{\multicolumn{1}{c|}{#1\tableHeaderVPhantom}}
\def\processcellNoLine#1{\multicolumn{1}{c}{#1\tableHeaderVPhantom}}
\def\mub#1{#1\,\micro b\xspace}
\def\nb#1{#1\,nb\xspace}
\def\pb#1{#1\,pb\xspace}
\def\fb#1{#1\,fb\xspace}
\def\ab#1{#1\,ab\xspace}
\def\belowfb{\fb{$< 1$}}
\def\belowab{\ab{$< 1$}}
\def\crossedOutMulticolumn#1{\multicolumn{#1}{c|}{---}}
\def\crossedOutMulticolumnNoLine#1{\multicolumn{#1}{c}{---}}
\renewcommand{\arraystretch}{1.2}
\def\HqT{H\protect\scalebox{0.8}{qT}\xspace}
\def\POWHEG{P\protect\scalebox{0.8}{OWHEG}\xspace}
\def\MINLO{MiN\protect\scalebox{0.8}{LO}\xspace}
\def\SMCatNLO{S--M\protect\scalebox{0.8}{C}@N\protect\scalebox{0.8}{LO}\xspace}
\def\MCatNLO{M\protect\scalebox{0.8}{C}@N\protect\scalebox{0.8}{LO}\xspace}
\def\MCatLO{M\protect\scalebox{0.8}{C}@L\protect\scalebox{0.8}{O}\xspace}
\def\LOPS{LO$\otimes$PS\xspace}
\def\MEPS{M\scalebox{0.8}{E}P\scalebox{0.8}{S}@LO\xspace}
\def\NLOPS{N\scalebox{0.8}{LO}P\scalebox{0.8}{S}\xspace}
\def\MENLOPS{ME\protect\scalebox{0.8}{NLO}PS\xspace}
\def\MEPSatNLO{M\protect\scalebox{0.8}{E}P\protect\scalebox{0.8}{S}@N\protect\scalebox{0.8{LO}\xspace}} 
\def\CKKW{CKKW\xspace}
\def\UNLOPS{UN\scalebox{0.8}{LO}P\scalebox{0.8}{S}\xspace}
\def\Vegas{V\protect\scalebox{0.8}{EGAS}\xspace}
\def\Herwig{H\protect\scalebox{0.8}{ERWIG}\xspace}
\def\Herwigpp{H\protect\scalebox{0.8}{ERWIG++}\xspace}
\def\Ariadne{A\protect\scalebox{0.8}{RIADNE}\xspace}
\def\Alpgen{A\protect\scalebox{0.8}{LPGEN}\xspace}
\def\Madgraph{M\protect\scalebox{0.8}{AD}G\protect\scalebox{0.8}{RAPH}\xspace}
\def\Madevent{M\protect\scalebox{0.8}{AD}E\protect\scalebox{0.8}{VENT}\xspace}
\def\Blackhat{B\protect\scalebox{0.8}{LACK}H\protect\scalebox{0.8}{AT}\xspace}
\def\Pythia{P\protect\scalebox{0.8}{YTHIA}\xspace}
\def\Jetset{J\protect\scalebox{0.8}{ETSET}\xspace}
\def\Vincia{V\protect\scalebox{0.8}{INCIA}\xspace}
\def\aMCatNLO{aM\protect\scalebox{0.8}{C}@N\protect\scalebox{0.8}{LO}\xspace}
\def\PowhegBox{P\protect\scalebox{0.8}{OWHEG}B\protect\scalebox{0.8}{OX}\xspace}
\def\FeynRules{F\protect\scalebox{0.8}{EYN}R\protect\scalebox{0.8}{ULES}\xspace}
\def\SpringBases{S\protect\scalebox{0.8}{PRING/}B\protect\scalebox{0.8}{ASES}\xspace}
\def\Professor{P\protect\scalebox{0.8}{ROFESSOR}\xspace}
\def\Rivet{R\protect\scalebox{0.8}{IVET}\xspace}
\def\Jetphox{J\protect\scalebox{0.8}{ET}P\protect\scalebox{0.8}{HOX}\xspace}
\def\Diphox{D\protect\scalebox{0.8}{I}P\protect\scalebox{0.8}{HOX}\xspace}
\def\Resbos{R\protect\scalebox{0.8}{ES}B\protect\scalebox{0.8}{OS}\xspace}
\def\BlackHat{B\protect\scalebox{0.8}{LACK}H\protect\scalebox{0.8}{AT}\xspace}
\def\GoSam{G\protect\scalebox{0.8}{O}S\protect\scalebox{0.8}{AM}\xspace}
\def\Golem{G\protect\scalebox{0.8}{OLEM}95\xspace}
\def\Samurai{S\protect\scalebox{0.8}{AMURAI}\xspace}
\def\MCFM{M\protect\scalebox{0.8}{CFM}\xspace}
\def\OpenLoops{O\protect\scalebox{0.8}{PEN}L\protect\scalebox{0.8}{OOPS}\xspace}
\def\SherpaOpenLoops{S\protect\scalebox{0.8}{HERPA}+O\protect\scalebox{0.8}{PEN}L\protect\scalebox{0.8}{OOPS}\xspace}
\def\Collier{C\protect\scalebox{0.8}{OLLIER}\xspace}
\def\VBFNLO{V\protect\scalebox{0.8}{BF}N\protect\scalebox{0.8}{LO}\xspace}
\def\FastJet{F\protect\scalebox{0.8}{ASTJET}\xspace}
\def\Sherpa{S\protect\scalebox{0.8}{HERPA}\xspace}
\def\sherpa{\Sherpa}
\def\Comix{C\protect\scalebox{0.8}{OMIX}\xspace}
\def\Apacic{A\protect\scalebox{0.8}{PACIC++}\xspace}
\def\Amegic{A\protect\scalebox{0.8}{MEGIC++}\xspace}
\def\CSS{C\protect\scalebox{0.8}{SS}\xspace}
\def\Ahadic{A\protect\scalebox{0.8}{HADIC++}\xspace}
\def\Adicic{A\protect\scalebox{0.8}{DICIC++}\xspace}
\def\Hadrons{H\protect\scalebox{0.8}{ADRONS++}\xspace}
\def\Photons{P\protect\scalebox{0.8}{HOTONS++}\xspace}
\def\LEP{LEP\xspace}
\def\LHC{LHC\xspace}
\def\Tevatron{Tevatron\xspace}
\def\Hera{HERA\xspace}
\def\Aleph{ALEPH\xspace}
\def\Delphi{DELPHI\xspace}
\def\Opal{OPAL\xspace}
\def\DO{D\O\ }
\def\CDF{CDF\xspace}
\def\ATLAS{ATLAS\xspace}
\def\CMS{CMS\xspace}
\def\RunII{Run II\xspace}
\def\SecRef#1{Section~\ref{#1}}
\def\EqRef#1{Equation~(\ref{#1})}
\def\EqRefs#1#2{Equations~(\ref{#1})-~(\ref{#2})}
\def\TabRef#1{Table~\ref{#1}}
\def\FigRef#1{Figure~\ref{#1}}
\def\FigRefs#1#2{Figuress~\ref{#1}-~\ref{#2}}
\def\abs#1{\left| #1\right|}
\def\rbr#1{\left( #1\right)}
\def\abr#1{\langle #1\rangle}
\def\cbr#1{\left\{ #1\right\}}
\def\sbr#1{\left[ #1\right]}
\def\im{\imath}
\def\jm{\jmath}
\def\args#1{\{\vec{#1}\}}
\def\argc#1#2{\{\vec{#1}\}_{\rm #2}}
\def\ptmin{p_{\perp,{\rm min}}}
\def\done{{\rm d}}
\def\order{\mathcal{O}}
\def\mc#1{\mathcal{#1}}
\def\mr#1{\mathrm{#1}}
\def\mb#1{\mathbb{#1}}
\def\dst{\displaystyle}
\def\sst{\scriptstyle}
\def\bmap#1#2#3{b_{#1,#2}(#3)}
\def\rmap#1#2#3{r_{\widetilde{#1},\tilde{#2}}(#3)}
\def\sbmap#1#2#3#4{b^{\rm(#1)}_{#2,#3}(#4)}
\def\srmap#1#2#3#4{r^{\rm(#1)}_{\widetilde{#2},\tilde{#3}}(#4)}
\def\nnb{\nonumber}
\def\bea{\begin{eqnarray}}
\def\eea{\end{eqnarray}}
\def\bi{\begin{itemize}}
\def\ei{\end{itemize}}
\def\hl{\vphantom{$\int_A^B$}}
\def\Hl{\vphantom{$\int\limits_a^b$}}
\def\reserrs#1#2#3{$#1_{-#2}^{+#3}$ pb}

\def\trule{\rule[-1.5mm]{0mm}{6mm}}

\def\todo#1{\textbf{\color{red}#1}}
\def\changed#1{\textbf{\color{blue}#1}}

\makeatletter
\def\thefontsize#1{{#1 The current font size is: \f@size pt\par}}
\makeatother

\section{$V$+jets\footnote{Editors: F.~Febres~Cordero and F.~Krauss}}
\label{ref:Vnjets}

In this section we study the associated production of a weak vector
boson and jets at a proton-proton collider with $\sqrt{s}=100$~TeV and
an expected accumulated total integrated luminosity of several
$ab^{-1}$.  Such a collider will allow to explore extreme kinematical
configurations for processes like $V$+many jets ($V=W^\pm,Z)$, giving
yet newer ways to test the Standard Model of particle physics at
scales significantly above the TeV scale.  Even more, many new physics
scenarios predict enhancements in the production of vector bosons and
jets, and so a clear understanding of SM model predictions is
important.

We present here general properties of total and differential cross
sections in order to obtain a first characterisation of the collision
environment.  Two broad classes of kinematical cuts are explored,
called `democratic' and `hierarchical' below.  The `democratic' cuts
are characterized by imposing a single minimum jet $p_{\rm T}$ cut on
all jets, while 'hierarchical' cuts impose a very large $p_{\rm T}$
cut on the hardest jet in the event and keep a softer cut for all
other jets.  These choices are known to affect the perturbative
behaviour of QCD, and we explore it now in this new high-energy
environment.  In particular we will be interested in regions of phase
space where the various cuts imply large scale ratios and thereby
induce correspondingly large logarithms.

Because uncertainties largely cancel in ratios of observables, we
devote our attention to scaling properties of jet ratios -- for
example the behaviour of cross sections in dependence on jet
multiplicities, and ratios between different types of vector bosons.
We also explore a number of differential cross sections, such as
integrated $p_{\rm T}$ spectra.  Finally, we record
parton-distribution function uncertainties on the processes' inclusive
cross sections.

The predictions are obtained employing a number of current methods.
These include primarily as fixed-order predictions at leading order
(LO) and next-to-leading order (NLO) in QCD, but for some sensitive
observables we also establish the impact of parton-shower effects.

\subsection{Setup}

In our discussion of $V$+jets results we consider only vector bosons
decaying to leptons of the first generation. Thus the final-state
signatures include electrons, electron neutrinos and jets. For
$Z$-bosons the decay products are explicitly specified being either
pairs of electrons or neutrinos, mimicking the missing signature,
while $W$ bosons decay to $e\nu_e$ pairs.  We consider in detail five
distinct phase-space regions for these processes, which are defined by
`basic', `low-democratic', `high-democratic', `low-hierarchical' and
`high-hierarchical' sets of cuts, given in eqs.~(\ref{eqn:Vjets-cuts})
and in Table~\ref{tab_ntps_Vjets}.  The `basic' cuts treat all jets on
equal terms with a minimum jet transverse momentum $p_T^{\rm min}$
that it is varied between $50$~GeV and $1000$~GeV.  The labels
`low'/`high' refer to the low and high transverse momentum ($p_T$)
cuts on all final state objects, whereas the labels
`democratic'/`hierarchical' refer to a uniform $p_T$ cut on all jets
or requiring a distinguished jet with high $p_T$.  For simplicity
identical $p_T$ cuts are applied to charged leptons and missing
neutrinos, which are measured as missing energy.  We denote the
transverse energy of the jets by $p_T^{\rm jet}$ and $p_T^{\rm
  lead-jet}$ for the jet with the largest transverse momentum
$p_T$. The transverse momentum cut of the charged leptons and single
neutrinos (missing energy) will be uniformly denoted by $p_T^{\rm
  lepton}:=p_T^{e}=p_T^{\nu}$.  In general, jets are reconstructed
with the anti-$k_T$ algorithm with a radius parameter of $R=0.4$,
using the \FastJet package~\cite{Cacciari:2008gp,Cacciari:2011ma}.
%
\begin{table}[h]
\begin{center}
\small
\setlength\tabcolsep{2.pt}
\begin{tabular}{ccccc}
\hline
 \multicolumn{5}{c}{ Phase-space cuts for $pp\rightarrow Z/W+jets+X$}  \\
 \hline
 basic  & low-democratic & high-democratic  & low-hierarchical &
 high-hierarchical \\ \hline\hline
 $p_T^{\rm jet}>p_{T}^{\rm min}$ &
 $p_T^{\rm jet}>50$~GeV &
 $p_T^{\rm jet}>500$~GeV &
 $p_T^{\rm jet}>50$~GeV &
 $p_T^{\rm jet}>500$~GeV \\
 --- &
 $p_T^{\rm lead-jet}>10^2$~GeV &
 $p_T^{\rm lead-jet}>10^3$~GeV &
 $p_T^{\rm lead-jet}>2\cdot 10^3$~GeV &
 $p_T^{\rm lead-jet}>10^4$~GeV \\\hline
 $p_T^{\rm lepton}>30$~GeV &
 $p_T^{\rm lepton}>30$~GeV &
 $p_T^{\rm lepton}>50$~GeV &
 $p_T^{\rm lepton}>30$~GeV &
 $p_T^{\rm lepton}>50$~GeV \\
\hline \end{tabular}
\caption{ The five phase-space regions considered. For the `basic' set
  of cuts $p_{T}^{\rm min}$ will be varied from 50 GeV to 1
  TeV. \label{tab_ntps_Vjets} }
\end{center} \end{table}
%

The following cuts on jet- and lepton-pseudo rapidities $\eta^{{\rm
    jet},e}$ and on $Z$ mass ($M_{ee}$) and $W$ transverse mass
($M_T^W$) are common to all five kinematical regions:
\begin{eqnarray} 
\mbox{rapidity cuts:}&& |\eta^{\rm jet}|<5\,,\quad
|\eta^e|<4\,\nonumber\\ 
\mbox{W-bosons:} && M_{\rm T}^W>40~GeV\label{eqn:Vjets-cuts}\\ 
\mbox{Z-bosons:} && 
Z\rightarrow e^+e^-:\quad 66~GeV < M_{ee} < 116~GeV \nonumber\ ,\\
&&Z\rightarrow \nu_e\bar\nu_e:\quad E_{\rm T, miss}> 100~GeV \nonumber\ ,
\end{eqnarray}
where the missing transverse energy $E_{\rm miss}$ is given by the sum
of all transverse (anti-)neutrino momenta $E_{\rm T,miss}=|\,\vec
p_T^{\,\nu}+\vec p_T^{\,\bar\nu}\,|$.

\subsubsection{Computational setup}
\label{sec:Vjets-setup}

For the fixed-order results at leading order (LO), the \Sherpa
framework~\cite{Gleisberg:2003xi,Gleisberg:2008ta} has been used, in
particular the \Comix matrix-element
generator~\cite{Gleisberg:2008fv}.  For calculations at
next-to-leading order (NLO) accuracy in the strong-coupling expansion,
the combination of the \Blackhat \cite{Berger:2008sj} and \Sherpa
packages are used.  The virtual matrix elements are provided by the
\Blackhat library. For $V$+4-jet production we have employed a
leading-color approximation of the one-loop matrix elements \footnote{
  For details of the calculation, the reader is referred to the
  corresponding articles for the cases of
  $W$+jets~\cite{Berger:2009zg,Berger:2009ep,Berger:2010zx,Bern:2013gka}
  and $Z$+jets~\cite{Ita:2011wn,Berger:2010vm}.}. The remaining
Born-level, real radiation corrections as well as integration
framework is provided by \Sherpa.  Infrared subtraction is
consistently treated by the Catani-Seymour
method~\cite{Catani:1996jh}, automated in \Sherpa~\cite{Gleisberg:2007md}. \\
For parton-level results including parton-shower effects the multi-jet
merging technology of \cite{Catani:2001cc,Hoeche:2009rj} is used, with
the parton shower built on Catani-Seymour subtraction kernels as
proposed in ref.~\cite{Nagy:2005aa} and implemented in
ref.~\cite{Schumann:2007mg }.  Higher-order accuracy is included based
on the \MCatNLO method~\cite{Frixione:2002ik} in the version
implemented in \Sherpa~\cite{Hoche:2010kg,Hoeche:2011fd} and the
multi-jet merging at NLO described in
refs.~\cite{Gehrmann:2012yg,Hoeche:2012yf} are employed. The zero-jet
inclusive cross section is obtained in NLO accuracy with the
higher-jet multiplicities being leading order in strong-coupling
expansion.  All calculations employ the CT14nlo parton-distribution
functions (PDF) for NLO results, and CT14llo for the reference LO
results.  The PDFs are accessed through the LHAPDF
interface~\cite{Buckley:2014ana}. The PDFs also provides the strong
coupling $\alpha_S(\mu)$ throughout.

In the fixed-order calculations, the renormalisation scale ($\mu_{R}$)
and the factorisation scale ($\mu_{F}$) are chosen identical and
defined through,
\begin{eqnarray} \label{Eq::barHT_VBPlusJets} 
    \mu_{R}\,=\mu_{F}=\,\bar{H}_T\,:=\, E_{\rm
    T}^V+\frac{1}{N_J}\,\sum\limits_{j=1}^{N_J}\,p^{\rm jet}_{T,j}\,,
\end{eqnarray} 
where, $E_{\rm T}^V=\sqrt{m_V^2+p_{T,V}^2}$ and $N_{J}$ is the overall
number of jets in the process.  Transverse momenta of the jets are
denoted by $p^{\rm jet}_{T,j}$.  For the fixed order LO and NLO QCD
results we employ the total partonic transverse energy,
\begin{equation}  \label{eqn:NLO-scale}
\mu_{R}\,= \mu_{F}\,= \,\hat{H}^\prime_{\rm T} \,:=\,
E_{\rm T}^V+\sum_{i=1}^{N_{\rm P}} p^{\rm parton}_{T,i}\,, 
\end{equation}
where $N_{\rm P}$ denotes the total number of final-state partons. The
parton momenta are denoted by $p^{\rm parton}_{T,i}$.  We set the
renormalization and factorization scales equal and vary them according
to $\mu=\mu_R=\mu_F=c_s \hat H_{\rm T}^\prime$, with $c_s\in
\{1/2,1/\sqrt{2},1,\sqrt{2},2\}$, to obtain the conventional estimate
of the size due to the truncation of the perturbative series.

Standard Model input parameters are defined through the $G_\mu$ scheme
with
\begin{align} 
m_Z &= 91.188\,{\rm GeV}\,, & \Gamma_Z &=2.49\,{\rm GeV}\,, \nonumber\\ 
m_W &= 80.419\,{\rm GeV}\,, & \Gamma_W &=2.06\, {\rm GeV}\,, \nonumber\\ 
m_H &= 125\,{\rm GeV}\,, & \Gamma_H &=0.00407\, {\rm GeV}\,, \nonumber\\ 
m_t &= 175\,{\rm GeV}\,, & \Gamma_t &=1.5\, {\rm GeV}\,, \nonumber\\ 
G_\mu &= 1.16639\times 10^{-5}\,{\rm GeV}^{-2}\,,& \sin^2\theta_W &=
1-\tilde{m}^2_W/\tilde{m}^2_Z \,, 
\end{align}
where $\tilde{m}_V^2 = m_V^2+i\Gamma_Vm_V$.  Unstable particles are
consistently treated through the complex mass
scheme~\cite{Denner:1999kn} in all but the NLO calculations, in which
the decay products are distributed according to a Breit--\-Wigner
distribution and real values for all coupling constants are
maintained.

\subsection{Inclusive cross sections}
\label{sec:Vjets-xs}

\subsubsection{Leading-order cross sections}
\label{sec:Vjets-LO-xs}
\begin{figure}[thpb]
  \begin{center}
    \includegraphics[width=0.80\linewidth]{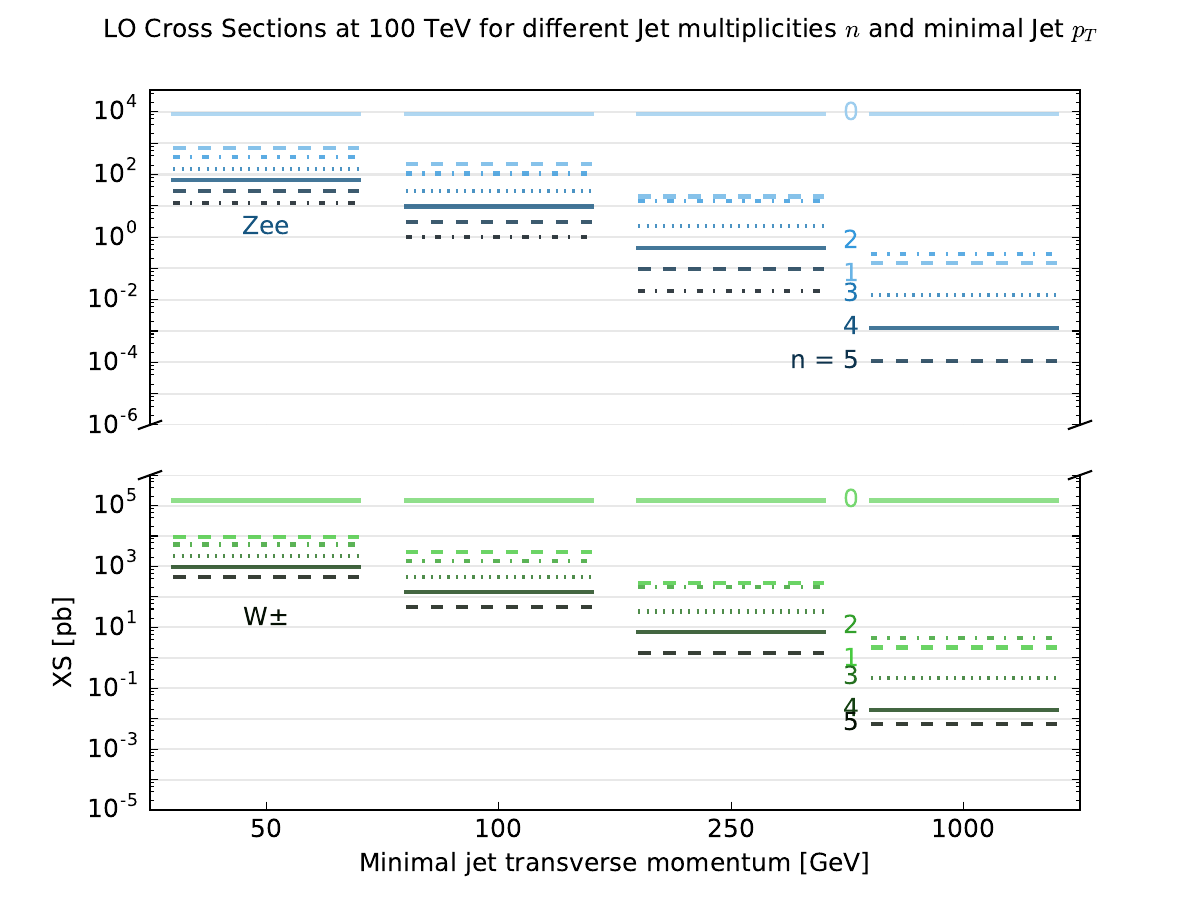}
    \caption{The leading-order cross sections against $p_{T}^{\rm min}$ 
      given by 50~GeV, 100~GeV, 250~GeV and 1000~GeV for the associated
      production of jets and a $Z$ or $W$
      boson decaying into leptons.
      \label{Figure:Table_minpt_XS}}
  \end{center}
\end{figure}
In Tables~\ref{Table:Table_WJetInclusive} and
~\ref{Table:Table_ZPlusJets} leading-order cross sections for the
production of a weak vector boson $V$ ($V=W^\pm$ or $Z$), which decays
into a massless lepton pair, in conjunction with up to six jets are
shown, employing the `basic' type of kinematical cuts.  The production
cross sections are displayed with four distinct values of $p_{T}^{\rm
  min}$ varied over the values $50$~GeV, $100$~GeV, $250$~GeV and
$1000$~GeV.  As a function of $p_{T}^{\rm min}$ total cross sections
are reduced by up to four orders of magnitude, but they still reach a
few attobarns for the highest multiplicities.  The cross sections
range over about 9 orders of magnitude from a few to a few dozen
nanobarns for inclusive production to a few attobarns when the vector
bosons are accompanied by six TeV jets.  Even for relatively soft jets
with a minimal transverse momentum of 50~GeV, the cross sections for
$V$+6 jets are still of the order of tens of picobarns.  Irrespective
of potentially large higher-order corrections, these first few numbers
already indicate that a future $\sqrt{s}=100$ TeV collider will
provide a very busy environment.  An obvious result of this is that
very large QCD backgrounds, even at high scales, will render this a
challenging environment for searches that involve signatures with many
jets.  These findings are condensed in
Figure~\ref{Figure:Table_minpt_XS}, which exhibits the cross section
$W^\pm$ and $Z$ production in association with jets, using
`democratic' cuts, and in Figure~\ref{Figure:Wpnj_pt_cumulative},
displaying the cumulative cross sections, including parton shower
effects in a simulation invoking also parton showering effects, based
on multi-jet merging technology.  In
Fig.~\ref{Figure:Wpnj_pt_distribution} the $p_\perp$ distribution of
the few first jets -- if existent -- and the $W$ boson is shown, based
on the same simulation.  Focusing on the regime of transverse momenta,
this figure suggests that for leading jets with transverse momenta
above around a TeV the recoil is mainly provided by a second jet
rather than by the $W$ boson.  Such kinematical situations are
therefore probably better identified as a (real) weak correction to
QCD dijet production rather than the real QCD correction to
$Vj$-associated production.
\begin{table}[th]
  \begin{center}
    \small
    \begin{tabular}{|l||c|c|c|c||c|c|c|c|}
      \hline
      & 
      \multicolumn{4}{c||}{ $pp\rightarrow$\Wpjn-jet+X }  &
      \multicolumn{4}{c|}{ $pp\rightarrow$\Wmjn-jet+X } \\ \hline
      $n/p^{\rm min}_{T} $ &
      50 GeV & 100 GeV & 250 GeV & 1000 GeV &
      50 GeV & 100 GeV & 250 GeV & 1000 GeV 
      \\ \hline\hline 
      $0$ &
      \multicolumn{4}{c||}{ \nb{40.51(5)} } &
      \multicolumn{4}{c|}{ \nb{34.29(4)} } \\\hline
      $1$ & \pb{2617(5)} & \pb{847(1)}  & \pb{80.3(1)}  & \fb{673(1)}   & \pb{2202(4)} & \pb{699(1)}  & \pb{62.5(1)} & \fb{443(1)}  \\
      $2$ & \pb{1482(8)} & \pb{427(2)}  & \pb{60.9(2)}  & \fb{1368(6)}  & \pb{1199(6)} & \pb{339(1)}  & \pb{45.8(1)} & \fb{886(3)}  \\
      $3$ & \pb{626(3)}  & \pb{125(1)}  & \pb{9.94(9)}  & \fb{71.2(6)}  & \pb{461(4)}  & \pb{94.6(9)} & \pb{6.75(6)} & \fb{39.9(3)} \\
      $4$ & \pb{286(1)}  & \pb{42.6(2)} & \fb{2166(9)}  & \fb{6.65(2)}  & \pb{208(1)}  & \pb{29.8(1)} & \fb{1390(6)} & \fb{3.51(1)} \\
      $5$ & \pb{128(1)}  & \pb{14.1(1)} & \fb{461(3)}   & \ab{592(3)}   & \pb{89.9(7)} & \pb{9.09(7)} & \fb{276(1)}  & \ab{289(1)}  \\
      $6$ & \pb{54.9(5)} & \pb{4.67(4)} & \fb{100.3(9)} & \ab{53.3(4)}  & \pb{37.2(3)} & \pb{2.94(2)} & \fb{57.4(5)} & \ab{24.8(1)}  \\
      \hline
    \end{tabular}
    \caption{Leading-order cross sections for the production of a
      leptonically decaying $W^+$ or $W^-$ in association with $n$
      jets. `Basic' cuts have been employed, with transverse momentum
      cuts ranging from $p_{T}^{\rm min}=50$~GeV to $p_{T}^{\rm
        min}=1$~TeV.
    \label{Table:Table_WJetInclusive}} \end{center}
\end{table}
\begin{table}[th]
  \begin{center}
    \small
    \begin{tabular}{|l||c|c|c|c||c|c|c|c|}
      \hline
      & 
      \multicolumn{4}{c||}{ $pp\rightarrow Z(\rightarrow e\bar{e})+n$-jet+X }  &
      \multicolumn{4}{c|}{ $pp\rightarrow Z(\rightarrow \nu\bar{\nu})+n$-jet+X } \\ \hline
      $n/p^{\rm min}_{T} $ &
      50 GeV & 100 GeV & 250 GeV & 1000 GeV &
      50 GeV & 100 GeV & 250 GeV & 1000 GeV 
      \\ \hline\hline 
      $0$ &
      \multicolumn{4}{c||}{\pb{8921(8)}  } &
      \multicolumn{4}{c|}{ \pb{17619(18)}} \\\hline
      $1$ & \pb{696(2)}  & \pb{213.8(4)} & \pb{20.04(4)} & \fb{151.7(3)} & \pb{1372(3)}  & \pb{421.8(9)} & \pb{39.56(8)} & \fb{300.3(7)} \\
      $2$ & \pb{378(2)}  & \pb{106.7(5)} & \pb{14.57(6)} & \fb{293(2)}   & \pb{745(4)}   & \pb{212(1)}   & \pb{28.9(2)}  & \fb{580(3)}   \\
      $3$ & \pb{151(2)}  & \pb{29.0(3)}  & \pb{2.24(2)}  & \fb{14.2(2)}  & \pb{293(3)}   & \pb{58.5(6)}  & \pb{4.37(4)}  & \fb{28.1(3)}  \\      
      $4$ & \pb{66.8(3)} & \pb{9.54(4)}  & \fb{463(2)}   & \ab{1280(5)}  & \pb{132.1(6)} & \pb{18.7(1)}  & \fb{905(4)}   & \ab{2509(9)}  \\
      $5$ & \pb{28.4(3)} & \pb{3.11(3)}  & \fb{95.3(6)}  & \ab{109.0(7)} & \pb{56.4(5)}  & \pb{6.07(6)}  & \fb{186(2)}   & \ab{213(2)}   \\
      $6$ & \pb{12.1(2)} & \pb{0.98(1)}  & \fb{19.4(2)}  & \belowfb & \pb{24.4(3)}  & \pb{1.95(2)}  & \fb{40.2(4)}  & \belowfb      \\
      \hline
    \end{tabular}
    \caption{Leading-order cross sections for the production of a $Z$ boson
      decaying either into a lepton or neutrino pair in association with $n$
      jets. `Basic' cuts have been employed with transverse momentum cut ranging
      from $p_{T}^{\rm min}=50$~GeV to $p_{T}^{\rm min}=1$~TeV.
      \label{Table:Table_ZPlusJets}}
  \end{center}
\end{table}

\begin{figure}[thpb]
\begin{center}
\includegraphics[width=0.70\linewidth]{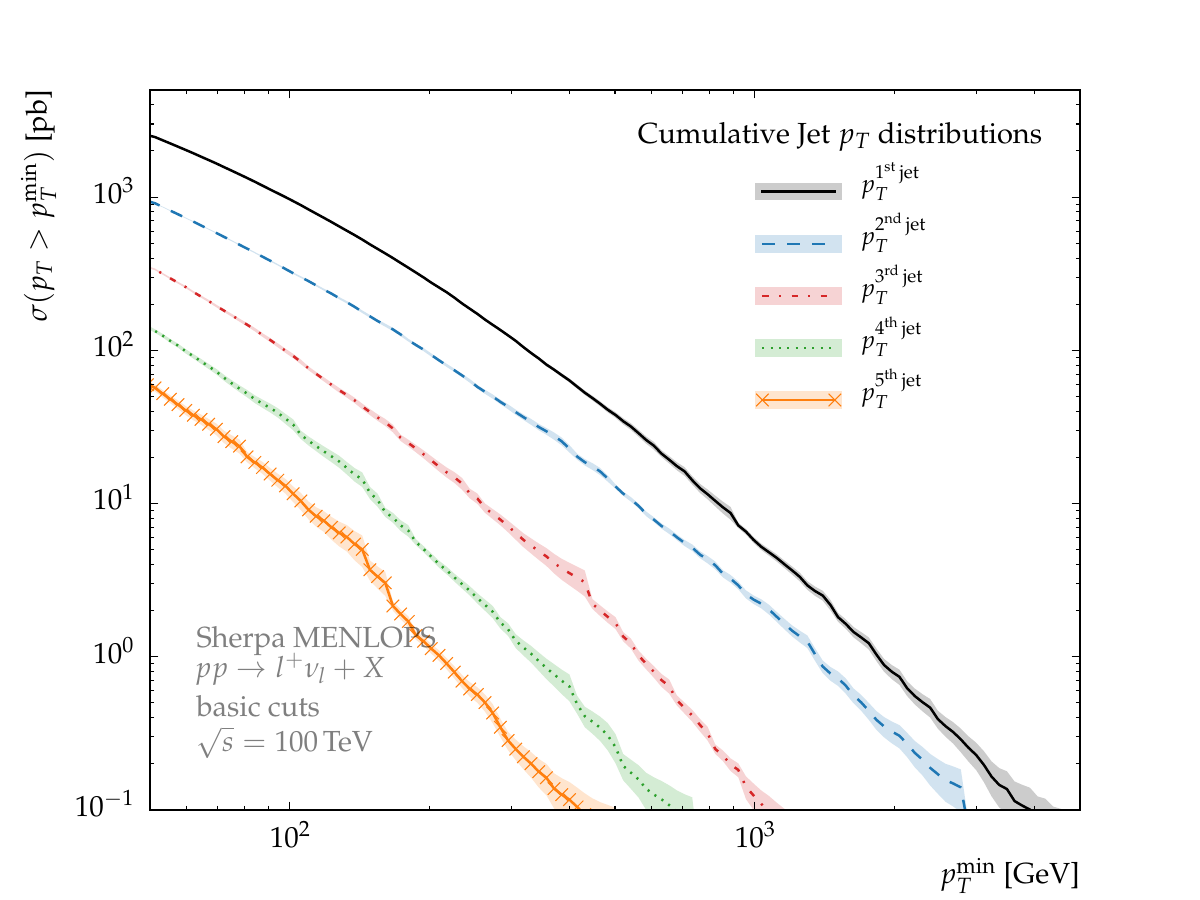}
    \caption{The total cross sections for the production of \Wpjn-jets
($n=1,...,5$) as a function of the $p_T^{\rm min}$, with the `basic' set of cuts. 
\label{Figure:Wpnj_pt_cumulative}} 
\end{center} 
\end{figure}
\begin{figure}[thpb]
\begin{center}
\includegraphics[width=0.70\linewidth]{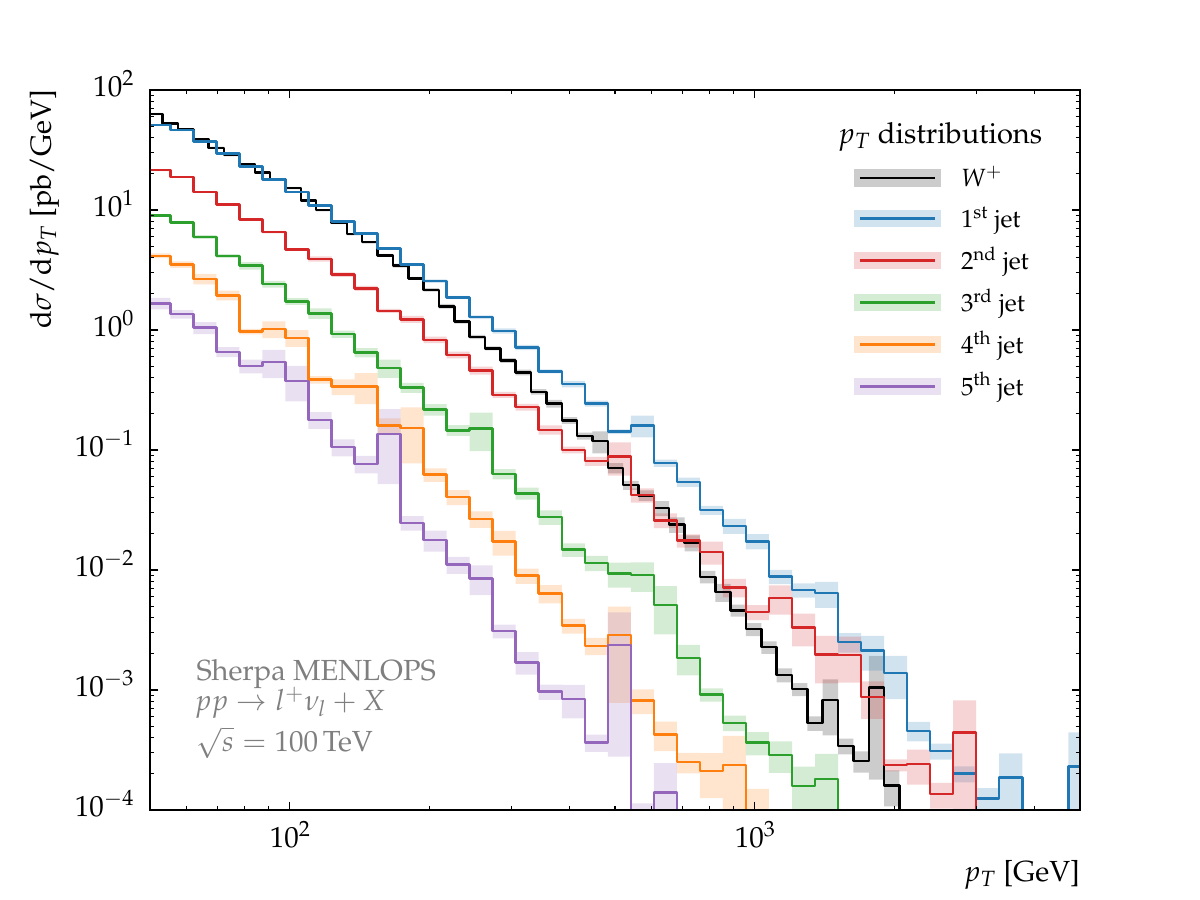}
\caption{Jet-$p_T$ distributions in an inclusive
sample of $W^+$ production. Also shown is the $p_T^{W^+}$ distribution.
\label{Figure:Wpnj_pt_distribution}}
\end{center} 
\end{figure}

\subsubsection{Next-to-leading order QCD corrections}
\label{sec:Vjets-NLO-xs}

In Tables \ref{tab_Wmj100TeV_total_xs}-\ref{tab_Zj100TeV_total_xs} we
give LO and NLO total inclusive cross sections for vector boson
production in association with 1 to 5 jets. We show sensitivity to
renormalization and factorization scales as superscripts and
subscripts, respectively. In parenthesis we quote the associated
statistical integration error for each total cross section.  The cross
sections at 100~TeV range over several orders of magnitude, reaching
cross sections of only few attobarns in the case of high-hierarchical
cuts.  The theoretical control over the cross section predictions is
estimated with a number of indicators: the scale variation dependence,
jet ratios and PDF-uncertainties. In what follows, we discuss the
scale variation dependence, but postpone the discussion of jet ratios
and PDF uncertainties to Sections \ref{sec:Vjets-ratios} and
\ref{sec:Vjets-validation}.

Perturbative calculations depend on the unphysical renormalization and
factorization scales due to the truncation of the perturbative series
for the scattering processes.  As commonly done, we estimate the size
truncated higher-order terms by varying the renormalization and
factorization scales.

In Tables~\ref{tab_Wmj100TeV_total_xs}-\ref{tab_Zj100TeV_total_xs} the
upper and lower scale variation is given super/sub
scripts. Renormalization and factorization are set to equal values and
varied simultaneously (see Section \ref{sec:Vjets-setup}).  We observe
the expected increase in scale dependence with growing jet
multiplicities due to the higher powers in the strong coupling
$\alpha_S(\mu)$. The linear growth of the scale dependence at LO is
significantly reduced at NLO.  The systematic of the scale variation
dependence is comparable for the different types of vector bosons.  To
summarize the results in the tables: the scale variation dependence
reduces at $100$~TeV from between 20\% to 50\% at LO to about 10\% at
NLO for all non-hierarchical cuts.  The case of hierarchical cuts is
perturbatively unstable, as can be seen from cross sections increasing
with jet multiplicity at LO, from the large difference of LO and NLO
cross sections, and also from the scale dependence at NLO which is not
as much reduced as in the non-hierarchical cases.  Such behavior,
however, is not unexpected, as LO hard matrix element will over
estimate rates of soft radiation, which are common in the hierarchical
environment. Nevertheless NLO results give a better description, which
can be compared to the jet-ratio results from the shower predictions
results presented in section~\ref{sec:Vjets-ratios}.

\begin{table}[th]
\begin{center}
\small
{\centering
\begin{tabular}{|c||c|c||c|c|}
        \hline \multicolumn{5}{|c|}{ $pp\rightarrow$ \Wmjn-jet+X } \\\hline 
        & 
\multicolumn{2}{c||}{{ low-democratic (100~TeV)[pb]}} &  
\multicolumn{2}{c|}{{ high-democratic (100~TeV)[fb]}}  \\
\hline
 $n$ & LO & NLO & LO & NLO \\
\hline
1 & $481.2(4)^{+0.0}_{-2.5}$ & $811(4)^{+38}_{-31}$ & $258.9(4)^{+25.7}_{-22.5}$ & $1139(30)^{+160}_{-131}$ \\
\hline
2 & $526.2(7)^{+68.6}_{-59.0}$ & $524(10)^{+2}_{-7}$  & $749(2)^{+146}_{-116}$ & $885(10)^{+34}_{-50}$ \\
\hline
3 & $253.5(7)^{+68.0}_{-50.8}$ & $212(7)^{+1}_{-15}$ & $151.0(6)^{+46.2}_{-33.5}$ & $164(4)^{+4}_{-11}$ \\
\hline
4 & $101.1(7)^{+41.0}_{-27.5}$ & $92(5)^{+2}_{-9}$ & $21.3(1)^{+8.8}_{-5.9}$ & $21.2(9)^{+1.7}_{-2.8}$ \\
\hline
5 & $36.4(5)^{+19.7}_{-12.1}$ & --- & $2.81(4)^{+1.48}_{-0.92}$ & --- \\
\hline
&
\multicolumn{2}{c||}{{   low-hierarchical (100~TeV)[pb]}} &  
\multicolumn{2}{c|}{{  high-hierarchical (100~TeV)[fb]}}     \\
\hline
1 & $0.01394(1)^{+0.00174}_{-0.00148}$ & $0.1003(3)^{+0.0173}_{-0.0139}$ & $0.001330(1)^{+0.000266}_{-0.000210}$ & $0.01730(6)^{+0.00393}_{-0.00304}$ \\
\hline
2 & $0.1117(2)^{+0.0236}_{-0.0185}$ & $0.127(1)^{+0.004}_{-0.007}$ & $0.01880(2)^{+0.00484}_{-0.00365}$ & $0.0230(2)^{+0.0017}_{-0.0020}$ \\
\hline
3 & $0.212(1)^{+0.065}_{-0.047}$ & $0.103(8)^{+0.013}_{-0.037}$ & $0.01363(3)^{+0.00471}_{-0.00333}$ & $0.0143(2)^{+0.0006}_{-0.0012}$ \\
\hline
4 & $0.240(2)^{+0.099}_{-0.066}$ & $0.08(2)^{+0.02}_{-0.06}$ & $0.00559(2)^{+0.00245}_{-0.00162}$ & $0.0056(2)^{+0.0004}_{-0.0007}$ \\
\hline
5 & $0.204(3)^{+0.106}_{-0.066}$ & --- & $0.00165(2)^{+0.00089}_{-0.00055}$ & --- \\
\hline
\end{tabular}
}
\caption{Fixed order \Wmjn-jet~+~X cross sections.  The setup is
    specified by the low/high/democratic/hierarchical phase-space regions
    described in Section~\ref{sec:Vjets-setup}.  Scale dependence variation is
    given in lower and upper limits and the statistical integration errors is
    given by the number in parenthesis next to the central value.
    \label{tab_Wmj100TeV_total_xs} }
\end{center}
\end{table}

\begin{table}[th]
\begin{center}
\small
{\centering
\begin{tabular}{|c||c|c||c|c|}
        \hline \multicolumn{5}{|c|}{ $pp\rightarrow$\Wpjn-jet+X  } \\\hline 
        & 
\multicolumn{2}{c||}{{ low-democratic (100~TeV)[pb]}} &  
\multicolumn{2}{c|}{{ high-democratic (100~TeV)[fb]}}  \\
\hline
 $n$ & LO & NLO & LO & NLO \\
\hline
1 & $563.1(5)^{+0.0}_{-2.8}$ & $926(7)^{+45}_{-36}$ & $405.3(8)^{+39.8}_{-34.8}$ & $1714(20)^{+238}_{-195}$  \\
\hline
2 & $622.5(9)^{+81.9}_{-70.2}$ & $593(10)^{+4}_{-10}$ & $1148(2)^{+223}_{-177}$ & $1362(20)^{+58}_{-82}$ \\
\hline
3 & $314(1)^{+84}_{-63}$ & $279(8)^{+0}_{-11}$ & $247(2)^{+75}_{-55}$ & $256(10)^{+7}_{-17}$ \\
\hline
4 & $127.2(9)^{+51.4}_{-34.5}$ & $98(8)^{+0}_{-8}$ & $37.3(3)^{+15.4}_{-10.3}$ & $37(2)^{+2}_{-4}$ \\
\hline
5 & $49.3(7)^{+26.5}_{-16.3}$ & --- & $5.03(6)^{+2.62}_{-1.64}$ & --- \\
\hline
&
\multicolumn{2}{c||}{{   low-hierarchical (100~TeV)[pb]}} &  
\multicolumn{2}{c|}{{  high-hierarchical (100~TeV)[fb]}}     \\
\hline
1 & $0.02499(1)^{+0.00304}_{-0.00259}$ & $0.1673(8)^{+0.0284}_{-0.0228}$ & $0.004191(2)^{+0.000802}_{-0.000639}$ & $0.0445(1)^{+0.0099}_{-0.0077}$ \\
\hline
2 & $0.1922(4)^{+0.0402}_{-0.0315}$ & $0.208(3)^{+0.005}_{-0.010}$ & $0.05128(6)^{+0.01295}_{-0.00982}$ & $0.0584(3)^{+0.0040}_{-0.0048}$ \\
\hline
3 & $0.371(2)^{+0.113}_{-0.082}$ & $0.19(1)^{+0.02}_{-0.06}$ & $0.0393(1)^{+0.0134}_{-0.0095}$ & $0.0354(7)^{+0.0012}_{-0.0030}$ \\
\hline
4 & $0.437(9)^{+0.178}_{-0.120}$ & $0.13(2)^{+0.04}_{-0.12}$ & $0.0168(1)^{+0.0072}_{-0.0048}$ & $0.0144(5)^{+0.0010}_{-0.0018}$ \\
\hline
5 & $0.39(1)^{+0.20}_{-0.12}$ & --- & $0.0052(1)^{+0.0028}_{-0.0017}$ & --- \\
\hline
\end{tabular}
}
\caption{Fixed order \Wpjn-jet~+~X cross sections.  The setup is
    specified by the low/high/democratic/hierarchical phase-space regions
    described in Section~\ref{sec:Vjets-setup}.  Scale dependence variation is
    given in lower and upper limits and the statistical integration errors is
    given by the number in parenthesis next to the central value.
    \label{tab_Wpj100TeV_total_xs} }
\end{center}
\end{table}

\begin{table}[th]
\begin{center}
\small
{\centering
\begin{tabular}{|c||c|c||c|c|}
        \hline \multicolumn{5}{|c|}{ $pp\rightarrow$\Zjn-jet+X  } \\\hline 
        & 
\multicolumn{2}{c||}{{ low-democratic (100~TeV)[pb]}} &  
\multicolumn{2}{c|}{{ high-democratic (100~TeV)[fb]}}  \\
\hline
 $n$  & LO & NLO & LO & NLO \\
\hline
1 & $186.0(2)^{+0.2}_{-2.0}$ & $290(1)^{+11}_{-9}$ & $127.4(2)^{+12.3}_{-10.8}$ & $504(2)^{+67}_{-55}$ \\
\hline
2 & $188.0(3)^{+23.6}_{-20.5}$ & $181(3)^{+0}_{-2}$ & $337.1(7)^{+64.8}_{-51.6}$ & $396(6)^{+16}_{-22}$ \\
\hline
3 & $92.3(1)^{+24.3}_{-18.2}$ & $79(2)^{+0}_{-3}$ & $69.6(1)^{+21.1}_{-15.3}$ & $73(1)^{+3}_{-6}$ \\
\hline
4 & $37.1(1)^{+14.8}_{-10.0}$ & $27(1)^{+0}_{-2}$ & $10.36(4)^{+4.28}_{-2.87}$ & $10.3(4)^{+1.0}_{-1.6}$ \\
\hline
5 & $13.8(1)^{+7.4}_{-4.5}$ & --- & $1.38(1)^{+0.72}_{-0.45}$ & --- \\
\hline
&
\multicolumn{2}{c||}{{   low-hierarchical (100~TeV)[pb]}} &  
\multicolumn{2}{c|}{{  high-hierarchical (100~TeV)[fb]}}     \\
\hline
1 & $0.007193(3)^{+0.000875}_{-0.000745}$ & $0.0462(1)^{+0.0077}_{-0.0062}$ & $0.0009173(4)^{+0.0001777}_{-0.0001412}$ & $0.01017(2)^{+0.00227}_{-0.00176}$ \\
\hline
2 & $0.05284(8)^{+0.01101}_{-0.00864}$ & $0.0574(6)^{+0.0013}_{-0.0026}$ & $0.01149(1)^{+0.00292}_{-0.00221}$ & $0.01337(5)^{+0.00093}_{-0.00113}$ \\
\hline
3 & $0.0989(2)^{+0.0302}_{-0.0219}$ & $0.054(3)^{+0.006}_{-0.017}$ & $0.00861(1)^{+0.00294}_{-0.00209}$ & $0.0080(2)^{+0.0003}_{-0.0007}$ \\
\hline
4 & $0.1140(3)^{+0.0465}_{-0.0313}$ & $0.049(6)^{+0.012}_{-0.031}$ & $0.003631(8)^{+0.001576}_{-0.001046}$ & $0.00336(9)^{+0.00025}_{-0.00043}$ \\
\hline
5 & $0.096(1)^{+0.050}_{-0.031}$ & --- & $0.001095(9)^{+0.000582}_{-0.000362}$ & --- \\
\hline
\end{tabular}
}
\caption{Fixed order \Zjn-jet~+~X cross sections for
    production.  The setup is
    specified by the low/high/democratic/hierarchical phase-space regions
    described in Section~\ref{sec:Vjets-setup}.  Scale dependence variation is
    given in lower and upper limits and the statistical integration errors is
    given by the number in parenthesis next to the central value.
    \label{tab_Zj100TeV_total_xs} }
\end{center}
\end{table}

\begin{figure}[thpb]
\begin{center}
\includegraphics[width=0.50\linewidth]{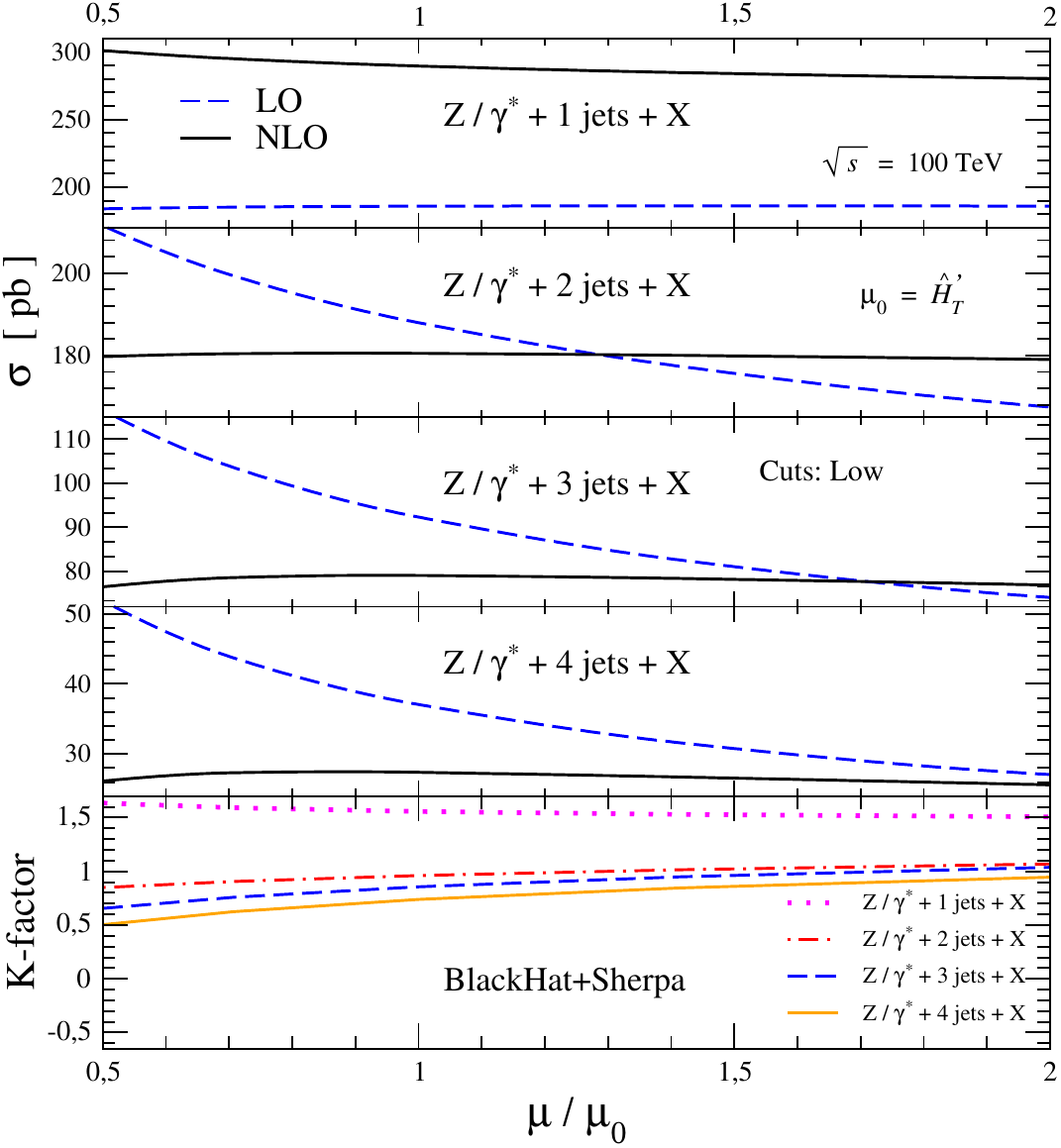}
\hspace{-2mm}
\includegraphics[width=0.50\linewidth]{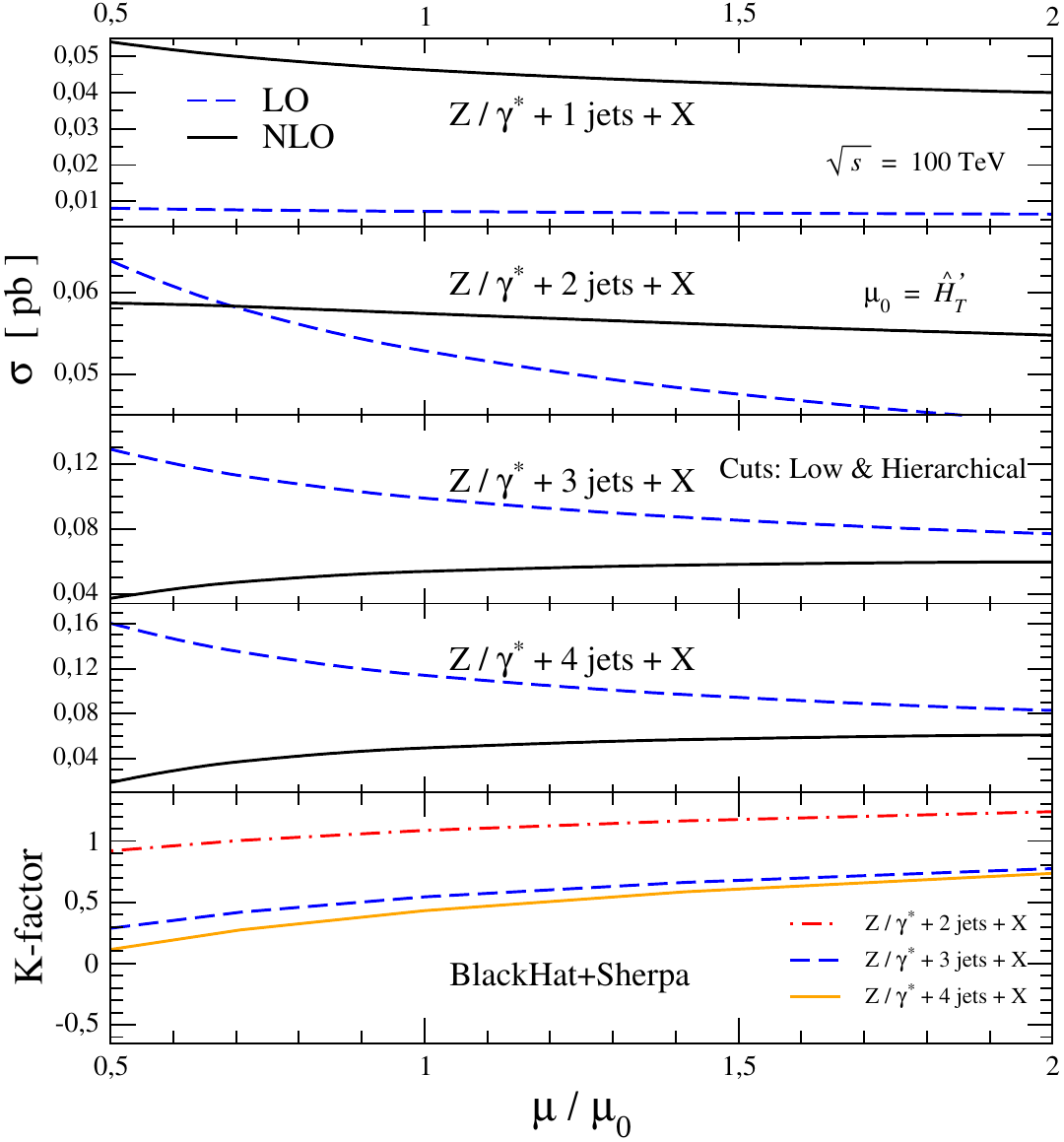}
\caption{Scale sensitivity for total cross sections with `low' cuts in $Z(\rightarrow ee)+$n-jet
production at LO and NLO.  K-factors are shown in the bottom panels. On the left
we show cases with `low-democratic' cuts and on the right with `low-hierarchical'.
\label{Figure:Zjets_Scale_Dependence}}
\end{center}
\end{figure}

\begin{figure}[thpb]
\begin{center}
\includegraphics[width=0.50\linewidth]{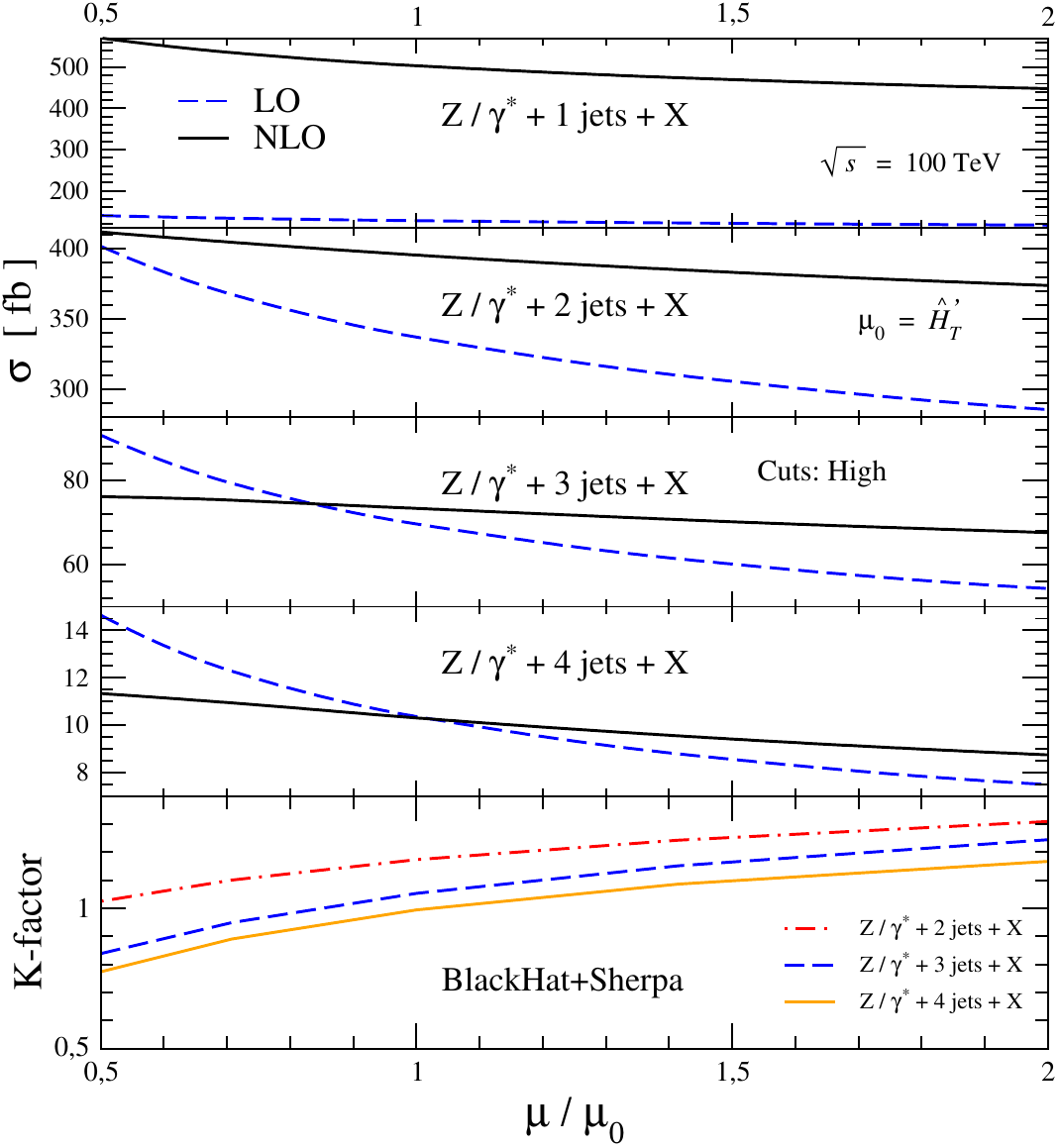}
\hspace{-2mm}
\includegraphics[width=0.50\linewidth]{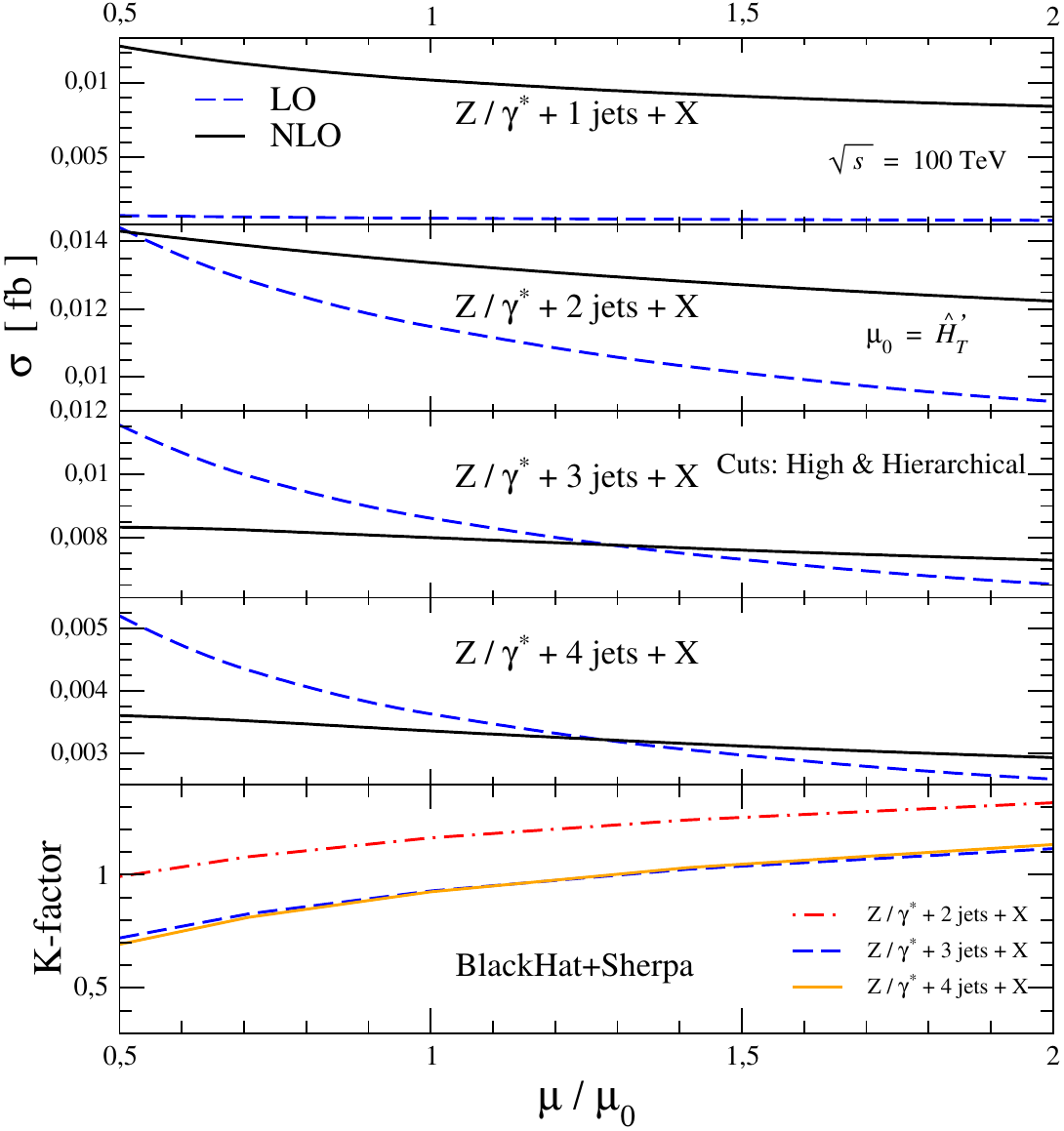}
\caption{Scale sensitivity for total cross sections with High cuts in $Z+$jets
production at LO and NLO.  K-factors are shown in the bottom panels. On the left
we show cases with `high-democratic' cuts and on the right with high-hierarchical.
\label{Figure:Zjets_High_Scale_Dependence}}
\end{center}
\end{figure}

To illustrate the stability of the NLO QCD results, in
Figures~\ref{Figure:Zjets_Scale_Dependence} and
\ref{Figure:Zjets_High_Scale_Dependence} we show a full set of scale
dependence plots for all kinematical regimes in $Z+n$-jet
production. It is clear that the dynamical scale choice $\mu=\hat
H_{\rm T}^\prime$ represents a natural scale for all the cuts
considered, and even does a good job over phase space. In
Figure~\ref{fig_Wm_jet_pt} we actually show differential cross
sections for the $p_{\rm T}$ of the $n$-th jet in inclusive $W^-$
production. In the bottom panel we show differential LO/NLO ratios
together with scale bands. Except for the well known giant K-factor in
the 1 jet bin, all perturbative corrections appear as mild for
configuration with jets with $p_{\rm T}$ of up to 10 TeV.  Notice that
in the highest bins, cross sections per bin are at the order of few
attobarns.

\begin{figure}[thpb]
\begin{center}
\includegraphics[clip,scale=0.58]{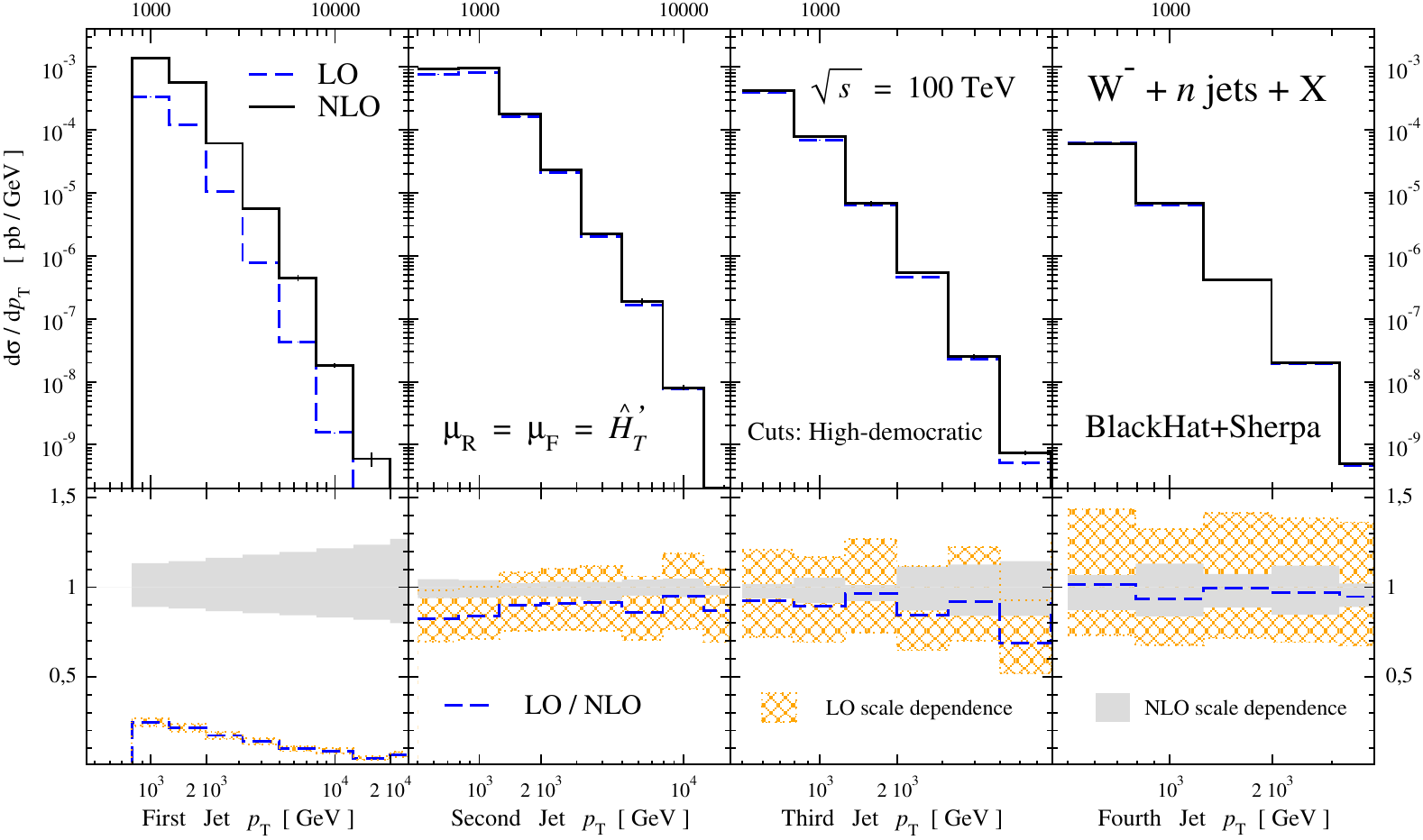}
\end{center}
\caption{Differential cross sections for inclusive $W^-$ production in
  the $n^{\rm th}$-jet $p_{T}$. Results are shown employing
  `high-democratic' cuts. The bottom panels show LO/NLO ratios as well
  as scale sensitivity.  }
\label{fig_Wm_jet_pt}
\end{figure}

\begin{figure}[thpb]
\begin{center}
\includegraphics[clip,scale=0.58]{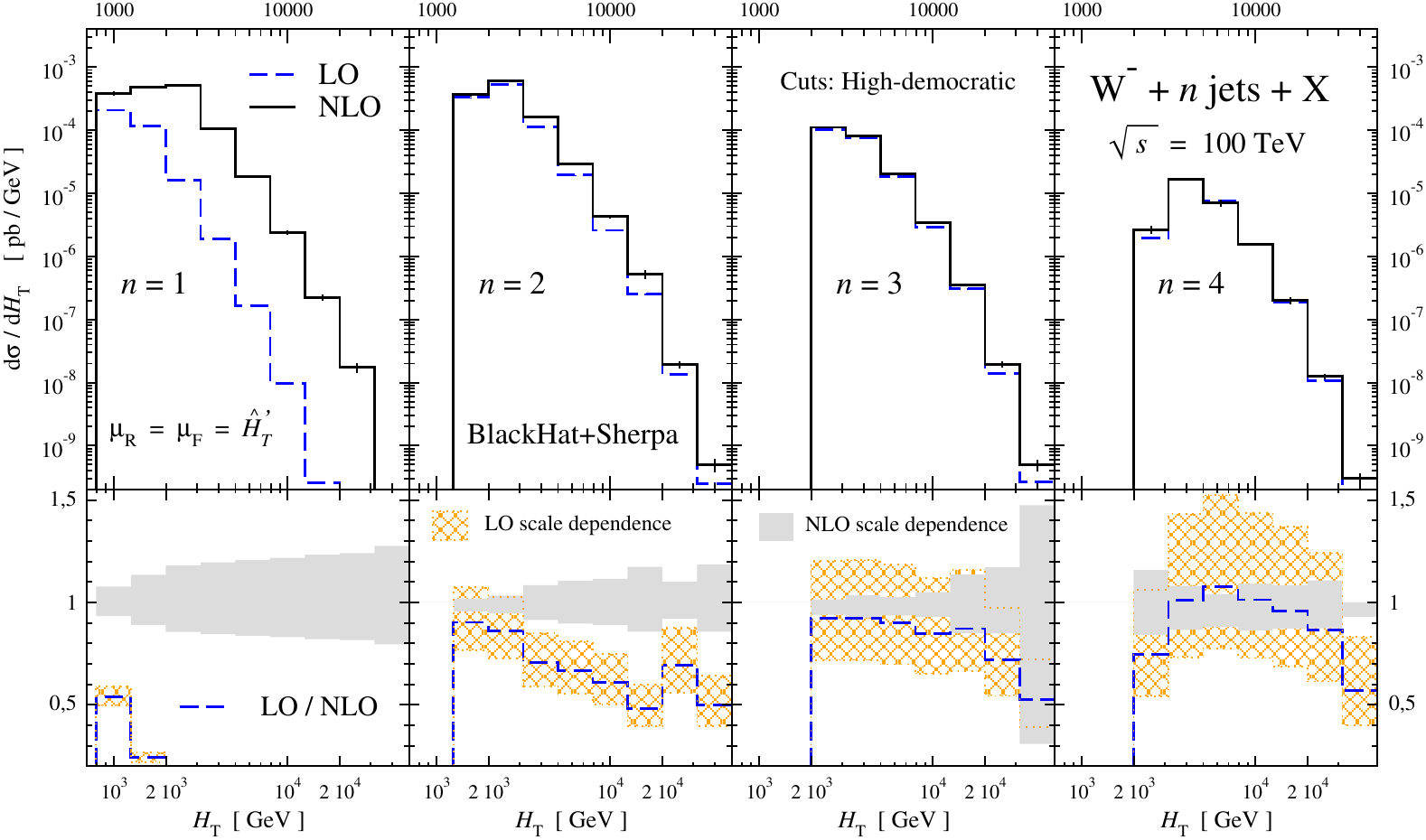}
\end{center}
\caption{Hadronic $H_T$ distribution in samples of \Wmjn-jets
  ($n=1,2,3,4$).  Results are shown employing `high-democratic'
  cuts. The bottom panels show LO/NLO ratios as well as scale
  sensitivity.  }
\label{fig_Wm_jet_HT}
\end{figure}

\begin{figure}[thpb]
\begin{center}
\includegraphics[clip,scale=0.58]{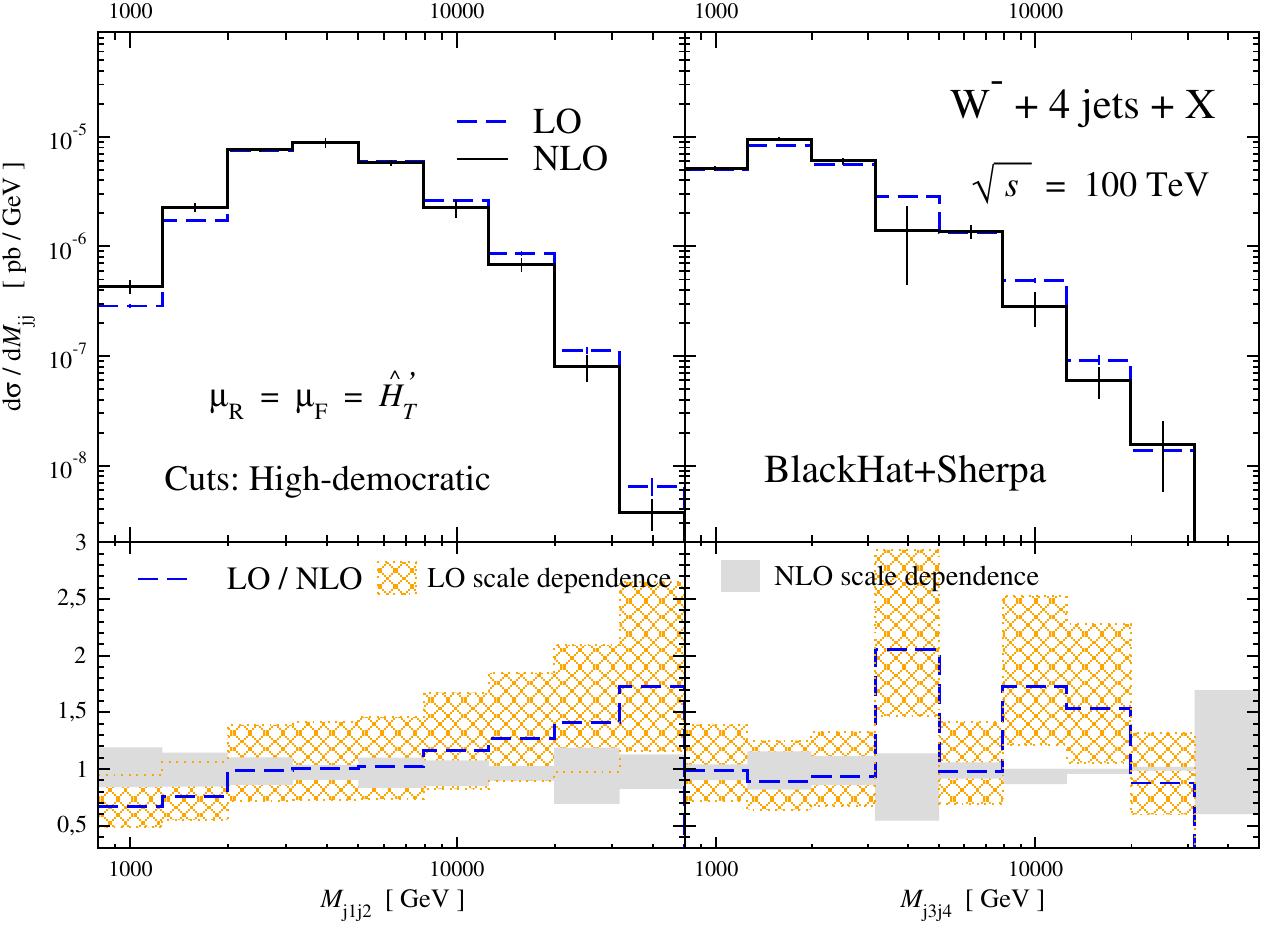}
\end{center}
\caption{Jet pair invariant masses $M_{j_1j_2}$ and $M_{j_3j_4}$ in of
  \Wmjjjj-jet production.  Results are shown employing
  `high-democratic' cuts. The bottom panels show LO/NLO ratios as well
  as scale sensitivity.  }
\label{fig_Wm4j_Mjj}
\end{figure}

\begin{figure}[thpb]
\begin{center}
\includegraphics[clip,scale=0.45]{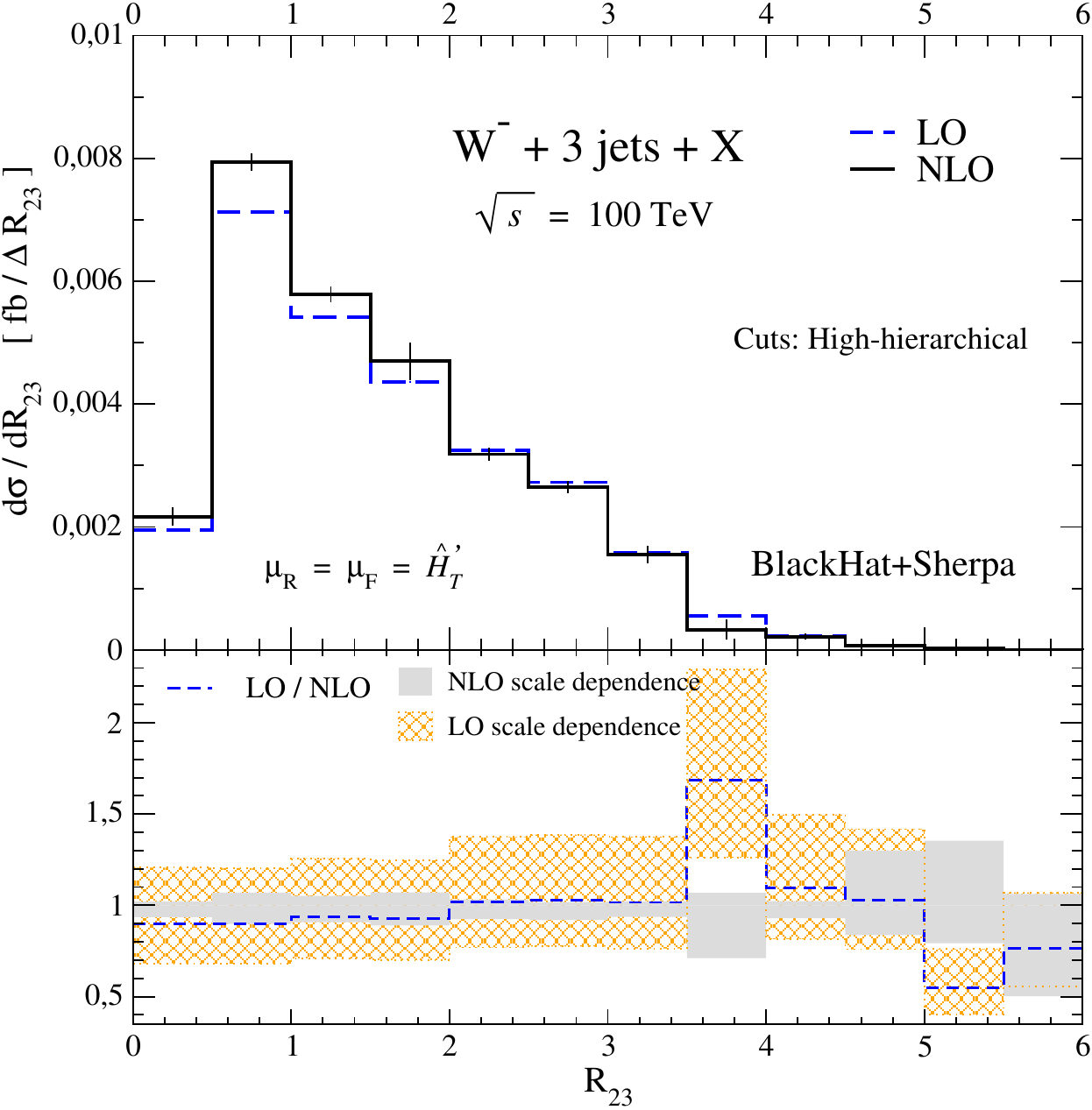}
\end{center}
\caption{$\Delta R$ separation between the sub-leading jets in
  \Wmjjj-jet production. Results are shown employing
  `high-hierarchical' cuts. The bottom panel shows LO/NLO ratio as
  well as scale sensitivity.  }
\label{fig_Wm3j_dRjj}
\end{figure}

The NLO QCD predictions for the $n^{th}$-jet $p_T$ shown in
Figure~\ref{fig_Wm_jet_pt} allow to explore the accessibility of very
hard jets at the $\sqrt{s}=100$ TeV machine. The threshold for
producing a few events with a single hard jet (considering an
integrated luminosity of several inverse attobarns) is around
$20$~TeV. Not surprisingly, all these events will be accompanied with
a second hard jet, as we can see from inspecting the tail of the
second jet $p_T$. On the other hand, few events will be recorded with
three jets (and a weak vector boson) with more than $5$~TeV of $p_T$,
and for four jets the threshold is around $3$ ~TeV.

An interesting picture emerges from the hadronic $H_T$ distributions
shown in Figure~\ref{fig_Wm_jet_HT}. The very large NLO corrections in
the \Wmj jet process is understood by the release of a kinematical
constraint that basically allows the vector boson to be soft in events
with large $H_T$. But here we can also see that quantum corrections
tend to increase the $H_T$ distributions for samples with two or more
jets. Extra jet radiation is favored in high $H_T$ environments, again
not surprisingly. This effect is such that for the larger
multiplicities we see that the differential cross sections are quite
similar for the $n=2,3$ and $4$ in the very high--$H_T$ tails. One
should then expect a sizable set of events with very large numbers of
jets.  In Figure~\ref{fig_Wm4j_Mjj} we show the di-jet mass
distributions for the pairs $(j_1,j_2)$ and $(j_3,j_4)$ in \Wmjjjj-jet
production. For both distributions corrections are generally mild, but
shape changes are clear for $M_{j_1j_2}$.  The radiation steepens the
slope of the $M_{j_1j_2}$ spectrum, but events with invariant masses
larger than $30$~TeV will be abundant.  In Figure~\ref{fig_Wm3j_dRjj}
we present the $R$ separation of the second and third-hardest jet in a
high-hierarchical configuration. Those are the hardest jets below the
very hard jet required. As can be seen these jets are produced in a
collimated fashion, with the potential singularity cut by the jet
algorithm (with $R=0.4$ for us). Extra radiation push the jets even
more close, as can be inferred from the change in shape of the
distribution.

\subsection{Cross-section ratios}
\label{sec:Vjets-ratios}

We present ratios of cross sections: Jet-production ratios,
\begin{equation}
R_n=\frac{\sigma^{V+n-\rm jet}}{\sigma^{V+(n-1)-\rm jet}} \label{eqn:Vjets-jetratio}
\end{equation}
are considered, giving the probability for the emission of an
additional jet.  The resulting ratios are displayed in Table
\ref{tab_jr_Wmj100TeV_total_xs}.  Theoretical uncertainties tend to be
reduced in these ratios, as many common features (like PDF's, alphas,
scale dependence) largely cancel in the ratios.  This renders them
particularly helpful for comparisons with experimental measurements.
For the present study we are interested in the systematic behaviour of
the ratios $R_n$ for two reasons.
\begin{itemize}
\item On the one hand, the understanding of the systematics of the
  ratio as a function of jet-multiplicity ($n$) allows to extrapolate
  from low to high jet multiplicities. This gives a handle to explore
  the collision environment.  Higher jet multiplicities are required
  for a definitive statement. We refer to `staircase'-behaviour when a
  convergence of the jet ratios to a fixed value $R_n\rightarrow R_s$
  for increasing $n$ is observed.  'Poisson-scaling', meaning that the
  emission of additional jets follows a Poisson distribution and thus
  a decreasing probability $R_n\rightarrow \bar n/(n+1)$ for
  intermediate jet multiplicities $n$ (with $\bar n$ a constant).
  Based on the predictions in Table \ref{tab_jr_Wmj100TeV_total_xs} we
  expect `staircase' in the `democratic' setup at $100$~TeV. The
  asymptotic jet emission probability depends on the phase-space
  configuration.  For the high-hierarchical setup the presented ratios
  suggest a Poisson scaling, which is expected for the statistical
  character of an additional soft-jet emission given the high-$p_T$
  jet enforced by the cuts.  For reasons explained in Section
  \ref{sec:Vjets-validation} the ratios $R_2$ including the
  predictions for single-jet production require the addition of even
  higher QCD corrections~\cite{Ridder:2015dxa,Boughezal:2015ded}.
\item On the other hand the ratios give a tool to probe the validity
  of the perturbative computations in the respective phase-space
  regions (see Section \ref{sec:Vjets-validation}).
\end{itemize}
The picture just described can be explored much more deeply by the
results from the shower calculation. Indeed, in
Figures~\ref{fig_jets_ratios_staircase}-\ref{fig_jets_ratios_poisson}
we show the scaling properties of the jet production ratios employing
our MENLOPS results.  With this we are able to look at production of
up to 14 jets and we clearly see the staircase and Poissonian
scaling. We have added fits to these scalings by fitting the ratios
with $n=1,...,4$, and it appears that the extrapolations work
remarkably well, making them a useful tool for further studies.

\begin{table}[th]
\begin{center}
\begin{tabular}{|c||c|c||c|c|}
\hline
 \multicolumn{5}{|c|}{ Ratios $R_n$ for \quad $pp\rightarrow$\Wmjn-jet+X \quad over\quad  $pp\rightarrow$\Wmjnm-jet+X} \\
\hline
& \multicolumn{2}{c||}{low-democratic (100~TeV)}
& \multicolumn{2}{c|}{high-democratic (100~TeV)}\\
\hline
 $n$  & LO & NLO & LO & NLO \\
\hline
$2$ & $1.094(2)$ & $0.65(1)$ & $2.894(8)$ & $0.78(2)$ \\
\hline
$3$ & $0.481(1)$ & $0.40(2)$ & $0.201(1)$ & $0.186(5)$ \\
\hline
$4$ & $0.398(3)$ & $0.43(3)$ & $0.141(1)$ & $0.129(6)$ \\
\hline
$5$ & $0.361(5)$ & --- & $0.132(2)$ & --- \\
\hline
& \multicolumn{2}{c||}{low-hierarchical (100~TeV)}
& \multicolumn{2}{c|}{high-hierarchical (100~TeV)}\\
\hline
$2$ & $8.01(2)$ & $1.27(2)$ & $14.13(2)$ & $1.32(1)$ \\
\hline
$3$ & $1.90(1)$ & $0.81(7)$ & $0.725(2)$ & $0.62(1)$ \\
\hline
$4$ & $1.13(1)$ & $0.7(2)$  & $0.410(2)$ & $0.39(1)$ \\
\hline
$5$ & $0.85(2)$ & --- & $0.295(5)$ & --- \\
\hline
\end{tabular}
\caption{Jet-production ratios for the $pp\rightarrow$\Wmjn-jet~+~X processes
are given. The numbers are based on Table~\ref{tab_Wmj100TeV_total_xs} .
\label{tab_jr_Wmj100TeV_total_xs} } 
\end{center}
\end{table}

\begin{figure}[thpb]
\begin{center}
\includegraphics[clip,scale=0.90]{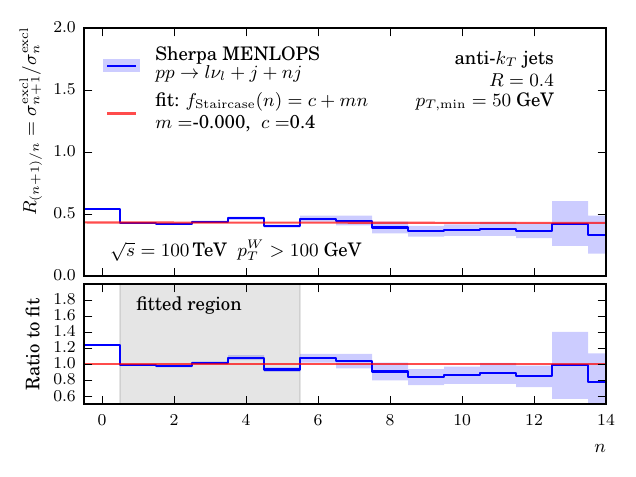}
\end{center}
\caption{Scaling properties of jet-production ratios for
  `low-democratic' configurations. Corresponding fits are shown in
  solid (red) lines, as extracted from fitting the shaded regions.  }
\label{fig_jets_ratios_staircase}
\end{figure}

\begin{figure}[thpb]
\begin{center}
\includegraphics[clip,scale=0.90]{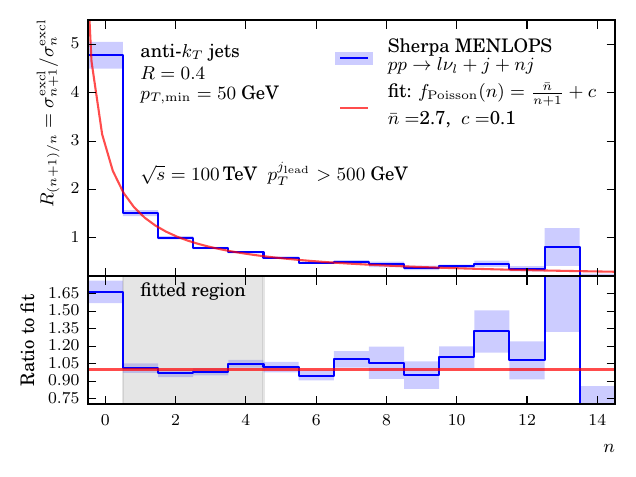}
\end{center}
\caption{Scaling properties of jet-production ratios for
  `low-hierarchical' (right) configurations. Corresponding fits are
  shown in solid (red) lines, as extracted from fitting the shaded
  regions.  }
\label{fig_jets_ratios_poisson}
\end{figure}

Tables
\ref{tab_Wm_Wp_j100TeV_total_xs}-\ref{tab_Z_Wp_j100TeV_total_xs}
display ratios of the inclusive cross section of different
vector-boson types,
\begin{equation} 
R^{V/V'}_n=\frac{\sigma^{V+n-\rm jet}}{\sigma^{V^\prime+n-\rm jet}}\,. 
\label{eqn:Vjets-chargeratio} \end{equation}
The ratios point to the dominant productions channels and the respective parton
luminosities in the respective phase-space regions. 
The monotonically decreasing ratios can be attributed to the increasing
up-quark to down-quark ratio with increasing Bjorken-$x$ values
\cite{Kom:2010mv}.  Thus for increasing collision energies, lower $x$-values are
probed leading to a reduction of the up-down ratio. This leads to a relative
increase of the $W^+$ production as compared to $W^-$ and $Z$-production. In
contrast, harder cuts enforce higher x-values, thus reducing $W^+$ production
compared to the other heavy vector bosons. Similarly, the production of
additional final state jets requires higher partonic initial-state energies
increasing the x-value and thus reducing the relative size of $W^+$ production.
These mechanisms explain the monotonicity systematic of the charge asymmetry
ratios in Tables
\ref{tab_Wm_Wp_j100TeV_total_xs}-\ref{tab_Z_Wp_j100TeV_total_xs}. \\
%
\begin{table}[thpb]
\begin{center}
\begin{tabular}{|c||c|c||c|c|}
\hline
 \multicolumn{5}{|c|}{ Ratios \quad $R^{W^-/W^+}_n$\quad  for\quad $pp\rightarrow$\Wmjn-jet+X \quad over\quad  $pp\rightarrow$\Wpjn-jet+X} \\
\hline
& \multicolumn{2}{c||}{low-democratic (100~TeV)}
& \multicolumn{2}{c|}{high-democratic (100~TeV)}\\
\hline
 $n$  & LO & NLO & LO & NLO \\
\hline
1 & $0.8545(0.0010)$ & $0.8765(0.0083)$ & $0.6388(0.0016)$ & $0.6647(0.0188)$ \\
\hline
2 & $0.8454(0.0017)$ & $0.8836(0.0261)$ & $0.6528(0.0019)$ & $0.6501(0.0125)$ \\
\hline
3 & $0.8073(0.0035)$ & $0.7575(0.0332)$ & $0.6106(0.0055)$ & $0.6416(0.0348)$ \\
\hline
4 & $0.7948(0.0082)$ & $0.9385(0.0931)$ & $0.5701(0.0052)$ & $0.5689(0.0342)$ \\
\hline
5 & $0.7388(0.0145)$ & --- & $0.5574(0.0103)$ & --- \\
\hline
& \multicolumn{2}{c||}{low-hierarchical (100~TeV)}
& \multicolumn{2}{c|}{high-hierarchical (100~TeV)}\\
\hline
1 & $0.5580(0.0005)$ & $0.5993(0.0032)$ & $0.3174(0.0002)$ & $0.3887(0.0016)$ \\
\hline
2 & $0.5812(0.0015)$ & $0.6118(0.0125)$ & $0.3666(0.0005)$ & $0.3937(0.0035)$ \\
\hline
3 & $0.5712(0.0042)$ & $0.5365(0.0531)$ & $0.3471(0.0012)$ & $0.4033(0.0098)$ \\
\hline
4 & $0.5492(0.0117)$ & $0.5990(0.2074)$ & $0.3336(0.0024)$ & $0.3877(0.0175)$ \\
\hline
5 & $0.5283(0.0170)$ & --- & $0.3160(0.0090)$ & --- \\
\hline
\end{tabular}
\caption{Ratios of the \Wpjn-jet production cross sections divided by
  the \Wmjn-jet production cross section
  (Table~\ref{tab_Wmj100TeV_total_xs} with respect to
  Table~\ref{tab_Wpj100TeV_total_xs}).\label{tab_Wm_Wp_j100TeV_total_xs}
}
\end{center}
\end{table}
%
%
\begin{table}[thpb]
\begin{center}
\begin{tabular}{|c||c|c||c|c|}
\hline
 \multicolumn{5}{|c|}{ Ratios\quad $R^{Z/W^+}_n$\quad for \quad $pp\rightarrow$\Zjn-jet+X \quad over\quad  $pp\rightarrow$\Wpjn-jet+X} \\
\hline
& \multicolumn{2}{c||}{low-democratic (100~TeV)}
& \multicolumn{2}{c|}{high-democratic (100~TeV)}\\
\hline
 $n$  & LO & NLO & LO & NLO \\
\hline
1 & $0.3303(0.0004)$ & $0.3128(0.0028)$ & $0.3143(0.0008)$ & $0.2939(0.0043)$ \\
\hline
2 & $0.3020(0.0006)$ & $0.3044(0.0084)$ & $0.2938(0.0009)$ & $0.2904(0.0062)$ \\
\hline
3 & $0.2941(0.0011)$ & $0.2830(0.0107)$ & $0.2816(0.0023)$ & $0.2864(0.0151)$ \\
\hline
4 & $0.2913(0.0023)$ & $0.2803(0.0274)$ & $0.2779(0.0023)$ & $0.2765(0.0162)$ \\
\hline
5 & $0.2790(0.0049)$ & --- & $0.2737(0.0041)$ & --- \\
\hline
& \multicolumn{2}{c||}{low-hierarchical (100~TeV)}
& \multicolumn{2}{c|}{high-hierarchical (100~TeV)}\\
\hline
1 & $0.2879(0.0002)$ & $0.2764(0.0014)$ & $0.2189(0.0001)$ & $0.2284(0.0007)$ \\
\hline
2 & $0.2749(0.0007)$ & $0.2754(0.0055)$ & $0.2241(0.0003)$ & $0.2289(0.0015)$ \\
\hline
3 & $0.2663(0.0015)$ & $0.2807(0.0224)$ & $0.2193(0.0007)$ & $0.2260(0.0079)$ \\
\hline
4 & $0.2611(0.0053)$ & $0.3902(0.0881)$ & $0.2167(0.0013)$ & $0.2331(0.0103)$ \\
\hline
5 & $0.2498(0.0075)$ & --- & $0.2097(0.0054)$ & --- \\
\hline
\end{tabular}
\caption{Ratios of the \Zjn-jet production cross sections divided by
  the \Wpjn-jet production cross section
  (Table~\ref{tab_Zj100TeV_total_xs} with respect to
  Table~\ref{tab_Wpj100TeV_total_xs}).\label{tab_Z_Wp_j100TeV_total_xs}
}
\end{center}
\end{table}

\subsection{Scaling behaviour: jet multiplicities or transverse momenta}
\label{sec:Vjets-scaling}

In this section we consider cross sections as a function of input
parameters or jet-multiplicities. Considering the behaviour of cross
sections as a function of initial state energies or transverse
momentum cuts allows one to understand the discovery potential as a
function of collision parameters.  For `basic' sets of cuts, we
display the dependence of the inclusive, fixed-order, NLO cross
sections as a function of the jet transverse-momentum cuts in
Figures~\ref{fig_ptscan_100TeV} and \ref{fig_ptscan_14TeV}, for
$\sqrt{s}=100$ and $14$ TeV respectively.  Comparing these two figures
we see that for the largest multiplicity shown ($n=4$) a variation in
$p_{T}^{\rm min}$ from $50$~GeV to $200$~GeV reduces cross sections by
one extra order of magnitude at $\sqrt{s}=14$~TeV.  Larger decreases
are of course expected for larger multiplicities, as the energy
available for extra radiation is increased by more than a factor of 7.

A very important observation extracted from
Figures~\ref{fig_ptscan_100TeV} and \ref{fig_ptscan_14TeV} is the
stability of the quantum corrections, related in part to the dynamical
scale choice $\mu=\hat H_{\rm T}^\prime$ that helps the LO predictions
to remain close to the more scale-independent NLO predictions. This
trend is associated with also the stability of total cross sections
explored in all kinematical regimes in
Tables~\ref{tab_Wmj100TeV_total_xs}-\ref{tab_Zj100TeV_total_xs}.

\begin{figure}[thpb]
\begin{center}
\includegraphics[clip,scale=0.45]{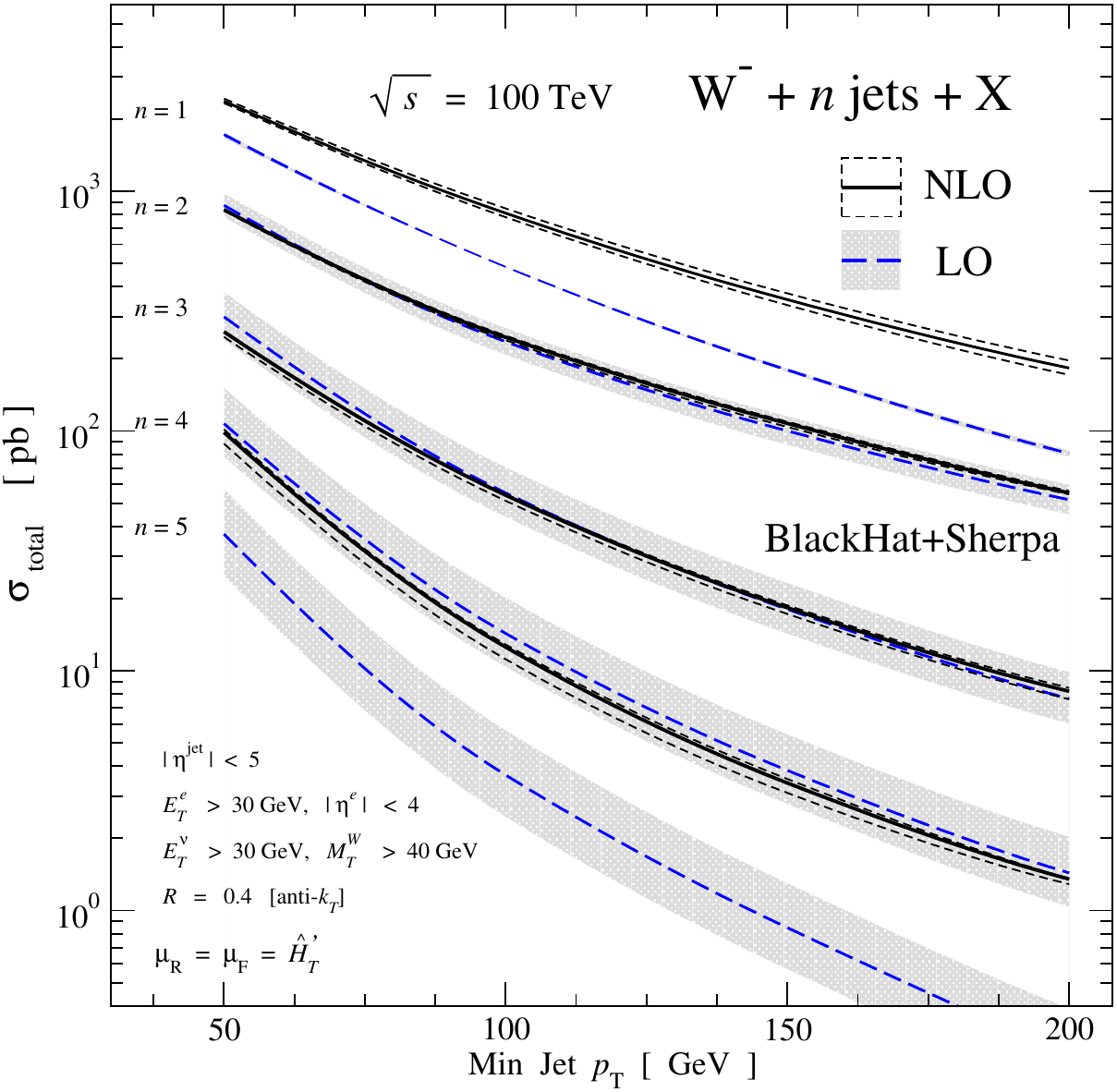}
\end{center}
\caption{Cross sections for \Wmjn-jets production as a function of the
  $p_{T}^{\rm min}$, as part of the `basic' set of cuts, with
  $\sqrt{s}=100$~TeV.  Top lines are for $n=1$ and bottom for $n=5$.
  Solid-black (dashed-blue) lines show corresponding NLO (LO) results,
  and we include also scale dependence bands with the small-dashed
  lines (shade).  }
\label{fig_ptscan_100TeV}
\end{figure}

\begin{figure}[thpb]
\begin{center}
  \includegraphics[clip,scale=0.45]{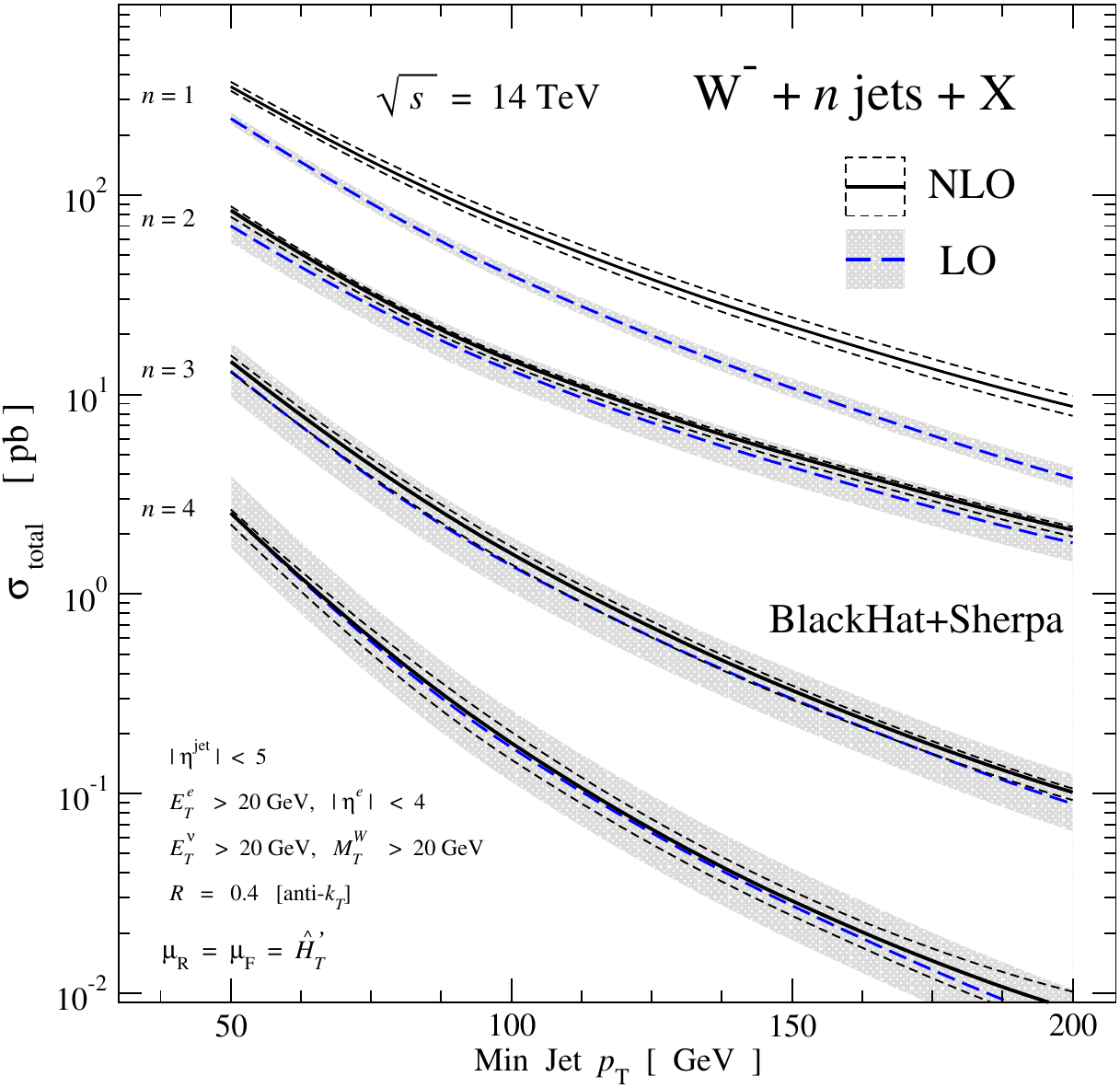}
\end{center}
\caption{Cross sections for \Wmjn-jets production as a function
of the $p_{T}^{\rm min}$, with $\sqrt{s}=14$~TeV.
The phase-space regions of the final-state objects is adjusted to
the initial-state energies as indicated in the figure.
Top lines are for $n=1$ and bottom lines for $n=4$.  Solid-black
(dashed-blue) lines show corresponding NLO (LO) results, and we
include also scale dependence bands with the small-dashed lines
(shade).  }
\label{fig_ptscan_14TeV}
\end{figure}

\subsection{Perturbative stability}
\label{sec:Vjets-validation}

Finally in this section, we validate the reliability of the
perturbative description of the scattering processes.  We explore both
the relative size of the quantum corrections ($K$-factors) and the
uncertainties associated to PDF's, as extracted from the error sets
provided by CT14nlo.

For convenience in Table \ref{tab_K_Wmj100TeV_total_xs} we give
explicit tables of the relative size of NLO corrections compared to
the fixed order LO predictions. The K-factors, defined by
$K_V^n=\sigma^{\rm V\!+\!n\!-\!jet}_{\rm NLO}/\sigma^{\rm
  V\!+\!n\!-\!jet}_{LO}$ are given in Table
\ref{tab_K_Wmj100TeV_total_xs}.  For our scale choice
(\ref{eqn:NLO-scale}) the corrections are modest for the `democratic'
jet cuts ranging from a K-factor 0.7 to 1.1 of associated production
of 2 to 4 jets. The K-factors increase with jet multiplicity.  The
case of single jet production behaves in a different way reaching
K-factors of 1.7\,. This increase is attributed to the known
phenomenon that additional production channels open in the real
radiation at NLO as well as the increasing phase space of hadrons
recoiling against the heavy vector boson
\cite{Catani:1997xc,Rubin:2010xp}, and it can be seen even more
markedly on the first jet differential distribution in
Figure~\ref{fig_Wm_jet_pt} (notice that NNLO QCD corrections stabilize
the perturbative
prediction~\cite{Ridder:2015dxa,Boughezal:2015ded}). NLO corrections
for $\sqrt{s}=14$~TeV and $\sqrt{s}=100$~TeV are comparable in size.
For hierarchical phase space cuts K-factors increase to the ranges 1.2
- 0.8 for $\sqrt{s}=14$~TeV and spread out to 1.1 - 0.3 at
$\sqrt{s}=100$~TeV for 2,3 and 4 associated jets. Single jet
production receives large corrections by a factor of 3.5 at
$\sqrt{s}=14$~TeV and up to factor of 7 at $\sqrt{s}=100$~TeV.
Reliable predictions for the associated production of heavy vector
bosons and a single jet require the inclusion of further corrections.
The hierarchical cuts introduce an additional scale as compared to the
`democratic' setup. The NLO prediction gives a better description of
the multi-scale processes as compared to the LO computation. The scale
setting, providing a renormalization and factorization scales
depending on the event kinematics, leads to a reliable predictions for
inclusive cross sections at LO.

\begin{table}[ht]
\begin{tabular}{|c||c||c||c|}
\hline \multicolumn{4}{|c|}{ K-factors for \quad $pp\rightarrow$ \Wmjn-jet+X   } \\\hline 
$n$ & low-democratic (14~TeV) & low-democratic (100~TeV)  & high-democratic (100~TeV) \\
\hline
1 & $1.78(1)$ & $1.686(9)$ & $4.4(1)$ \\
\hline
2 & $1.12(2)$ & $1.00(2)$ & $1.18(2)$ \\
\hline
3 & $1.01(2)$ & $0.84(3)$ & $1.09(3)$ \\
\hline
4 & $0.96(3)$ & $0.91(5)$ & $1.00(4)$ \\
\hline
 & low-hierarchical (14~TeV) & low-hierarchical (100~TeV)  & high-hierarchical (100~TeV) \\
\hline
1 & $3.49(1)$ & $7.19(2)$ & $13.01(4)$ \\
\hline
2 & $1.16(1)$ & $1.14(1)$ & $1.223(9)$ \\
\hline
3 & $0.90(2)$ & $0.48(4)$ & $1.05(1)$ \\
\hline
4 & $0.81(4)$ & $0.32(9)$ & $1.00(3)$ \\
\hline
\end{tabular}
\caption{K-factors for \Wmjn-jet~+~X predictions for
  Table~\ref{tab_Wmj100TeV_total_xs}. The size of the NLO corrections is
  representative for all types of heavy vector bosons
  discussed.}\label{tab_K_Wmj100TeV_total_xs} \end{table}

A further indicator for the validity of the perturbative predictions
are the jet ratios (\ref{eqn:Vjets-jetratio}). In the perturbative
regime an additional hard emission is dressed with a factor of the
strong coupling constant. The emission of additional jets should thus
be suppressed and jet ratios are expected to be small, i.e.\ smaller
than one.  In Table~\ref{tab_jr_Wmj100TeV_total_xs} we give the jet
ratios for the predictions
Tables~\ref{tab_Wmj100TeV_total_xs}-\ref{tab_Zj100TeV_total_xs}.  For
$n>2$ we observe ratios of the order one pointing to the consistency
of the fixed-order predictions. Further discussions of the scaling of
the inclusive cross sections can be found in Section
\ref{sec:Vjets-ratios}.  As a final probe of our current ability to
make meaningful quantitative predictions for proton-proton collisions
at a $100$~TeV machine, we collect the variance of the cross sections
induced by the PDF uncertainties. In table (\ref{tab_PDF_total_xs}) we
present the one-sigma relative uncertainties induced by the PDF's for
the NLO predictions of
Tables~\ref{tab_Wmj100TeV_total_xs}-\ref{tab_Zj100TeV_total_xs}.\\
We observe a comparable PDF uncertainties in $\sqrt{s}=14$~TeV and
$\sqrt{s}=100$~TeV predictions from $1-2 \%$ for moderate cuts
(`low-democratic'). The uncertainties rise in more extreme regions of
phases space; in the `high-democratic' region $1-3\%$ uncertainty
intervals are observed, in low-hierarchical $2-3\%$ and in the
high-hierarchical uncertainty intervals up to $7\%$ are observed for
the highest jet multiplicities.  \Wpjn-jet predictions are performing
best as expected for the proton-proton initial state. We also notice
slightly larger PDF errors for $W^-$ cross sections as compared to
$W^+$, pointing to larger contributions from better-constrained
valence quarks in the latter. Of course, with the data collected
finally at the LHC, better understanding shall follow, and these PDF
uncertainties will then be further reduced.

\begin{table}[ht]
\begin{tabular}{|c||c|c|c||c|c|c||c|c|c|}
\hline
& \multicolumn{3}{c||}{{  $pp\rightarrow$\Wmjn-jet+X }} 
& \multicolumn{3}{c||}{{  $pp\rightarrow$\Wpjn-jet+X }} 
& \multicolumn{3}{c|}{{  $pp\rightarrow $\Zjn-jet+X }} \\\hline
& \multicolumn{9}{c|}{{  democratic} } \\\hline
& \multicolumn{2}{c|}{{  low} } 
& \multicolumn{1}{c||}{{  high} } 
& \multicolumn{2}{c|}{{  low} } 
& \multicolumn{1}{c||}{{  high} } 
& \multicolumn{2}{c|}{{  low} } 
& \multicolumn{1}{c|}{{  high} } \\
\hline
$n/\sqrt{s}$  & 14~TeV & 100~TeV & 100~TeV  & 14~TeV & 100~TeV & 100~TeV & 14~TeV & 100~TeV & 100~TeV  \\
\hline
1 & $1.42\%$ & $1.89\%$  
& $1.42\%$ 
& $1.17\%$ & $1.76\%$ 
& $1.17\%$
& $1.19\%$ & $1.85\%$
& $1.36\%$ \\
\hline
2 & $1.42\%$ & $1.60\%$ 
& $1.55\%$  
& $1.13\%$ & $1.56\%$ 
& $1.19\%$
& $1.25\%$ & $1.52\%$ 
& $1.43\%$ \\
\hline
3 & $1.65\%$ & $1.44\%$  
& $2.08\%$
& $1.23\%$ & $1.48\%$ 
& $1.76\%$ 
& $1.48\%$ & $1.46\%$ 
& $2.03\%$ \\
\hline
4 & $2.25\%$ & $1.45\%$ 
& $2.76\%$ 
& $1.59\%$ & $1.43\%$ 
& $2.02\%$ 
& $2.05\%$ & $1.49\%$ 
& $2.69\%$ \\
\hline
& \multicolumn{9}{c|}{{ hierarchical} }\\ 
\hline
1 & $2.88\%$ & $2.18\%$ 
& $5.43\%$ 
& $2.05\%$ & $1.62\%$ & $3.30\%$ 
& $2.51\%$ & $1.99\%$ 
& $3.80\%$ \\
\hline
2 & $2.90\%$ & $2.14\%$ & $5.83\%$ 
& $2.09\%$ & $1.61\%$ 
& $3.11\%$
& $2.55\%$ & $2.01\%$
& $3.63\%$ \\
\hline
3 & $3.14\%$ & $2.10\%$ 
& $5.83\%$ 
& $2.31\%$ & $1.57\%$ 
& $3.32\%$ 
& $2.88\%$ & $2.06\%$ 
& $4.14\%$ \\
\hline
4 & $3.57\%$ & $3.03\%$ 
& $6.80\%$  
& $2.61\%$ & $1.26\%$ 
& $4.01\%$ 
& $3.36\%$ & $2.30\%$ 
& $4.70\%$ \\
\hline
\end{tabular}
\caption{Percentage PDF errors for one-sigma error bands for NLO
  $pp\rightarrow$\Vjn-jet+X cross sections in
  Tables~\ref{tab_Wmj100TeV_total_xs}-\ref{tab_Zj100TeV_total_xs}.
\label{tab_PDF_total_xs} } \end{table}

\clearpage
\section{Vector boson and heavy flavours\footnote{Editor: J. Campbell}}
\label{sec:vqq}

\def\zzz{zzz}

\subsection{Overview}

The production of a weak vector boson, $V=W^\pm,Z$. with a pair of
heavy quarks is important for a number of reasons.  These processes
admit the study of mechanisms of heavy flavor production for events
that can be more easily controlled experimentally, due to the presence
of the weak boson, particularly if it decays leptonically.  These
channels therefore open the possibility of constraining, for instance,
hypotheses of intrinsic heavy quark contributions to the proton
distribution functions.  For the case of bottom quarks the resulting
final states provide important backgrounds for many studies that will
be of high interest at a 100~TeV collider.  For example, $Vb\bar b$
production is an irreducible background to associated Higgs production
with subsequent Higgs boson decay, $H \to b \bar b$, and the case
$V=W^\pm$ represents backgrounds to several top quark production
processes.

Since the top quark is relatively short-lived and decays before
hadronizing, these cases are qualitatively different from $c$- and
$b$-quark production; the top production processes will therefore be
considered instead in Section~\ref{ref:top}.  For definiteness, here
we focus on the case of bottom quarks.  For the case of two
identified, well-separated heavy quarks at transverse momenta of order
$20$~GeV or higher, there is essentially no difference between the
rates for $Wb\bar b$ and $Wc\bar c$ production.  The $Zc\bar c$
production rate differs from that for $Zb\bar b$ due to the change
from a down-type to up-type coupling to quarks.  Due to the dominance
of the $gg \to Zb\bar b$ channel at 100~TeV, the $Zc\bar c$ rate is
very well-approximated from a rescaling by the ratio,
\begin{equation}
\frac{V_c^2 + A_c^2}{V_b^2 + A_b^2} \approx 0.78 \,,
\end{equation} 
where $V_q$ and $A_q$ are the vector and axial couplings of the $Z$
boson to quarks.
 
\begin{figure}[htb]
\begin{center}
\includegraphics[width=0.49\textwidth]{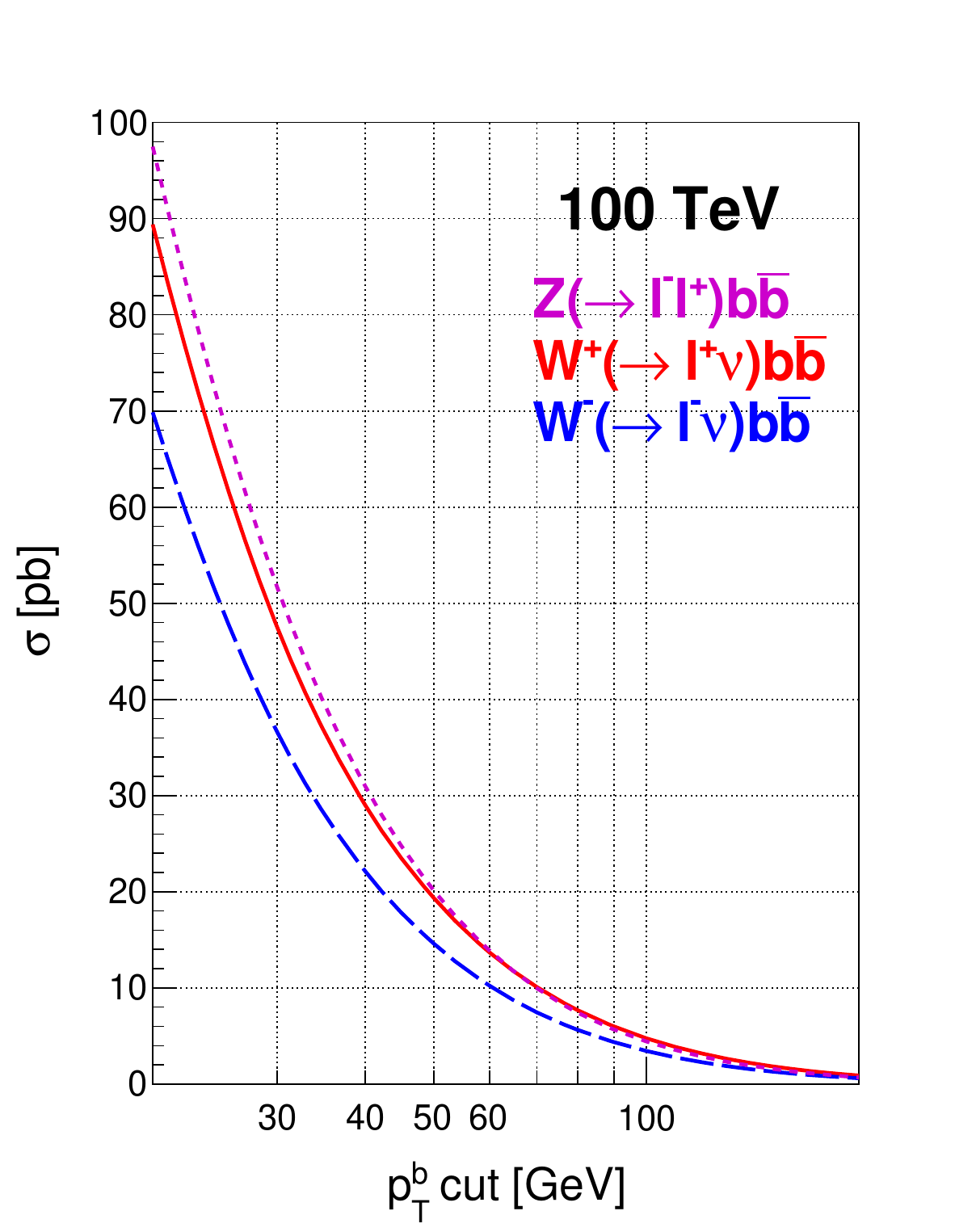}
\includegraphics[width=0.49\textwidth]{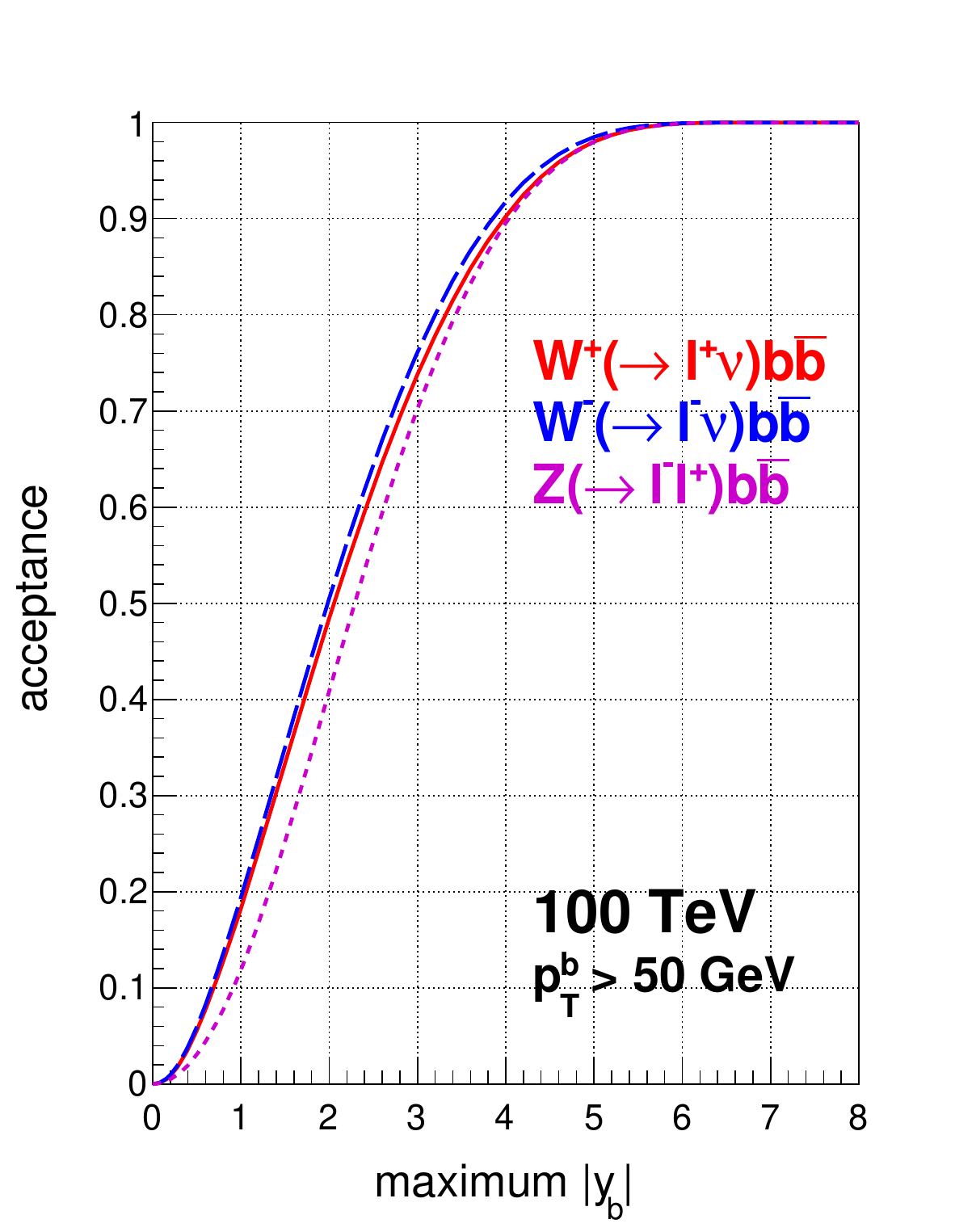}
\end{center}
\caption{Left: cross-sections for $Vb\bar b$ processes as a function
  of the minimum $b$-jet transverse momentum.  Right: the fraction of
  events accepted for a given maximum $b$-jet rapidity, for the case
  $p_T^b > 50$~GeV.}
\label{fig:VQQptyrap} 
\end{figure}
The $Vb{\bar b}$ rates for representative processes at a $pp$ collider
operating at 100 TeV are indicated in
figure~\ref{fig:VQQptyrap}.~\footnote{Cross-sections have been
  computed at NLO in
  MCFM~\cite{Campbell:1999ah,Campbell:2011bn,Campbell:2015qma}, using
  default parameters and the NLO CT14 pdf set with the scale choice
  $\mu_r = \mu_f = m_V$.}  Decays of the vector bosons into the
cleanest leptonic modes is assumed, accounting for a single family of
leptons only, i.e. $W \to e\nu$ and $Z \to e^- e^+$.  No acceptance
cuts are placed on the vector boson decay products while the bottom
quarks are clustered into $b$-jets using the anti-$k_T$ jet algorithm
with $R=0.4$. Events are only accepted if they contain at least two
$b$-jets, that are initially subject to only very loose cuts,
\begin{equation}
p_T^b, p_T^{\bar b} > 20~\mbox{GeV} , \qquad
|y^b|, |y^{\bar b}| < 10 \,.
\end{equation}
The impact of stricter cuts on the transverse momentum and rapidity of
the $b$-jets is also assessed.  In figure~\ref{fig:VQQptyrap} (left)
the cross-section is shown as a function of the minimum transverse
momentum of the $b$-jets.  Over the range shown all of the
cross-sections are of similar size.  This is purely coincidental since
the branching ratio for $Z \to e^- e^+$ is much smaller than for $W
\to e \nu$; before the vector boson decay the $Zb\bar b$ process is
much larger since it proceeds through LO diagrams with two gluons in
the initial state.  The reduction of the cross-section due to
more-realistic cuts on the $b$-jet rapidities can be gauged from
figure~\ref{fig:VQQptyrap} (right).  This shows the acceptance,
defined as $\sigma(|y_b| < |y_b^{max})/ \sigma(|y_b| < 10)$, for a
possible operating point represented by the cut $p_T^b > 50$~GeV.  The
acceptance is rather similar for all $Wb\bar b$ and $Zb \bar b$ cases,
although the somewhat broader $b$-jet rapidity distribution for the
$Zb \bar b$ process results in the smallest acceptance for a given
rapidity. Efficient $b$-tagging to rapidities of around 3 would
capture approximately $70$\% of the cross-section, while the $90$\%
level is only attained at 4 units of rapidity.  A summary of the
cross-sections at a few representative working points is shown in
table~\ref{table:Vbbxsec}.
\begin{table}[h!]
\begin{center}
\begin{tabular}{|c|c|c|c|}
  \hline
  Process & $p_T^b > 50$~GeV & $p_T^b > 50$~GeV, $|y_b|<3$ &  $p_T^b > 100$~GeV \\
  \hline
  $W^+ (\to \ell^+ \nu) b \bar b     $ & 19.4 & 14.3 & 4.76 \\
  $W^- (\to \ell^- \bar\nu) b \bar b $ & 14.7 & 11.2 & 3.45 \\
  $Z (\to \ell^- \ell^+ ) b \bar b   $ & 20.3 & 14.3 & 4.44 \\
  \hline
\end{tabular}
\caption{Cross-sections (in pb) for $Vb\bar b$ processes under various $b$-jet acceptance cuts.}
\label{table:Vbbxsec}
\end{center}
\end{table}

\begin{figure}[htb]
\begin{center}
\includegraphics[width=0.8\textwidth]{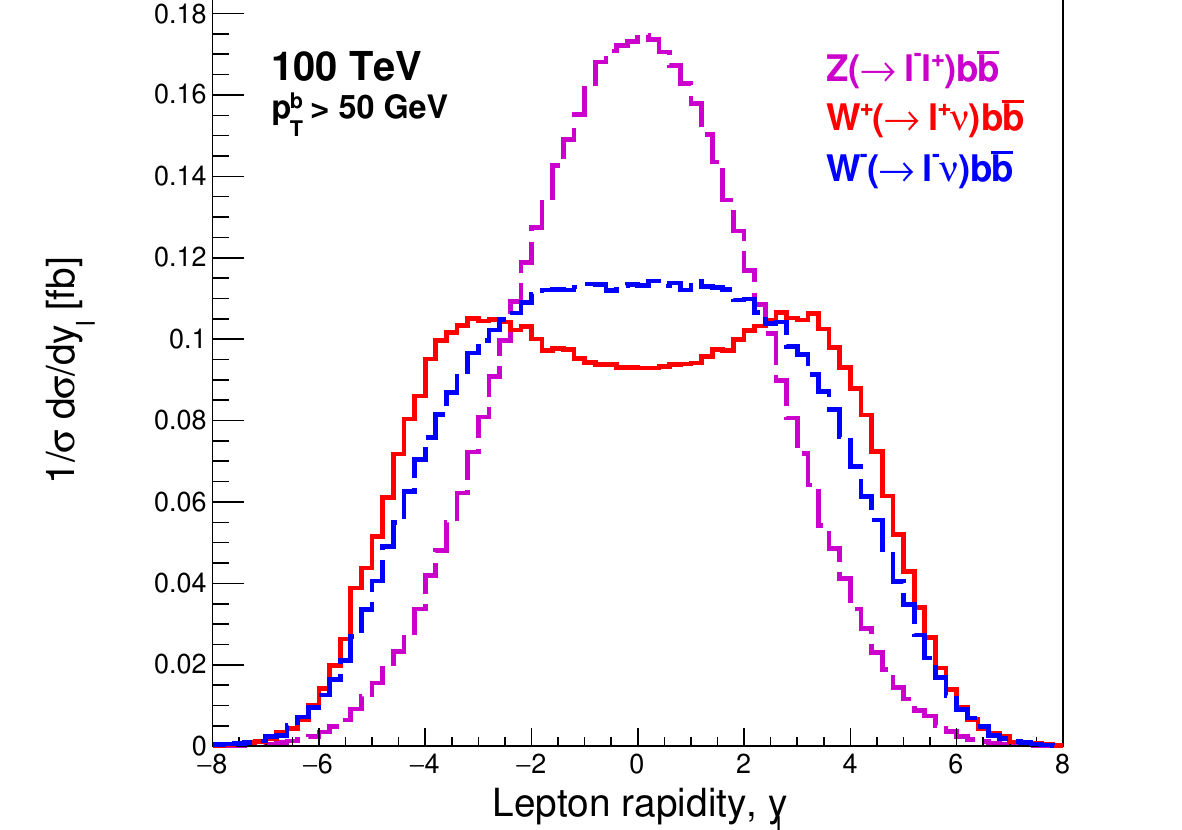}
\end{center}
\caption{The normalized rapidity distributions for the charged leptons produced in the $Vb\bar b$
processes, for the $b$-jet transverse momentum cut $p_T^b > 50$~GeV.}
\label{fig:VQQleprap} 
\end{figure}
Finally, figure~\ref{fig:VQQleprap} shows the shapes of the rapidity
distributions of the charged leptons in each process.  For the $Zb\bar
b$ process the dominance of the gluon pdf contributions leads to a
rather central distribution, with most leptons produced in the region
$|y| \lesssim 3$.  For $Wb\bar b$ production there is still a
significant valence-quark contribution that leads to a wider central
plateau, with a substantial number of events produced out to four
units of rapidity.

\subsection{Fully differential $Wb\bar b+X$ production}

We now turn to a careful investigation of $Wb\bar b+X$ production
using a fully differential calculation of the process in which a $W$
boson is produced in association with two $b$ jets and a further light
jet. Through the use of the \MINLO{}
prescription~\cite{Hamilton:2012np} this calculation can be used to
describe not only the emission of additional light jets, but also
inclusive $Wb\bar b$ production.  We will use this calculation to
study these final states under three sets of selection cuts, that are
appropriate for studies of $Wb\bar{b}+X$ production itself, or as a
background to $HW$ or single-top searches, respectively.

\subsubsection{Computational setup}
The computation is performed using the \Wbb{} and \Wbbj{} generators
available in the \POWHEGBOX{} framework~\cite{Nason:2004rx,
  Frixione:2007vw, Alioli:2010xd} and developed
in~\cite{Luisoni:2015mpa}. The tree-level amplitudes, which include
Born, real, spin- and colour-correlated Born amplitudes, were
automatically generated using an interface~\cite{Campbell:2012am} to
\MG{}~\cite{Stelzer:1994ta,Alwall:2007st}, whereas the one-loop
amplitudes were generated with
\GOSAM{}~\cite{Cullen:2011ac,Cullen:2014yla} via the
Binoth-Les-Houches (BLHA)
interface~\cite{Binoth:2010xt,Alioli:2013nda}, presented for the
\POWHEGBOX{} and \GOSAM{} in~\cite{Luisoni:2013cuh}.
\noindent
The version of \GOSAM{}~\cite{Cullen:2014yla} that was run is the 2.0: it
uses \QGRAF{}~\cite{Nogueira:1991ex}, \FORM~\cite{Kuipers:2012rf} and
\SPINNEY{}~\cite{Cullen:2010jv} for the generation of the Feynman
diagrams. These diagrams are then computed at running time with
\NINJA{}~\cite{vanDeurzen:2013saa,Peraro:2014cba}, which is a reduction
program based on the Laurent expansion of the
integrand~\cite{Mastrolia:2012bu}, and using {\tt
  OneLOop}~\cite{vanHameren:2010cp} for the evaluation of the scalar one-loop
integrals.
\noindent
For unstable phase-space points, the reduction automatically switches
to \GOLEM{}~\cite{Cullen:2011kv}, that allows to compute the same
one-loop amplitude evaluating tensor integrals.
\noindent
The \Wbb{} and \Wbbj{} generators include bottom-mass effects and spin
correlations of the leptonic decay products of the $W$ boson.  Despite
the fact that the computation is performed with massive quarks in the
decoupling scheme~\cite{Collins:1978wz}, where $\as$ is running with 4
light flavours, a switch to allow for a running with 5 light flavours
and the usage of pdfs with 5 flavours, as proposed
in~\cite{Cacciari:1998it}, has been implemented. The details technical
for the switch in the case at hand can be found in the Appendix of the
ref.~\cite{Luisoni:2015mpa}.

All the results have been obtained setting the bottom mass to
$m_{b}=4.75$~\GeV{} and using the
MMHT2014nlo68cl~\cite{Harland-Lang:2014zoa} pdf set. Jets have been
clustered with the \FASTJET{} package~\cite{Cacciari:2005hq,
  Cacciari:2011ma}, with radii which depend on the type of analysis
performed.  The renormalization and factorization scales have been set
according the \MINLO{} prescription~\cite{Hamilton:2012np}, as
described in ref.~\cite{Luisoni:2015mpa}.  The results presented in
the following sections have been computed at fixed next-to-leading
order level, plus \MiNLO.  Parton-shower effects have not been taken
into account.  The errors in the plots and in the tables have a
statistic origin and come from the numeric integration of the
results. No scales or pdf variations have been studied in this
contribution.

\subsubsection{$\boldsymbol{Wb\bar{b}}$ selection cuts}
\label{sec:wbbj_sel}
We begin by presenting results for the production of a $W$ boson in
association with two hard $b$ jets in the final state.  For this
analysis we use the anti-$\kt$ jet algorithm~\cite{Cacciari:2008gp}
with jet radius set to $R=0.4$.  We require the presence of exactly
two $b$ jets with transverse momentum $\pt^{\sss b}>50$~\GeV{} and we
apply three different cuts on the transverse momentum of additional
light jets, i.e.~$\pt^{\sss j}>1$, 100 or 500~\GeV.  This allows to
investigate the fully inclusive $Wb\bar{b}$ production, where the
light jet can become unresolved, as well as final states where
additional light jets are present. We stress that the former case can
be explored only due to the use of the \MINLO{} prescriptions, where
appropriate Sudakov form factors damp the soft and collinear regions
associated with the extra light jet. We also show some comparisons
with the NLO predictions obtained with the \Wbb{} generator, in which
the renormalization and factorization scales have been set to
$\mu=\sqrt{\hat{s}}/4$, where $\hat{s}$ is the square of the partonic
center-of-mass energy, as suggested in
ref.~\cite{Luisoni:2015mpa}. For exclusive kinematic regions, where
the jet is resolved and has high transverse momentum, the \Wbb{} code
describes the jet at LO, while the \Wbbj{} one gives a description at
NLO.

Results for the fiducial cross sections are reported in
Table~\ref{table:wbb_fiducial_XS} for the \Wbb{} generator, and in the
top rows of Table~\ref{table:wbbjets_fiducial_XS}, for the \Wbbj{}
one, where we also report the corresponding values computed at
$\sqrt{s}=14$~\TeV. The increase in the cross section from 14 to
100~\TeV{} is much larger than the relative increase in the
center-of-mass energy (roughly a factor of $7$), and it becomes larger
by sharpening the cuts on the transverse momenta of $b$ and light
jets. Furthermore, there is a 20\% difference between the NLO
$Wb\bar{b}$ cross section and the one for $Wb\bar{b}+1$ jet with
$\pt^{\sss j}>1$~\GeV. Instead, we note that the $100$~TeV result for
\Wbbj{} for the most inclusive case ($\pt^{\sss j}>1$~GeV) and two
$b$-jets with $\pt^{\sss b}>50$~\GeV{} ($34.0 \pm 0.6$pb) agrees
extremely well with the pure NLO prediction of $34.1$pb computed at
$\mu=m_{W}$ that has been presented in
Table~\ref{table:Vbbxsec}.\footnote{Apart from the difference in the
  method of calculation, there is also a small mismatch in the choice
  of PDFs.}

\begin{figure}[htb]
\begin{center}
\includegraphics[width=0.49\textwidth]{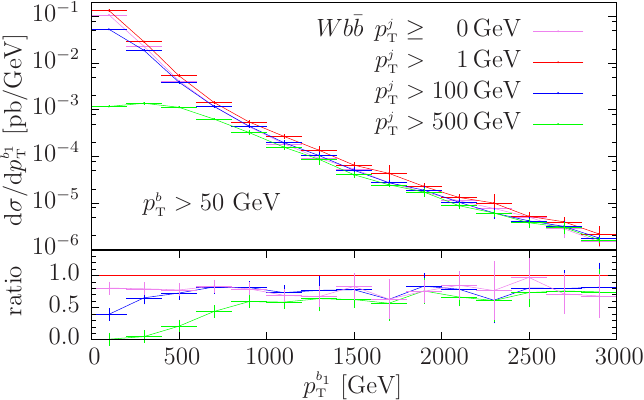}
\includegraphics[width=0.49\textwidth]{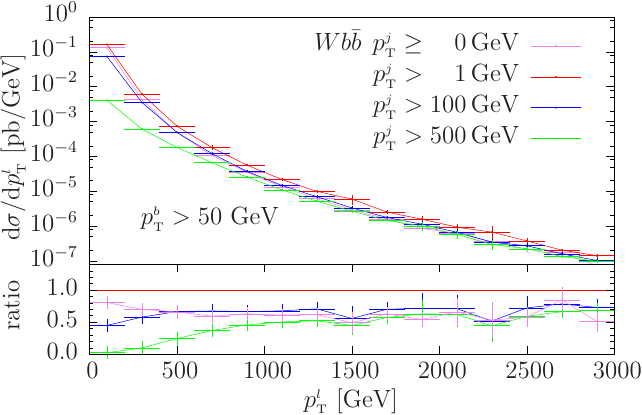}
\end{center}
\caption{Transverse-momentum distributions of the hardest $b$ jet~(left) and
  of the charged lepton~(right) for $Wb\bar{b}j$ production at
  $\sqrt{s}=100$~\TeV. The results using the NLO \Wbb{} code are shown too.}
\label{fig:b1-lept-pt}
\end{figure}
\begin{figure}[htb]
\begin{center}
\includegraphics[width=0.49\textwidth]{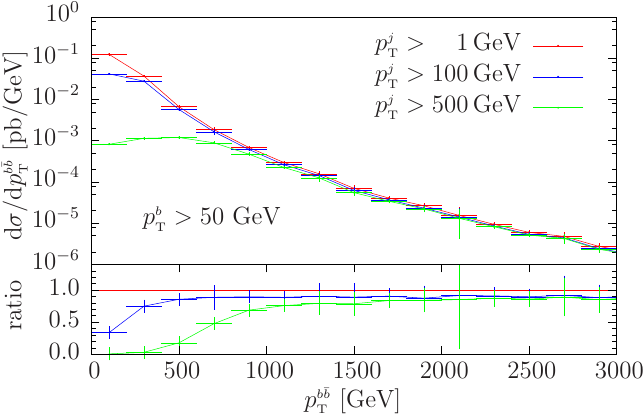}
\includegraphics[width=0.49\textwidth]{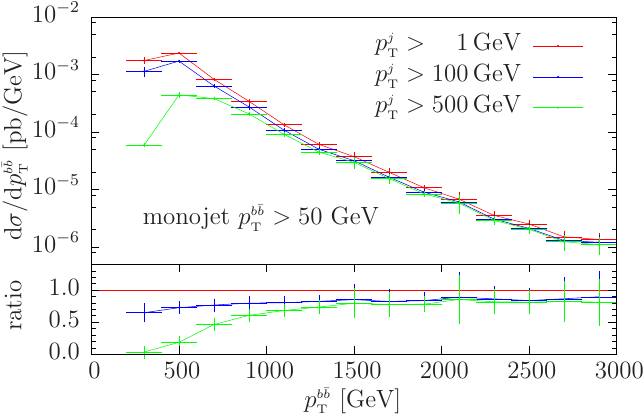}
\end{center}
\caption{Transverse-momentum distributions of the two-$b$ jet
  system~(left) and of the $b\bar{b}$-monojet~(right) for $Wb\bar{b}j$
  production at $\sqrt{s}=100$~\TeV. Details of the jet algorithm
  employed in the two cases are reported in the text.}
\label{fig:bb-pt}
\end{figure}
\begin{figure}[htb]
\begin{center}
\includegraphics[width=0.49\textwidth]{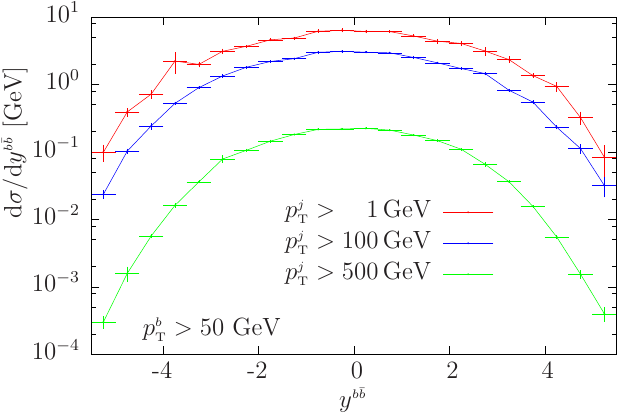}
\includegraphics[width=0.49\textwidth]{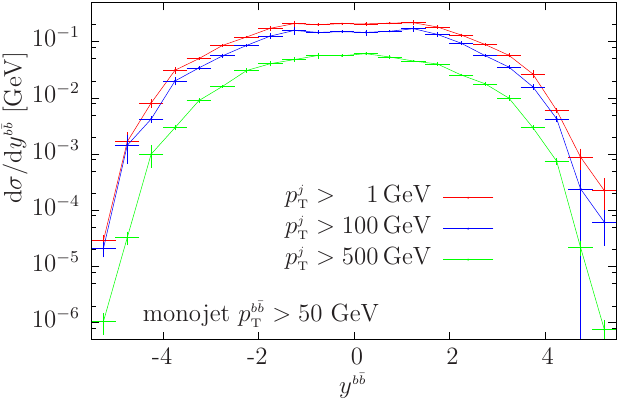}
\end{center}
\caption{Rapidity distributions of the two-$b$ jet system~(left) and
  of the $b\bar{b}$-monojet~(right) for $Wb\bar{b}j$ production at
  $\sqrt{s}=100$~\TeV. Details of the jet algorithm employed in the
  two cases are reported in the text.}
\label{fig:bb-y}
\end{figure}

Figure~\ref{fig:b1-lept-pt} shows the transverse-momentum distribution
of the hardest $b$ jet and of the charged lepton, respectively. These
observables can be described also by the \Wbb{} generator, and we plot
the corresponding curves for comparison.  Figures~\ref{fig:bb-pt}
and~\ref{fig:bb-y} display, on the left panels, the transverse
momentum and the rapidity distribution of the two-$b$ jet system,
respectively.  The right panels in the figures will be discussed in
Sec.~\ref{sec:mono_sel}. In all these plots the different $\pt$ cuts
on the light jets induce differential ratios which vary only in the
low transverse-momentum regions, while being almost constant
elsewhere. For the transverse-momentum distributions the differences
between $\pt^{\sss j}>100$~\GeV{} and $\pt^{\sss j}>500$~\GeV{} are
restricted to the region below 1 \TeV, where the bulk of the cross
section sits. For harder transverse momenta the cut on the light jet
loses its importance, leading to ratios of order one.  The impact of
the transverse-momentum cut on the light jet on the rapidity
distributions of the two-$b$ jet system has instead larger effects, as
can be seen in Fig.~\ref{fig:bb-y}, with constant ratios in the
considered rapidity range.

\begin{figure}[htb]
\begin{center}
\includegraphics[width=0.49\textwidth]{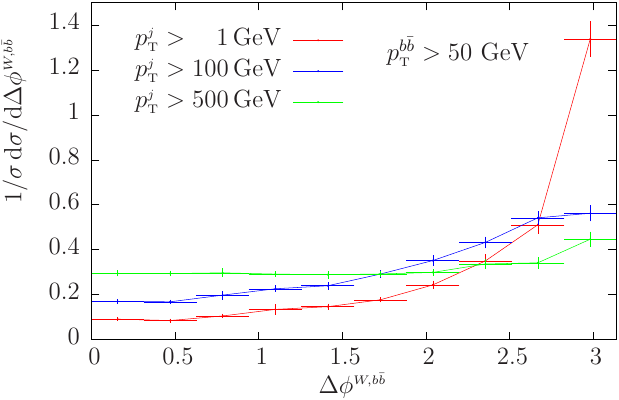}
\includegraphics[width=0.49\textwidth]{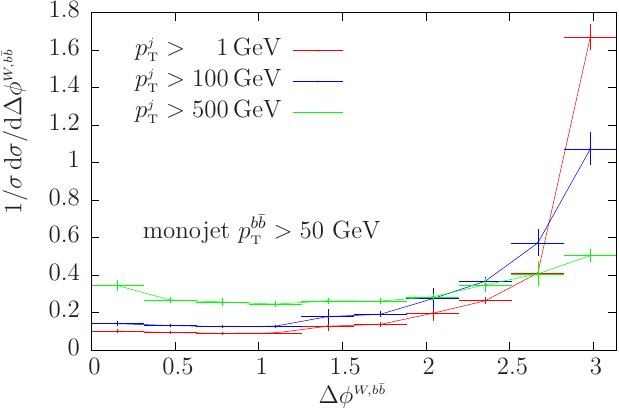}
\end{center}
\caption{Normalized differential cross-section distribution as a
  function of the azimuthal angle separation between the $W$ boson and
  the two-$b$ jet system, on the left, and the $b\bar{b}$-monojet, on
  the right, respectively, at $\sqrt{s}=100$~\TeV. Details of the jet
  algorithm employed in the two cases are reported in the text.}
\label{fig:W-bb-dphi}
\end{figure}
\begin{figure}[htb]
\begin{center}
\includegraphics[width=0.49\textwidth]{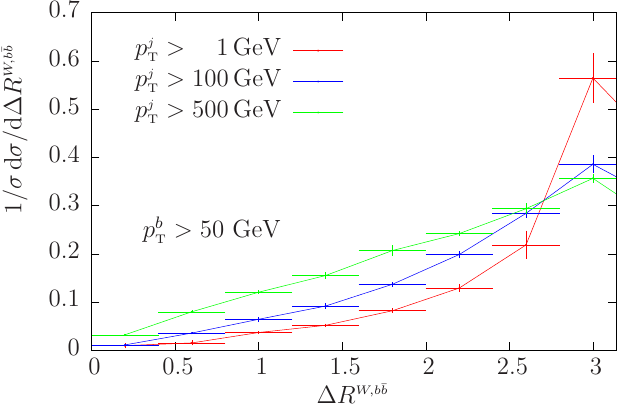}
\includegraphics[width=0.49\textwidth]{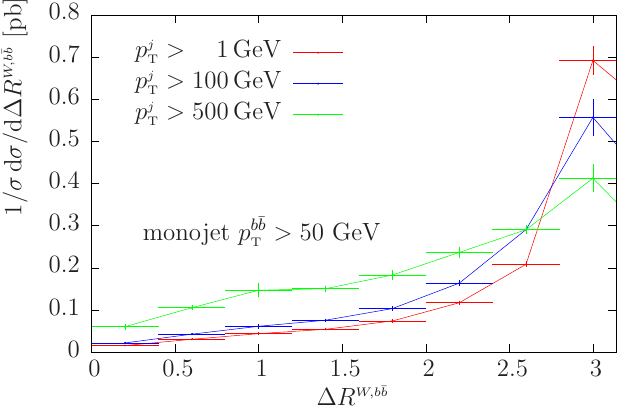}
\end{center}
\caption{Normalized differential cross-section distribution as a
  function of the radial distance between the $W$ boson and the
  two-$b$ jet system, on the left, and the $b\bar{b}$-monojet, on the
  right, respectively, at $\sqrt{s}=100$~\TeV. Details of the jet
  algorithm employed in the two cases are reported in the text.}
\label{fig:W-bb-dr}
\end{figure}

Figures~\ref{fig:W-bb-dphi} and~~\ref{fig:W-bb-dr} show the normalized
distribution of the azimuthal angle $\Delta\phi^{W,b\bar{b}}$ and the
radial distance $\Delta R^{W,b\bar{b}}$ between the $W$ boson and the
two-$b$ jet system respectively. In the most inclusive case
($\pt^{\sss j}>1$~\GeV), the $W$ boson and the $b\bar{b}$-system are
preferably produced back-to-back in azimuth and tend to have a large
radial distance. In addition, when extra hard jet radiation is
required, the distributions become flatter as the hardness of the
additional jet is increased.

\begin{figure}[htb]
\begin{center}
\includegraphics[width=0.49\textwidth]{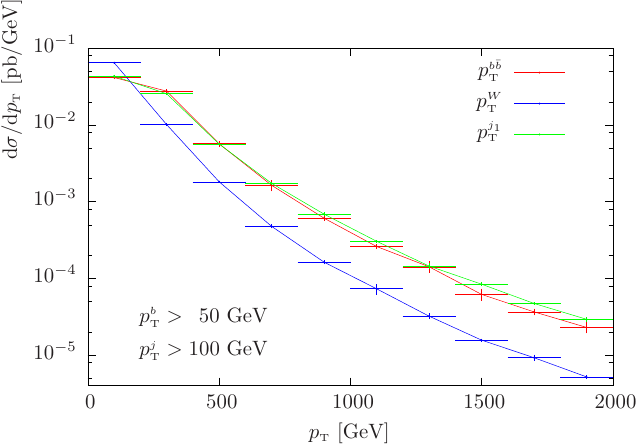}
\includegraphics[width=0.49\textwidth]{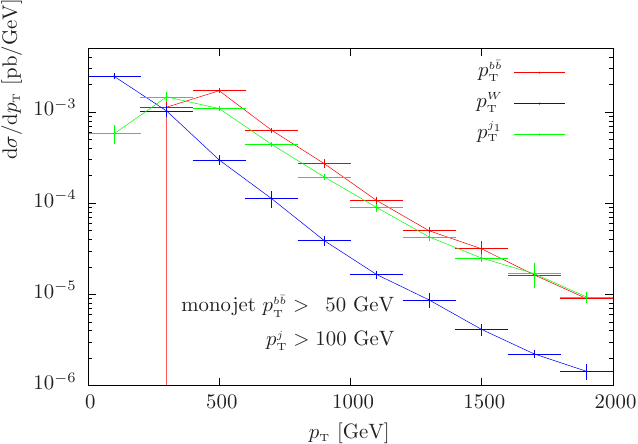}
\end{center}
\caption{Transverse-momentum distribution of the two-$b$ jet system,
  of the $W$ boson and of the hardest light jet. Ratio plots are shown
  too. The differential cross section has a cut on the $b$ jet, in the
  left panels, and on the monojet in the right ones.}
\label{fig:bbvsW-pt}
\end{figure}
\begin{figure}[htb]
\begin{center}
\includegraphics[width=0.49\textwidth]{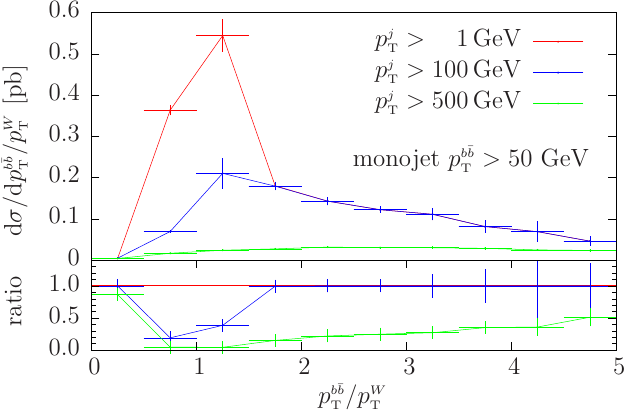}
\includegraphics[width=0.49\textwidth]{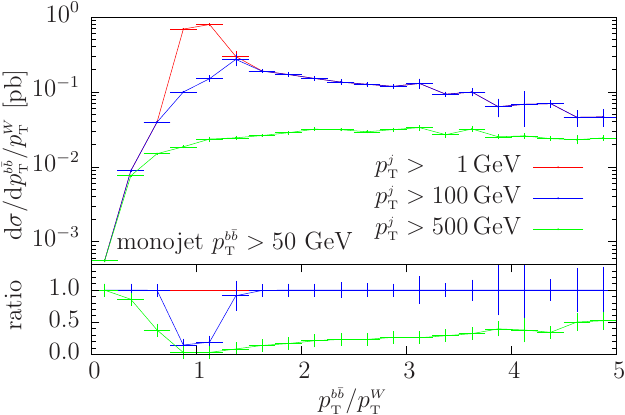}
\end{center}
\caption{Differential cross sections as a function of the ratio of the
  transverse momentum of the two-$b$ jet system over the $W$ boson one.}
\label{fig:bbvsW-ptratio}
\end{figure}
Finally, in Figure~\ref{fig:bbvsW-pt} the transverse momentum spectra
of the $W$ boson, the two-$b$ jet system and the extra light tagged
jet are compared. A clear difference in the distribution of the vector
boson with respect to the other twos can be seen, the former being
much softer at high transverse momentum. In the high-$\pt$ tail, we
note that the jet tends to be slightly harder than the two-$b$ jet
system.

\subsubsection{Higgsstrahlung selection cuts}
\label{sec:mono_sel}
In this section we investigate $Wb\bar{b}+X$ production as irreducible
background for the associated production of a Higgs boson and a $W$,
where the Higgs boson decays into a $b\bar{b}$ pair.
It is well known that, for boosted-boson kinematics, the signal to
background ratio for Higgs detection improves
considerably~\cite{Butterworth:2008iy}. In fact, in this case, there
is a high probability that the two $b$ quarks are clustered into a
single fat jet. We study then the level of background to this
associated Higgs production channel, by looking at events where the
$W$ boson is produced in association with a fat $b$-flavoured monojet,
containing the $b\bar{b}$ quark pair. These events are likely to
become very frequent at center-of-mass energies of the order of
hundreds of~\TeV.  In this analysis, jets are reconstructed using the
Cambridge/Aachen algorithm~\cite{Dokshitzer:1997in} with a jet radius
$R=0.7$. Furthermore we require the invariant mass of the monojet
$m^{b\bar{b}}$ to be between 100 and 150~\GeV , and a minimum
transverse momentum of $\pt^{b\bar{b}}>50$~\GeV.  As done in
Sec.~\ref{sec:wbbj_sel}, we impose three different transverse-momentum
cuts on the extra light jets, i.e.~$\pt^{\sss j}>1$, 100 or 500~\GeV.
The fiducial cross sections computed at 14 and 100~\TeV{} are
presented in the central rows of
Table~\ref{table:wbbjets_fiducial_XS}.

Coming to the differential distributions, in the right panels of the
Figures~\ref{fig:bb-pt}--\ref{fig:bbvsW-pt} we plot the same kinematic
distributions as plotted in the left panels, this time considering the
monojet instead of the two $b$ jets.  Due to the presence of the
additional cut on the invariant mass of the $b\bar{b}$ system, these
distributions are two order of magnitude smaller than the
corresponding ones in the left panels, but present similar shapes.
The right panel of Fig.~\ref{fig:W-bb-dr} shows the differential cross
section as a function of the azimuthal angle
$\Delta\phi^{W,b\bar{b}}$. This distribution is almost insensitive to
a cut on the transverse momentum of the light jet of 100~\GeV, while
it shows larger deviations with respect to the most inclusive case,
when the cut is increased to 500~\GeV. In the latter case, the
distribution becomes nearly flat over the whole kinematical range.

Dedicated analyses are needed to compare directly signal and
background, in order to assess the effectiveness of these cuts.

As far as the differential cross section as a function of the radial
distance $\Delta R^{W,b\bar{b}}$ is concerned,
Figure~\ref{fig:W-bb-dr} shows that only the events separated by a
large $\Delta R^{W,b\bar{b}}$ are more affected by the harder cut on
the additional jet transverse momentum.

\begin{figure}[htb]
\begin{center}
\includegraphics[width=0.49\textwidth]{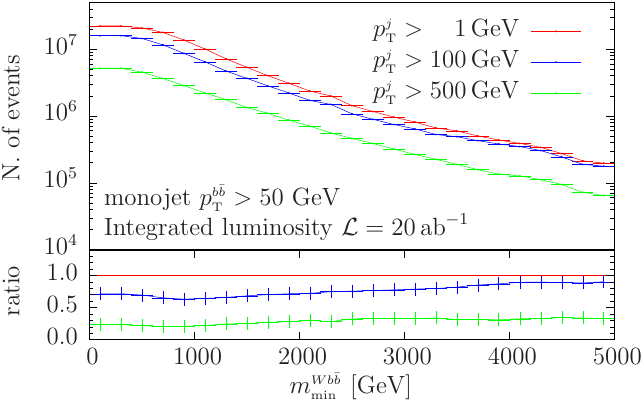}
\includegraphics[width=0.49\textwidth]{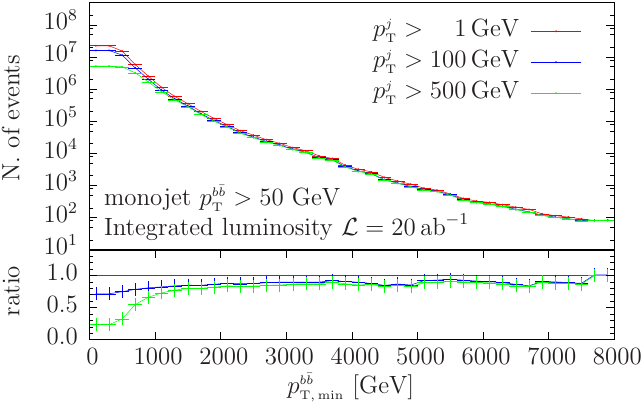}
\end{center}
\caption{Invariant-mass distribution of the $b\bar{b}$ system (left)
  and number of events as a function of the minimum transverse
  momentum of the $b\bar{b}$ system (right) in the monojet search at
  $\sqrt{s}=100$~\TeV. The number of events is computed assuming an
  integrated luminosity of $\mathcal{L}=20$~ab$^{-1}$.}
\label{fig:bb-m}
\end{figure}
The behavior of the ratios of the transverse momentum spectra of the
$W$ boson, of the two-$b$ jet system and of the extra light tagged
jet, shown in the right panels of Fig.~\ref{fig:bbvsW-pt}, for the
monojet search, is similar to the ones in the left panels.

In the study of the monojet selection cuts, it is interesting to study
the differential cross section as a function of the ratio
$\pt^{b\bar{b}}/\pt^{W}$, which is shown in
Figure~\ref{fig:bbvsW-ptratio}.  The two panels show the same
distribution on a linear~(left panel) and logarithmic scale~(right
panel).  While for the most inclusive case, the bulk of the cross
section is given by events where the ratio of the transverse momenta
is close to one, as the cut on the extra jet gets harder, the
distributions flatten, showing that events where the $W$ boson is
softer than the two-$b$ jet system clearly prevail.

In Figure~\ref{fig:bb-m}, on the left panel, the number of events as
function of the minimal invariant mass of the $Wb\bar{b}$ system is
shown. The right panel of Figure~\ref{fig:bb-m} shows instead the
number of events as a function of the minimum transverse momentum of
the monojet. In both cases, an integrated luminosity of
$\mathcal{L}=20$~ab$^{-1}$ is assumed. It is striking that, with the
aforementioned cuts, the number of background events induced by
$Wb\bar{b}+X$ is around $10^{6}$, with a transverse-momentum cut on
the monojet of the order of $1$~\TeV, even in the case where the
light-jet transverse momentum is required to be above 500~\GeV. This
fact should be kept in mind in order to asses the experimental
sensitivity, in searches of massive particles decaying into a pair of
boosted bottom quarks, in association with a hard lepton and missing
transverse energy.

\subsubsection{Single-top selection cuts}
\label{sec:stop_sel}
The last scenario we consider is single-top production. To estimate
the size and shape of the background induced by $Wb\bar{b}+X$
production on single-top searches, we require the presence of exactly
two resolved jets in the final state, one of which must be a $b$ jet,
while the other hast to be a light jet.
\noindent
We have recombined the partons using the anti-$\kt$ algorithm with
$R=0.4$ and have not distinguished between jets containing only one
$b$ quark, one $b$ anti-quark or a $b\bar{b}$ pair, considering them
on the same footing as a $b$ jet.
We have computed kinematic distributions applying the following cuts
on the transverse momentum of the $b$ jet and of the light jet
\begin{equation}
\pt^{\sss j/b}> 50, \ 100~\GeV. \nonumber
\end{equation}
Furthermore, we have imposed a cut on the transverse momentum of the
sum of the momenta of the $W$ and of the $b$ jet, to simulate the
effect of a cut on a reconstructed top-quark momentum $p^{\sss t}$, in
single-top production
\begin{equation}
\pt^{\sss t} > 0, \  500, \ 1000~\GeV.\nonumber
\end{equation}
We refer to the reconstructed $Wb$ system as ``top'' jet, in the following.
\begin{figure}[htb]
\begin{center}
\includegraphics[width=0.49\textwidth]{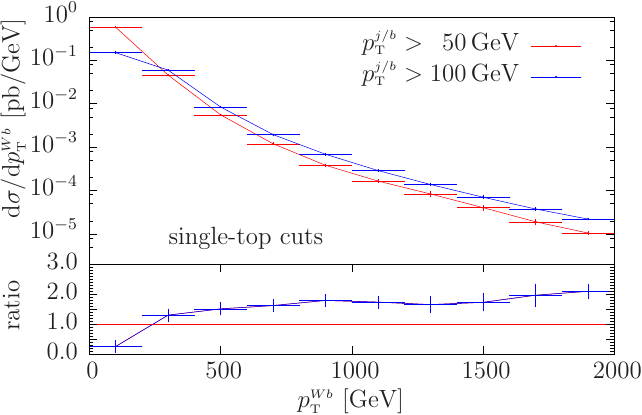}
\includegraphics[width=0.49\textwidth]{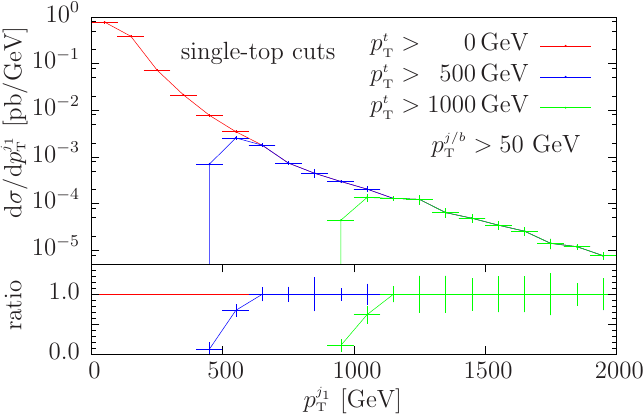}
\end{center}
\caption{Transverse-momentum distributions of the $Wb$ system~(left) and of
  the light jet~(right) at $\sqrt{s}=100$~\TeV.}
\label{fig:st-Wb-pt}
\end{figure}
\begin{figure}[htb]
\begin{center}
\includegraphics[width=0.49\textwidth]{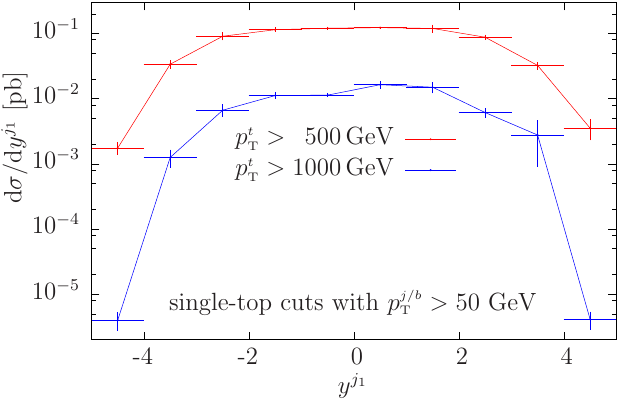}
\includegraphics[width=0.49\textwidth]{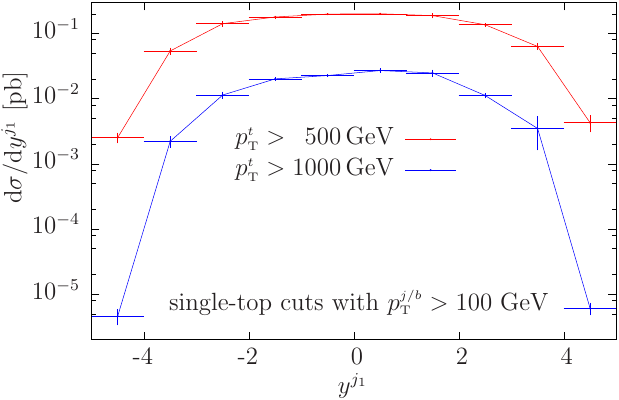}
\end{center}
\caption{Rapidity distribution of the light jet for two different
  transverse-momentum cuts on the reconstructed jets and on the ``top'', i.e.~the
  $Wb$ system, at $\sqrt{s}=100$~\TeV.}
\label{fig:st-j1-y}
\end{figure}
\begin{figure}[htb]
\begin{center}
\includegraphics[width=0.49\textwidth]{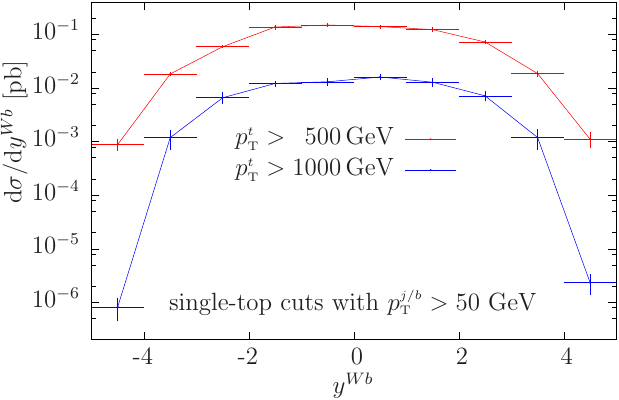}
\includegraphics[width=0.49\textwidth]{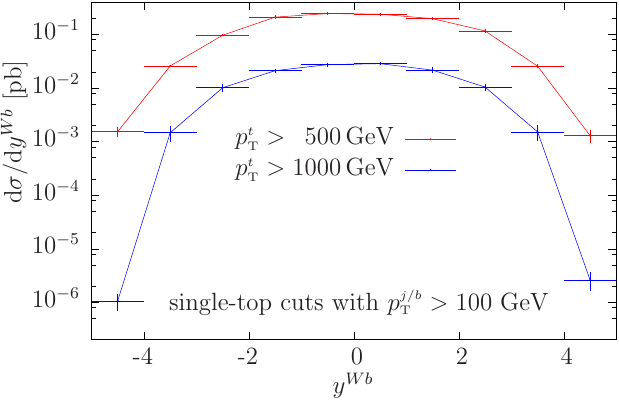}
\end{center}
\caption{Rapidity distribution of the ``top'' system for two different
  transverse-momentum cuts on the reconstructed jets, at
  $\sqrt{s}=100$~\TeV.}
\label{fig:st-Wb-y}
\end{figure}
\begin{figure}[htb]
\begin{center}
\includegraphics[width=0.49\textwidth]{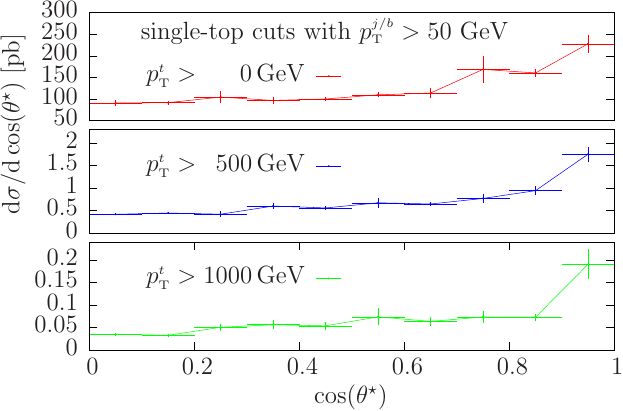}
\includegraphics[width=0.49\textwidth]{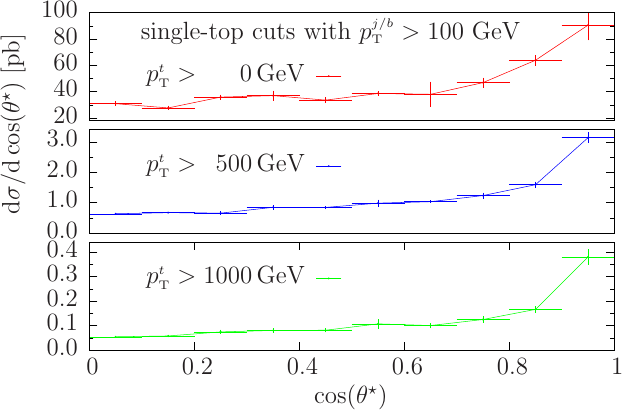}
\end{center}
\caption{Differential cross section as a function of the cosine of the angle
  $\theta^*$ between the charged lepton and the light jet, in the ``top''
  rest frame, at $\sqrt{s}=100$~\TeV.}
\label{fig:st-costhstar}
\end{figure}

\noindent
In the last rows of Table~\ref{table:wbbjets_fiducial_XS} we give the
fiducial cross sections computed within the acceptance cuts reported
above. We observe an inversion when comparing the effect of the
transverse-momentum cut on the $b$ jet for $\pt^{\sss t}>0$~\GeV{} and
$\pt^{\sss t}>500$, 1000~\GeV. In the former case, the fiducial cross
section decreases both at 14 and at 100~\TeV, whereas in the latter
twos, the cross sections increase when applying harder cuts. This is
due to the peculiarity of the adopted event selection: in fact,
requiring only one resolved $b$ jet induces automatically a veto on
the second $b$ jet present at LO. By hardening the cut on $\pt^{b}$, a
wider kinematic region opens up for the additional unresolved $b$ jet,
leading to the observed increase.  It would be interesting in the
future to study the sensitivity of the single-top signal to this cut,
and to compare it to the one we are studying here.  The inversion is
also clearly visible in the first bin in the left panel of
Figure~\ref{fig:st-Wb-pt}, which shows the transverse-momentum
distribution of the ``top'' jet for $\pt^{\sss j/b}>50$~\GeV{} and
$\pt^{\sss j/b}>100$~\GeV. On the right panel of
Figure~\ref{fig:st-Wb-pt}, we plot the transverse-momentum
distributions of the light tagged jet for different cuts on the
``top'' jet transverse momentum. Finally, in
Figures~\ref{fig:st-j1-y}--\ref{fig:st-costhstar} we plot the
differential cross sections as a function of the rapidity of the light
jet, of the rapidity of the ``top'' jet and of the cosine of the angle
$\theta^*$ between the charged lepton and the light jet in the ``top''
rest frame. In the left panels of these figures we consider a cut on
the $b$ and light jet of $\pt^{\sss j/b}>50$~\GeV, while in the right
panels, this cut has been set to $\pt^{\sss j/b}>100$~\GeV. While the
shape of the curves in the two panels are very similar, increasing the
cut on the ``top'' jet decreases the distributions by more than one
order of magnitude.

\begin{table}[h!]
\centering
\small
\begin{tabular}{| r @{$\,\pm\,$} l | r @{$\,\pm\,$} l || r @{$\,\pm\,$} l | r @{$\,\pm\,$} l |}
  \hline
  \multicolumn{4}{|c||}{$\sigma^{Wb\bar{b}}_{\mathrm{NLO}}$ [pb] @ 14 \TeV} &
  \multicolumn{4}{|c|}{$\sigma^{Wb\bar{b}}_{\mathrm{NLO}}$ [pb] @ 100 \TeV\phantom{\Big|}}     \\
  \hline
  \multicolumn{8}{|c|}{$Wb\bar{b}$ selection cuts\phantom{\Big|}}\\
  \hline
  \multicolumn{2}{|c|}{$\pt^{\sss b}> 0$~\GeV}
  &\multicolumn{2}{|c||}{$\pt^{\sss b}> 50$~\GeV}
  &\multicolumn{2}{|c|}{$\pt^{\sss b}> 0$~\GeV}
  &\multicolumn{2}{|c|}{$\pt^{\sss b}> 50$~\GeV\phantom{\Big|}} \\
  \hline
    $102.83$ & $0.07$ & $1.399$ & $0.001$ & $988$ & $11$ & $27.28$ & $0.03$ \\
  \hline
\end{tabular}
\caption{$Wb\bar{b}$ fiducial cross sections in pb at NLO accuracy for the
  scenario considered in Sec.~\ref{sec:wbbj_sel}, for a proton-proton
  collider at 14 and 100~\TeV, computed with the \Wbb{} code.}
\label{table:wbb_fiducial_XS}
\end{table}

\begin{table}[h!]
\centering
\small
\begin{tabular}{|l|| r @{$\,\pm\,$} l | r @{$\,\pm\,$} l || r @{$\,\pm\,$} l
    | r @{$\,\pm\,$} l |}
\hline
  \ & \multicolumn{4}{|c||}{$\sigma_{\mathrm{NLO+MiNLO}}$ [pb] @ 14
    \TeV} & \multicolumn{4}{|c|}{$\sigma_{\mathrm{NLO+MiNLO}}$ [pb] @ 100 \TeV\phantom{\Big|}}     \\
  \hline
  \multicolumn{9}{|c|}{$Wb\bar{b}$ selection cuts\phantom{\Big|}}\\
  \hline
  & \multicolumn{2}{|c|}{$\pt^{\sss b}> 0$~\GeV}
  & \multicolumn{2}{|c||}{$\pt^{\sss b}> 50$~\GeV}
  & \multicolumn{2}{|c|}{$\pt^{\sss b}> 0$~\GeV}
  & \multicolumn{2}{|c|}{$\pt^{\sss b}> 50$~\GeV\phantom{\Big|}} \\
  \hline
  $\pt^{\sss j}>\phantom{10}1$ \GeV &  $  96.0$ & $   6.7$ & $   1.78$ & $   0.13$ & $1179$ & $  46$ & $34.0$ & $ 0.6$ \\
  $\pt^{\sss j}>100$           \GeV &  $  5.84$ & $  0.09$ & $  0.416$ & $  0.008$ & $ 149$ & $ 4.0$ & $15.5$ & $ 0.1$ \\
  $\pt^{\sss j}>500$           \GeV &  $0.0355$ & $0.0003$ & $0.00764$ & $0.00004$ & $3.80$ & $0.17$ & $1.00$ & $0.01$ \\
  \hline
  \multicolumn{9}{|c|}{Higgsstrahlung selection cuts\phantom{\Big|}}\\
  \hline
  & \multicolumn{4}{|c||}{$\pt^{\sss b}> 50$~\GeV}
  & \multicolumn{4}{|c|}{$\pt^{\sss b}> 50$~\GeV\phantom{\Big|}} \\
  \hline
  $\pt^{\sss j}>\phantom{10}1$  \GeV & \multicolumn{2}{|r @{$\;\pm$}}{$0.0215$} & \multicolumn{2}{l||}{\hspace{-0.1cm}$0.0003$}
                                   & \multicolumn{2}{|r @{$\;\pm$}}{$  1.11$} & \multicolumn{2}{l|}{\hspace{-0.2cm}$0.022$} \\
  $\pt^{\sss j}>100$            \GeV & \multicolumn{2}{|r @{$\;\pm$}}{$0.0122$} & \multicolumn{2}{l||}{\hspace{-0.1cm}$0.0002$}
                                   & \multicolumn{2}{|r @{$\;\pm$}}{$ 0.794$} & \multicolumn{2}{l|}{\hspace{-0.2cm}$0.021$} \\
  $\pt^{\sss j}>500$            \GeV & \multicolumn{2}{|r @{$\;\pm$}}{$0.00237$} & \multicolumn{2}{l||}{\hspace{-0.1cm}$0.00002$}
                                   & \multicolumn{2}{|r @{$\;\pm$}}{$  0.259$} & \multicolumn{2}{l|}{\hspace{-0.2cm}$0.005$} \\
  \hline
  \hline
 \multicolumn{9}{|c|}{Single-top selection cuts\phantom{\Big|}}\\
  \hline
  & \multicolumn{2}{|c|}{$\pt^{\sss j/b}>50$~\GeV}
  & \multicolumn{2}{|c||}{$\pt^{\sss j/b}>100$~\GeV}
  & \multicolumn{2}{|c|}{$\pt^{\sss j/b}>50$~\GeV}
  & \multicolumn{2}{|c|}{$\pt^{\sss j/b}>100$~\GeV \phantom{\Bigg|}} \\
  \hline
  $\pt^{\sss t}>\phantom{100}0$  \GeV & $6.00   $ & $0.18   $ & $1.62   $ & $0.06   $ & $126   $ & $ 4     $ & $44.3 $ & $ 1.6  $ \\
  $\pt^{\sss t}>\phantom{1}500$  \GeV & $0.009  $ & $0.001  $ & $0.12   $ & $0.001  $ & $0.72  $ & $ 0.02  $ & $1.16  $ & $ 0.03 $ \\
  $\pt^{\sss t}>1000$            \GeV & $0.0005 $ & $0.0001 $ & $0.0006 $ & $0.0001 $ & $0.070 $ & $ 0.004 $ & $0.123 $ & $ 0.005  $ \\
  \hline

  \hline
\end{tabular}
\caption{$Wb\bar{b}j$ fiducial cross sections in pb at NLO+MiNLO accuracy for
  the different scenarios considered in Secs.~\ref{sec:wbbj_sel},
  \ref{sec:mono_sel} and~\ref{sec:stop_sel} for a proton-proton collider at
  14 and 100~\TeV, computed with the \Wbbj{} code.}
\label{table:wbbjets_fiducial_XS}
\end{table}

\clearpage
\section{Gauge boson pair production\footnote{Editor: D. Rathlev}} 
\label{sec:VV}

\def\Matrix{{\sc Matrix}}
\def\Munich{{\sc Munich}}
\def\OpenLoops{{\sc OpenLoops}}
\def\Collier{{\sc Collier}}
\def\CutTools{{\sc CutTools}}
\def\OneLOop{{\sc OneLOop}}

\def\zz{\ensuremath{ZZ}}
\def\z{\ensuremath{Z}}

\def\abbrev{}
\def\llog{\text{\abbrev LL}}
\def\nll{\text{\abbrev NLL}}
\def\nnll{\text{\abbrev NNLL}}
\def\lo{\text{\abbrev LO}}
\def\nlo{\text{\abbrev NLO}}
\def\nnlo{\text{\abbrev NNLO}}
\def\nlonll{\nlo\plus\nll}
\def\nnlonnll{\nnlo\plus\nnll}
\def\qcd{{\abbrev QCD}}

\def\pppbar{pp}
\def\ppbar{p\bar{p}}
\def\pp{\textrm{pp}}
\def\qqbar{q\bar{q}}
\def\qqbarprime{q{\bar q^\prime}}
\def\bbbar{b\bar{b}}
\def\ccbar{c\bar{c}}
\def\ttbar{t\bar{t}}
\def\cbar{\bar{c}}

\def\citere#1{Ref.~\cite{#1}}
\def\citeres#1{Refs.~\cite{#1}}
\def\reffi#1{Fig.~\ref{#1}}
\def\reffis#1{Figs.~\ref{#1}}
\def\refta#1{Tab.~\ref{#1}}

\def\Tstrut{\rule{0pt}{3.0ex}}         
\def\Bstrut{\rule[-1.5ex]{0pt}{0pt}}   

\subsection{ZZ production}
\label{sec:VV_Vpair}
All numerical results in this section and the next have been produced
with the {{\rmfamily \scshape Matrix}\xspace} code \footnote{\Matrix{}
  is the abbreviation of ``\Munich{} Automates qT subtraction and
  Resummation to Integrate Cross Sections'', by M.~Grazzini,
  S.~Kallweit, D.~Rathlev, M.~Wiesemann.  In preparation.}. For the SM
parameters we use $m_W=80.399$ GeV, $m_Z=91.1876$ GeV,
$\Gamma_W=2.1054$ GeV, $\Gamma_Z=2.4952$ GeV and $G_F=1.6639\cdot
10^{-5}$ GeV$^{-2}$.

We start by considering the rapidity acceptance of $ZZ\to
e^+e^-\mu^+\mu^-$ production. We apply basic $ZZ$ selection cuts of
$66$ GeV $<m_{\ell\ell}<116$ GeV on the invariant mass of oppositely
charged leptons of the same flavour and consider two different $p_T$
thresholds of $20$ and $100$ GeV on the leptons. Renormalization and
factorization scales are set to the sum of transverse energies of the
two $Z$ bosons, $\mu_R=\mu_F=\mu=E_T^{Z,1}+E_T^{Z,2}$, with
$E_T^{Z}=\sqrt{m_Z^2+\left(p_T^{Z}\right)^2}$, and we use LO, NLO and
NNLO MMHT2012 sets ~\cite{Harland-Lang:2014zoa} at the LO, NLO and
NNLO respectively. Tab. \ref{tab:ZZ_total} shows the fiducial cross
section corresponding to this setup at LO, NLO and NLO+gg, in which
the finite and gauge invariant gluon-fusion contribution is
included. For comparison, Tab. \ref{tab:ZZ_total} also provides the
inclusive cross section without any transverse-momentum cut at 100
TeV, for which we also provide the NNLO cross section. It can be seen
that at 100 TeV the gluon-fusion contribution provides roughly 70\% of
the full NNLO correction, consistent with Ref.\cite{Cascioli:2014yka}.

\begin{table}[ht]
  \begin{center}
    \begin{tabular}{|c| c| c| c| c|}
      \hline
      {\rule{0pt}{3.0ex}}
      $\sqrt{s}$ (TeV) & $\sigma_{\textrm{LO}}$ (fb) &
      $\sigma_{\textrm{NLO}}$ (fb) & $\sigma_{\textrm{NLO+gg}}$ (fb) & $\sigma_{\textrm{NNLO}}$ (fb)\\ [0.5ex]
      \hline
      {\rule{0pt}{3.0ex}}
      14 $\left(p_T^\ell>20\, \textrm{GeV}\right)$ & 15.51 & 21.63 & 23.71 &{\rule[-1.5ex]{0pt}{0pt}}\\
      \hline
      {\rule{0pt}{3.0ex}}
      100 $\left(\mathrm{incl.}\right)$ & 284.7  & 361 & 430 & 460 {\rule[-1.5ex]{0pt}{0pt}}\\
      \hline
      {\rule{0pt}{3.0ex}}
      100 $\left(p_T^\ell>20\, \textrm{GeV}\right)$ & 181.8 & 230.2 & 269.2 &{\rule[-1.5ex]{0pt}{0pt}}\\
      \hline
      {\rule{0pt}{3.0ex}}
      100 $\left(p_T^\ell>100\, \textrm{GeV}\right)$ & 0.4778 & 0.888  & 1.514 &{\rule[-1.5ex]{0pt}{0pt}}\\
      \hline
    \end{tabular}
  \end{center}
  \renewcommand{\baselinestretch}{1.0}
  \caption{Fiducial cross section for $ZZ$ production at the LHC at
    LO, NLO and NLO+gg. Leptonic branching ratios included.} 
  \label{tab:ZZ_total}
\end{table}

Fig. \ref{fig:ZZ_eta} shows the rapidity acceptance
$\sigma(|\eta^\ell|<\eta_{cut})/\sigma$ for the final-state leptons as
a function of the maximum rapidity cut. For a cut on the minimal
lepton transverse momentum of $20$ GeV, a rapidity cut with
$\eta_{cut}\approx 3$ removes around 50\% of the total cross
section. If the lepton transverse momentum cut is increased to $100$
GeV, the leptons are forced to be more transverse, and a rapidity cut
of $\eta_{cut}\approx 2$ retains 50\% of the cross section.

\begin{figure}[htpb]
        \centering
        \includegraphics[width=0.8\textwidth]{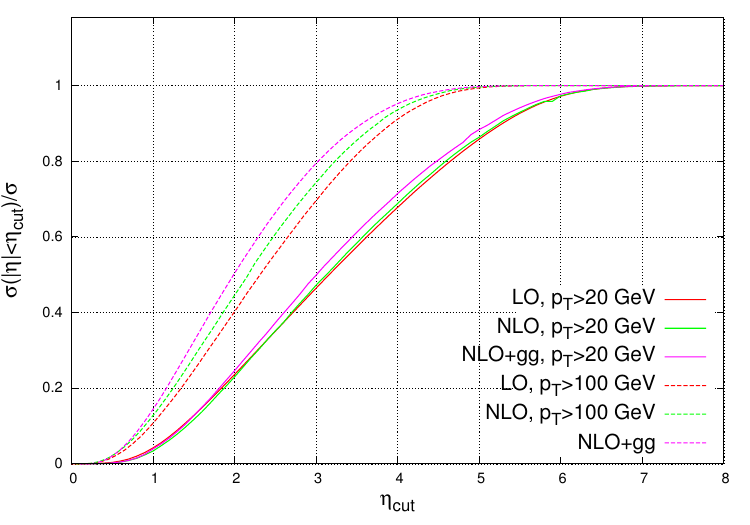}
        \caption{Rapidity acceptance of 4 lepton production at 100 TeV
          as a function of the maximum lepton rapidity at LO (red),
          NLO (green) and NNLO (blue) with a $p_T^{\ell}>20$ GeV
          (solid) and a $p_T^{\ell}>100$ GeV cut (dashed).}
    \label{fig:ZZ_eta}
\end{figure}

For comparison, Fig. \ref{fig:ZZ_eta_14} shows the rapidity acceptance
at a center of mass energy of $14$ TeV and a minimum lepton transverse
momentum of $20$ GeV. Compared to the situation at $100$ TeV, the
events are more central and a rapidity cut of $\eta_{cut}\approx 3$
retains more than 70\% of the cross section.

\begin{figure}[htpb]
        \centering
        \includegraphics[width=0.8\textwidth]{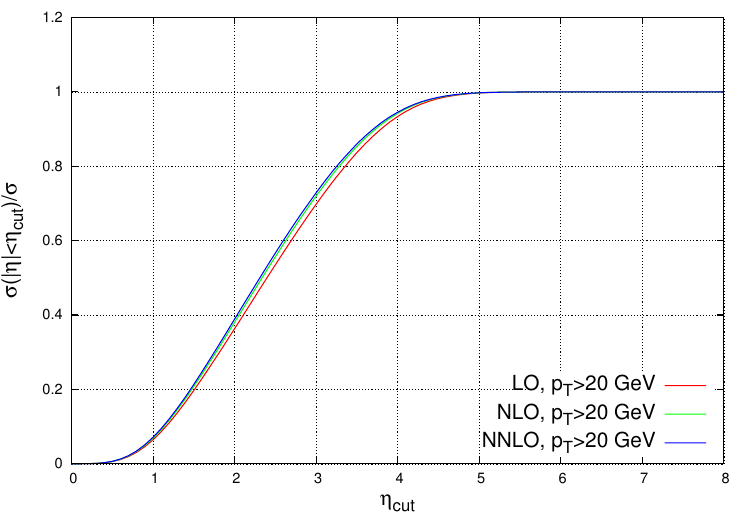}
        \caption{Rapidity acceptance of 4 lepton production at 14 TeV
          as a function of the maximum lepton rapidity at LO, NLO and
          NNLO for $p_T^{\ell}>20$ GeV.}
    \label{fig:ZZ_eta_14}
\end{figure}

Fig. \ref{fig:ZZ_pT} shows the acceptance as a function of the minimal
lepton transverse momentum. The cross section is rapidly falling when
increasing the transverse momentum requirements on the leptons, and a
cut of $100$ GeV leads to a reduction of the cross section of more
than a factor of 200 when compared to the original cut of $20$ GeV.
\begin{figure}[htpb]
        \centering
        \includegraphics[width=0.8\textwidth]{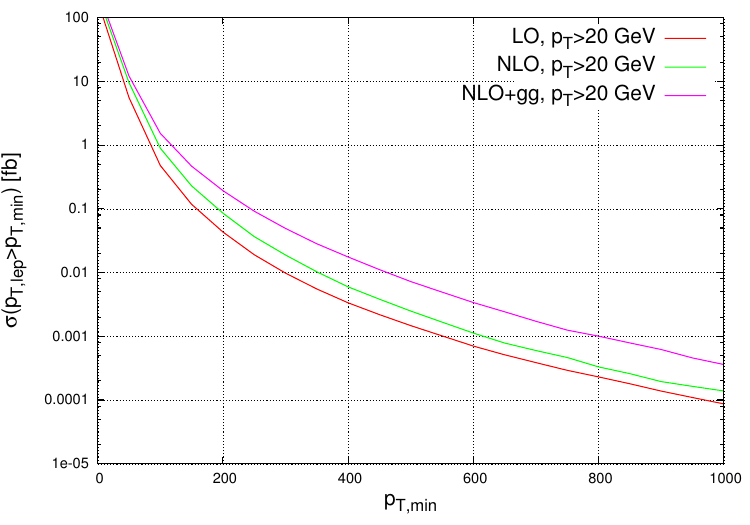}
        \caption{4 lepton production cross section at 100 TeV as a
          function of the minimum lepton transverse momentum at LO,
          NLO and NNLO.}
    \label{fig:ZZ_pT}
\end{figure}

New physics at very high energies can be described by an effective
field theory at lower energies, where heavy particles running in loops
might modify the couplings between SM particles. The effective
operators are suppressed by the scale of new physics, and are
therefore most pronounced in the high-energy tales of
distributions. Fig. \ref{fig:ZZ_m} shows the 4 lepton cross section
above a minimal cut on the invariant mass of the final-state
system. Although it drops off rapidly, even at very high invariant
masses of $\sim 2$ TeV a cross section of around 1 fb remains.

\begin{figure}[htpb]
        \centering
        \includegraphics[width=0.8\textwidth]{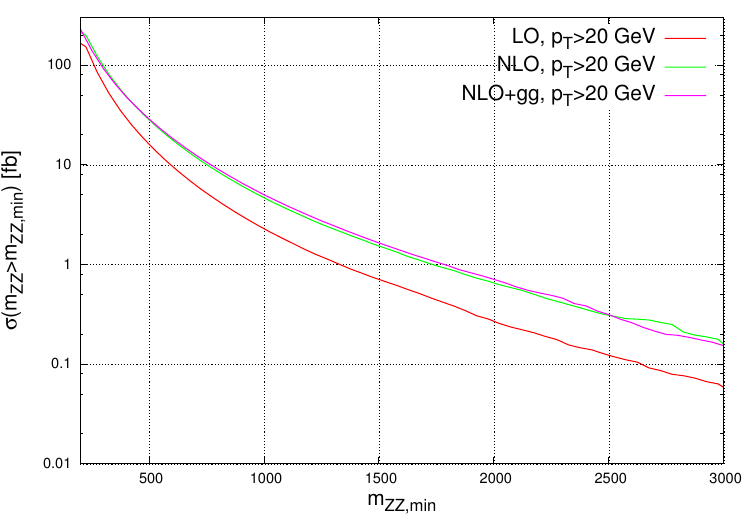}
        \caption{4 lepton production cross section at 100 TeV as a
          function of the minimum invariant mass of the final-state
          system at LO and NLO.}
    \label{fig:ZZ_m}
\end{figure}

\subsection{WW production}
\subsubsection{Top-contamination issues}
We now move to $W^+W^-$ production. Compared to ZZ production,
$W^+W^-$ production comes with the additional complication that the
inclusive cross section is not straight-forwardly defined in
perturbation theory. The reason lies in the contamination by $Wt$ and
$t\bar{t}$ production entering at NLO and NNLO, respectively, if the
bottom-quark is considered massless
\cite{Gehrmann:2014fva}. \reffi{fig:ww_diags_singlet} shows diagrams
contributing to the single-real correction to $W^+W^-$
production. While the non-resonant diagrams (left) are part of the
genuine QCD corrections, also resonant $Wt$ diagrams appear in the
same partonic channel. Resonant $Wt$ production amounts to around 30\%
of the LO $W^+W^-$ cross section. The problem is even more severe at
NNLO, where diagrams as the one shown in \reffi{fig:ww_diags_doublet}
start to contribute in the double-real emission correction. Besides
QCD corrections to $W^+W^-$ production (left), the same channel also
contains diagrams from resonant $\ttbar$ production, leading to an
increase of the cross section of around 400\%.

\begin{figure*}[ht]
\begin{center}
\hspace{-4mm}
\includegraphics[scale=1.3,trim=1.5cm 26.0cm 15cm 0.5cm]{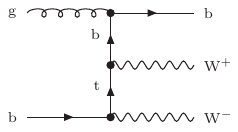}
\hspace{8mm}
    \includegraphics[scale=1.3,trim=1.5cm 26.0cm 15cm 0.5cm]{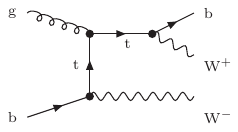} 
\end{center}
\vspace{-4mm}
  \caption{Feynman diagrams contributing to the $gb\to W^+W^-b$ subprocess.}
  \label{fig:ww_diags_singlet}
\end{figure*}

\begin{figure*}[ht]
\begin{center}
\hspace{-4mm}
\includegraphics[scale=1.3,trim=1.5cm 26.0cm 15cm 0.5cm]{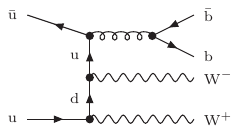}
\hspace{8mm}
    \includegraphics[scale=1.3,trim=1.5cm 26.0cm 15cm 0.5cm]{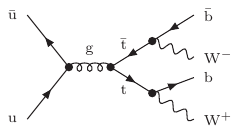} 
\end{center}
\vspace{-4mm}
  \caption{Feynman diagrams contributing to the $u\bar{u}\to W^+W^-\bbbar$ subprocess.}
  \label{fig:ww_diags_doublet}
\end{figure*}

While the top-contamination only affects partonic channels involving
b-quarks in the external states, these channels cannot
straight-forwardly be neglected in the computation, as they are
crucial to the cancellation of collinear divergences. However, they
can be rendered IR finite by specifying a finite b-quark mass, i.e.\
by working in a 4-flavour scheme (4FS). In a 4FS, all partonic
channels with external b-quarks and thus the top-contamination can be
removed from the computation, resulting in a sensible definition of
the $W^+W^-$ cross section. This procedure leads to additional
theoretical uncertainties on the level of $2\%$ for LHC collider
energies, which is well below the remaining scale dependence even at
NNLO (see \citere{Gehrmann:2014fva} for more details).

However, using a finite b-quark at a 100 TeV collider is much less
justified, and one might worry about missing significant contributions
from $\bbbar$ initial states. To obtain a rough estimate of the size
of these effects, one can compare the LO cross sections obtained in
the 4FS and in the 5FS. We find that with NNPDF3.0 sets, the
difference is negligible at 14 TeV, and amounts to $\sim-5\%$ at a 100
TeV proton-proton collider. We conclude that while the
top-contamination problem cannot be considered solved at 100 TeV, a
4FS computation can be used to obtain a useful estimate of the cross
sections for $W^+W^-$ production at a future 100 TeV collider.

\subsubsection{Predictions at 100 TeV}
We apply a lower cut of $10$ GeV on the invariant mass of the
electron-muon pair and consider two different $p_T$ thresholds of $20$
and $100$ GeV. In both setups we require a minimal missing transverse
momentum equal to the lepton $p_T$ threshold. Renormalization and
factorization scales are set to the sum of transverse energies of the
two $W$ bosons, $\mu=E_T^{W^+}+E_T^{W^-}$, with
$E_T^{W^\pm}=\sqrt{m_W^2+\left(p_T^{W^\pm}\right)^2}$. As the full
NNLO calculation including the leptonic decay is not available yet, we
limit the discussion to the NLO results. We do however include the
gluon-fusion contribution. \refta{tab:WW_total} shows the fiducial
cross sections at 14 and 100 TeV. The scale uncertainties are on the
level of $\pm 15\%$ at LO and $\pm 4\%$ at NLO. The PDF uncertainties
are estimated to be $\pm 7\%$ at LO and reduce to $\pm 1\%$ at NLO. We
note that there are huge NLO corrections when applying a strict
transverse-momentum cut of 100 GeV. This is in contrast to the
analogous results for $ZZ$ production in \refta{tab:ZZ_total}. The
difference is due to the fact that the missing transverse-momentum cut
(instead of the cut on individual leptons as in the $ZZ$ case)
suppresses configurations with back-to-back neutrinos, and favors
final states with a large total transverse momentum of the $W^+W^-$
system. Non-vanishing $W^+W^-$ transverse momenta only arise at the
next-to-leading order.

\begin{table}[ht]
  \begin{center}
    \begin{tabular}{|c| c| c| c| c|}
      \hline
      {\rule{0pt}{3.0ex}}
      $\sqrt{s}$ (TeV) & $\sigma_{\textrm{LO}}$ (fb) & $\sigma_{\textrm{NLO}}$ (fb)& $\sigma_{\textrm{NLO+gg}}$ (fb) & $\sigma_{\textrm{NNLO}}$ (fb)\\ [0.5ex]
      \hline
      {\rule{0pt}{3.0ex}}
      14 $\left(p_T^\ell>20\, \textrm{GeV}\right)$ & 509 & 759 & 805 & {\rule[-1.5ex]{0pt}{0pt}}\\
      \hline
      {\rule{0pt}{3.0ex}}
      100 $\left(\mathrm{incl.}\right)$ & 8162  & 12877 & 13992 & 15362 {\rule[-1.5ex]{0pt}{0pt}}\\
      \hline
      {\rule{0pt}{3.0ex}}
      100 $\left(p_T^\ell>20\, \textrm{GeV}\right)$ & 4685 & 8027 & 8738 & {\rule[-1.5ex]{0pt}{0pt}}\\
      \hline
      {\rule{0pt}{3.0ex}}
      100 $\left(p_T^\ell>100\, \textrm{GeV}\right)$ & 18.09 & 89.6 & 98.3 & {\rule[-1.5ex]{0pt}{0pt}}\\
      \hline
    \end{tabular}
  \end{center}
  \renewcommand{\baselinestretch}{1.0}
  \caption{Fiducial cross section for $W^+W^-$ production at the LHC
    at LO, NLO and NLO+gg. Leptonic branching ratios included.} 
  \label{tab:WW_total}
\end{table}

Fig. \ref{fig:WW_eta} shows the rapidity acceptance
$\sigma(|\eta^\ell|<\eta_{cut})/\sigma$ for the final-state leptons as
a function of the maximum rapidity cut. For a cut on the minimal
lepton transverse momentum of $20$ GeV, a rapidity cut with
$\eta_{cut}\approx 3$ removes around 45\% of the total cross
section. If the lepton transverse momentum cut is increased to $100$
GeV, the leptons are forced to be more transverse, and a rapidity cut
of $\eta_{cut}\approx 2$ retains 50\% of the cross section.

\begin{figure}[htpb]
        \centering
        \includegraphics[width=0.8\textwidth]{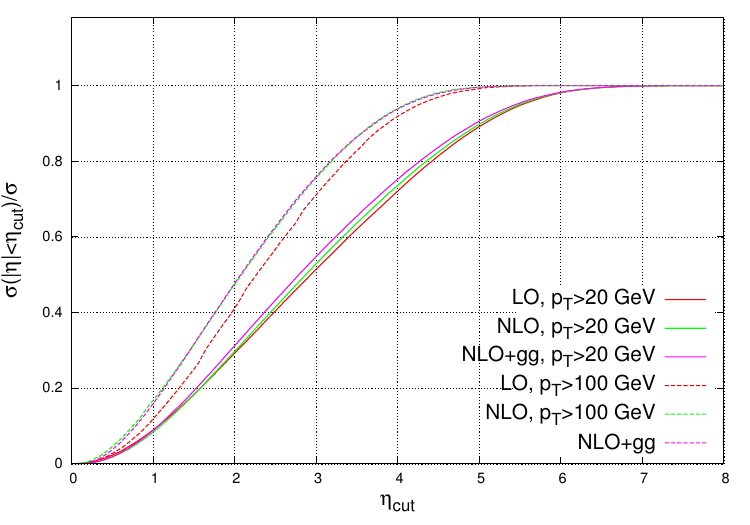}
        \caption{Rapidity acceptance of $W^+W^-$ production at 100 TeV
          as a function of the maximum lepton rapidity at LO (red) and
          NLO (green) with a $p_T^{\ell}>20$ GeV (solid) and a
          $p_T^{\ell}>100$ GeV cut (dashed).}
    \label{fig:WW_eta}
\end{figure}

Fig. \ref{fig:WW_pT} shows the cross section as a function of the
minimal lepton transverse momentum. Transverse momentum cuts higher
than $\sim150$ GeV cut away more than $99\%$ of the fiducial cross
section.
\begin{figure}[htpb]
        \centering
        \includegraphics[width=0.8\textwidth]{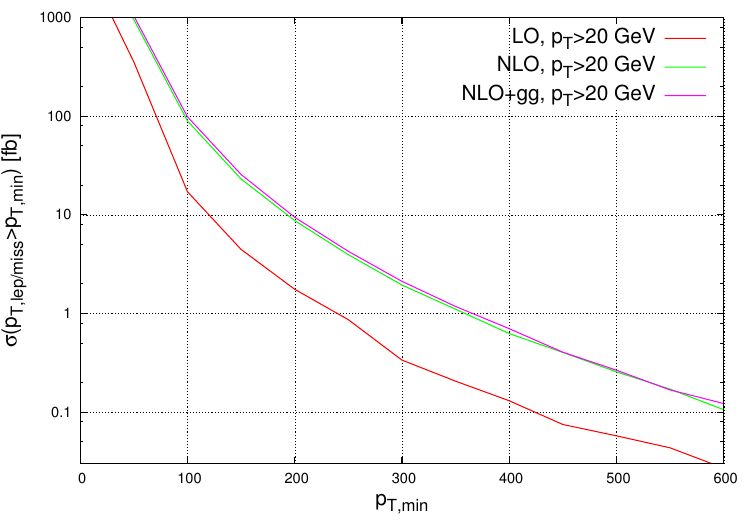}
    \caption{$W^+W^-$ cross section at 100 TeV as a function of the minimum lepton transverse momentum at LO and NLO.}
    \label{fig:WW_pT}
\end{figure}

\clearpage
\subsection{$\gamma\gamma$ production} 
\label{sec:VV_phph}
Diphoton production at hadronic colliders is a very relevant process,
both from the point of view of testing the SM
predictions~\cite{Chatrchyan:2011qt,Aaltonen:2011vk,Chatrchyan:2014fsa,Abazov:2013pua,Aad:2011mh,Aad:2012tba}
as for new physics searches.  \emph{Direct} or \emph{prompt} photons
provide an ideal test to QCD since they constitute a theoretically and
experimentally clean final state: on the theory side, because they do
not have QCD interactions with other final state particles;
experimentally, because photon energies and momenta can be measured
with high precision by modern detectors.

Besides purely QCD-related considerations, diphoton final states have
played a crucial role in the recent discovery of the Higgs boson at
the LHC \cite{cha:2012gu,aad:2012gk}. They are also important in many
new physics scenarios \cite{:2012afa,:2012mx}, in particular in the
search for extra-dimensions~\cite{Aad:2012cy} or
supersymmetry~\cite{CMS:2012un}. And,
recently~\cite{CMS:2015dxe,ATLAS:2015750}, an excess in the diphoton
invariant mass spectrum was observed in searches for new physics in
high mass diphoton events in $pp$ collisions at 13 TeV.

We are interested in the process $pp\rightarrow \gamma \gamma X$.  The
lowest-order process ($\mathcal{O}(\alpha_S^0)$) occurs via the quark
annihilation subprocess $q\bar{q}\rightarrow \gamma\gamma$. The
next-to-leading order (NLO) QCD corrections have been computed and
implemented in the fully-differential Monte Carlo codes
\texttt{DIPHOX}~\cite{Binoth:1999qq}, \texttt{2gammaMC}
\cite{Bern:2002jx} and \texttt{MCFM}~\cite{Campbell:2011bn}. A
calculation that includes the effects of transverse-momentum
resummation is implemented in \texttt{Resbos}~\cite{Balazs:2007hr}.

At next-to-next-to-leading order (NNLO), all the
($\mathcal{O}(\alpha_S^2)$) contributions were put together in a
complete and consistent calculation in the \texttt{2$\gamma$NNLO} code
\cite{Catani:2011qz} for first time.  The next-order gluonic
corrections to the box contribution $gg\rightarrow \gamma\gamma$
(which are part of the N$^3$LO QCD corrections to diphoton production)
were also computed in ref.~\cite{Bern:2002jx} and found to have a
moderate quantitative effect.

The transverse momentum $p_T$ spectrum of the diphoton pair has been
calculated in fully-differential Monte Carlo codes at
LO~\cite{Binoth:1999qq,Bern:2002jx,Campbell:2011bn,Balazs:2007hr} and
at NLO~\cite{Catani:2011qz,DelDuca:2003uz,Gehrmann:2013aga}. Recently,
first calculations for diphoton production in association with
two~\cite{Bern:2014vza,Gehrmann:2013bga,Badger:2013ava} and
three~\cite{Badger:2013ava} jets at NLO became available. The
transverse momentum resummation for diphoton production at NNLL + NNLO
was recently presented in Ref.~\cite{Cieri:2015rqa} and implemented in
the $2\gamma${\tt Res} numerical code.

Besides direct photon production from the hard subprocess, photons can
also be produced from the fragmentation of QCD partons. The
computation of fragmentation subprocesses requires (the poorly known)
non-perturbative information, in the form of parton fragmentation
functions of the photon (the complete single- and double-fragmentation
contributions are implemented in \texttt{DIPHOX} \cite{Binoth:1999qq}
for diphoton production at the first order in $\alpha_S$).  However,
the effect of the fragmentation contributions is sizebly reduced by
the {photon isolation} criteria that are necessarily applied in hadron
collider experiments to suppress the very large irreducible background
(\textit{e.g.}, photons that are faked by jets or produced by hadron
decays). Two such criteria are the so-called ``standard'' cone
isolation and the ``smooth'' cone isolation proposed by Frixione
\cite{Frixione:1998jh}. The standard cone isolation is easily
implemented in experiments, but it only suppresses a fraction of the
fragmentation contribution.  By contrast, the smooth cone isolation
(formally) eliminates the entire fragmentation contribution.  All the
results presented in this section were obtained with the smooth
isolation prescription, which, for the parameters used in the
experimental analysis reproduces the standard result within a $1\%$
accuracy~\cite{Andersen:2014efa} at NLO.

In this section we present some benchmark results on diphoton
production at $\sqrt{s}=100$~TeV, of possible relevance to Higgs boson
studies as well as to BSM searches. We compute the NLO and NNLO QCD
radiative corrections at the fully-differential level. In all the NLO
results presented in this section we consider also the box
contribution at the lowest order in the strong coupling constant
($\mathcal{O}(\alpha_S^2)$).

The acceptance criteria used in the numerical results presented in
this section are the following: $ p_T^{\gamma} \geq 30$~GeV and the
rapidity of both photons has to satisfy $|y_\gamma|<2.5$.  We use the
MSTW2008 \cite{Martin:2009iq} sets of parton distributions, with
densities and $\alpha_S$ evaluated at each corresponding order (i.e.,
we use $(n+1)$-loop $\alpha_S$ at N$^n$LO, with $n=0,1,2$), and we
consider $N_f=5$ massless quarks/antiquarks and gluons in the initial
state. The default renormalization ($\mu_R$) and factorization
($\mu_F$) scales are set to the value $\mu_R=\mu_F =\sqrt{
  M_{\gamma\gamma}^2 + p_{T\gamma\gamma}^2}$. The QED coupling
constant $\alpha$ is fixed to $\alpha=1/137$.

The smooth cone isolation prescription is as follows: we consider a
cone of radius\\ \mbox{$r=\sqrt{(\Delta \eta)^2+(\Delta \phi)^2}$}
around each photon and we require that the total amount of hadronic
(partonic) transverse energy $E_T$ inside the cone is smaller than
$E_{T\, max}(r)$,
\begin{equation}
E_{T\, max}(r) \equiv  \epsilon_\gamma \,p_T^\gamma \left(\frac{1-\cos r}{1- \cos R}\right)^n \, ,
\end{equation}
where $p_T^\gamma$ is the photon transverse momentum; the isolation
criterion $E_T < E_{T\, max}(r)$ has to be fulfilled for all cones
with $r\leq R$.  The isolation parameters are set to the values
$\epsilon_\gamma=0.05$, $n=1$ and $R=0.4$ in all the numerical results
presented in this section. In Ref.~\cite{Andersen:2014efa} it was shown
that implementing $\epsilon_\gamma=0.05$ the effects of the
fragmentation contribution are under control, in the sense that the
NLO cross section obtained with the smooth cone isolation criterion
coincides with the corresponding NLO cross section obtained with the
standard cone isolation criterion at the percent level.
 
\begin{figure}[htb]
\begin{center}
\begin{tabular}{cc}
  \includegraphics[width=0.47\textwidth]{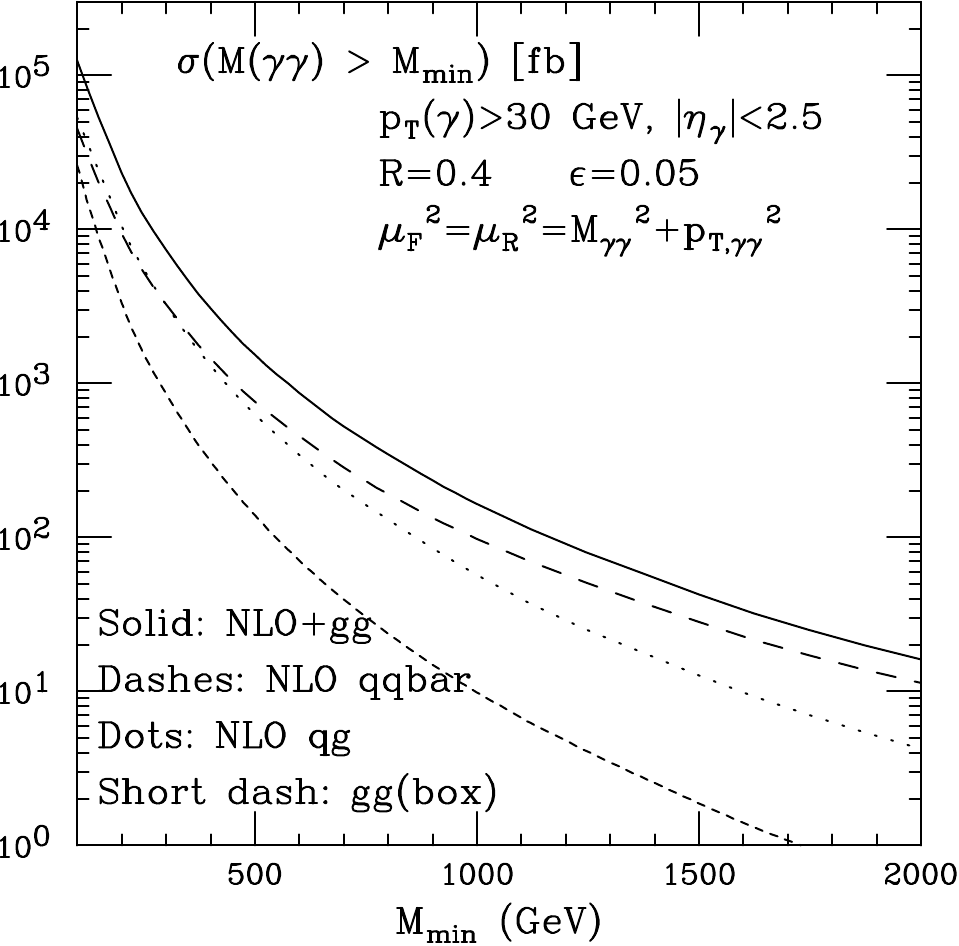} &
  \includegraphics[width=0.49\textwidth]{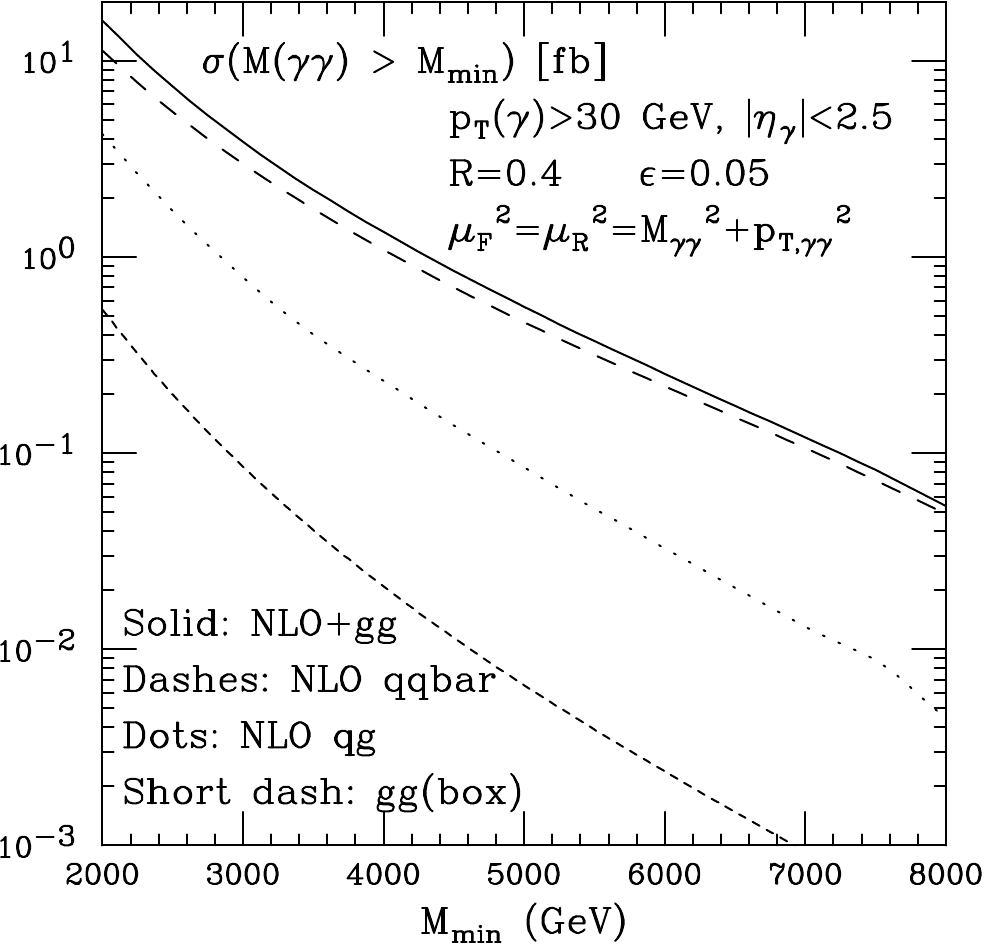}\\
\end{tabular}
\end{center}
\caption{\label{fig:diph_1} { Integrated diphoton invariant mass
    distribution, over different mass ranges. We display the full NLO
    cross-section, inclusive of the box contribution at the lowest order
    ($\mathcal{O}(\alpha_S^2)$), with the different partonic channels
    present at this perturbative level.}}
\end{figure}

In Fig.~\ref{fig:diph_1} we present our results for the integrated
invariant mass distribution.  While for low values of
$M_{\gamma\gamma}$ the box contribution (formally
$\mathcal{O}(\alpha_S^2)$) is of the same order of the LO $q\bar{q}$
contribution ($\mathcal{O}(\alpha_S^0)$), for large values of the
invariant mass, the LO cross section is at least one order larger than
the box contribution. Moreover, notice that for large values of the
lower cut in the diphoton invariant mass ($M_{\gamma\gamma}^{{\rm
    min}}>$ 400~GeV), the contribution to the cross section due to
partonic channels containing at least a gluon (in the initial state)
is negligible with respect to the $q\bar{q}$ channel. This is mostly
due to the greater impact of the isolation cut, which affects directly
processes like $qg\to qg \gamma\gamma$, where, to have a large
$M_{\gamma\gamma}$, one of the two photons is preferentially radiated
by the final state quark.

We note that the cross-section is of the order of a several tens of ab
for $M_{\gamma\gamma}\gsim 8$~TeV, meaning of order 1000 events for
the expected integrated luminosity (20-30~ab$^{-1}$).

\begin{figure}[htb]
\begin{center}
\begin{tabular}{cc}
  \includegraphics[width=0.47\textwidth]{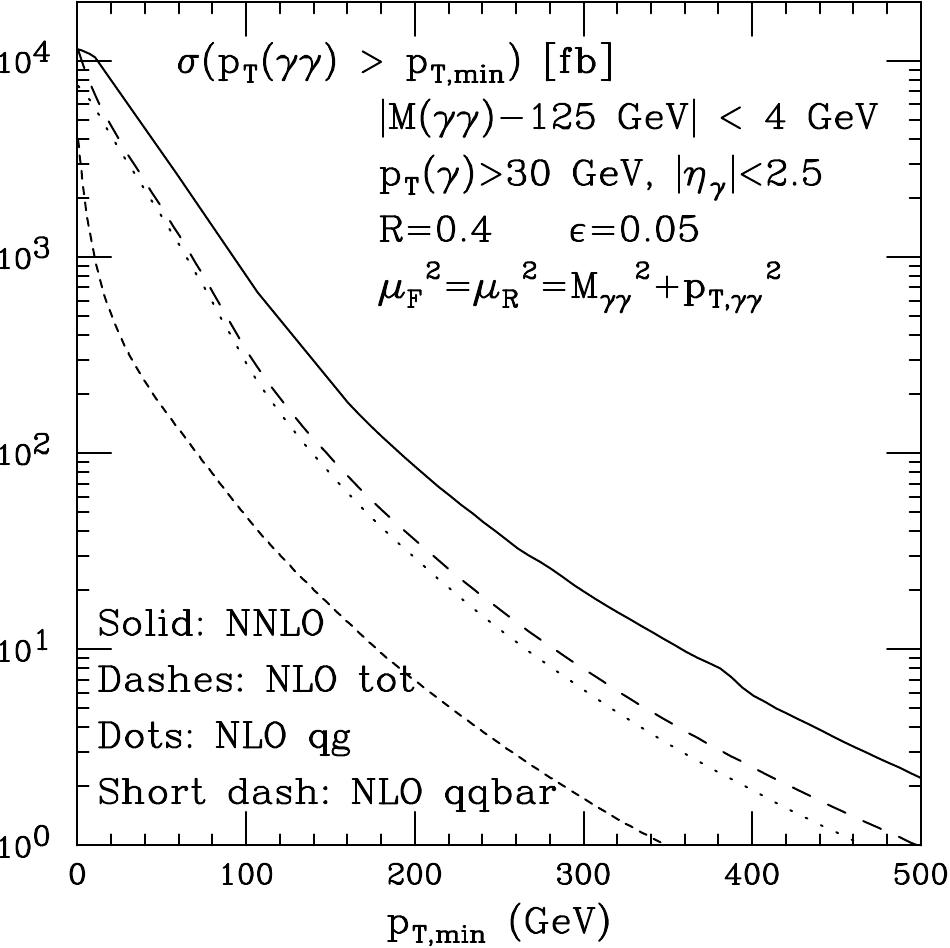} &
  \includegraphics[width=0.49\textwidth]{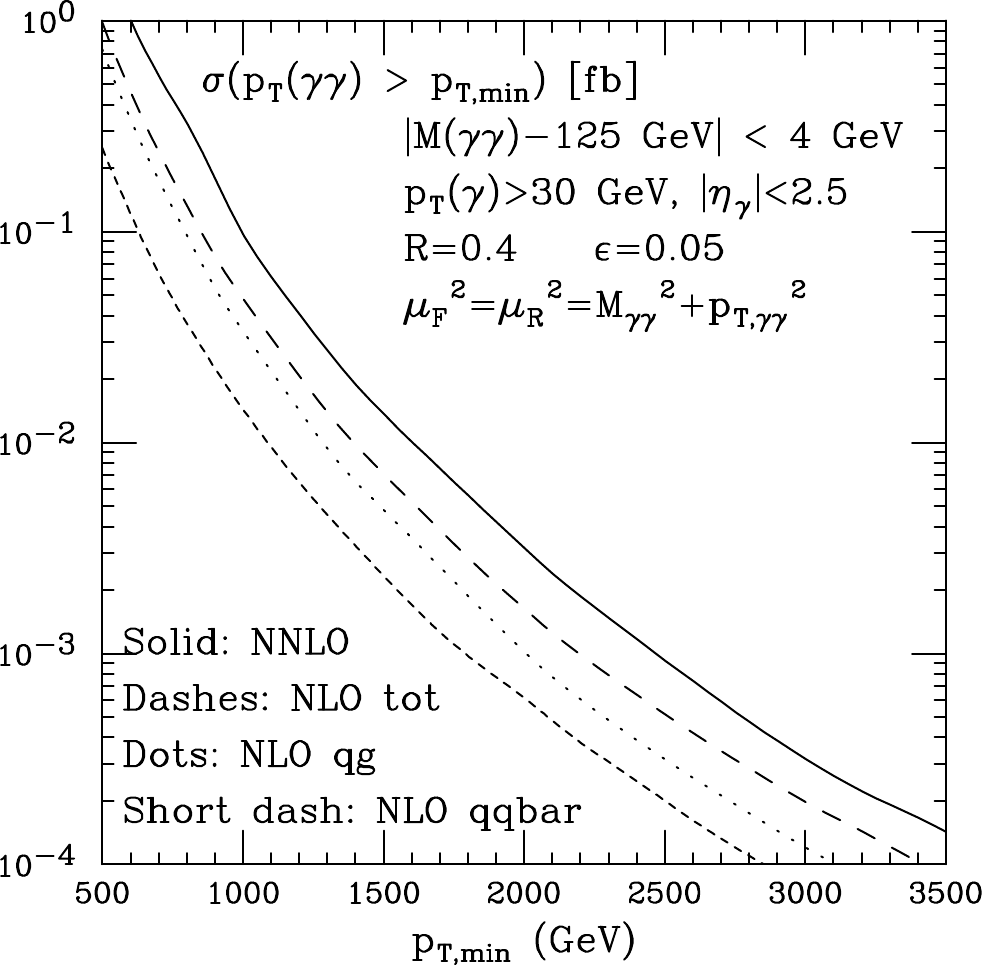}\\
\end{tabular}
\end{center}
\caption{\label{fig:diph_2} { Integrated diphoton transverse momentum
    distribution, subject to the constraint $|M_H - M_{\gamma\gamma}|<
    4$~GeV ($M_H =125$~GeV). We compare the NNLO and the NLO
    cross-sections, and the relative contributions of the $qg$ and
    $q\bar{q}$ processes at NLO.}}
\end{figure}

In Fig.~\ref{fig:diph_2} we show the integrated diphoton transverse
momentum distribution requiring $|M_H - M_{\gamma\gamma}|< 4$~GeV
($M_H =125$~GeV). 
The restriction of the diphoton invariant mass to this interval is
kept in all plots of interest for Higgs physics .
The notation NLO vs NNLO refers here to the order at which the
inclusive $\gamma\gamma+X$ process is evaluated, namely $O(\alpha_S)$
and $O(\alpha_S^2)$, respectively. In this language, NLO is actually
the first order at which the photon pair develops a transverse
momentum, and NNLO is the first genuine radiative correction to the
$p_T$ distribution.  Notice that at $O(\alpha_S^2)$ the $gg$ box
contribution does not generate a transverse momentum for the diphoton
pair, this will only arise at $O(\alpha_S^3)$. The tree-level $gg$
contributions of $O(\alpha_S^2)$ are small and, while they are
included in the NNLO, they are not shown separately in the plot.

In the left panel we compare the NNLO contribution with the NLO
cross-section.  We are not considering here transverse momentum
resummation (as implemented in $2\gamma${\tt Res} or {\tt Resbos}),
since the $p_T$ range of interest in these plots is well above the
values where Sudakov effects are relevant.

While for the integrated invariant mass distribution
(Fig.~\ref{fig:diph_1}) the $q\bar{q}$ partonic channel dominates the
cross section, in the diphoton integrated transverse momentum
distribution the $q\bar{q}$ and $qg$ channels are at the same order
over the whole transverse momentum range. It is easy to see that the
invariant mass cut on the diphoton pair forces the two photons to be
close to each other, and thus the $qg$ initial state process does not
need to be penalized by the isolation requirement which suppresses
this channel in the large-mass spectrum.

\begin{figure}[htb]
\begin{center}
\begin{tabular}{cc}
  \includegraphics[width=0.47\textwidth]{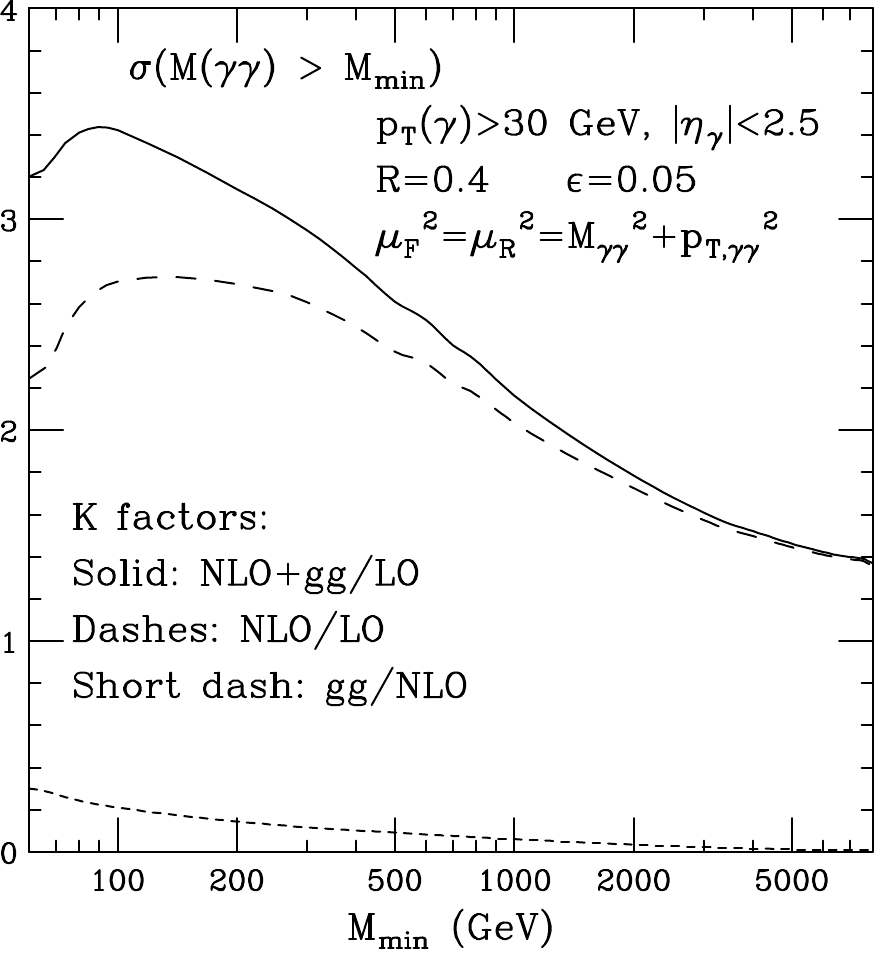} &
  \includegraphics[width=0.49\textwidth]{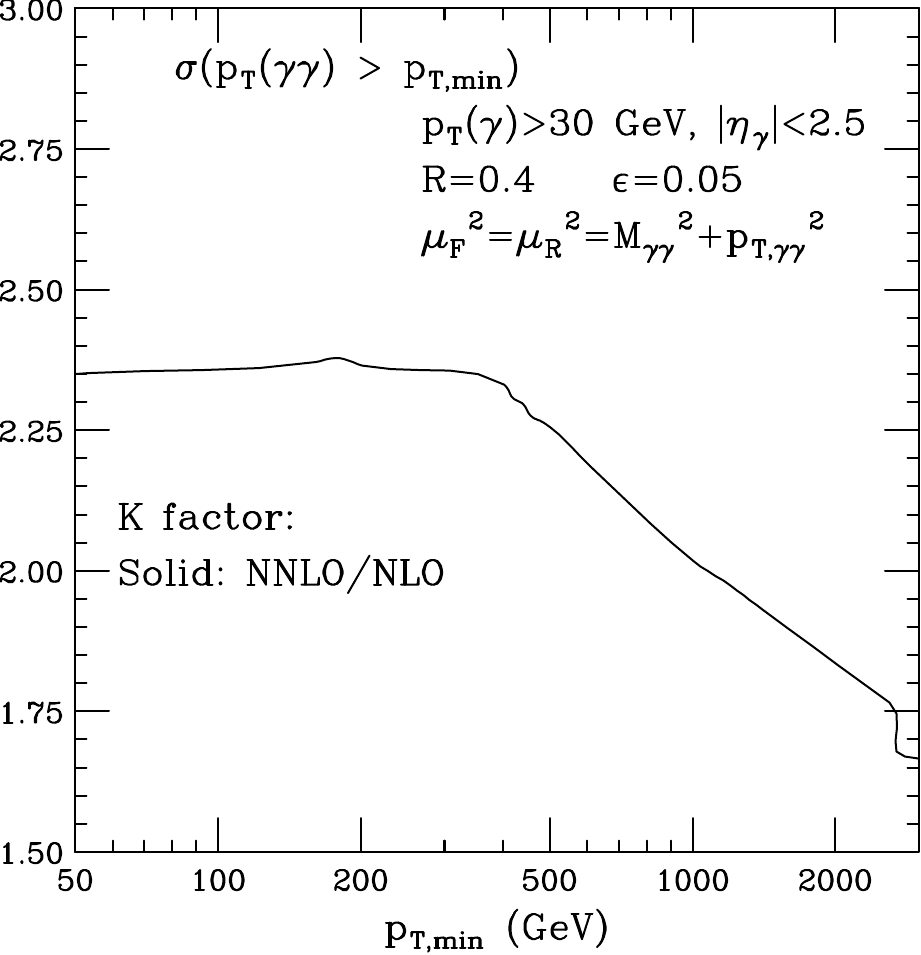}\\
\end{tabular}
\end{center}
\caption{\label{fig:diph_3} {Left panel: K factors for the diphoton
    invariant mass distribution from Fig.~\ref{fig:diph_1}.
    Right panel: K factors for the diphoton
    $p_T$  distribution from Fig.~\ref{fig:diph_2}.}}
\end{figure}

In Fig.~\ref{fig:diph_3} we show the K factors for the diphoton mass
and $p_T$ spectra, calculated for Fig.~\ref{fig:diph_1} and
Fig.~\ref{fig:diph_2}.

We observe that the NLO contributions introduce very large corrections
to the cross-section mainly for low and moderate values of the
invariant mass distribution $M_{\gamma\gamma}^{{\rm min}} <$ 1~TeV. At
high mass the K factors (NLO+box)/LO or (NLO)/LO tend to 1.4 (at
$\sqrt{s}=14$~TeV, K=NNLO/NLO $\simeq$ 1.2 for $M_{\gamma\gamma}^{{\rm
    min}} \simeq$ 3~TeV)). Likewise the $K$ factor of the diphoton
$p_T$ spectrum tends to be larger than 2 up to $p_T\sim 400$~GeV, and
to diminish after that.

\clearpage
\subsection{Anomalous couplings from $WW$ and $W\gamma$ production} 
\label{sec:VV_tgc}

In this section we explore the potential of the FCC to constraint or
measure anomalous triple-gauge couplings.
As an example, we consider $W^+W^-$ with $W$ bosons decaying into
electrons or muons and $W^+\gamma$ production with $W^+$ decaying into
a positron and a neutrino.

We consider an extension of the SM Lagrangian which includes up to
dimension six operators
\begin{equation}
{\cal L} = {\cal L}_{\rm SM} + \sum_i \frac{c_i}{\Lambda^2} {\cal O}_i
+ \ldots \,.
\end{equation}
In particular, we consider the effect of the following, CP-conserving,
dimension six operators~\cite{Zeppenfeld:1987ip}
\begin{eqnarray}
&{\cal O}_{WWW} = {\rm Tr}[W_{\mu\nu} W^{\nu \rho} W_\rho^\mu]\,, \quad\nonumber\\
&{\cal O}_{W} = (D_\mu \Phi)^\dagger W^{\mu \nu} (D_\nu \Phi)\,,\quad \nonumber\\
& {\cal O}_{B} = (D_\mu \Phi)^\dagger B^{\mu \nu} (D_\nu \Phi)\,,
\end{eqnarray}
with $\Phi$ being the Higgs doublet field and 
\begin{eqnarray}
D_\mu &=& \partial_\mu +\frac{i}{2} g \tau^I W_\mu^I +\frac{i}{2}g' B_\mu\,,\nonumber \\
W_{\mu\nu} &=& \frac{i}{2}g \tau^I\left(\partial_\mu W_\nu^I
- \partial_\nu W_\mu^I + g \epsilon_{IJK} W_\mu^J W_\nu^K\right)\,,\nonumber\\
B_{\mu\nu} &=& \frac{i}{2} g'\left(\partial_\mu B_\nu-\partial_\nu B_\mu\right)\,.
\end{eqnarray}
We remark that since the higher dimensional operators
can be seen as low energy remnants of some new heavy degrees of
freedom integrated out at scale 
$\Lambda$, we are implicitly assuming the scale of new physics $\Lambda$ to
be larger than the energy range we are probing.

The effect of these operators can also be equivalently expressed
in terms of anomalous couplings. The corresponding
modification of the SM Lagrangian is written as
\begin{eqnarray}
&{\cal L}_{TGC} = i g_{WWV} \left(
g_1^V(W_{\mu\nu}^+ W^{-\nu} - W^{+\mu} W^{-}_{\mu\nu}) V_\nu +
\kappa_V W_\mu^+W_\nu^- V^{\mu\nu} +\right.\nonumber\\ 
&\left.\frac{\lambda_V}{m_W^2} W_{\mu}^{+\nu}W_{\nu}^{-\rho}V_{\rho}^{\mu}
\right)\,, 
\end{eqnarray}
where $V=\gamma,Z$, $W_{\mu\nu}^\pm = \partial_\mu W_\nu^\pm
- \partial_\nu W_\mu^\pm$, $V_{\mu\nu}^\pm = \partial_\mu V_\nu
- \partial_\nu V_\mu$, $g_{WW\gamma} = -e$ and $g_{WWZ}=-e\cot\theta_w$.
At tree level the anomalous couplings can be related
to the coefficients of the dimension six operators via the following relations 

\begin{eqnarray}
&g_1^Z = 1+ c_{\rm w} \frac{m_Z^2}{2\Lambda^2}\,,\nonumber \\
&\kappa_\gamma = 1+ (c_{\rm w}+c_{\rm b})  \frac{m_W^2}{2\Lambda^2}\,,\nonumber\\
&\kappa_Z = 1+ (c_{\rm w}-c_{\rm b}\tan^2\theta_W)  \frac{m_W^2}{2\Lambda^2}\,,\nonumber\\
&\lambda_\gamma = \lambda_Z = c_{\rm www} \frac{3 g^2 m_W^2}{2\Lambda^2}\,.
\end{eqnarray}

For $W^+W^-$, we consider predictions at Les Houches event level
obtained with the {\tt POWHEG} $WW$
code~\cite{Melia:2011tj,Nason:2013ydw}.  
In this way NLO corrections are included together with Sudakov 
effects associated with the hardest
radiation, but the effects of the subsequent parton shower,
hadronization, or underlying event corrections are not included. We
remark that we do not include here loop-induced gluon-gluon channels.
We consider the following minimal set of cuts on the charged leptons
\mbox{$p_{t,l} > 20$~GeV}, $|\eta_l| < 2.5$, and a cut of $20$~GeV on
the missing transverse momentum. Jets are reconstructed using the
anti-$k_t$ algorithm~\cite{Cacciari:2008gp} with $R=0.6$. Furthermore,
in order to reduce the top background, we veto events where the
invariant mass of any charged lepton combined with any jet is below
$200$~GeV.  We have verified that, in the distributions that we have
considered, lowering this cut does not lead to significant changes,
and that the top contribution is negligible.

For $W^+\gamma$, when NLO QCD corrections are taken into account, real
gluon induced diagrams arise. For these contributions we need a
strategy for the treatment of photon fragmentation contribution,
namely the infrared divergent configurations where the photon becomes
soft or collinear to the emitting quark.  Since in the \texttt{POWHEG
  BOX} approach~\cite{Barze:2014zba} for the treatment of photon
fragmentation contribution there are two underlying Born
configurations at LHE level, $W^+\gamma$ and $W^+j$, the analysis at
event level would be highly inefficient because of the $W^+j$
contribution which would largely dominate.  Therefore we consider
predictions at NLO accuracy with smooth isolation
prescription~\cite{Frixione:1998jh} applied at generation stage.  In
our analysis following cuts are applied: $p_{T}^l > 20$ GeV,
$|\eta_{i}| < 2.5$ with $i = e^+,\gamma$ and $\Delta R_{l\gamma} >
0.7$ . Moreover, in order to improve the efficiency, since we are
interested in the $p_{t,\gamma}$ distribution only, we have put a
generation cut at 100 GeV, after checking that the effect of the cut
is negligible in the high $p_t$ region.

All the results have been obtained using
NNPDF30\_nlo\_as\_0118~\cite{Ball:2014uwa}. In order to understand the
sensitivity to the different operators we turn on the coefficient of
one operator at a time. We examined several observables and find that,
as well-known, the sensitivity to dimension six operators appears in
the region of large transverse momenta or invariant masses. As an
example, we consider the invariant mass $m_{ll}$ of the dilepton pair
for $W^+W^-$ and the photon's transverse momentum for $W^+\gamma$.

Our results, presented in Fig.~\ref{fig:wgamma} for $W^+\gamma$ and in
Figs.~\ref{fig:cwww-cw}, ~\ref{fig:cb-lambda} (left) for $W^+W^-$ are
shown in terms of integrated rates.  In particular, in the upper
panels, we show the number of events assuming 10~ab$^{-1}$ of
integrated luminosity for different values of the coefficients of the
operators.
In the lower panels we quantify the significance of the excess by
showing the ratio of the number of events in excess of the SM
prediction divided by the squared-root of the number of events
predicted in the SM, $(N_{\rm C} - N_{\rm SM})/\sqrt{N_{\rm
    SM}}$. Under the assumption that SM backgrounds can be measured
and predicted precisely, the above quantity gives a rough indication
of the significance that can be reached with an integrated luminosity
of 10~ab$^{-1}$.
The two horizontal lines in the lower panels indicate the $3\sigma$
and $5\sigma$ significance. For each operator, we show the
distributions corresponding to three choices of the coefficients of
the operators, that envelope the $3\sigma$ and $5\sigma$ lines.  For
$W^+W^-$ the maximal sensitivity has a peak for given values of
$m_{ll}$. This corresponds to a value where the departure from the SM
predictions are big enough, but the statistics remains
significant. This is not the case for $W^+\gamma$ distributions when
we consider the departures from the SM prediction due to
$\mathcal{O}_W + \mathcal{O}_B$. In this case the positive effect due
to the presence of the anomalous coupling $\kappa_\gamma$ is not
sufficient to compensate the drop in the number of events in the
distributions' tails.  Therefore the sensitivity does not peak in the
region of large transverse momenta. We also remark that, in order to
achieve a significance around $5\sigma$ for the $W^+\gamma$ process,
we have taken values of $c_{\rm w}$ and $c_{\rm b}$ which are roughly
two orders of magnitude higher with respect to the $W^+W^-$ case.  In
other words, the $W^+ W^-$ process is more effective in constraining
the coefficients $c_{\rm w}$ and/or $c_{\rm b}$ than the $W^+\gamma$
process.

We see that, compared to current bounds from 8 TeV
LHC~\cite{Khachatryan:2015sga,Aad:2014mda}, bounds improve by more
than two orders of magnitude.

For $W^+ W^-$ production, the right plot in Fig.~\ref{fig:cb-lambda}
shows the value of the scale $\Lambda$ such that we enter the strong
coupling regime, according to the rules of dimensional analysis given
in ref.~\cite{Gavela:2016bzc}.  Looking for example at the $c_{\rm
  www}/\Lambda^2=0.04$~TeV$^{-2}$ case, we see that the non
perturbative region is at scales of several tens of TeV, where our
optimal region for the determination of the anomalous couplings is
below 10~TeV. The situation is somewhat worse for the $c_{\rm
  W}/\Lambda^2=0.08$ and especially for the $c_{\rm b}/\Lambda^2=0.2$
cases, where the non-perturbative region is reached near 10 TeV, and
the optimal scale for the determination of the anomalous couplings is
near 8~TeV.

\begin{figure}[htb]
  \centering
    \includegraphics[width=0.49\textwidth]{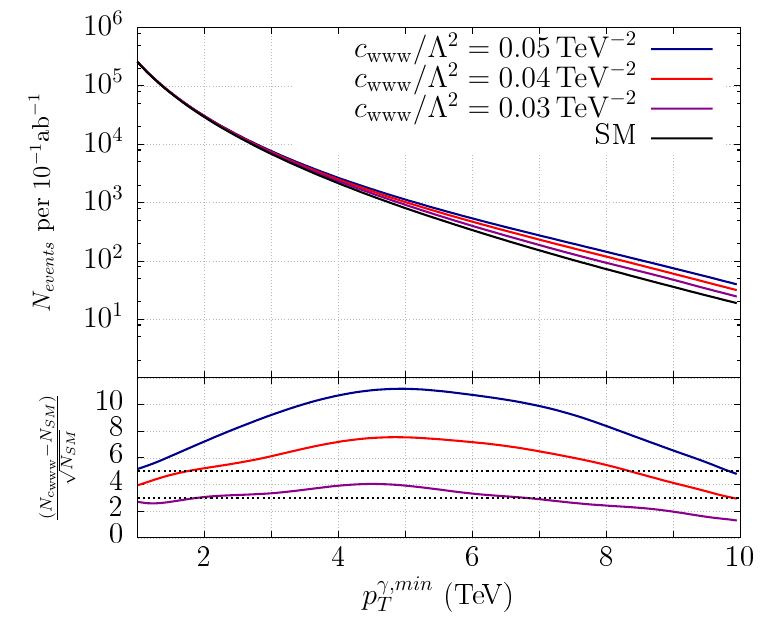}
    \includegraphics[width=0.49\textwidth]{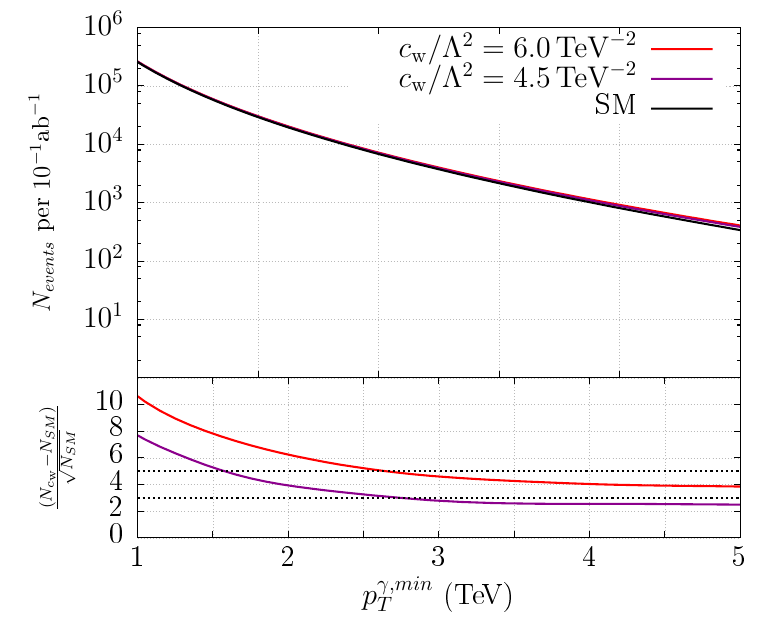}
  \caption{Upper panels: number of events with $p_{\rm t,\gamma} > p_{\rm
      t,\gamma}^{\rm min}$ in the SM and for various values of $c_{\rm
      www}$ (left)
    and $c_{\rm w}$ (right) for an accumulated luminosity of 10
    ab$^{-1}$ (the effect of $c_{\rm b}$ is the same of $c_{\rm w}$). 
    Lower panels: significance computed as
    $(N_{c_{i}}-N_{\rm SM})/\sqrt{N_{\rm SM}}$.}
  \label{fig:wgamma}
\end{figure}

\begin{figure}[htb]
  \centering
    \includegraphics[width=0.49\textwidth]{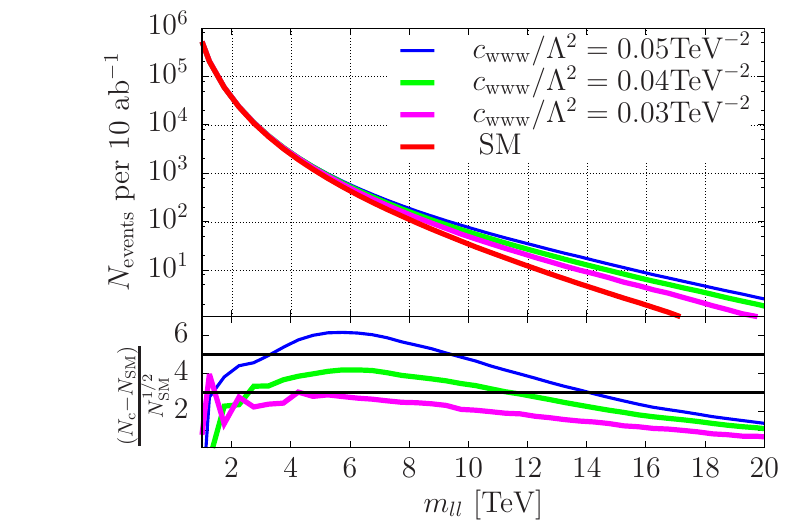}
    \includegraphics[width=0.49\textwidth]{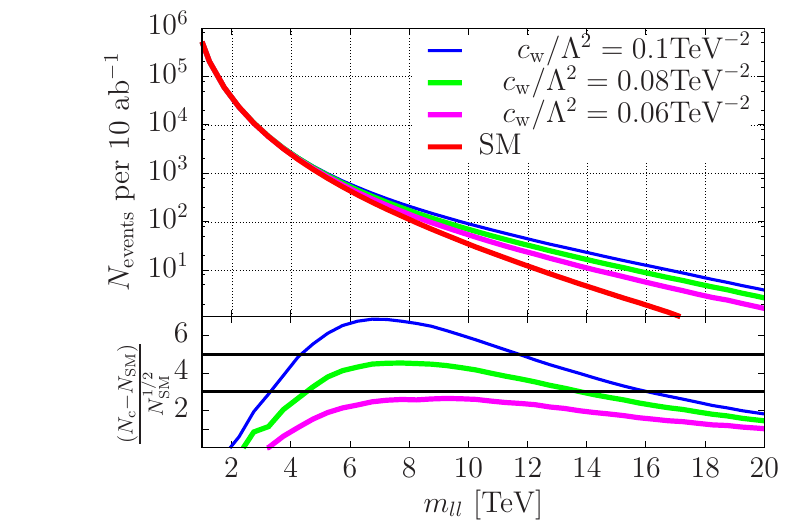}
    \caption{Upper panels: number of events with $m_{\rm ll} > m_{\rm
        ll}^{\rm min}$ in the SM and for various values of $c_{\rm
        www}$ (left) and $c_{\rm w}$ (right) for an accumulated
      luminosity of 10 ab$^{-1}$. Lower panels: significance computed
      as $(N_{c_{i}}-N_{\rm SM})/\sqrt{N_{\rm SM}}$.}
  \label{fig:cwww-cw}
\end{figure}

\begin{figure}[htb]
  \centering
    \includegraphics[width=0.49\textwidth]{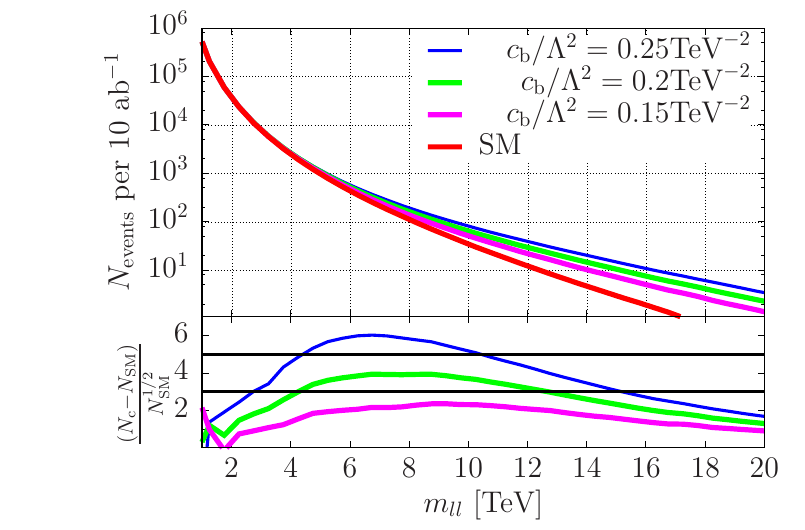}
    \includegraphics[width=0.49\textwidth,height=5.0cm]{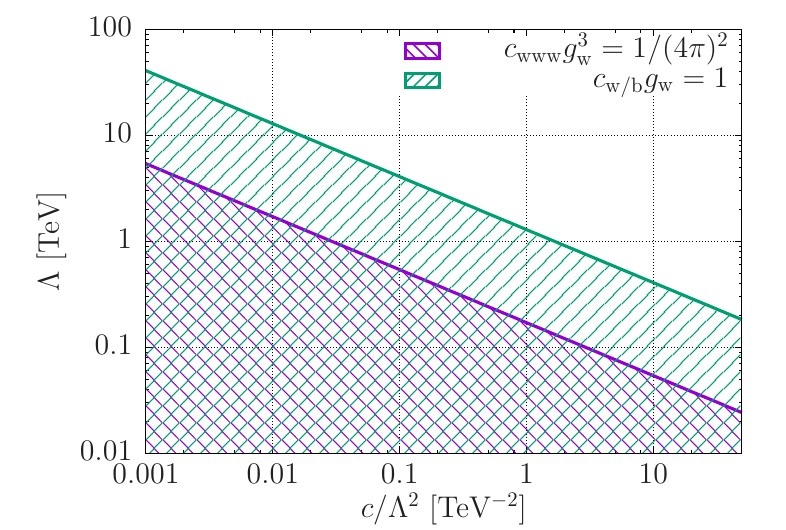}
    \caption{Left: same as fig.~\ref{fig:cwww-cw} for $c_{\rm
        b}$. Right: the perturbative region (hatched regions) in the
      $c/\Lambda^2$--$\Lambda$ plane.}
  \label{fig:cb-lambda}
\end{figure}

\clearpage
\subsection{$VV$+jet production} 
\label{sec:VVjet}
\def\mO{\mathcal{O}}
\def\rG{r_\Gamma}
\def\e{\epsilon}
\def\vev{\langle v \rangle}
\def\beq{\begin{equation}}
\def\eeq{\end{equation}}
\def\beqn{\begin{eqnarray}}
\def\eeqn{\end{eqnarray}}
\def\al{\alpha}
\def\cN{{\cal N}}
\def\ga{\gamma}
\def\eps{\epsilon}
\def\nn{\nonumber}
\def\pu{p_{1}}
\def\pd{p_{2}}
\def\pt{p_{3}}
\def\ro{\rho}
\def\si{\sigma}
\def\Tria{\mathrm{Tria}}
\def\BubD{\mathrm{BubD}}
\def\arccosh{\mathrm{arccosh}}
\def\Mbar{\overline{M}}

\def\pb{{\ensuremath\rm pb}}
\def\fb{{\ensuremath\rm fb}}
\def\ora{\overrightarrow}
\def\GeV{\ifmmode {\mathrm{\ Ge\kern -0.1em V}}\else
                   \textrm{Ge\kern -0.1em V}\fi}%
\def\TeV{\ifmmode {\mathrm{\ Te\kern -0.1em V}}\else
                   \textrm{Te\kern -0.1em V}\fi}%

\def\htr#1{\color{red} #1}
\def\htb#1{\color{blue} #1}
\def\htg#1{\color{green} #1}
\def\htrd#1{\htr{\st{#1}}}
\def\htbd#1{\htb{\st{#1}}}
\def\htgd#1{\htg{\st{#1}}}

\def\leftb{\left(}
\def\rightb{\right)}
\def\wt{\widetilde}
\def\eqn{equation}\def\beq{\begin{equation}}
\def\eeq{\end{equation}}
\def\beqn{\begin{eqnarray}}
\def\eeqn{\end{eqnarray}}
\def\nn{\nonumber}
\def\spa#1.#2{\left\langle#1\,#2\right\rangle}
\def\spb#1.#2{\left[#1\,#2\right]}
\def\spaa#1.#2.#3{\langle\mskip-1mu{#1} 
                  | #2 | {#3}\mskip-1mu\rangle}
\def\spbb#1.#2.#3{[\mskip-1mu{#1}
                  | #2 | {#3}\mskip-1mu]}
\def\spab#1.#2.#3{\langle\mskip-1mu{#1} 
                  | #2 | {#3}\mskip-1mu]}
\def\spba#1.#2.#3{[\mskip-1mu{#1} 
                  | #2 | {#3}\mskip-1mu\rangle}
\def\spaba#1.#2.#3.#4{\langle\mskip-1mu{#1} 
                  | #2 | #3 | {#4}\mskip-1mu\rangle}
\def\q{{\vphantom{\qb}{q}}}
\def\l{{\vphantom{\lb}{l}}}
\def\Q{{\vphantom{\Qb}{Q}}}
\def\qb{{\bar q}}
\def\Qb{{\bar Q}}
\def\s#1{s_{#1}}

\def\bentarrow{\:\raisebox{1.3ex}{\rlap{$\vert$}}\!\rightarrow}
\def\bothdk#1#2#3#4#5{
\begin{array}{r c l}
#1 & \rightarrow & #2#3 \\
 & & \:\raisebox{1.3ex}{\rlap{$\vert$}}\raisebox{-0.5ex}{$\vert$}
\phantom{#2}\!\bentarrow #4 \\
 & & \bentarrow #5
\end{array}
}

\subsubsection{Overview}
We here consider the hadronic production of $W$ pairs in association
with a single jet at next-to-leading order (NLO) in QCD at a proton
collider with a center-of-mass energy of 100 \TeV. The $W$ bosons
decay leptonically, with all spin correlations included.  At tree
level this process corresponds to the partonic reaction,
\begin{equation}
\label{WWjetprocess}
\bothdk{q+\bar q}{W^+ +}{W^-+g}{\mu^-+\nu_\mu}{\nu_e+e^+}
\end{equation}
with all possible crossings of the partons between initial and final
states. Tree level diagrams for this process are shown in
Fig.~\ref{fig:LOdiags}.

\begin{figure}
\begin{center}
\includegraphics[scale=0.3]{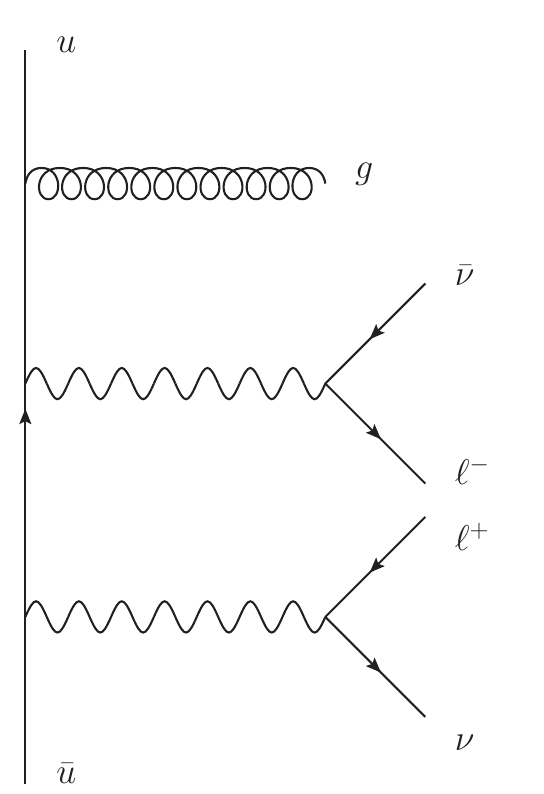} \hspace*{1.5cm}
\includegraphics[scale=0.3]{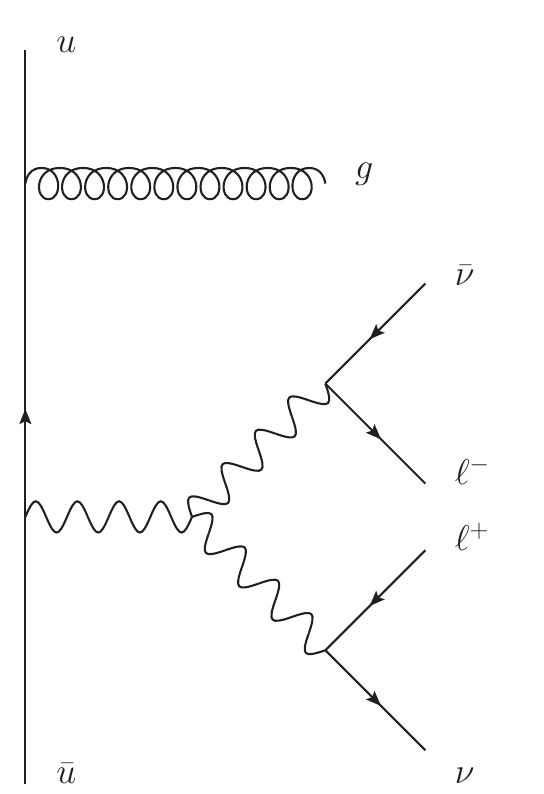} \\ ~~ \\
(a) \hspace*{5cm} (b)
\end{center}
\caption{Sample diagrams entering the calculation of the leading order amplitude
for the $WW+$jet process, corresponding to (a) $W$ emission from the quark line and (b) emission
from an intermediate $Z$ boson or photon.\label{fig:LOdiags}}
\end{figure}

At next-to-leading order we must include the emission of an additional
parton, either as a virtual particle to form a loop amplitude, or as a
real external particle.  Sample diagrams for virtual NLO contributions
are shown in Fig.~\ref{fig:loopdiags}; in addition, one-loop
corrections to Fig.~\ref{fig:LOdiags} (b) must be included. All
results presented in the following have been obtained using the
calculation of Ref.~\cite{Campbell:2015hya}\footnote{See also
  \cite{Campbell:2016fph}.}, where virtual corrections have been
obtained using generalized unitarity methods
\cite{Britto:2004nc,Britto:2005ha,Britto:2006sj,Forde:2007mi,Mastrolia:2009dr,Badger:2008cm}. The
combination of the virtual contributions with born and real emission
diagrams has implemented into MCFM
\cite{Campbell:1999ah,Campbell:2015qma}. Note that we do not include
the effects of any third-generation quarks, either as external
particles or in internal loops.
\begin{figure}
\begin{center}
\includegraphics[scale=0.3]{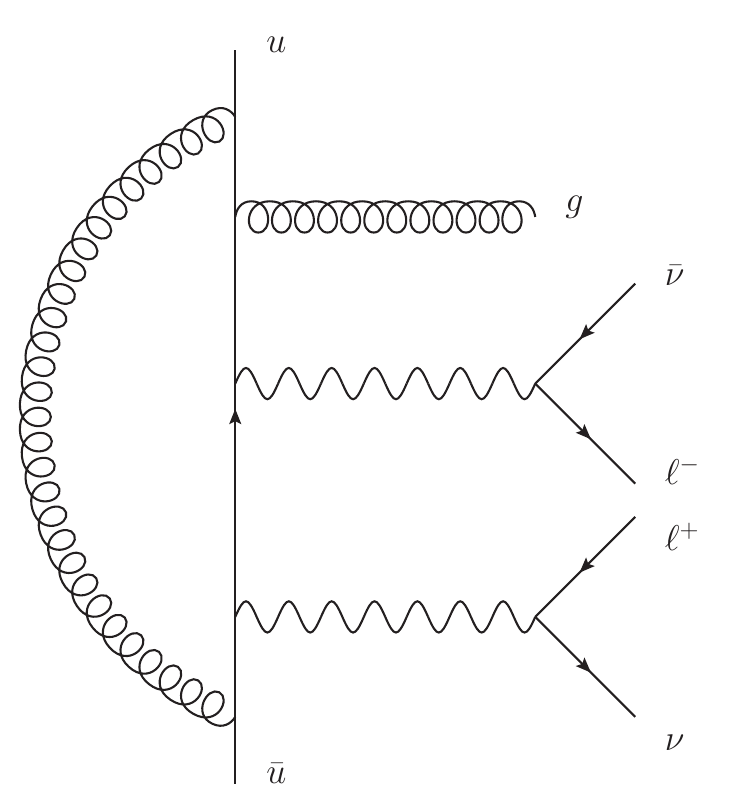} \hspace*{1.5cm}
\includegraphics[scale=0.3]{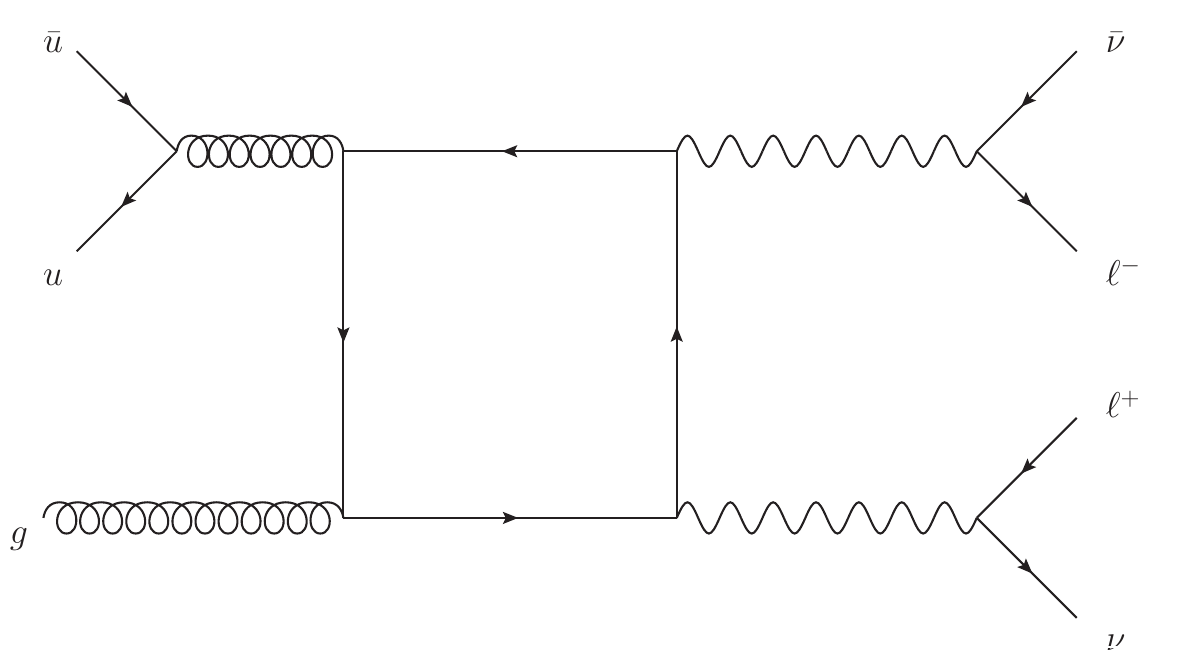}
\end{center}
\caption{Sample diagrams entering the calculation of the one-loop amplitude
for the $WW+$jet process.  The one-loop diagrams can be categorized according
to whether a gluon dresses a leading-order amplitude (left), or whether the
diagram includes a closed fermion loop (right).\label{fig:loopdiags}}
\end{figure}

\subsubsection{Total cross sections}
\begin{table}
\begin{center}
\begin{tabular}{|c|c|c|c|}
\hline
$m_W$               & 80.385 GeV           & $\Gamma_W$ & 2.085 GeV \\
$m_Z$               & 91.1876 GeV          & $\Gamma_Z$ & 2.4952 GeV \\
$e^2$               & 0.095032             & $g_W^2$    & 0.42635 \\ 
$\sin^2\theta_W$    & $0.22290$            & $G_F$      & $0.116638\times10^{-4}$ GeV$^{-2}$\\
\hline
\end{tabular}
\caption{The values of the mass, width and electroweak parameters used to produce
the results in this subsection.
\label{parameters}}
\end{center}
\end{table}
The results presented in this section have been obtained using the
parameters shown in Table~\ref{parameters}.  In calculations of LO
quantities we employ the CTEQ6L1 PDF set~\cite{Pumplin:2002vw}, while
at NLO we use CT10~\cite{Lai:2010vv}.  The renormalization and
factorization scales are usually chosen to be the same, $\mu_R = \mu_F
= \mu$, with our default scale choice $\mu = \mu_0$ given by,
\begin{equation}
\mu_0 \equiv \frac{H_T}{2} = \frac{1}{2} \sum_i p_{\perp}^i \;.
\end{equation}
The sum over the index $i$ runs over all final state leptons and
partons.  Jets are defined using the anti-$k_T$ algorithm with
separation parameter $R=0.5$ and must satisfy,
\begin{equation}
p_{\perp}^\text{jet} > p_{\perp, \text{cut}}^{\text{jet}} \;, \qquad
|\eta^\text{jet}| <4.5 \;.
\label{eq:jetcuts}
\end{equation}
The cross-sections predicted at LO and NLO are shown in
Fig.~\ref{fig:ptjet}, as a function of $p_{\perp,
  \text{cut}}^{\text{jet}}$ and for values as large as $20$~TeV.
The cross-sections at NLO are significantly larger than those at LO,
by as much as an order of magnitude at $10$~TeV and beyond.
\begin{figure}
\begin{center}
\includegraphics[width=0.35\textwidth,angle=-90]{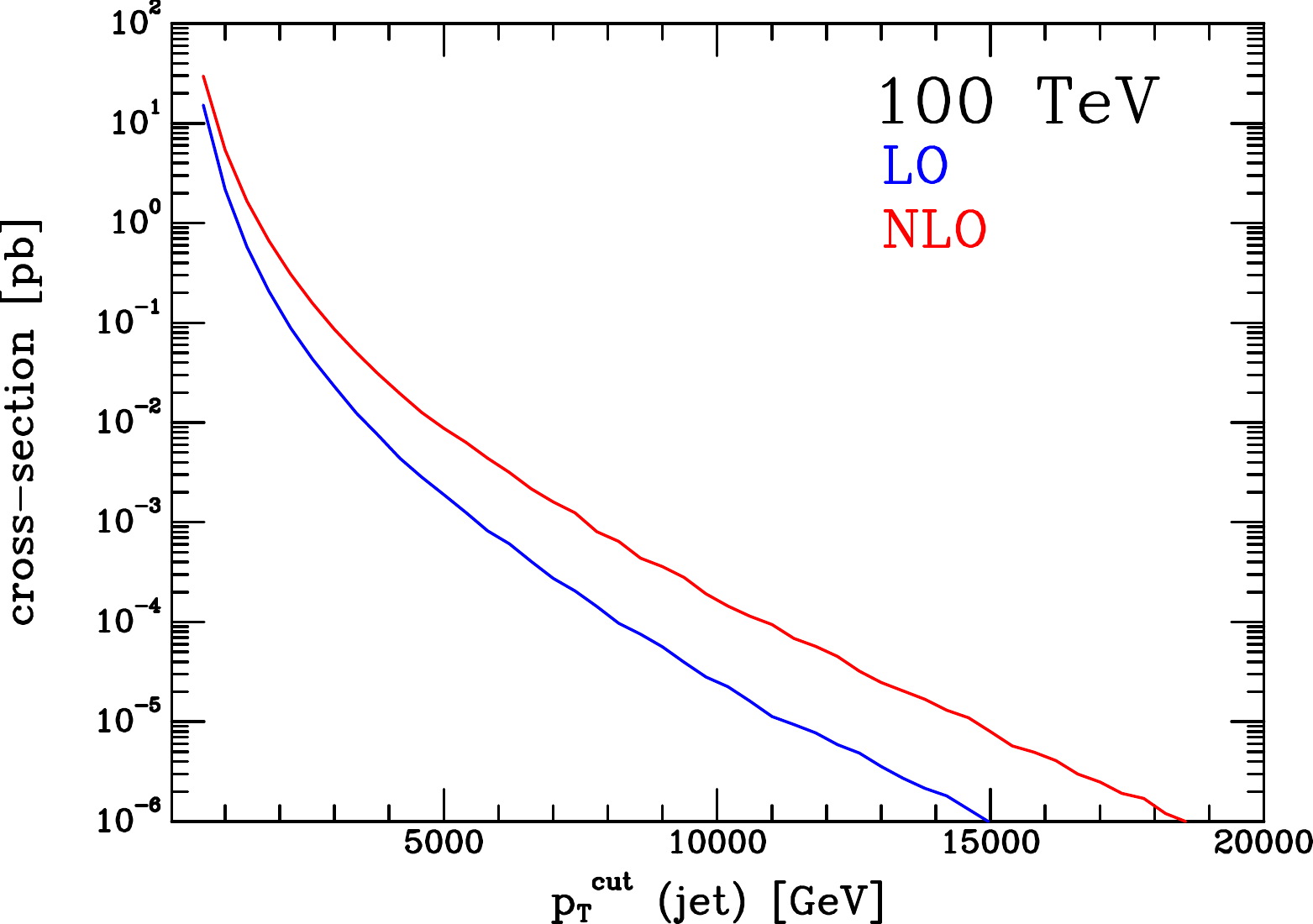}
\end{center}
\caption{Cross-sections at  $100$~TeV,
as a function of the transverse momentum
cut on the jet.  
\label{fig:ptjet}}
\end{figure}

As useful operating points, we use $p_{\perp,
  \text{cut}}^{\text{jet}}=25~\mbox{GeV}$ and also choose to study the
additional case $p_{\perp, \text{cut}}^{\text{jet}}=300~\mbox{GeV}$,
which we will label 100 \TeV * in the following.  The cross-sections
for $WW$+jet production, together with the corresponding values for
the 14 \TeV~ LHC and under the basic jet cuts of
Eq.~(\ref{eq:jetcuts}), are collated in Table~\ref{xsecs}
\footnote{Note that there is a minor typographical error in
  Ref.~\cite{Campbell:2015hya} in the relative uncertainty due to
  scale variations for the LO cross section at 100 \TeV, which we have
  corrected here.}.  Note that the effect of the decays of the $W$
bosons is not included. At the 100 TeV machine, the jet cut of 300~GeV
has been chosen so that the cross section is similar in size to the 14
TeV cross section, as can be seen from Table~\ref{xsecs}.  This cut
provides a useful benchmark in a different kinematic regime that may
be more appropriate at that collider energy.
\renewcommand{\baselinestretch}{1.5}
\begin{table}
\begin{center}
\begin{tabular}{|r|r|c|c|}
\hline
$\sqrt s$~~~~ & $p_{\perp, \text{cut}}^{\text{jet}}$ & $\sigma_{LO}$~[pb] & $\sigma_{NLO}$~[pb] \\
\hline
$14$~TeV      &  25 GeV & $39.5_{-11.0\%}^{+11.7\%} $  & $48.6_{-4.0\%}^{+3.8\%}$ \\
$100$~TeV     &  25 GeV & $648_{-19.3\%}^{+22.3\%}   $  & $740_{-9.3\%}^{+4.5\%} $ \\
$100$~TeV     & 300 GeV & $30.3^{+11.22\%}_{-10.56\%}$&$53.7^{+8.0\%}_{-7.6\%}$ \\
\hline
\end{tabular}
\renewcommand{\baselinestretch}{1.0}
\caption{Cross-sections for the process $p p \to WW$+jet at proton-proton colliders
of various energies, together with estimates of the theoretical uncertainty
from scale variation as described in the text. 
Monte Carlo uncertainties are at most a
single unit in the last digit shown shown in the table.
\label{xsecs}}
\end{center}
\end{table}
\renewcommand{\baselinestretch}{1.0}

An interesting feature of the higher order corrections to processes
such as the one at hand is the existence of so-called ``giant
K-factors''~\cite{Frixione:1992pj,Frixione:1993yp,Rubin:2010xp}.  An
observable that exemplifies this effect is $H_T^\text{jets}$, which is
defined to be the scalar sum of all jet transverse momenta in a given
event.  At NLO, real radiation contributions arise in which two hard
partons are produced approximately back-to-back, with the $W^+W^-$
system relatively soft.  Such configurations are not captured at all
by the LO calculations, in which the parton and $W^+W^-$ system are
necessarily balanced in the transverse plane.  This results in the by
now well-known feature of huge NLO corrections at large
$H_T^\text{jets}$, as shown in Fig.~\ref{fig:HTjets}.
\begin{figure}
\begin{center}
\includegraphics[width=0.35\textwidth,angle=-90]{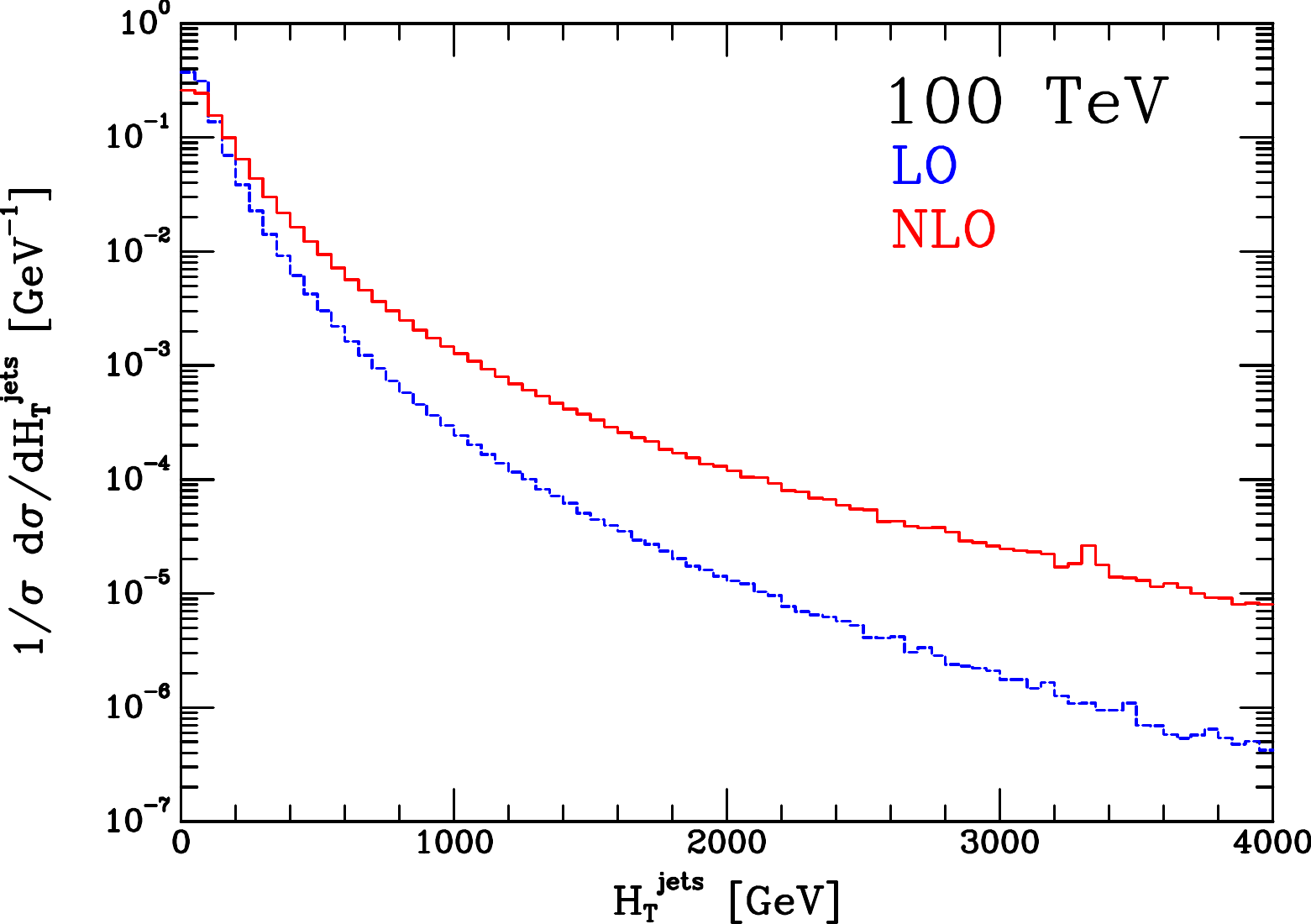}
\end{center}
\caption{The distribution of the observable
$H_T^\text{jets}=\sum_\text{jets} p_\perp^\text{jet}$ at LO and NLO. 
\label{fig:HTjets}}
\end{figure}

We see that the NLO predictions are at least an order of magnitude
larger than their LO counterparts in the tails of the
distributions\footnote{This effect also appears at a 14 \TeV~ LHC,
  cfr. e.g. \cite{Campbell:2016fph}.}. This onset occurs well before
the interesting multi-TeV region.

Another interesting topic to investigate is the {\sl total} number of
events for selection cuts, i.e.
\begin{\eqn}\label{eq:sigint}
  \sigma_{\text{tot}}\left( \text{cut}
  \right)\,=\,\int\,d\sigma\,\Theta\left( \text{cut} \right),
\end{\eqn}
where the cuts for dimensionful quantities can reach
$\mathcal{O}\left( \TeV \right)$. Figure \ref{fig:xss} displays
similar distributions for the quantities $H_{T,\text{jets}}$ and
$p^{WW}_{T}$.
\begin{figure}
\begin{center}
\includegraphics[width=0.33\textwidth,angle=-90]{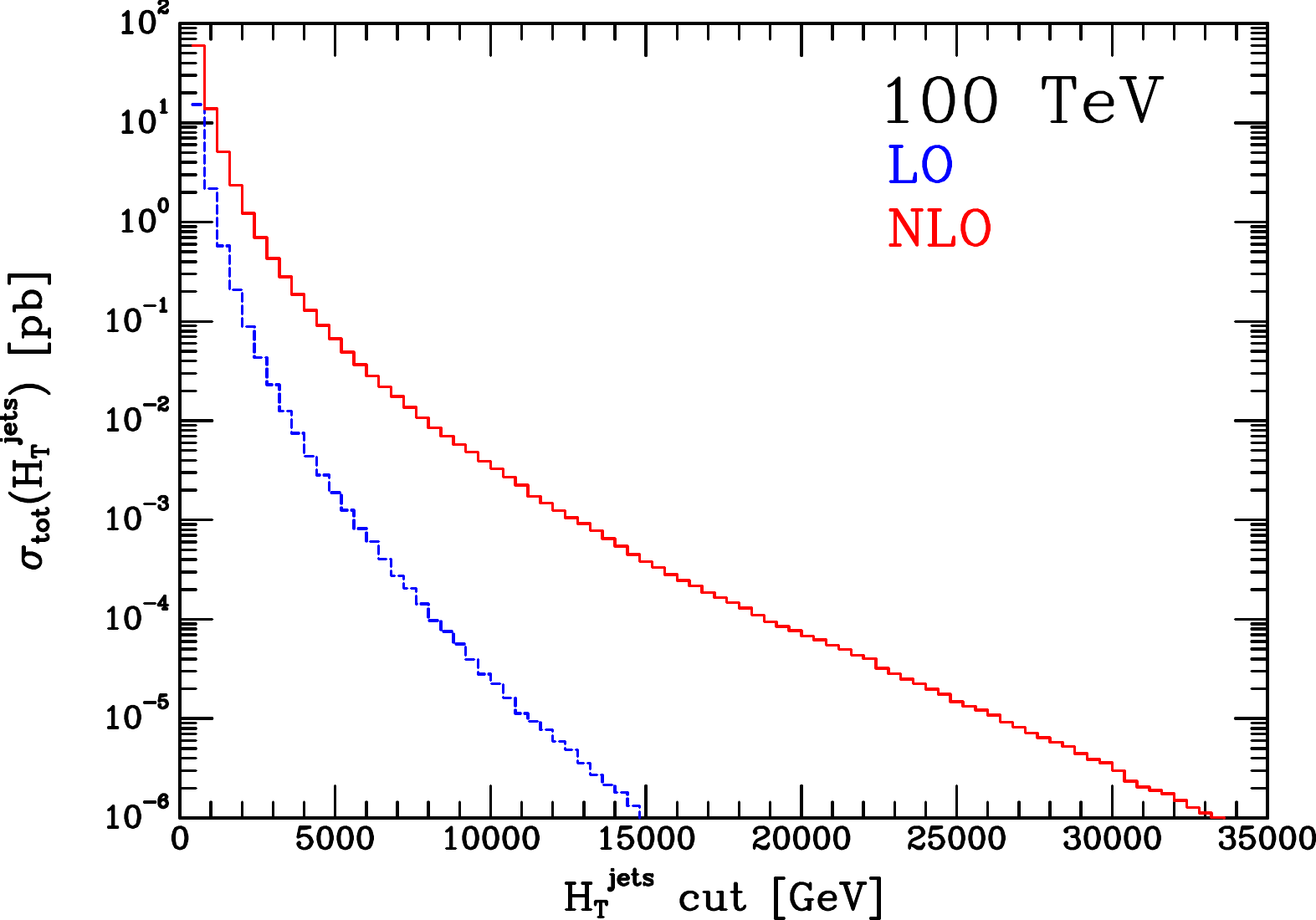} \hspace*{0.5cm}
\includegraphics[width=0.33\textwidth,angle=-90]{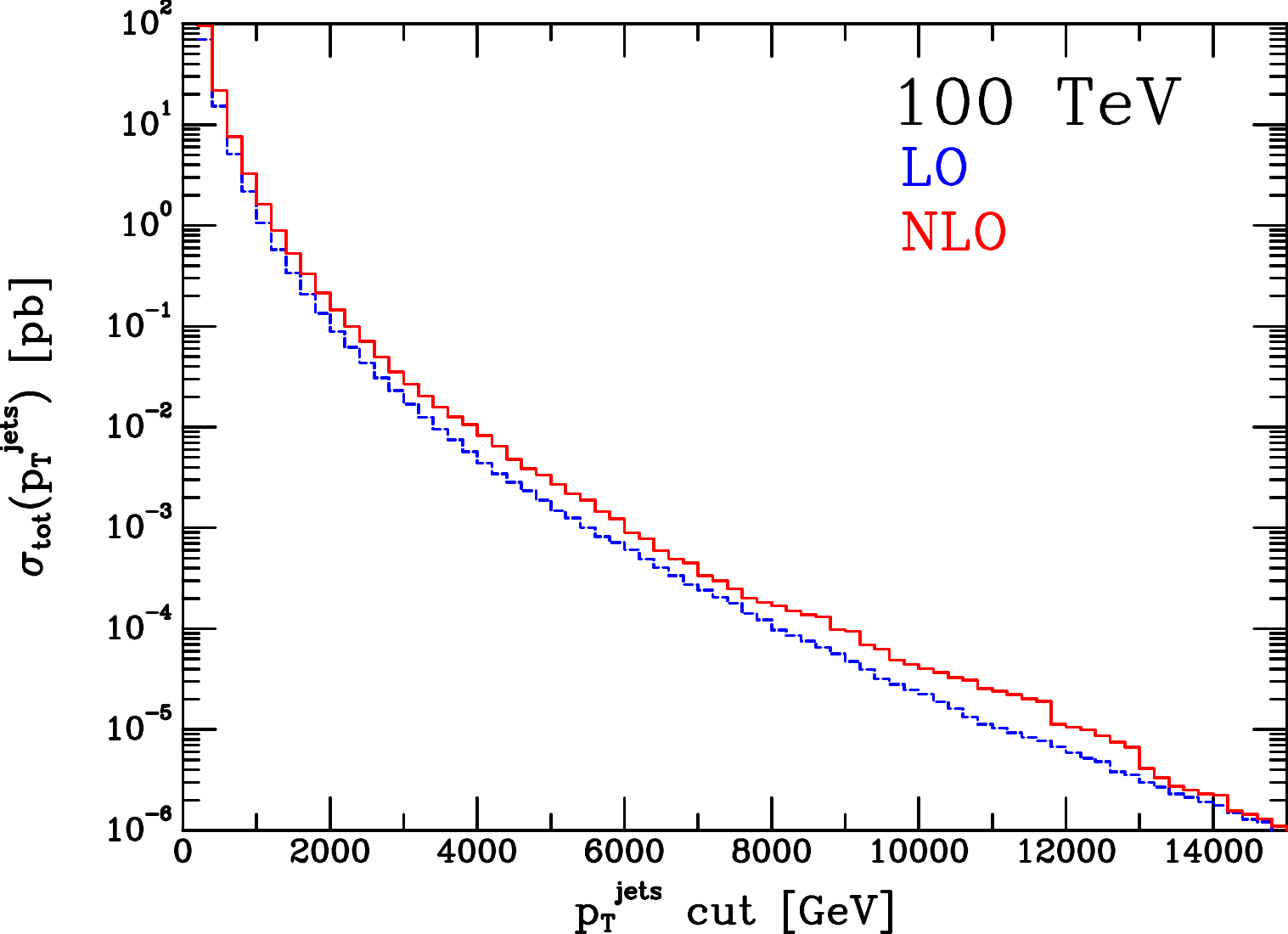}
\end{center}
\caption{Total integrated cross sections (cf. Eq.~ (\ref{eq:sigint})) at LO and NLO, for the quantity 
$H_T^\text{jets}=\sum_\text{jets} p_\perp^\text{jet}$ (left) and $|p_T^{WW}|\,\equiv\,|p_T^\text{jets}|$, the transverse momentum of the {\sl complete} jet system (right), with $p_\perp^\text{cut}\,=\,25\,\GeV$ in both cases. 
\label{fig:xss}}
\end{figure}

\subsubsection{Differential distributions}

To illustrate some of the key differences between the predictions for
$WW$+jet production at the two collider energies, we now examine NLO
predictions for a number of kinematic distributions.  For this study
we consider leptonic decays of the $W^+$ and $W^-$ bosons, but do not
apply any cuts on the decay products. We also show the respective
distributions at the 14 \TeV~ LHC for comparison.
Fig.~\ref{fig:comp1} shows two quantities that characterize the
overall nature of this process, the transverse momentum of the leading
jet and the scalar sum of all jet and lepton transverse momenta,
$H_T$.  All histograms have been normalized to the total NLO
cross-sections given earlier, in order to better compare their shapes.
At $100$~TeV the leading jet is significantly harder than at $14$~TeV.
The $H_T$ distribution is also harder at $100$~TeV with, of course, a
significant shift in the peak once the jet cut is
raised.~\footnote{This observable is also frequently used as a cut
  variable in searches for physics beyond the SM, for example in
  Refs.~\cite{Aad:2014gka,Aad:2015wqa}, where cuts are placed in the
  range $\sim 0.6$--$2\TeV$ depending on the details of the search
  strategy.}
\begin{figure}
\begin{center}
\includegraphics[width=0.33\textwidth,angle=-90]{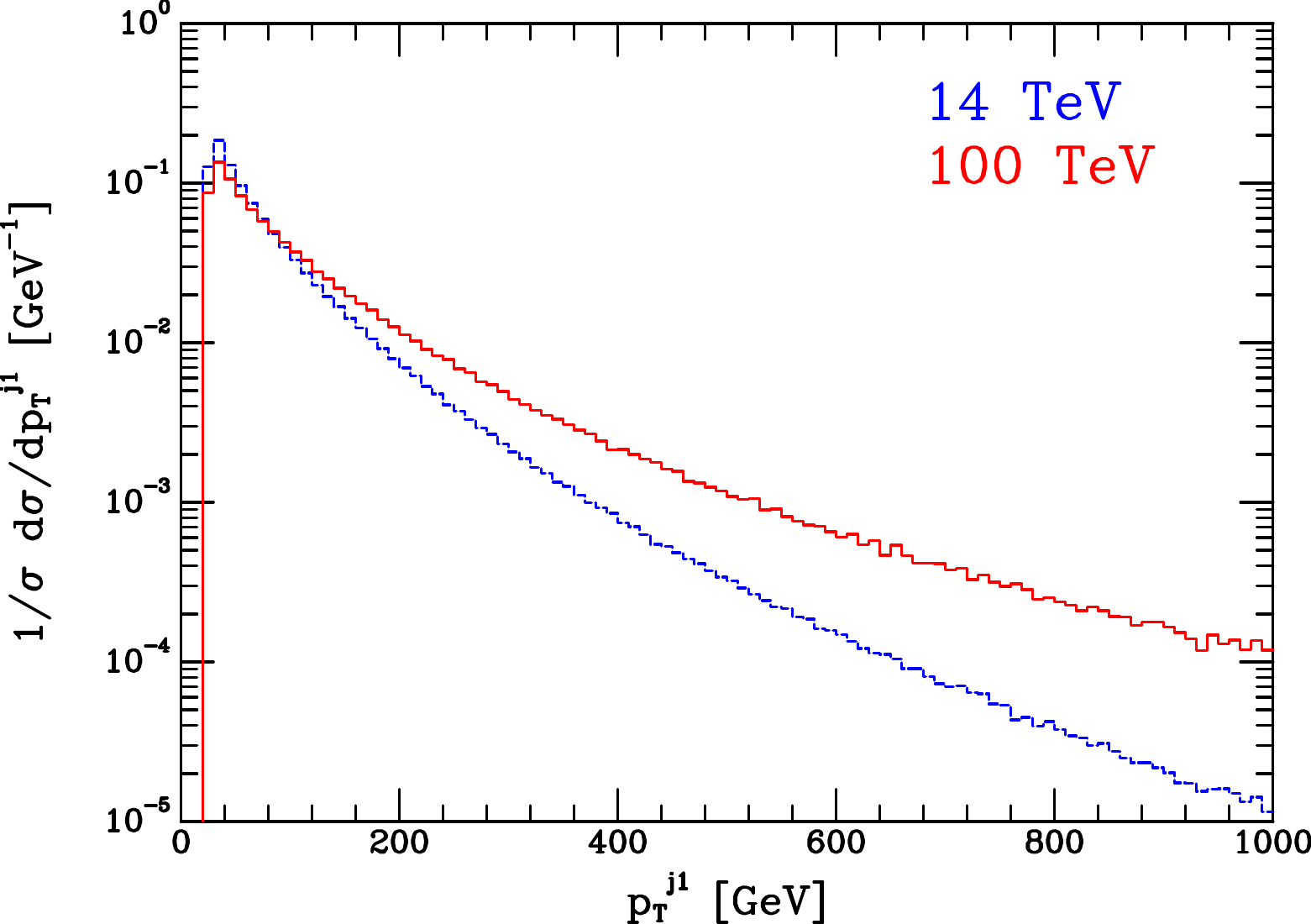} \hspace*{0.5cm}
\includegraphics[width=0.33\textwidth,angle=-90]{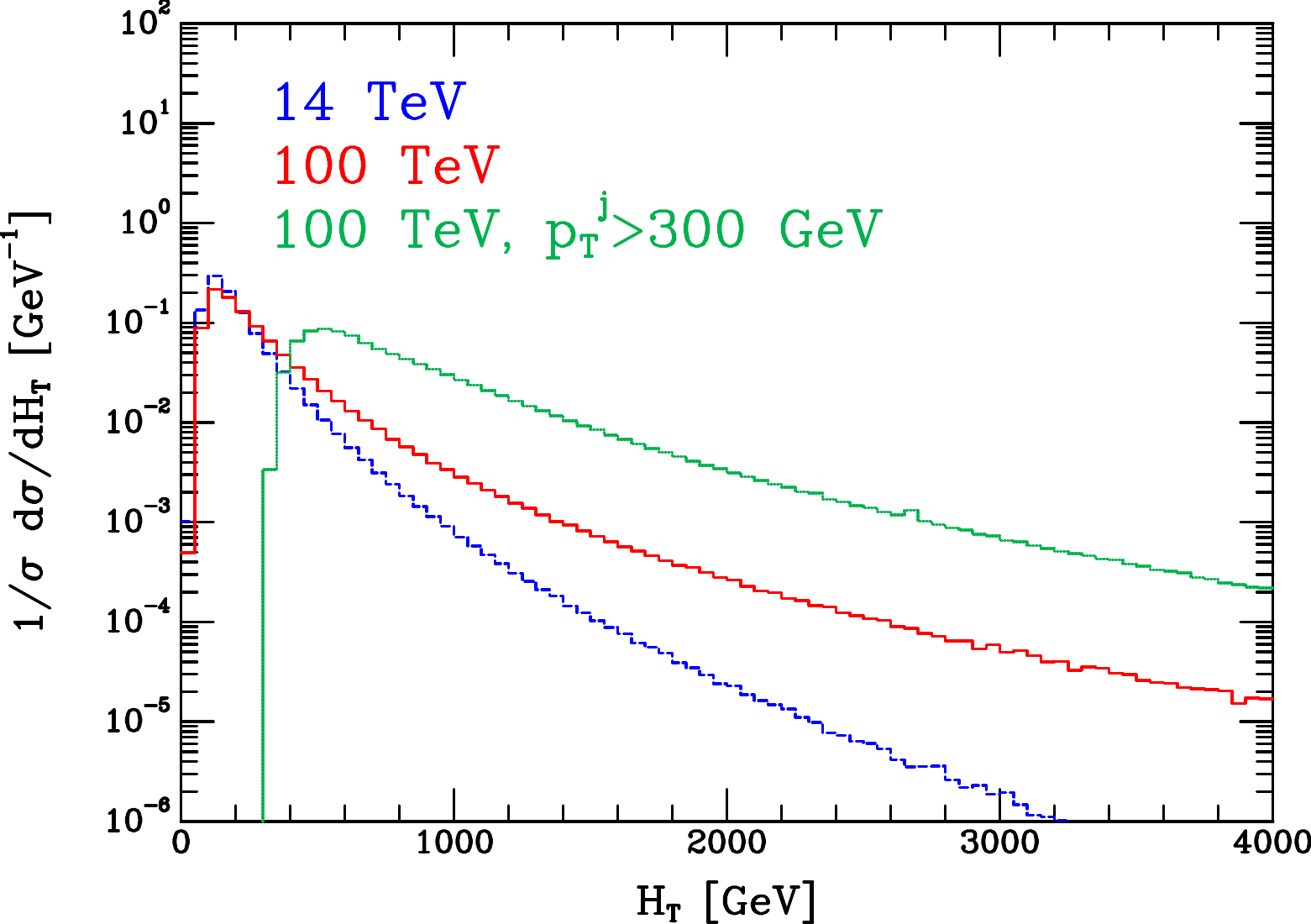}
\end{center}
\caption{NLO $p_{\perp, j}$ (left) and $H_T$ (right) distributions, normalized by the respective
total cross sections, for 14 \TeV (red), 100 \TeV (blue), and 100 \TeV * (green)
\label{fig:comp1}}
\end{figure}

Turning to leptonic observables, Fig.~\ref{fig:comp2} shows the
transverse momentum and rapidity of the positron from the $W^+$ decay.
The transverse momentum spectrum of the positron falls much less
steeply at $100$~TeV, and even less so with a higher jet cut.  The
rapidity distribution of the positron is also changed non-trivially,
with the broader peak at $100$~TeV reflecting the fact that the
process is probing a much smaller parton fraction.  When the jet cut
is raised to $300$~GeV the required parton fraction is again larger so
that the shape is a little closer to the one found at
$14$~TeV.~\footnote{Although not shown here, the jet rapidity exhibits
  a similar behaviour.}
\begin{figure}
\begin{center}
\includegraphics[width=0.33\textwidth,angle=-90]{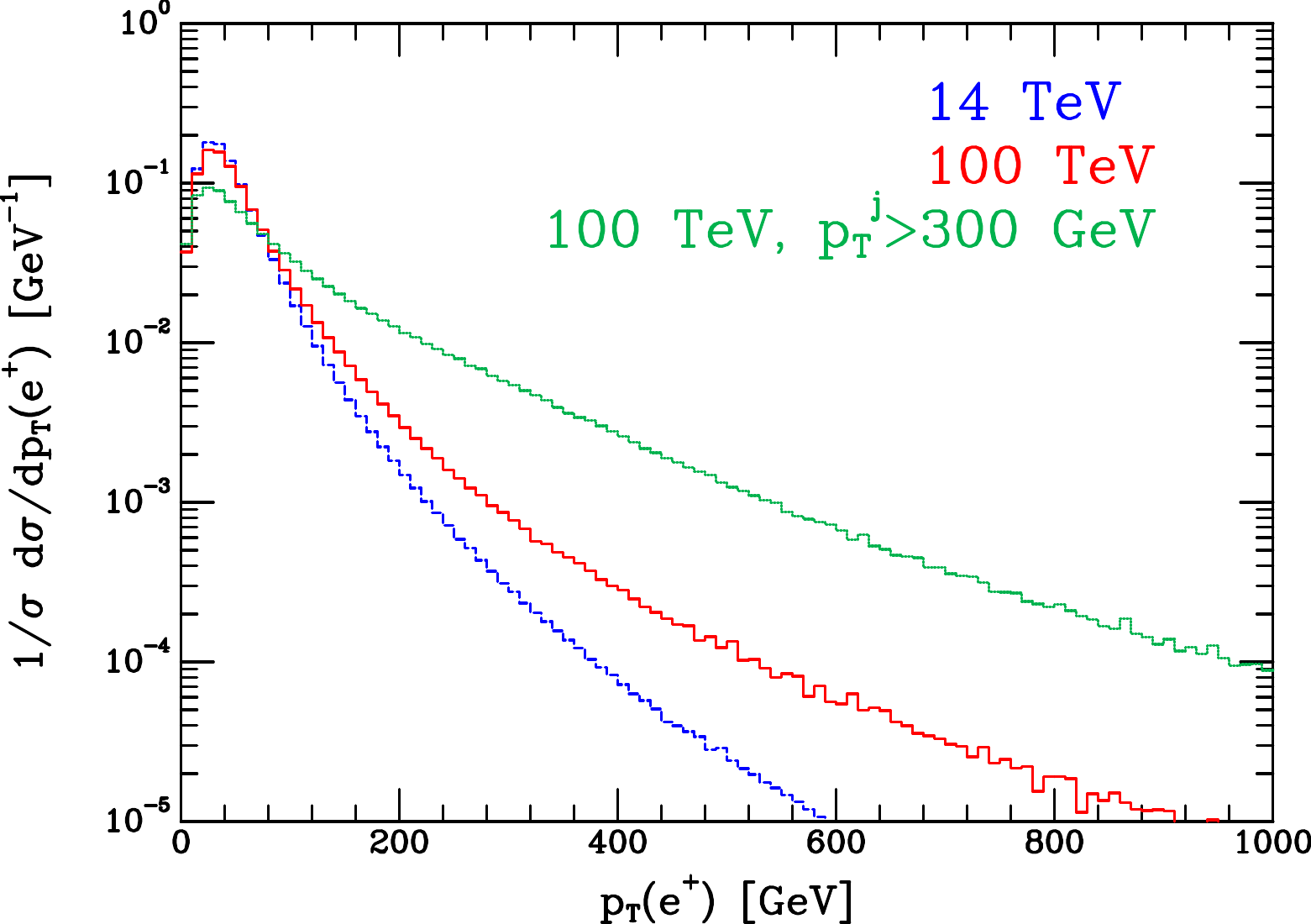} \hspace*{0.5cm}
\includegraphics[width=0.33\textwidth,angle=-90]{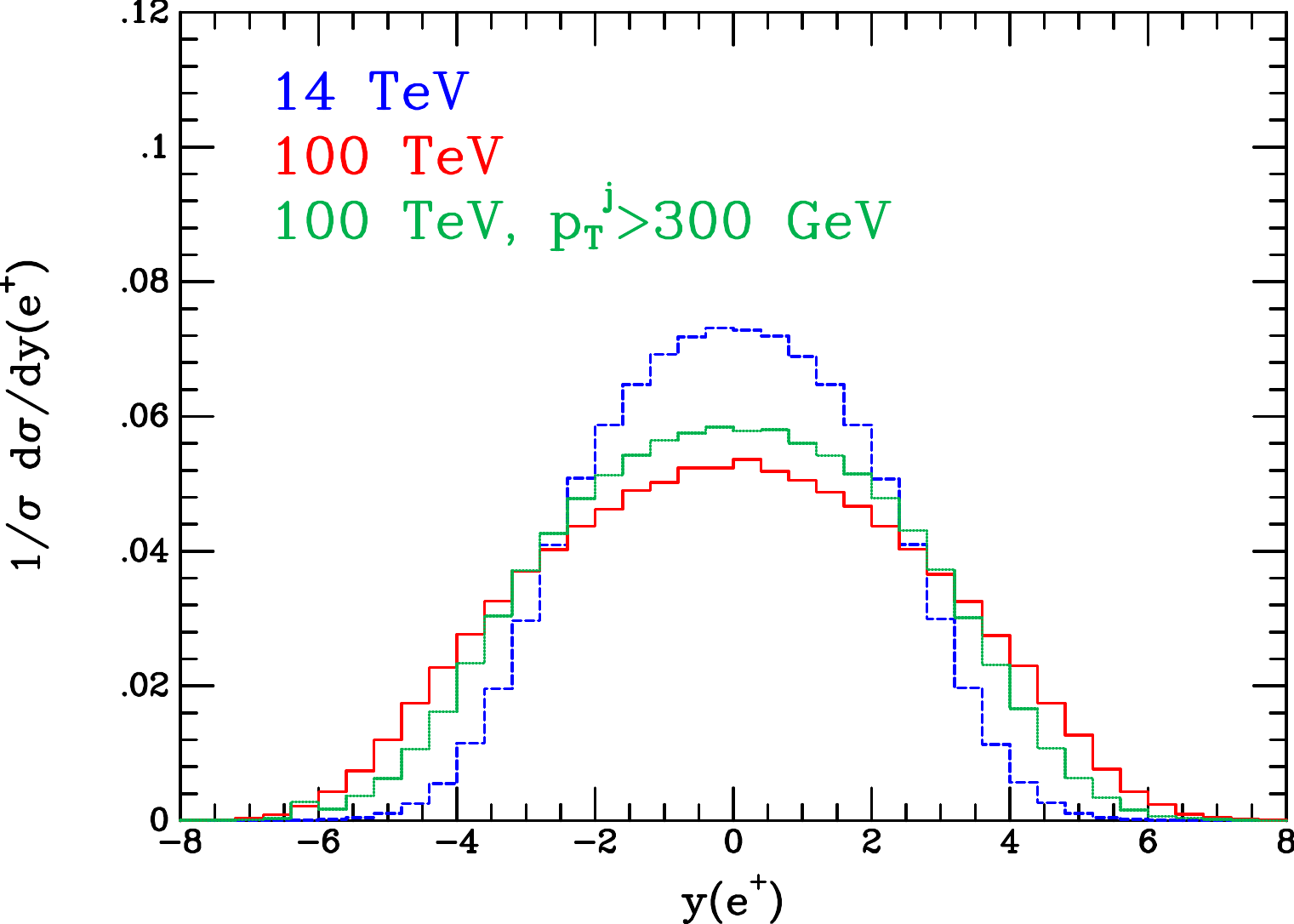}
\end{center}
\caption{NLO  $p_{\perp, \ell}$ (left) and $\eta_\ell$ (right)
distributions, normalized by the respective total cross sections, for 14 \TeV (red),
100 \TeV (blue), and 100 \TeV * (green) 
\label{fig:comp2}}
\end{figure}
An observable that is particularly interesting for this process is the
azimuthal angle between the electron and the positron, which can be
used to isolate contributions to this final state from Higgs boson
decays.  As shown in Fig.~\ref{fig:comp3}, under the usual jet cuts at
$14$~TeV, this distribution is peaked towards $\Delta \Phi_{\ell
  \ell}=\pi$, a feature which persists at $100$~TeV using the same jet
cut.  Once the jet cut is raised significantly, the recoil of the
$W^+W^-$ system results in the two leptons instead being
preferentially produced closer together, i.e. in the region $\Delta
\Phi_{\ell \ell} \to 0$.  This is the same region of $\Delta
\Phi_{\ell \ell}$ that is favoured by events produced via the Higgs
boson decay.  Even if the jet threshold at a $100$~TeV collider were
not as high as $300$~GeV, such a shift in this distribution could be
an important consideration in optimizing Higgs-related analyses in the
$W^+W^-$ decay channel.
\begin{figure}
\begin{center}
  \includegraphics[width=0.33\textwidth,angle=-90]{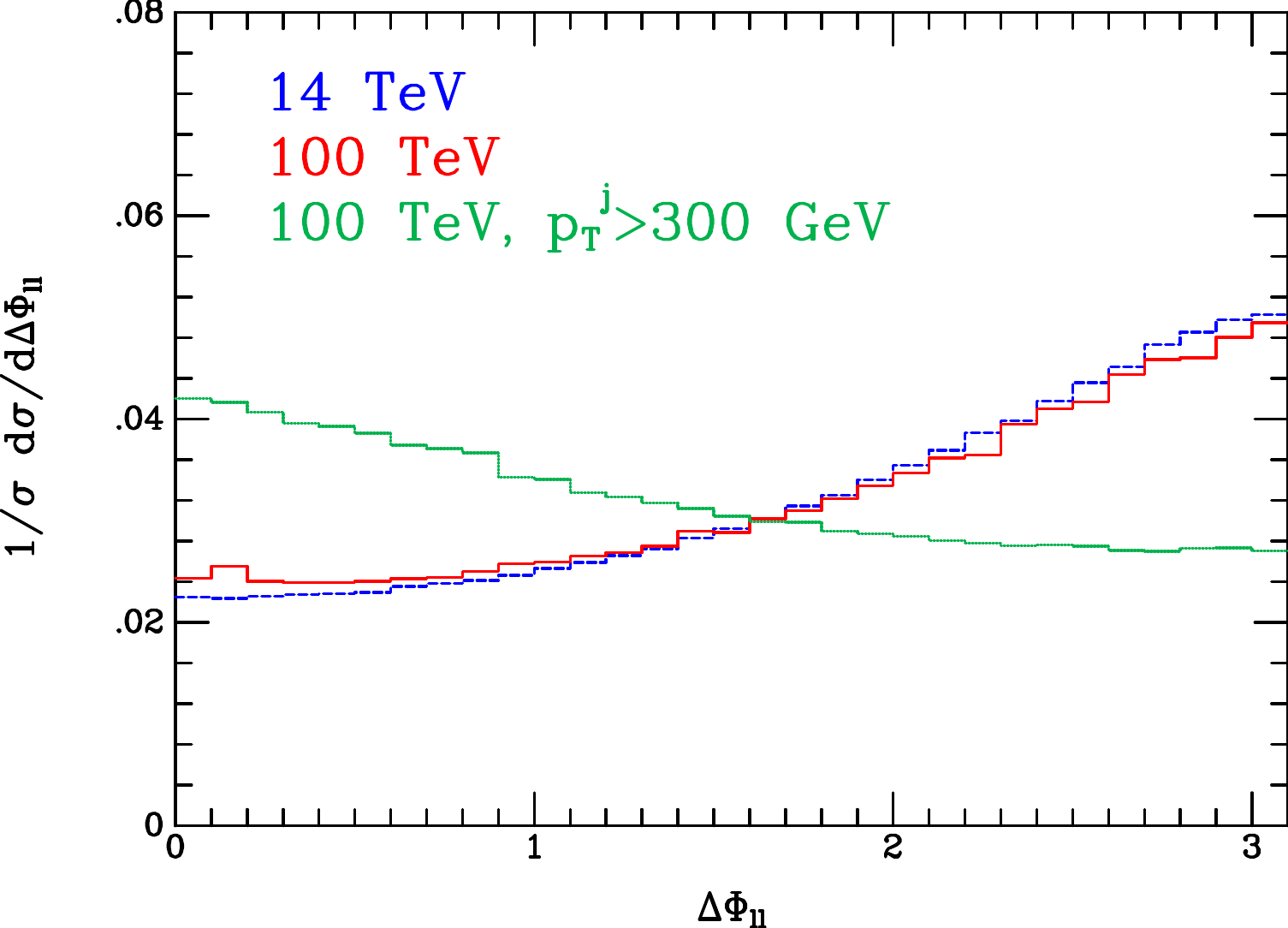}
  \hspace*{0.5cm}
  \includegraphics[width=0.33\textwidth,angle=-90]{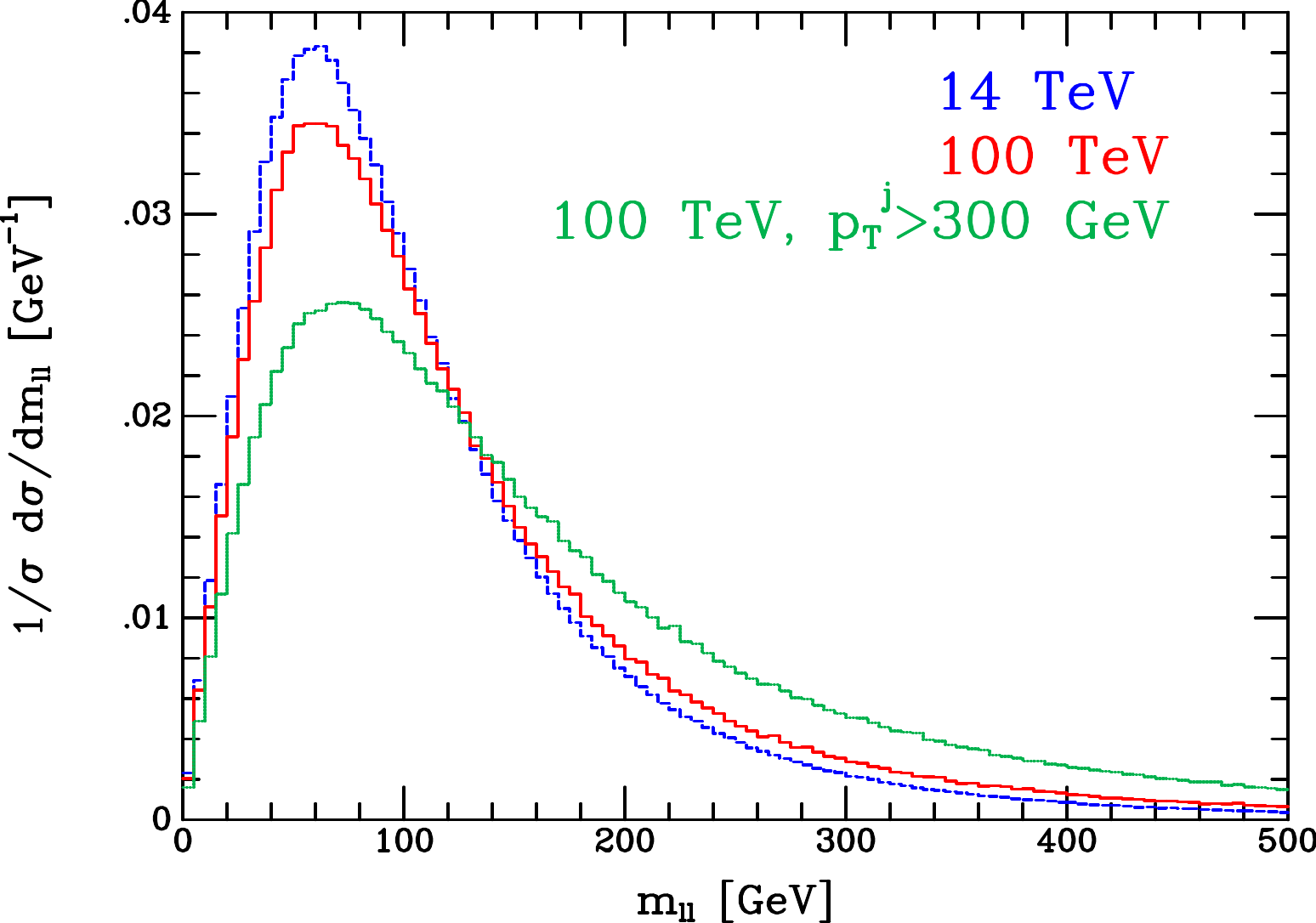}
\end{center}
\caption{NLO $\Delta \Phi_{\ell \ell}$ (left) and $m_{\ell \ell}$ (right)
distributions, normalized by the respective total cross sections, for 14 \TeV (red),
100 \TeV (blue), and 100 \TeV * (green) 
\label{fig:comp3}}
\end{figure}
Despite this shift to smaller $\Delta \Phi_{\ell \ell}$, the
combination of this effect with the change in the $p_{\perp, \ell}$
distribution shown earlier results in a relatively similar
distribution for $m_{\ell \ell}$, albeit with a longer tail.

Finally, we show the distribution of the transverse momentum for the
dilepton system $p_\perp^{\ell\ell}$, after cutting on the dilepton
invariant mass. The corresponding cross section values are given in
table \ref{mll}, while distributions are shown in figure
\ref{fig:etll}.

\renewcommand{\baselinestretch}{1.5}
\begin{table}
\begin{center}
\begin{tabular}{|r|c|c|}
\hline
 $m_{\ell\ell}^\text{max}$ & $\sigma_{LO}$~[pb] & $\sigma_{NLO}$~[pb] \\
\hline
125 \GeV &4.76 &5.34\\
50 \GeV &1.48 &1.64\\
\hline
\end{tabular}
\renewcommand{\baselinestretch}{1.0}
\caption{Cross-sections for the process $p p \to WW$+jet at a $100$~TeV proton-proton collider, for two different
cuts on the dilepton invariant mass. The listed values include leptonic branching ratios.
Monte Carlo uncertainties are at most a
single unit in the last digit shown shown in the table.
\label{mll}}
\end{center}
\end{table}
\renewcommand{\baselinestretch}{1.0}
\begin{figure}
\begin{center}
\includegraphics[width=0.33\textwidth,angle=-90]{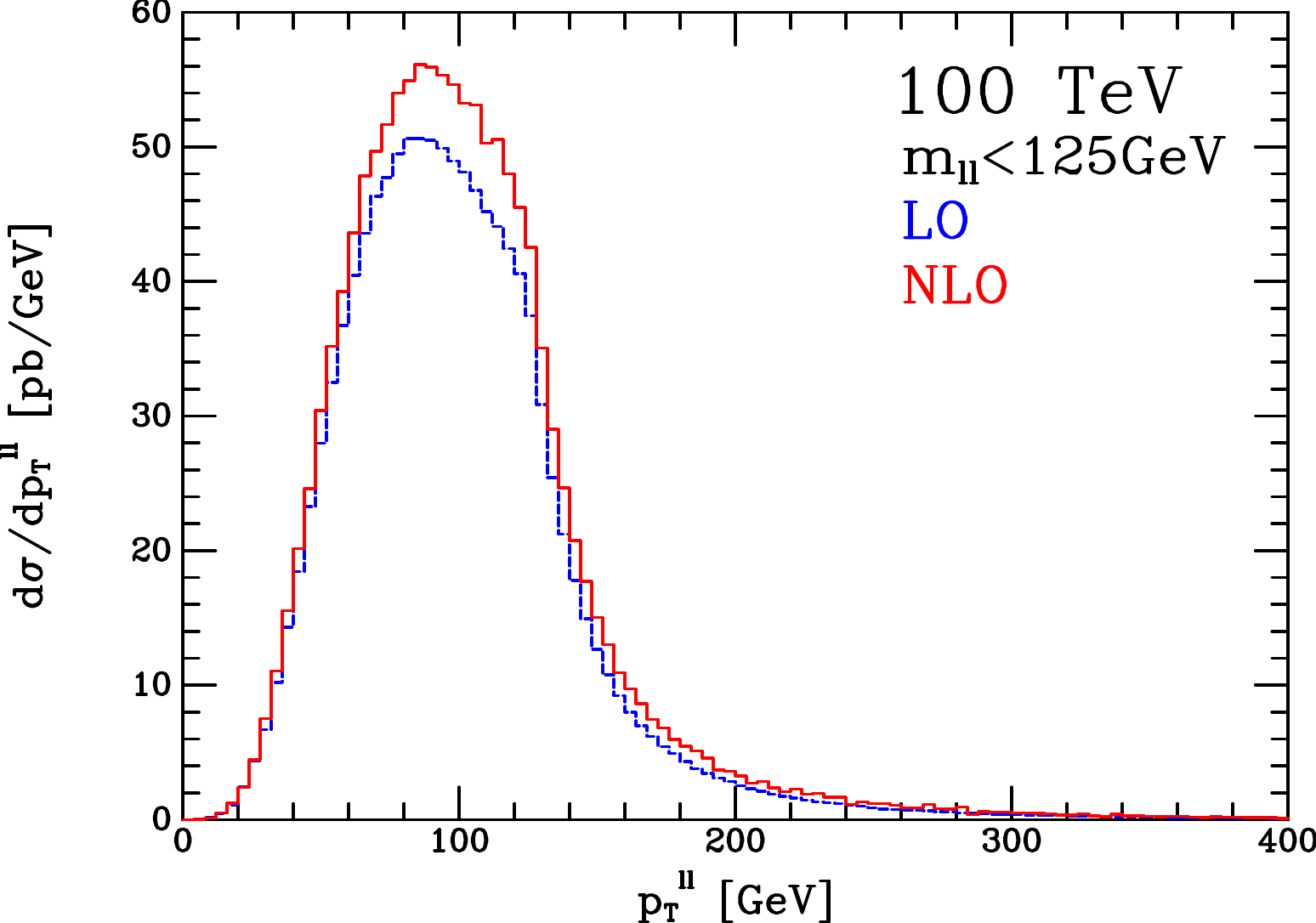} \hspace*{0.5cm}
\includegraphics[width=0.33\textwidth,angle=-90]{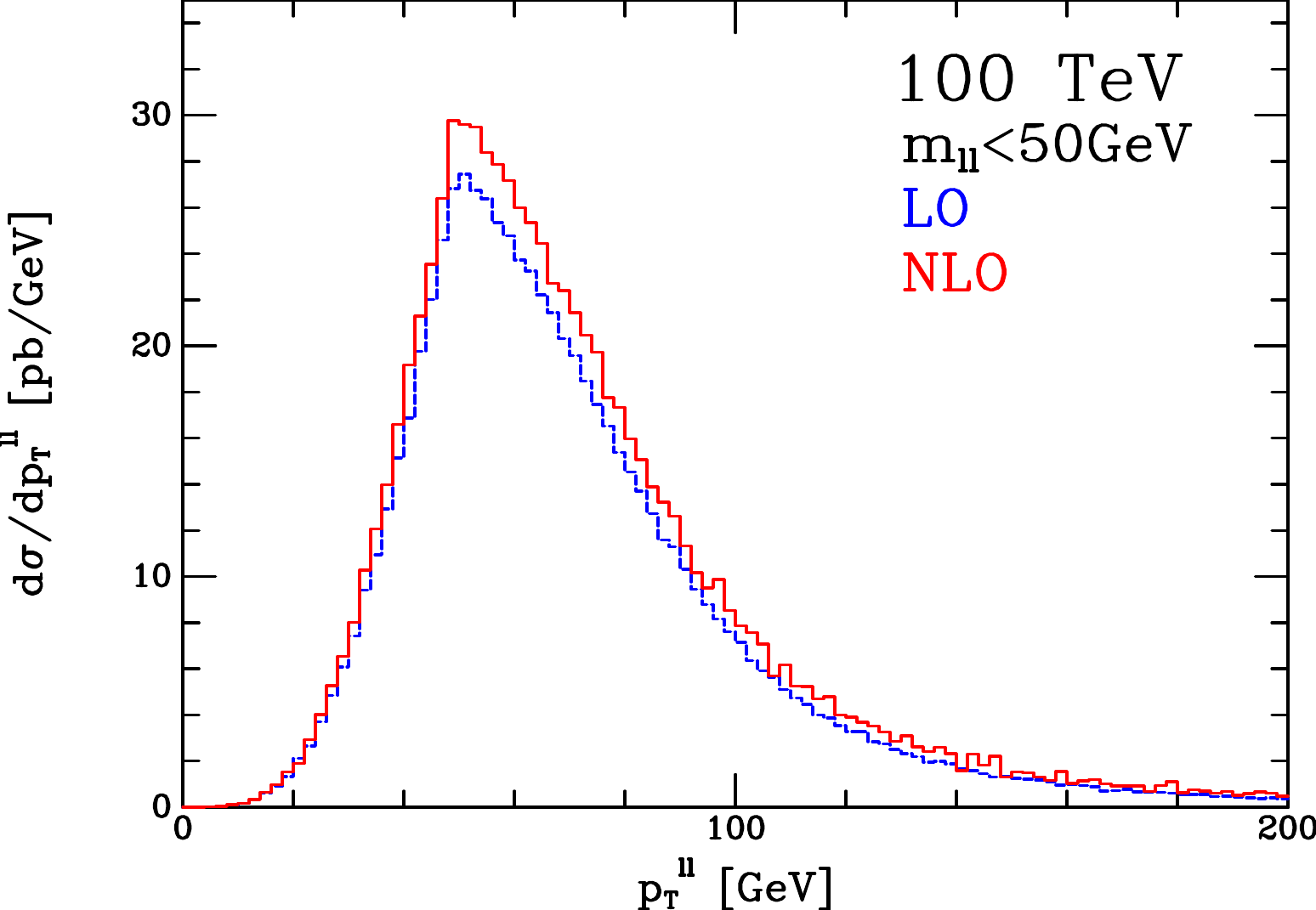}
\end{center}
\caption{ Transverse momentum of the dilepton system at LO and NLO, for $m_{\ell\ell}<125\,\GeV$ (left)
 and $m_{\ell\ell}<50\,\GeV$ (right).
\label{fig:etll}}
\end{figure}

\subsubsection{Summary}
Of course, at 100~\TeV~dimensionful variables, such as $p_\perp$ and
$m_{\ell\ell}$, exhibit longer tails in the distributions than at
14~\TeV.  This simply reflects the increased center-of-mass energy of
the system.  However this increase of the center-of-mass energy also
leads to broader rapidity distributions.  Furthermore, applying a
higher $p_\perp$ cut significantly changes distributions for the
dilepton azimuthal angle $\Delta \Phi_{\ell\ell}$ as well as the total
transverse momentum of the visible system $H_T$, which are frequently
used for background suppression for Higgs measurements or BSM
searches, respectively. In case such an increased cut is applied, this
needs to be taken into account when devising the respective search
strategies at a 100 \TeV~machine.

\clearpage
\section{Electroweak production of gauge bosons in VBF and VBS
  processes\footnote{Editor: B.J\"ager}}
\label{sec:vbf}
Vector boson fusion (VBF) and vector boson scattering (VBS) processes
provide particularly promising means for probing the mechanism of
electroweak symmetry breaking. At hadron colliders, this class of
reactions proceeds via the scattering of (anti-)quarks by the exchange
of weak gauge bosons in the $t$-channel with subsequent emission of
weak gauge bosons, i.e. the purely electroweak (EW) reactions $pp\to
Vjj$ and $pp\to VVjj$, respectively (with $V$ denoting a $W^\pm$ or a
$Z$ boson). In this report, we focus on leptonic decays of the weak
bosons. The jets emerging from the quarks in VBF and VBS reactions are
typically located in the forward and backward regions of the
detector. Little QCD activity is encountered in the central region of
rapidity. These characteristic features can be exploited for a
powerful suppression of a priori large QCD backgrounds. In the
following, we will consider EW $W^+jj$, $Zjj$, $W^+W^+jj$, $W^+Zjj$,
$W^+W^-jj$, and $ZZjj$ production at a 100~TeV proton-proton collider
in the context of the Standard Model. We will devise selection cuts
tailored for an optimization of the respective signal processes in the
presence of the most abundant QCD backgrounds, in particular
QCD-induced $VVjj$ processes and, in the case of $W^+W^-jj$ final
states, backgrounds constituted by $t\bar t$ production in association
with up to two jets.  For VBS reactions, we assume, for simplicity,
that each gauge boson is decaying into a different type of lepton
pair, and neglect interference effects that in principle could arise
from final states involving same-type leptons. Off-shell and
non-resonant contributions to the respective 2-lepton+2-jet or
4-lepton+2-jet final states are fully taken into account in all signal
channels.

After a description of the general setup and input parameters of our
study in Sec.~\ref{ssec:input}, we will discuss various VBS-induced
double and single gauge-boson production processes in
Secs.~\ref{ssec:wpwpjj}--\ref{ssec:wpwmjj}, and Sec.~\ref{sec:vbf_V},
respectively. In Sec.~\ref{ssec:bench} benchmark cross sections for
the various VBS signal processes are provided.

\subsection{Input parameters and setup}
\label{ssec:input}
Our numerical calculations are performed with the {\tt VBFNLO} Monte
Carlo package
\cite{Arnold:2008rz,Oleari:2003tc,Jager:2006zc,Jager:2006cp,Bozzi:2007ur,Jager:2009xx,Campanario:2013qba,Campanario:2013gea,Campanario:2014ioa}
for all $Vjj$ and $VVjj$ processes apart from the QCD-induced
$W^+W^-jj$ mode, and the {\tt Madgraph} code package
\cite{Alwall:2011uj} for the remaining processes, including the
top-induced backgrounds. In principle, the public {\tt
  POWHEG-BOX}~package
\cite{Nason:2004rx,Frixione:2007vw,Alioli:2010xd} provides
implementations for several VBS signal and background processes
including NLO-QCD corrections matched with parton showers
\cite{Jager:2011ms,Jager:2012xk,Schissler:2013nga,Jager:2013mu,Jager:2013iza,Melia:2011gk,Re:2012zi,Campbell:2013vha,Frixione:2007nw,Alioli:2011as}. However,
since the major goal of this study is to explore the capabilities of a
future high-energy collider facility rather than to perform a
precision analysis, we will refrain from using this tool here.

For the results presented in this section we use the SM masses and
widths,
\begin{equation}
\label{eq:SMinput}
\begin{array}{clrccl}
M_W&=80.385~\mathrm{GeV}, &&  \Gamma_W &= 2.097547~\mathrm{GeV}\,,
\\
M_Z &=91.1876~\mathrm{GeV}, && \Gamma_Z &= 2.508827~\mathrm{GeV}\,,
\\
M_H &=125.09~\mathrm{GeV}, & & \Gamma_H &= 0.004066~\mathrm{GeV}\,,
\\
m_\mathrm{top}&=172.5~\mathrm{GeV},& & \Gamma_\mathrm{top} &= 1.340488~\mathrm{GeV}\,. 
\end{array}
\end{equation}
The EW coupling constant is computed in the $G_\mu$ scheme from the
above input parameters and the Fermi constant $G_\mu =
1.1663787\times10^{-5}~\mathrm{GeV}^{-2}$, via
\begin{equation}
  \alpha_{G_\mu} = \frac{\sqrt{2}\,G_\mu M_W^2}{\pi}\left(1-\frac{M_W^2}{M_Z^2}\right)\,.
\end{equation}
External $b$-and $t$-quark contributions are disregarded throughout in
all matrix elements.  For the parton distribution functions (PDFs) of
the proton, we use the MMHT2014lo/nlo68cl sets
\cite{Harland-Lang:2014zoa} at LO and NLO, respectively, and the
corresponding values of $\alpha_s$ as provided by the LHAPDF
repository \cite{Buckley:2014ana}.
As factorization scale, $\mu_F$, and renormalization scale, $\mu_R$, for the EW $VVjj$ processes we use
\begin{equation}
\mu_F=\mu_R = Q_i\,, 
\end{equation}
where the $Q_i$ denote the momentum transfer of the incoming to the
outgoing quark on the upper and lower fermion lines, respectively. For
the QCD induced $VVjj$ processes, we use
\begin{equation}
\mu_F=\mu_R = \frac{1}{2}H_T\,, 
\end{equation}
with 
\begin{equation}
H_T = \sum_{i} p_{T,i}+E_T(V_1)+E_T(V_2)\,,\end{equation}
where the summation is running over all final-state partons in an
event, and the transverse energy of each weak boson is determined by
its transverse momentum, $p_{T,V}$, and mass, $M_V$, via
\begin{equation}
E_T(V)=\sqrt{p_{T,V}^2+M_V^2}\,.
\end{equation}
%

%

For our numerical analysis, we use a set of minimal selection cuts.
For the reconstruction of jets, we use the anti-$k_\mathrm{T}$
algorithm with $R=0.4$, and demand a minimum transverse momentum,
\begin{equation}
\label{eq:jet-def}
p_\mathrm{T,jet} \ge 50~\mathrm{GeV}\,.
\end{equation}
The two hardest jets fulfilling the cut of Eq.~(\ref{eq:jet-def}) are
called ``tagging jets''. These two jets are required to reside in
opposite hemispheres of the detector,
\begin{equation}
\label{eq:ysign}
y_{j_1}\times y_{j_2} < 0\,. 
\end{equation}

For charged leptons we impose cuts on transverse momenta, rapidities,
and jet-lepton separations in the rapidity-azimuthal angle plane,
\begin{equation}
\label{eq:ptl-cut}
p_{T,\ell} \ge 20~\mathrm{GeV}\,,\qquad
|y_{\ell}| \le 5~\,,\qquad  
\Delta R_\mathrm{jet,\ell} \ge 0.4\,\,.
\end{equation}
A very powerful tool for the suppression of background processes is
provided by requiring the charged leptons to be located in between the
two tagging jets in rapidity,
\begin{equation}
\label{eq:lgap}
 y_{j,min}^{tag} < y_{\ell}  < y_{j,max}^{tag}\,. 
\end{equation}

For the $ZZjj$, $W^\pm Z jj$, and $Zjj$ processes, to suppress
contributions from photons of very small virtuality we furthermore
require a minimal invariant mass for all pairs of oppositely charged
leptons,
\begin{equation}
\label{eq:ll-cuts}
M_{\ell^+\ell^-} > 15~\mathrm{GeV}\,.  
\end{equation}
In addition to these minimal cuts, process-specific selection cuts are
devised for each channel.

\subsection{$W^+W^+jj$}
\label{ssec:wpwpjj}
For the $W^+W^+jj$ channel, we consider the representative $\nu_e
e^+\nu_\mu \mu^+ jj$ final state.  We found that the EW signal in the
presence of QCD-induced $W^+W^+jj$ production can be improved by a set
of selection cuts that are imposed in addition to the minimal cuts of
Eqs.~(\ref{eq:jet-def})--(\ref{eq:lgap}).
Because of the absence of gluon-induced contributions in the
QCD-induced production mode a very large signal-to-background ($S/B$)
ratio of 29.35 can be achieved by rather moderate customized cuts on
the separation of the two tagging jets,
\begin{equation}
\label{eq:jjcuts_wpp}
m_{jj}>500~\mathrm{GeV}\,,\quad
\Delta y_{jj}>1.5\,.
\end{equation}
With this set of cuts, we obtain cross sections of $\sigma^\mathrm{EW}
= 49.335(8)$~fb and $\sigma^\mathrm{QCD} = 1.681(2)$~fb for EW- and
QCD-induced $W^+W^+jj$ production, respectively, at LO. The NLO-QCD
corrections to the EW signal process are small, resulting in a cross
section of $\sigma^\mathrm{EW}_\mathrm{NLO} = 52.56(2)$~fb. We note,
however, that the NLO-QCD corrections are not flat, but affect bulk
and tail of distributions in a non-trivial manner. To illustrate this
effect, we depict the transverse momentum distribution of the hardest
jet at LO and NLO~QCD in Fig.~\ref{fig:ptjet-nloqcd}.
%
%
\begin{figure}
\begin{center}
\includegraphics[width=0.5\textwidth]{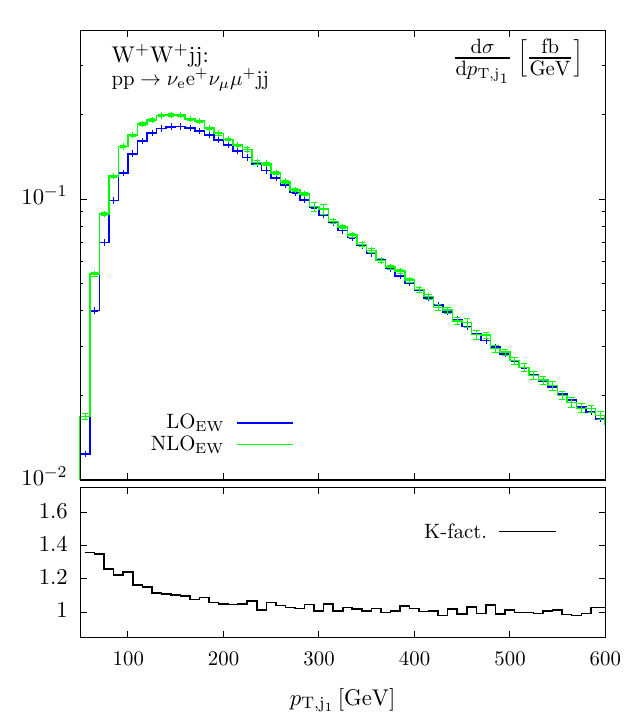}
\caption{Transverse-momentum distribution of the hardest jet in  $pp\to \nu_e e^+\nu_\mu \mu^+ jj$ via VBS, within the selection cuts of Eqs.~(\ref{eq:jet-def})--(\ref{eq:lgap}) and Eq.~(\ref{eq:jjcuts_wpp}). The upper panel shows the LO (blue line) and the NLO-QCD results (green line) for the EW process, while the lower panel displays the K-factor that is defined as the ratio of the NLO to the LO result.  
\label{fig:ptjet-nloqcd}
}
\end{center}
\end{figure}
%
Despite the non-negligible impact of NLO-QCD corrections, in the
following we restrict our analysis to LO, since at this time details
of a possible experimental setup represent the dominant source of
uncertainties.

Figure~\ref{fig:wpwpjj-plots} shows the EW signal and the QCD
background for the same distribution and, in addition, for the
transverse mass distribution of the gauge-boson system.
%
%
\begin{figure}
\includegraphics[width=0.5\textwidth]{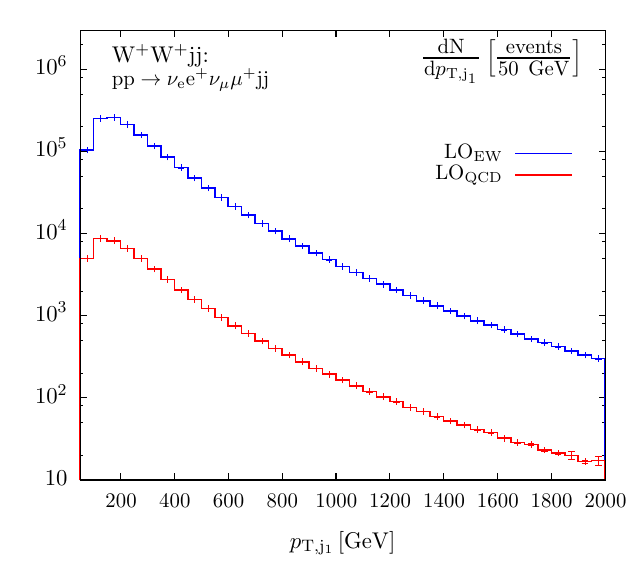}
\includegraphics[width=0.5\textwidth]{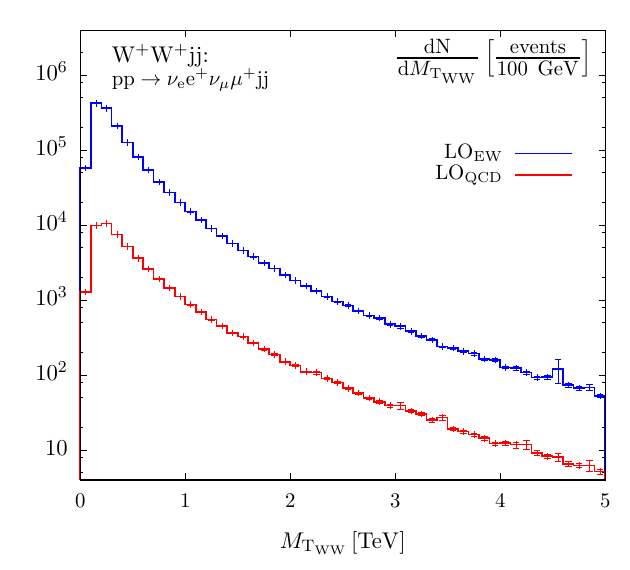}
\caption{Transverse-momentum distribution of the hardest jet~(l.h.s.) and transverse-mass distribution of the gauge-boson system~(r.h.s) for the EW-induced (blue line) and QCD-induced (red line) contributions to  $pp\to \nu_e e^+\nu_\mu \mu^+ jj$, within the selection cuts of Eqs.~(\ref{eq:jet-def})--(\ref{eq:lgap}) and Eq.~(\ref{eq:jjcuts_wpp}) for an integrated luminosity of 30~ab$^{-1}$. 
\label{fig:wpwpjj-plots}
}
\end{figure}
%
%
In order to spot new physics that mostly impacts the tails of
invariant-mass and transverse-momentum distributions, searches
typically focus on the kinematic region of large invariant masses of
the gauge-boson system. In the presence of two neutrinos, this
quantity is not fully reconstructible.  In this case, the transverse
mass of the $W^+W^+$ system is considered instead, that is defined by
\begin{equation}
M_{T_{WW}} =
 \sqrt{
           \left(E_{T}^{\ell\ell}+E_{T}^{miss}\right)^2
          -\left(\vec{p}_T^{\ell\ell}+\vec{p}_T^\mathrm{miss}\right)^2
          }\,,
\end{equation}
where 
\begin{equation}
E_{T}^{\ell\ell} = \sqrt{(\vec{p}_T^{\ell\ell})^2+M_{\ell\ell}^2}\,,\quad 
E_{T}^{miss} = |\vec{p}_T^\mathrm{miss}|\,.
\end{equation}
Here, $\vec{p}_T^{\ell\ell}$ denotes the transverse momentum of the
charged-lepton system, and $\vec{p}_T^\mathrm{miss}$ the total
transverse momentum of the neutrino system.

The transverse-mass distribution depicted in
Fig.~\ref{fig:wpwpjj-plots} clearly exhibits that the EW signal is
dominating over the entire kinematic range. Thus, even after the
application of a severe cut on $M_{T_{WW}}$ that might be necessary in
new physics searches, the impact of the QCD-induced background on the
VBS signal will remain small.  In order to quantify the number of
events per bin we are assuming an integrated luminosity of
30~ab$^{-1}$.

In Fig.~\ref{fig:wpwpjj-nevts} we show the number of events above a
specific value of the tagging jets' transverse momentum and the
gauge-boson system's transverse mass, respectively, assuming an
integrated luminosity of 30~ab$^{-1}$.
%
%
\begin{figure}
\includegraphics[width=0.5\textwidth]{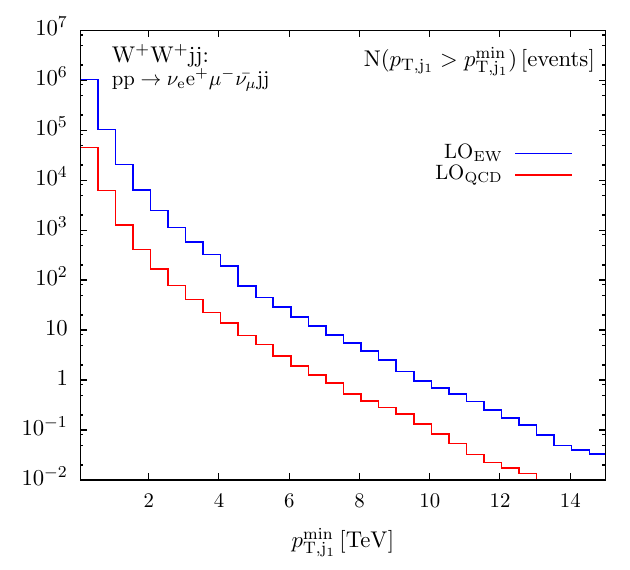}
\includegraphics[width=0.5\textwidth]{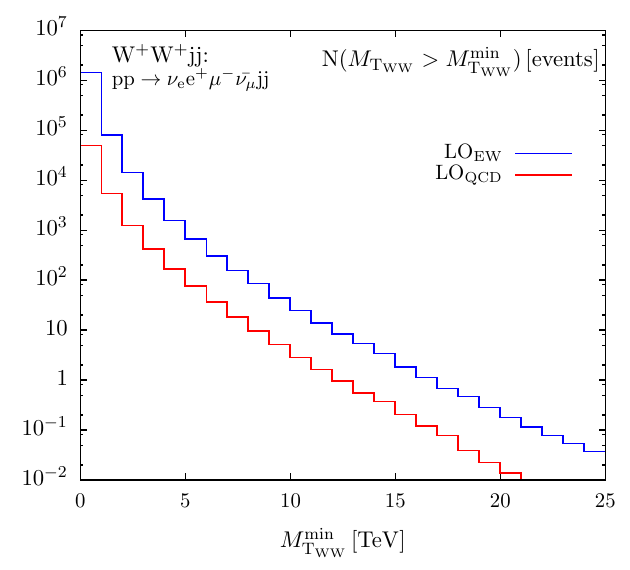}
\caption{Total number of events produced with $p_{T,j_1}>p_{T,j_1}^\mathrm{min}$ (l.h.s.) and  with  $M_{T_{WW}}>M_{T_{WW}}^\mathrm{min}$~(r.h.s.) for the EW-induced (blue line) and QCD-induced (red line) contributions to  $pp\to \nu_e e^+\nu_\mu \mu^+ jj$, within the selection cuts of Eqs.~(\ref{eq:jet-def})--(\ref{eq:lgap}) and Eq.~(\ref{eq:jjcuts_wpp}) for an integrated luminosity of 30~ab$^{-1}$. 
\label{fig:wpwpjj-nevts}
}
\end{figure}
%
%

%
\subsection{$W^+Z jj$}
\label{ssec:wpzjj}
For the $W^+Zjj$ channel, we consider the representative $\nu_e
e^+\mu^-\mu^+ jj$ final state.  An optimization of the $S/B$ ratio in
the $W^+Z jj$ channel can be achieved when in addition to the cuts of
Eqs.~(\ref{eq:jet-def})--(\ref{eq:ll-cuts}) the following
process-specific cuts are imposed:
\begin{equation}
\label{eq:jjcuts_wpz}
m_{jj}>2500~\mathrm{GeV}\,,\quad
\Delta y_{jj}>5\,.
\end{equation}
With the cuts of Eqs.~(\ref{eq:jet-def})--(\ref{eq:ll-cuts}) and
Eq.~(\ref{eq:jjcuts_wpz}), we obtain a cross section of
$\sigma^\mathrm{EW} = 5.0547(7)$~fb and $\sigma^\mathrm{QCD} =
2.801(1)$~fb for EW- and QCD-induced $W^+Zjj$ production,
respectively, at LO, resulting in an $S/B$ ratio of 1.80.
%
%
\begin{figure}
\includegraphics[width=0.5\textwidth]{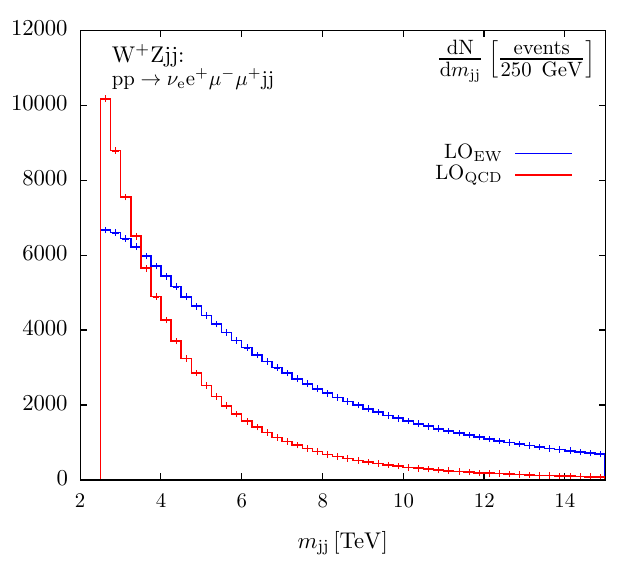}
\includegraphics[width=0.5\textwidth]{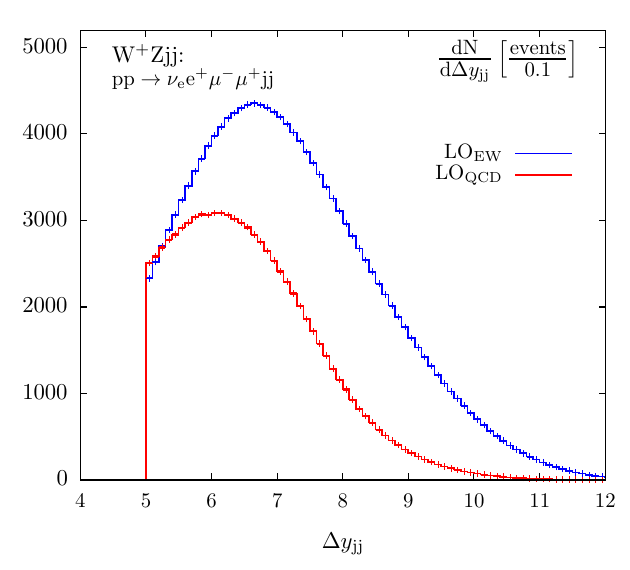}
\caption{Invariant-mass distribution (l.h.s.) and rapidity separation of the two tagging jets (r.h.s.) for the EW-induced (blue line) and QCD-induced (red line) contributions to  $pp\to \nu_e e^+\mu^-\mu^+ jj$, within the selection cuts of Eqs.~(\ref{eq:jet-def})--(\ref{eq:ll-cuts}) and Eq.~(\ref{eq:jjcuts_wpz})  for an integrated luminosity of 30~ab$^{-1}$.  
\label{fig:wpzjj-plots}
}
\end{figure}
%
%
For this setup, the invariant mass distribution and the rapidity
separation of the two tagging jets are shown in
Fig.~\ref{fig:wpzjj-plots}. Obviously, in the QCD-induced production
mode the two jets tend to be closer, which is essential for the design
of cuts for the improvement of the $S/B$ ratio.

In contrast to $W^+W^+jj$ and $W^+W^-jj$ final states where the
invariant mass of the two-gauge-boson system cannot be determined in
the fully leptonic decay modes, such a reconstruction is possible in
the $W^+Zjj$ channel using kinematical constraints to estimate the
longitudinal component of the neutrino momentum.  The distribution of
the invariant mass computed from these reconstructed momenta is
depicted in Fig.~\ref{fig:wpzjj-mvv} together with the number of
events above a specific value of $M_{WZ}$, assuming an integrated
luminosity of 30~ab$^{-1}$.
\begin{figure}
\includegraphics[width=0.5\textwidth]{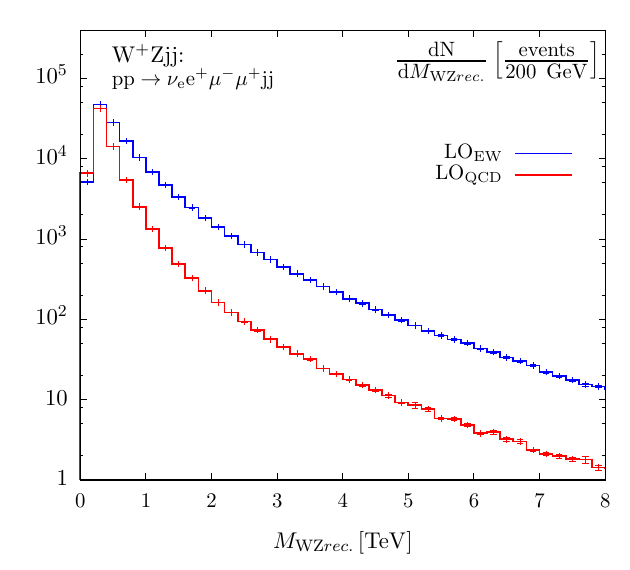}
\includegraphics[width=0.5\textwidth]{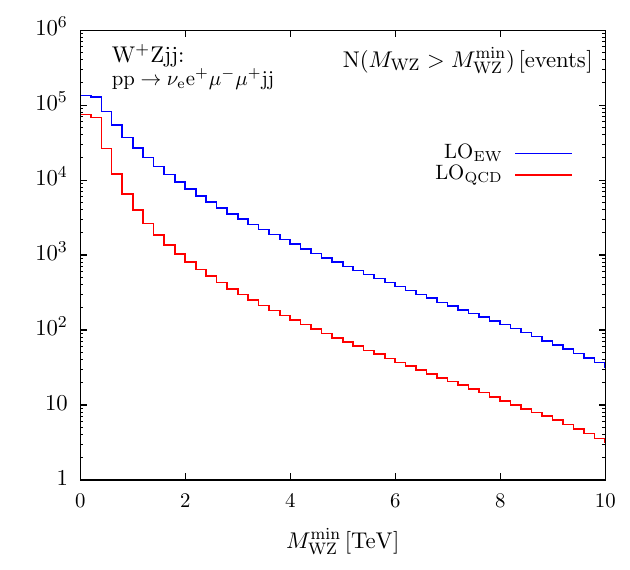}
\caption{Invariant-mass distribution of the $WZ$ system reconstructed from the lepton momenta~(l.h.s.) and total number of events produced with $M_{WZ}>M_{WZ}^\mathrm{min}$~(r.h.s) for the EW-induced (blue line) and QCD-induced (red line) contributions to  $pp\to \nu_e e^+\mu^-\mu^+ jj$, within the selection cuts of Eqs.~(\ref{eq:jet-def})--(\ref{eq:ll-cuts}) and Eq.~(\ref{eq:jjcuts_wpz}).  An integrated luminosity of 30~ab$^{-1}$ is assumed. 
\label{fig:wpzjj-mvv}
}
\end{figure}
%

%
\subsection{$ZZ jj$}
\label{ssec:zzjj}

The $ZZjj$ channel is of particular phenomenological relevance, both,
as VBS process that is sensitive, for instance, to new scalar
resonances in the TeV regime, and as background to Higgs production
via vector boson fusion in the $H\to ZZ$ decay mode. Here, we focus on
the fully leptonic final state where each $Z$ boson decays into a
lepton pair of different type, i.e.\ the process $pp\to
e^-e^+\mu^-\mu^+ jj$.

Proceeding in the same manner as for the $W^+W^+ jj$ and $W^+Z jj$
processes, we devise a set of selection cuts enhancing the impact of
the EW production mode with respect to QCD-induced $ZZjj$
production. To this end, we impose the basic selection cuts of
Eqs.~(\ref{eq:jet-def})--(\ref{eq:ll-cuts}), amended by the extra cuts
of
\begin{equation}
\label{eq:jjcuts_zz}
m_{jj}>2000~\mathrm{GeV}\,,\quad
\Delta y_{jj}>3\,.
\end{equation}
With these cuts, we find a LO cross section of $\sigma^\mathrm{EW} =
2.1506(7)$~fb and $\sigma^\mathrm{QCD} = 0.2533(2)$~fb for EW- and
QCD-induced $ZZjj$ production, respectively, resulting in an $S/B$
ratio of 8.49.

The invariant mass of the $ZZ$ system can be fully reconstructed from
the momenta of the final-state charged leptons. Figure~\ref{fig:mzz}
shows the four-lepton invariant-mass distribution in two different
ranges.  At low values of $M_{ZZ}$, an interesting structure can be
observed that is due to the $Z$ peak around 91~GeV and, for the EW
production mode, the Higgs resonance at 125~GeV. Both channels exhibit
a broad continuum contribution above the $Z$-pair production threshold
with the QCD contribution decreasing slightly faster than the EW
contribution.
%
%
\begin{figure}
\includegraphics[width=0.5\textwidth]{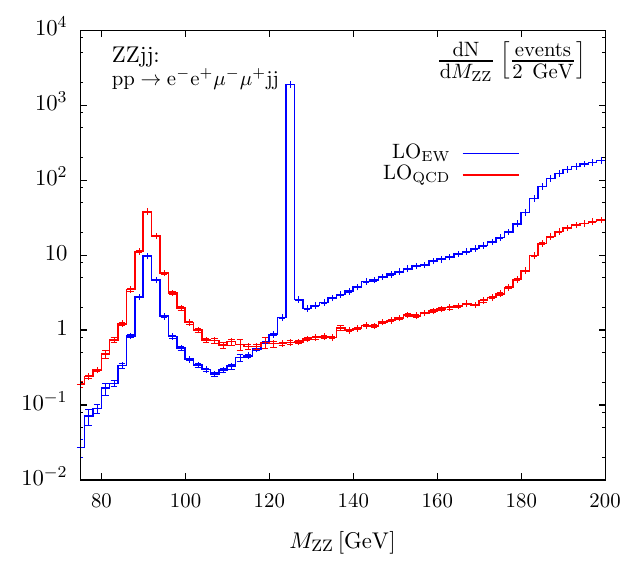}
\includegraphics[width=0.5\textwidth]{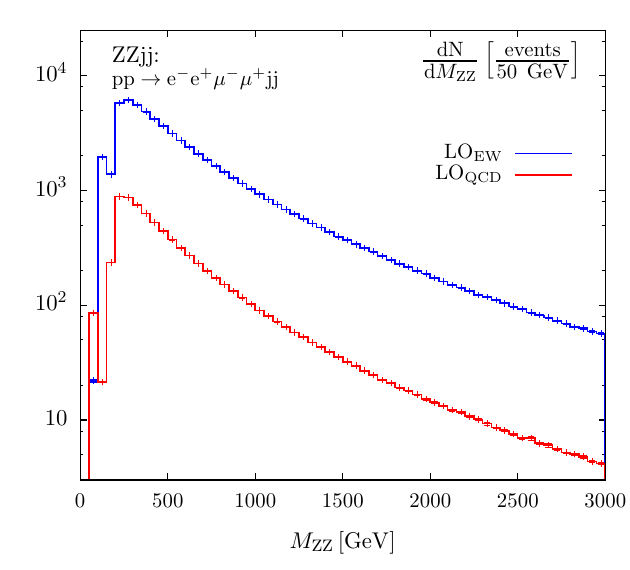}
\caption{Invariant-mass distribution of the four-lepton system for two different ranges of the EW-induced (blue line) and QCD-induced (red line) contributions to  $pp\to e^-e^+\mu^-\mu^+jj$, within the selection cuts of Eqs.~(\ref{eq:jet-def})--(\ref{eq:ll-cuts}) and Eq.~(\ref{eq:jjcuts_zz}). An integrated luminosity of 30~ab$^{-1}$ is assumed. 
\label{fig:mzz}
}
\end{figure}
%
%
In Fig.~\ref{fig:zzjj-nevts} we show the number of events above a
specific value of the tagging jets' transverse momenta and invariant
mass, respectively, assuming an integrated luminosity of 30~ab$^{-1}$.
%
\begin{figure}
\includegraphics[width=0.5\textwidth]{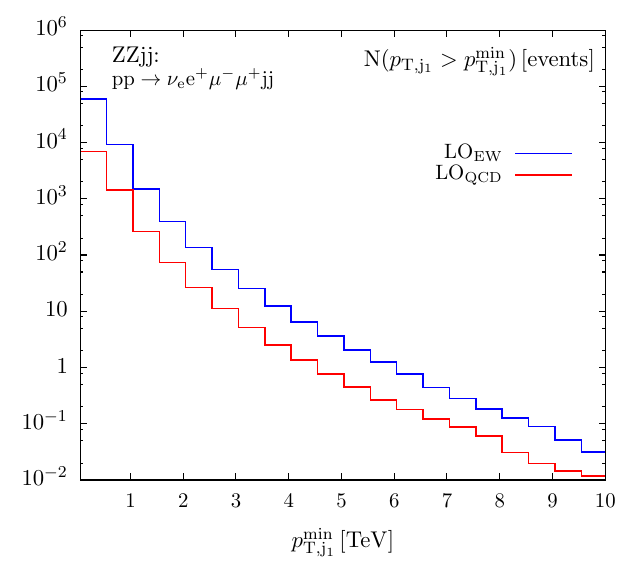}
\includegraphics[width=0.5\textwidth]{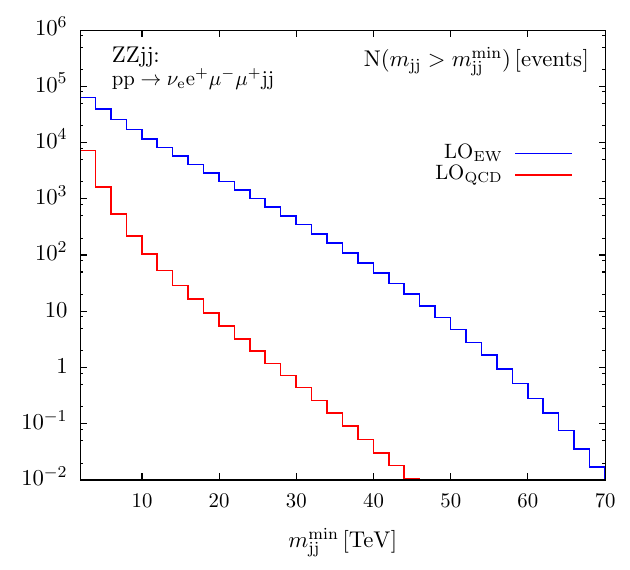}
\caption{Total number of events produced with $p_{T,j_1}>p_{T,j_1}^\mathrm{min}$ (l.h.s.) and  with $m_{jj}>m_{jj}^\mathrm{min}$ (r.h.s.) for the EW-induced (blue line) and QCD-induced (red line) contributions to  $pp\to e^-e^+\mu^-\mu^+jj$, within the selection cuts of Eqs.~(\ref{eq:jet-def})--(\ref{eq:ll-cuts}) and Eq.~(\ref{eq:jjcuts_zz}). An integrated luminosity of 30~ab$^{-1}$ is assumed. 
\label{fig:zzjj-nevts}
}
\end{figure}
%
%

\subsection{$W^+W^-jj$}
\label{ssec:wpwmjj}
The strategy applied to the $W^+W^-jj$ channel differs from the
respective analyses of other channels, as in this case the dominant
source of background to the VBS signal is provided not by QCD-induced
$W^+W^-jj$ production, but by top-pair production in association with
jets. In the $t\bar t$ channel, when the dominant decay modes of the
top quarks into $W$~bosons and bottom quarks are considered, the
bottom quarks can be misidentified as light-flavor tag jets.  Even
more problematic are modes where a $t\bar t$~pair is produced in
association with one or two jets that may mimic the tag jets of a VBS
event. Because of the large event rates, despite the application of
efficient $b$-veto techniques it is difficult to reduce the background
associated with these various $t\bar t$ processes below the level of
the signal cross section with cut-based techniques.  In order to find
an optimal set of selection cuts for EW $W^+W^-jj$ production, we
therefore take $t\bar t$, $t\bar t+$1~jet, $t\bar t+$2~jet, and
QCD-induced $W^+W^-jj$ production processes into account. We use {\tt
  MadGraph5} for the simulation of the top backgrounds that we
generically refer to as $t\bar t +$jets. We focus on final states with
different types of leptons, $e^+\nu_e\mu^-\bar\nu_{\mu} jj$.

An optimal $S/B$ ratio is obtained with the basic selection cuts of
Eqs.~(\ref{eq:jet-def})--(\ref{eq:lgap}) and additional cuts on the
separation of the two tagging jets,
\begin{equation}
\label{eq:jjcuts_wpwm}
m_{jj}>2000~\mathrm{GeV}\,,\quad
\Delta y_{jj}>5\,.
\end{equation}
For the suppression of the $t\bar t +$jets backgrounds, we veto any
events with an identified $b$~quark, assuming the $b$-tagging
efficiencies listed in Tab.~\ref{tab:btag}.
\begin{table}
\begin{center}
\begin{tabular}{|c|c|c|}
\hline
& &\\
$p_{Tj}^{veto}$ [GeV] & $1.4 < |\eta_j^{veto}|$  & $|\eta_j^{veto}|<1.4$
  \\
& &\\
\hline
& & \\
20 - 50   & $60\%$ & $70\%$ \\
50 - 80   & $65\%$ & $75\%$ \\
80 - 120  & $70\%$ & $80\%$ \\
120 - 170 & $70\%$ & $80\%$ \\
$>$ 170   & $65\%$ & $75\%$ \\
& &\\
\hline
\end{tabular}
\caption{
\label{tab:btag}
Assumed $b$-tagging efficiencies as functions of the transverse
momentum of the jet for different rapidity ranges (adapted from
Ref.~\cite{Englert:2008tn}).}
\end{center}
\end{table}
Events passing the $b$-veto are rejected, if they exhibit any jet in
the rapidity interval between the two tagging jets,
\begin{equation}
\label{eq:yveto}
 y_{j,min}^{tag} < y_j^{veto} < y_{j,max}^{tag}\,. 
\end{equation}
Note that in our LO calculation the VBS signal and the QCD-induced
$W^+W^-jj$ background never exhibit more than two jets and thus always
pass the cut of Eq.~(\ref{eq:yveto}).
With the full set of selection cuts and the $b$-veto procedure we
apply, we find cross sections of $\sigma^\mathrm{EW} = 58.28(2)$~fb,
$\sigma^\mathrm{QCD} =17.1(1)$~fb, and $\sigma^{t\bar t+\mathrm{jets}}
=5.2(4)$~fb.

\subsection{Single gauge-boson production via VBF}
\label{sec:vbf_V}
The efficient suppression of QCD backgrounds is much more challenging
for single gauge-boson production via VBF than in the case of
gauge-boson pair production via VBS. A simple cut-based analysis is
not capable of yielding $S/B$ ratios much larger than one. More
advanced techniques will be necessary for a clean isolation of the VBF
signal in these cases. We nonetheless report our results for a simple
cut-based study here to convey which orders of magnitude are to be
expected for signal and background cross sections after VBF-specific
selection cuts are imposed. We consider the representative $e^-e^+jj$
and $\nu_e e^+ jj$ final states for the $Zjj$ and $W^+jj$ processes,
respectively.

We impose the cuts of Eqs.~(\ref{eq:jet-def})--(\ref{eq:lgap}).
Furthermore, the tagging jets are required to exhibit a large
invariant mass and be well-separated in rapidity,
\begin{equation}
\label{eq:jjcuts_v}
m_{jj}>2000~\mathrm{GeV}\,,\quad
\Delta y_{jj}>5\,.
\end{equation}
For the $Zjj$ production process, in addition the cut of
Eq.~(\ref{eq:ll-cuts}) is applied to the decay leptons.

The cut on the on the lepton rapidity relative to the tagging jets,
Eq.~(\ref{eq:lgap}), is particularly important for the suppression of
the QCD backgrounds that typically feature leptons not located in
between the tagging jets. The impact of this cut is illustrated by
Fig.~\ref{fig:ystar_lep}, where for $pp\to \nu_e e^+ jj$ we show the
distribution of the $y_\ell^\star$ variable, defined as
\begin{equation}
y_\ell^\star = y_\ell -\frac{y^{tag}_{j_1}+y^{tag}_{j_2}}{2}\,,
\end{equation}
without and with the cut of Eq.~(\ref{eq:lgap}). 
%
%
\begin{figure}
\includegraphics[width=0.5\textwidth]{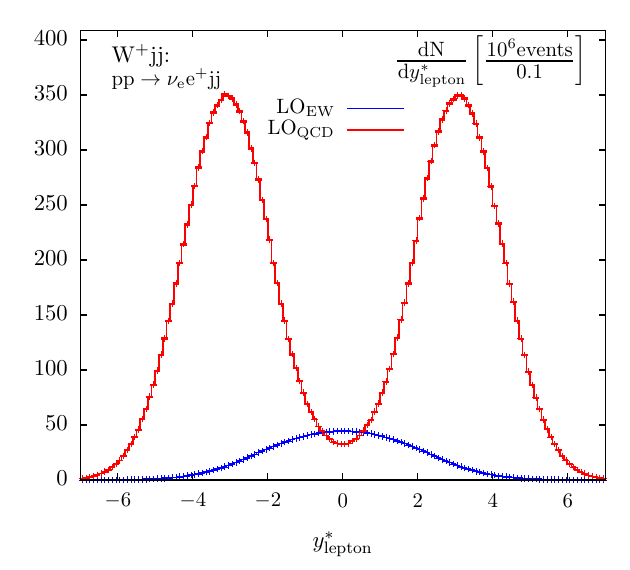}
\includegraphics[width=0.5\textwidth]{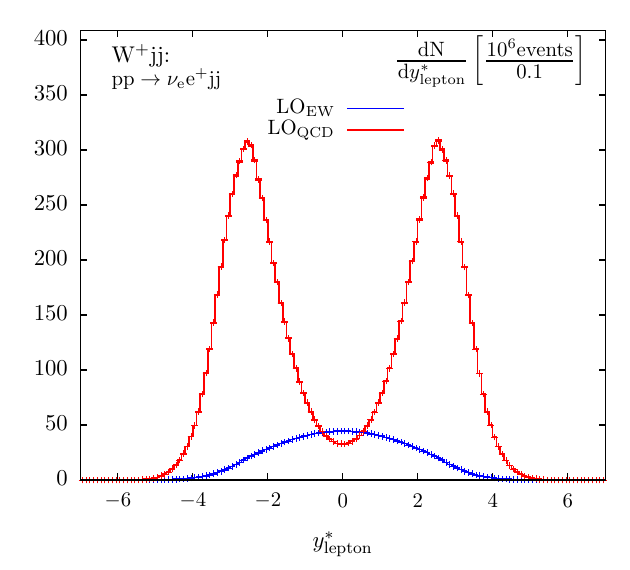}
\caption{Distribution of the $y_\ell^\star$ variable for the EW-induced (blue line) and QCD-induced (red line) contributions to  $pp\to  \nu_e  e^+  jj$, within the selection cuts of Eqs.~(\ref{eq:jet-def})--(\ref{eq:ptl-cut}) and Eq.(\ref{eq:jjcuts_v}),  without~(l.h.s.) and with~(r.h.s.) the lepton rapidity-gap cut of Eq.~(\ref{eq:lgap}). An integrated luminosity of 30~ab$^{-1}$ is assumed.  
\label{fig:ystar_lep}
}
\end{figure}
%
%
The cut has an impact of about 40\% on the QCD background, while it
reduces the EW signal cross section only marginally.

With the above-listed cuts (including the requirement on the lepton
rapidity), the cross sections given in Tab.~\ref{tab:vbf-v} are
obtained for the EW signal and the respective QCD background processes
in the $W^+jj$ and $Zjj$ modes when decays of the gauge bosons into a
specific lepton pair are considered.
%
%
\renewcommand{\arraystretch}{1.2}
\begin{table}
\begin{center}
\begin{tabular}{|c|c|c|c|}
\hline
 &EW production&QCD production&$S/B$\\
\hline
$ \sigma^\mathrm{LO}(W^+jj)$  &   6980.1(8)  & 41324(10) &$0.17$\\
\hline
$\sigma^\mathrm{LO}(Zjj)$  &   1079.5(3)  & 5164(1)  &$0.21$\\
\hline
\end{tabular}
\caption{
\label{tab:vbf-v}
Cross sections for the EW-induced $Vjj$ production processes together with the irreducible QCD background and the signal-to-background ratio, $S/B$,  within the default cuts for $Vjj$ processes discussed in the text. Decays of the weak bosons into a specific leptonic final state are included as detailed in the text. All cross sections are given in [fb]. 
} 
\end{center}
\end{table}
%
%

The larger $S/B$ ratios given in Tab.~\ref{tab:vbf-v2} can be
obtained, if the more severe cuts
\begin{equation}
\label{eq:vcuts}
m_{jj}>3000~\mathrm{GeV}\,,\quad
\Delta y_{jj}>6\,, \quad 
 |y_\ell| \le 1\,, 
\end{equation}
are imposed on the tagging jets and the charged leptons. 
%
%
%
\begin{table}
\begin{center}
\begin{tabular}{|c|c|c|c|}
\hline
 &EW production&QCD production&$S/B$\\
\hline
$\sigma^\mathrm{LO}(W^+jj)$&   1488.1(4)  & 1227.8(8) & $1.21 $ \\
\hline
$\sigma^\mathrm{LO}(Zjj)$ &   154.4(1) & 138.0(1) & $ 1.12$ \\
\hline
\end{tabular}
\caption{
\label{tab:vbf-v2}
Cross sections for the EW-induced $Vjj$ production processes together with the irreducible QCD background and the signal-to-background ratio, $S/B$,  within the default cuts for $Vjj$ processes discussed in the text and the additional cuts of Eq.~(\ref{eq:vcuts}). Decays of the weak bosons into a specific leptonic final state are included as detailed in the text. 
All cross sections are given in [fb]. 
} 
\end{center}
\end{table}
%
%

In Figs.~\ref{fig:zjj-nevts} and~\ref{fig:wjj-nevts}, for the $Zjj$
and $W^+jj$ production modes we show the number of events above a
specific value of the tagging jets' transverse momenta and invariant
mass, respectively, assuming an integrated luminosity of 30~ab$^{-1}$.
%
%
\begin{figure}
\includegraphics[width=0.5\textwidth]{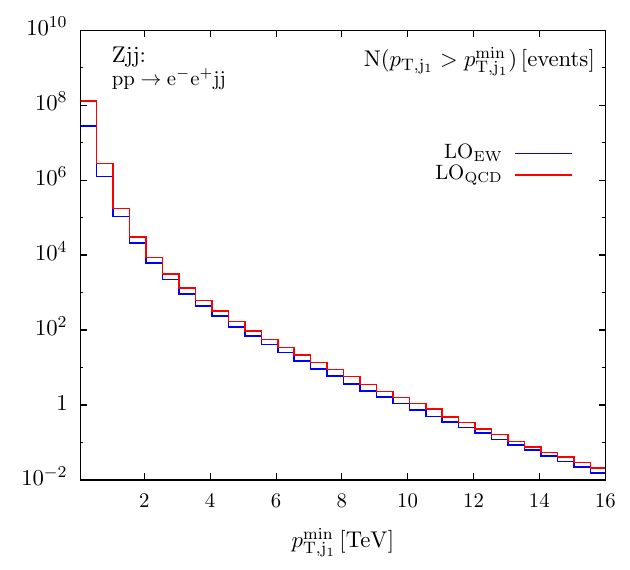}
\includegraphics[width=0.5\textwidth]{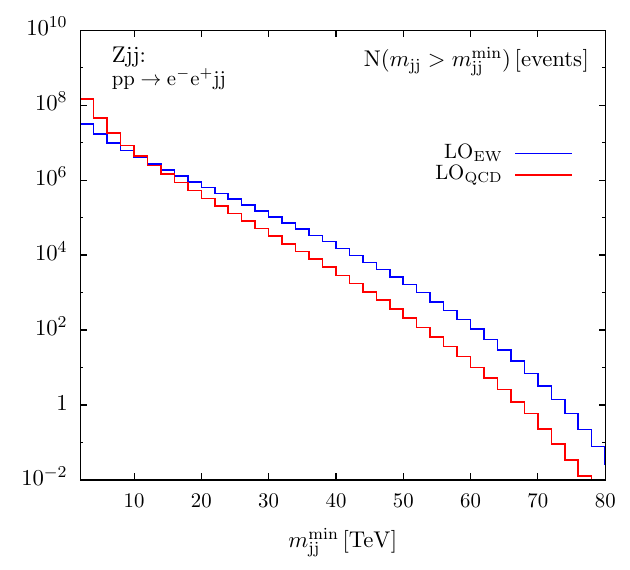}
\caption{Total number of events produced with $p_{T,j_1}>p_{T,j_1}^\mathrm{min}$ (l.h.s.) and  with $m_{jj}>m_{jj}^\mathrm{min}$ (r.h.s.)  for the EW-induced (blue line) and QCD-induced (red line) contributions to  $pp\to e^-e^+jj$, within the selection cuts of Eqs.~(\ref{eq:jet-def})--(\ref{eq:ll-cuts}),  and Eq.~(\ref{eq:jjcuts_v}).  An integrated luminosity of 30~ab$^{-1}$ is assumed. 
\label{fig:zjj-nevts}
}
\end{figure}
%
%
%
%
\begin{figure}
\includegraphics[width=0.5\textwidth]{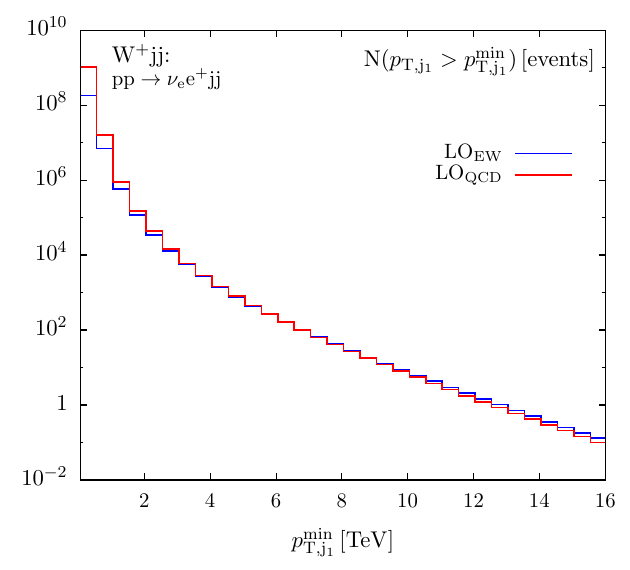}
\includegraphics[width=0.5\textwidth]{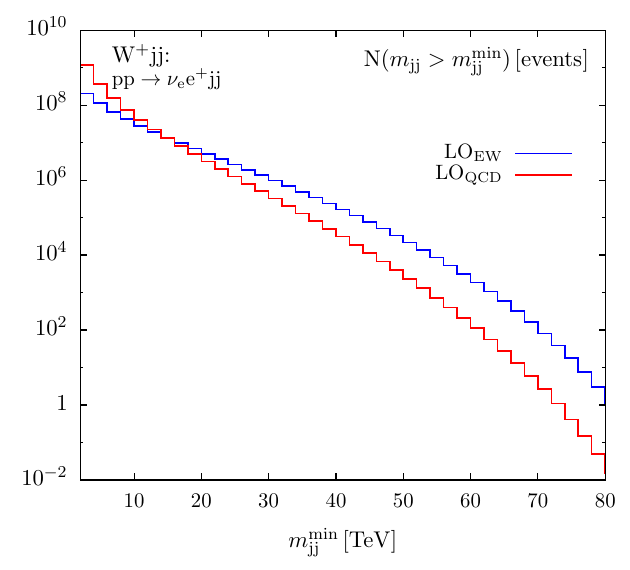}
\caption{Total number of events produced with $p_{T,j_1}>p_{T,j_1}^\mathrm{min}$ (l.h.s.) and  with $m_{jj}>m_{jj}^\mathrm{min}$ (r.h.s.)  for the EW-induced (blue line) and QCD-induced (red line) contributions to  $pp\to \nu_e e^+jj$, within the selection cuts of Eqs.~(\ref{eq:jet-def})--(\ref{eq:lgap}),   and Eq.~(\ref{eq:jjcuts_v}).  An integrated luminosity of 30~ab$^{-1}$ is assumed. 
\label{fig:wjj-nevts}
}
\end{figure}
%
%

\subsection{Benchmark cross sections}
\label{ssec:bench}
As we have shown above, dedicated sets of selection cuts are essential
for obtaining optimal signal-to-background ratios in the environment
of a high-energy hadron collider. Nonetheless, we here provide cross
sections for the various VBS processes within simple cut scenarios to
facilitate comparisons among the various channels.

In Tab.~\ref{tab:xsec1} we list numbers for an inclusive setup where
we only impose the transverse-momentum cuts of Eq.~(\ref{eq:jet-def})
on the two tagging jets reconstructed via the anti-$k_\mathrm{T}$
algorithm with $R=0.4$.  For processes with final-state $Z$~bosons we
additionally require
\begin{equation}
M_{\ell^+\ell^-} > 66~\mathrm{GeV}
\label{eq:mll=mz}
\end{equation}
%
%
for all oppositely-signed lepton pairs to suppress contributions from
photons splitting into lepton pairs.
%
%
\begin{center}
\begin{table}
\begin{center}
\begin{tabular}{|r|r|}
\hline
VBS channel & cross section [fb]\\
\hline\hline
 \PWp jj         & 41 200   \\
\hline
 Z jj            &  7 215 \\
\hline
\hline
\PWp \PWm jj           & 245.7       \\
\hline
\PWp \PWp jj           & 104.8     \\
\hline
\PWp Z jj              & 19.64    \\
\hline
ZZ jj                 & 5.372     \\
\hline
\end{tabular}
\caption{
\label{tab:xsec1}
Cross sections for various VBS processes within the cuts of
Eq.~(\ref{eq:jet-def}). For processes with $Z$~bosons, additionally
the cut of Eq.~(\ref{eq:mll=mz}) is imposed. Decays of the weak bosons
into a specific leptonic final state are included as detailed in the
text. Statistical errors are at the permille level in each case.  }
\end{center}
\end{table}
\end{center}
%
%
In Tab.~\ref{tab:xsec2} we additionally impose VBS-specific cuts on the tagging jets, 
\begin{equation}
\label{eq:jet-cut-tab2}
y_{j1}\times y_{j2} < 0\,,\qquad m_{jj}> 2000~\mathrm{GeV}\,,  \qquad \Delta y_{jj}>5\,. 
\end{equation}
%
%
\begin{center}
\begin{table}
\begin{center}
\begin{tabular}{|r|r|}
\hline
VBS channel & cross section [fb]\\
\hline\hline
 \PWp jj         & 8 670  \\
\hline
 Z jj            &  1 461   \\
\hline
\hline
\PWp \PWm jj           & 93.27       \\
\hline
\PWp \PWp jj           &   48.35       \\
\hline
\PWp Z jj              &  8.312    \\
\hline
ZZ jj                 &2.419     \\
\hline
\end{tabular}
\caption{
\label{tab:xsec2}
Cross sections for various VBS processes within the cuts of
Eqs.~(\ref{eq:jet-def}) and (\ref{eq:jet-cut-tab2}). For processes
with $Z$~bosons, additionally the cut of Eq.~(\ref{eq:mll=mz}) is
imposed. Decays of the weak bosons into a specific leptonic final
state are included as detailed in the text. Statistical errors are at
the permille level in each case.  }
\end{center}
\end{table}
\end{center}
%
%
Cross sections with realistic cuts on the decay leptons as given in
Eqs.~(\ref{eq:ptl-cut})--(\ref{eq:lgap}) are listed in
Tab.~\ref{tab:xsec3}.
%
%
\begin{center}
\begin{table}
\begin{center}
\begin{tabular}{|r|r|}
\hline
VBS channel & cross section [fb]\\
\hline\hline
 \PWp jj         & 6 979   \\
\hline
 Z jj            &   1 050 \\
\hline
\hline
\PWp \PWm jj           & 58.30        \\
\hline
\PWp \PWp jj           & 32.36     \\
\hline
\PWp Z jj              &  4.875    \\
\hline
ZZ jj                 & 1.415     \\
\hline
\end{tabular}
\caption{
\label{tab:xsec3}
Cross sections for various VBS processes within the cuts of
Eqs.~(\ref{eq:jet-def}), (\ref{eq:jet-cut-tab2}),
(\ref{eq:ptl-cut})--(\ref{eq:lgap}). For processes with $Z$~bosons,
additionally the cut of Eq.~(\ref{eq:mll=mz}) is imposed.  Decays of
the weak bosons into a specific leptonic final state are included as
detailed in the text. Statistical errors are at the permille level in
each case.  }
\end{center}
\end{table}
\end{center}

\clearpage
\section{Jets\footnote{Editors: A.~Larkoski, M.~Pierini, M.~Selvaggi}}
\label{sec:jets}
The production of jets is the process that by far dominates, at all
distance scales, the final states emerging from hard collisions among the
proton constituents.

\subsection{Spectroscopy with high-mass dijets}
\label{sec:jet_spect}

A central goal of the 100 TeV collider would be the discovery of new
states with multi-TeV masses.  If these states are able to be produced
at a $pp$ collider, then they must decay to light quarks and gluons.
Additionally, these states may decay to electroweak-scale objects,
especially if they are related to the (solution of the) hierarchy
problem.  The dominant decay modes of electroweak bosons and the top
quark is to hadronic final states.  Therefore, we should generically
expect that final states with jets are among the most sensitive to new
physics signals.  In this section, we will study resonances that decay
to pairs of QCD jets or electroweak objects and the sensitivity of jet
algorithm parameters to reconstructing invariant mass spectra.

In this section, and the following sections, we simulate events as
follows.  Narrow color-singlet resonances with masses of 10, 20, 30,
and 40 TeV that decay to pairs of top quarks, $W$ bosons, light
quarks, or gluons in $pp$ collision events at 100 TeV are generated
with MadGraph\_aMC@NLO v2.3.2.2 \cite{Alwall:2014hca}.  The top quark
and $W$ boson final states are decayed fully hadronically.  The
parton-level events are then showered with Pythia v8.2
\cite{Sjostrand:2014zea} or Herwig++ v2.7.1 \cite{Bahr:2008pv}.  The
resulting jets are clustered with the anti-$k_T$ algorithm
\cite{Cacciari:2008gp} using FastJet v3.1.3 \cite{Cacciari:2011ma}.
Only particles with pseudorapidity $|\eta|<2.5$ are included in the
jet clustering and only jets with transverse momentum $p_T$ larger
than 20\% of the mass of the mother resonance are included.  This
latter cut effectively imposes a cut on the pseudorapidity of the jets
$|\eta_J|\lesssim 1.5$.  For this analysis, we are most interested in
the required performance and resolution of the detector to reconstruct
the jets and the resonance, and so this cut will not directly affect
that.  It is to guarantee that the jets we are studying are indeed
those that originated from the resonance decay.

In Figs.~\ref{fig:mres_tt}-\ref{fig:mres_ww}, we plot the invariant
mass distribution of the two highest $p_T$ jets from events with a 20
TeV resonance.  We scan over the jets' radii ranging from $R = 0.05$
to $R = 0.5$.  Because the resonance is almost always produced at
rest, the total invariant mass of these events will be about 20 TeV.
As the radius of the jets increases, more radiation in the final state
is captured in the jets.  The long tail of the mass distributions
extending below 20 TeV indicates that there is some amount of
radiation from the decay of the resonance that is not being captured
in the two hardest jets.  This tail decreases as the jet radius
increases and is essentially absent for hadronically decaying $W$
bosons, for the range of $R$ considered.  $W$ bosons are color
singlets, and so do not radiate at wide angles.  Therefore, once the
jet radius is large enough to capture the $W$ decay products, then
essentially all of the radiation in the final state is in the jets.

\begin{figure}[h!]
\centering
\includegraphics[width=0.45\linewidth]{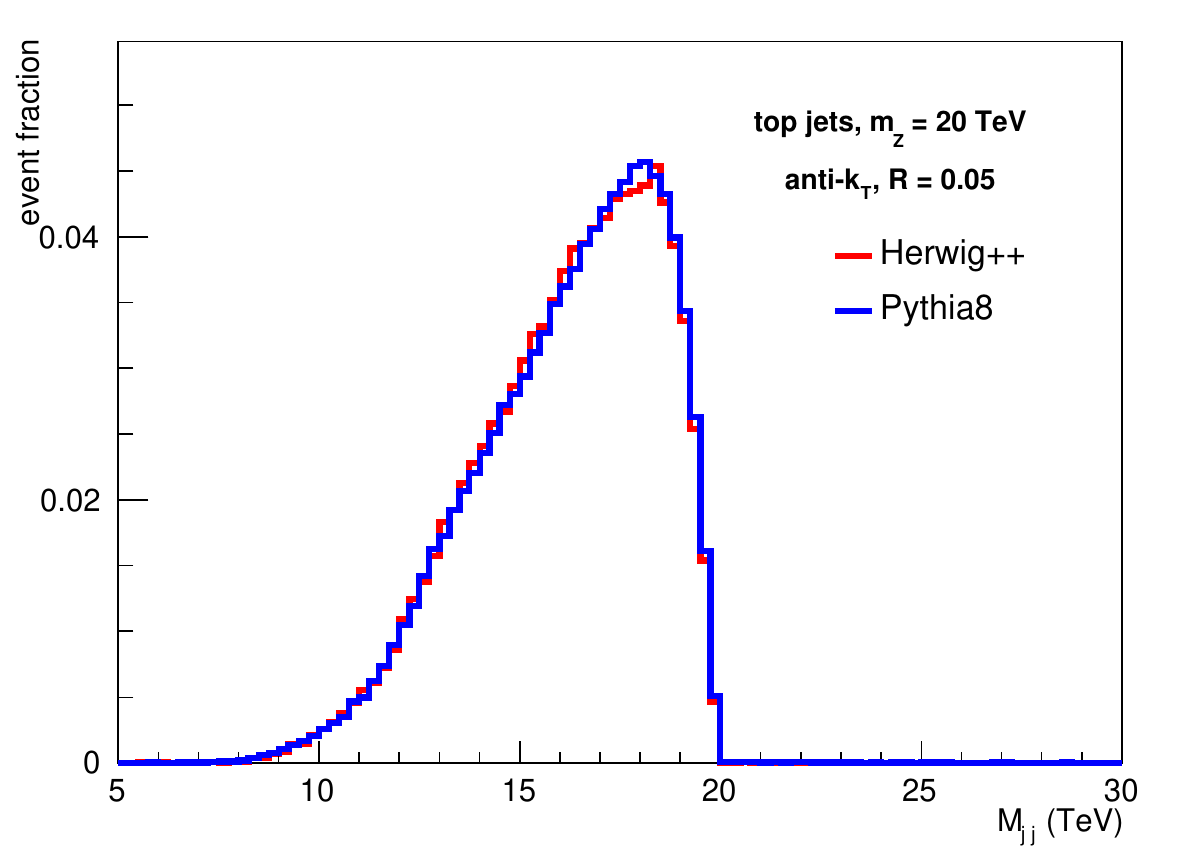}
\includegraphics[width=0.45\linewidth]{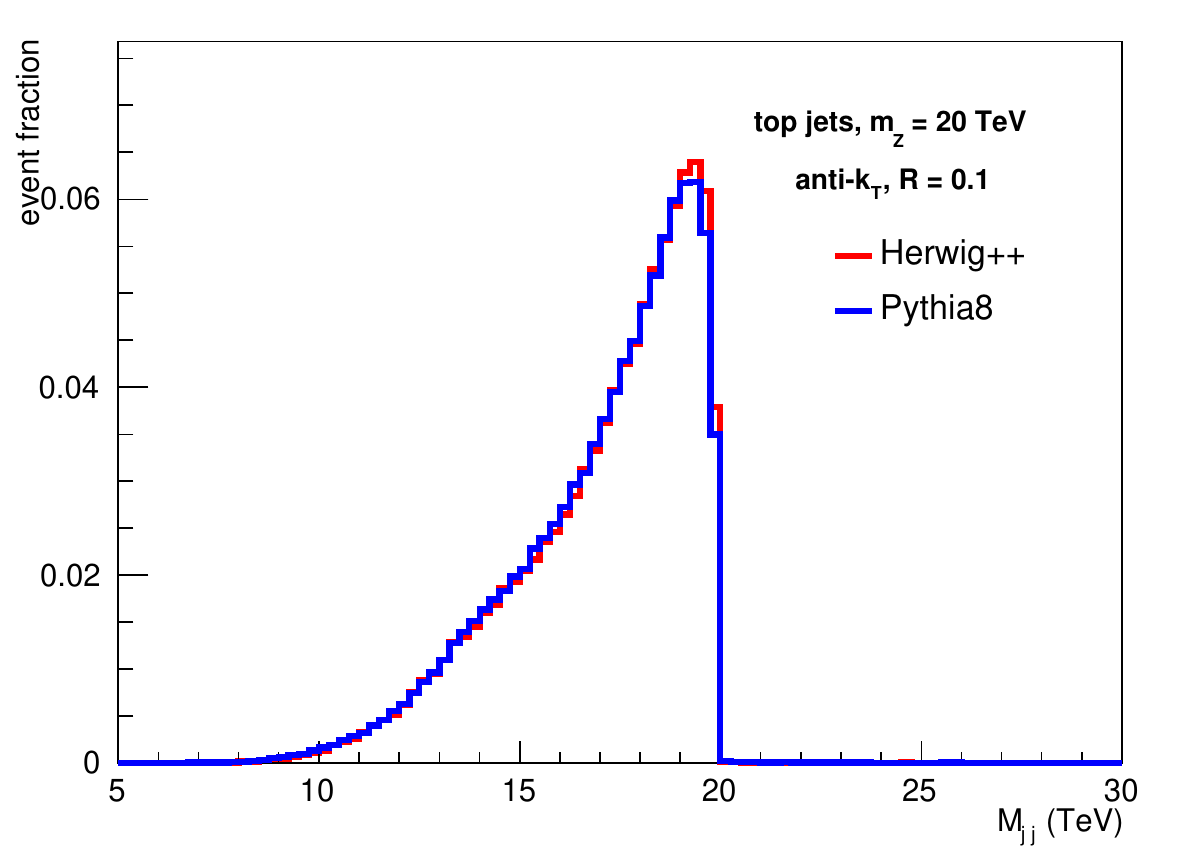}
\\
\includegraphics[width=0.45\linewidth]{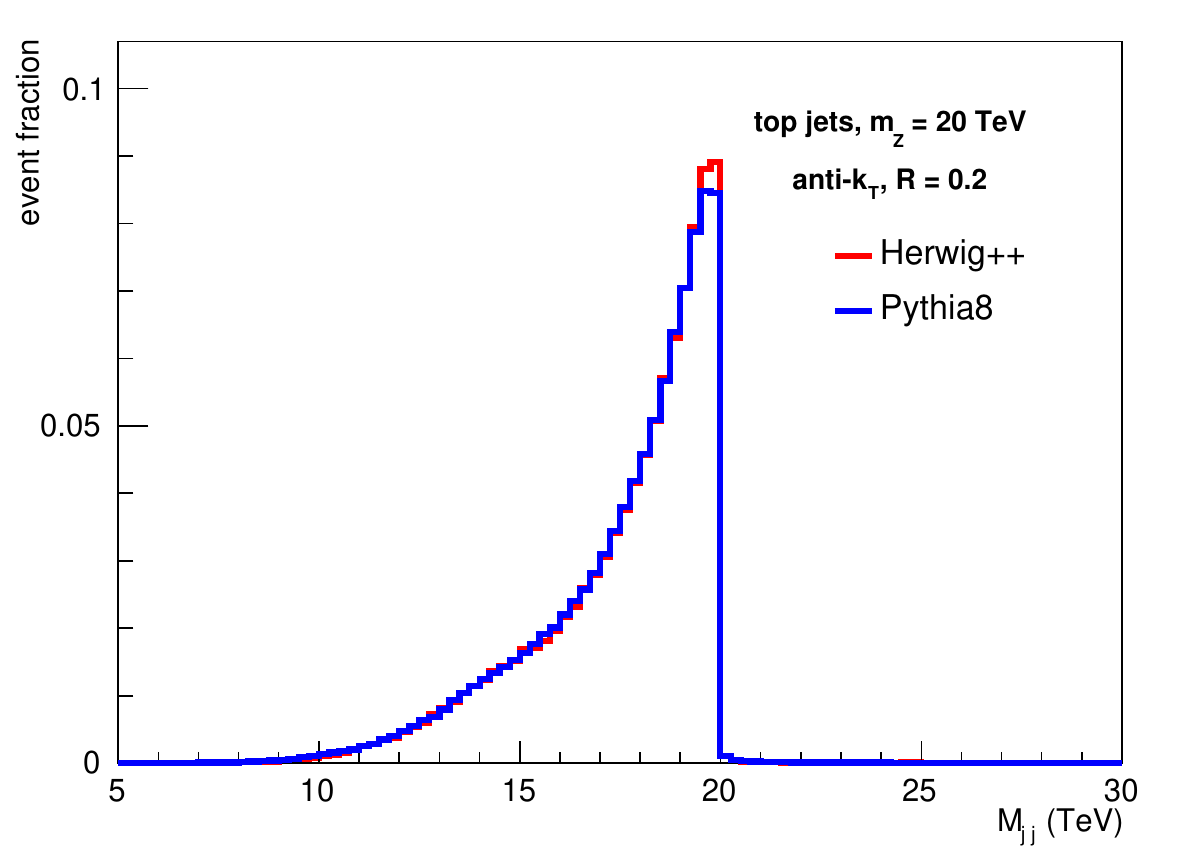}
\includegraphics[width=0.45\linewidth]{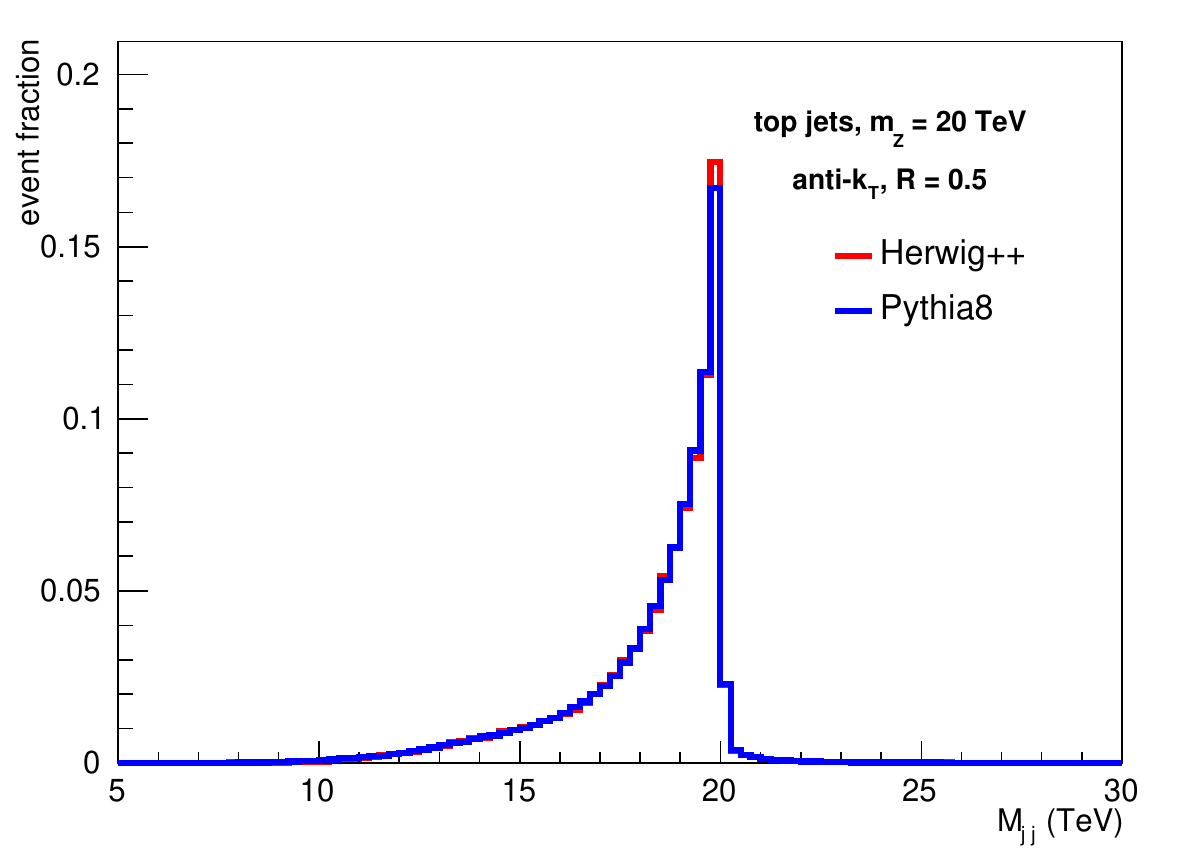}
\caption{ Dijet invariant mass spectrum of boosted top quarks produced
  from the decay of a 20 TeV resonance with jet radii ranging from
  $R=0.05$ to $R = 0.5$.  }
\label{fig:mres_tt}
\end{figure}

\begin{figure}[h!]
\centering
\includegraphics[width=0.45\linewidth]{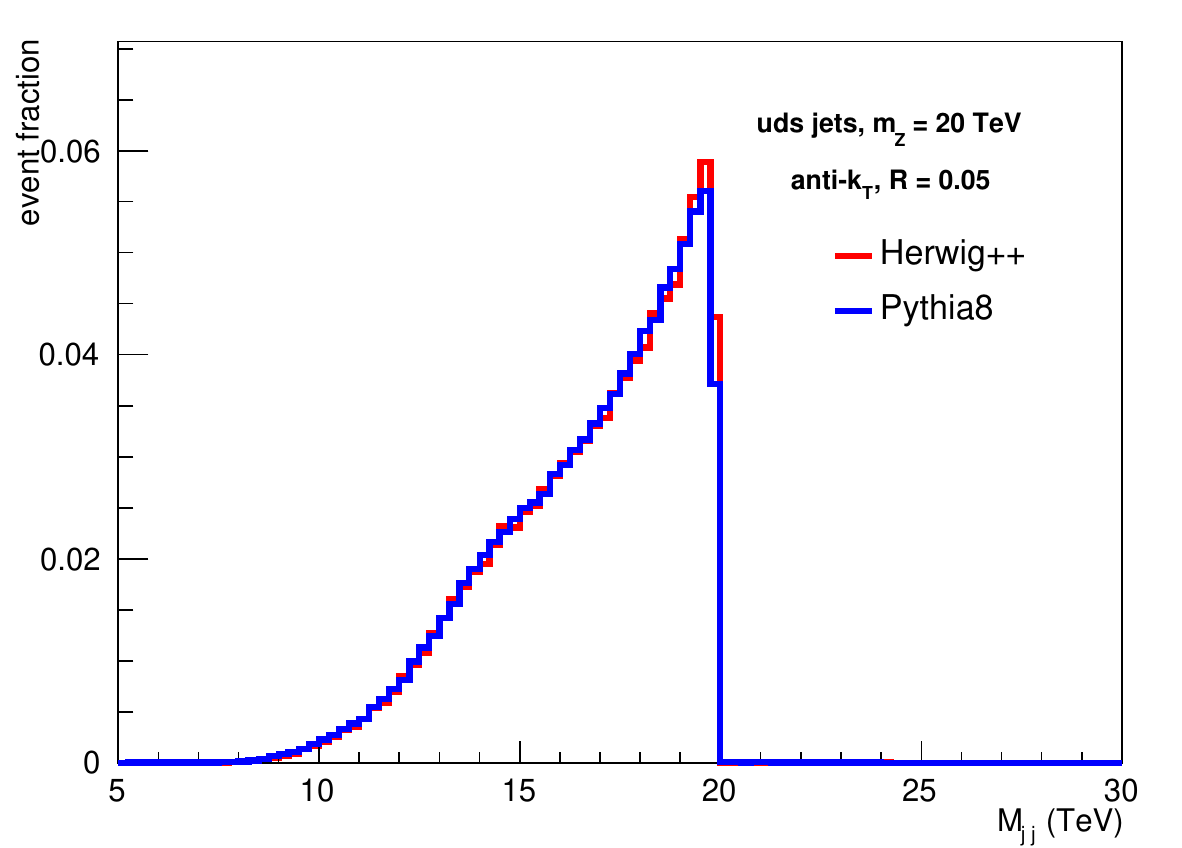}
\includegraphics[width=0.45\linewidth]{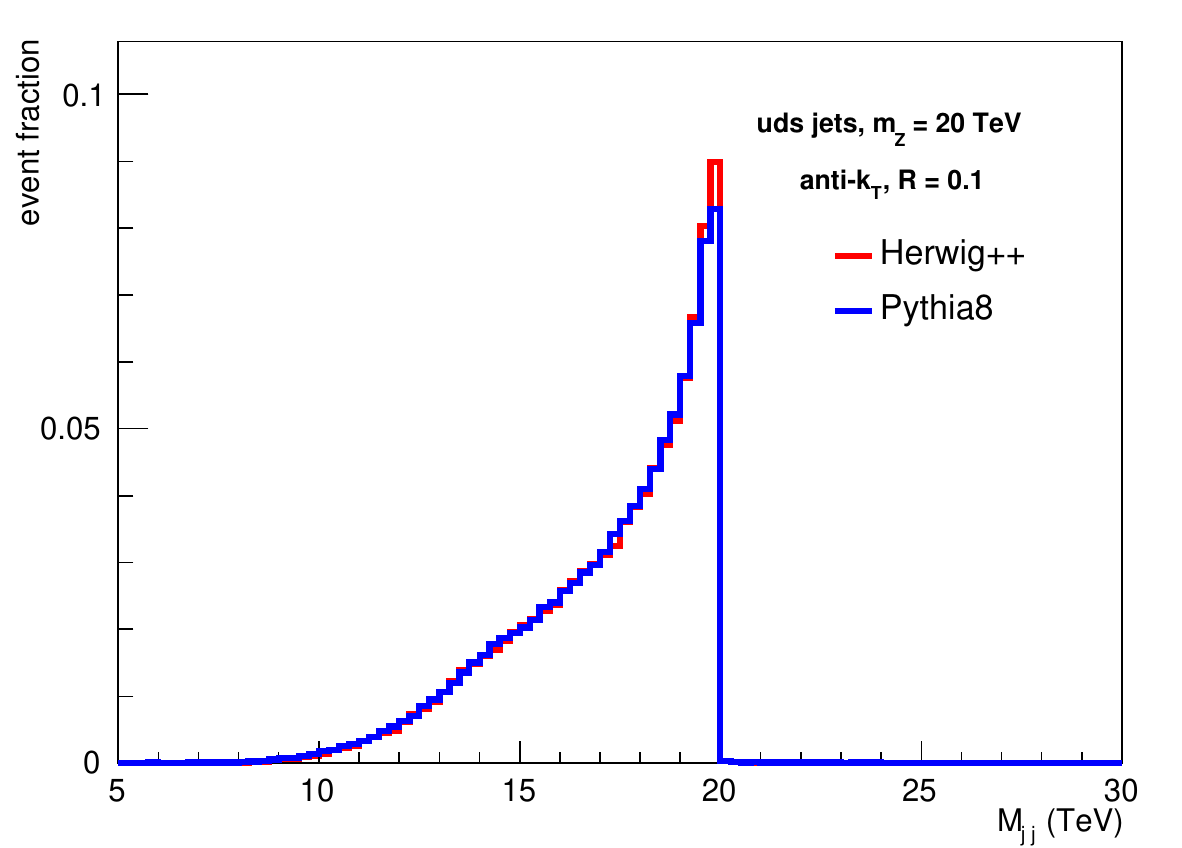}
\\
\includegraphics[width=0.45\linewidth]{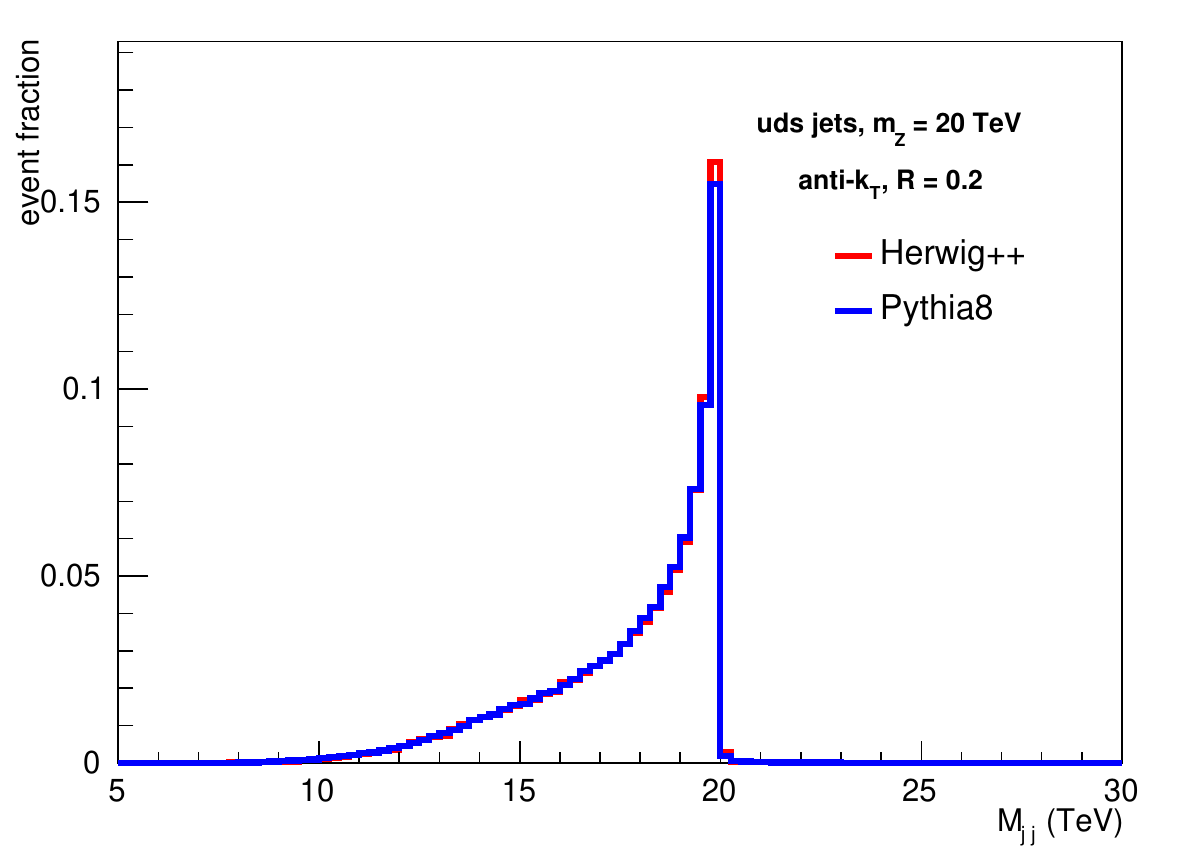}
\includegraphics[width=0.45\linewidth]{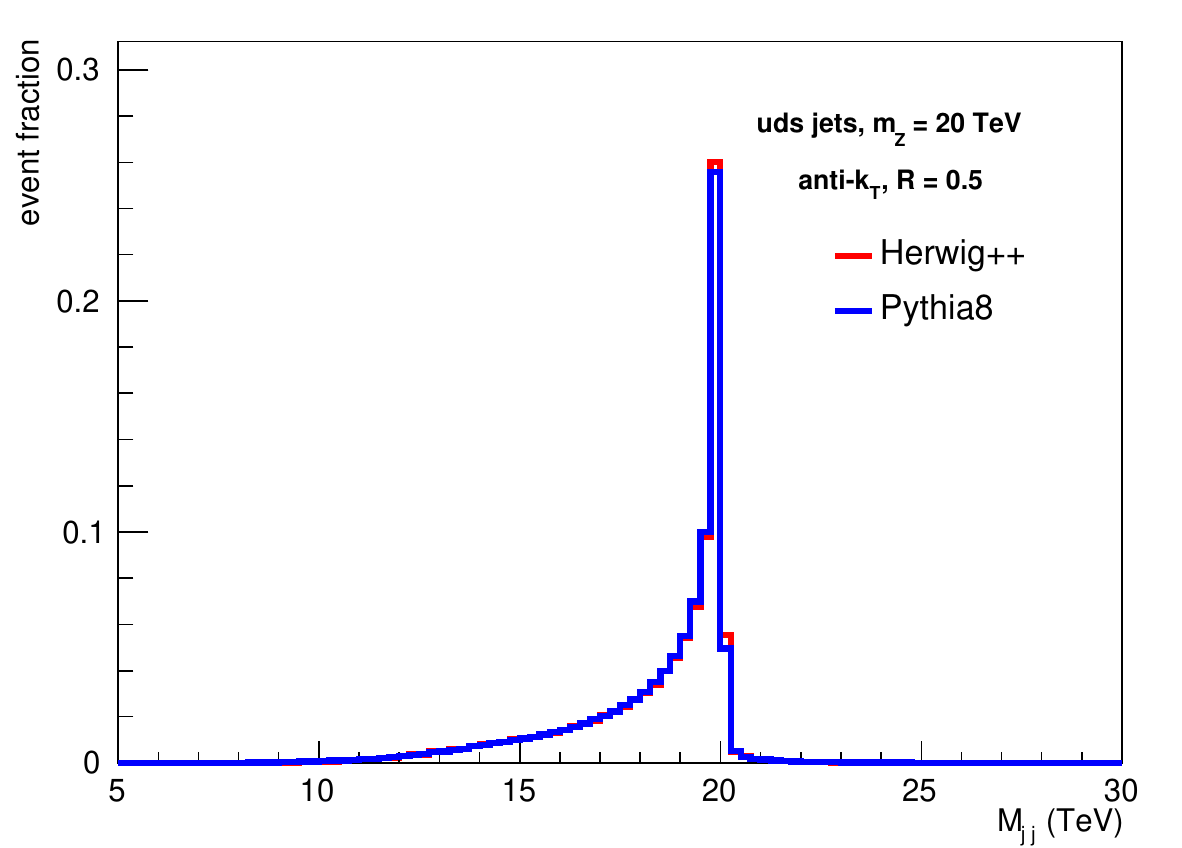}
\caption{ Light QCD quark dijet invariant mass spectrum produced from
  the decay of a 20 TeV resonance with jet radii ranging from $R=0.05$
  to $R = 0.5$.  }
\label{fig:mres_jj}
\end{figure}

\begin{figure}[h!]
\centering
\includegraphics[width=0.45\linewidth]{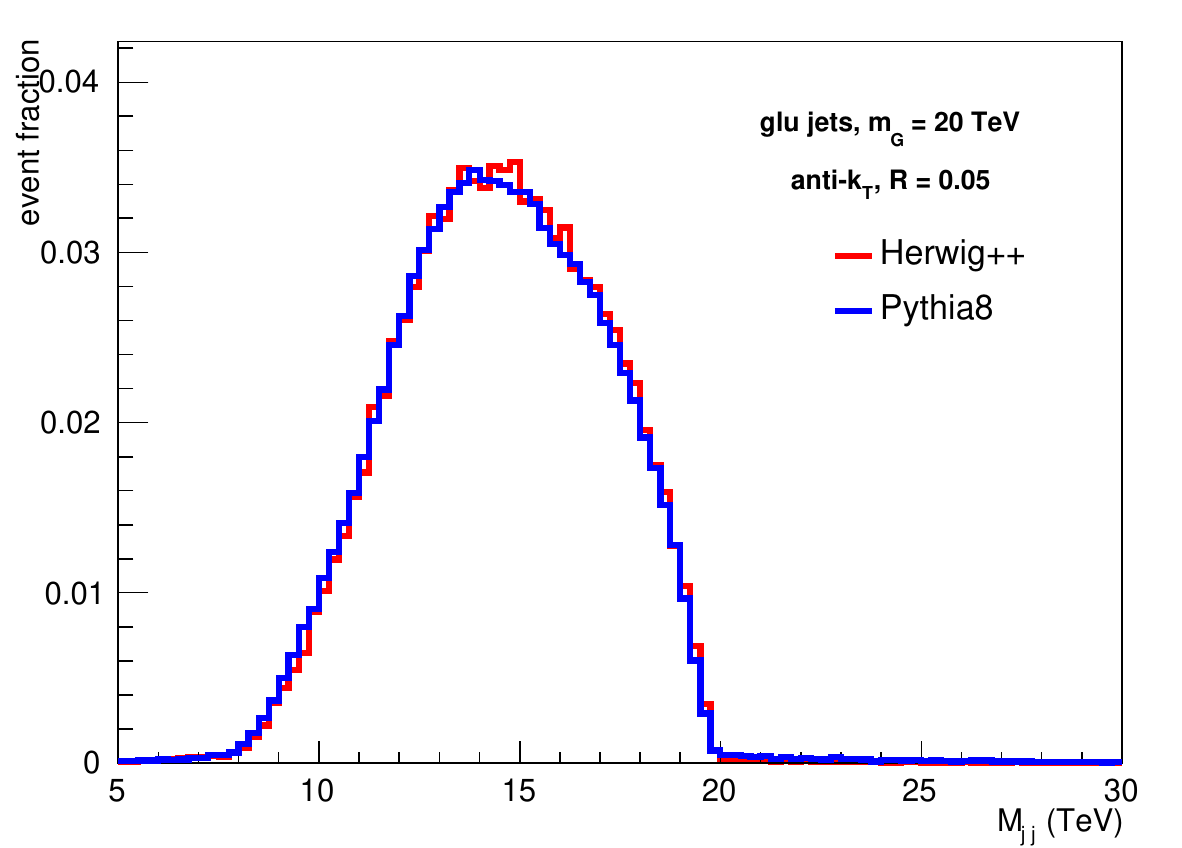}
\includegraphics[width=0.45\linewidth]{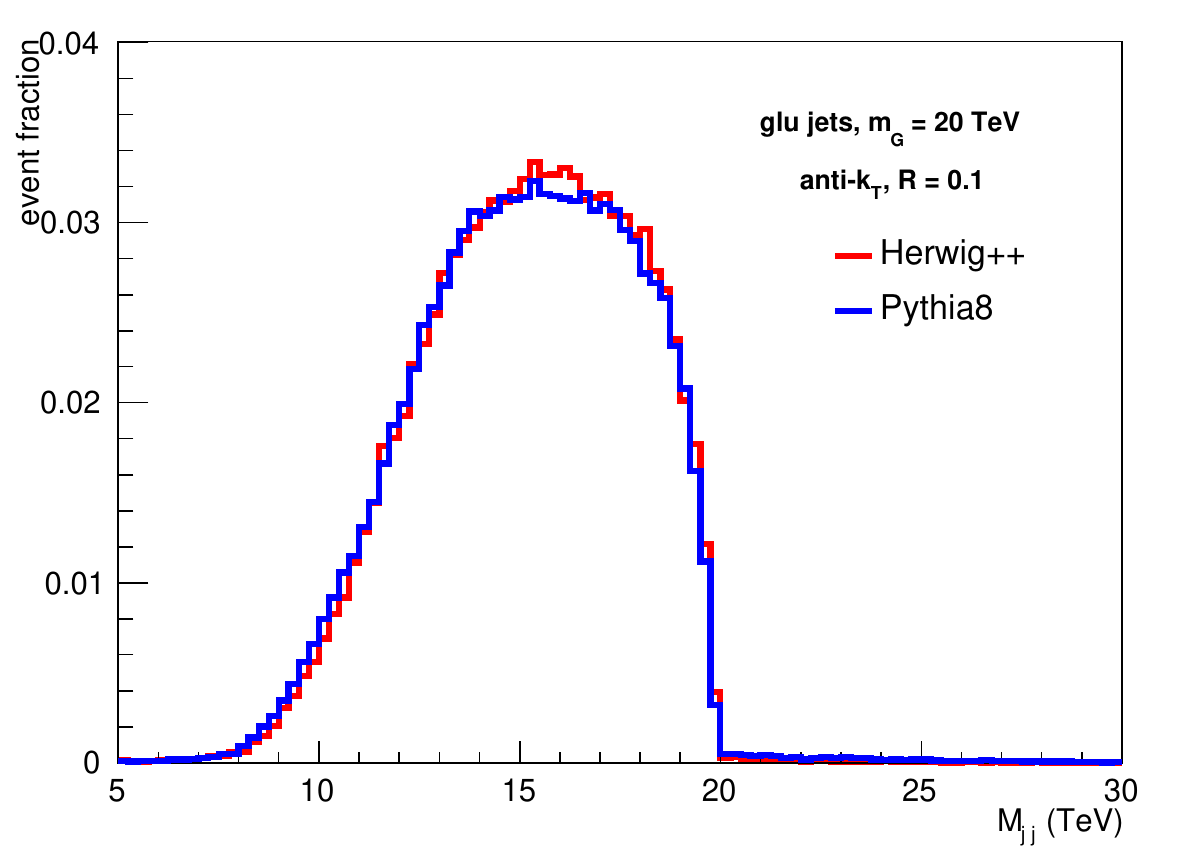}
\\
\includegraphics[width=0.45\linewidth]{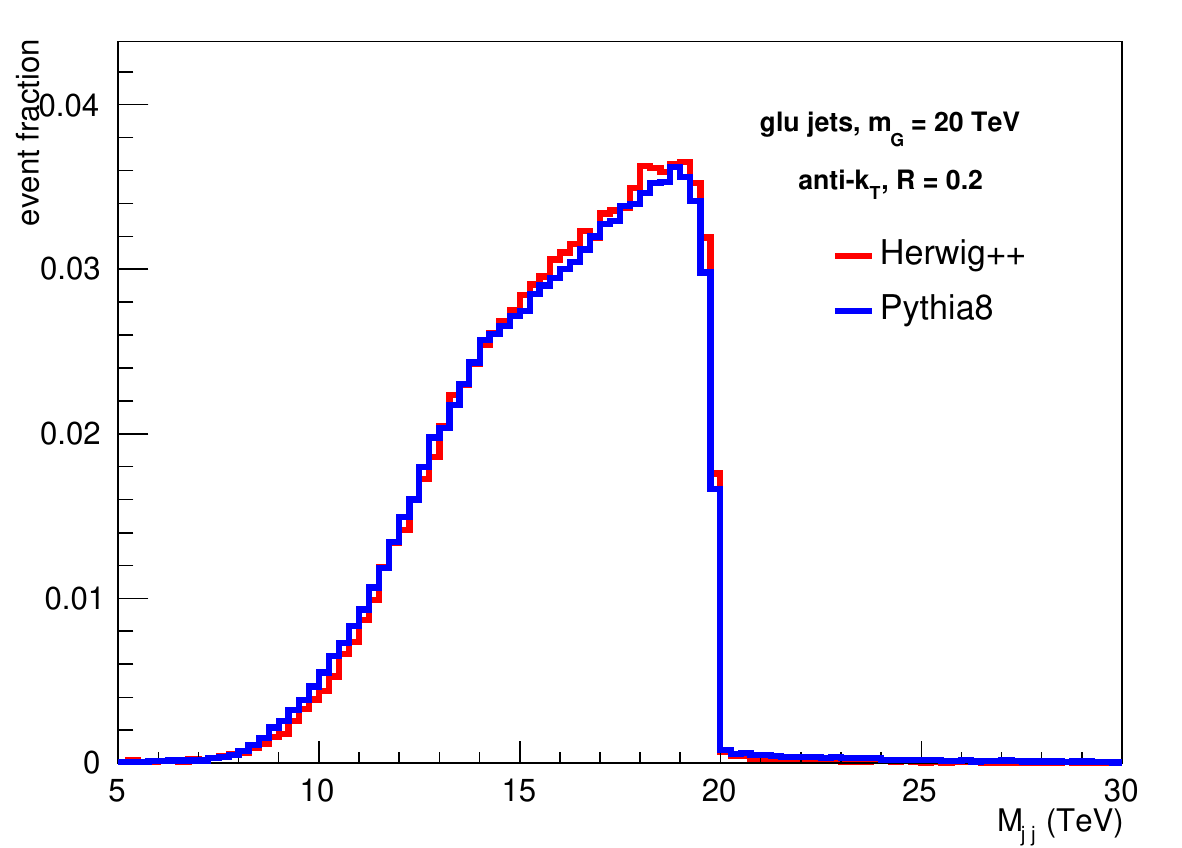}
\includegraphics[width=0.45\linewidth]{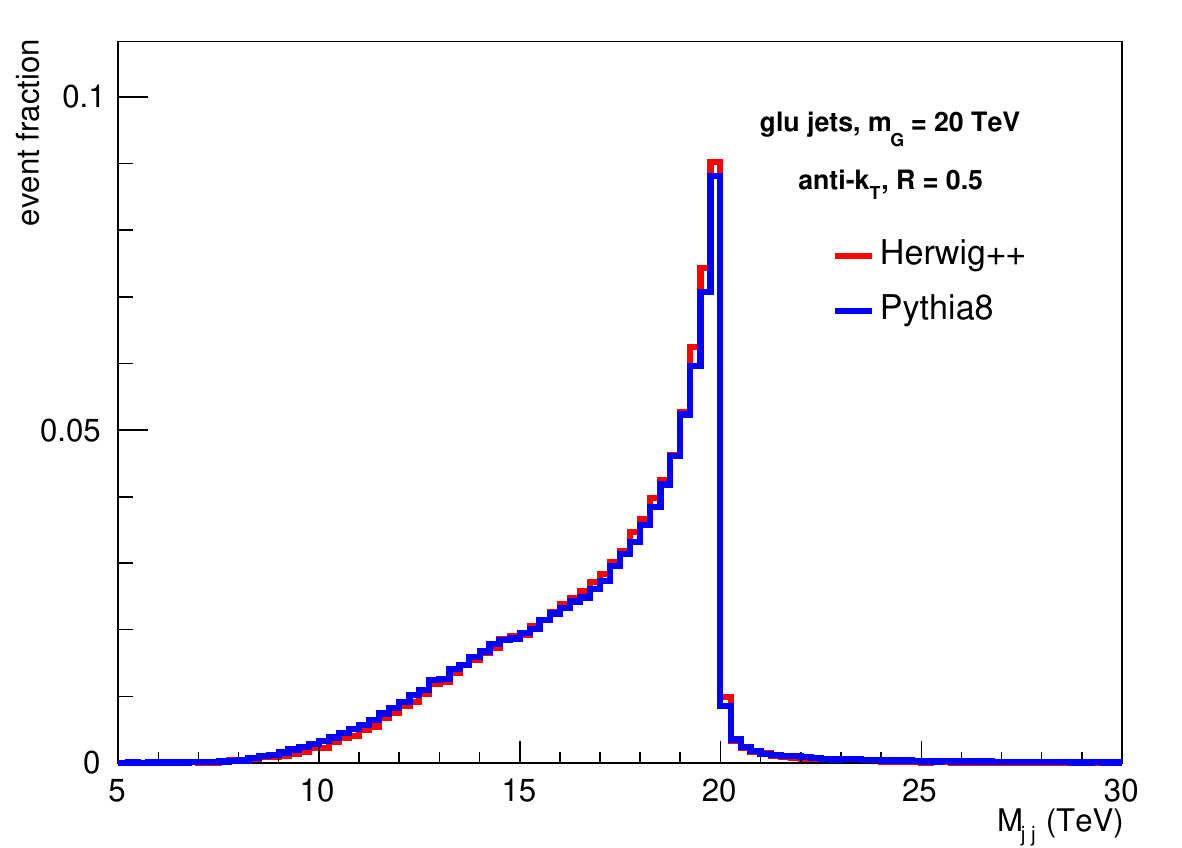}
\caption{ Gluon dijet invariant mass spectrum produced from the decay
  of a 20 TeV resonance with jet radii ranging from $R=0.05$ to $R =
  0.5$.  }
\label{fig:mres_gg}
\end{figure}

\begin{figure}[h!]
\centering
\includegraphics[width=0.45\linewidth]{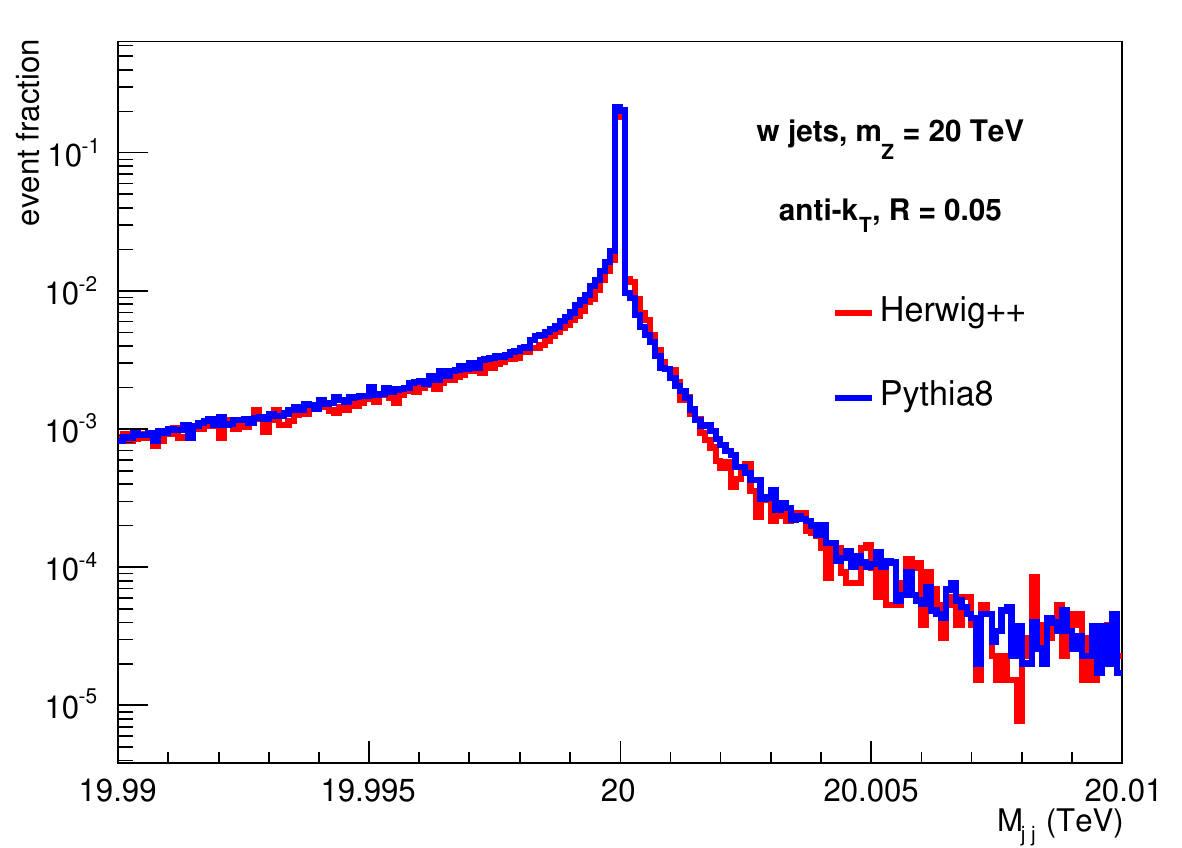}
\includegraphics[width=0.45\linewidth]{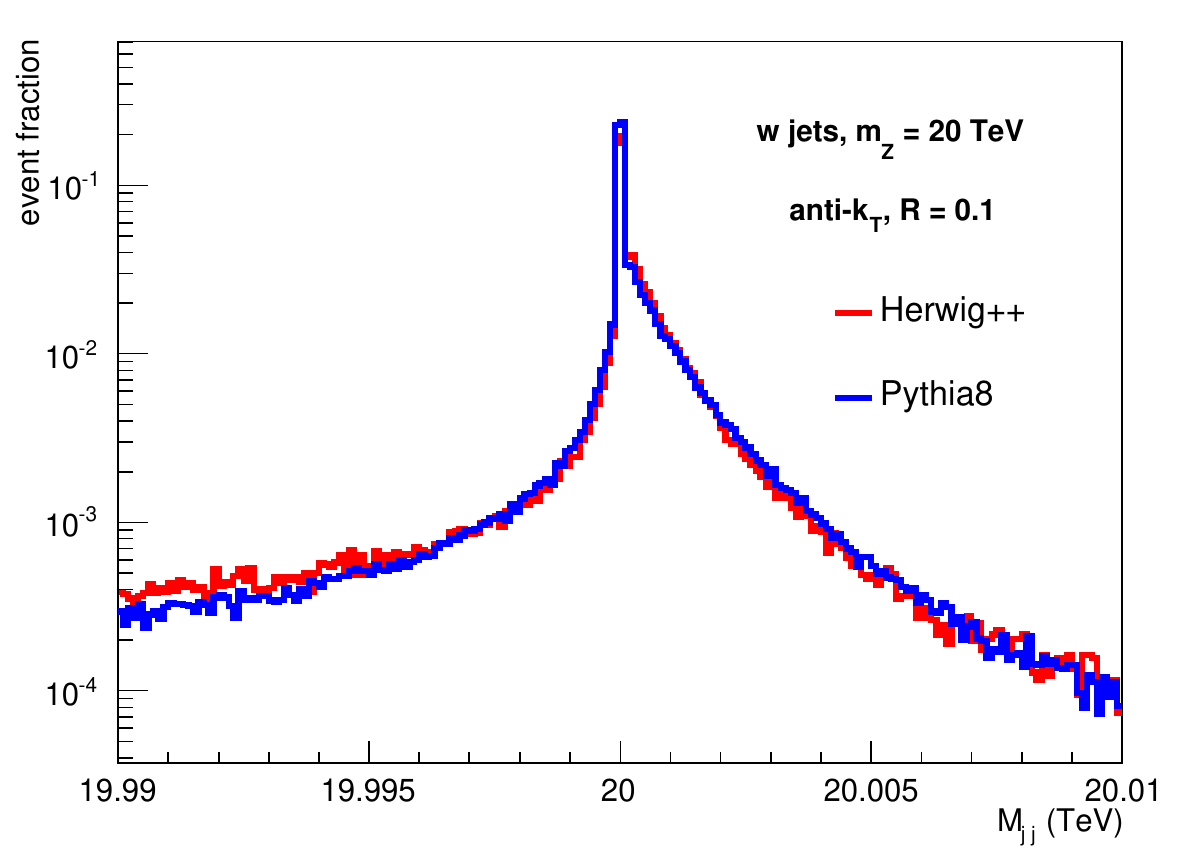}
\caption{ Dijet invariant mass spectrum of boosted $W$ bosons produced
  from the decay of a 20 TeV resonance with jet radii of $R=0.05$ and
  $R = 0.1$.  }
\label{fig:mres_ww}
\end{figure}

\begin{figure}[h!]
\centering
\includegraphics[width=0.45\linewidth]{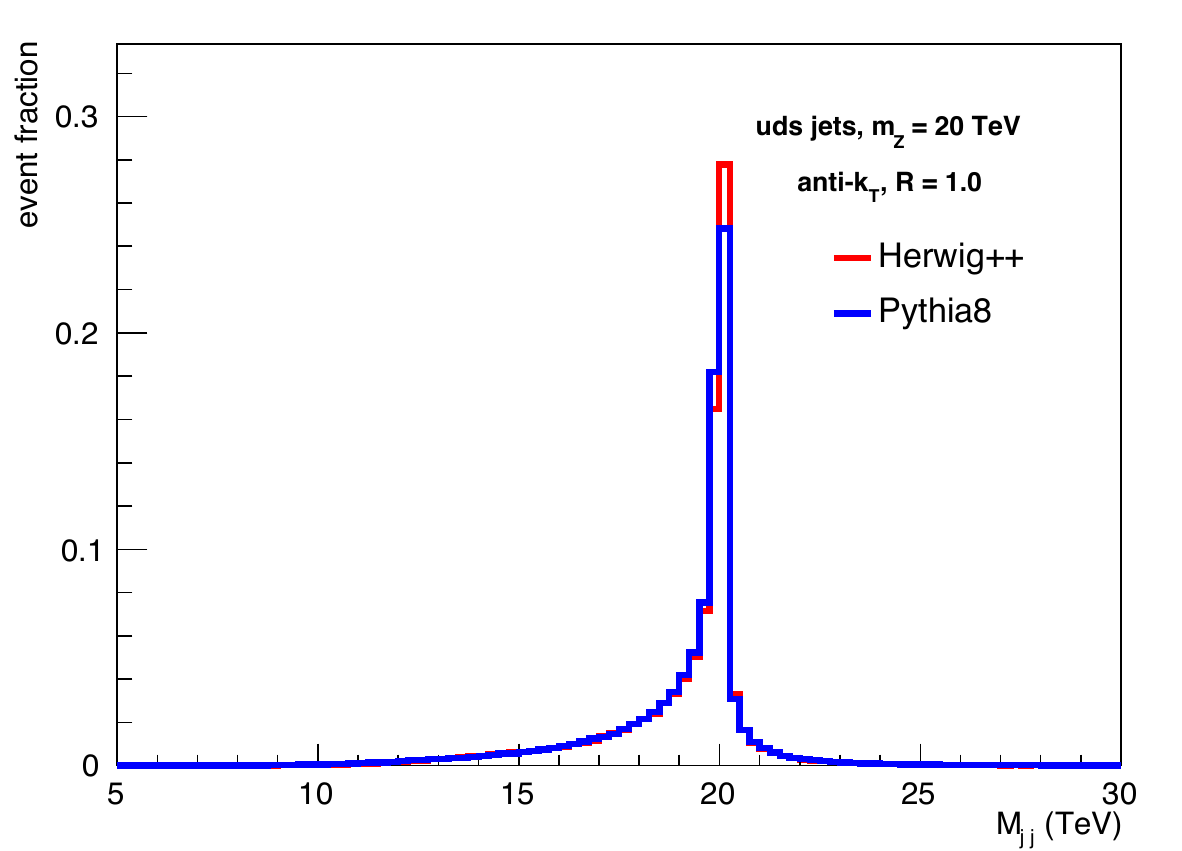}
\includegraphics[width=0.45\linewidth]{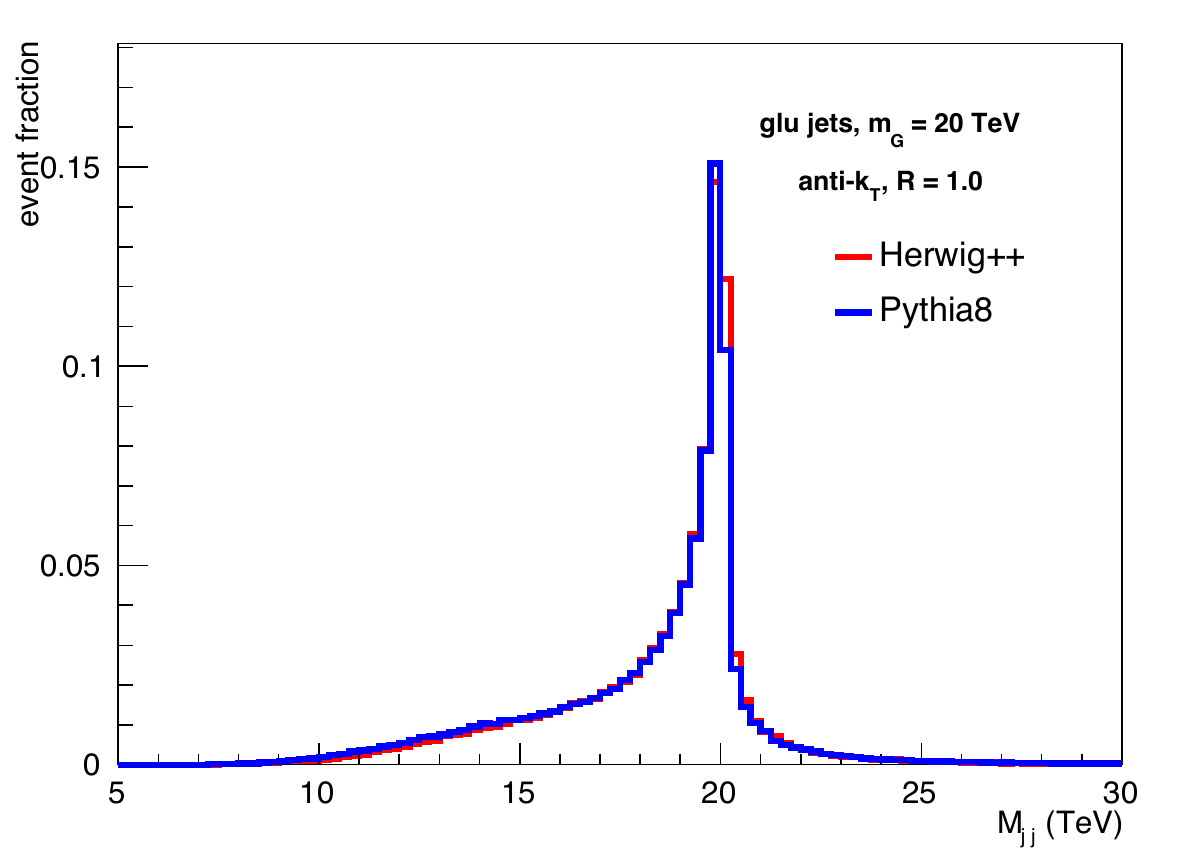}
\caption{ Light quark dijet invariant mass spectrum (left) and gluon
  dijet invariant mass spectrum (right) produced from the decay of a
  20 TeV resonance with jet radius of $R = 1.0$.  }
\label{fig:qg_R1}
\end{figure}

For colored top quarks, light quarks, and gluons the tail is never
completely removed, as long as there is radiation in the event not
captured in the jet.  There is always a non-zero probability that a
colored parton will emit radiation outside of the jet and therefore
will effectively lose energy.  By increasing the jet radius, the tail
of the resonance mass distribution extending to small masses can be
reduced.  In Fig.~\ref{fig:qg_R1}, we plot the dijet invariant mass
for resonances decaying to gluons and light quark jets with jet radius
$R=1.0$.  As compared to earlier plots, where the jet radius extended
to only $R=0.5$, the mass distribution is much more symmetric and the
tail extending to small masses is nearly eliminated.

This effect on the $p_T$ of the jet can be estimated in the small jet
radius $R$ limit.  The average $p_T$ loss $\langle \delta p_T\rangle$
due to perturbative radiation is \cite{Dasgupta:2007wa,Salam:2009jx}
\begin{equation}
\langle \delta p_T \rangle = \frac{\alpha_s}{\pi} L_i \log R + {\cal O}(\alpha_s)\,.
\end{equation}
$L_i$ is a constant that depends on the flavor of the jet:
\begin{align}
L_q  &= \left(
2\log 2 -\frac{3}{8}
\right) C_F \,,\\
L_g &= \left(
2\log 2 -\frac{43}{96}
\right) C_A +\frac{7}{48}n_f T_R \,.
\end{align}
For resonances that decay to two jets, this $p_T$ loss can be
translated into the average difference between the true resonance mass
and the dijet invariant mass, $\langle \delta m \rangle$.  To lowest
order in the small jet radius limit, assuming that the resonance is
produced at rest, this mass difference is approximately
\begin{equation}\label{eq:massdiff}
\langle \delta m \rangle \simeq -m\frac{\alpha_s}{2\pi} L_i \log R + {\cal O}(\alpha_s)\,,
\end{equation} 
where $m$ is the mass of the resonance.

\begin{figure}[h!]
\centering
\includegraphics[width=0.45\linewidth]{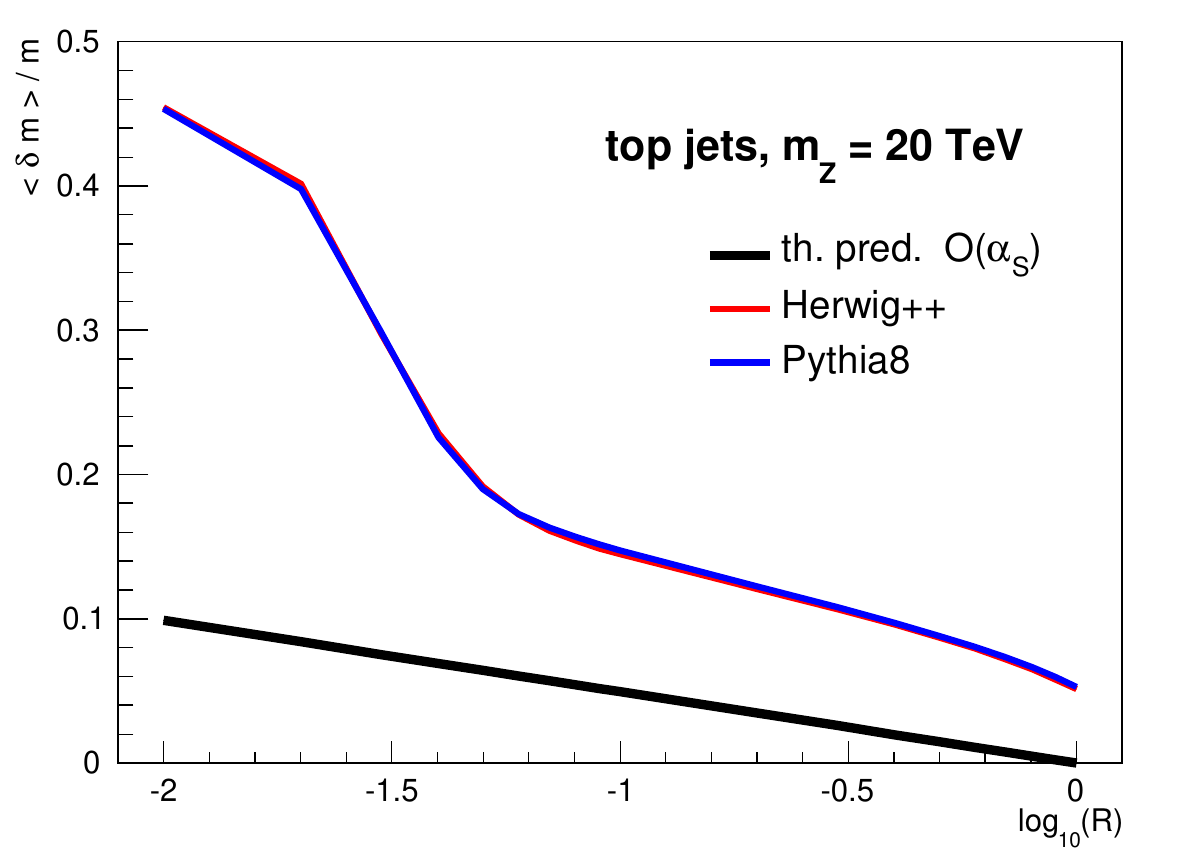}
\includegraphics[width=0.45\linewidth]{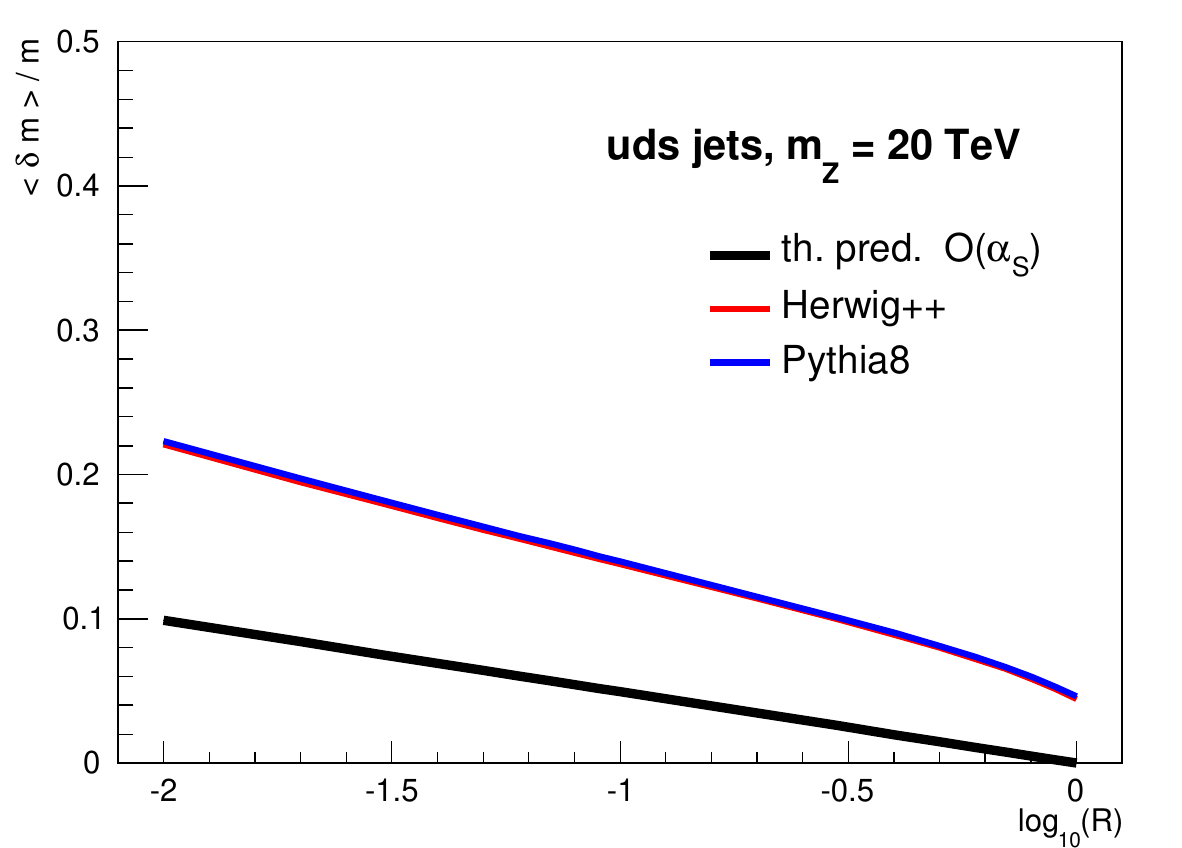}
\\
\includegraphics[width=0.45\linewidth]{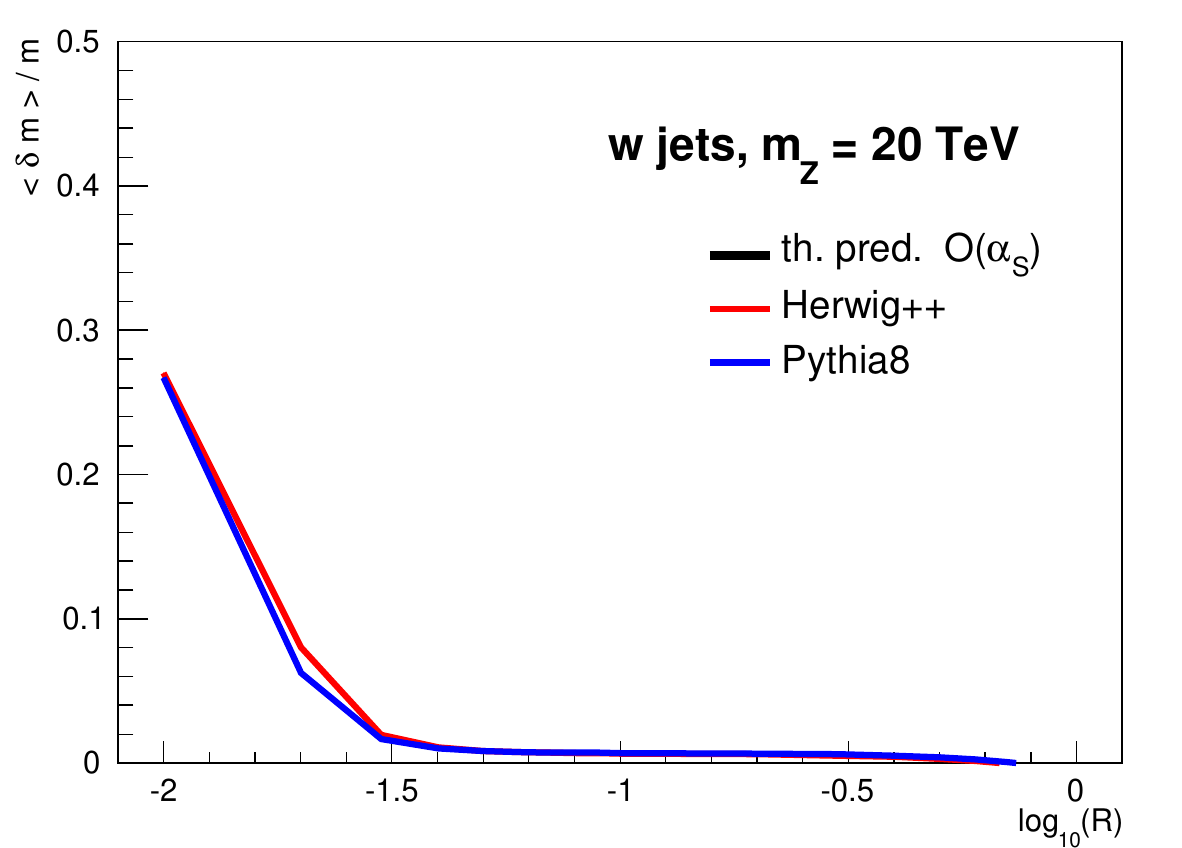}
\includegraphics[width=0.45\linewidth]{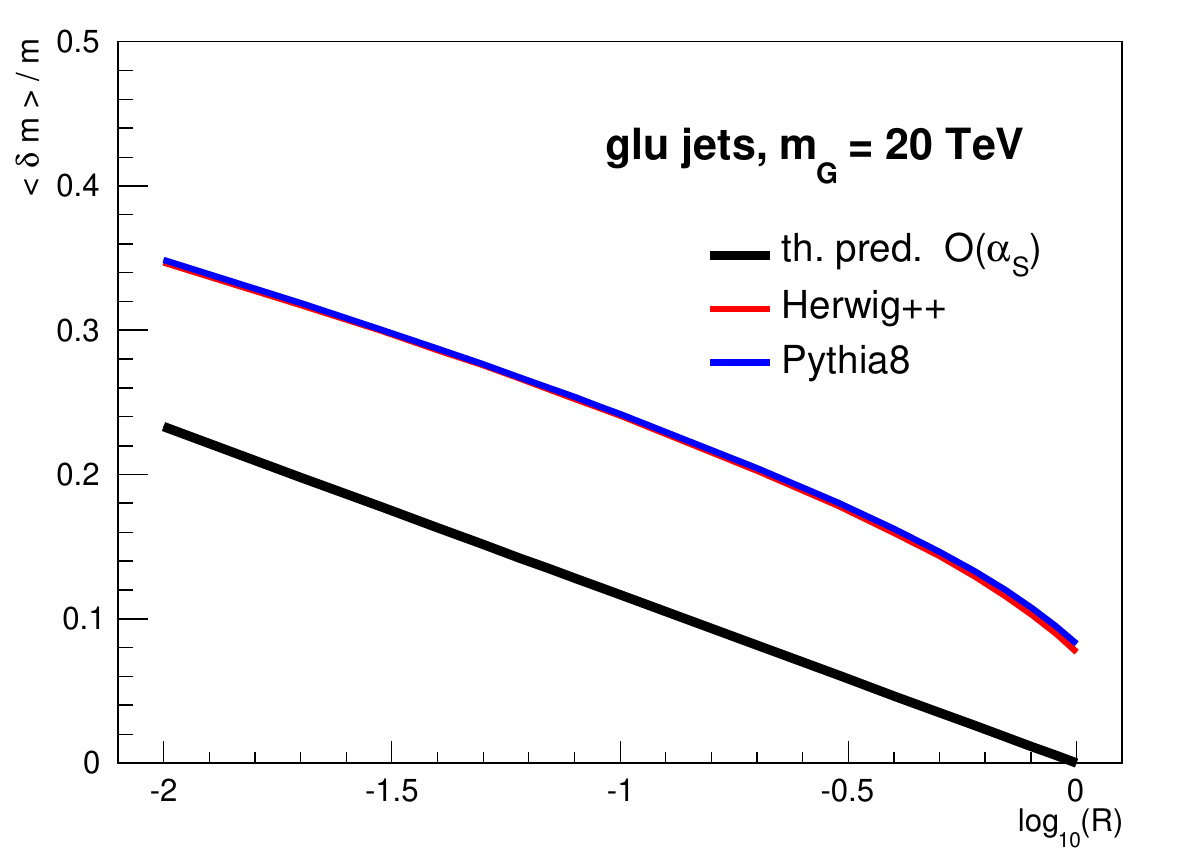}
\caption{ Average fractional difference between dijet invariant mass
  from anti-$k_T$ jets with various radii and the true 20 TeV
  resonance that decays to tops, light QCD quarks, $W$s, and gluons.
  The prediction of Eq.~\ref{eq:massdiff} is shown for reference.  }
\label{fig:ave_mres}
\end{figure}

In Fig.~\ref{fig:ave_mres}, we plot the average difference between the
dijet invariant mass and the true resonance mass $\langle \delta
m\rangle$ as a function of the jet radius $R$.  On these plots, we
have also included the analytic prediction of Eq.~\ref{eq:massdiff}
for reference.  Once the jet radius is large enough to capture all of
the $W$ decay products (above about $R = 0.03$), the di-$W$ invariant
mass is very close to the true resonance mass, as expected because it
is a color-singlet.  For light quark and gluon jets, the prediction in
Eq.~\ref{eq:massdiff} agrees very well with the slope of the curve
from the Monte Carlos.  The offset differs, but is affected by
$R$-independent ${\cal O}(\alpha_s)$ corrections that we have not
included.  Like for $W$s, if the jet radius is too small, then all of
the decay products of the top quark will not be captured in the jet.
However, once the jet radius is above about $R = 0.06$, the top quark
emits radiation outside of the jet in the same manner as a light
quark.

As a quantitative measure of the optimal precision to which resonance
masses can be reconstructed, in Fig.~\ref{fig:fwhm} we plot the
fractional full-width half-maximum of the reconstructed 20 TeV
resonance that decays to $W$ bosons and gluons as a function of the
jet radius.  As illustrated in Fig.~\ref{fig:mres_ww}, the width of
the resonance decaying to $W$ bosons is exceptionally small, and
appears to only be limited by the intrinsic width of the resonance.
If the jet radius is too large, however, then more contamination
radiation will be captured by the jet, smearing out the resonance
peak.  For resonances decaying to gluons, the opposite is true.  If
the jet radius is too narrow, then a significant amount of final state
radiation will exit the jet, greatly reducing the resolution of the
resonance peak.  However, as the jet radius increases, more of this
radiation is captured in the jet, improving the resolution.  Note,
however, that even with the largest jet radius, the resolution of the
resonance mass for gluons is at the percent level, as compared to less
than a part per mille for $W$s.

\begin{figure}[h!]
\centering
\includegraphics[width=0.45\linewidth]{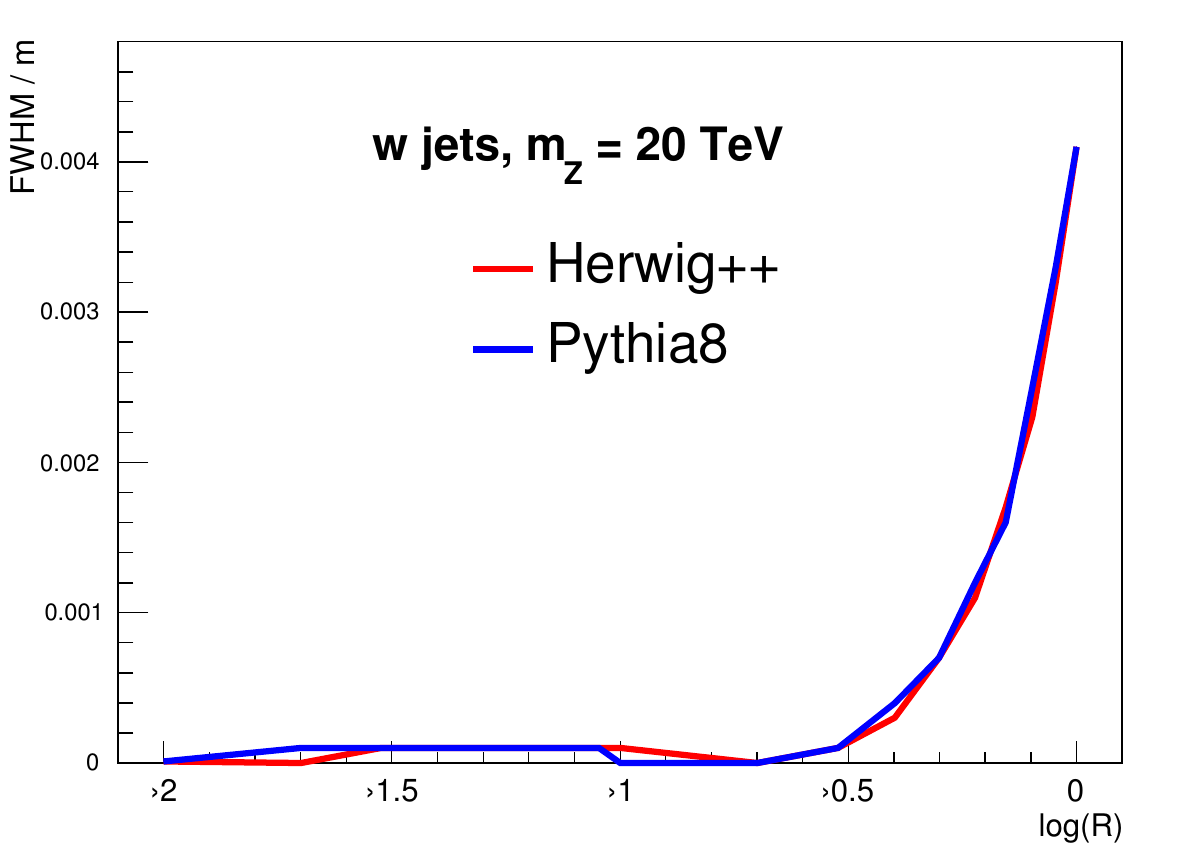}
\includegraphics[width=0.45\linewidth]{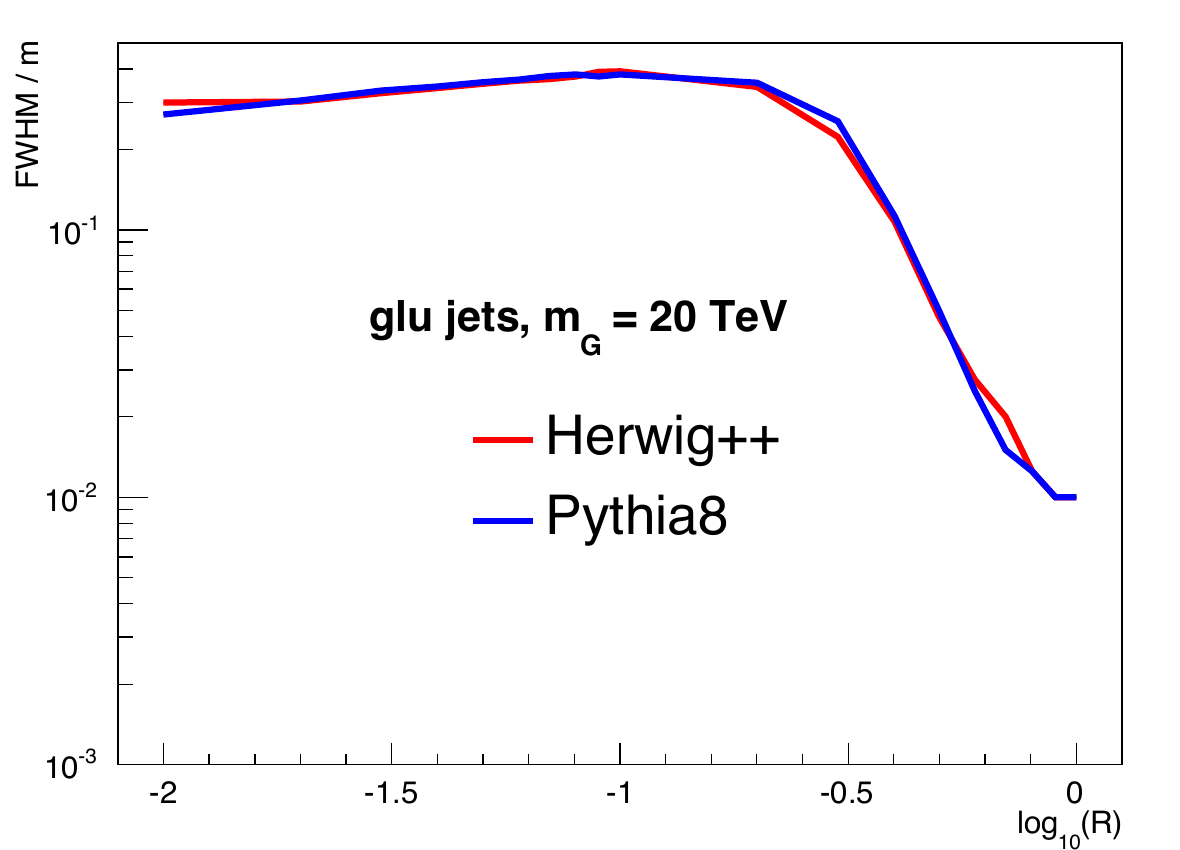}
\caption{
Full-width half maximum of the 20 TeV resonance that decayed to boosted $W$ bosons (left) and gluon jets (right).
} 
\label{fig:fwhm}
\end{figure}

\subsection{SM physics of boosted objects}
\label{sec:jet_boost}

Given that jets or hadronically decaying electroweak objects may be
the most powerful probe into new, high scale physics, it is necessary
to efficiently identify their origin.  For electroweak particles, the
most sensitive single observable is the mass and jets with masses
around 100 GeV are evidence of electroweak origin.  Jets initiated by
light QCD partons, on the other hand, have no intrinsic high-energy
scale.  Depending on the cuts made on the jets as imposed by the jet
algorithm, the mass spectra of QCD jets will be correspondingly
sculpted and may peak in the electroweak mass window.  More detailed
analyses can be performed for identifying specific jets; see for
example the studies in
Refs.~\cite{Auerbach:2014xua,Aguilar-Saavedra:2014iga,Larkoski:2015yqa},
dedicated to top quarks.  We will review the conclusions of some of
these top quark studies in Sec.~\ref{sec:toptagsec}.

Jet mass distributions are plotted in
Figs.~\ref{fig:mjet_10_005}-\ref{fig:mjet_10_05}.  Here, we plot the
masses of the jets from the resonance decays studied in the previous
section with the same cuts imposed.  The mass of the resonance is
fixed to 10 TeV, and the jet radius is varied from $R=0.05$ to
$R=0.5$.  For quark and gluon jets, the mass distributions increase as
the jet radius increases.  For these jets, the peak of the mass
distribution is located approximately at
\begin{equation}
m_\text{peak}^\text{QCD}\simeq \frac{\alpha_s}{\pi} C_i R p_T \,,
\end{equation}
where $C_i$ is the color of the jet and $p_T$ is its transverse momentum.  As illustrated in the plots, the peak of these QCD jets is in the electroweak mass range for $R \gtrsim 0.2$.  Therefore, by decreasing the jet radius, we reduce the number of QCD jets that look like jets from hadronic decays of electroweak objects.

The mass of jets initiated by hadronic decays of top quarks or $W$
bosons have very different dependence on jet radius.  At the smallest
jet radius studied, $R=0.05$, there is a significant amount of
radiation in the decays that are not captured in the jet.  For $W$
bosons, while there is a pronounced peak at the $W$ mass, there is a
tail at small masses indicating that a fraction of the jets do not
contain both prongs of the $W$ decay.  For top quarks, there actually
is no peak at the top mass whatsoever.  Some jets do consist of the
$W$ from the decay, but the bulk is a smooth, falling distribution.
As the radius is increased more of the decay products are included in
the jets, and so for $R\gtrsim 0.1$, most of the top and $W$ jets
exhibit dominant peaks at their expected masses.  As a rule of thumb,
the critical jet radius necessary to capture all of the decay products
for a resonance of mass $m$ is approximately
\begin{equation}\label{eq:rcrit}
R_\text{crit}\simeq \frac{2m}{p_T}\,.
\end{equation}

When the jet radius is increased to $R=0.5$, however, the mass
distribution is significantly deformed.  This is due to the inclusion
of more contamination radiation in the jet, that is uncorrelated with
the decay.  This radiation may come from the initial state or
underlying event and is approximately uniformly distributed over the
area of the jet.  Therefore, its contribution to the transverse
momentum of the jet scales like the area of the jet, $R^2$, while its
contribution to the mass of the jet scales like $R^4$.  Roughly, in
changing the jet radius from $R=0.2$ to $R=0.5$, the effect of
contamination radiation on the jet mass increased by a factor of
almost 40.  This illustrates that, to accurately reproduce the
resonance peak, to reduce QCD backgrounds, and to eliminate
contamination, a jet radius close to the critical radius
$R_\text{crit}$ in Eq.~\ref{eq:rcrit} should be used.

\begin{figure}[h!]
\centering
\includegraphics[width=0.45\linewidth]{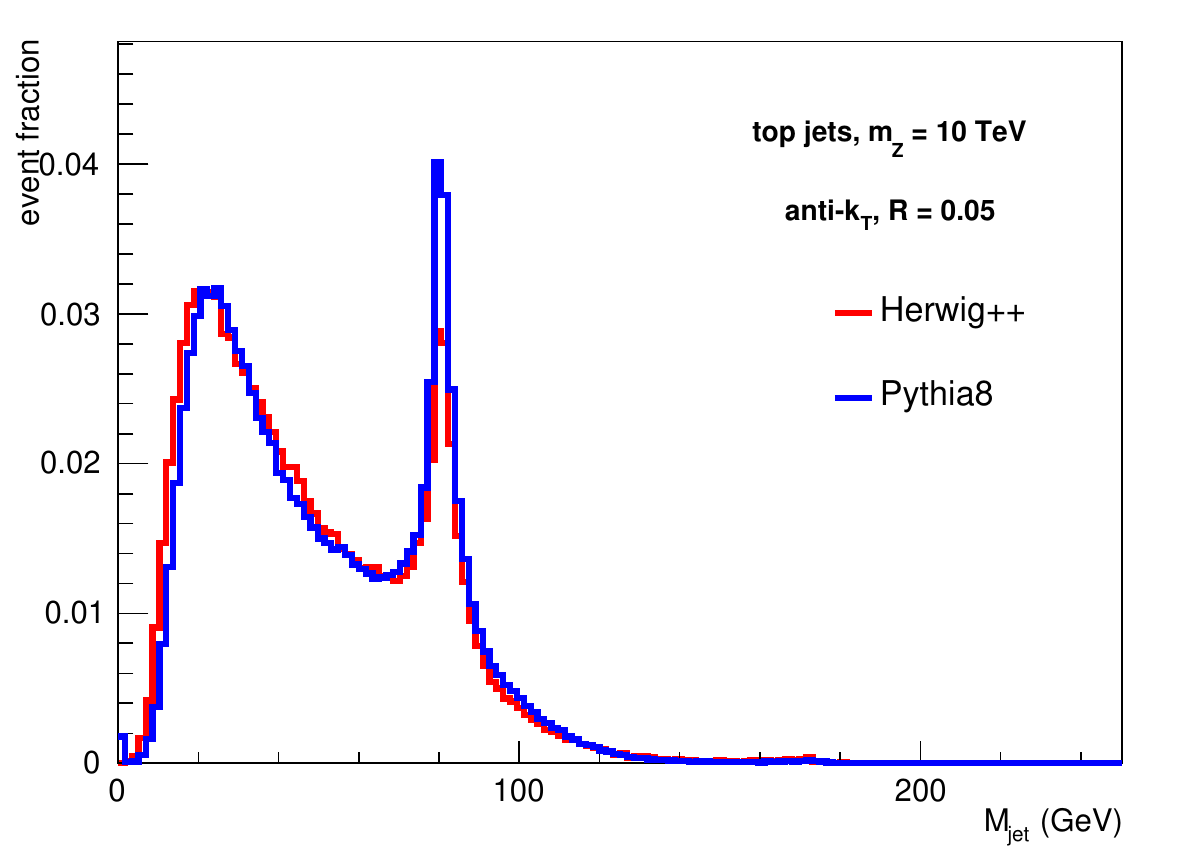}
\includegraphics[width=0.45\linewidth]{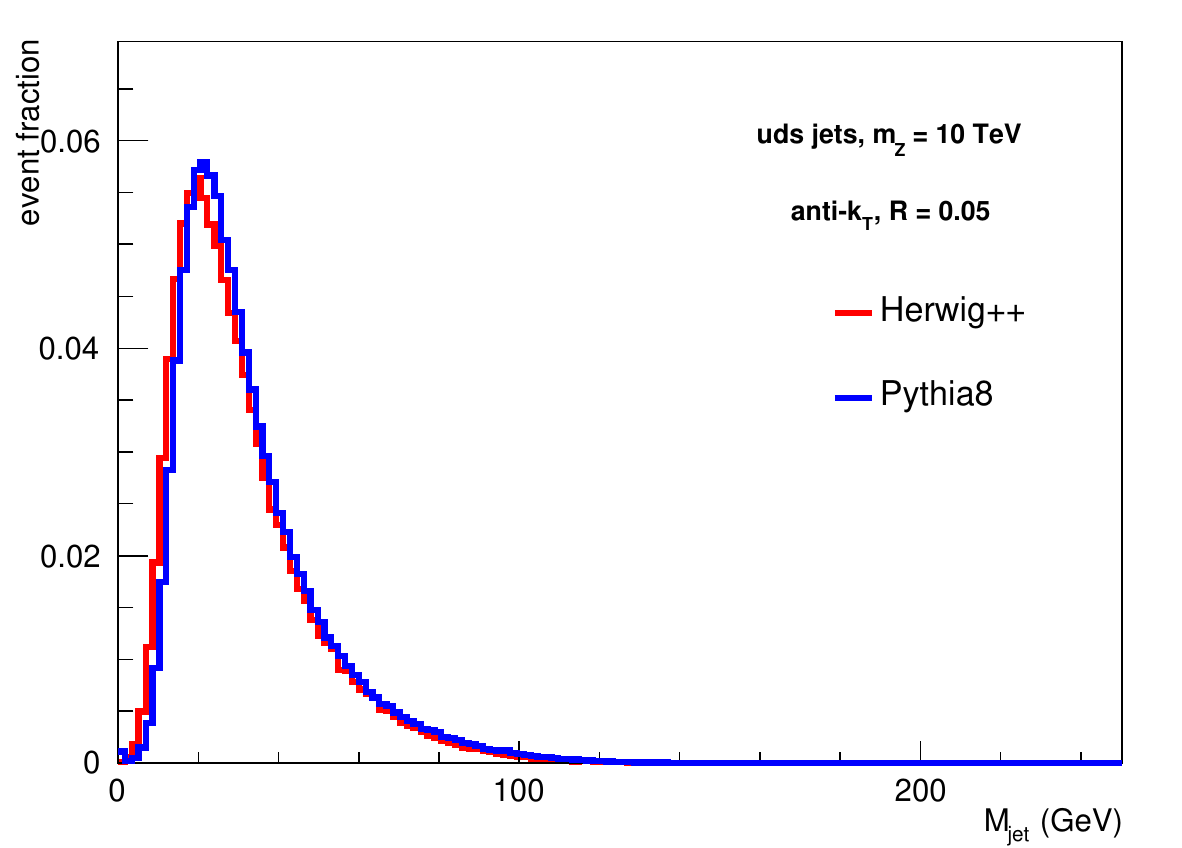}
\\
\includegraphics[width=0.45\linewidth]{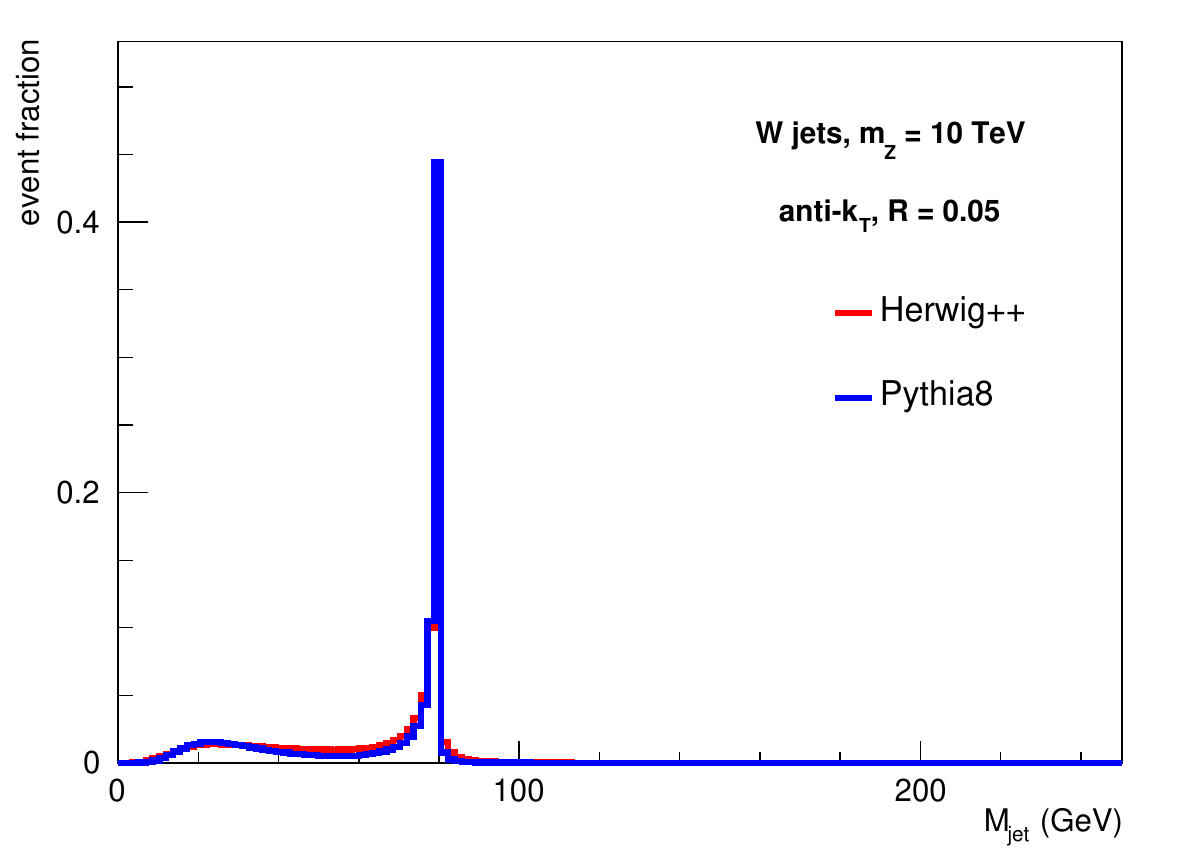}
\includegraphics[width=0.45\linewidth]{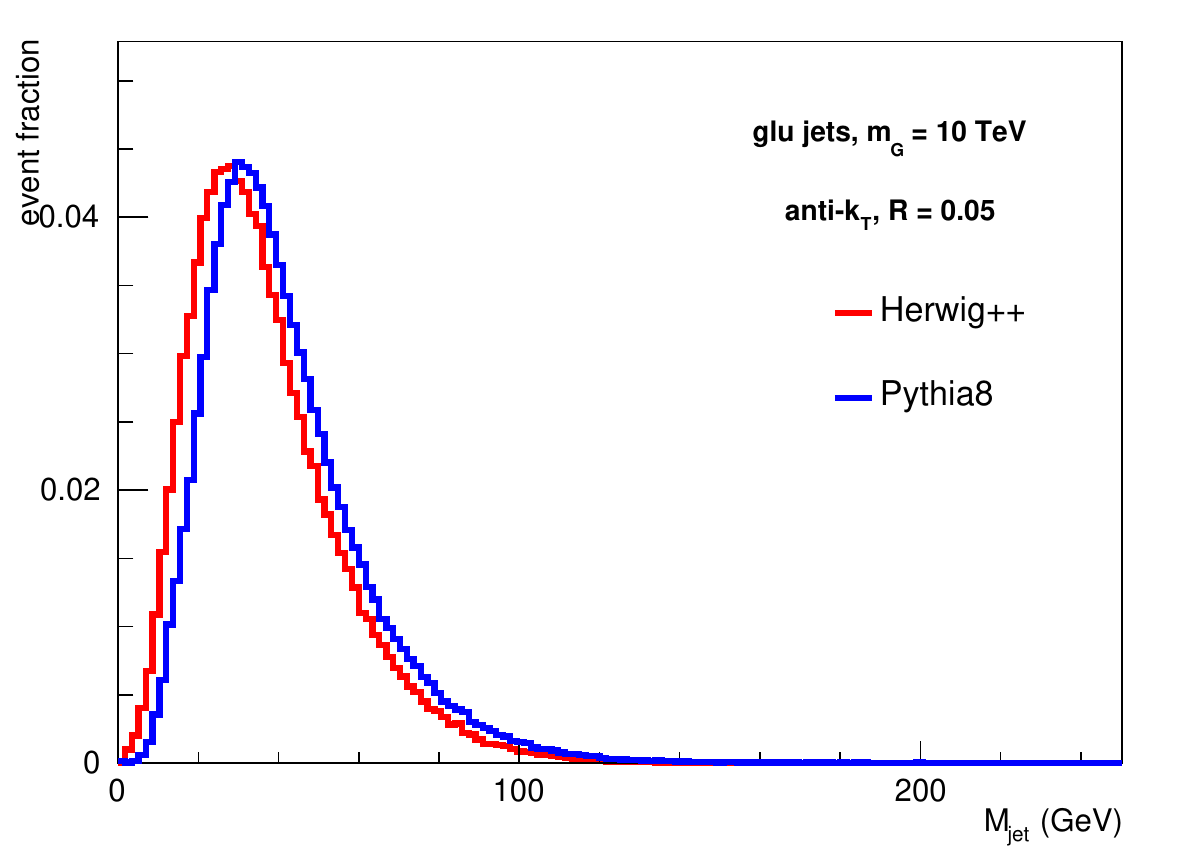}
\caption{ Jet mass distribution of $R=0.05$ jets produced from 10 TeV
  resonance decays to tops, light QCD quarks, $W$s, and gluons.  }
\label{fig:mjet_10_005}
\end{figure}

\begin{figure}[h!]
\centering
\includegraphics[width=0.45\linewidth]{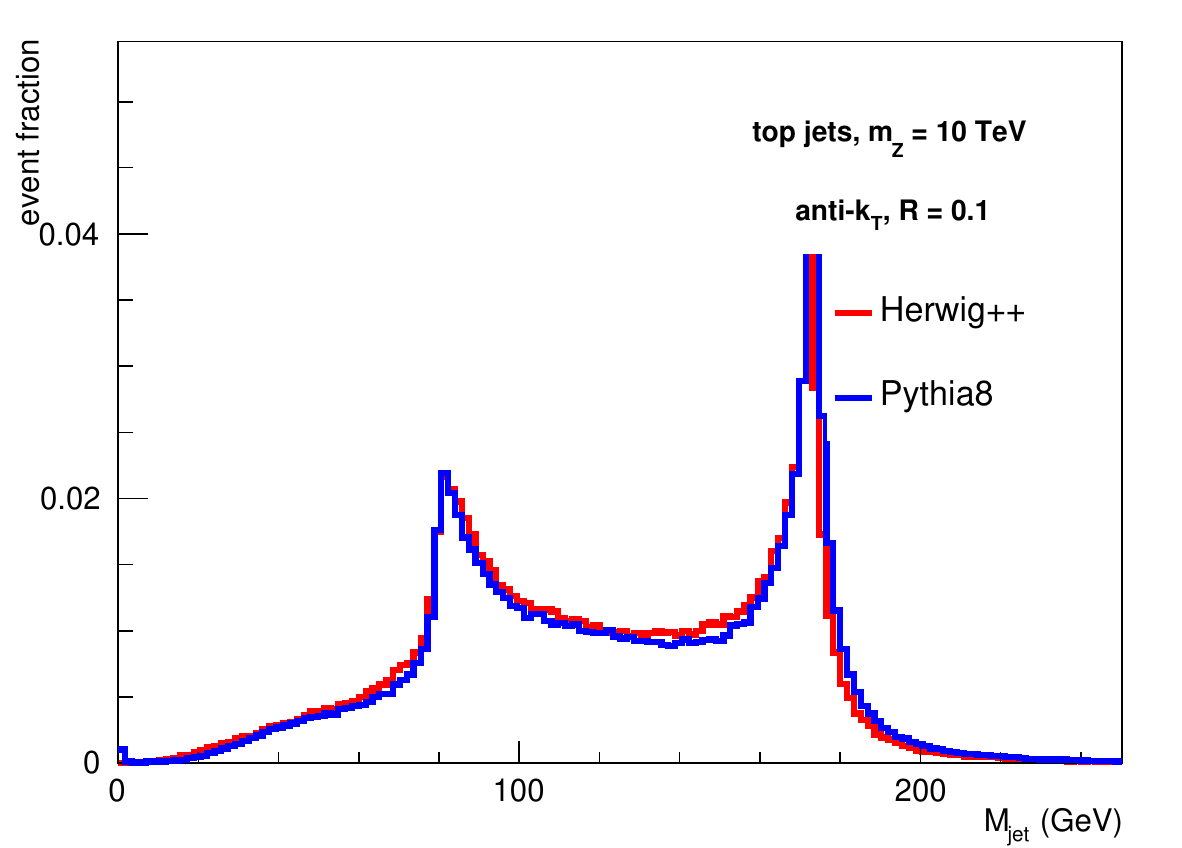}
\includegraphics[width=0.45\linewidth]{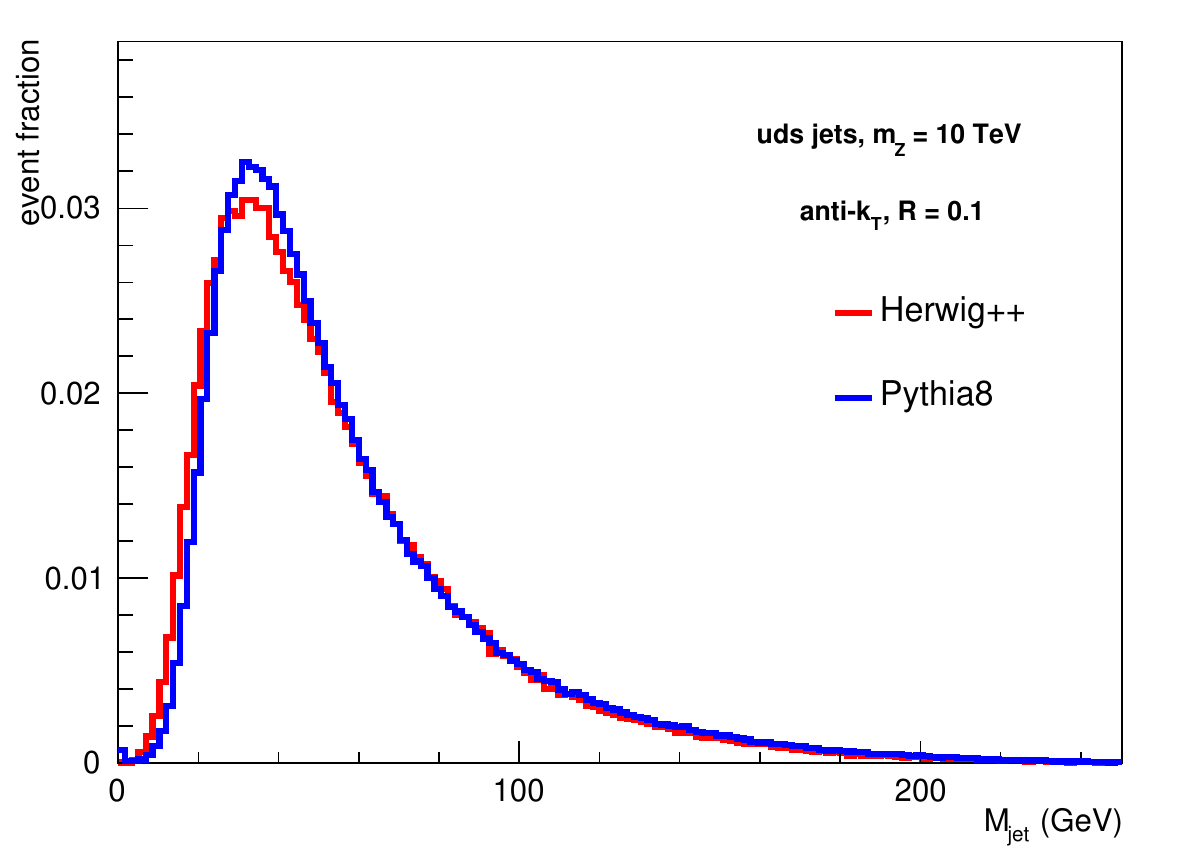}
\\
\includegraphics[width=0.45\linewidth]{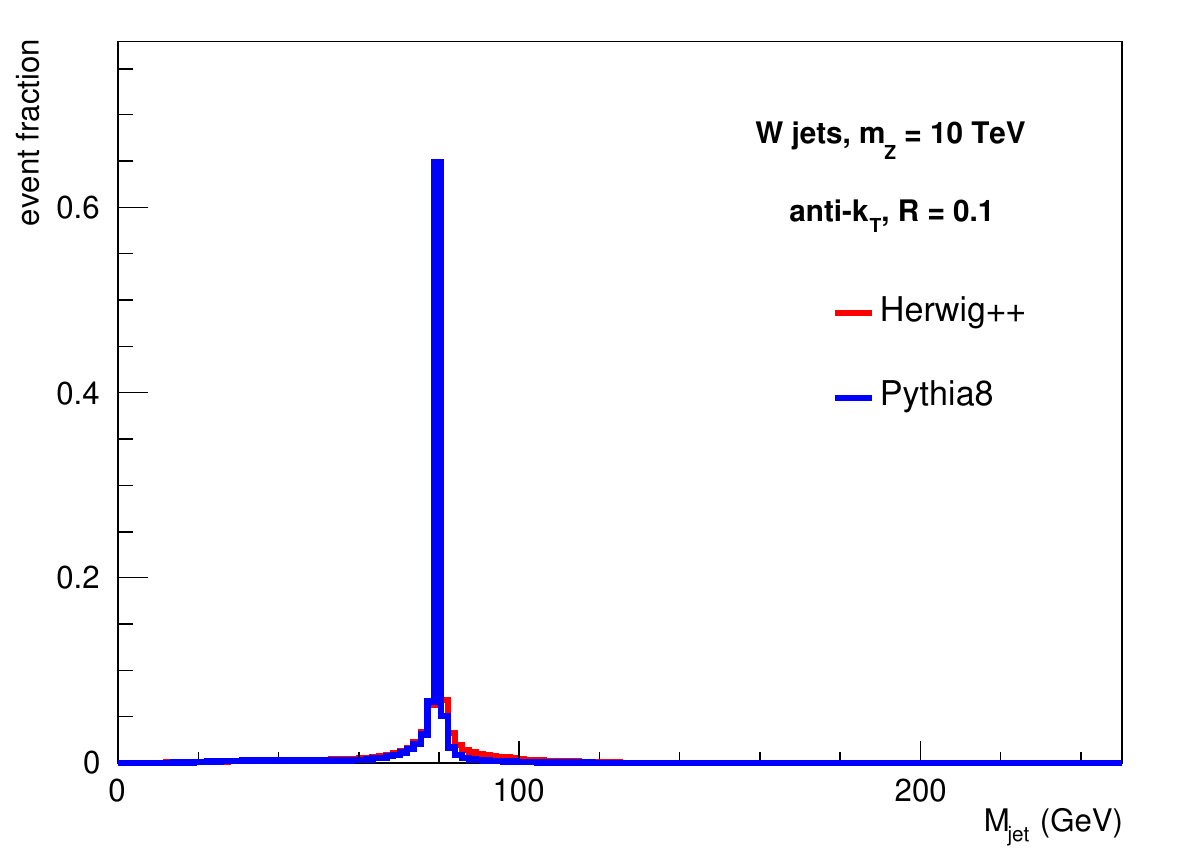}
\includegraphics[width=0.45\linewidth]{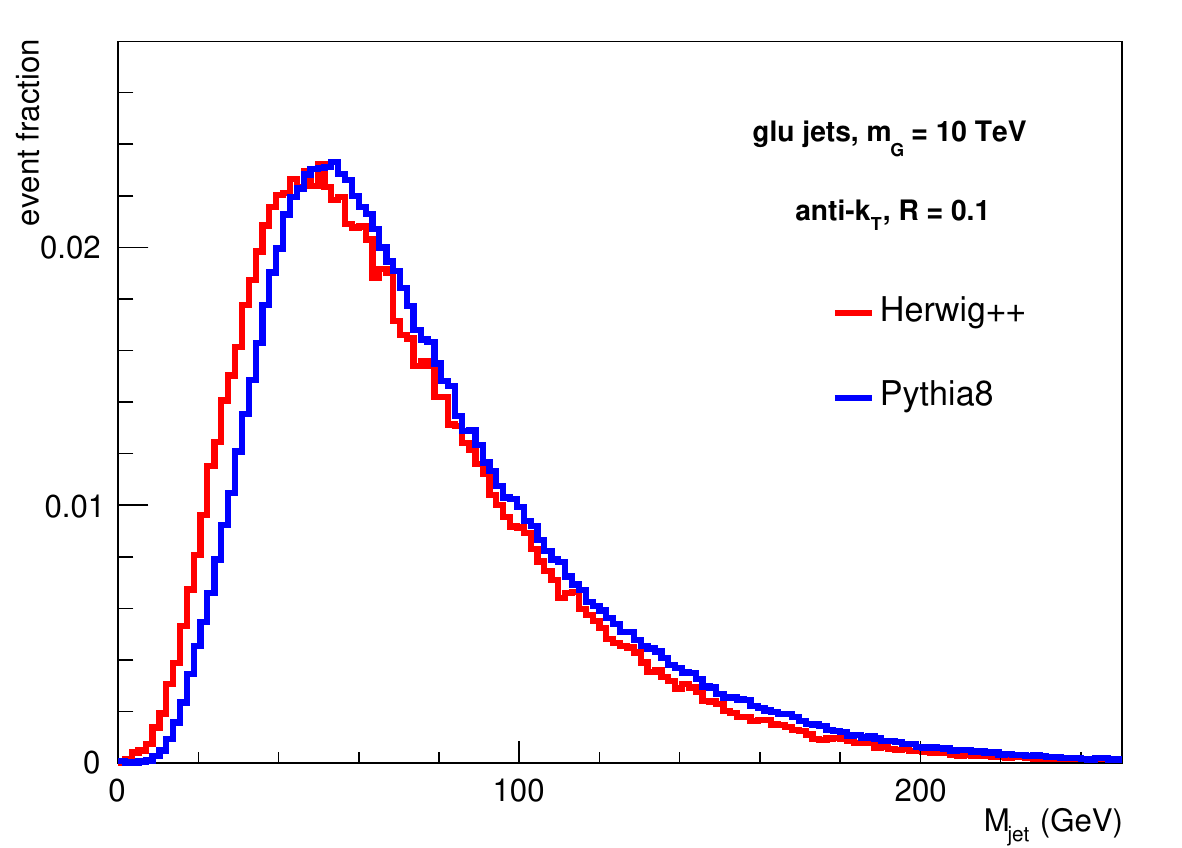}
\caption{ Jet mass distribution of $R=0.1$ jets produced from 10 TeV
  resonance decays to tops, light QCD quarks, $W$s, and gluons.  }
\label{fig:mjet_10_01}
\end{figure}

\begin{figure}[h!]
\centering
\includegraphics[width=0.45\linewidth]{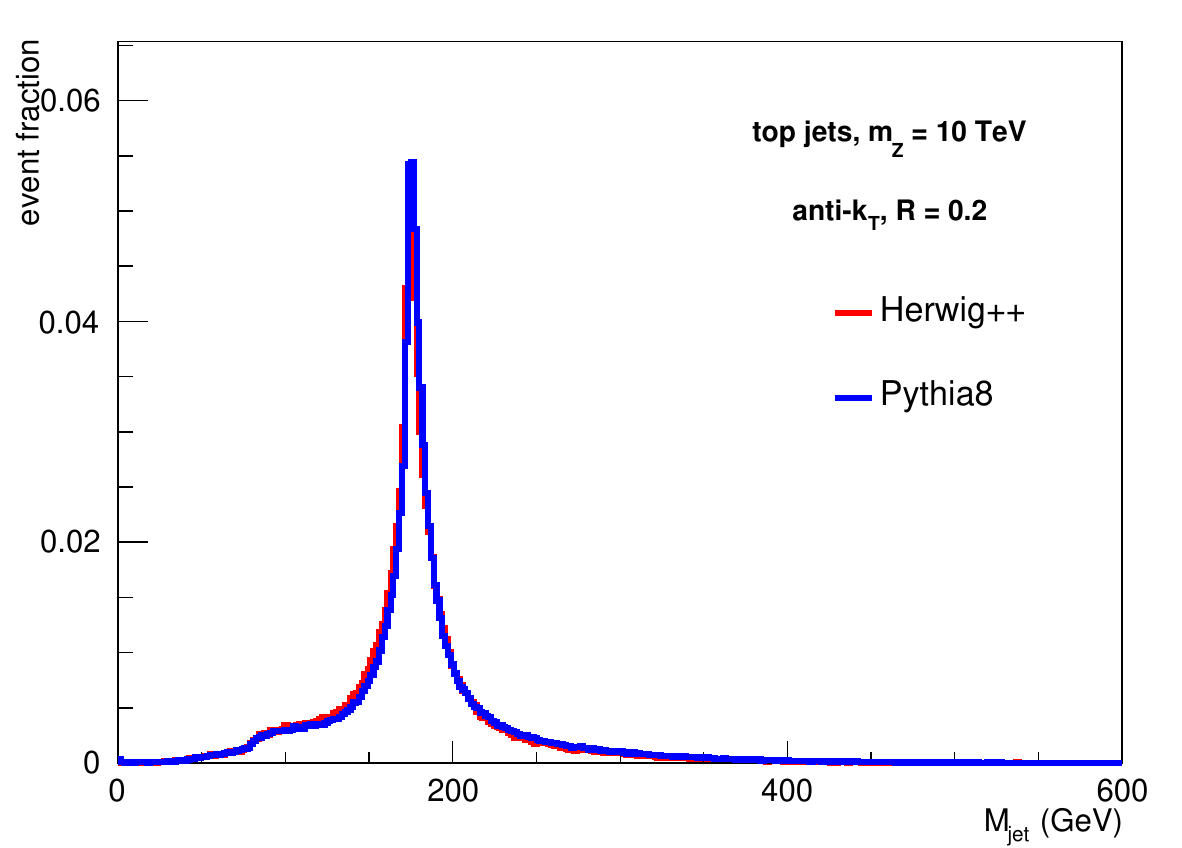}
\includegraphics[width=0.45\linewidth]{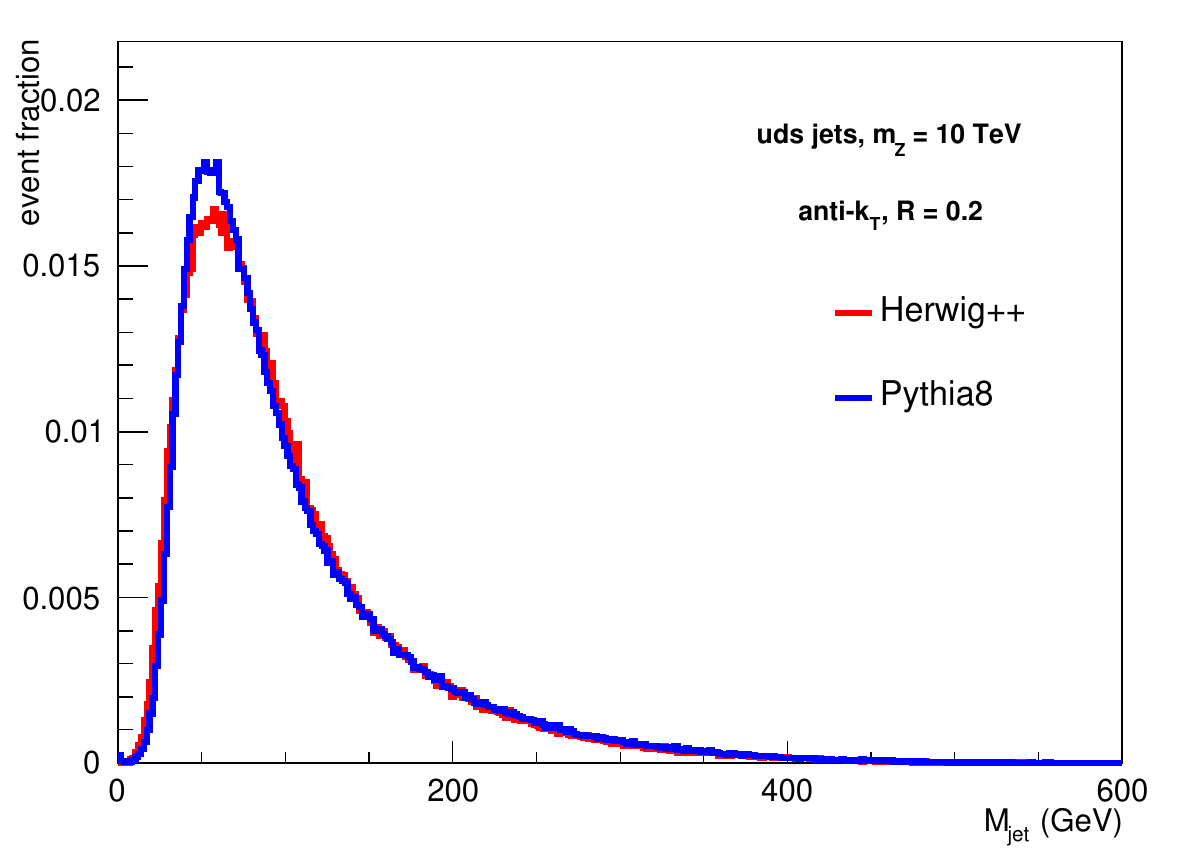}
\\
\includegraphics[width=0.45\linewidth]{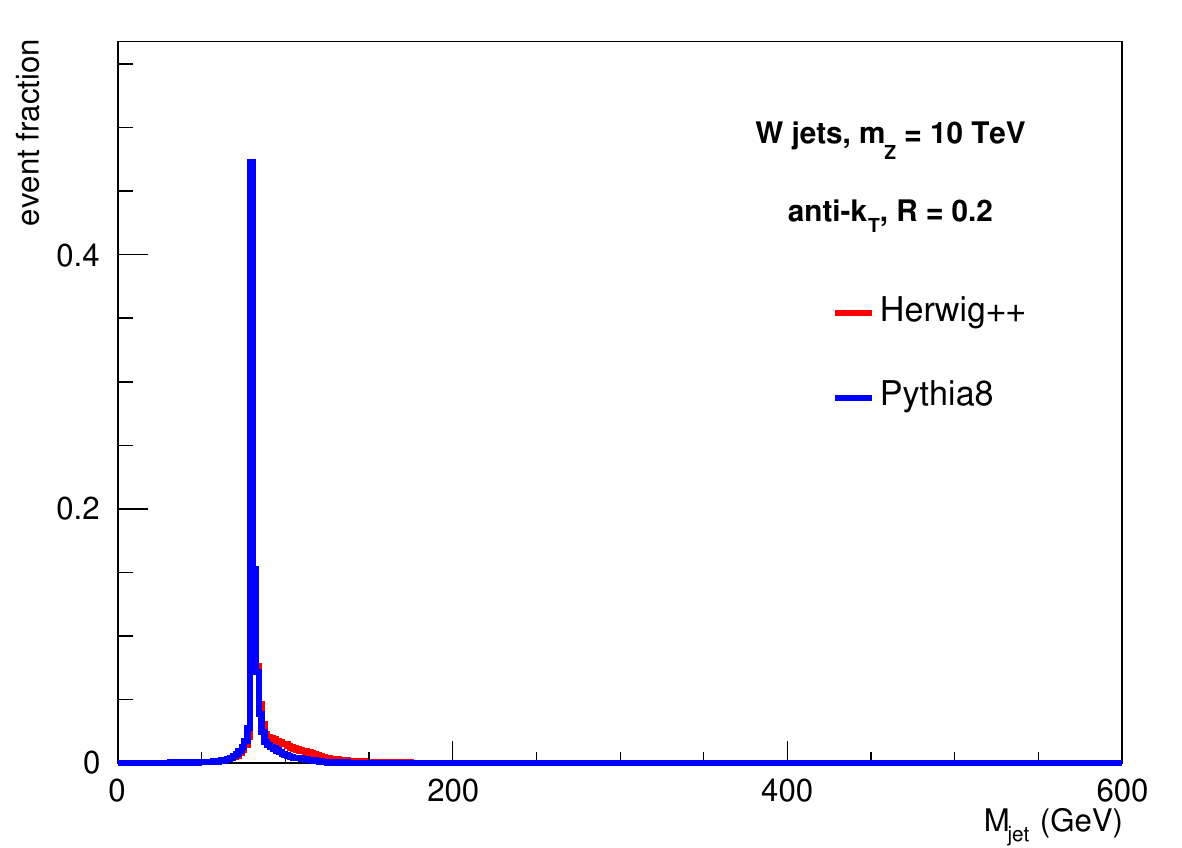}
\includegraphics[width=0.45\linewidth]{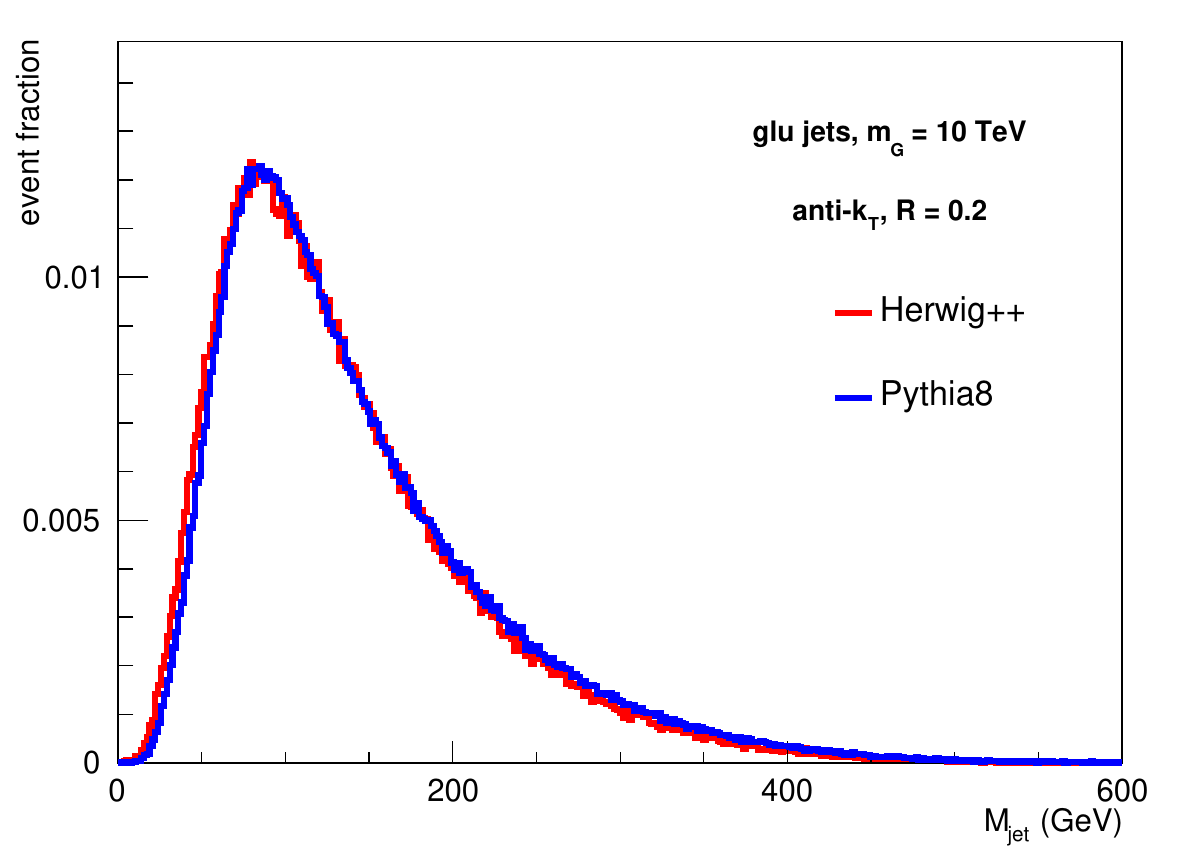}
\caption{ Jet mass distribution of $R=0.2$ jets produced from 10 TeV
  resonance decays to tops, light QCD quarks, $W$s, and gluons.  }
\label{fig:mjet_10_02}
\end{figure}

\begin{figure}[h!]
\centering
\includegraphics[width=0.45\linewidth]{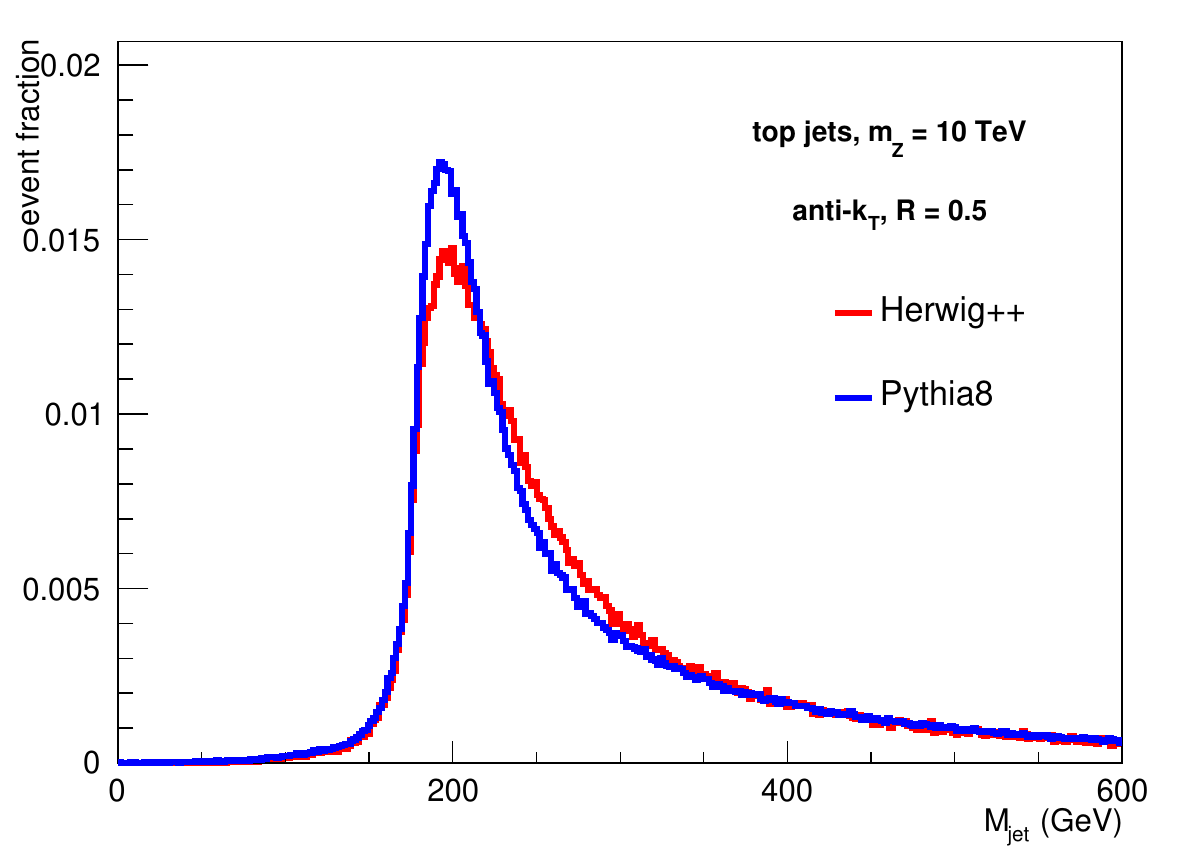}
\includegraphics[width=0.45\linewidth]{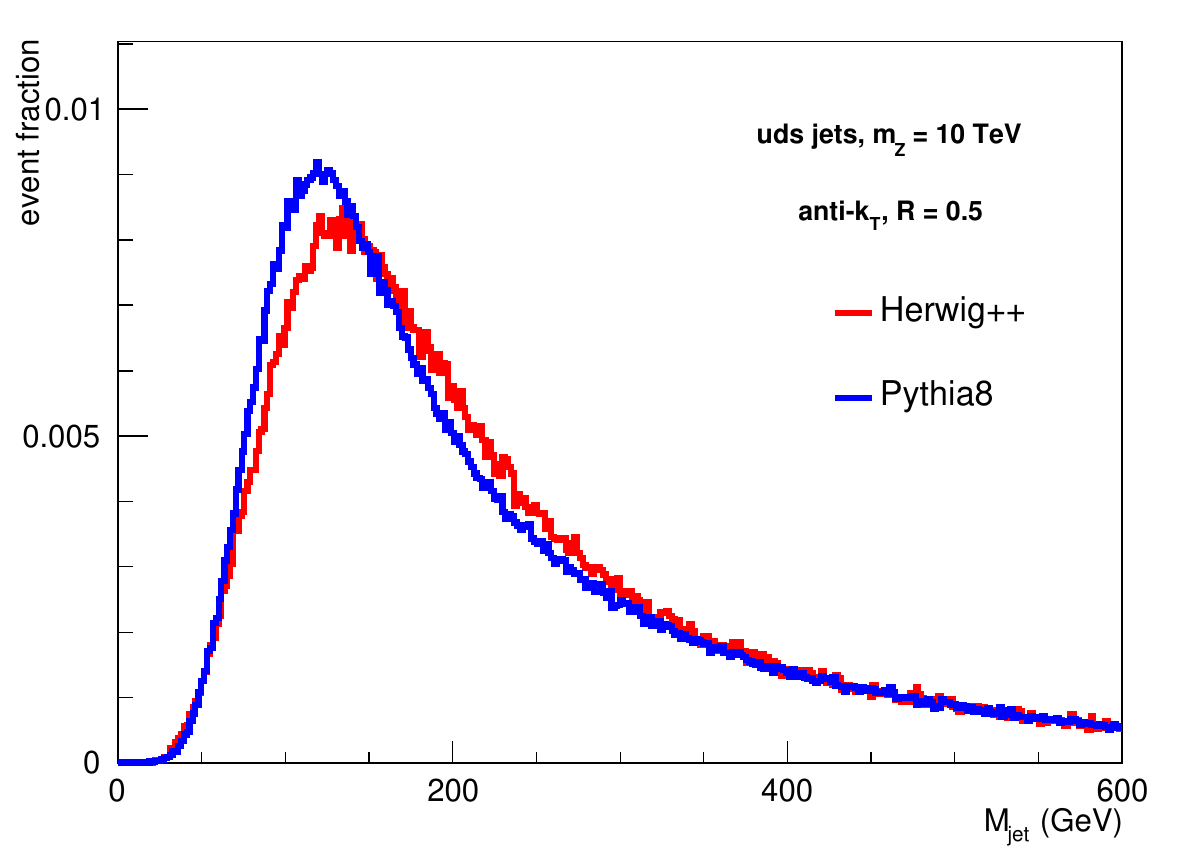}
\\
\includegraphics[width=0.45\linewidth]{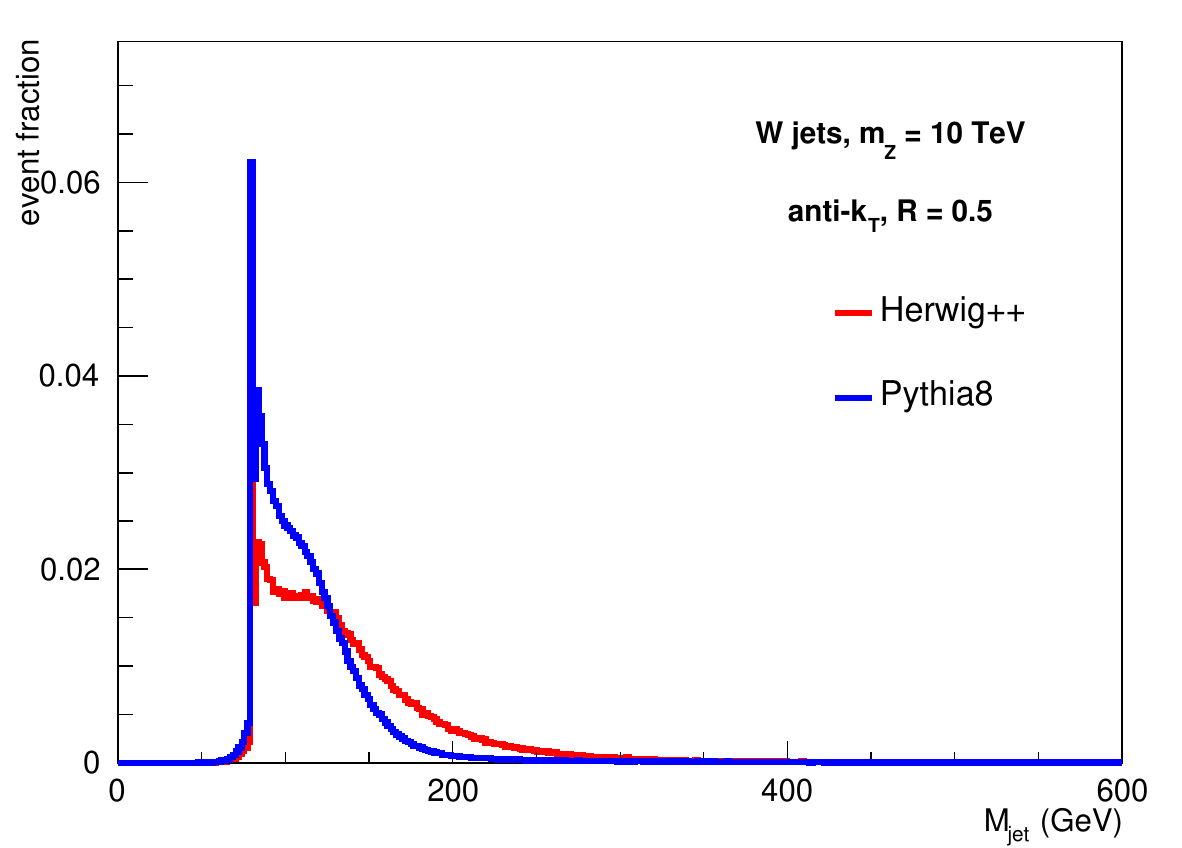}
\includegraphics[width=0.45\linewidth]{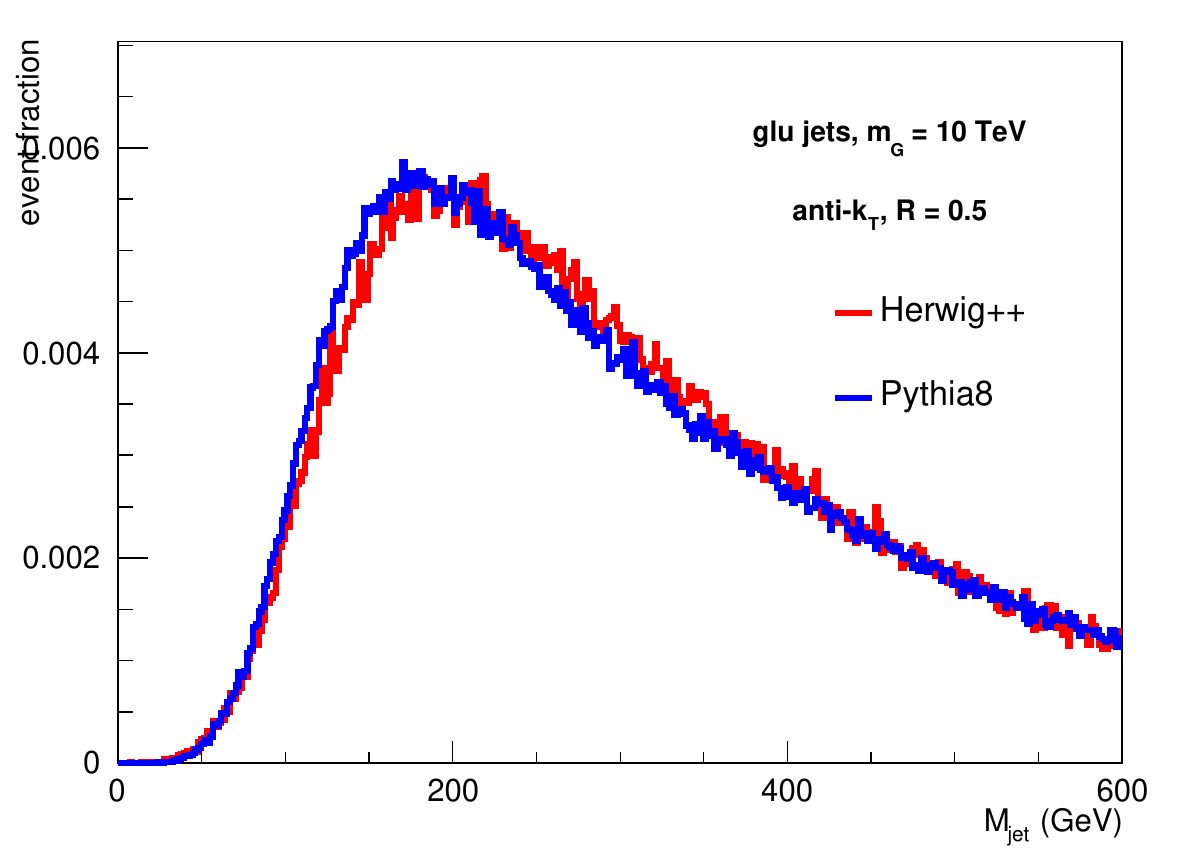}
\caption{
Jet mass distribution of $R=0.5$ jets produced from 10 TeV resonance decays to tops, light QCD quarks, $W$s, and gluons.
  } 
\label{fig:mjet_10_05}
\end{figure}

These observations are further illustrated in
Figs.~\ref{fig:efrac_10}-\ref{fig:efrac_40}.  Here, we have plotted
the average energy fraction located within an angle $\Delta R$ from
the jet center.  The jet radius is fixed to be $R=0.5$ and the mass of
the resonance that decays to the jets ranges from 10 to 40 TeV.  As
expected from the approximate scale invariance of QCD, the average
energy curves for quark and gluon jets is essentially independent of
the $p_T$ of the jet.  The top mass jets exhibit small $p_T$
dependence between the 10 and 20 TeV resonance mass samples, but are
independent for the higher mass samples.  For sufficiently high $p_T$
jets, top quarks are just light quarks.  For the boosted $W$ bosons,
on the other hand, almost all of the jet $p_T$ is contained within the
critical radius $R_\text{crit}$.  $W$ bosons are color singlets and so
there is no scale above which their radiation pattern looks like light
QCD jets.  This suggests that isolation requirements, similar to that
used for $\tau$ identification at the LHC, could be used to purify a
sample with boosted, hadronically-decaying $W$ bosons.

\begin{figure}[h!]
\centering
\includegraphics[width=0.45\linewidth]{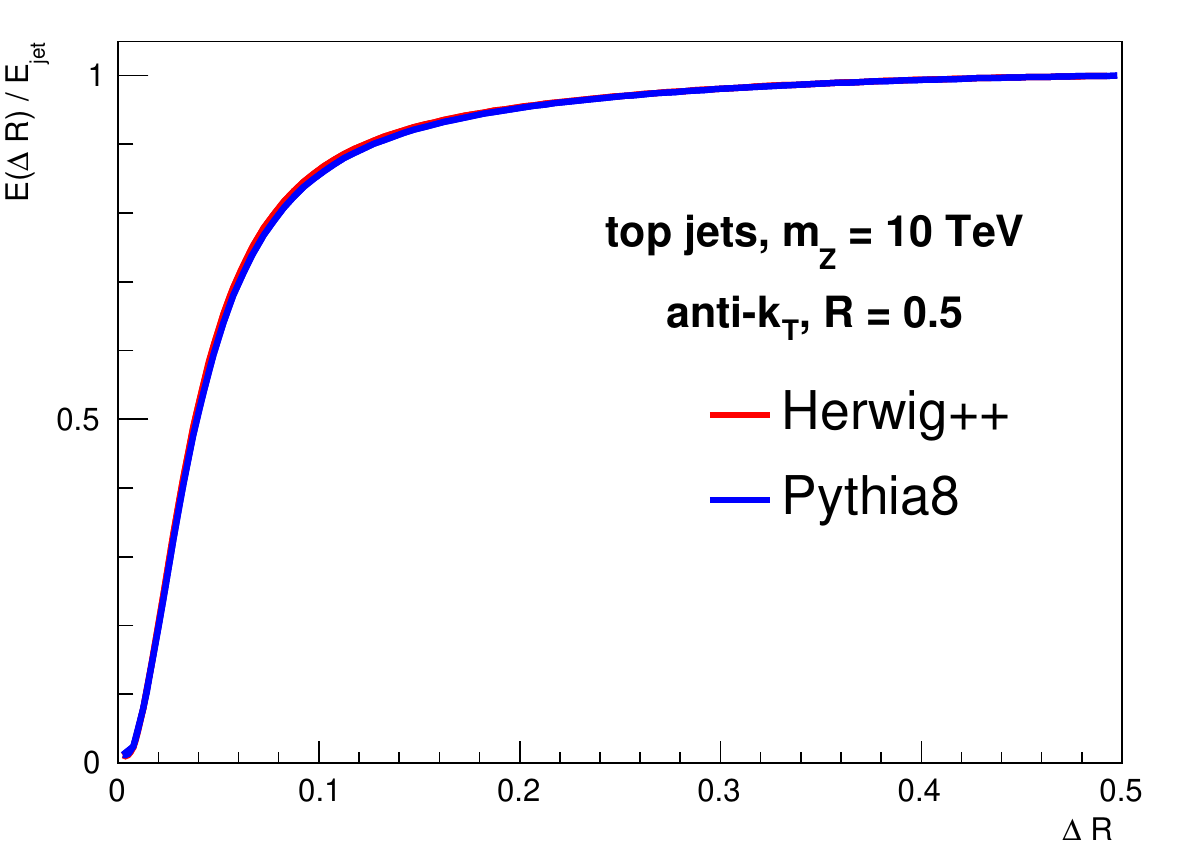}
\includegraphics[width=0.45\linewidth]{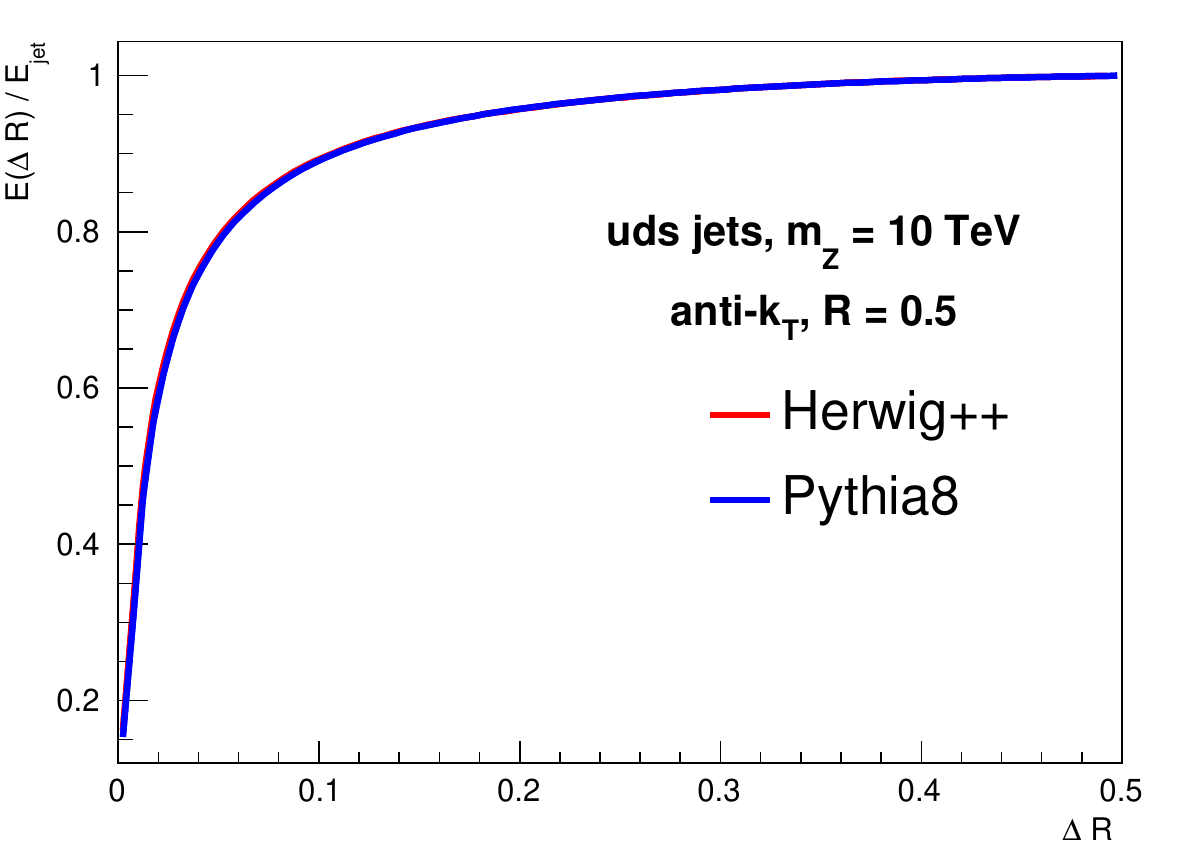}
\\
\includegraphics[width=0.45\linewidth]{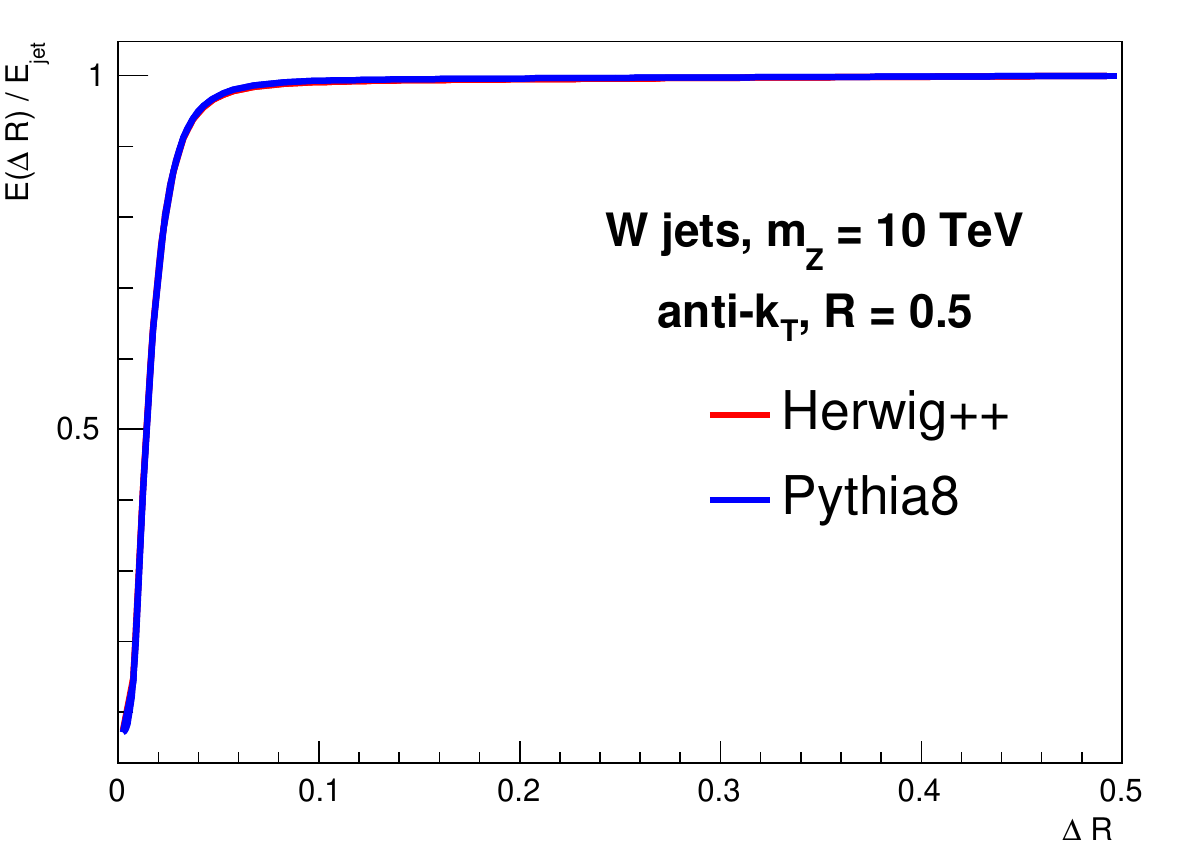}
\includegraphics[width=0.45\linewidth]{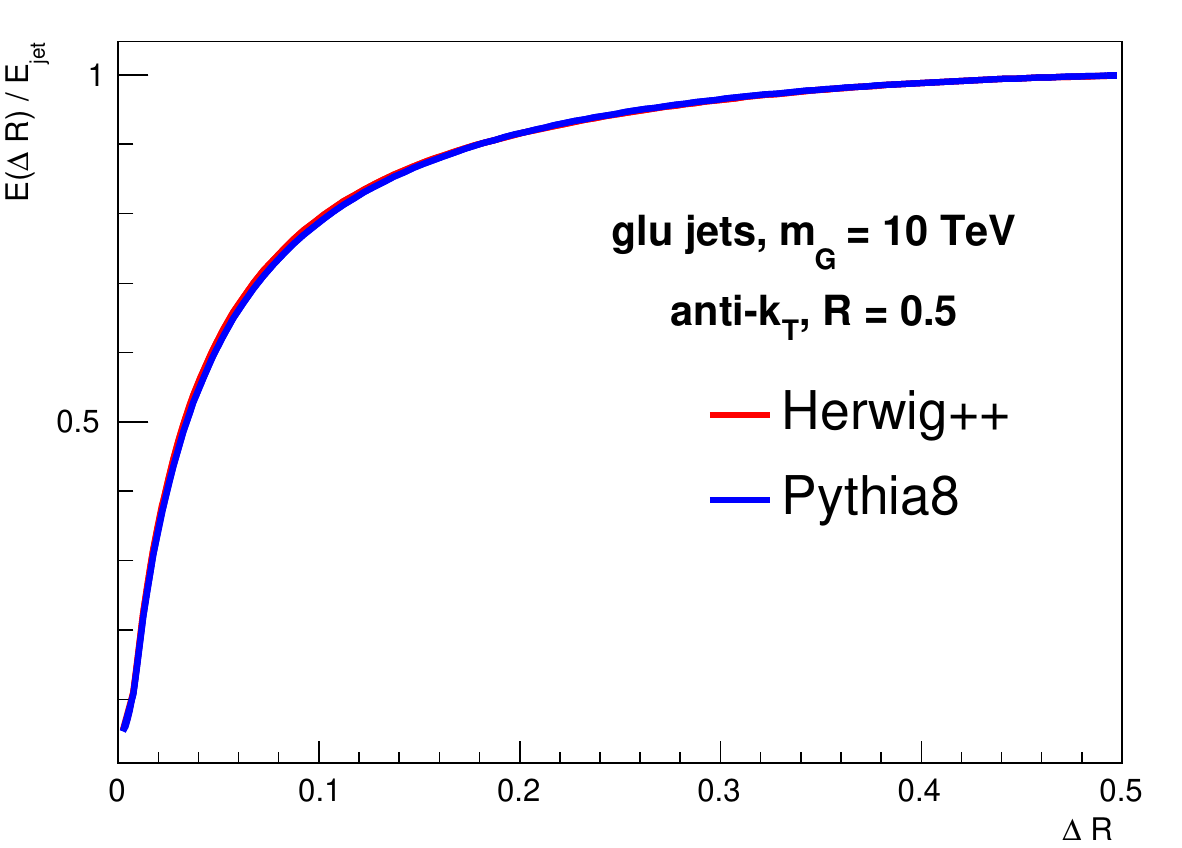}
\caption{ Average energy fraction contained within and angular scale
  $\Delta R$ of jets produced from 10 TeV resonance decays to tops,
  light QCD quarks, $W$s, and gluons.  }
\label{fig:efrac_10}
\end{figure}

\begin{figure}[h!]
\centering
\includegraphics[width=0.45\linewidth]{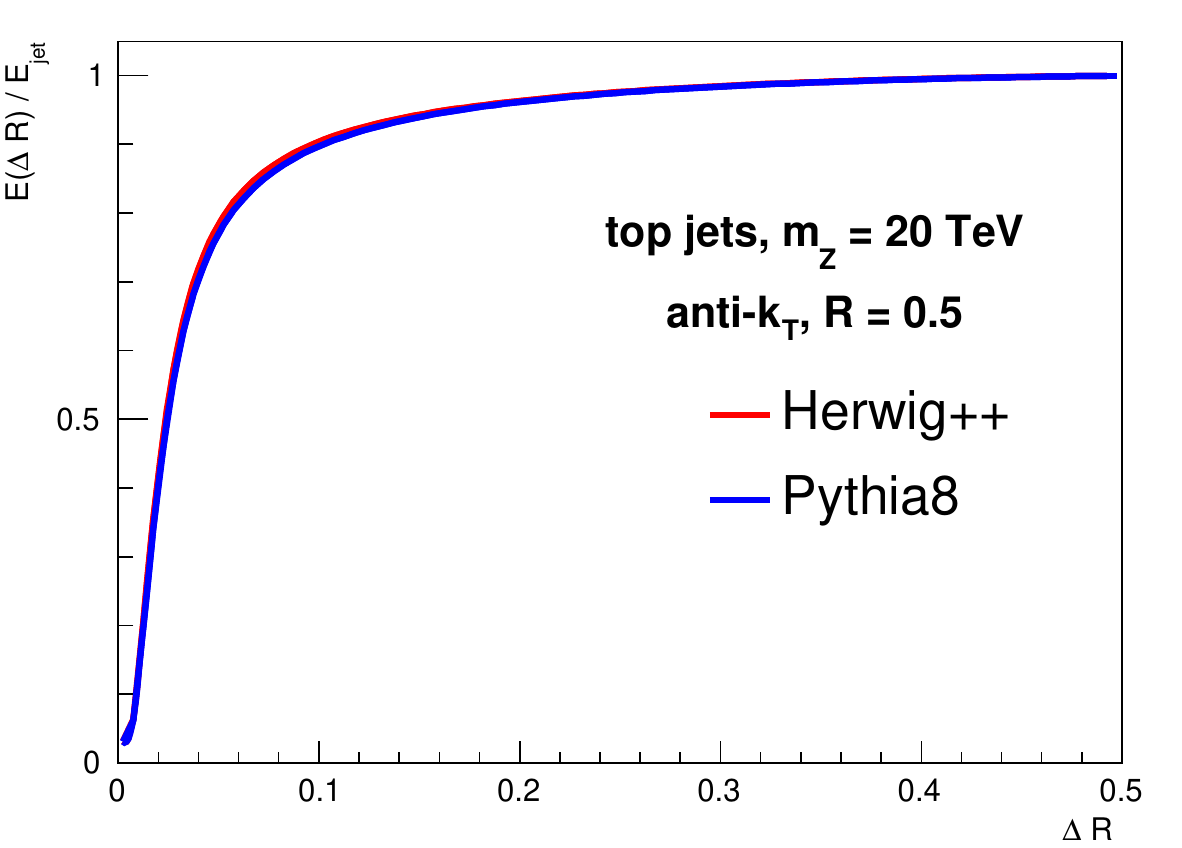}
\includegraphics[width=0.45\linewidth]{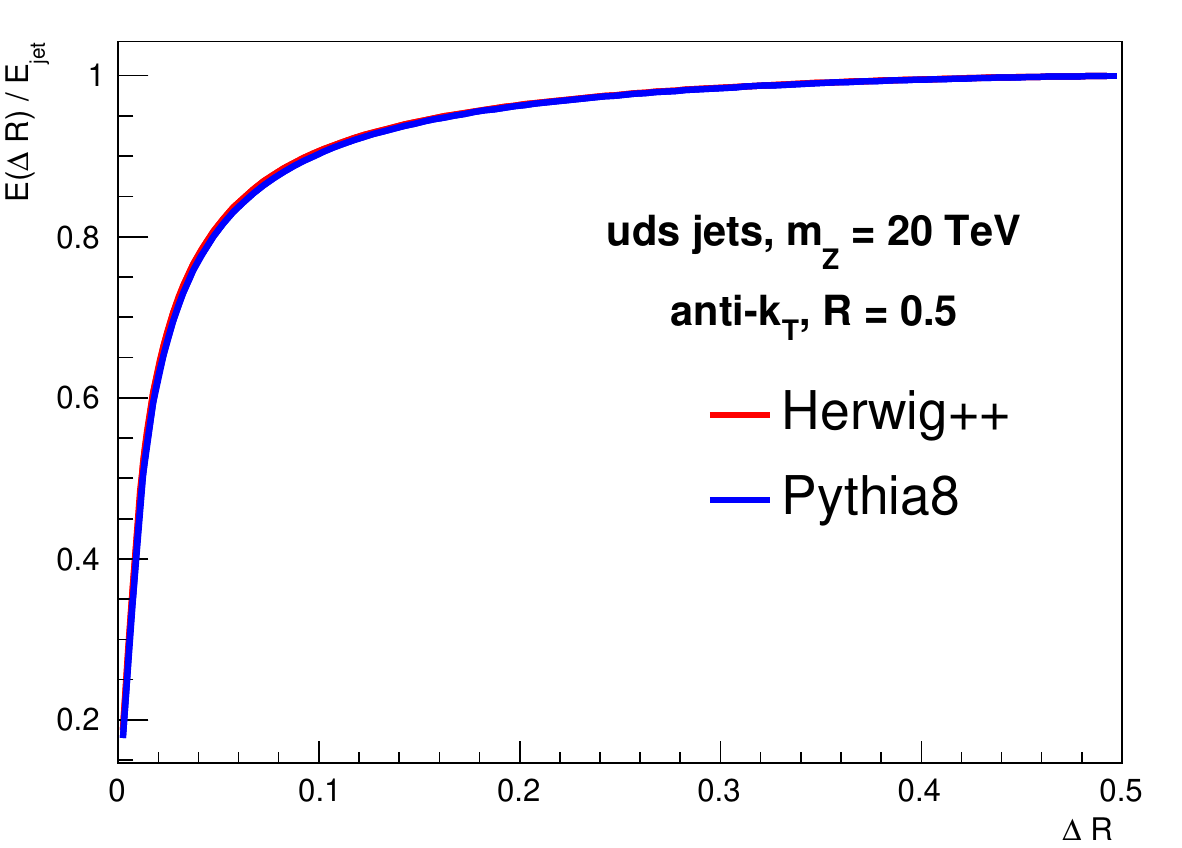}
\\
\includegraphics[width=0.45\linewidth]{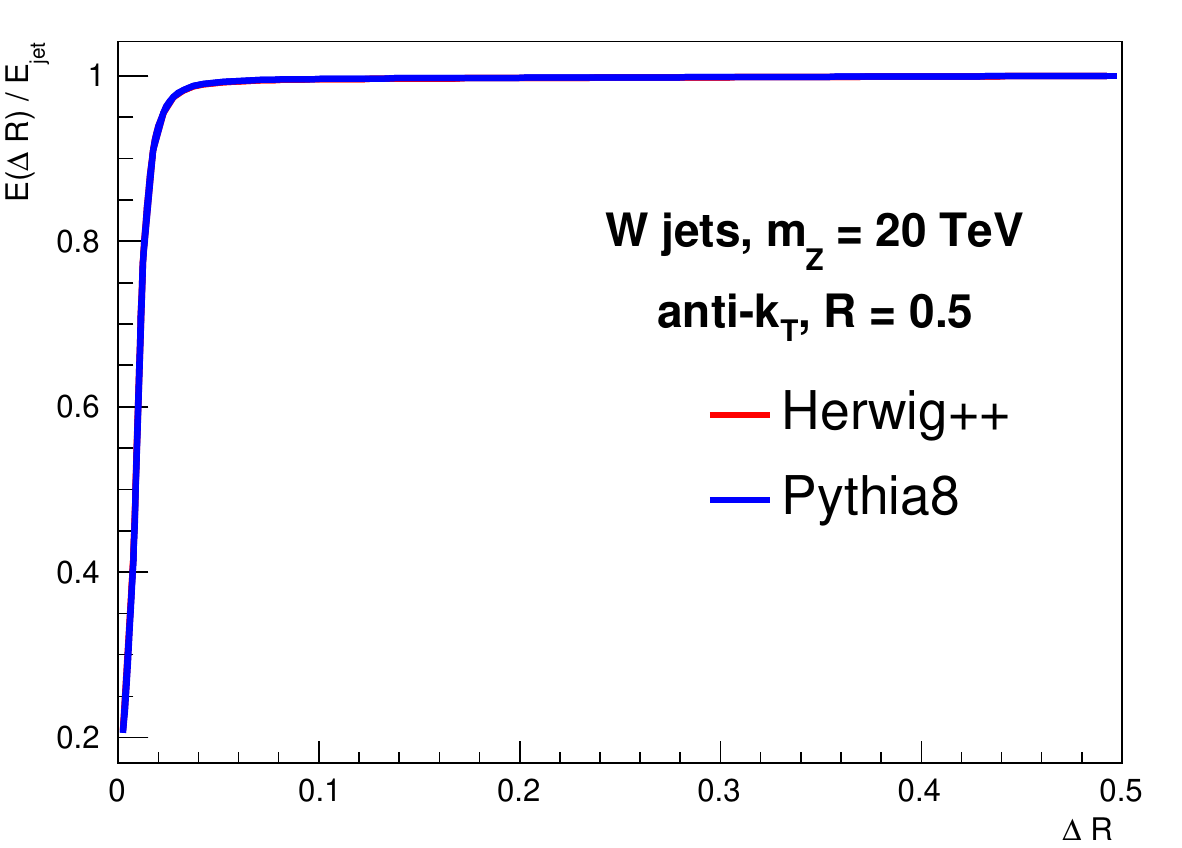}
\includegraphics[width=0.45\linewidth]{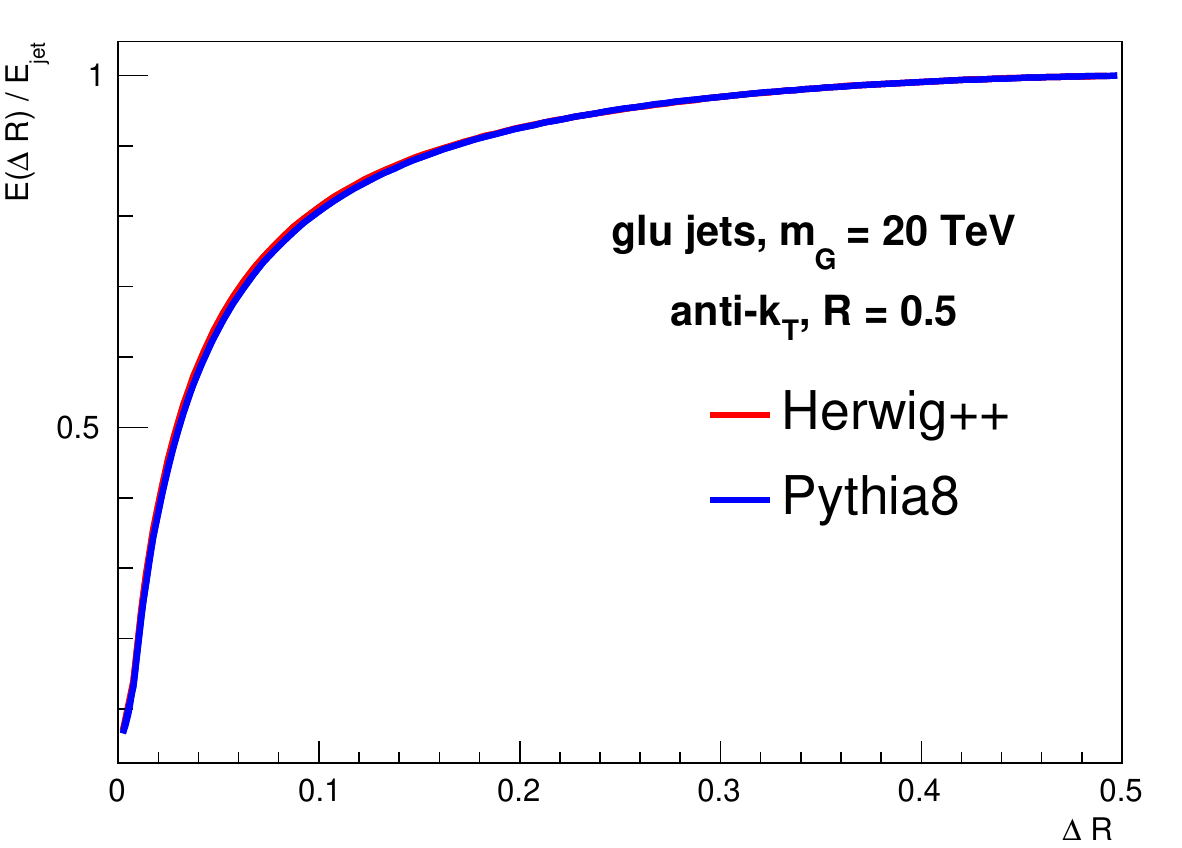}
\caption{ Average energy fraction contained within and angular scale
  $\Delta R$ of jets produced from 20 TeV resonance decays to tops,
  light QCD quarks, $W$s, and gluons.  }
\label{fig:efrac_20}
\end{figure}

\begin{figure}[h!]
\centering
\includegraphics[width=0.45\linewidth]{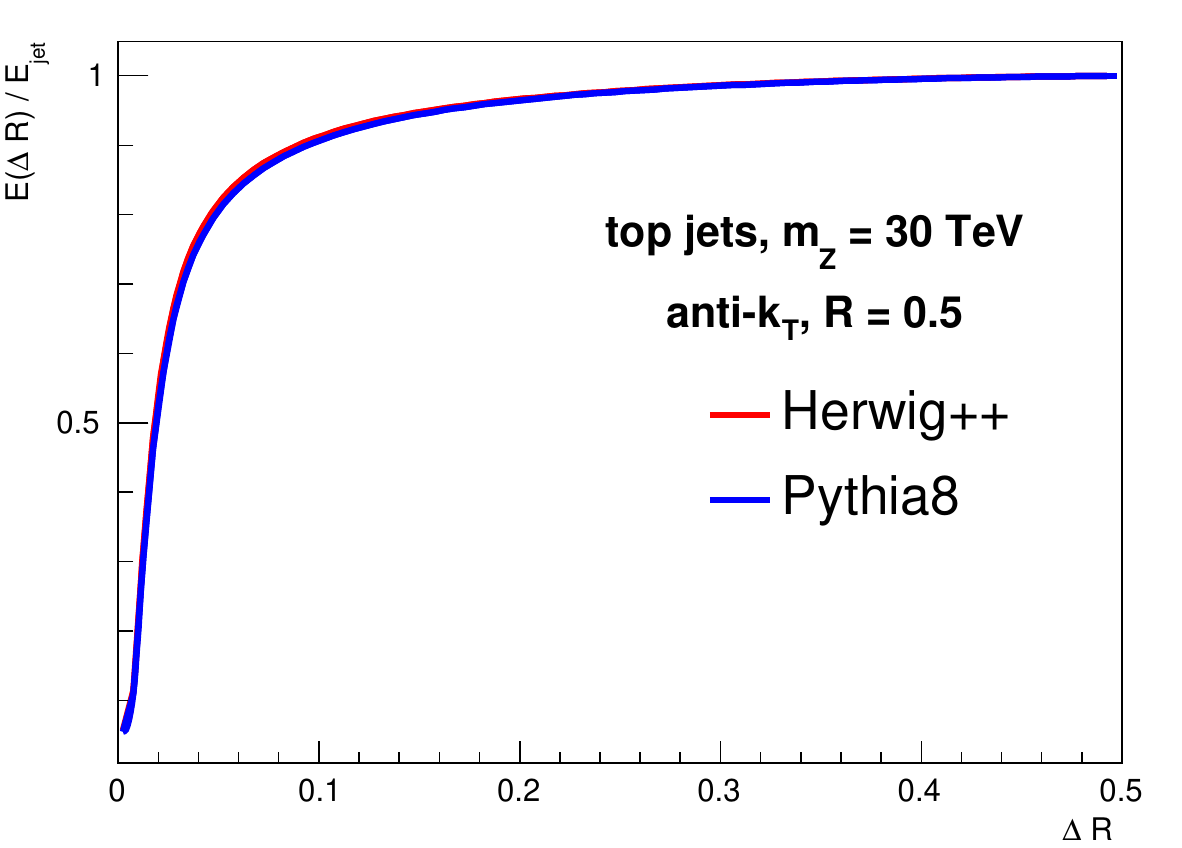}
\includegraphics[width=0.45\linewidth]{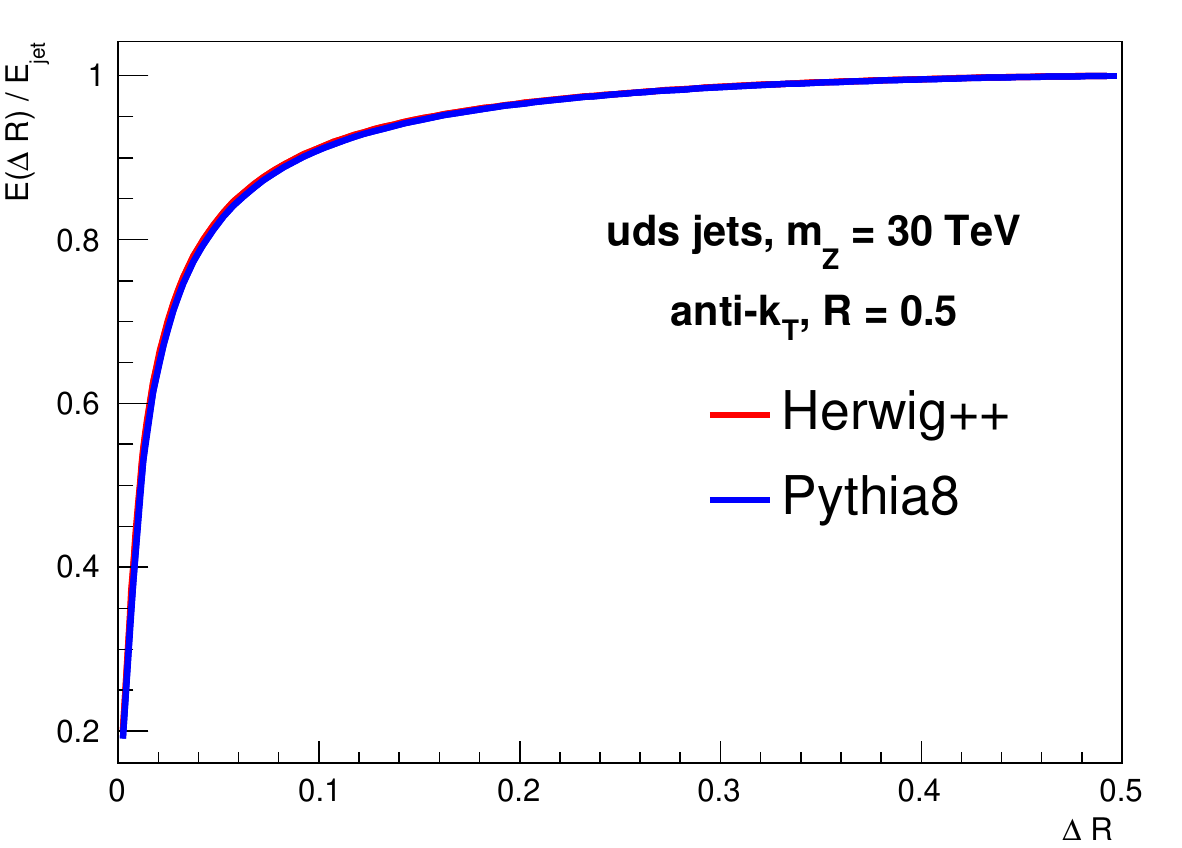}
\\
\includegraphics[width=0.45\linewidth]{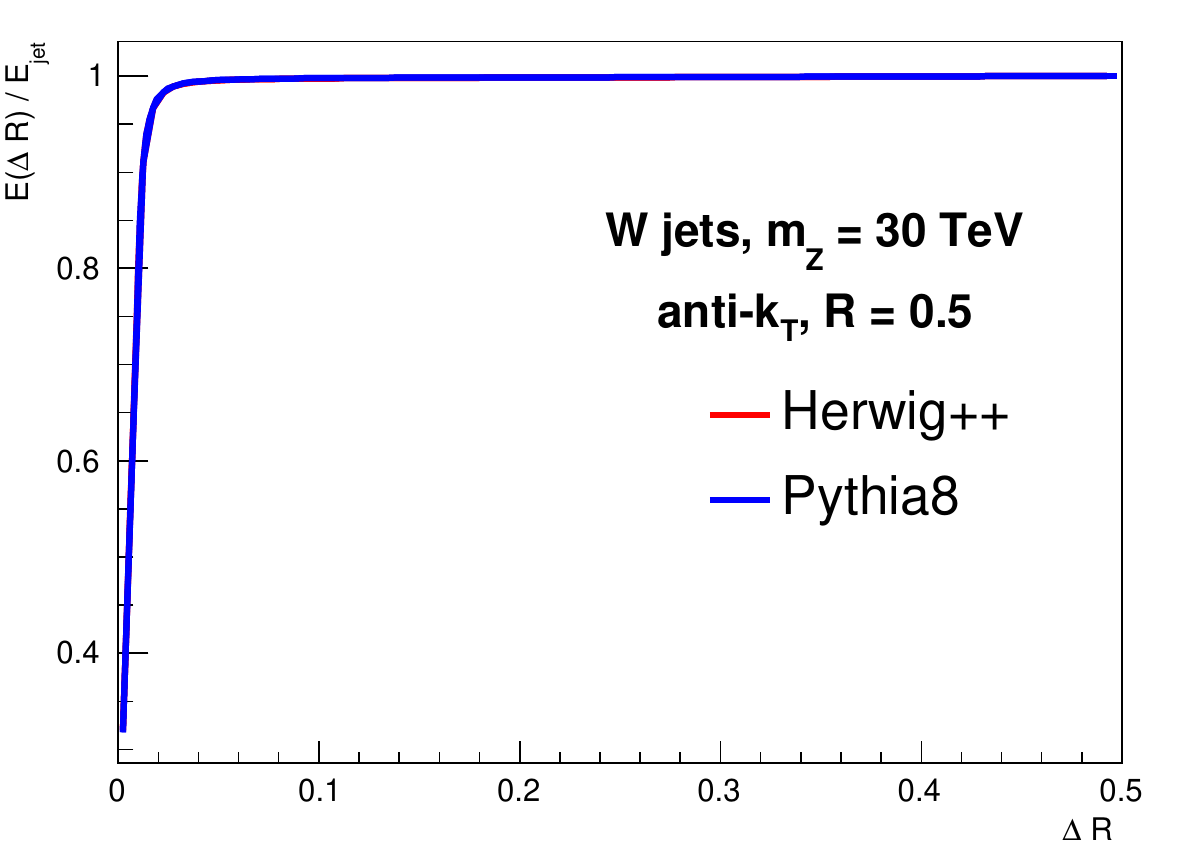}
\includegraphics[width=0.45\linewidth]{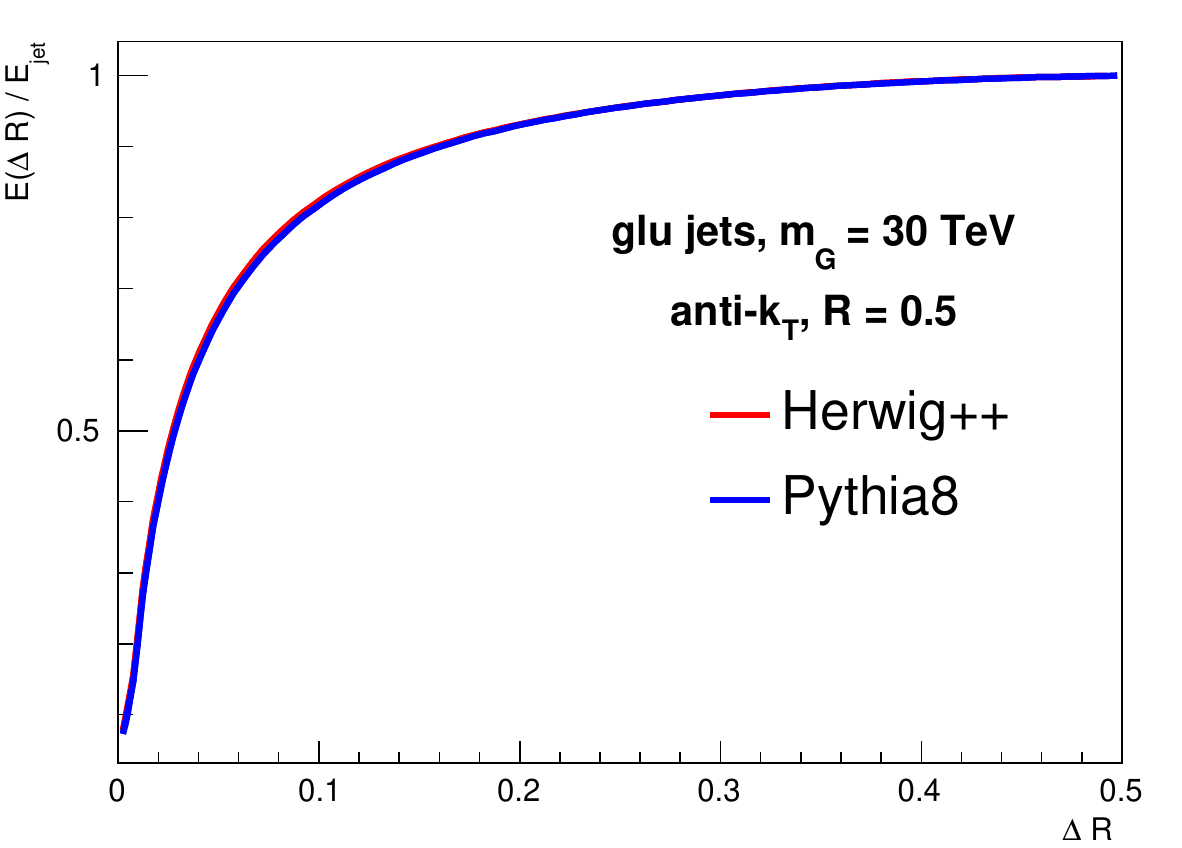}
\caption{ Average energy fraction contained within and angular scale
  $\Delta R$ of jets produced from 30 TeV resonance decays to tops,
  light QCD quarks, $W$s, and gluons.  }
\label{fig:efrac_30}
\end{figure}

\begin{figure}[h!]
\centering
\includegraphics[width=0.45\linewidth]{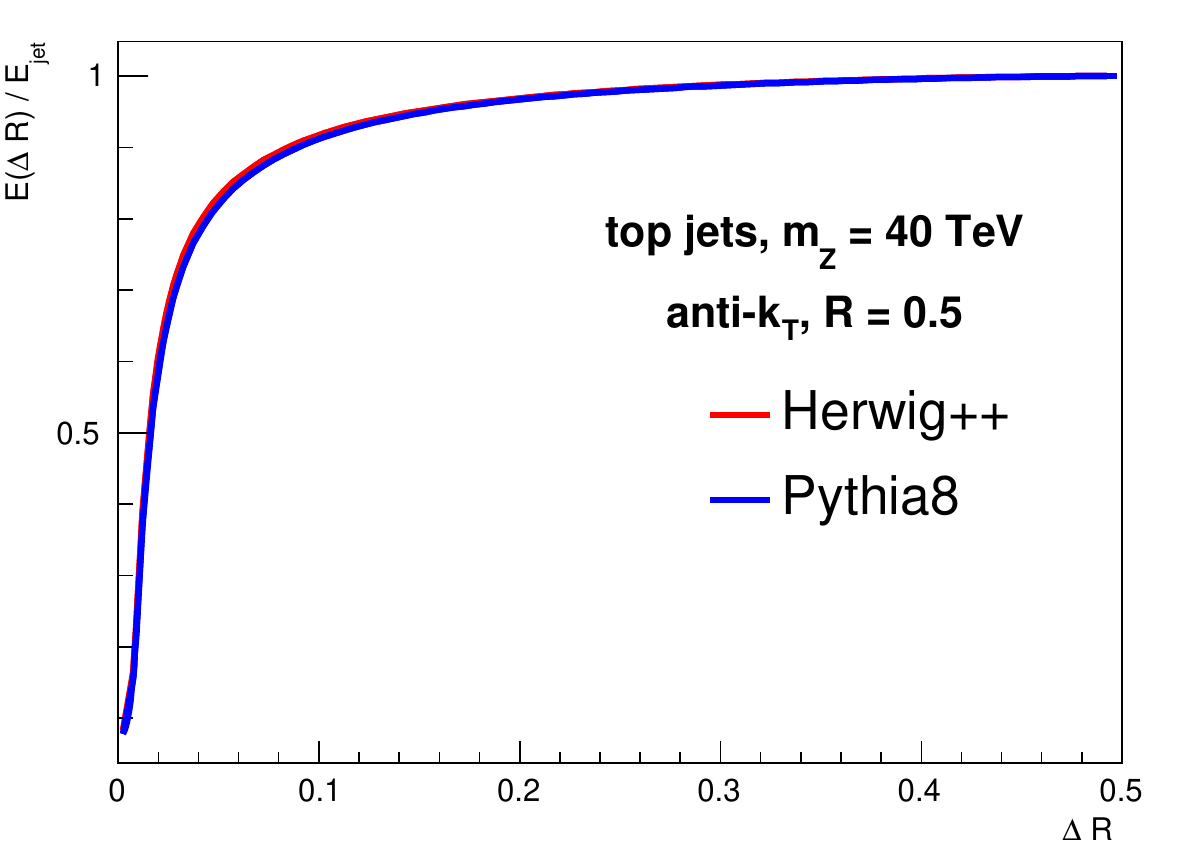}
\includegraphics[width=0.45\linewidth]{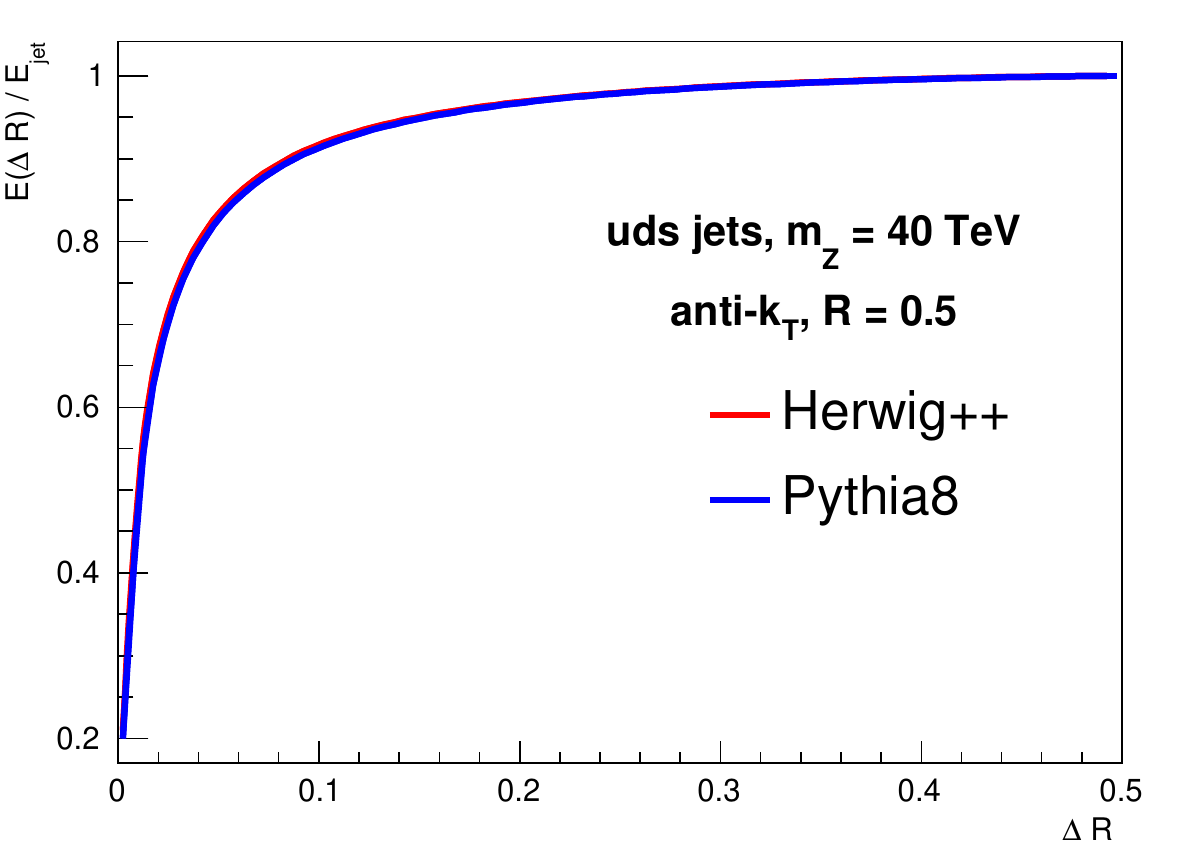}
\\
\includegraphics[width=0.45\linewidth]{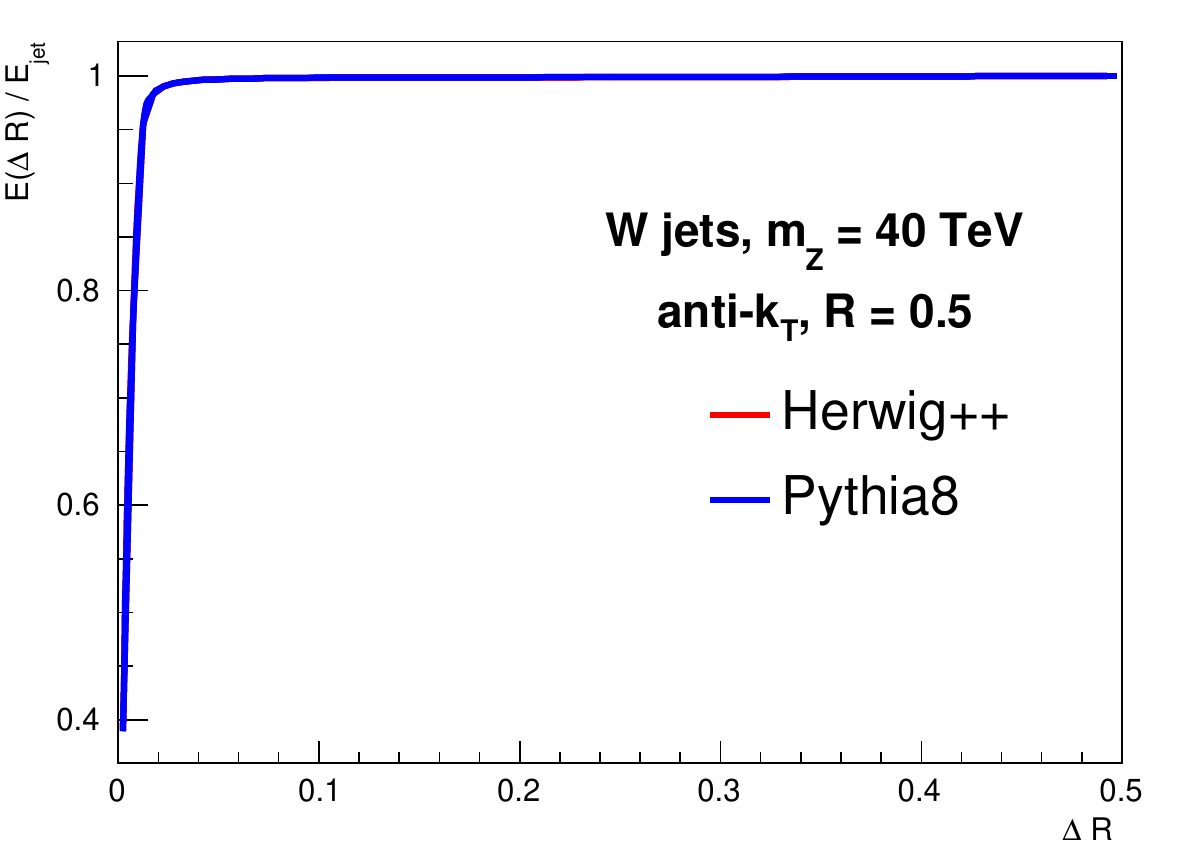}
\includegraphics[width=0.45\linewidth]{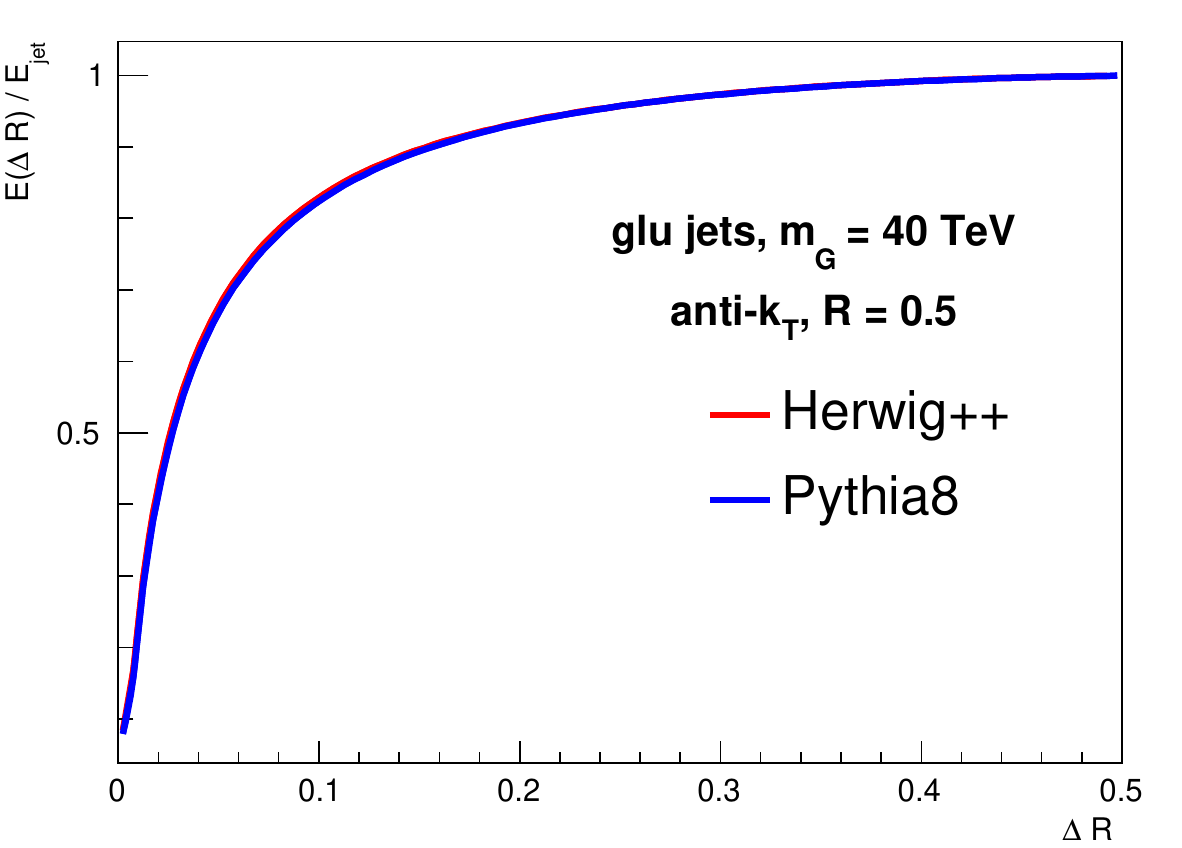}
\caption{ Average energy fraction contained within and angular scale
  $\Delta R$ of jets produced from 40 TeV resonance decays to tops,
  light QCD quarks, $W$s, and gluons.  }
\label{fig:efrac_40}
\end{figure}


\subsubsection{Top Quark Tagging at FCC}\label{sec:toptagsec}

Tagging hadronically-decaying boosted top quarks is a fundamental
problem at the LHC and will be important at the FCC also.  There has
been significant effort devoted to the development of observables and
algorithms for identification of top quarks at the LHC; see the
reviews~\cite{Abdesselam:2010pt,Plehn:2011tg,Altheimer:2012mn,Altheimer:2013yza,Adams:2015hiv}
and references therein.  In this section, we will review recent
studies of top quark identification at the FCC.

In the study of Ref.~\cite{Larkoski:2015yqa}, top quarks produced at
high boosts at the FCC were identified by measuring observables on
jets that are sensitive to the three-prong structure of the hadronic
top quark decay.  Due to the extreme hierarchy between possible $p_T$s
at the FCC and the top quark mass, there were several components of
the tagging algorithm proposed by Ref.~\cite{Larkoski:2015yqa}.  The
tagging procedure used there was the following:
\begin{enumerate}

\item Jets are first clustered using the anti-$k_T$ algorithm with a
  fixed jet radius of $R = 1.0$.  These $R = 1.0$ jets are then
  reclustered with a radius $R = 4m_\text{top}/p_T$, where $p_T$ is
  the transverse momentum of the original jet.  Only the hardest jet
  found from this reclustering is kept.  This procedure minimizes the
  effect of contamination radiation on the top quark mass measurement.

\item The invariant mass of the tracks $m_\text{tracks}$ in the
  resulting jet is measured.  To account for the bias of this mass
  measurement with respect to the total jet mass (at least on
  average), a rescaled track mass is defined as:
$$m_\text{res} = \frac{p_T}{p_T^\text{tracks}}m_\text{tracks}\,,$$
where $p_T$ is the total transverse momentum of the jet and $p_T^\text{tracks}$ is just the transverse momentum in tracks.  The rescaled jet mass was required to lie in the window $m_\text{res}\in [120,250]$ GeV around the top quark mass.

\item On these jets that passed the rescaled track mass cut, the
  substructure observables $N$-subjettiness
  \cite{Thaler:2010tr,Thaler:2011gf} and energy correlation functions
  \cite{Larkoski:2013eya} were measured exclusively on the tracks.
  Relevant for three-prong top quark jets, the $N$-subjettiness ratio
  $\tau_3/\tau_2$ and the energy correlation function observable $D_3$
  \cite{Larkoski:2014zma} were used.  Top quark signal jets take
  relatively small values for these observables while background jets
  initiated by light QCD partons have relatively large values, and so
  a cut can be applied to further discriminate boosted top quarks from
  background QCD jets.

\end{enumerate}
Depending on acceptance or purity criteria, the precise cut on the
observables $\tau_3/\tau_2$ and $D_3$ will change, so a useful way to
illustrate the discrimination power of an observable is with a signal
versus background efficiency curve, or ROC curve.  To generate the
results in this section only, we showered fixed-order events generated
with MadGraph\_aMC@NLO v2.2.2 with Pythia v6.4. Complete details of
event generation and the discrimination procedure are presented in
Ref.~\cite{Larkoski:2015yqa}.

\begin {table}
\begin{center}  
\begin{tabular}{l||c|c}
& CMS & FCC \\
  \hline
  \hline 
$B_z$ $(T)$ &  3.8 & 6.0 \\ 
  \hline
Length $(m)$ & 6 & 12 \\
 \hline
Radius $(m)$ & 1.3 &  2.6 \\
 \hline
 \hline
$\epsilon_0$ & 0.90 & 0.95\\
 \hline
 $R^*$ & 0.002 & 0.001  \\
 \hline
 \hline
$\sigma(p_T)/p_T $ &  $0.2\cdot p_T$ (TeV/c) &  $0.02\cdot p_T$ (TeV/c)\\
   \hline  
$\sigma(\eta,\phi)$ & 0.002 & 0.001 
\end{tabular}
\caption{Tracking-related parameters for the CMS and FCC setup in Delphes.}
\label{tab:trk_param}
\end{center}
\end{table}

\begin {table}
\begin{center}  
\begin{tabular}{l||c|c}
& CMS & FCC \\
  \hline
  \hline  
$\sigma(E)/E $ (ECAL)& 7\%/$\sqrt{E} \oplus 0.7\%$ &  3\%/$\sqrt{E} \oplus 0.3\%$\\
   \hline  
$\sigma(E)/E $ (HCAL)& 150\%/$\sqrt{E} \oplus 5\%$  &  50\%/$\sqrt{E} \oplus 1\%$\\
  \hline
  \hline
$\eta \times \phi$ cell size (ECAL)& $(0.02\times0.02)$ &  $(0.01\times0.01)$\\
  \hline  
 $\eta \times \phi$ cell size  (HCAL)& $(0.1\times0.1)$ & $(0.05\times0.05)$ 
\end{tabular}
\caption{Calorimeter parameters for the CMS and FCC setup in Delphes.}
\label{tab:cal_param}
\end{center}
\end{table}

To include at least a benchmark for detector effects,
Ref.~\cite{Larkoski:2015yqa} used the fast detector simulator
\textsc{Delphes} \cite{deFavereau:2013fsa}, with a hypothetical future
collider's detector modeled off of the CMS detector
\cite{Bayatian:2006zz}.  The detector simulation parameters of the
model CMS detector and FCC detector used in that study are summarized
in Tabs.~\ref{tab:trk_param} and \ref{tab:cal_param}.
Ref.~\cite{Larkoski:2015yqa} emphasized that the simulated detectors
are both quite conservative and would require a full GEANT-based
simulation~\cite{Agostinelli:2002hh} to truly accurately describe all
features of the FCC detector.

A few of the detector parameters were customized for the extreme
environment of the FCC, especially in the high density environment of
the tracking system.  The magnetic field strength $B$, the size of the
tracking radius $L$ and the single hit spatial resolution
$\sigma_{r\phi}$ are the main parameters that constrain the resolution
on the track transverse momentum:
 \begin{equation}
\frac{\sigma(p_T)}{p_{T}} \approx \frac{\sigma_{r\phi}}{B\cdot L^2}\,.
\end{equation}
The jet center has the highest density of charged particles, and so this should describe the dominant effect on the resolution. For tracks a distance $R$ from the jet center, we define the track resolution efficiency
\begin{equation}
\epsilon(R) = \frac{2\epsilon_0}{\pi} \arctan\left(\frac{R}{R^*}\right) \,.
\end{equation}
$R^*$ is a parameter that controls the angular resolution of the tracker, where we set $R^*=0.001$ for simulated FCC detector and $R^*=0.002$ for modeling the CMS detector.

\begin{figure}
\begin{center}
\includegraphics[width=0.45\linewidth]{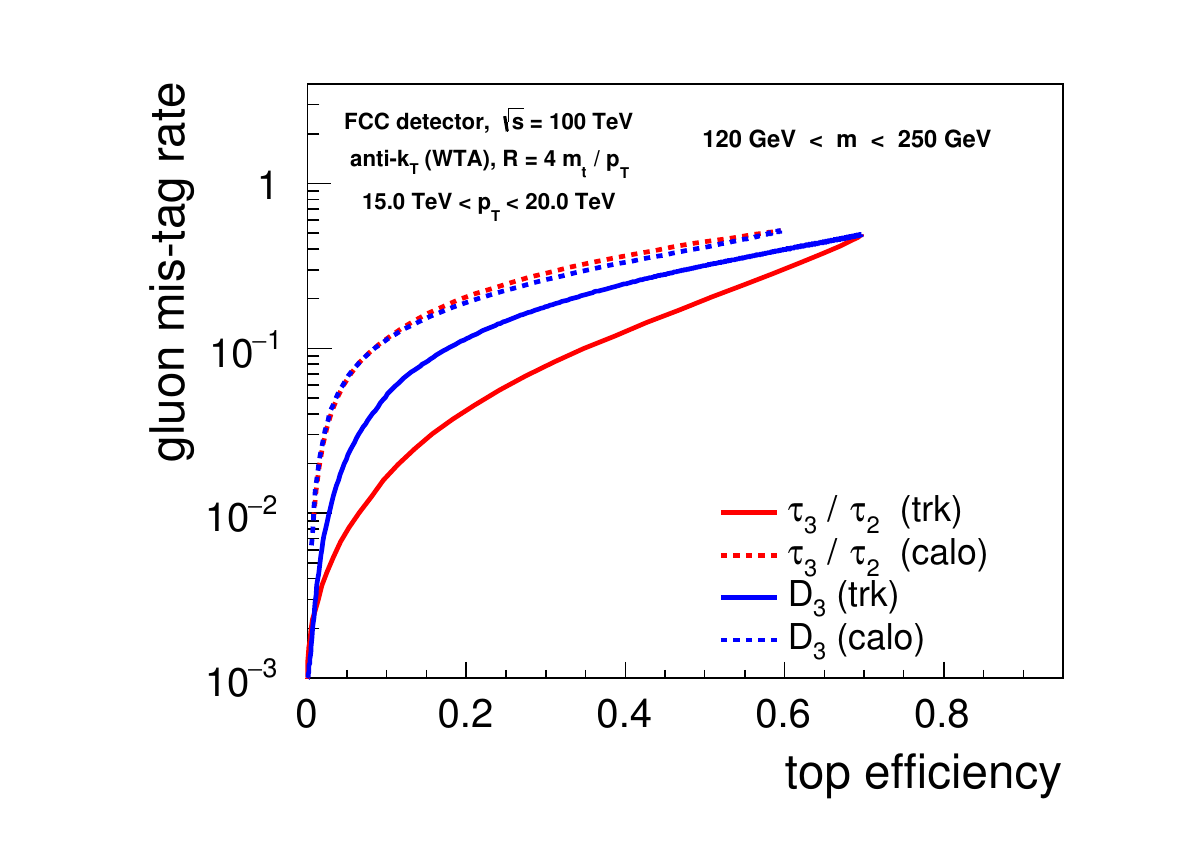}
\includegraphics[width=0.45\linewidth]{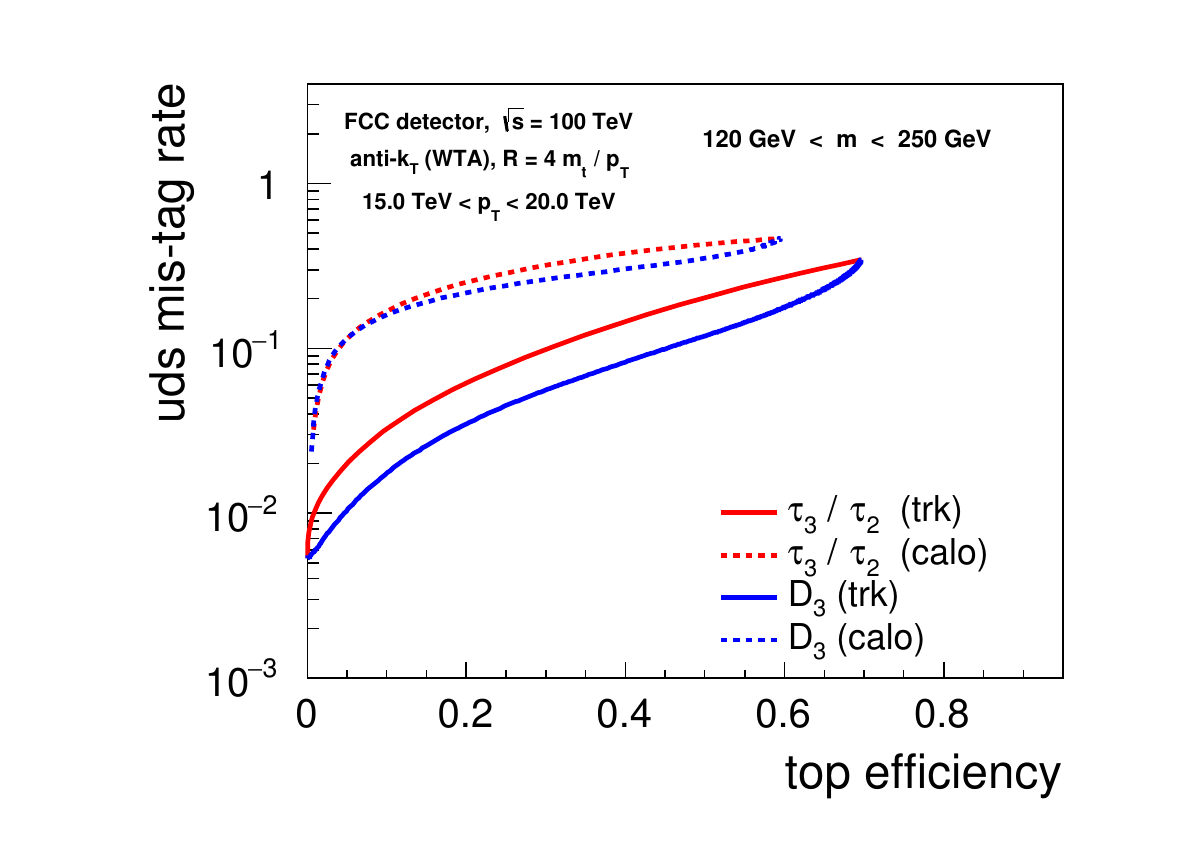}
\end{center}
\caption{ Signal vs.~background efficiency (ROC) curves for top quark
  identification from QCD background utilising $\tau_{3,2}$ and $D_3$
  with the FCC detector for $p_T \in[15,20]$ TeV. (left) top quarks
  vs.~gluon jets, (right) top quarks vs.~light quark jets.  The cut on
  the jet mass of $m\in[120,250]$ GeV is included in the efficiencies.
  Events were showered with Pythia v6.4.  }
\label{fig:roc_curves_FCC}
\end{figure}

Representative ROC curves are shown in Fig.~\ref{fig:roc_curves_FCC}
for discrimination of boosted top quarks from jets initiated by light
QCD partons at the FCC for jets with $p_T \in[15,20]$ TeV.  The quark
and gluon jet backgrounds have been separated and the ROC curves for
track- or calorimeter-based measurements are compared.  The effect of
the cut on the rescaled track jet mass is included in efficiencies.
Tab.~\ref{tab:rej_pt} lists background rejection rates in several jet
$p_T$ bins at fixed signal efficiencies of 20\%, 40\%, and 60\%.  The
performance of the simulated CMS and FCC detectors are also compared.

Note from Tab.~\ref{tab:rej_pt} that as the mass of the resonance
increases (corresponding to increasing jet $p_T$) the power to reject
light QCD jets decreases, at fixed top quark efficiency.  To have the
same top quark efficiency at multiple jet $p_T$s requires changing
observable cuts.  As the jet $p_T$ increases, one becomes more
sensitive to the finite angular resolution of the detector, which will
reduce the power to cleanly identify the three hard prongs of the
boosted top quark.  Therefore, as $p_T$ increases, the observable cuts
must become looser, which in turn means that more background quark or
gluon jets will also be included.

\begin{table}[t]
\begin{center}
\begin{tabular}{cc|ccccc}
\multicolumn{7}{c}{{\bf 20\% Top Efficiency}}\\
 &$p_T$ cut&$[2.5,5] $ TeV & $[5,7.5] $ TeV & $[7.5,10] $ TeV& $ [10,15] $ TeV&$ [15,20] $ TeV\\
\hline
{\multirow{2}{*}{gluons} } &CMS& 2\% & 3\%& 4\%& 5\%& 6\%\\
&FCC& 1\% & 2\%& 2\%& 3\%& 4\%\\
\cline{2-7}
{\multirow{2}{*}{quarks} } &CMS& 1\% & 2\%& 3\%& 5\%& 7\%\\
&FCC& 0.5\% & 1\%& 1.5\%& 2\%& 4\%\\
\end{tabular}
\begin{tabular}{cc|ccccc}
\multicolumn{7}{c}{{\bf 40\% Top Efficiency}}\\
 &$p_T$ cut&$[2.5,5] $ TeV & $[5,7.5] $ TeV & $[7.5,10] $ TeV& $ [10,15] $ TeV&$ [15,20] $ TeV\\
\hline
{\multirow{2}{*}{gluons} } &CMS& 7\% & 9\%& 10\%& 14\%& 17\%\\
&FCC& 5\% & 6\%& 7\%& 10\%& 12\%\\
\cline{2-7}
{\multirow{2}{*}{quarks} } &CMS& 3\% & 5\%& 7\%& 11\%& 17\%\\
&FCC& 1.5\% & 2.5\%& 4\%& 5\%& 8\%\\
\end{tabular}
\begin{tabular}{cc|ccccc}
\multicolumn{7}{c}{{\bf 60\% Top Efficiency}}\\
 &$p_T$ cut&$[2.5,5] $ TeV & $[5,7.5] $ TeV & $[7.5,10] $ TeV& $ [10,15] $ TeV&$ [15,20] $ TeV\\
\hline
{\multirow{2}{*}{gluons} } &CMS& 18\%  & 20\%& 24\%& 30\%& 38\%\\
&FCC& 13\% & 15\%& 20\%& 24\%& 25\%\\
\cline{2-7}
{\multirow{2}{*}{quarks} } &CMS& 7\% & 10\%& 15\%& 22\%& 30\%\\
&FCC& 4\% & 6\%& 8\%& 11\%& 15\%\\
\end{tabular}
\end{center}
\caption{  Table of background rejection rates at fixed signal efficiencies for jet $p_T$s ranging from $2.5$ TeV to $20$ TeV at the CMS or FCC detector.  For gluon (quark) jet backgrounds, efficiencies are determined from cuts on $\tau_{3,2}$ ($D_3$) measured on tracks. The cut on the rescaled track-based jet mass of $m_J\in[120,250]$ GeV is included in the efficiencies.  These results are from events showered with Pythia v6.4.
}
\label{tab:rej_pt}
\end{table}
\clearpage
\subsection{Boosted boson tagging}
\label{sec:jet_boost_MP}

A boson of mass $M$ decaying hadronically produces two quarks with
angular separation $\Delta R \approx 2 M /p_T$. At large momenta, the
separation becomes smaller than the jet size. Such a boson would be
seen in a detector as a single massive jet.

The identification of jets as hadronically decaying bosons opened new
perspectives at the LHC. The development of an effective tagging
algorithm for boosted vector bosons~\cite{Aad:2013gja,
  Khachatryan:2014vla} allowed to retain a good sensitivity to
resonances decaying to two bosons and heavier than $\approx
1$~TeV~\cite{Aad:2015owa,Aad:2014xka,Aad:2015ufa,Khachatryan:2015bma,Khachatryan:2015ywa,Khachatryan:2014hpa,Khachatryan:2014gha}.

The reconstruction of heavy jets needs a new detector design. A good
reconstruction of the boson mass requires both excellent energy and
angular resolution, since the jet mass depends on both the momenta of
the jet constituents and the angular separation among them. One can
then study the jet mass as a benchmark for calorimeter granularity.

As a reference, we take the case of Randall Sundrum (RS) graviton
$\rm{G_{RS}}$ decaying to two $Z$ bosons and study the reconstructed
mass resolution for different detector geometries.  Signal events are
generated with {\tt PYTHIA8}~\cite{Sjostrand:2007gs,Sjostrand:2014zea}
at a center-of-mass energy $\sqrt{s}=100$~TeV, for different values of
the $\rm{G_{RS}}$ mass. The jets reconstructed in these events are
compared to ordinary QCD jets, generated in $\rm{G_{RS}}\to q \bar q$
and $\rm{G_{RS}} \to gg$ decays. These samples have the same
kinematic features (e.g., $p_T$ and $\eta$ distributions) as the
corresponding jets from $Z$ bosons, as long as the $p_T$ is much
larger than the $Z$ mass. Any difference observed in this study can then be
interpreted as related to the nature of the jet ($Z$ vs quarks and gluons).

Events are reconstructed with {\tt
  DELPHES3}~\cite{Ovyn:2009tx,deFavereau:2013fsa}, using the default
detector performances for the FCC detector, provided with the software
distribution.

Three detector scenarios are defined: (i) the baseline
detector geometry with calorimeter cells of size $\phi \times \eta =
0.5^o \times 0.01$ for ECAL and $2.5^o \times 0.05$ for HCAL. (ii)
twice the cell size both for ECAL and HCAL, keeping the same ECAL/HCAL
cell-size ratio. (iii) half the cell size for ECAL and HCAL, keeping the
same ECAL/HCAL size ratio.

Jets are clustered using the {\tt FASTJET}~\cite{Cacciari:2011ma}
implementation of the anti-$K_T$ algorithm~\cite{Cacciari:2008gp} with
jet-size parameter $R=0.25$, giving as input to the jet algorithm the
list of four-momenta for the particles reconstructed with the {\tt
  DELPHES} implementation of the particle-flow algorithm. The
performances of the tracking detector are fixed to the default
parametrization. Any difference observed is then genuinely related
to the change in the calorimeter geometry.
 
\begin{figure}[htbp]
\centering 
\includegraphics[width=.49\textwidth]{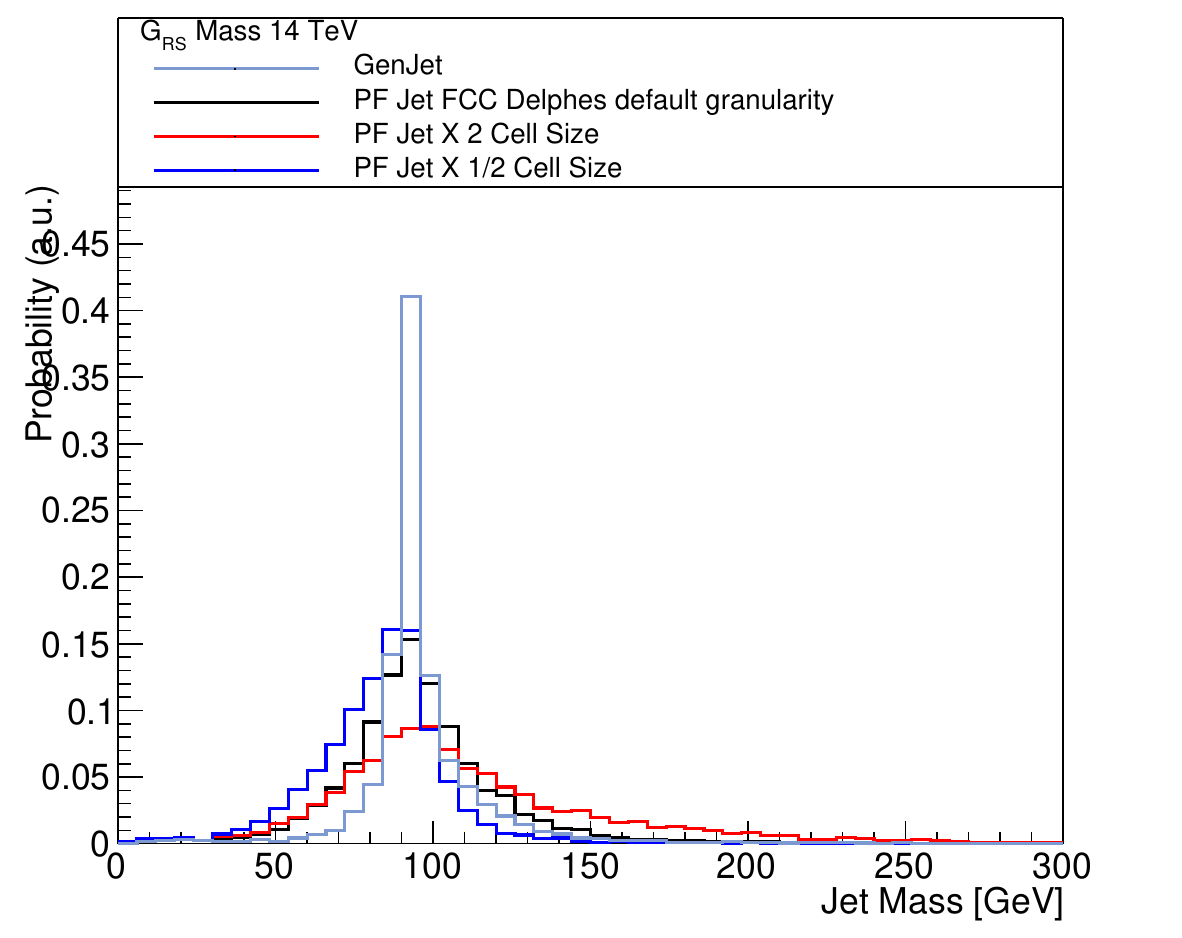}
\includegraphics[width=.49\textwidth]{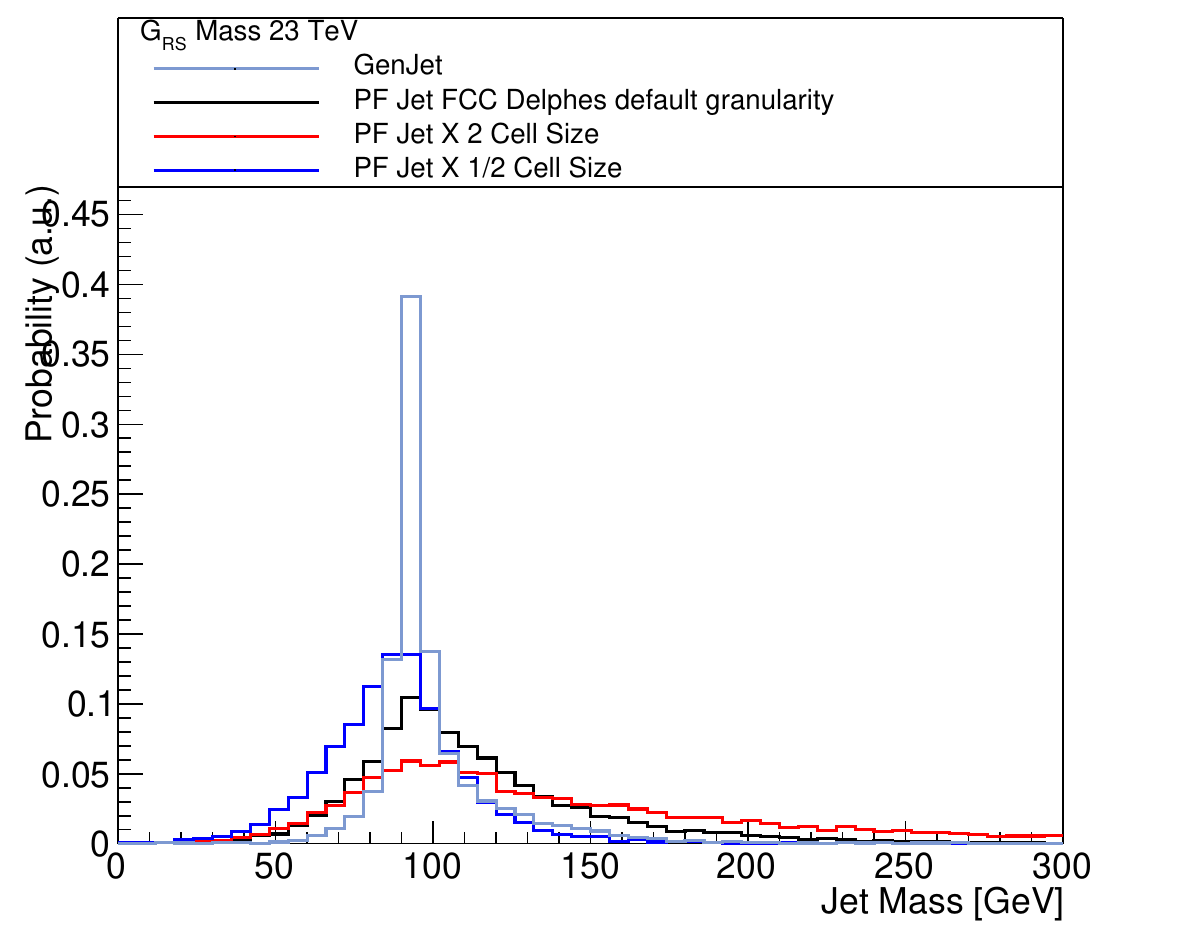} \\
\includegraphics[width=.49\textwidth]{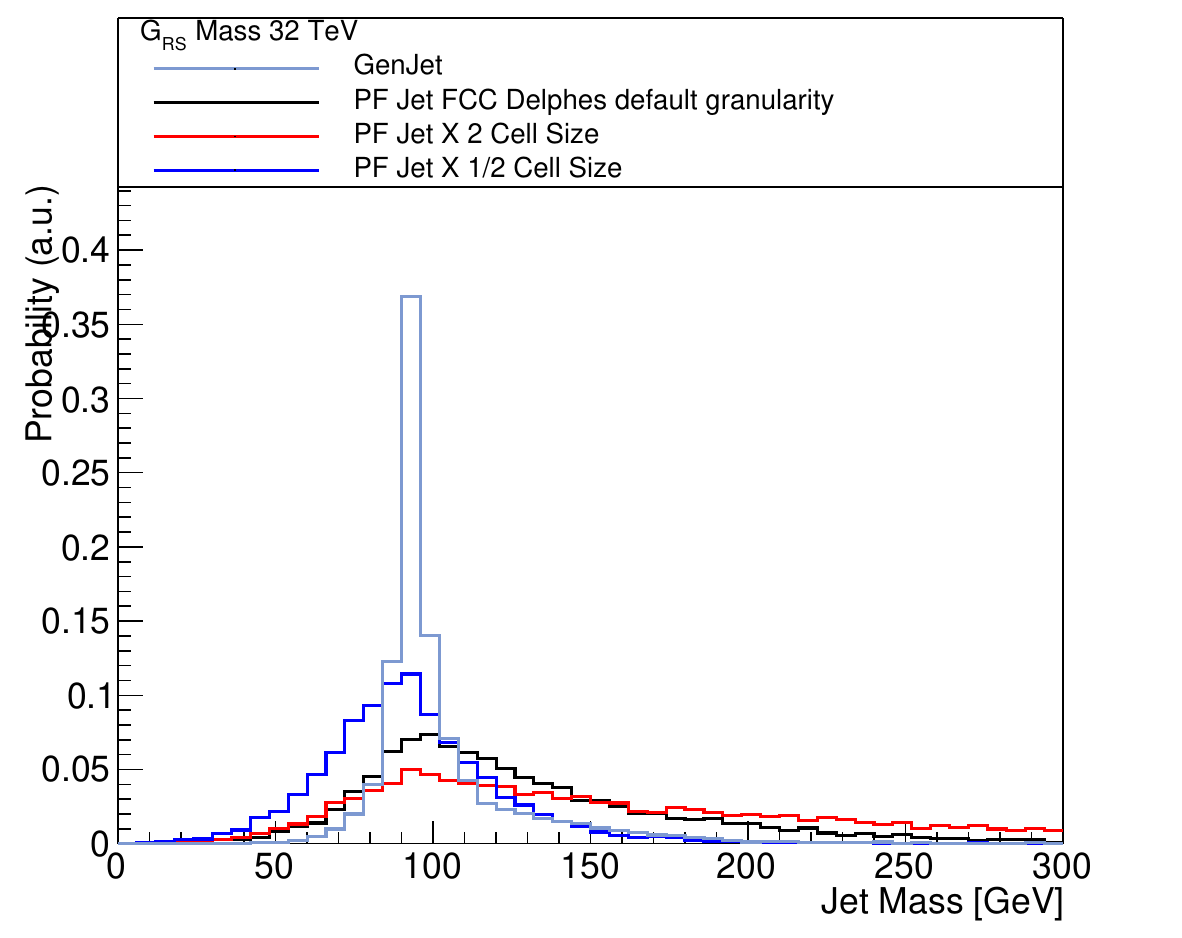}
\includegraphics[width=.49\textwidth]{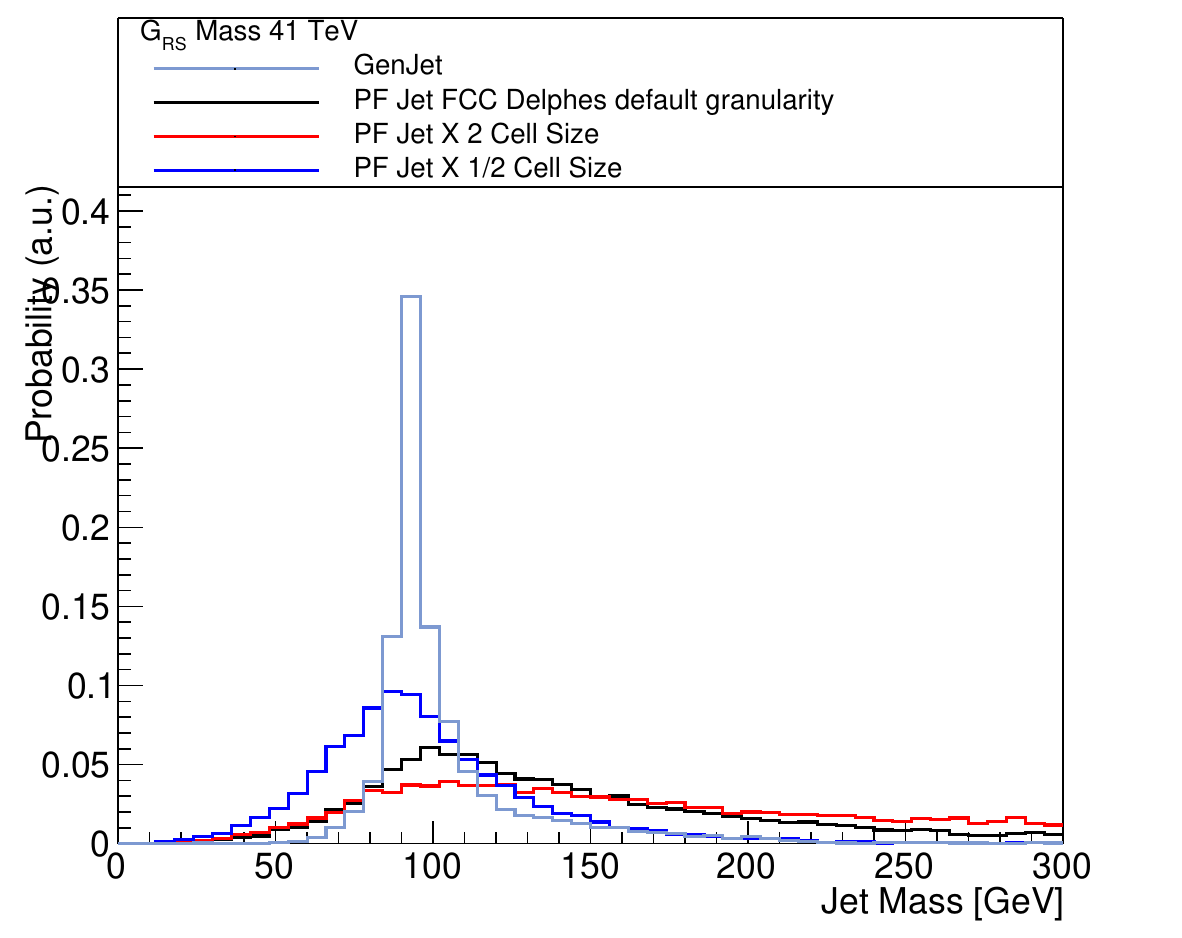} 
\caption{Jet mass distribution for $\rm{G_{RS}}$
  produced in pp collisions at $\sqrt{s}=100$~TeV and decaying to two
  $Z$ bosons. The $\rm{G_{RS}}$ mass is fixed to 14~TeV (top left),
  23~TeV (top right), 32~TeV (bottom left), and 41~TeV (bottom
  right). Different granularities are considered for the calorimeter
  cells. As a reference, the mass distribution for generator-level
  jets is also shown.\label{fig:jetMassRS}}
\end{figure}

Figure~\ref{fig:jetMassRS} shows the jet mass distribution for
different values of the $\rm{G_{RS}}$ mass, from 14 to 41 TeV. As a
comparison, the corresponding distribution obtained clustering
generated particles into jets (gen-jets) is shown, representing the
ideal case of a perfect detector
resolution. Table~\ref{tab:JetMassResol} summarises the resolution
corresponding to each granularity scenario. The resolution is
quantified with the $\sigma$ parameter of a Gaussian fit to the
distribution for mass values between 40 GeV and 140 GeV, scaled to the
mean value of the Gaussian. Besides the worsening of the resolution
with the coarser resolution, one should notice the increasing bias in
the peak position and the larger non-Gaussian tails

\begin {table}[htbp]
\begin{center}  
\begin{tabular}{l||c|c|c|c}
\hline
$\rm{G_{RS}}$ mass & gen-jets & baseline resolution & $\times 1/2$ granularity & $\times 2$ granularity \\
\hline
14 TeV & 5\% & 16\%  & 16\%  & 28\% \\
23 TeV & 6\%  & 22\%  & 22\%  & 29\% \\
32 TeV & 5\% & 24\%  & 25\%  & 33\% \\
41 TeV & 4\% & 28\%  & 26\%  & 36\% \\
\hline 
\end{tabular}
\caption{Relative resolution of the jet mass peak for $\rm{G_{RS}} \to
  ZZ$ events produced in pp collisions at $\sqrt{s}=100$~TeV. The
  resolution is quantified as the ratio between the $\sigma$ and $m$
  parameters of a Gaussian fit, in the jet mass range $[40,140]$
  GeV.\label{tab:JetMassResol}}
\end{center}
\end{table}

In Fig.~\ref{fig:MassGRS}, the mass distribution for the dijet system
is shown for the same values of $\rm{G_{RS}}$ mass. The events are
selected requiring $80<m_J \times m_Z/mode(m_J) <100$ GeV for each
jet.  The dijet-mass reconstruction exhibits poor scale and
resolution, induced by the small jet-size parameter $R$. In a
realistic search, this effect could be cured using a wide cone for
kinematic reconstruction and a narrow cone for jet tagging, similarly
to what is currently done in some LHC search. In
Fig.~\ref{fig:MassGRS}, the mass scale is partially compensated
applying a $m_Z/mode(m_J)$ rescale factor.

 \begin{figure}[htbp]
\centering 
\includegraphics[width=.49\textwidth]{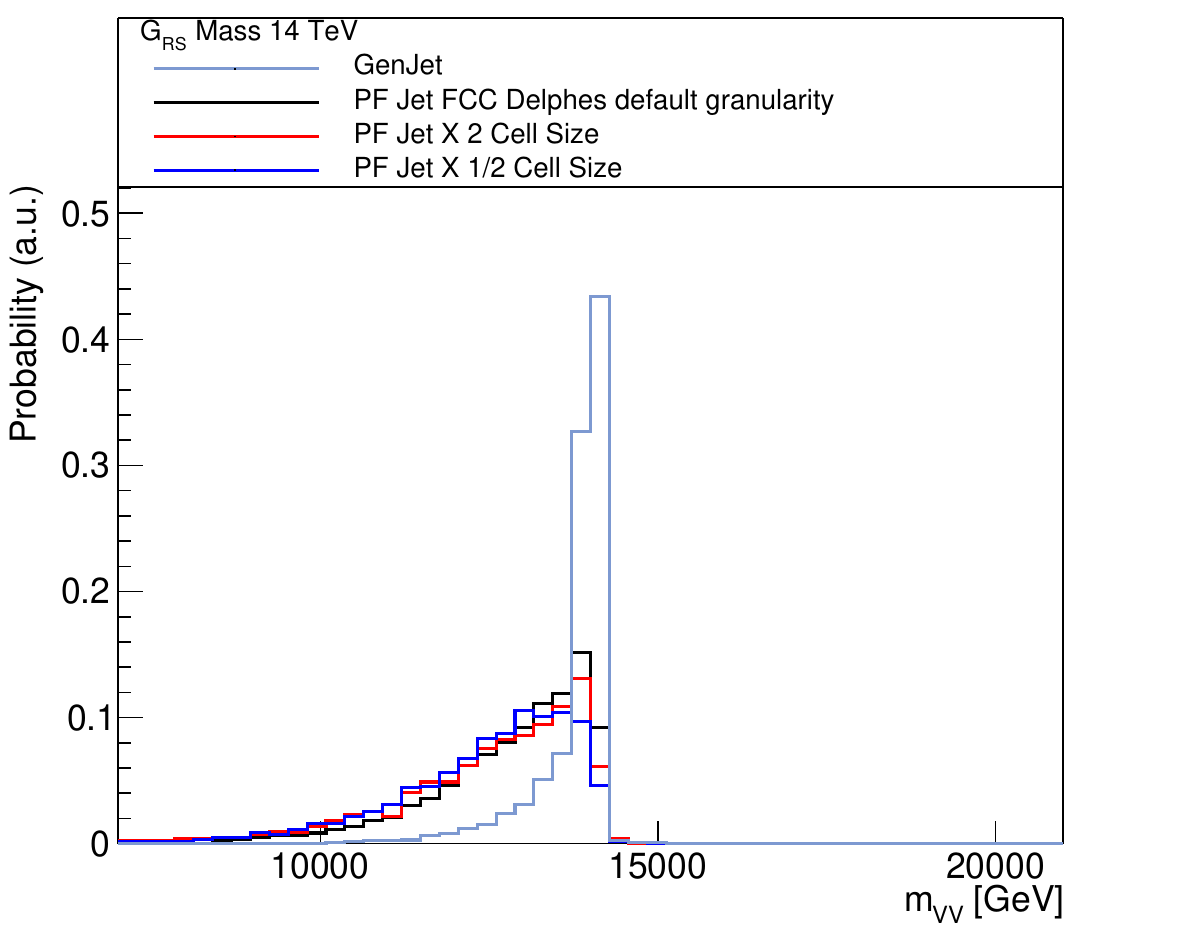}
\includegraphics[width=.49\textwidth]{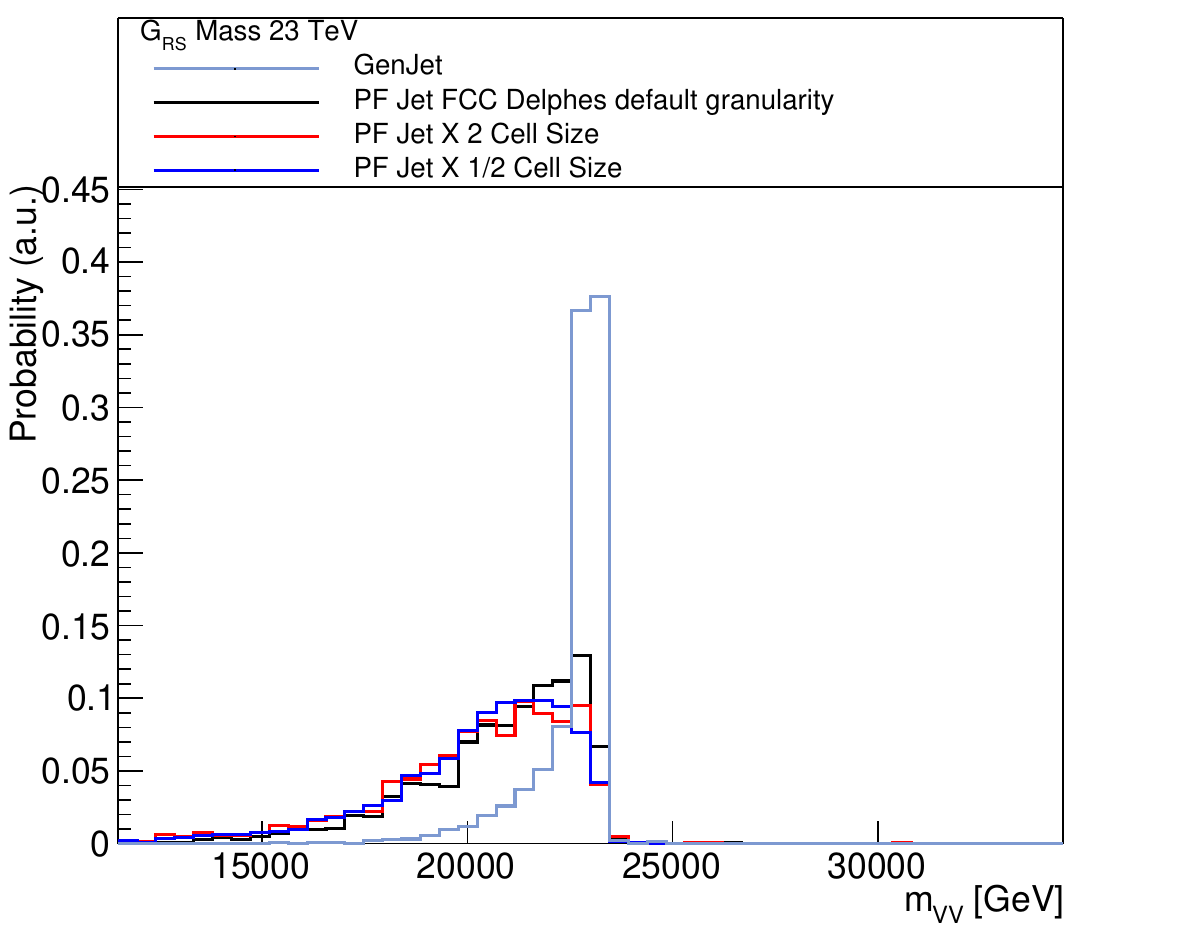}\\
\includegraphics[width=.49\textwidth]{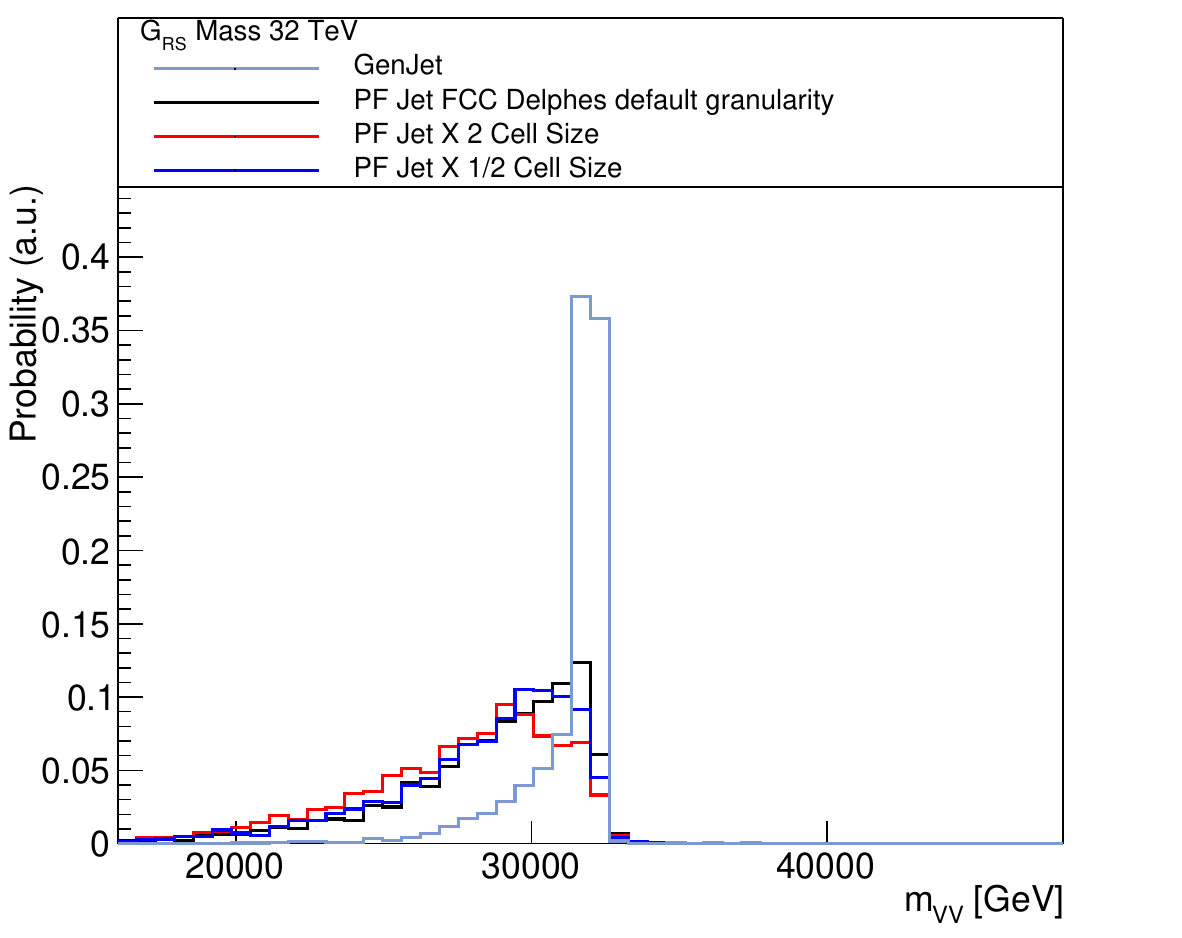}
\includegraphics[width=.49\textwidth]{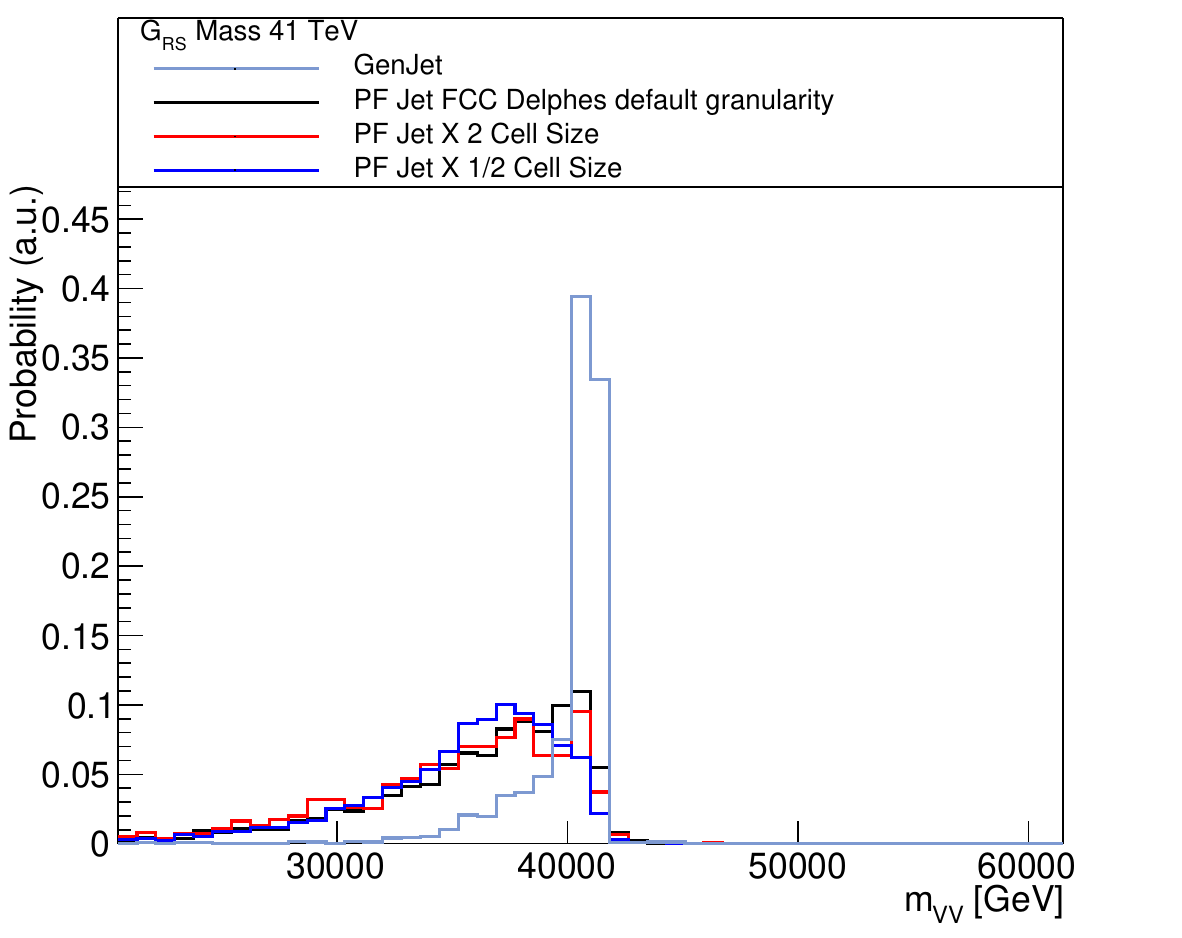}
\caption{\label{fig:MassGRS} Mass distribution for $\rm{G_{RS}}$
  produced in pp collisions at $\sqrt{s}=100$~TeV and decaying to two
  $Z$ bosons. The $\rm{G_{RS}}$ mass is fixed to 14~TeV (top left),
  23~TeV (top right), 32~TeV (bottom left), and 41~TeV (bottom
  right). Different granularities are considered for the calorimeter
  cells. As a reference, the mass distribution for generator-level
  jets is also shown.}
\end{figure}

Besides the jet mass, the identification of boosted bosons usually exploits
the so-called jet substructure, i.e. the study of the angular and
momentum distribution of the jet constituents in a massive jet. 

\begin{figure}[htbp]
\centering 
\includegraphics[width=.49\textwidth]{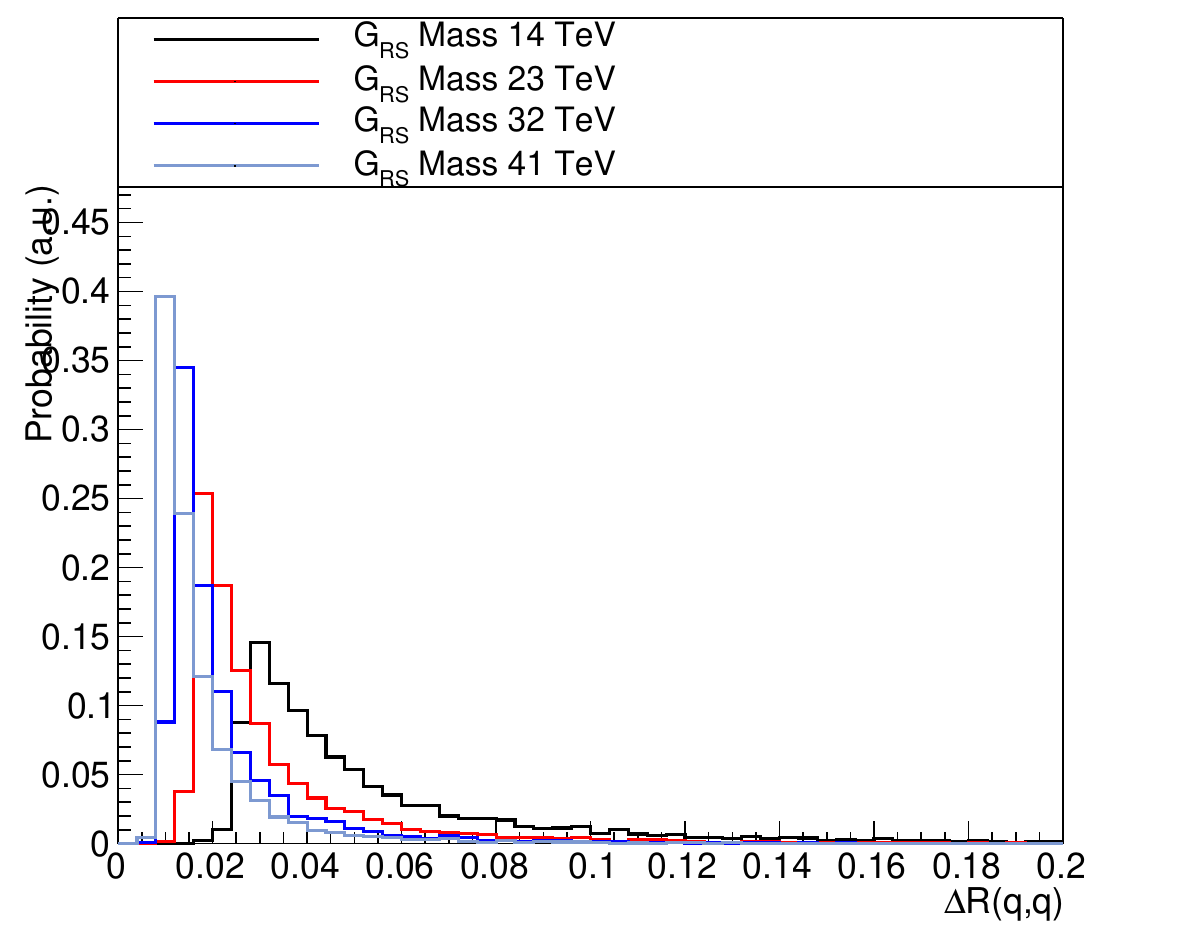}
\caption{\label{fig:DeltaRZZ} $\Delta R$ separation between the two
  quarks originating from the decay of a boosted $Z$ bosons in
  $\rm{G_{RS}}\to ZZ$ events, for different values of the
  $\rm{G_{RS}}$ mass.}
\end{figure}

In the $p_T$ range relevant for LHC searches, variables exploiting the
jet substructure typically aim to identify jets whose constituents can
be arranged into two subjets. At the FCC, the larger boost values
accessible in 100 TeV collisions change substantially the experimental
signature. The separation between the two subjets becomes very small
for large $\rm{G_{RS}}$ mass values, as shown in
Fig.~\ref{fig:DeltaRZZ}.

\begin{figure}[htbp]
\centering 
\includegraphics[width=0.95\textwidth]{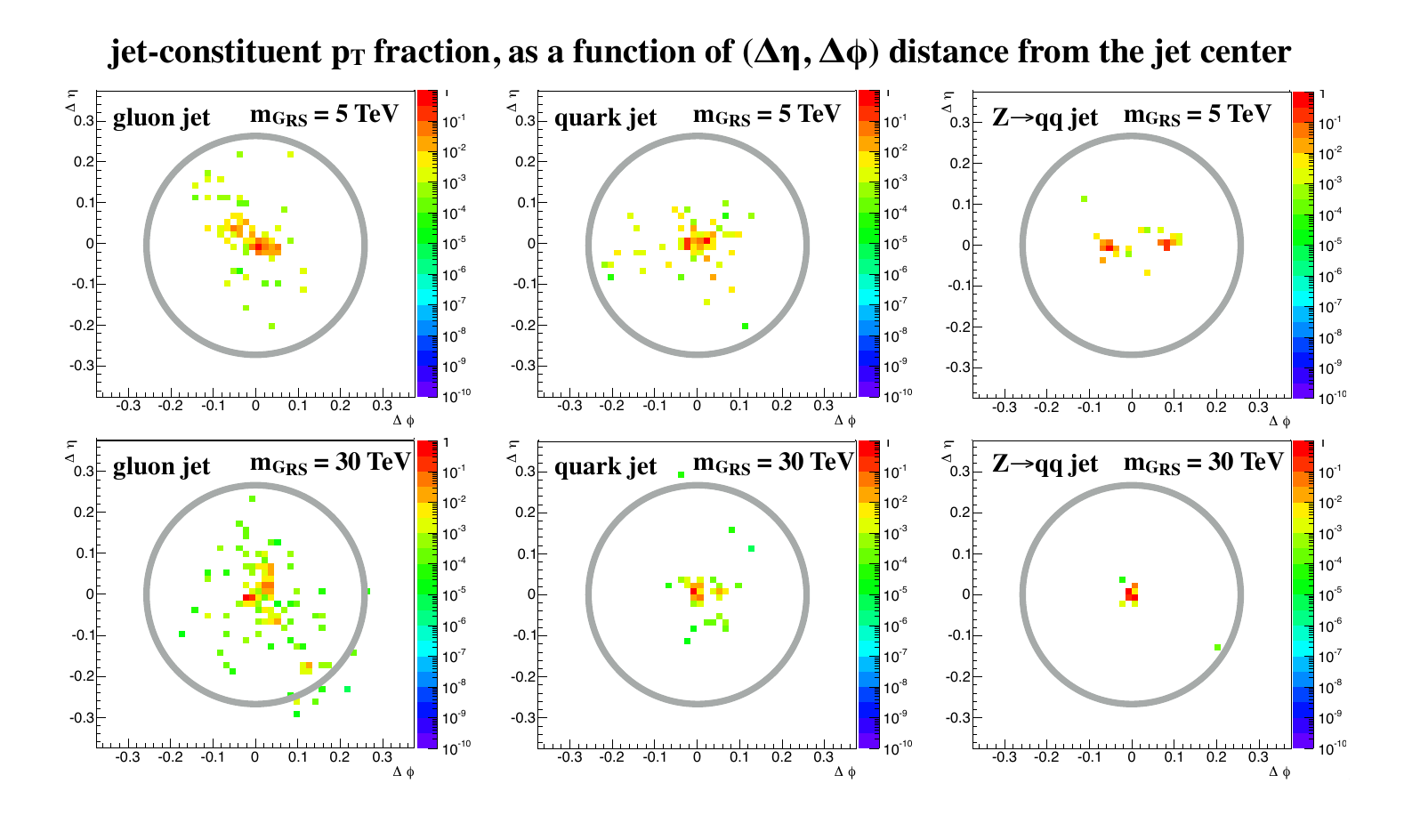}
\caption{\label{fig:pTflow} Ratio between the $p_T$ of jets
  constituents and jet $p_T$ for jets originating from gluon (left), a
  quark (centre), and $Z\to q \bar q$ (right), shown as a function of
  the $\eta$ and $\phi$ distance of each constituent from the jet
  centre. The top (bottom) plots refer to the highest-$p_T$ jet in a
  typical $\rm{G_{RS}}$ decay, for a mass value $m_{\rm{G_{RS}}}=
  5$~TeV ($m_{\rm{G_{RS}}}= 30$~TeV). An angular matching to the
  generated $Z$ boson is applied for $Z\to q \bar q$.}
\end{figure}

Consequently, the boosted boson is better identified as a single
narrow jet inside an otherwise empty jet, similarly to a $\tau$
lepton. This is represented in Fig.~\ref{fig:pTflow}, where the $p_T$
flow of typical boosted bosons and ordinary QCD jets is shown for a
small ($m_{\rm{G_{RS}}}= 5$~TeV) and a large ($m_{\rm{G_{RS}}}=
30$~TeV) value of $m_{\rm{G_{RS}}}$. As a function of the $\eta$ and
$\phi$ distance of each constituent from the jet centre, the
constituent $p_T$ is shown, normalized to the jet $p_T$. For small
$m_{\rm{G_{RS}}}$ two subjets in $Z$ jets are visible inside the
jet. For large $m_{\rm{G_{RS}}}$, the two subjets merge into a single
jet, while the rest of the jet is quite empty. For comparison, the
corresponding distributions are shown for typical jets from gluons and
quarks. No substantial change in the jet behavior is observed in this
case.

In view of this difference, a change in strategy could improve the
effectiveness of jet substructure as a tagging algorithm. As an
example, we consider the five quantities:
\begin{equation}
Flow_{n,5} = \sum_{p} \frac{|p_T^p|}{|p_T^{jet}|} 
\end{equation}
where $n=1,..,5$, $p_T^{jet}$ and $p_T^p$ are the jet and constituent
transverse momenta, respectively. The sum in the equation extends over
the jet constituents $p$ such that
\begin{equation}
\frac{n-1}{5} R \leq \Delta R(p,jet) < \frac{n}{5} R,
\end{equation}
where $R$ is the jet size and  $\Delta R(p,jet) = \sqrt{(\Delta \eta)^2 +
  (\Delta \phi)^2}$ is the angular distance between a given jet
constituent and the jet axis.

The five $Flow_{n,5}$ quantities are used together with the jet mass
as input features to train a boosted decision tree (BDT), using the
TMVA package~\cite{Hocker:2007ht}. The BDT is trained using as a
signal sample $\rm{G_{RS}} \to ZZ$ events with hadronically decaying
$Z$ bosons, while the background training sample is provided by jets
from $\rm{G_{RS}} \to q \bar q$ events ($q=u,d,c,s,b$). The training
is repeated for several values of $m_{\rm{G_{RS}}}$. 
For comparison, a BDT discriminant is trained with the same procedure,
using as input features the jet mass and the subjettiness ratio
$\tau_2/\tau_1$. The subjettiness variables~\cite{Thaler:2010tr} are
here used as a reference of the typical strategy followed for $V$-jet
tagging at the LHC. The distribution of the jet mass,
$Flow_{1,5}$, $Flow_{2,5}$, and the ROC curves for the two BDTs are
shown in Fig.~\ref{fig:features}.

\begin{figure}[tbp]
\centering 
\includegraphics[width=.49\textwidth]{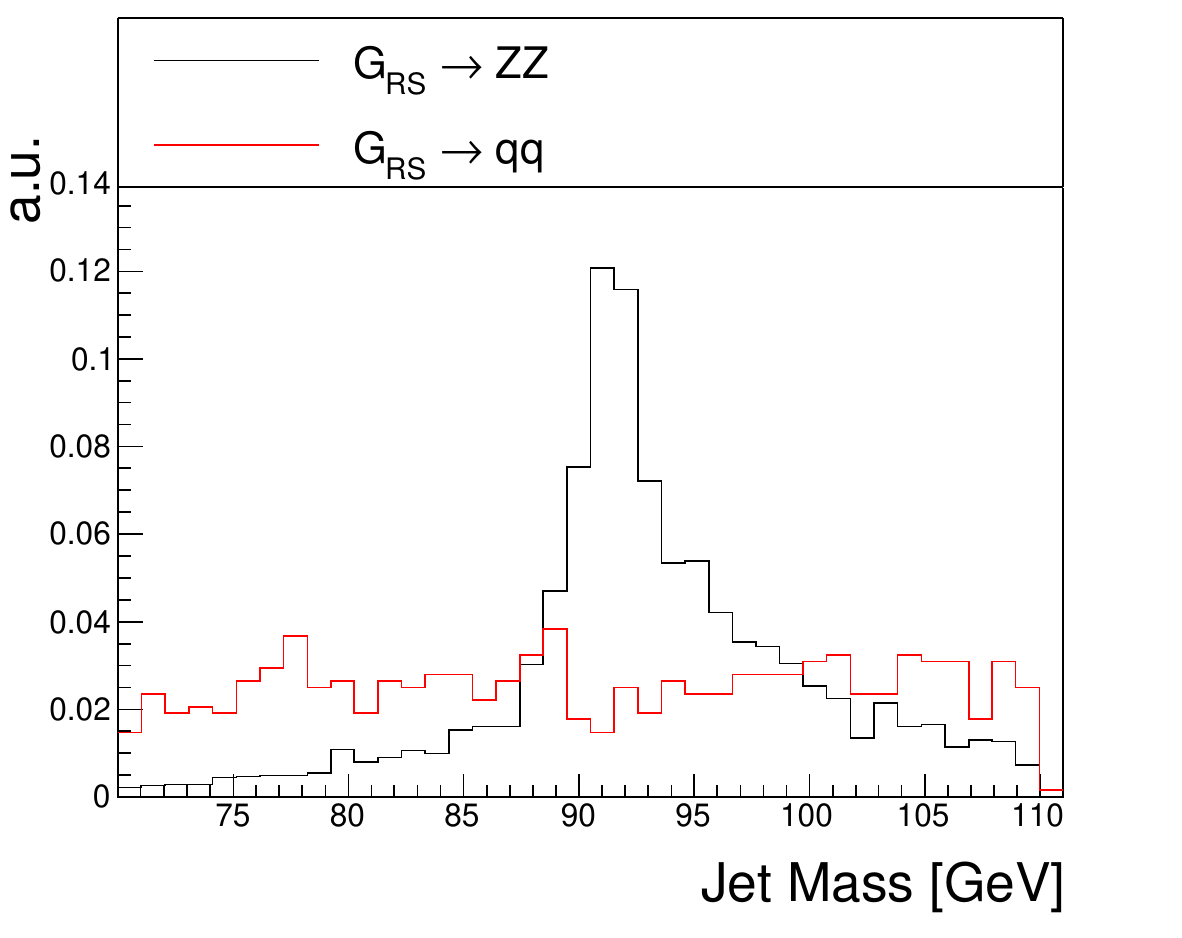}
\includegraphics[width=.49\textwidth]{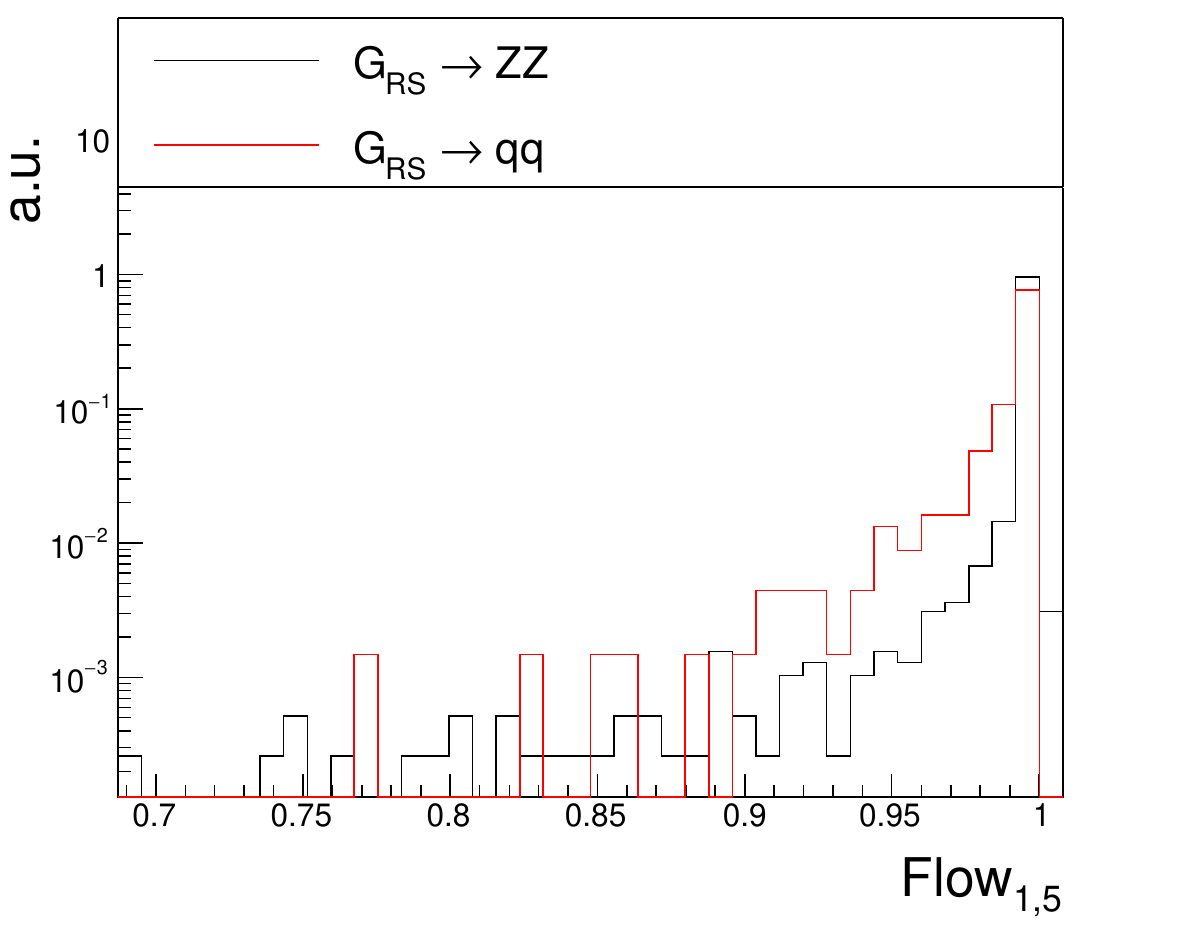}\\
\includegraphics[width=.49\textwidth]{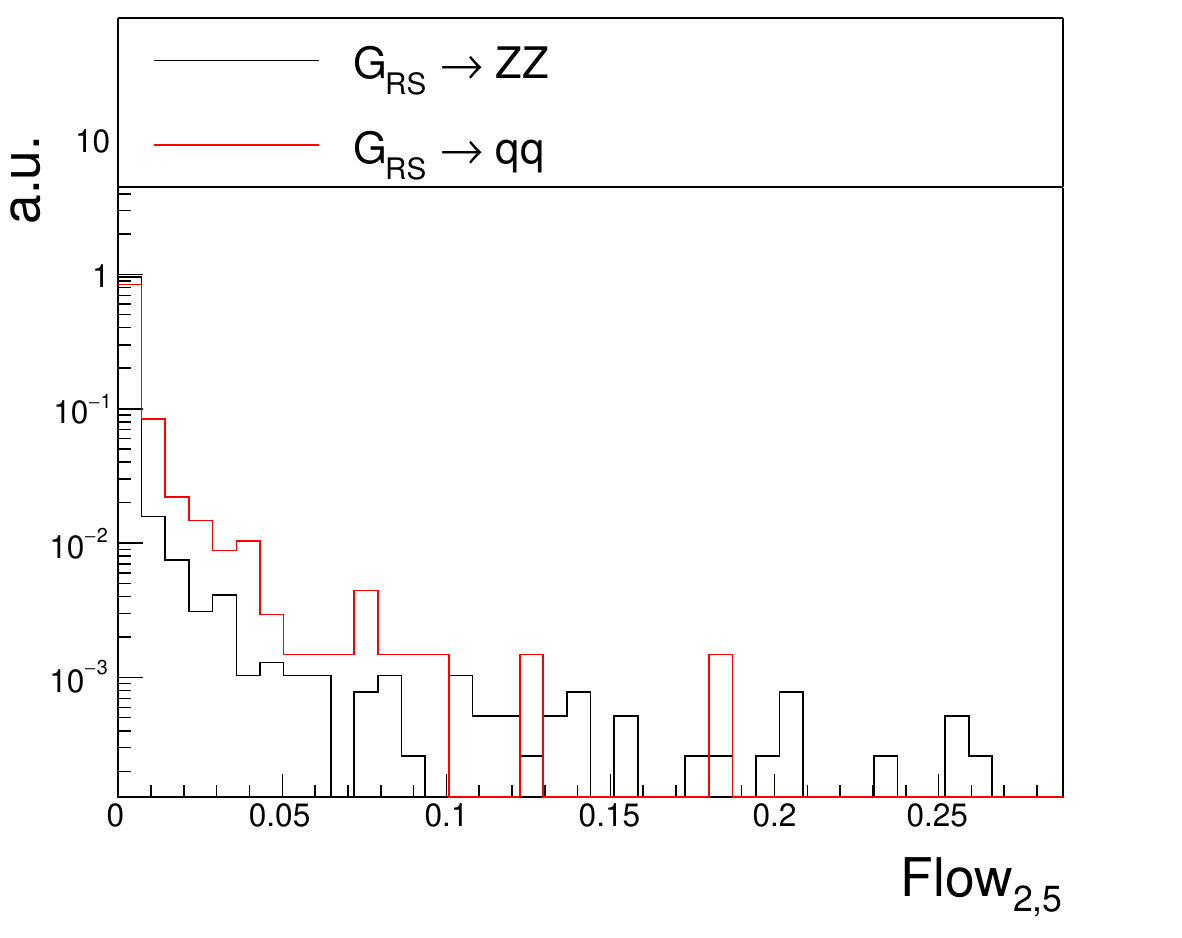}
\includegraphics[width=.49\textwidth]{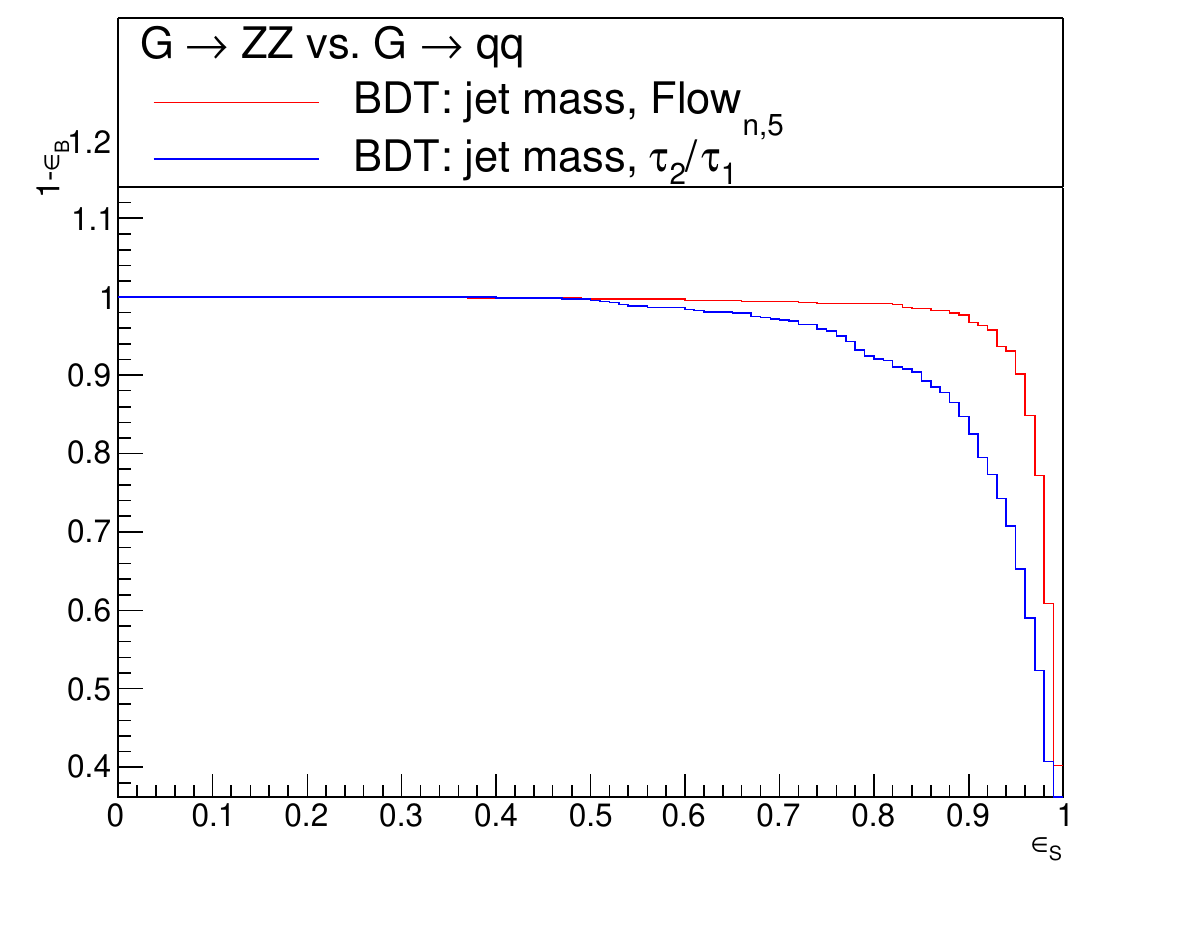}
\caption{\label{fig:features} Distribution of the Jet mass (top left),
  $Flow_{1,5}$ (top right), and $Flow_{2,5}$ (bottom left) for a
  signal hadronically decaying $Z$ bosons in $\rm{G_{RS}} \to ZZ$
  events and a background of jets in $\rm{G_{RS}} \to q \bar q$
  events. The ROC curve for a BDT trained from the five $Flow_{n,5}$
  and the jet mass is shown in the bottom-right plot, compared to the
  corresponding ROC curve trained from the jet mass and the
  subjettiness ratio $\tau_2/\tau_1$. The $\rm{G_{RS}}$ mass is fixed
  to 32 TeV.}
\end{figure}

The left plots in Fig.~\ref{fig:subjettinessXVV} show the tagging
efficiency obtained as a function of the $\rm{G_{RS}}$ mass for the
two discriminators when the false-positive rate (mistag) is fixed to
10\%, training the algorithm against quark and gluon jets. The right
plots in the same figure shows the mistag as a function of the
$\rm{G_{RS}}$ mass, when the tagging efficiency is fixed to
80\%. Similar results are obtained when $\rm{G_{RS}} \to gg$ events
are used as background.

\begin{figure}[tbp]
\centering 
\includegraphics[width=.49\textwidth]{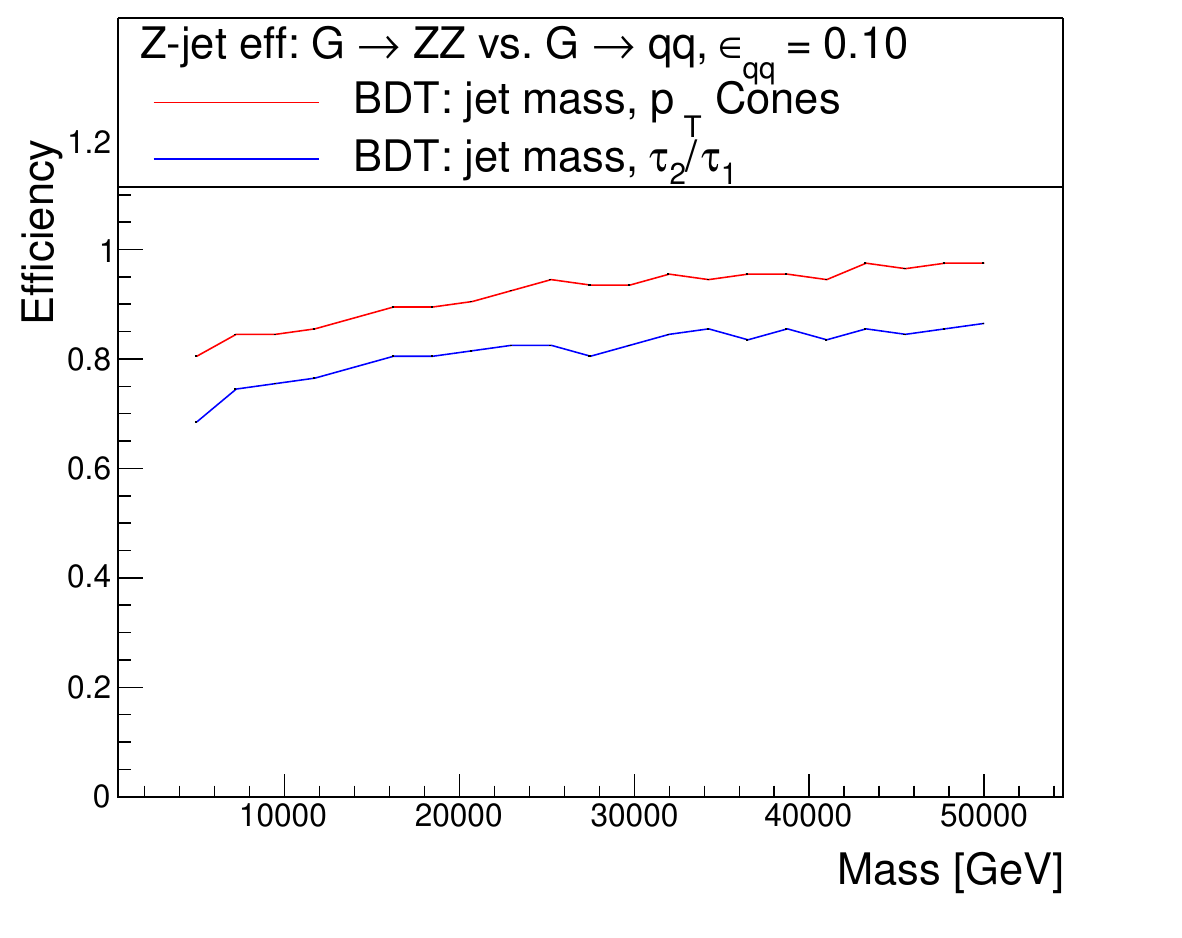}
\includegraphics[width=.49\textwidth]{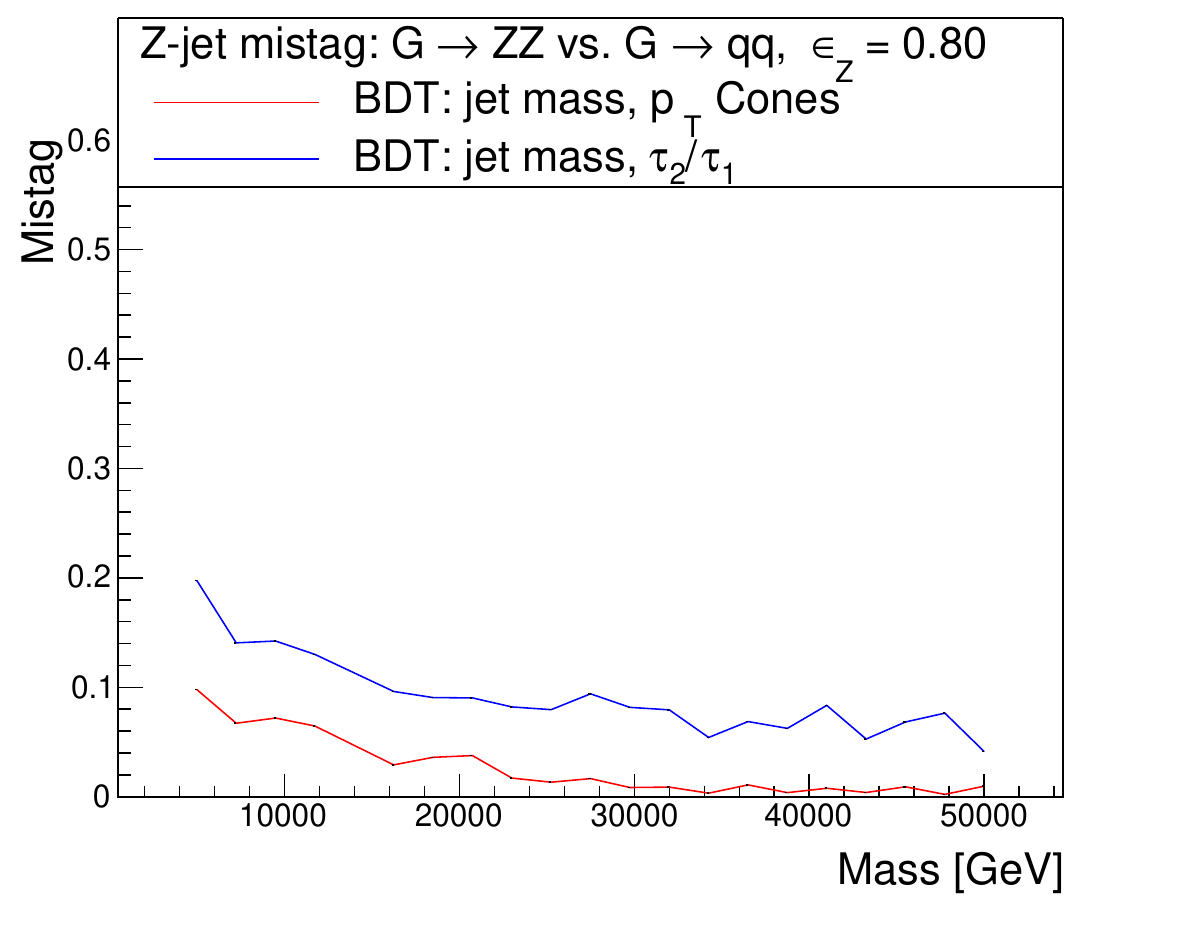} \\
\includegraphics[width=.49\textwidth]{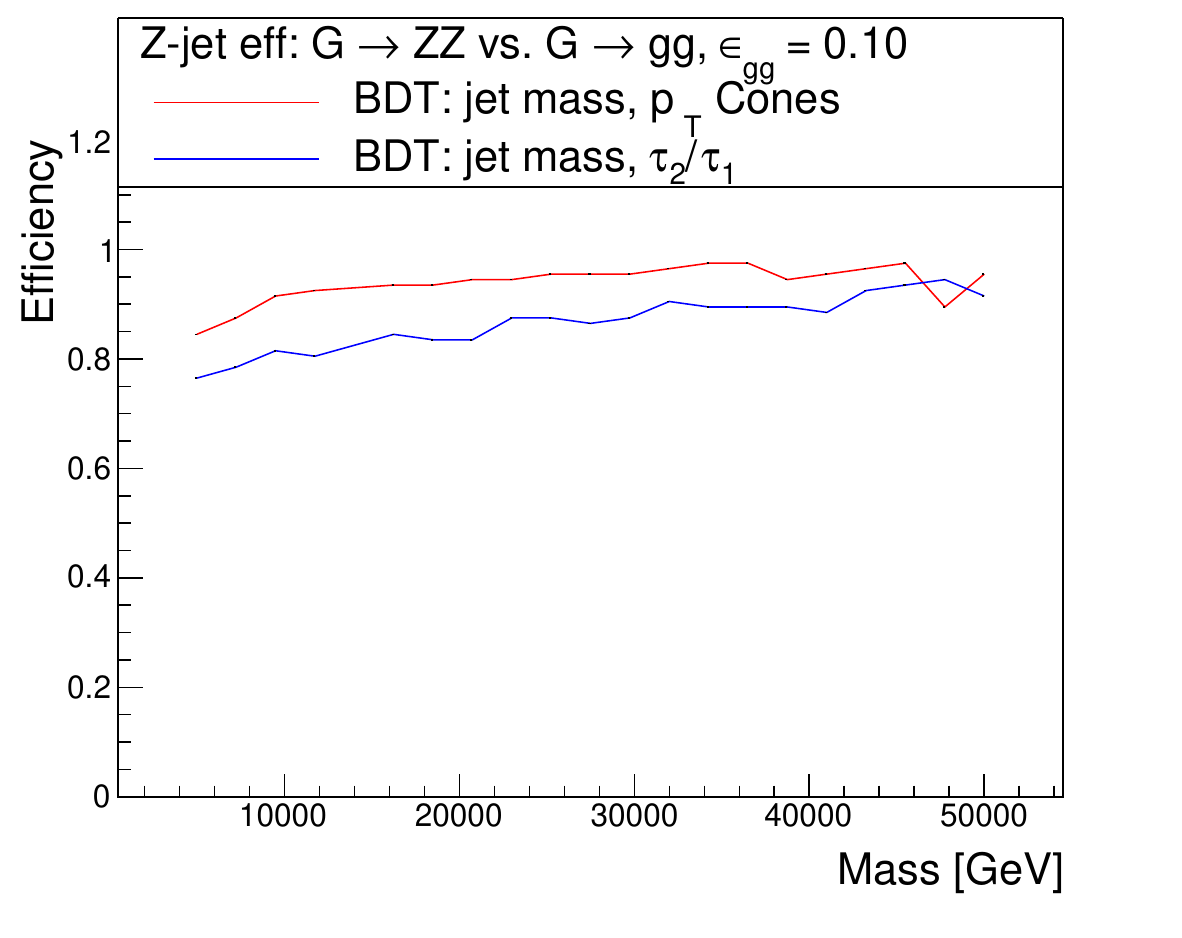}
\includegraphics[width=.49\textwidth]{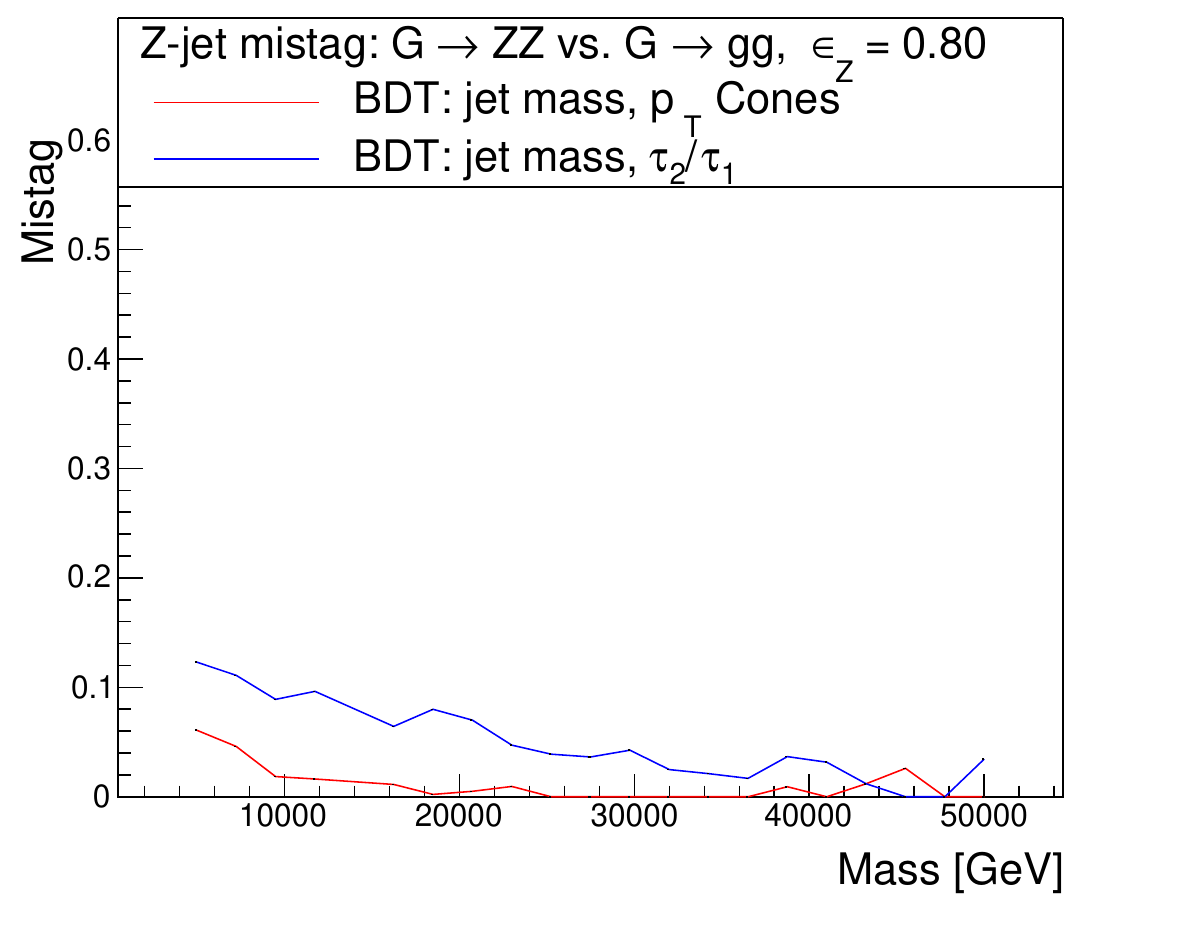} 
\caption{\label{fig:subjettinessXVV} Discrimination power of the
  V-tagging algorithms against quark (top) and gluon (bottom) jets:
  tagging efficiency as a function of the $\rm{G_{RS}}$ mass
  corresponding to a mistag rate of 10\% (left) and mistag rate as a
  function of the $\rm{G_{RS}}$ mass for a tagging efficiency of 80\%
  (right).}
\end{figure}

While this study highlights the importance of highly granular
calorimeters in with largely boosted vector-boson tagging, the
strategy discussed here is far from being an optimal exploitation of the
information provided by a granular calorimeter.  In this respect,
progresses made on image recognition and deep learning could have a
big impact on jet tagging in the future, as discussed in
Ref.~\cite{Almeida:2015jua,deOliveira:2015xxd}.

\subsection{Jet fragmentation at large $p_T$}
\label{sec:jet_frag}

The ability to tag a jet by measuring its mass or other properties
requires excellent resolution of its constituent particles.  For
optimal energy and angular resolution, finely segmented calorimetry
and precise tracking systems are required.  The resolution of both
will depend on the density of tracks and their momentum.  In a
high-density environment, it will be challenging to identify
individual tracks, thereby reducing mass resolution.  At extremely
high momenta, tracks will not bend substantially in the tracking
magnetic field and their charge and momentum may not be able to be
determined.  Designing tracking systems that can resolve both of these
issues will be required.

All plots in this section are generated using the same event and jet
criteria as discussed in Sec.~\ref{sec:jet_spect}.  In particular, we
require that all particles used in identifying jets have
pseudorapidity $|\eta| < 2.5$.

In Figs.~\ref{fig:ntrack_10}-\ref{fig:ntrack_40}, we plot the average
(mean) number of tracks with $p_T>p_T^{\min}$ in jets with radius
$R=0.5$ from the resonance decays studied throughout this section.
These plots demonstrate, for example, that a jet of any flavor from
the decay of 10 TeV resonance will have at least one track with
$p_T\gtrsim 500$ GeV.  For jets from the decays of 40 TeV resonances,
every jet will have at least one track with $p_T$ greater than about 2
TeV.  For precision measurements, it may be necessary to consider
rarer configurations; say, tracks that occur in 10\% or even only 1\%
of jets.  In this case, for jets from 10 TeV resonances, 1\% of jets
will have a track with $p_T\gtrsim 2$ TeV, while for jets from 40 TeV
resonances, this is increased to about 10 TeV.

While we have shown plots for jets with radius $R=0.5$, except at low
$p_T$, these plots are relatively insensitive to jet radius.  Because
these tracks carry such a large fraction of the total jet's transverse
momentum, they must be located very near the center of the jet.

\begin{figure}[h!]
\centering
\includegraphics[width=0.45\linewidth]{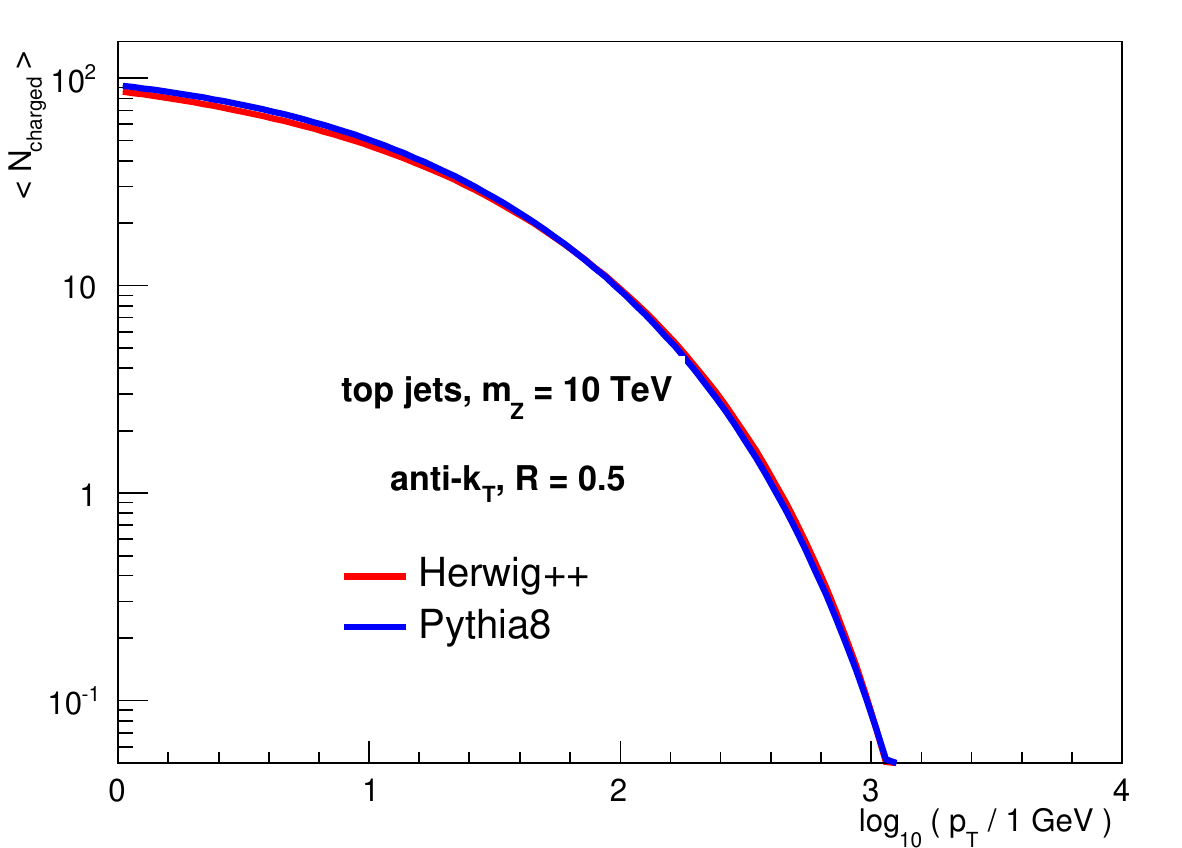}
\hspace{-0.02\textwidth}
\includegraphics[width=0.45\linewidth]{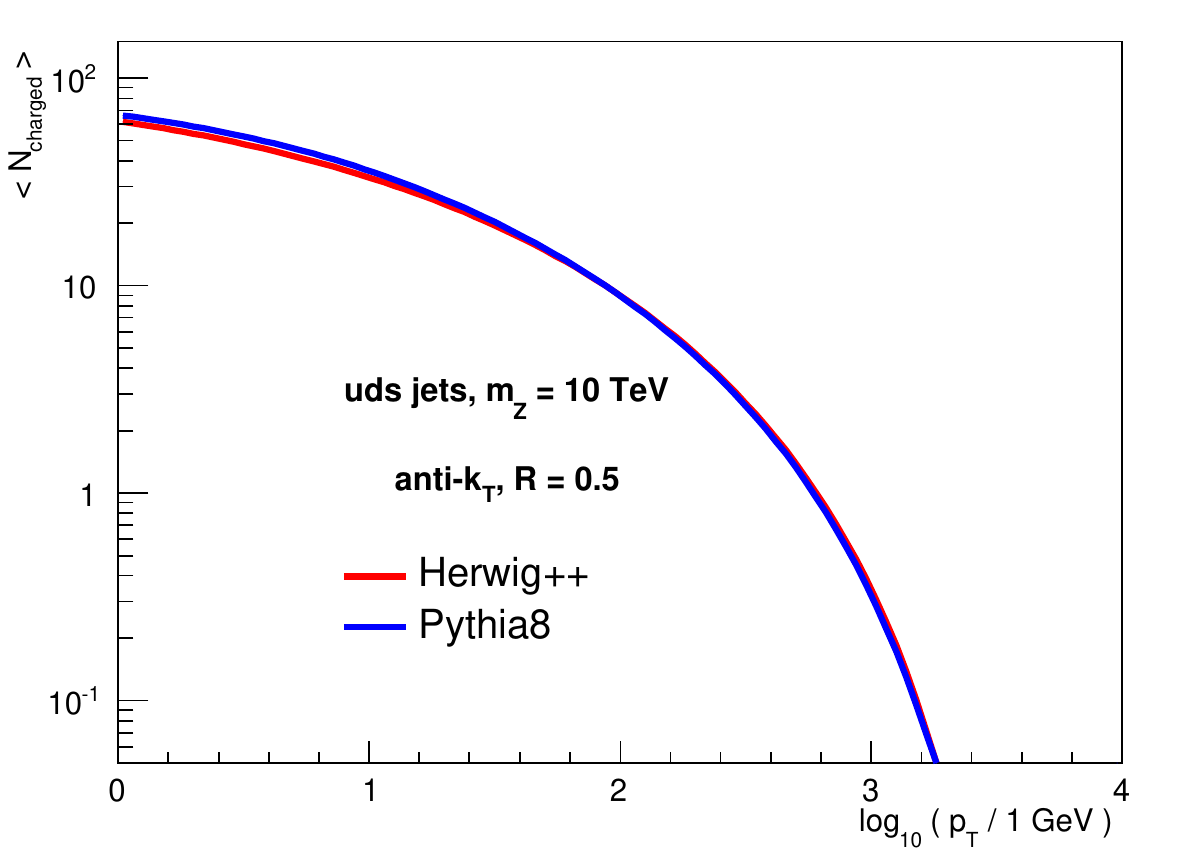}
\\
\includegraphics[width=0.45\linewidth]{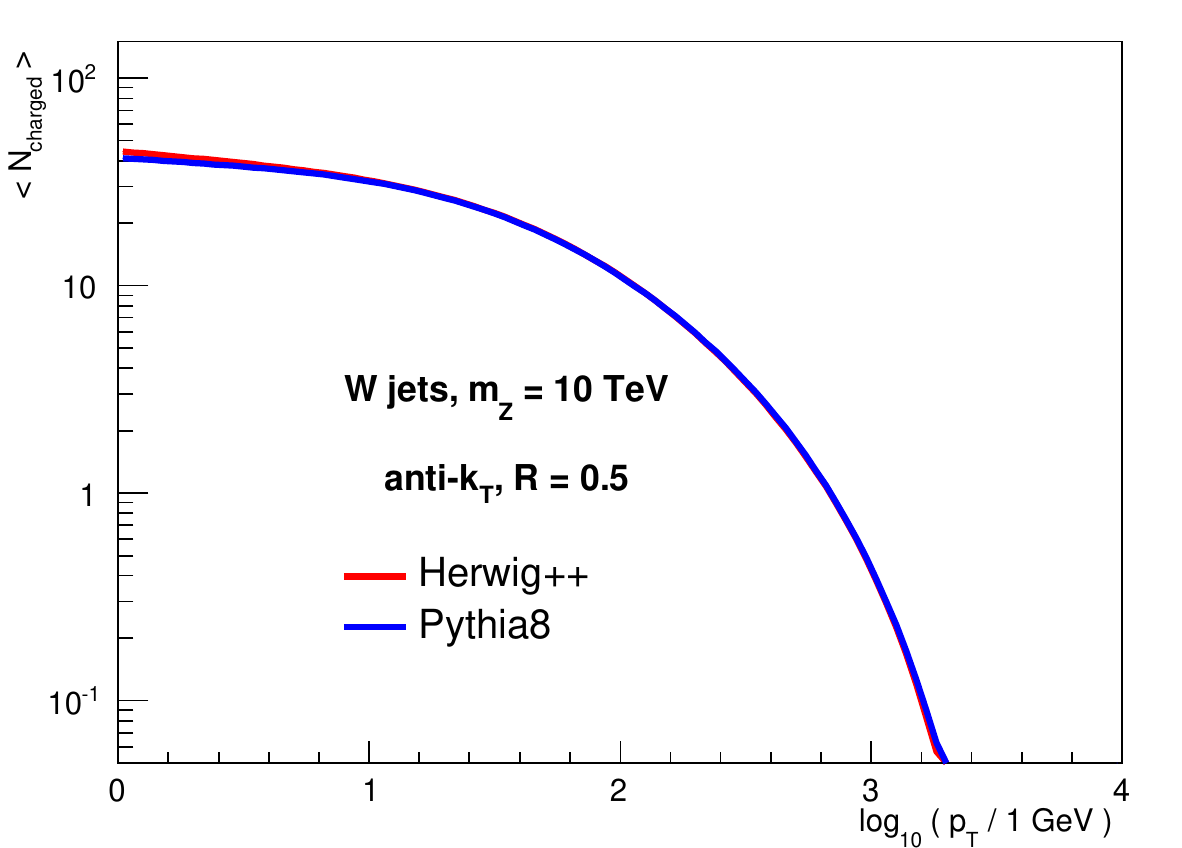}
\hspace{-0.02\textwidth}
\includegraphics[width=0.45\linewidth]{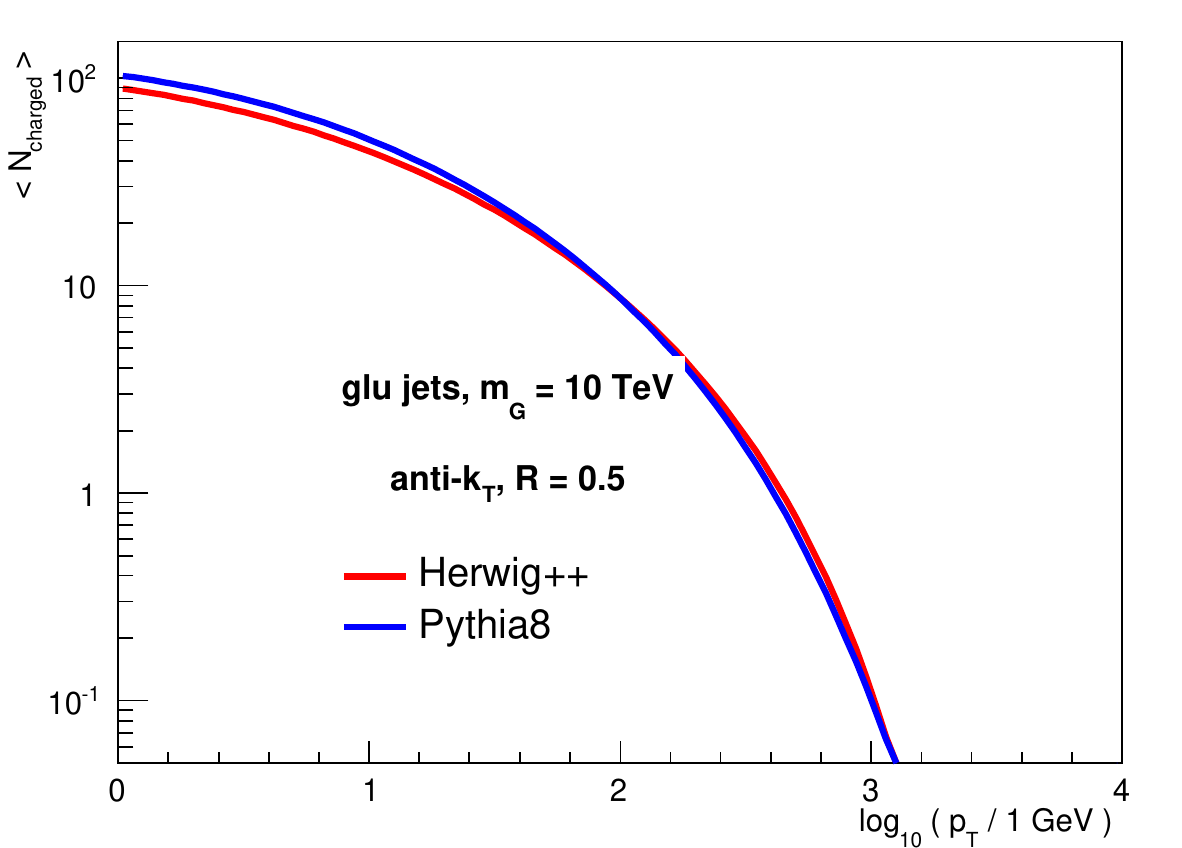}
\caption{ Average number of charged tracks with $p_T>p_T^{\min}$ in
  $R=0.5$ anti-$k_T$ jets produced from 10 TeV resonance decays to
  tops, light QCD quarks, $W$s, and gluons.  }
\label{fig:ntrack_10}
\end{figure}

\begin{figure}[h!]
\centering
\vspace{-2mm}
\includegraphics[width=0.45\linewidth]{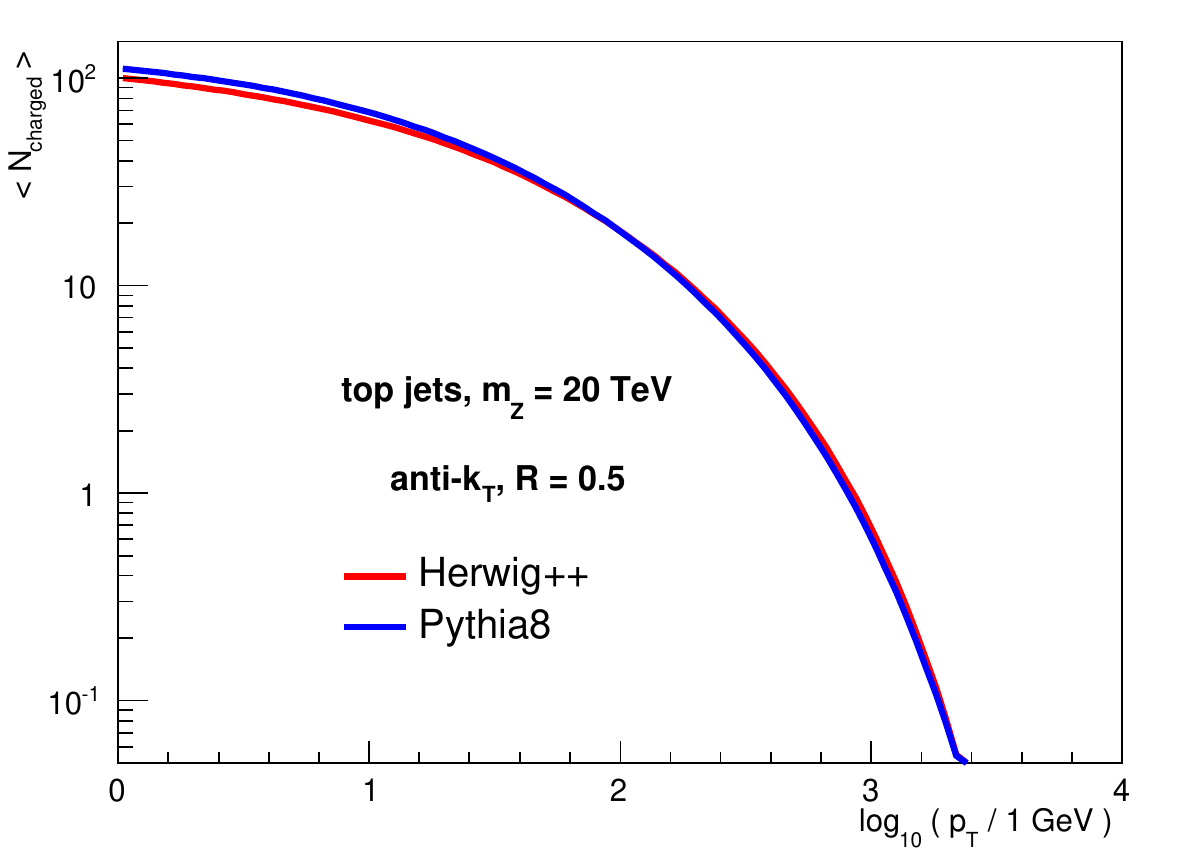}
\hspace{-0.02\textwidth}
\includegraphics[width=0.45\linewidth]{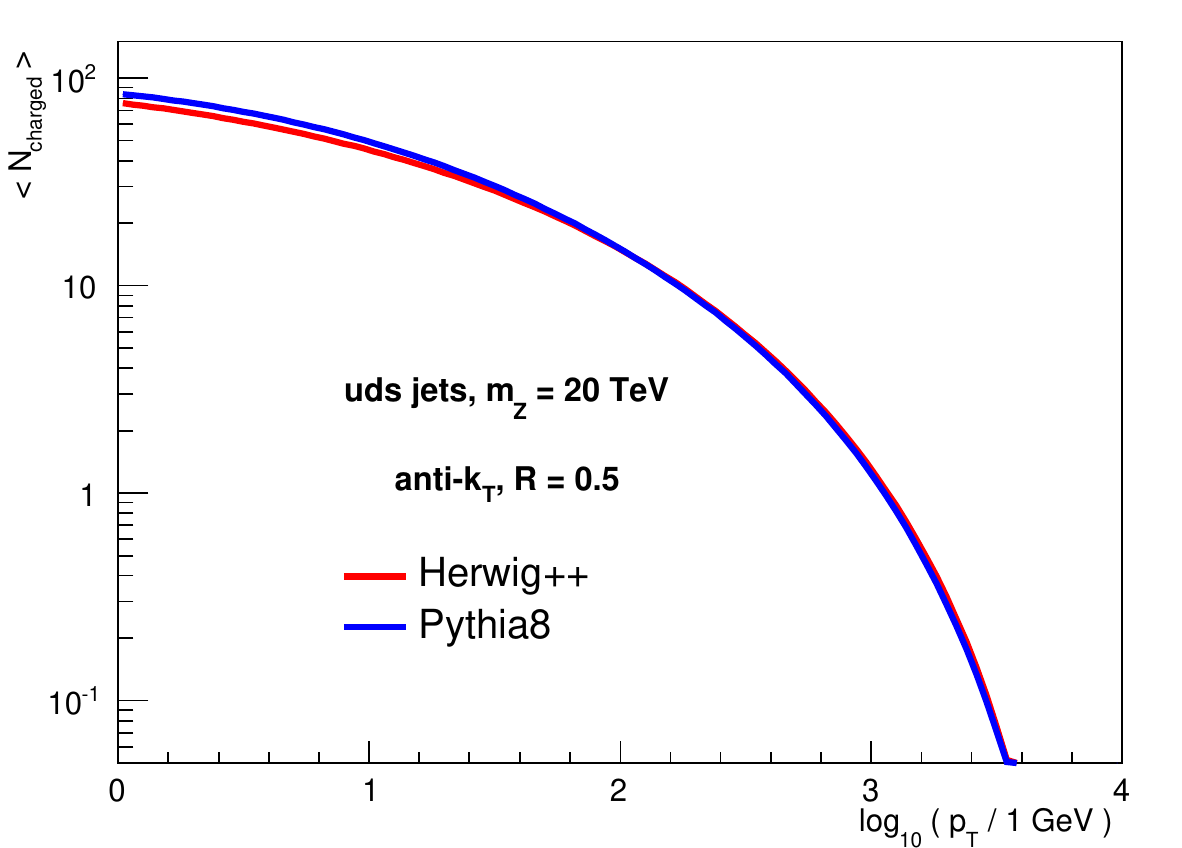}
\\ 
\includegraphics[width=0.45\linewidth]{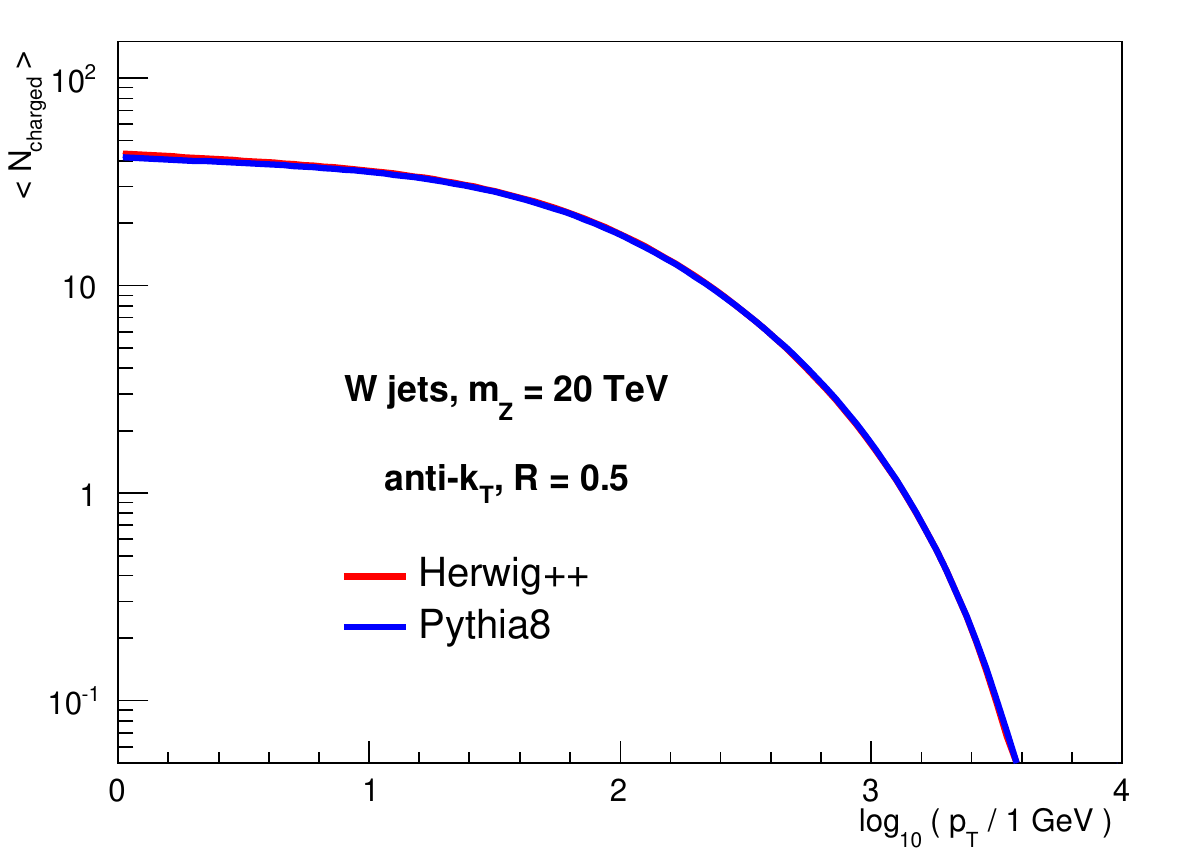}
\hspace{-0.02\textwidth}
\includegraphics[width=0.45\linewidth]{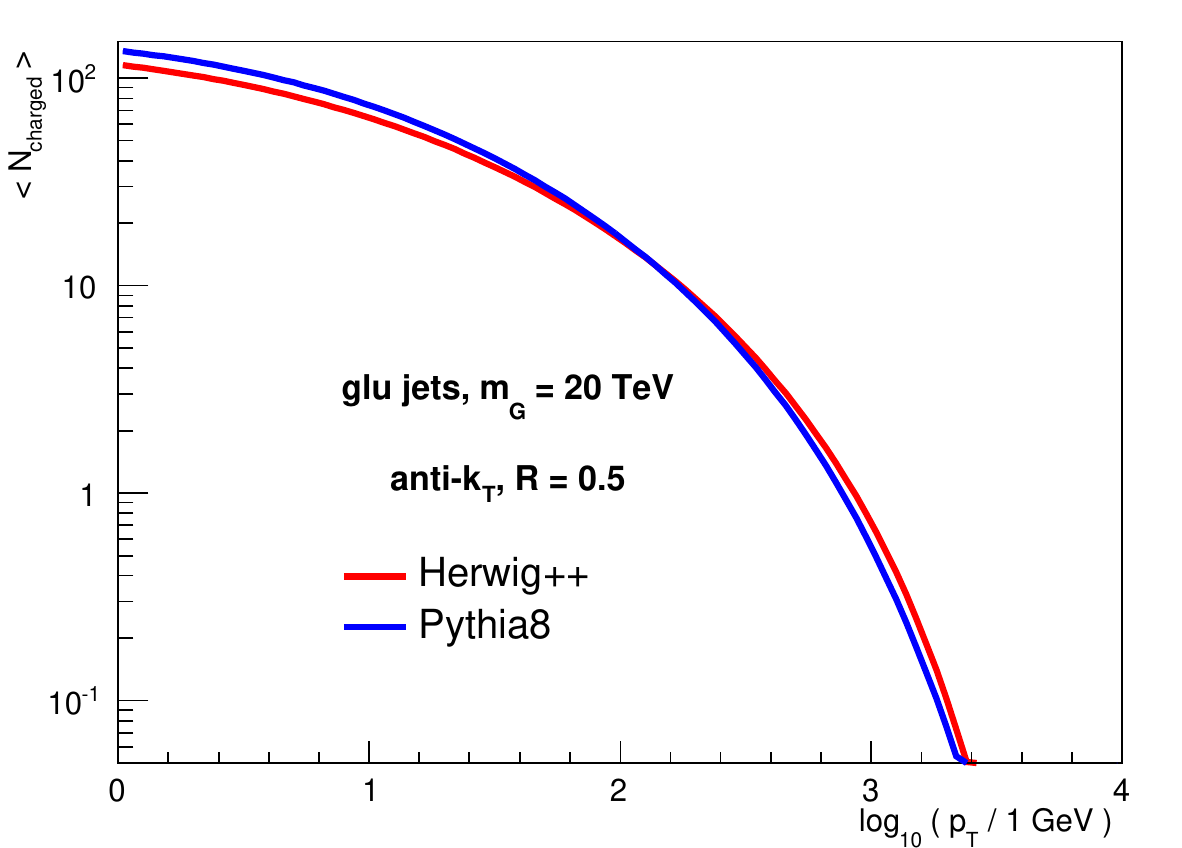}
\vspace{-2mm}
\caption{ Average number of charged tracks with $p_T>p_T^{\min}$ in
  $R=0.5$ anti-$k_T$ jets produced from 20 TeV resonance decays to
  tops, light QCD quarks, $W$s, and gluons.  }
\label{fig:ntrack_20}
\end{figure}

\begin{figure}[h!]
\centering
\includegraphics[width=0.45\linewidth]{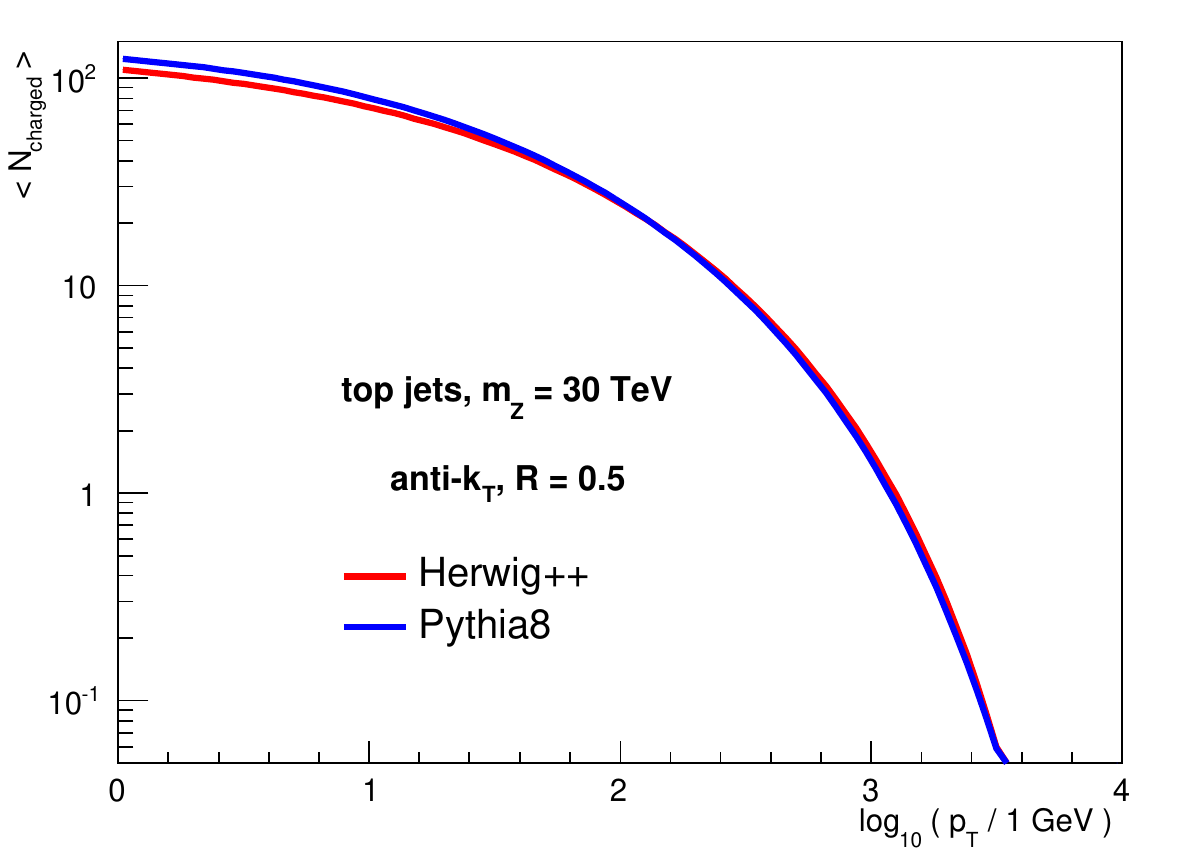}
\hspace{-0.02\textwidth}
\includegraphics[width=0.45\linewidth]{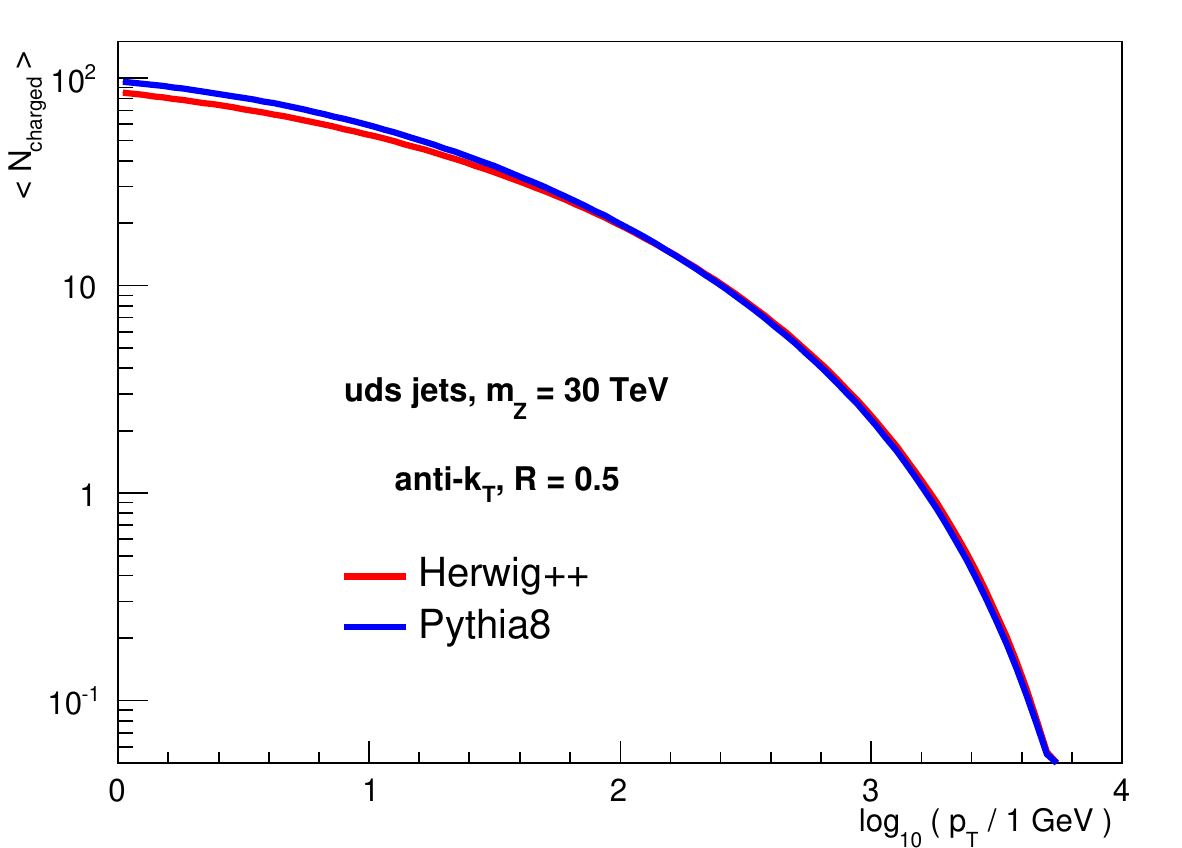}
\\
\includegraphics[width=0.45\linewidth]{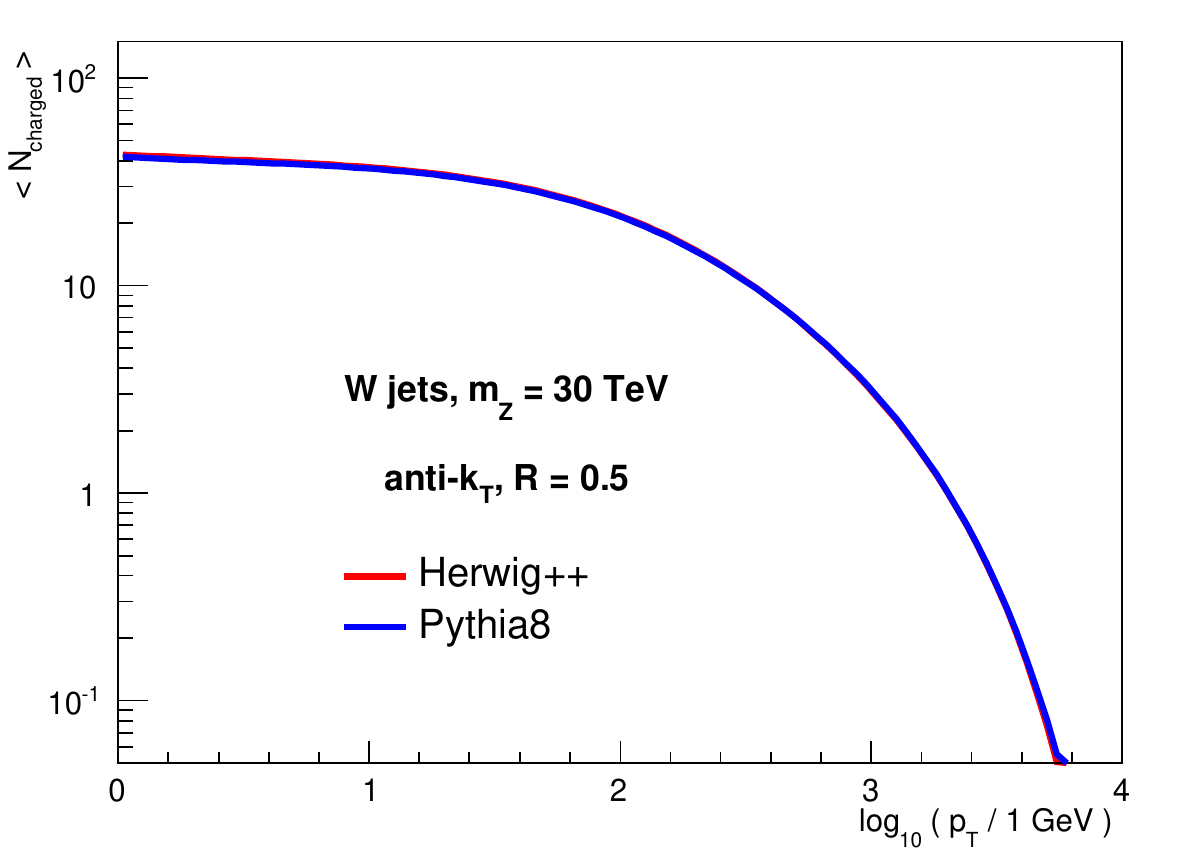}
\hspace{-0.02\textwidth}
\includegraphics[width=0.45\linewidth]{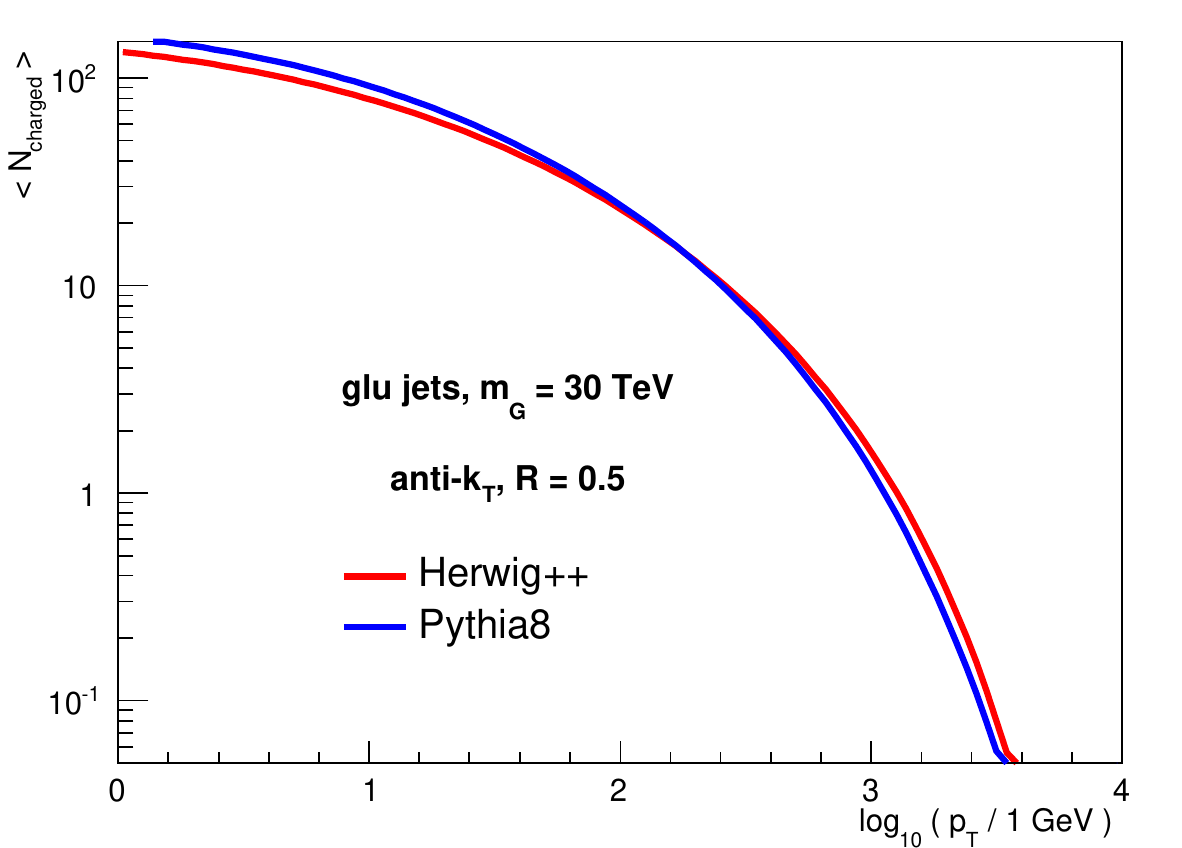}
\caption{ Average number of charged tracks with $p_T>p_T^{\min}$ in
  $R=0.5$ anti-$k_T$ jets produced from 30 TeV resonance decays to
  tops, light QCD quarks, $W$s, and gluons.  }
\label{fig:ntrack_30}
\end{figure}

\begin{figure}[h!]
\centering
\includegraphics[width=0.45\linewidth]{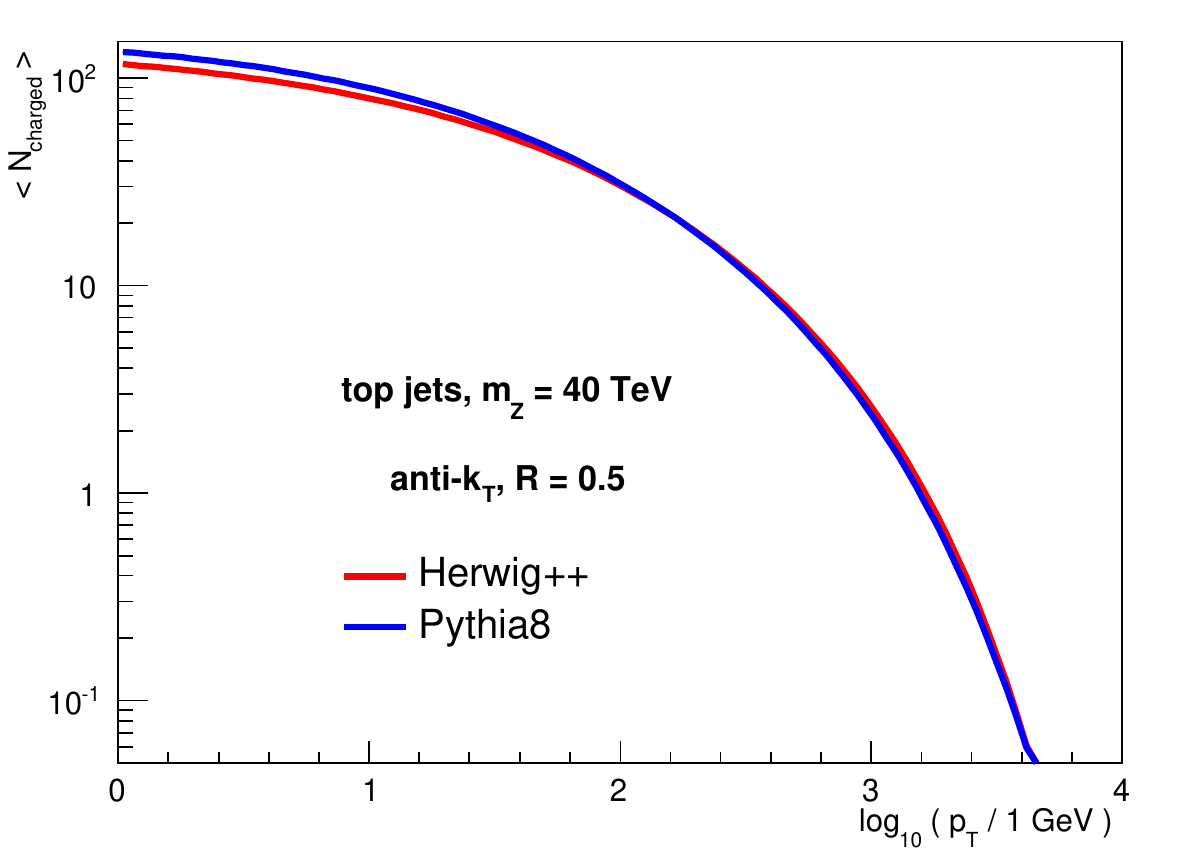}
\hspace{-0.02\textwidth}
\includegraphics[width=0.45\linewidth]{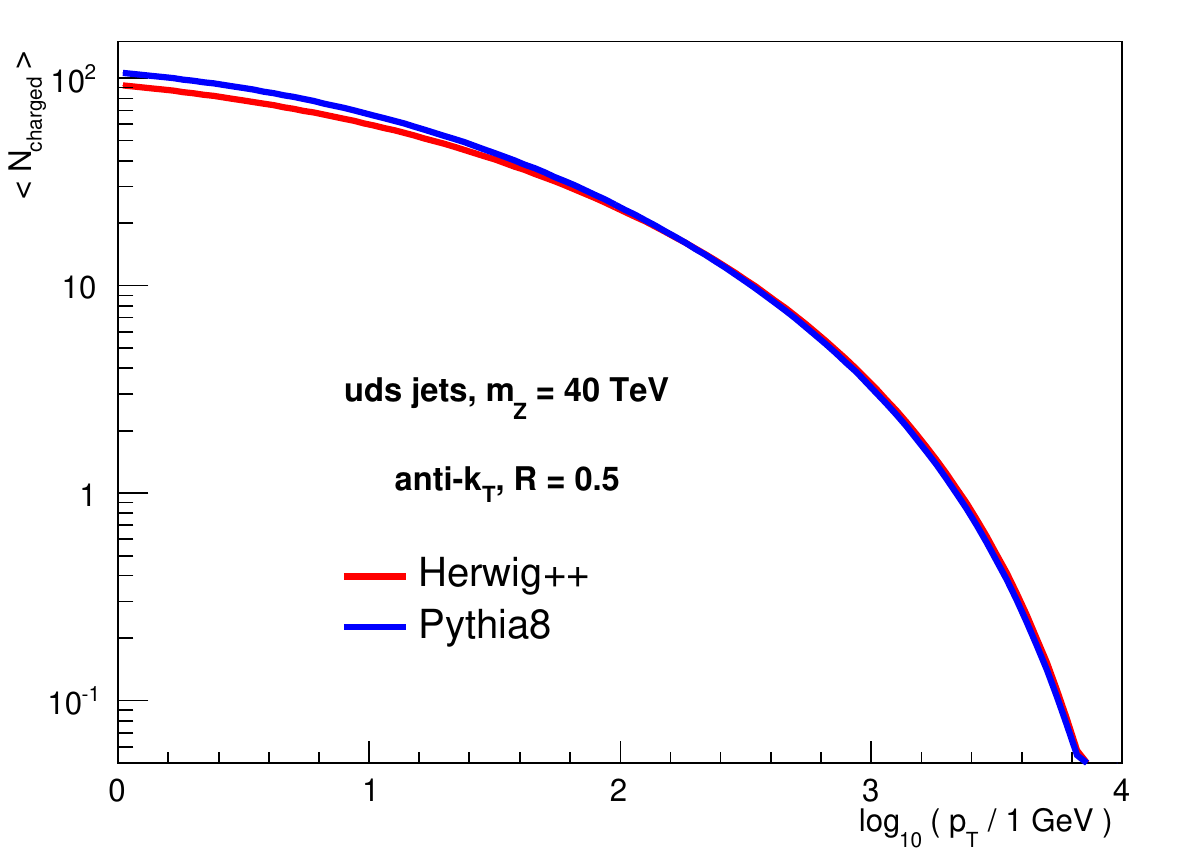}
\\
\includegraphics[width=0.45\linewidth]{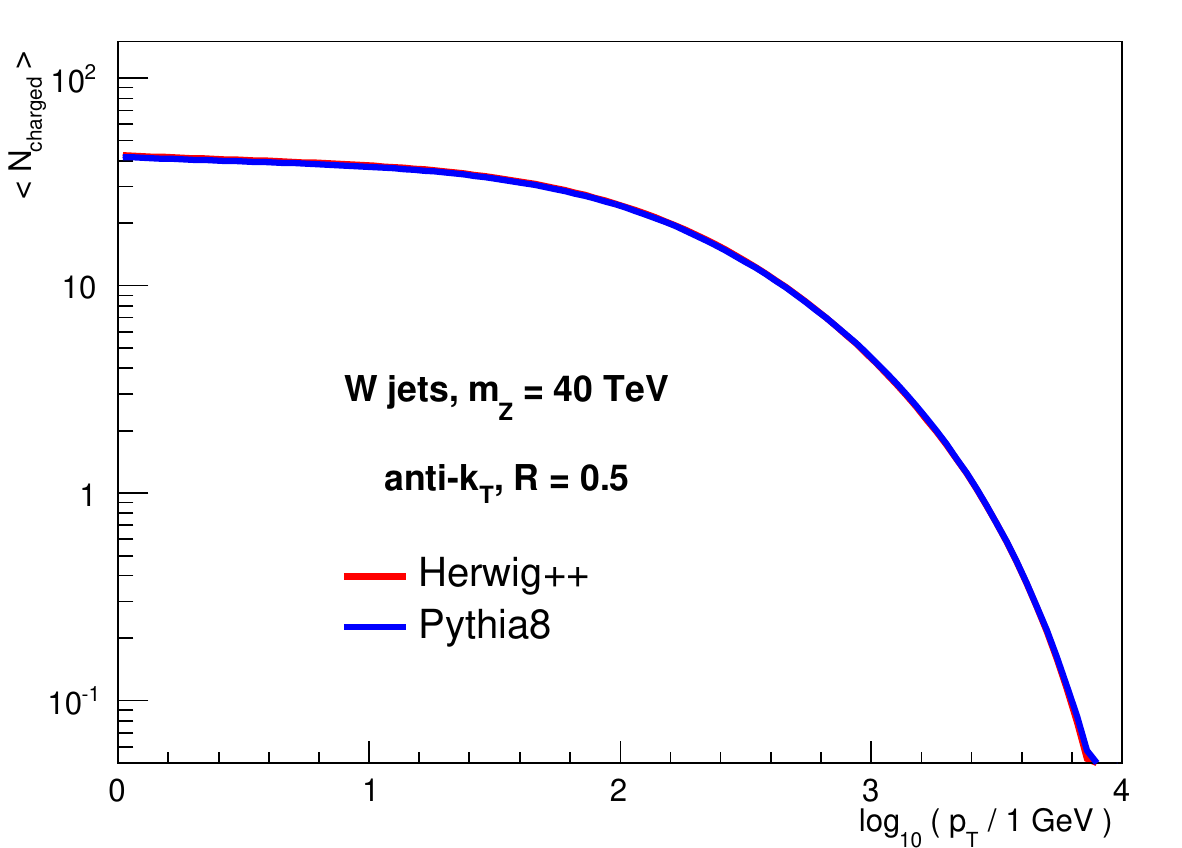}
\hspace{-0.02\textwidth}
\includegraphics[width=0.45\linewidth]{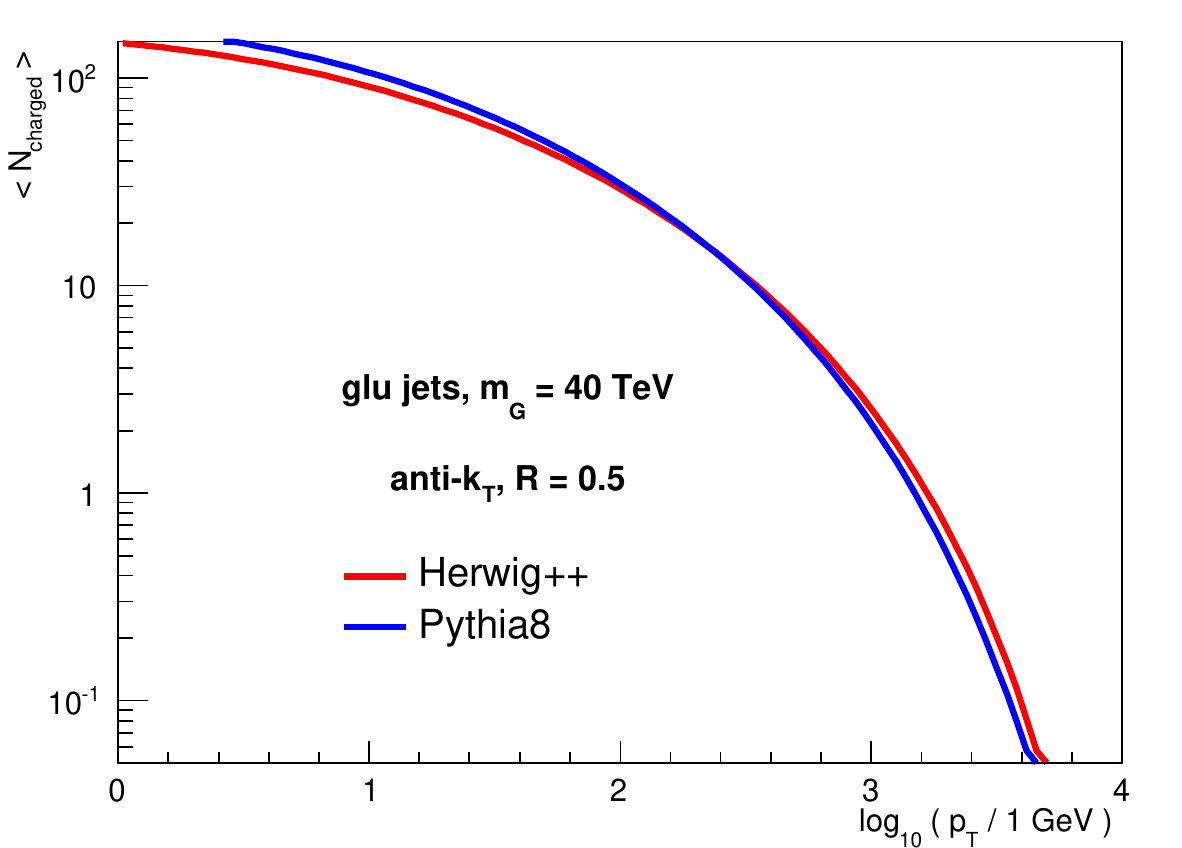}
\caption{ Average number of charged tracks with $p_T>p_T^{\min}$ in
  $R=0.5$ anti-$k_T$ jets produced from 40 TeV resonance decays to
  tops, light QCD quarks, $W$s, and gluons.  }
\label{fig:ntrack_40}
\end{figure}


In addition to having a sufficiently high magnetic field to measure
the momentum of high $p_T$ tracks, the tracking system must also be
able to resolve particles in a high density environment.  As a proxy
for tracker densities, in Figs.~\ref{fig:medR_10}-\ref{fig:medR_40},
we plot the median angle $\Delta \tilde R$ between two tracks with
$p_T$ greater than a minimum value.  The way that this median angle is
defined is as follows.  First, we take all tracks in a single jet with
$p_T$ greater than a minimum value and find the median distance
between pairs of those tracks.  Note that for a jet with two hard
prongs (like a boosted $W$), this median value will typically be
either close to 0 or $2m/p_T$, depending on the precise distribution
of tracks in the jet.  We take the median rather than the mean
pairwise track distance because the median is insensitive to outliers
and corresponds to the angular scale at which half of the pairs have a
larger angle and half a smaller angle.  Then, the median pairwise
angle of each jet is averaged over the ensemble.

These jets are produced from decays of resonances ranging from 10 to
40 TeV, and this median angle exhibits strong jet $p_T$ and flavor
dependence.  The distribution of this median angle for quark and gluon
flavor jets has no structure and the angle between tracks with the
same $p_T$ is approximately twice as large for gluons as compared to
quarks.
Because quark and gluon jets have no intrinsic high energy scale
associate with them, the distributions with different jet $p_T$s are
simply scaled by the ratio of their $p_T$s.  In the high mass tail,
the mass of the jet is determined by the relative angle of the hardest
particles in the jet.

The median angle between high $p_T$ tracks for top quark or $W$ jets
is very different.  These jets do have an intrinsic scale, and so this
median angle should manifest these scales.  For a jet with two-prongs
(like from a hadronically-decaying $W$) with mass $m$ and transverse
momentum $p_T$, the characteristic angle between the hard prongs is
$\theta=2m/p_T$.  Assuming that the prongs are very narrow and
otherwise approximately identical, when averaged over the jet
ensemble, the median pairwise angle will be roughly
$R^{\text{med}}\simeq m/p_T$.  For a jet with more hard prongs, like a
top quark jet, the median angle will be closer to the characteristic
jet angle.  The characteristic angular scales in top quark jets and
$W$ jets are
\begin{equation}
R_t^\text{med} \simeq \frac{2m_t}{p_T} \,, \qquad R_W^\text{med} \simeq \frac{m_W}{p_T}\,.
\end{equation}
Especially at the highest resonance masses, features are present in
the top and $W$ distributions near these angles.  Combining the
information in Fig.~\ref{fig:ntrack_40} and Fig.~\ref{fig:medR_40},
for instance, requires resolving angular scales of $\Delta \tilde
R\lesssim 10^{-3}$ to be able to reconstruct the substructure of
boosted $W$ bosons from 40 TeV resonances.

\begin{figure}[h!]
\centering
\includegraphics[width=0.45\linewidth]{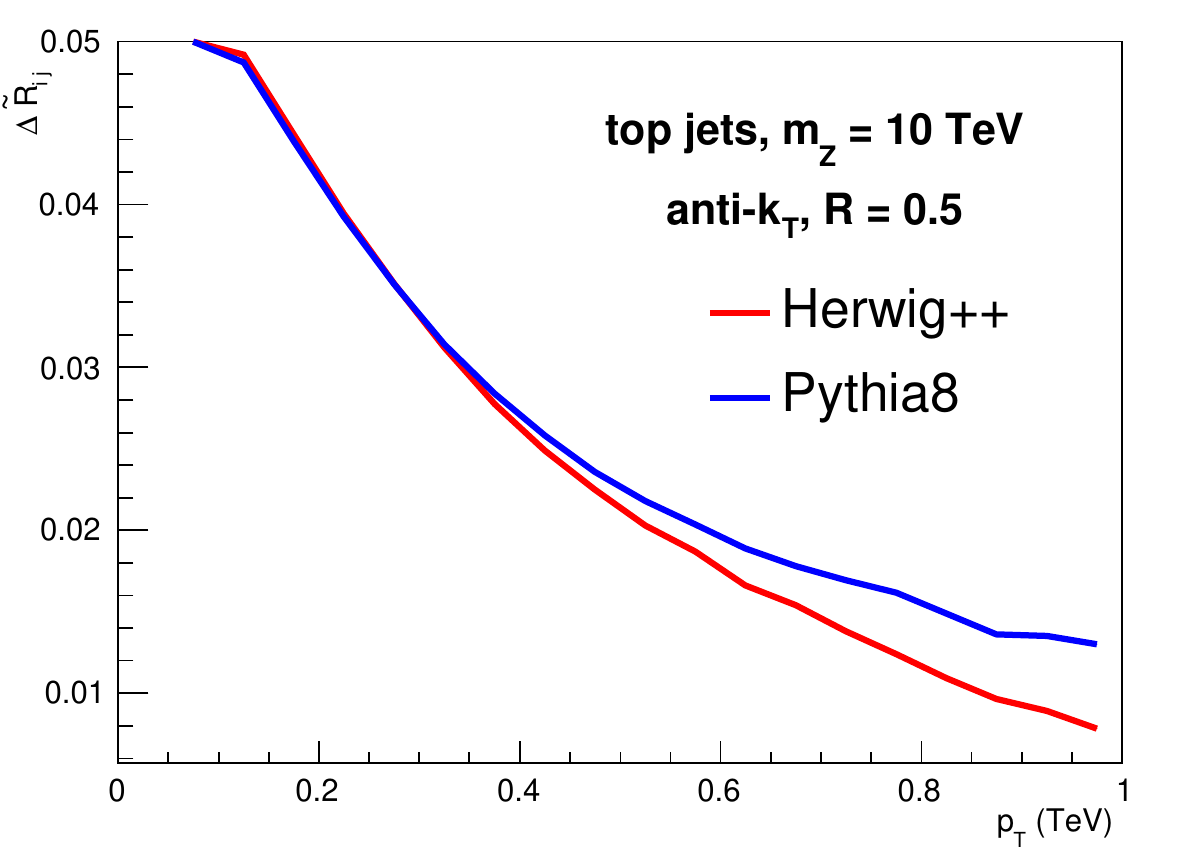}
\hspace{-0.02\textwidth}
\includegraphics[width=0.45\linewidth]{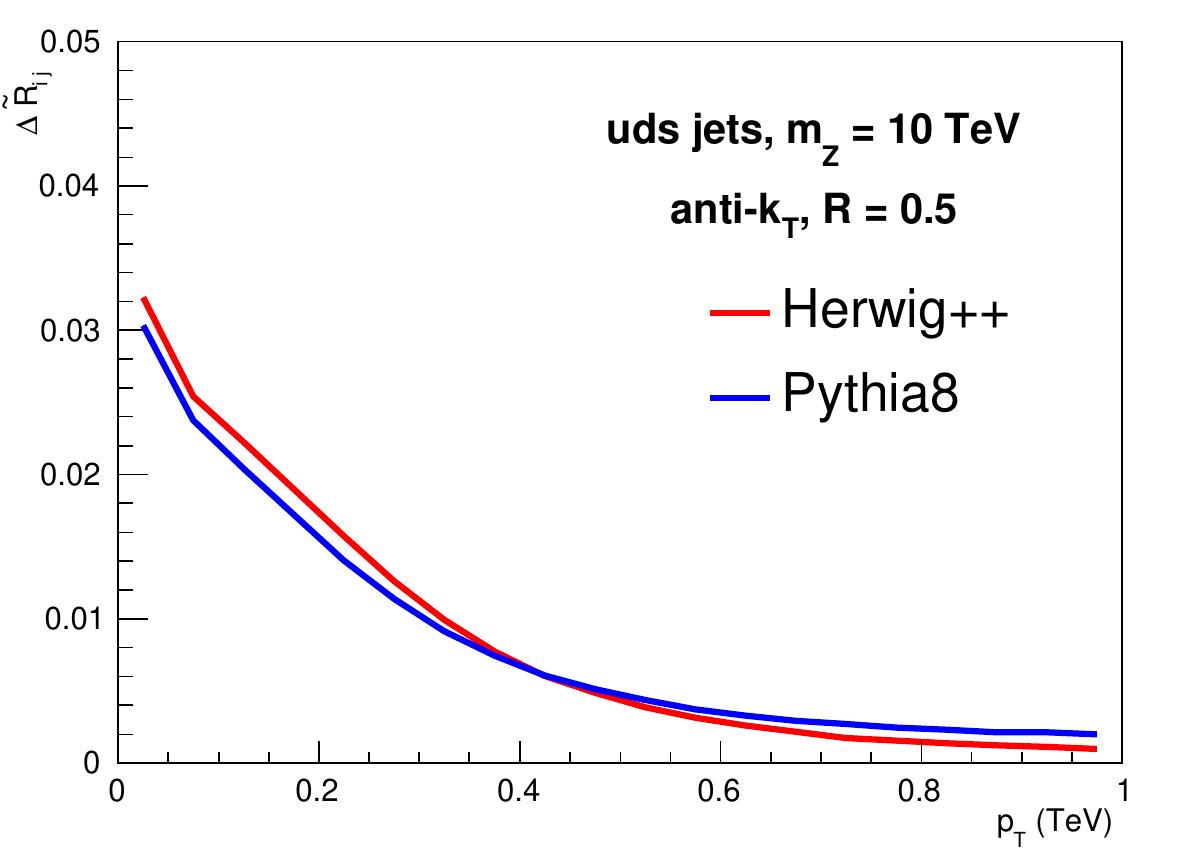}
\\
\includegraphics[width=0.45\linewidth]{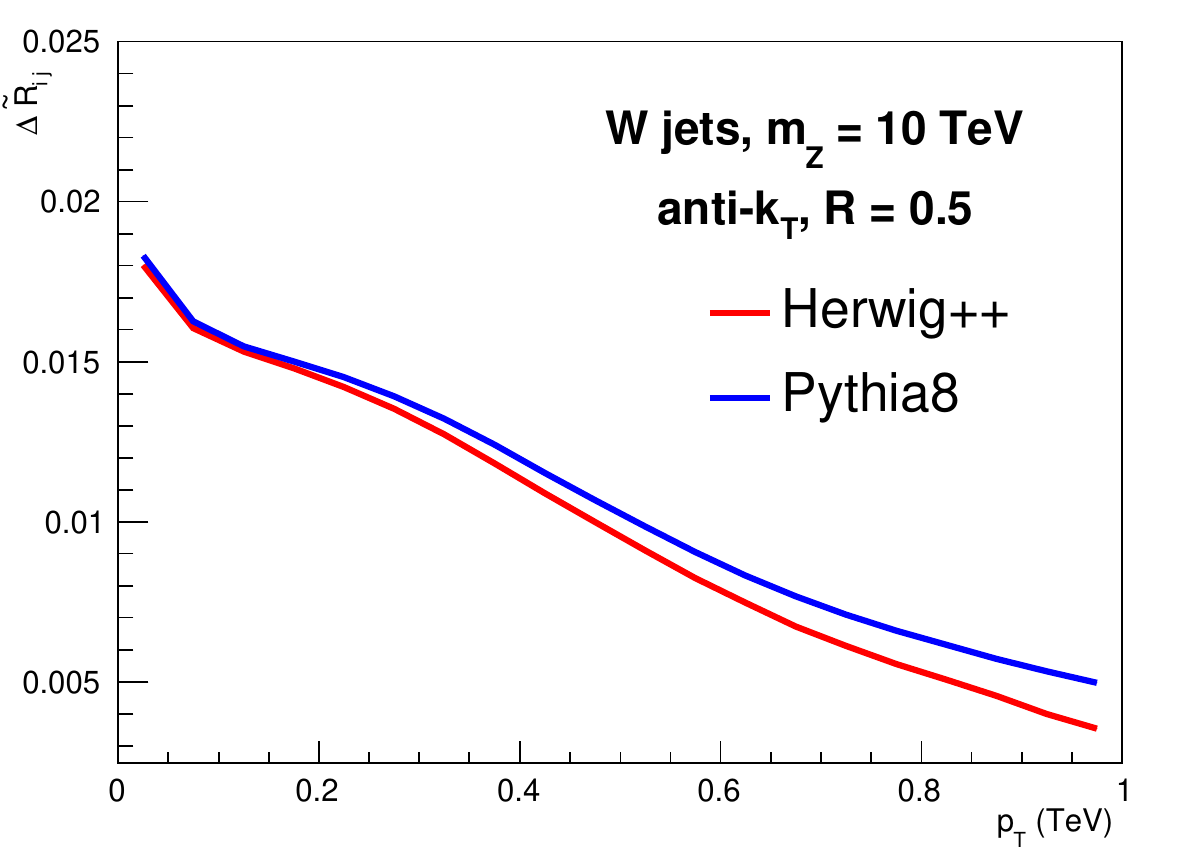}
\hspace{-0.02\textwidth}
\includegraphics[width=0.45\linewidth]{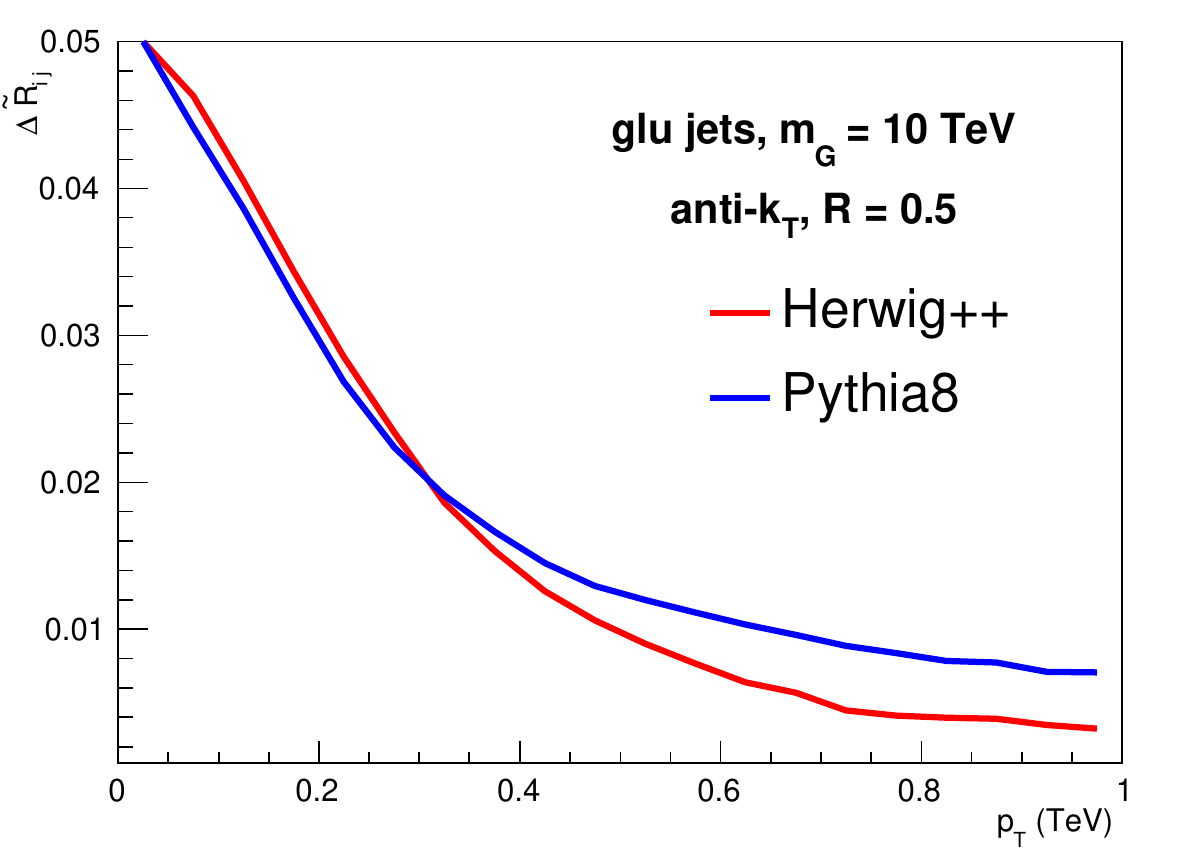}
\caption{
Median angular separation $\Delta \tilde R$ between charged tracks with $p_T>p_T^{\min}$ in $R=0.5$ anti-$k_T$ jets produced from 10 TeV resonance decays to tops, light QCD quarks, $W$s, and gluons.  
  } 
\label{fig:medR_10}
\end{figure}

\begin{figure}[h!]
\centering
\includegraphics[width=0.45\linewidth]{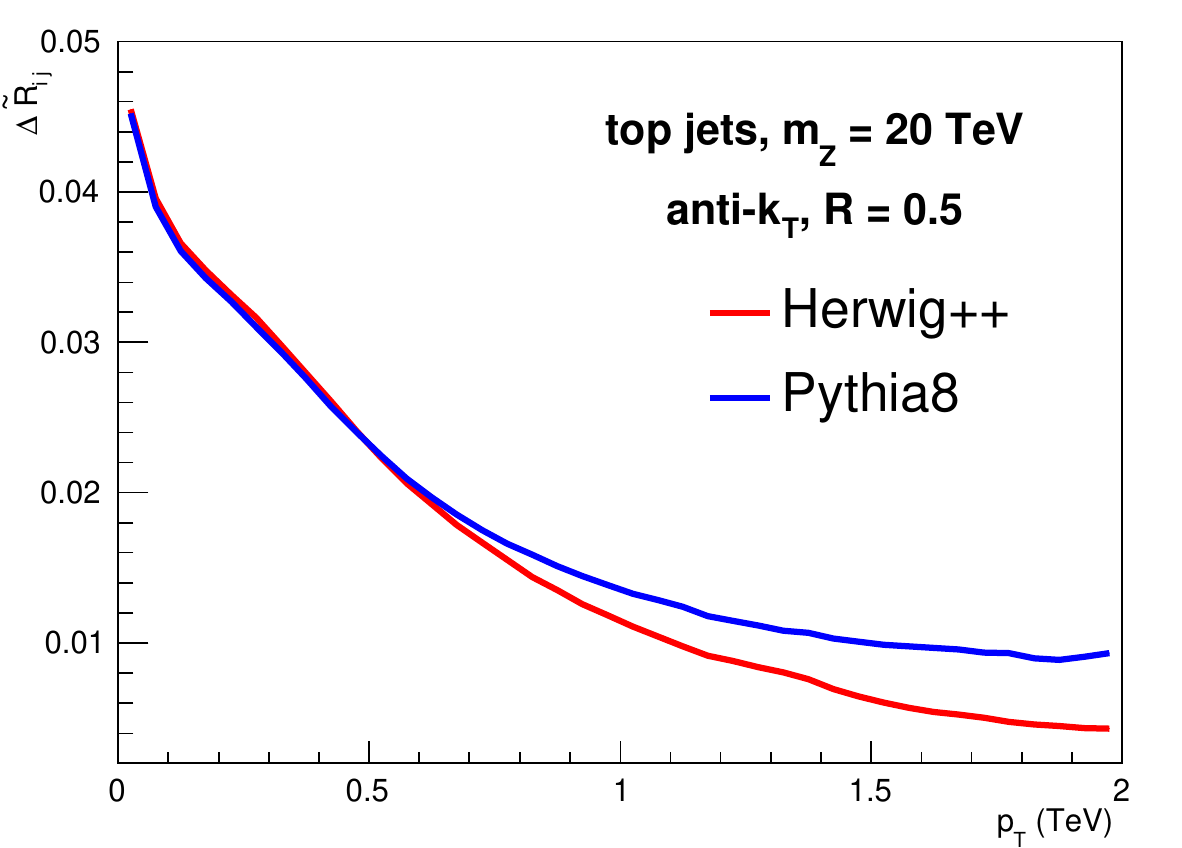}
\hspace{-0.02\textwidth}
\includegraphics[width=0.45\linewidth]{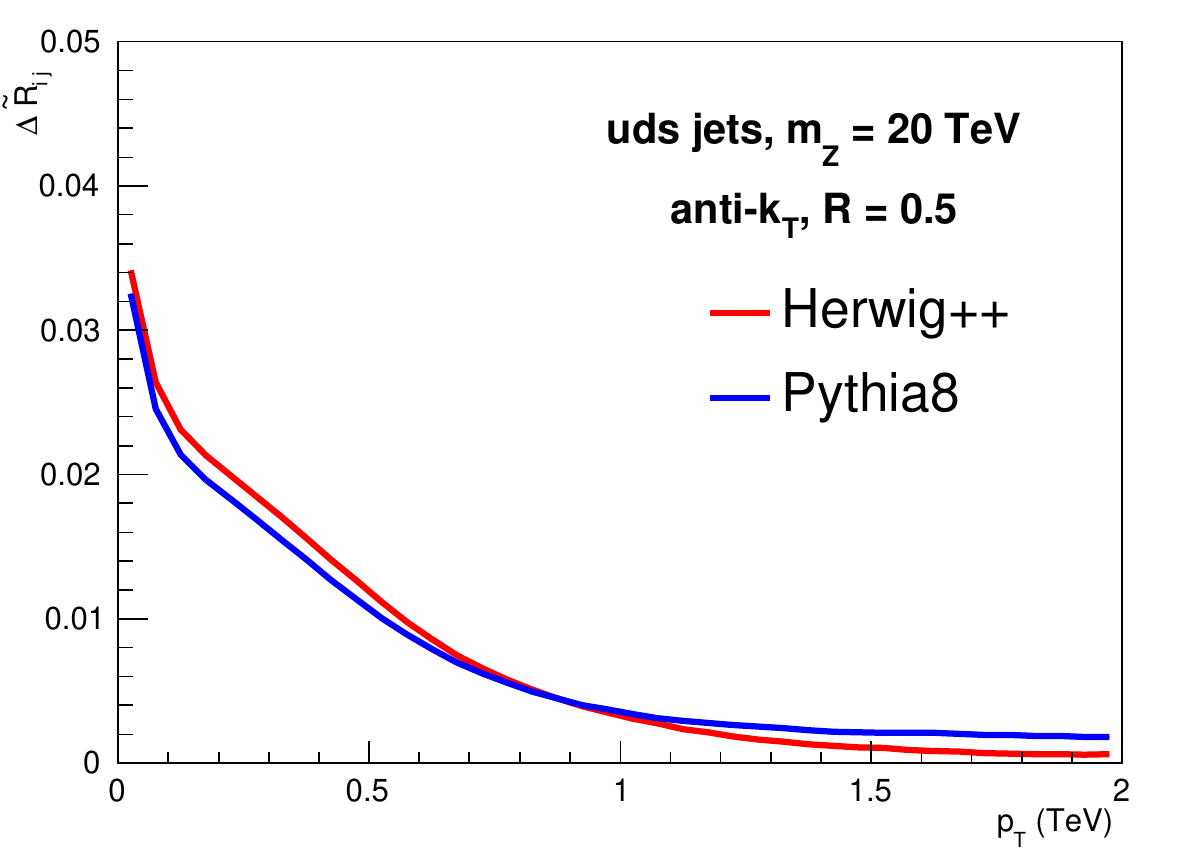}
\\
\vspace{-5mm}
\includegraphics[width=0.45\linewidth]{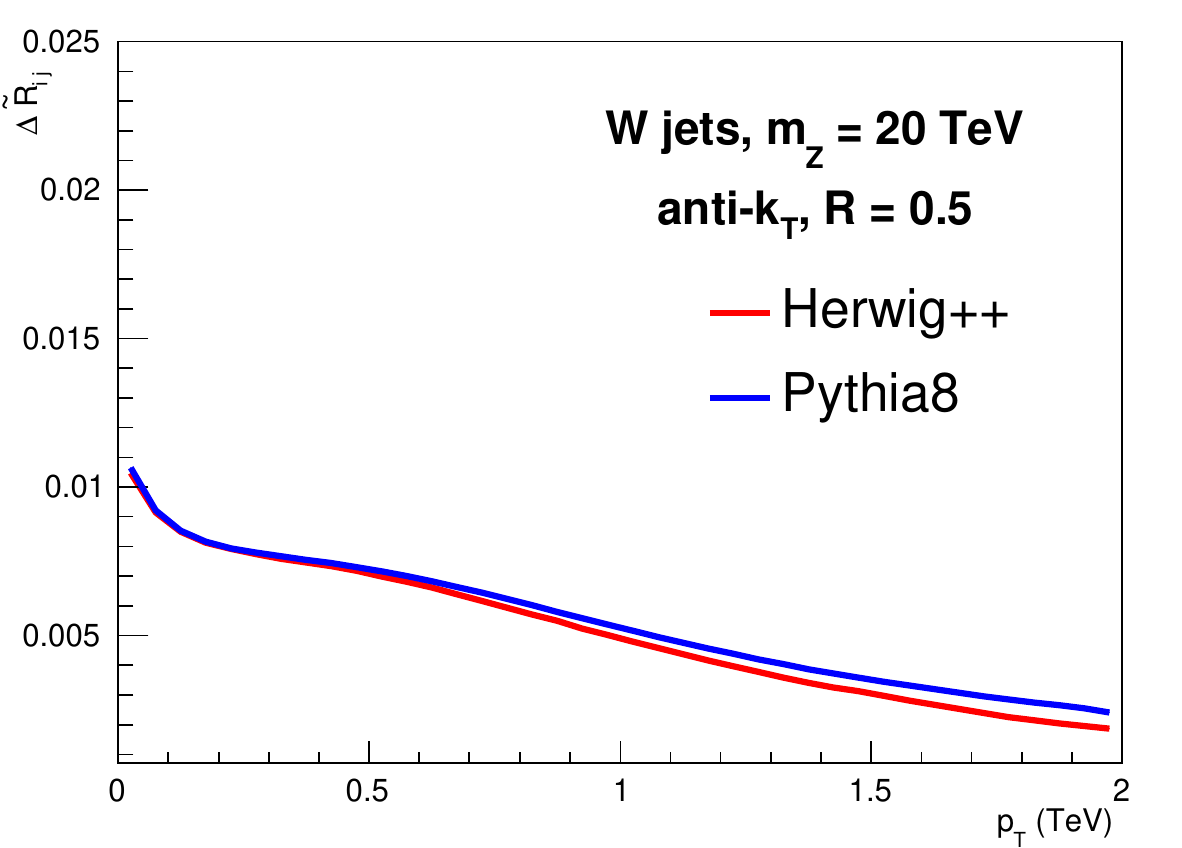}
\hspace{-0.02\textwidth}
\includegraphics[width=0.45\linewidth]{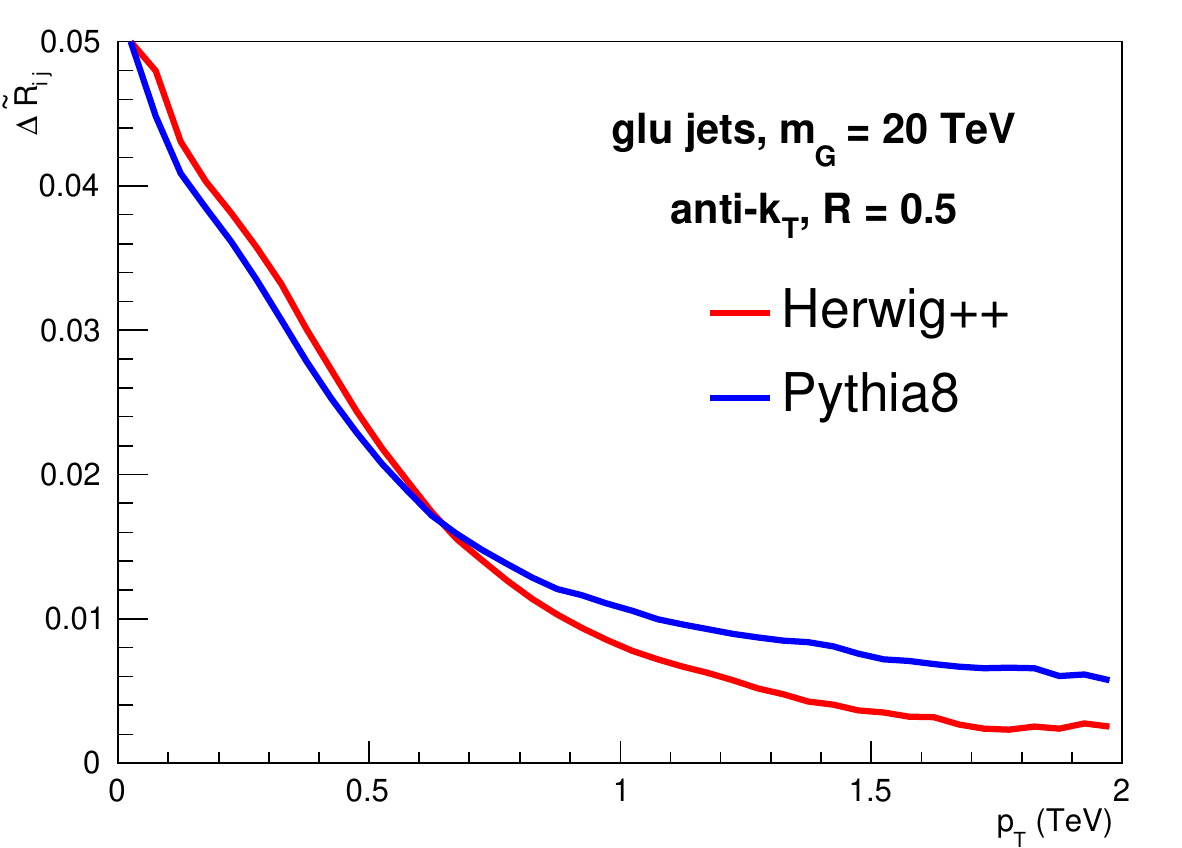}
\caption{
Median angular separation $\Delta \tilde R$ between charged tracks with $p_T>p_T^{\min}$ in $R=0.5$ anti-$k_T$ jets produced from 20 TeV resonance decays to tops, light QCD quarks, $W$s, and gluons.  
  } 
\label{fig:medR_20}
\end{figure}

\begin{figure}[h!]
\centering
\includegraphics[width=0.45\linewidth]{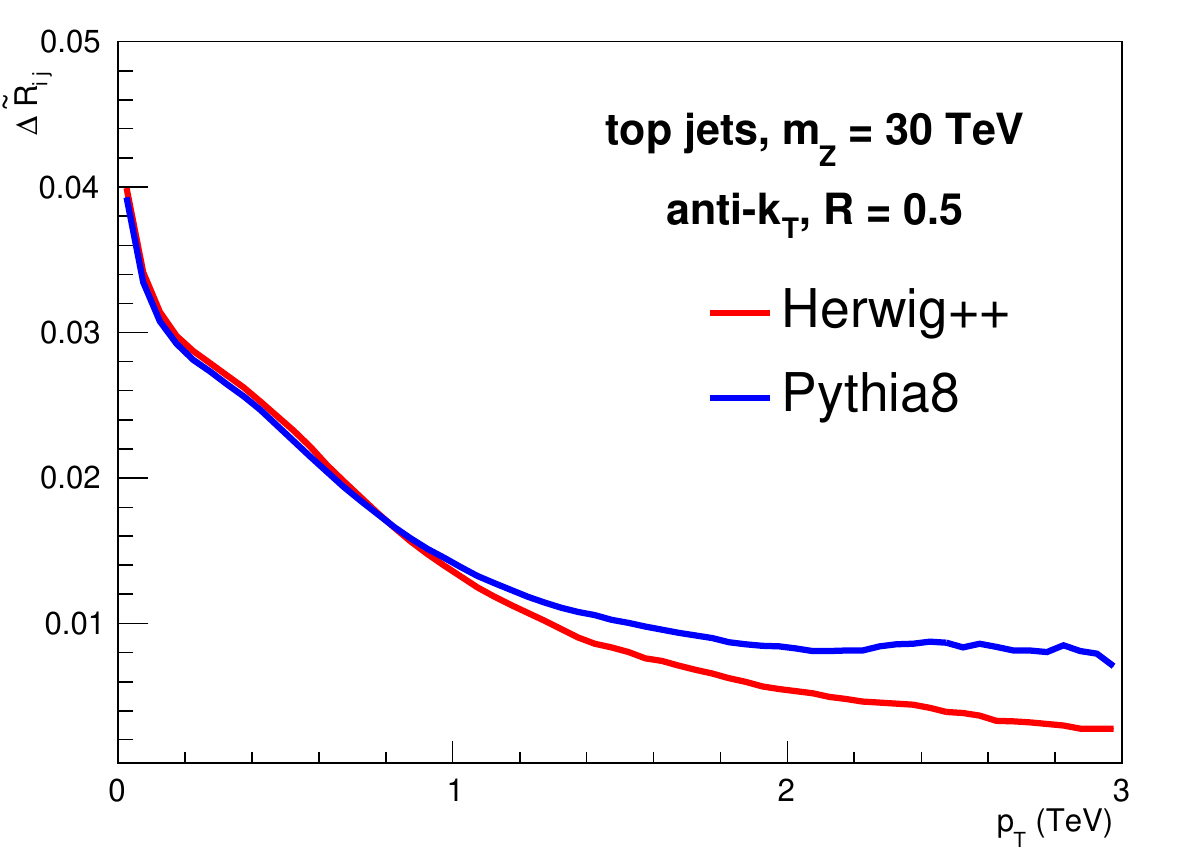}
\hspace{-0.02\textwidth}
\includegraphics[width=0.45\linewidth]{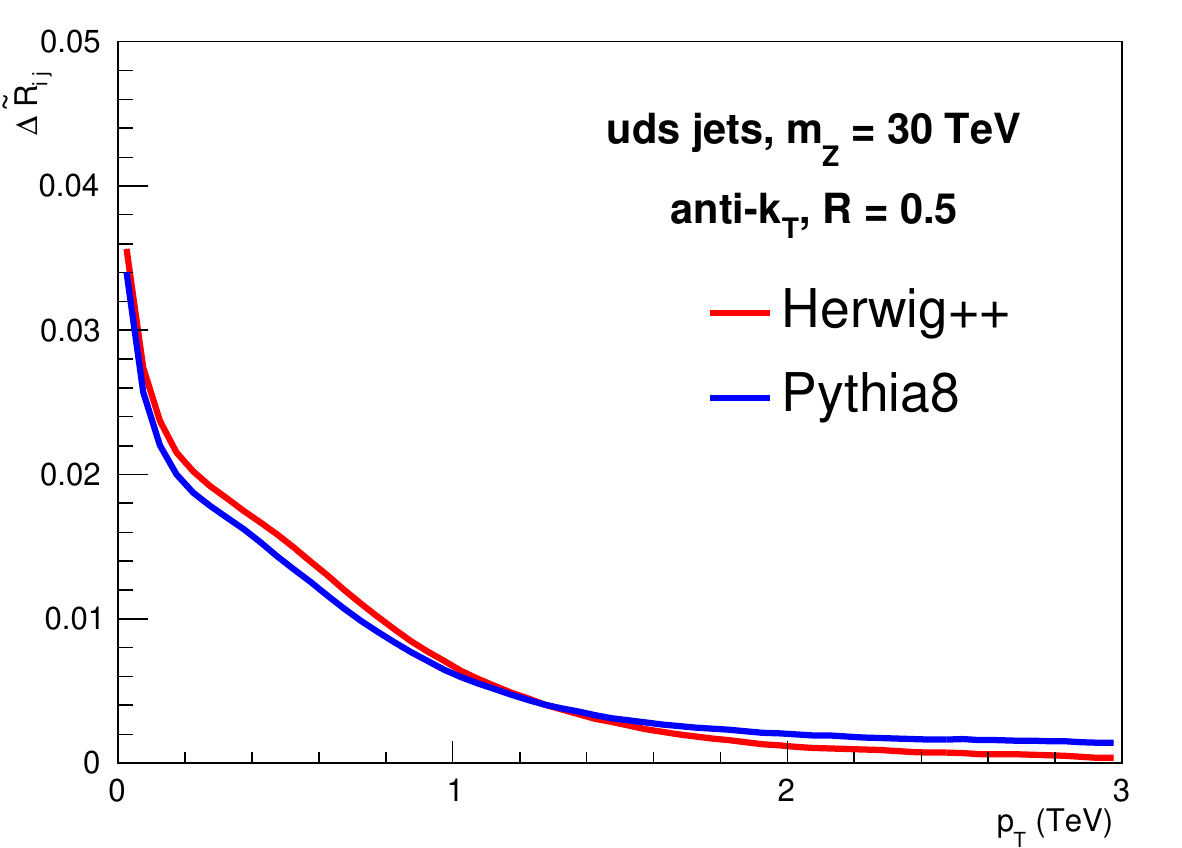}
\\
\includegraphics[width=0.45\linewidth]{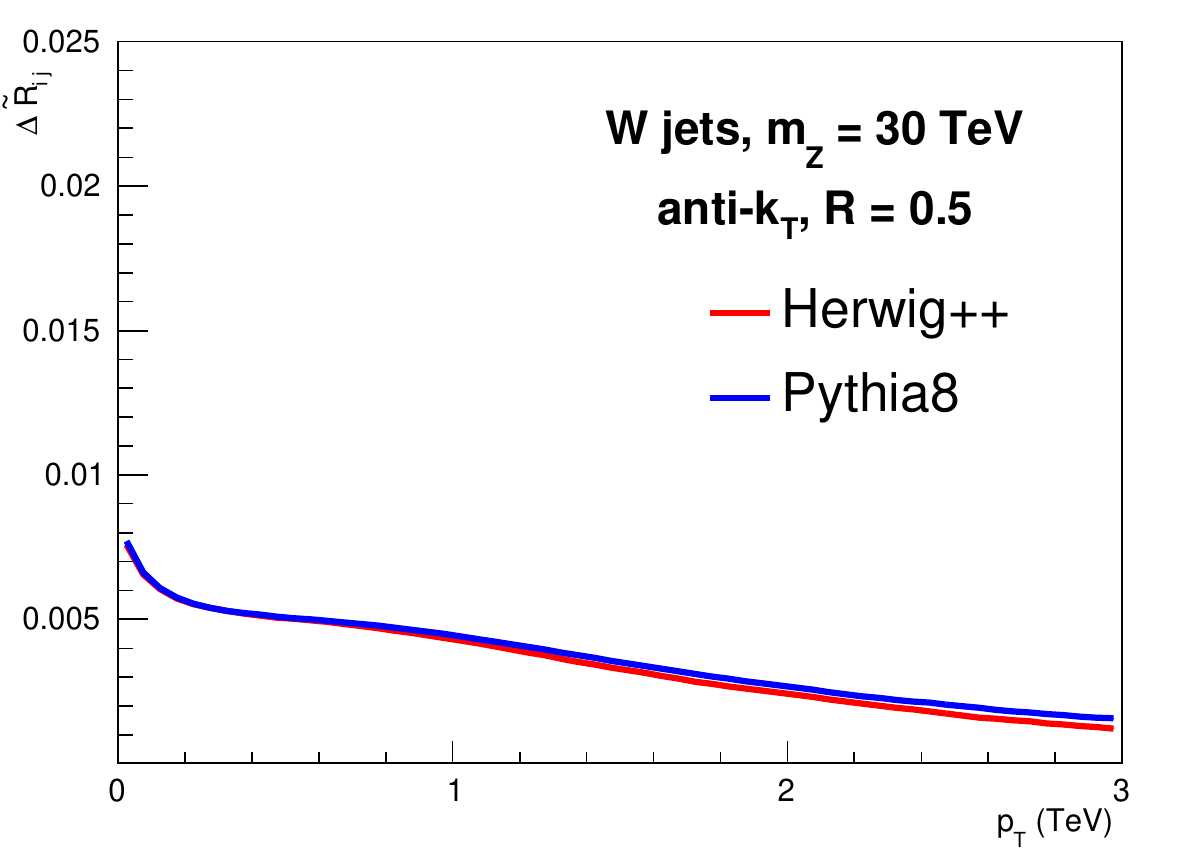}
\hspace{-0.02\textwidth}
\includegraphics[width=0.45\linewidth]{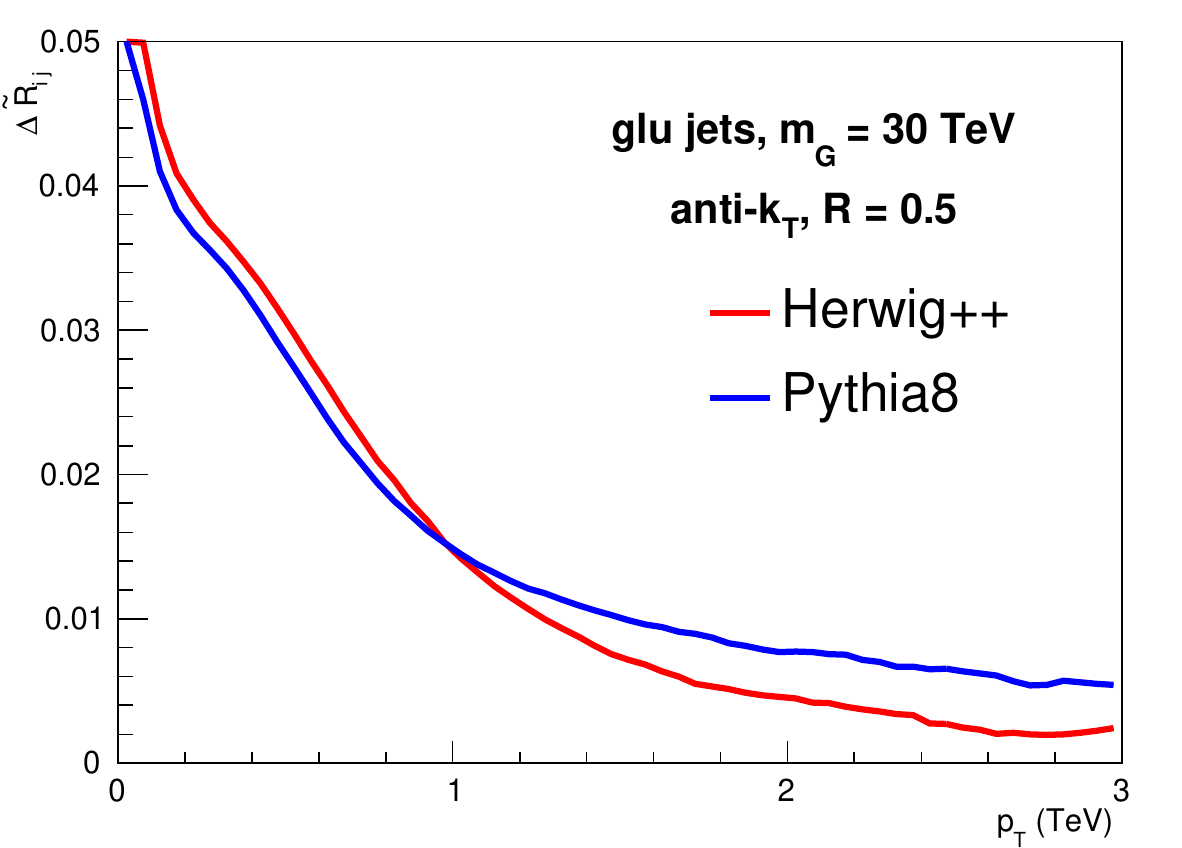}
\caption{
Median angular separation $\Delta \tilde R$ between charged tracks with $p_T>p_T^{\min}$ in $R=0.5$ anti-$k_T$ jets produced from 30 TeV resonance decays to tops, light QCD quarks, $W$s, and gluons.  
  } 
\label{fig:medR_30}
\end{figure}

\begin{figure}[h!]
\centering
\includegraphics[width=0.45\linewidth]{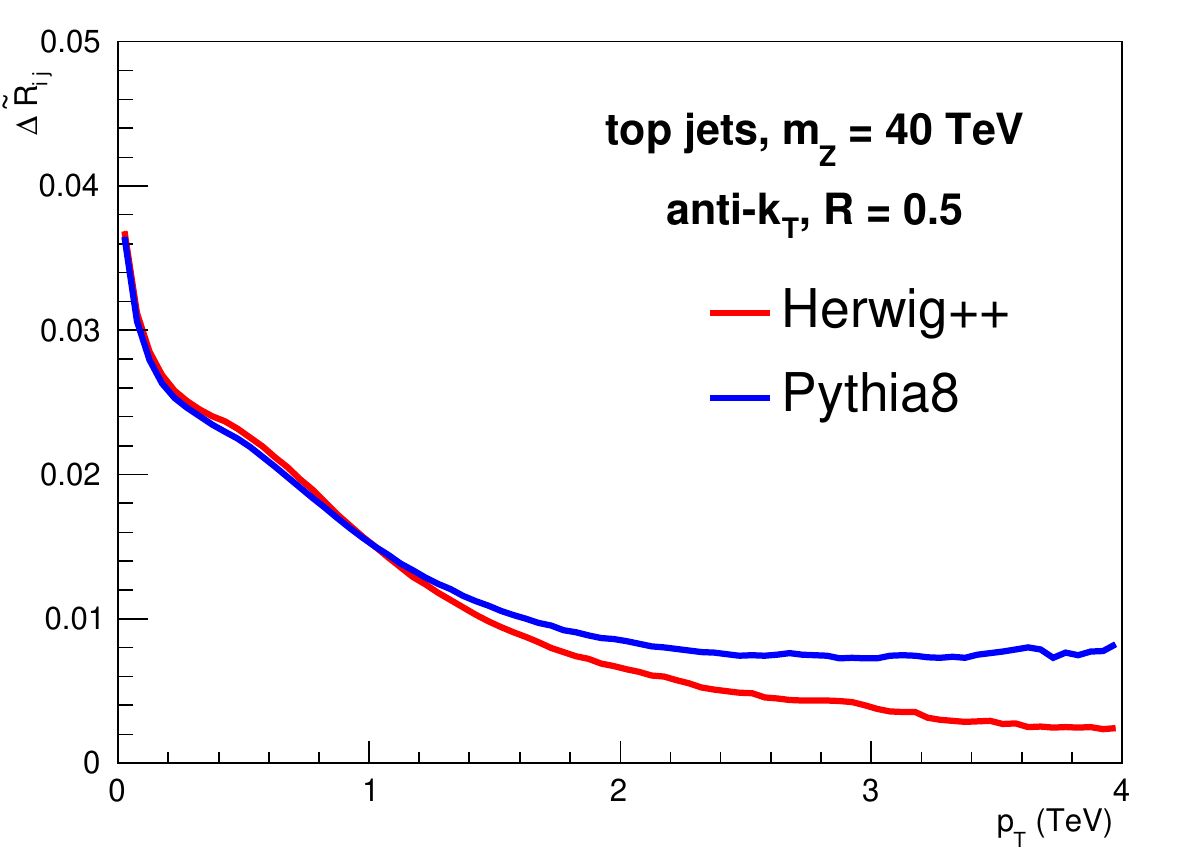}
\hspace{-0.02\textwidth}
\includegraphics[width=0.45\linewidth]{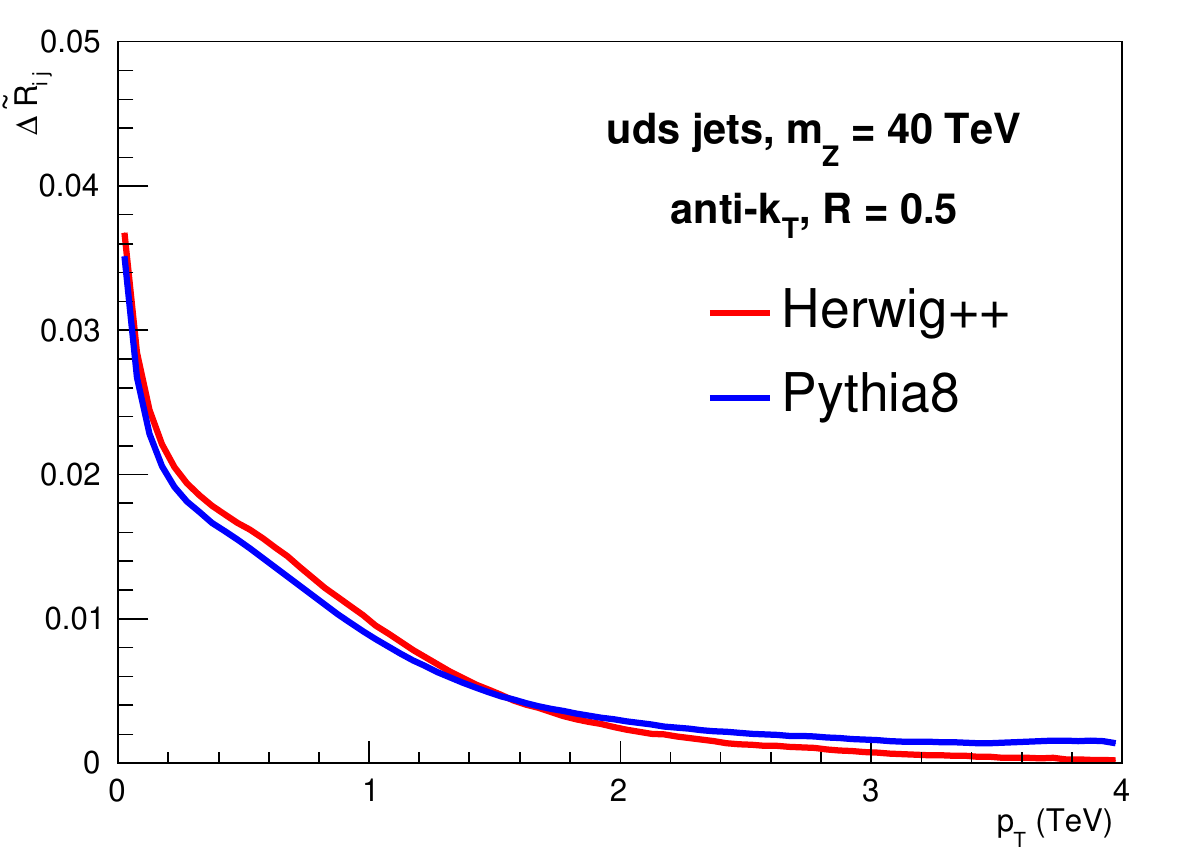}
\\
\includegraphics[width=0.45\linewidth]{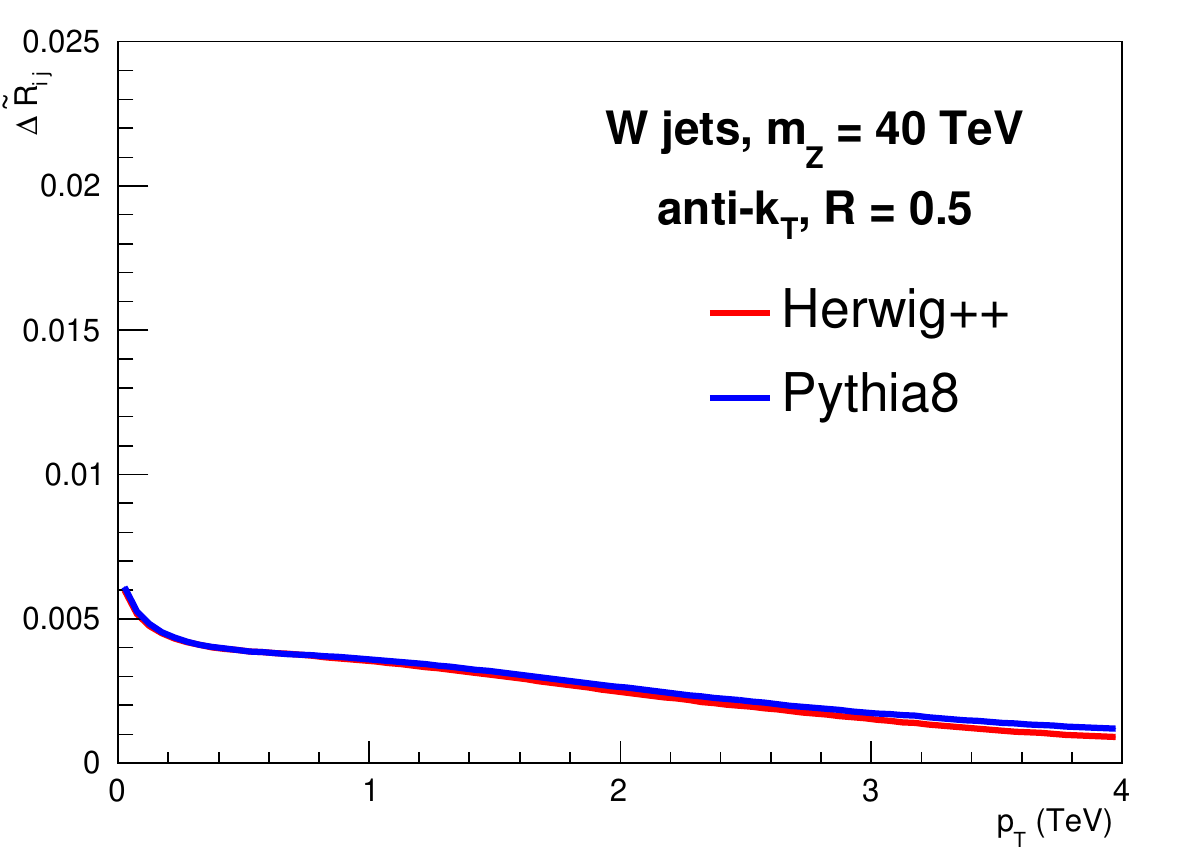}
\hspace{-0.02\textwidth}
\includegraphics[width=0.45\linewidth]{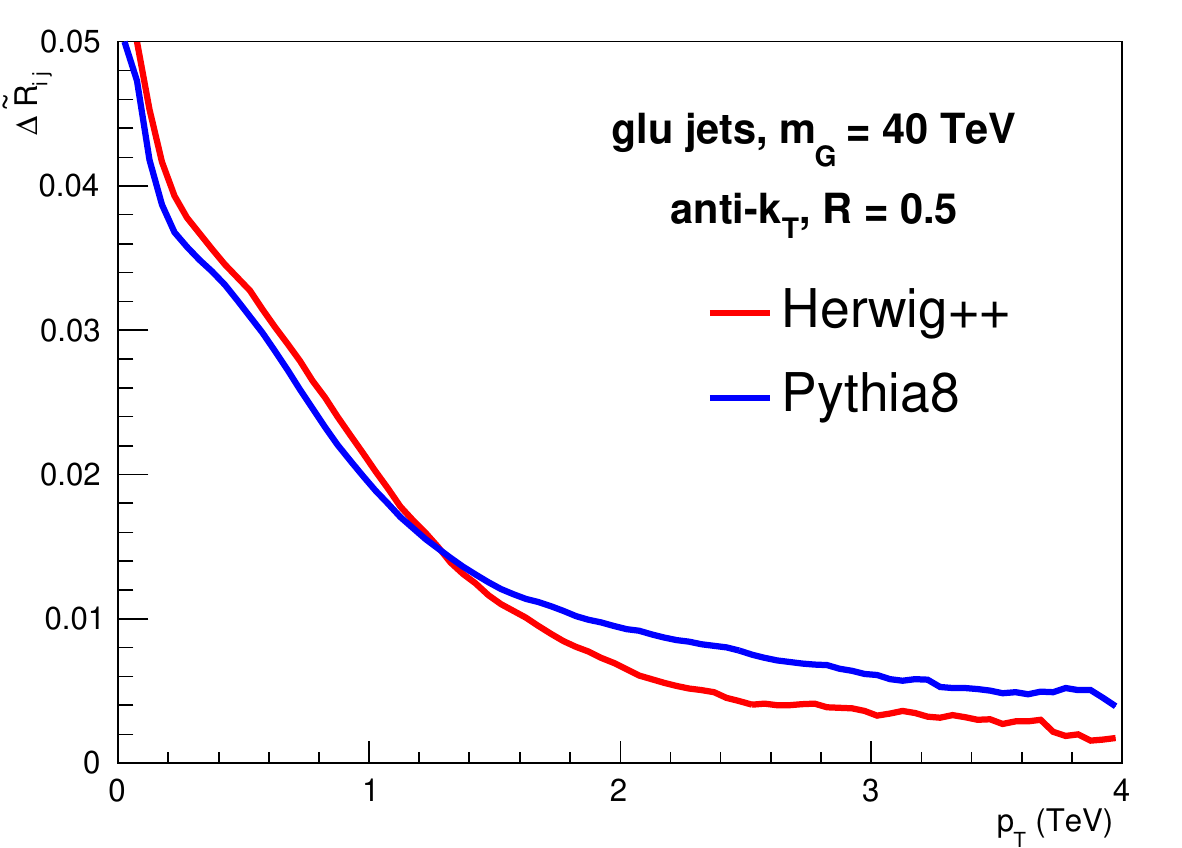}
\caption{
Median angular separation $\Delta \tilde R$ between charged tracks with $p_T>p_T^{\min}$ in $R=0.5$ anti-$k_T$ jets produced from 40 TeV resonance decays to tops, light QCD quarks, $W$s, and gluons.  
  } 
\label{fig:medR_40}
\end{figure}

\clearpage

\clearpage
\section{Multijets\footnote{Editors: S.~Badger and F.~Krauss}}
\label{ref:njets}

In this section we explore the total rates and distributions for final
states with multiple jets and photons. An overall feature of the
results presented here is the huge amount of multi-jet activity that
could be measured within the first few days of running. This opens up
many possibilities for searches of exotic physics beyond the Standard
Model such as black holes or instantaneous decaying into jets. A large
number or events containing systems with effective masses of 10 or
even 20 TeV would be observed which will also explore a region where
no prior experience of QCD exists.

A variety of different kinematic configurations are considered. These
can be broadly classified into two categories: \emph{democratic}, in
which cuts on the transverse momenta of all jets are treated equally,
and \emph{hierarchical}, in which harder cuts on the leading jet are
applied. The choices are known to affect the perturbative stability of
the observables which we investigate in sec.~\ref{secMJNLO}.

\subsection{Computational setup}

For the following studies, the \Sherpa event generation framework
\cite{Gleisberg:2003xi,Gleisberg:2008ta} has been used. Proton--proton
collisions at centre-of-mass energies of $100$ TeV are considered and,
in relevant cases, compared to collisions at LHC scale energies of 14
TeV to highlight interesting features of energy scaling.  Unless
stated otherwise, jets are reconstructed with the anti-$k_\perp$
algorithm with a radius parameter of $R=0.4$, using the \FastJet
package~\cite{Cacciari:2008gp,Cacciari:2011ma}. The Standard Model
input parameters are defined through the $G_\mu$ scheme. Unstable
fermions and bosons are treated through the complex mass
scheme~\cite{Denner:1999kn}.  All quarks apart from the top-quark are
assumed to be massless.  The effects of the top-quark are included in
the running of $\alpha_S$ for scales above the its mass.  For matrix
element generation and cross section calculations, the \Comix matrix
element generator~\cite{Gleisberg:2008fv} is employed. For the proton
PDFs the NNPDF3.0 NLO set~\cite{Ball:2014uwa} is used, which also
provides the strong coupling $\alpha_S$.  Renormalisation and
factorisation scales are defined in a process-specific way, and are
listed separately in the respective subsections.  For most
distributions the multijet merging technology
of~\cite{Catani:2001cc,Krauss:2002up,Hoeche:2009rj} \footnote{It is
  worth noting that other merging techniques exist, like for
  instance~\cite{Lonnblad:2001iq,Lonnblad:2011xx,Lonnblad:2012ng,
    Mangano:2001xp,Mangano:2006rw,Hamilton:2009ne}, which however by
  far and large have been shown to yield comparable results at lower
  energies, see for example~\cite{Alwall:2007fs}.} is employed, with
the parton shower built on Catani-Seymour subtraction kernels as
proposed in~\cite{Nagy:2005aa} and implemented
in~\cite{Schumann:2007mg}.

Next-to-leading order corrections are generated at fixed order using
\Sherpa together with the \textsc{NJet} one-loop matrix element
provider \cite{Badger:2012pg}. Real radiation is provided via the
Catani-Seymour subtraction method implemented in \Sherpa
\cite{Gleisberg:2007md} and the \Comix matrix element generator
\cite{Gleisberg:2008fv}. Root Ntuples are generated and analysed using
the CT14nlo PDF set \cite{Dulat:2015mca} which provides the strong
coupling $\alpha_S(m_Z) = 0.118$.

\subsubsection{Kinematic cuts}

Various cuts on the transverse momentum of the jets are considered and
specified in the later discussions. For runs with LO+PS/MEPS@LO no
additional kinematic requirements were taken for multi-jet
production. For processes involving photons, the additional constraint
that each photon should be at a radius of least $\Delta R \geq 0.4$
from every jet was imposed.

At NLO a mild rapidity cut on all jets and photons $|\eta_{j/\gamma}|
< 8$ was taken in addition to these requirements. At NLO care must be
taken to ensure photon final states are infrared safe. Accordingly, we
used the standard Frixione smooth cone isolation
\cite{Frixione:1998jh} with parameters $R=0.4$, $\varepsilon=0.1$ and
$n=1$.

\subsubsection{Scale choices}

In this section, we use a dynamic choice in general for the
factorisation and renormalisation scales, $\mu_{F/R}$, given by the
sum of transverse momenta
\begin{equation}
    \tfrac{1}{2} \hat{H}_T = \tfrac{1}{2}\left(\sum_{i} {p_{T,i}}\right)\,.
  \label{eq:HThat}
\end{equation}
For the fixed-order calculations in sec.~\ref{secMJNLO}, the sum runs
over final-state partons. This includes a single photon, if
present. For processes with two photons in the final state, we use
\begin{equation}
  \tfrac{1}{2} \hat{H}_T' = \tfrac{1}{2}\left(m_{T,\gamma\gamma} + \sum_{i\in \text{partons}} {p_{T,i}}\right)\,,
  \label{eq:mt2HThat}
\end{equation}
with $m_{T,\gamma\gamma}$ the transverse mass of the diphoton system.
For the leading-order \Sherpa setups considered in the following
section~\ref{sec:MJxsLO} the sum goes over anti-$k_T$ jets instead
(and photons), and is averaged:
\begin{equation}
    \bar{H}_T = \frac{1}{N_\text{jet} + N_\gamma} \sum_{i} {p_{T,i}}\,.
  \label{eq:mt2HThat2}
\end{equation}

\begin{table}[h!]
  \begin{center}
    \begin{tabular}{@{}l rrrr@{}}
      &
      \multicolumn{4}{c}{$j^n$ with $R\,=\,0.2$\tableHeaderVPhantom} \\
      \cline{2-5} \noalign{\smallskip}
      $n\;/\;\;p_{T,j} $ &
      \minptcellNoLine{50} & 
      \minptcellNoLine{100} & 
      \minptcellNoLine{250} & 
      \minptcellNoLine{1000}
      \\
      \hline \noalign{\smallskip}
      $2$ & \mub{315(1)}  & \mub{29.9(1)}  & \nb{1045(4)}   & \pb{3483(10)}\\
      $3$ & \mub{38.0(3)} & \mub{2.51(2)}  & \nb{54.1(3)}   & \pb{72.0(4)}\\
      $4$ & \mub{13.5(1)} & \nb{665(7)}    & \nb{10.0(1)}   & \pb{6.83(7)}\\
      $5$ & \mub{4.98(7)} & \nb{199(2)}    & \nb{2.02(2)}   & \fb{621(4)}\\
      $6$ & \mub{2.18(2)} & \nb{65.8(7)}   & \pb{456(5)}    & \fb{57.8(4)}\\
      $7$ & \mub{0.93(2)} & \nb{23.5(3)}   & \pb{112(1)}    & \fb{7.21(6)}\\
      $8$ & \mub{0.413(9)}   & \nb{8.1(2)}    & \pb{29.7(4)}   & \fb{0.832(8)}\\ \noalign{\smallskip}
      &
      \multicolumn{4}{c}{$j^n$ with $R\,=\,0.4$\tableHeaderVPhantom} \\
      \cline{2-5} \noalign{\smallskip}
      $n\;/\;\;p_{T,j} $ &
      \minptcellNoLine{50} & 
      \minptcellNoLine{100} & 
      \minptcellNoLine{250} & 
      \minptcellNoLine{1000}
      \\
      \hline \noalign{\smallskip}
      $2$ & \mub{315(1)}  & \mub{29.9(1)} & \nb{1045(4)}   & \pb{3483(10)}  \\
      $3$ & \mub{34.6(3)} & \mub{2.31(1)} & \nb{49.9(3)}   & \pb{66.7(3)}  \\
      $4$ & \mub{10.5(1)} & \nb{539(5)}   & \nb{7.82(8)}   & \pb{4.93(4)}  \\
      $5$ & \mub{3.40(4)} & \nb{130(1)}   & \nb{1.247(9)}   & \fb{358(2)}   \\
      $6$ & \mub{1.21(1)} & \nb{35.0(3)}  & \pb{229(2)}    & \fb{28.5(1)}  \\
      $7$ & \mub{0.406(6)}   & \nb{9.42(9)}  & \pb{42.0(4)}   & \fb{2.35(2)}  \\
      $8$ & \mub{0.154(2)}   & \nb{2.66(4)}  & \pb{8.12(9)}   & \fb{0.195(1)}   \\
    \end{tabular}
    \caption{Leading order cross sections for the production of anti-$k_T$ 
      jets with varying minimal $p_T$, ranging from 50~GeV to 1~TeV and 
      two different values of the jet algorithm radius parameter, $R=0.2$ 
      and $R=0.4$. For the calculation the scales $\mu_{F,R} = \bar{H}_T$
      have been used.
      \label{Table:Table_JetInclusive}}
  \end{center}
\end{table}

\subsection{Leading order inclusive cross sections and distributions
  \label{sec:MJxsLO}}

We performed the calculations in this section with \Sherpa at LO,
unless stated otherwise.  We begin with the inclusive multi-jet
production rates for up to 8 eight final state jets in
Table~\ref{Table:Table_JetInclusive}.  The rates are calculated with
varying a minimal $p_T$ cut, ranging from 50~GeV to 1~TeV and two
different values of the anti-$k_T$ radius parameter, $R=0.2$ and
$R=0.4$. Within the first days of running nanobarn cross-section
events with 3 or 4 jets of 250 GeV could be observed and final states
with up to eight 1 TeV jets will be observable with the order of a few
thousand events with the planned integrated luminosity.  We also show
the scaling behaviour of the ratio $\sigma(R=0.4)/\sigma(R=0.2)$ in
Figure \ref{Figure:Table_n_xs_R_ratios} using various minimum $p_T$
cuts. Assuming that jets are not overlapping, i.e.\ their distance
$\Delta R$ in $\eta$-$\phi$ being $\Delta R > 2R$, the total area $a$
they cover is given by
\begin{equation}
  a = N_J \pi R^2 \approx
  \begin{cases}
      0.5 \cdot N_\text{jet}  & \text{for}\;\;\; R = 0.4, \\
      0.13 \cdot N_\text{jet} & \text{for}\;\;\; R = 0.2.
  \end{cases}
  \label{eq:multijetarea}
\end{equation}
\begin{figure}[h!]
  \subfloat[Cross section comparison between collisions at
    $\sqrt{s}=\SI{14}{TeV}$ and $\sqrt{s}=\SI{100}{TeV}$.
    \label{fig:JetRatesLO14vs100}]{
    \includegraphics[width=0.46\linewidth]{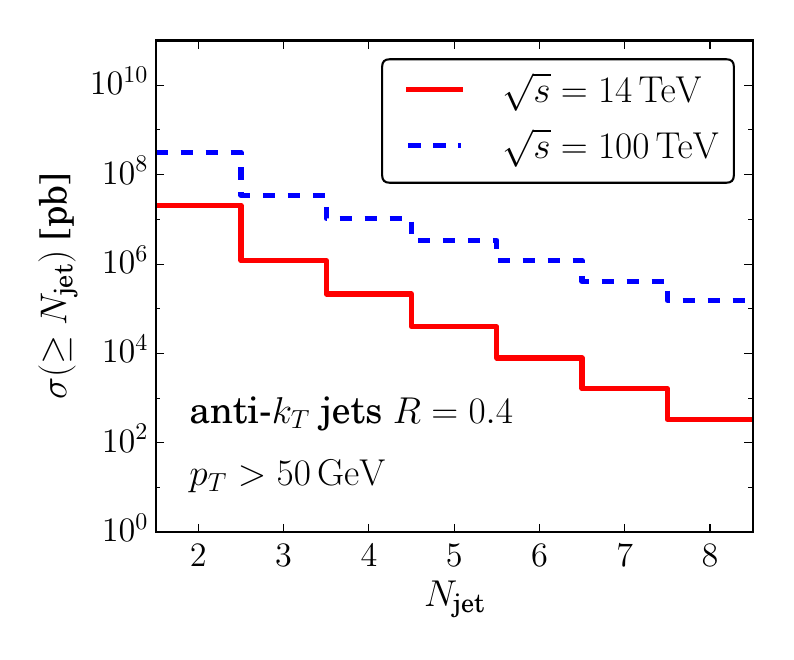}
  }
  \hfill
  \subfloat[Cross section ratios for different jet radii $R=0.4$ and $R=0.2$.
    Four different $p_T$ cuts are employed.\label{Figure:Table_n_xs_R_ratios}]{
    \includegraphics[width=0.46\linewidth]{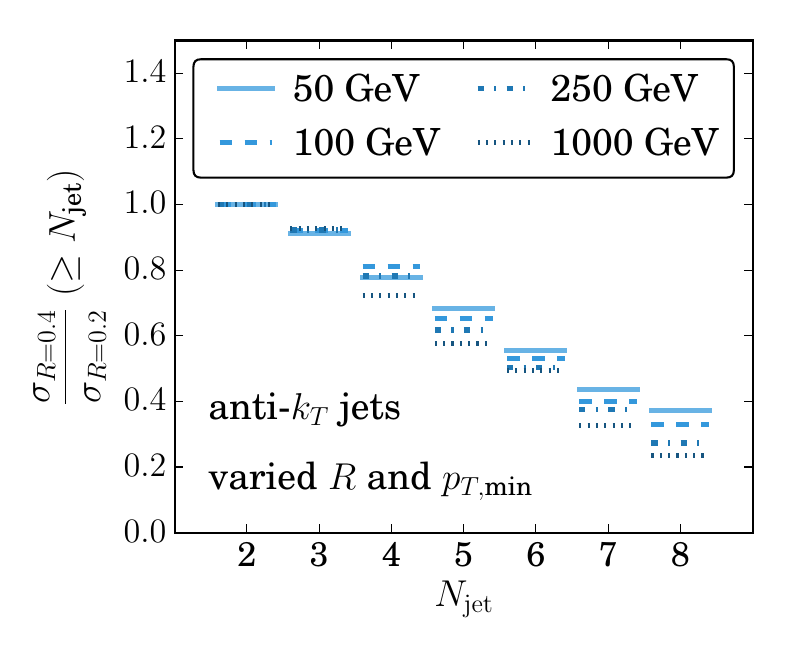}
  }
  \caption{Inclusive multiplicity cross sections for anti-$k_T$ jet production
    at leading order for $pp$ collisions. The scales are set to
    $\mu_{F,R}=\bar{H}_T$.
\label{figJetRatesLO14vs100}}
\end{figure}
For a detector with a coverage over 10 units in pseudo-rapidity,
similar to ATLAS or CMS at the LHC, the total acceptance region is
$2\pi \Delta \eta \approx 63$. In both cases of $R=0.4$ or $R=0.2$ the
total coverage is much greater than the area of the jets and so the
scaling behaviour is not driven by phase-space effects in acceptance
but expected to be defined through QCD dynamics. The total inclusive
cross sections are compared to the those at 14 TeV in Figure
\ref{fig:JetRatesLO14vs100} where one clearly sees the increasing
multiplicity of events at the higher centre-of-mass energy.

\begin{table}[h!]
  \begin{center}
    \begin{tabular}{@{}l rrrrr @{}}
      & 
      \multicolumn{5}{c}{$j^n$ with $p^{\rm min}_{T,j}\,\geq\,50$ GeV varying $p^{\rm lead}_{T,j}$\tableHeaderVPhantom}  \\
      \cline{2-6} \noalign{\smallskip}
      $n\;/\;\;p^{\rm lead}_{T,j} $ &
      \minptcellNoLine{500} &
      \minptcellNoLine{1000} & 
      \minptcellNoLine{2000} & 
      \minptcellNoLine{5000} & 
      \minptcellNoLine{10000} \\
      \hline \noalign{\smallskip}
      $2$ & \nb{67.4(2)} & \nb{3.48(1)} & \pb{139(1)} & \pb{1.06(1)} & \fb{11.3(1)}\\
      $3$ & \nb{178(1)} & \nb{11.0(1)} & \pb{485(3)} & \pb{3.91(2)} & \fb{39.3(1)}\\
      $4$ & \nb{214(2)} & \nb{16.9(1)} & \pb{864(8)} & \pb{7.39(7)} & \fb{74.6(6)}\\
      $5$ & \nb{191(1)} & \nb{18.7(1)} & \pb{1093(7)} & \pb{10.6(1)} & \fb{102(1)}\\
      $6$ & \nb{136(2)} & \nb{16.3(2)} & \pb{1133(1)} & \pb{11.9(1)} & \fb{113(1)} \\
    \end{tabular}
     \caption{Leading order cross sections for the
     production of anti-$k_T$ jets with minimal $p_T$ of $50$~GeV
     different values of leading-jet transverse momentum.
     For the calculation the scales $\mu_{F,R} = \bar{H}_T$
     have been used.
       \label{Table:Table_JetsStaggered}}
  \end{center}
\end{table}
\begin{figure}[h!]
  \begin{center}
    \includegraphics[width=0.5\linewidth]{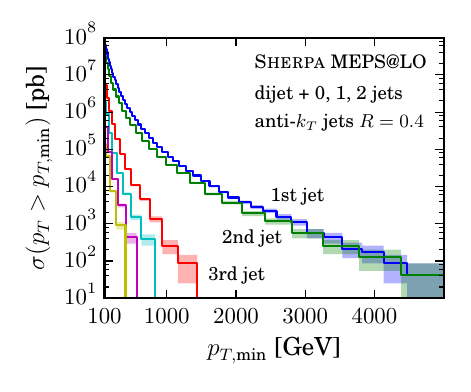}
    \caption{Cumulative leading order $p_T$ distributions for the first six highest $p_T$ jets ordered in $p_T$. The labels for the 4th through 6th jet are omitted,
    but follow the natural order.\label{fig:LOMJptdistr}}
  \end{center}
\end{figure}
In Table \ref{Table:Table_JetsStaggered} we consider inclusive cross
sections based on the corresponding leading order matrix elements for
multijet events with a minimum $p_T$ of $50$~GeV and different values
of minimum leading-jet transverse momentum. One sees again that
extreme kinematic configurations are clearly accessible, opening up
unexplored areas of QCD dynamics.  However, one can observe that some
of the leading order rate estimates do not decrease with increasing
final state multiplicity. Having a much harder cut on the leading jet
than on the subleading ones induces large scale hierarchies and thus
necessitates to consider higher order (logarithmic) corrections, e.g.\
through a parton shower simulations, cf.\ Secs.~\ref{secMJNLO} and
\ref{sec:MJscaling}. Turning our attention to the $p_T$ spectra, in
Fig.~\ref{fig:LOMJptdistr} we show cumulative distributions for a
democratic cut of 1 TeV on all jets for the first 6 jets ordered in
$p_T$. This sample now has been generated using a MEPS@LO setup with
matrix elements for up to two additional jets on top of the dijet core
process merged together and dressed with parton showers. Though the
energy distribution for the highest multiplicity jets fall quickly
many events where the 6th jet has still more than $400$--$500$ GeV
will be observed.  The leading jets are accessible at energies much
greater than 3 TeV, which we will explore further in the next section.

\subsubsection{Jet production in association with one or two photons}

Multijet events in association with photons are important backgrounds
to new physics searches. Ratios of $Z/\gamma$ production can be used
to estimate missing transverse energy from decays of the Z boson into
neutrinos. Di-photon signals are particularly important when studying
Higgs or potentially higher mass scalar resonances.

\begin{table}[h!]
  \begin{center}
    \begin{tabular}{@{}l rrrr@{}}
      &
      \multicolumn{4}{c}{$\gamma j^n$\tableHeaderVPhantom} \\
      \cline{2-5} \noalign{\smallskip}
      $n\;/\;\;p_{T,j} $ &
      \minptcellNoLine{50} & 
      \minptcellNoLine{100} & 
      \minptcellNoLine{250} & 
      \minptcellNoLine{1000}
      \\
      \hline \noalign{\smallskip}

	$1$ & \nb{75.19(8)} & \nb{9.38(2)} & \pb{479.0(9)} & \pb{3.045(6)}\\
	$2$ & \nb{27.3(1)}  & \nb{7.62(3)} & \pb{932(4)}   & \pb{14.31(5)}\\
	$3$ & \nb{14.8(2)}  & \nb{2.37(3)} & \pb{129(1)}   & \fb{573.(4)} \\
	$4$ & \nb{6.95(6)}  & \pb{757(6)}  & \pb{26.5(2)}  & \fb{52.1(5)} \\
	$5$ & \nb{3.20(3)}  & \pb{253(2)}  & \pb{5.61(4)}  & \fb{4.51(3)} \\
	$6$ & \nb{1.43(2)}  & \pb{82.7(8)} & \pb{1.20(2)}  & \fb{0.404(3)}\\
	$7$ & \nb{0.603(7)} & \pb{27.1(4)} & \pb{0.262(3)} & \belowfb     \\

      \noalign{\smallskip}
      &
      \multicolumn{4}{c}{$\gamma\gamma j^n$} \\  
      \cline{2-5} \noalign{\smallskip}
      $n\;/\;\;p_{T,j} $ &
      \minptcellNoLine{50} & 
      \minptcellNoLine{100} & 
      \minptcellNoLine{250} & 
      \minptcellNoLine{1000}
      \\
      \hline \noalign{\smallskip}
      $0$ & \pb{47.7(1)} & \pb{47.7(1)} & \pb{47.7(1)} &  \pb{47.7(1)} \\
      $1$ & \pb{29.74(7)} &  \pb{13.56(3)} & \pb{2.007(5)} &  \fb{35.1(1)}  \\
      $2$ & \pb{30.9(2)}  &  \pb{12.02(7)} & \pb{2.43(1)}  &  \fb{84.3(4)}  \\
      $3$ & \pb{21.0(8)}  &  \pb{5.35(4)}  & \fb{532(4)}   &  \fb{5.04(4)}  \\
      $4$ & \pb{12.9(1)}  &  \pb{2.25(1)}  & \fb{131.2(8)} &  \ab{519(2)}   \\
      $5$ & \pb{7.02(6)}  &  \fb{847(8)}   & \fb{30.6(3)}  &  \ab{49.2(3)}  \\
    \end{tabular}
     \caption{Leading order cross sections for the production of anti-$k_T$ jets
     in association with one or two photons. Democratic cuts on all jet $p_T$
     are taken at 50, 100, 250 and 1000 GeV. The photon transverse momenta
     must be larger than $p_{T,\gamma} > 50$ GeV and separated from all jets
     by at least $\Delta R \geq 0.4$.
     \label{Table:Table_PhotonJetInclusive}}
  \end{center}
\end{table}
Table~\ref{Table:Table_PhotonJetInclusive} shows leading order, i.e.\
pure tree-level, cross sections for the production of one or two
photons in association with jets with varying minimal $p_T$, ranging
from 50 GeV to 1 TeV and fixed $R=0.4$.  The transverse momentum of
the photon(s) must satisfy $p_{T,\gamma} \geq \SI{50}{GeV}$, and the
photon(s) must be separated from jets or other photons by at least
$\Delta R \geq 0.4$. Even though the additional powers of $\alpha$
lower the production rates considerably diphoton production with up to
2 or 3 TeV jets could be observed with the full integrated luminosity.

\begin{figure}[hbtp]
  \begin{center}
    \includegraphics[width=\linewidth]{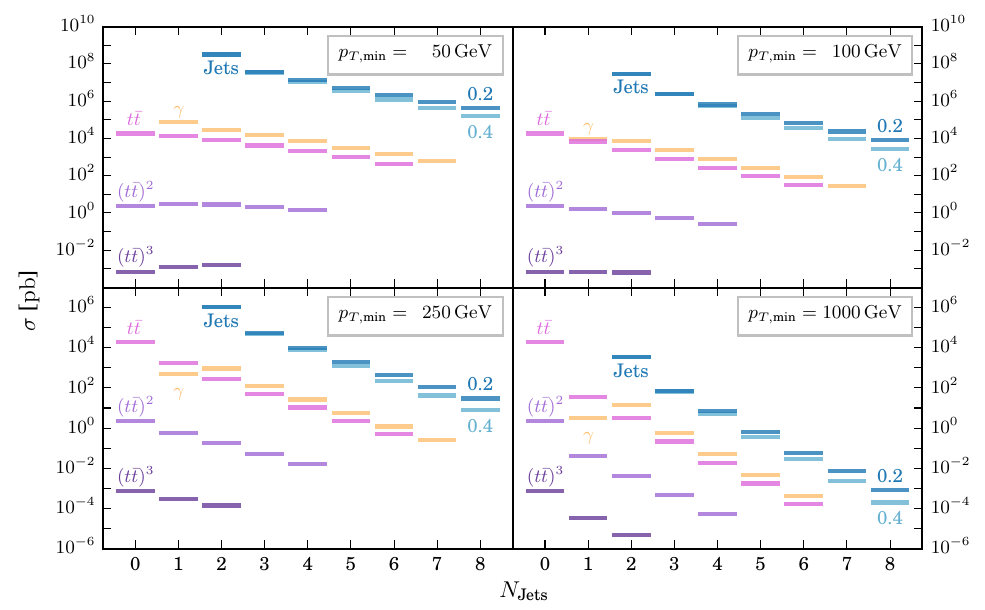}
    \caption{Inclusive cross section comparison between various QCD processes for different $p_{T,\text{min}}$. \label{fig:JetRatesLOQCD}}
  \end{center}
\end{figure}

To summarise the leading order results in this section we collect a
number of multi-jet QCD processes in
Figure~\ref{fig:JetRatesLOQCD}. For four different values of the
minimum $p_T$ we show pure jet productions with up to 8 jets and
single photon with up to 7 jets. As a comparison we also show top pair
production with up to 6 jets, two quark pairs with up to 4 jets and
three top pairs with up to two jets. The fact that the latter
processes are accessible with relatively high-$p_T$ jets impressively
demonstrates the degree to which QCD can be studied in the 100 TeV
environment, opening up huge amounts of phase-space for new physics
searches.

\subsection{NLO cross sections and K-factors}
\label{secMJNLO}

\subsubsection{Multijet production}

\begin{table}
\begin{center}
  \begin{tabular}[h]{cccccccc}
    & \multicolumn{7}{c}{$j^n$} \\
    \hhline{~-------}
    & \multicolumn{3}{c}{$p_T^{\rm min}> 50$ GeV} && \multicolumn{3}{c}{$p_T^{\rm min}> 500$ GeV} \\
    \hhline{~---~---}
    $n$ & LO & NLO & K && LO & NLO & K \\
    2 & $289.0^{+8.7}_{-13.6}$ $\mu$b & -- & -- && $66.0^{+12.2}_{-9.8}$ nb & -- & -- \\
    3 & $28.5^{+7.1}_{-5.4}$ $\mu$b & $15.1^{+3.1}_{-6.8}$ $\mu$b & $0.5^{+0.26}_{-0.30}$ && $1.7^{+0.6}_{-0.4}$ nb & $1.4^{+0.0}_{-0.3}$ nb & $0.8^{+0.27}_{-0.31}$ \\
    4 & $6.9^{+2.9}_{-1.9}$ $\mu$b & $2.2^{+1.3}_{-3.4}$ $\mu$b & $0.3^{+0.40}_{-0.45}$ && $153.2^{+68.3}_{-44.6}$ pb & $132.8^{+0.0}_{-27.9}$ pb & $0.9^{+0.34}_{-0.39}$
  \end{tabular}
\end{center}
\caption{Inclusive cross-sections for multijet production at NLO and LO using democratic cuts of $50$ and $500$ GeV.
  Renormalisation and factorisation scales are chosen equal with a central values $\mu_R = \mu_F = \hat{H}_T/2$
  with theoretical uncertainty estimated through variations over the range $[1/2,2]$. Cross-sections for massless $2\to 2$
  scattering are not well defined at NLO so the results are omitted (see footnote on page \pageref{footnote:dijet}). \label{tab:MJNLOXS}}
\end{table}
To study the impact of NLO correction representative NLO/LO
$K$-factors are presented with democratic cuts on all jets and
hierarchical cuts on the leading jet.  In Table \ref{tab:MJNLOXS} we
show the LO and NLO cross-sections for up to four jets with democratic
cuts on all jet transverse momenta of $50$ or $500$ GeV.  Since the
back-to-back configuration for di-jets cause the NLO phase-space to
become singular the cross-section is not well defined and the numbers
are not quoted%
\footnote{%
  This pathological behaviour of massless $2\rightarrow2$ scattering
  processes is well known (see for example equations (2.8) ad (2.9) of
  reference \cite{Frixione:1997ks}). In this case the unresolved
  contribution generates an additional singularity which is not
  cancelled by the virtual corrections in special back-to-back
  configurations.  For reference, we quote the values missing from
  Table \ref{tab:MJNLOXS}:
\begin{align*}
  \sigma_{pp\rightarrow \geq 2j}^{\rm NLO}(p_T^{\rm min} > 50 \,{\rm GeV}) &= -111.0^{+62.5}_{-66.0} \, \mu {\rm b} \\
  \sigma_{pp\rightarrow \geq 2j}^{\rm NLO}(p_T^{\rm min} > 500 \,{\rm GeV}) &= 10.4^{+11.3}_{-18.4} \,{\rm nb}
  \label{eq:NLO2jXS}
\end{align*}
\label{footnote:dijet}%
}. In Figure \ref{fig:MJ2jetpT} we show distributions for the 1st and
2nd jets ordered in $p_T$. Variations in the unphysical factorisation
and renormalisation scale choices at NLO leads to the expected
reduction in theoretical uncertainty - in this case around 10\% at
NLO.  The low $K$-factors for three and four jet production with a
$p_T$ cut at 50 GeV suggest this cut is too soft for fixed order
perturbation theory to work. With a the higher $p_T$ cut of $500$ GeV
the $K$-factors are much more reasonable and the slight decrease is in
agreement with previous computations at 7 and 8 TeV
\cite{Bern:2011ep,Badger:2012pf}.

\begin{figure}[hbtp]
  \begin{center}
    \begin{tabular}[h]{cc}
      \includegraphics[width=6cm]{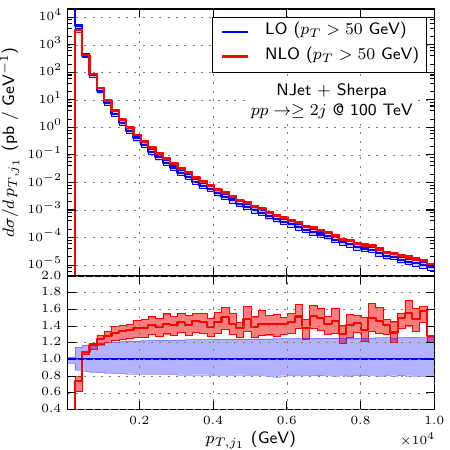} &
      \includegraphics[width=6cm]{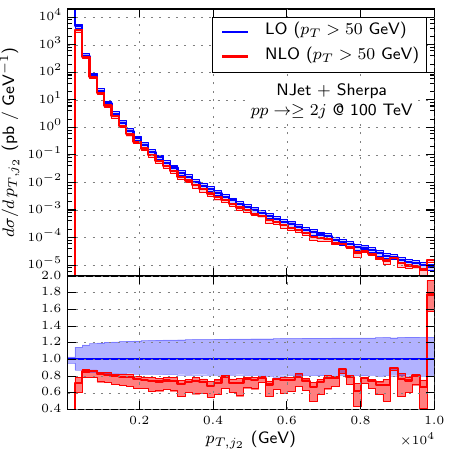} \\
      \includegraphics[width=6cm]{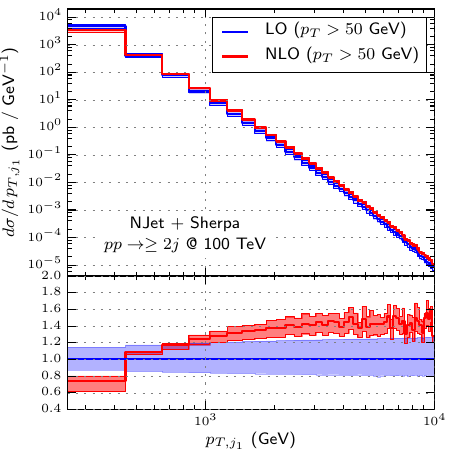} &
      \includegraphics[width=6cm]{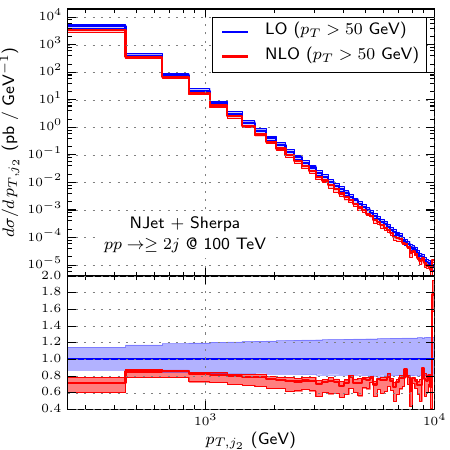}
    \end{tabular}
  \end{center}
  \caption{1st and 2nd leading jet $p_T$ for $pp\to\geq2j$ at 100
    TeV. LO and NLO scale variations in the range $[1/2,2]$ are shown
    around the central scale of $\mu_R=\mu_F=\hat{H}_T/2$. The top row
    shows a linear scale from $50$ GeV to $10$ TeV while the bottom
    row shows the same plot using a log scale over the range $250$ GeV
    to $10$ TeV in order to avoid the singularity which affects the
    first bin.  All plots use events generated using a minimum $p_T$
    cut of $50$ GeV.}
  \label{fig:MJ2jetpT}
\end{figure}

\begin{figure}[hbtp]
  \begin{center}
    \begin{tabular}[h]{ccc}
      \includegraphics[width=5cm]{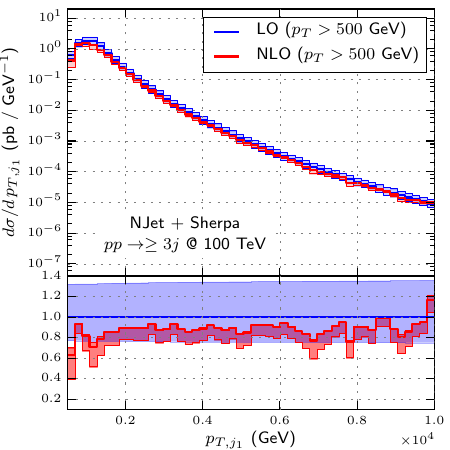} &
      \includegraphics[width=5cm]{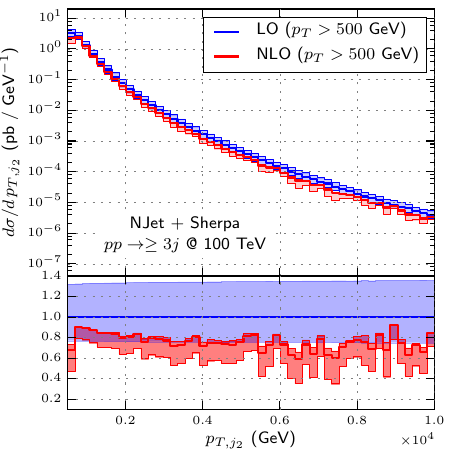} &
      \includegraphics[width=5cm]{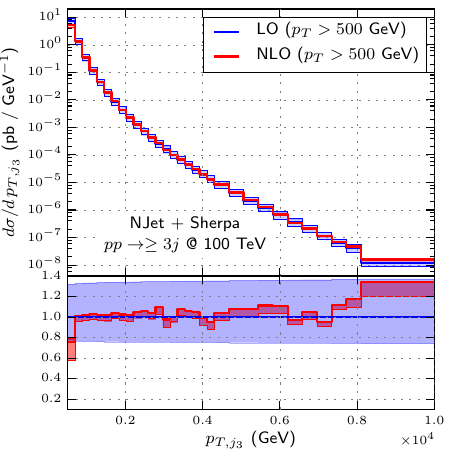} \\
      \includegraphics[width=5cm]{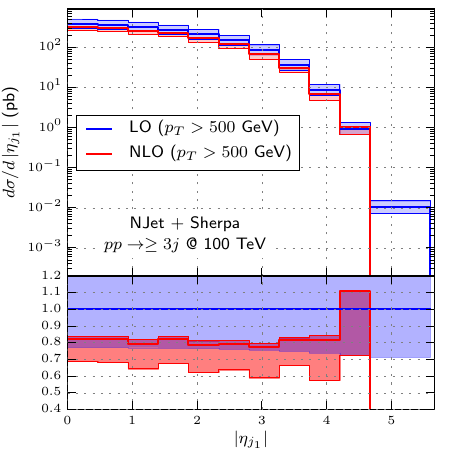} &
      \includegraphics[width=5cm]{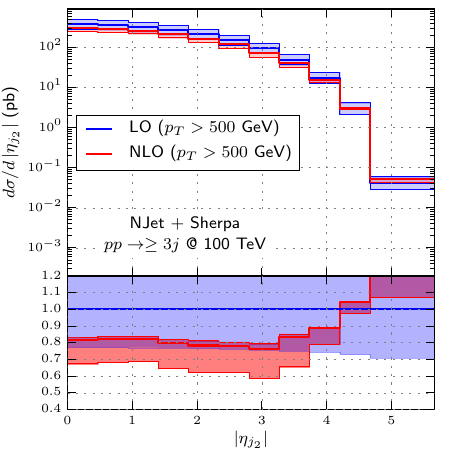} &
      \includegraphics[width=5cm]{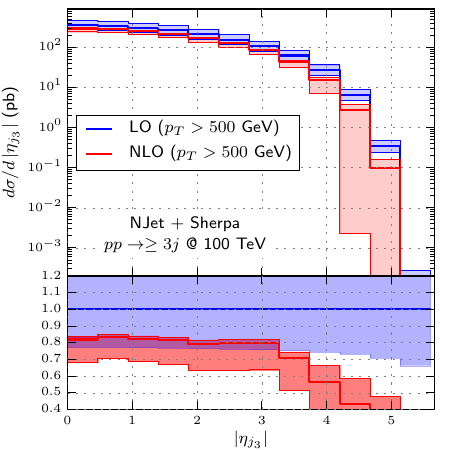}
    \end{tabular}
  \end{center}
  \caption{$p_T$ and rapidity distributions for the 1st, 2nd and 3rd
    leading jet ordered in $p_T$ for $pp\to\geq3j$ at 100 TeV with a
    minimum $p_T$ cut of $500$ GeV. LO and NLO scale variations in the
    range $[1/2,2]$ are shown around the central scale of
    $\mu_R=\mu_F=\hat{H}_T/2$.}
  \label{fig:MJ3jetpTrap}
\end{figure}

\begin{figure}[hbtp]
  \begin{center}
    \begin{tabular}[h]{cc}
      \includegraphics[width=6cm]{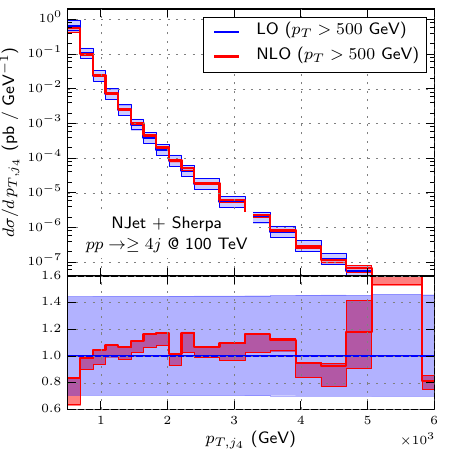} &
      \includegraphics[width=6cm]{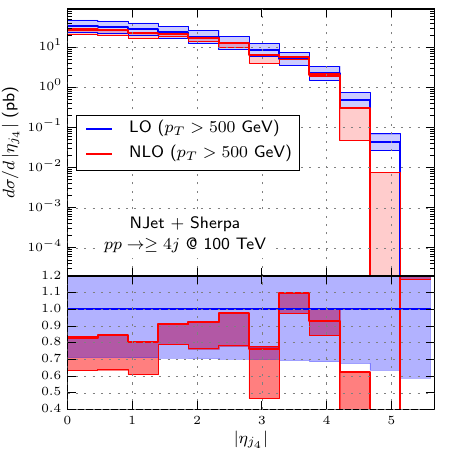}
    \end{tabular}
  \end{center}
  \caption{$p_T$ and rapidity distributions for the 4th jet ordered in
    $p_T$ for $pp\to\geq4j$ at 100 TeV. LO and NLO scale variations in
    the range $[1/2,2]$ are shown around the central scale of
    $\mu_R=\mu_F=\hat{H}_T/2$.}
  \label{fig:MJ4jetpTrap}
\end{figure}

In Figure \ref{fig:MJ3jetpTrap} we show distributions for the $p_T$
ordered jets from $500$ GeV to $10$ TeV in $pp\rightarrow 3j$ events
while in Figure \ref{fig:MJ4jetpTrap} we show distributions for 4th
leading jet from $500$ GeV to $10$ TeV in $pp\rightarrow 4j$ events.

\begin{figure}[hbtp]
  \begin{center}
    \begin{tabular}[h]{cc}
      \includegraphics[width=6cm]{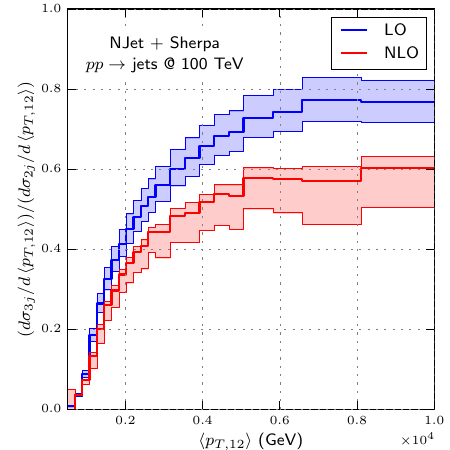} &
      \includegraphics[width=6cm]{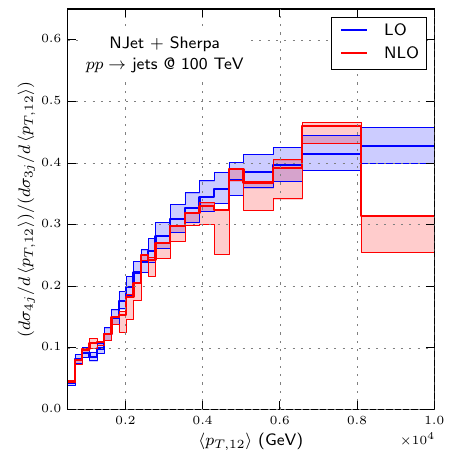}
    \end{tabular}
  \end{center}
  \caption{The $R_{3j/2j}$ and $R_{4j/3j}$ ratios as a function of average transverse momentum of the two leading jets at 100 TeV.
  $\langle p_{T,12} \rangle = \tfrac{1}{2}\left( p_{T,j_1}+p_{T,j_2} \right)$}
  \label{fig:MJRjpt}
\end{figure}

Figure \ref{fig:MJRjpt} shows two plots of multi-jet cross ratios as a
function of the leading jet $p_T$.  Though scaling behaviour of
multijet cross-sections will be covered in more detail in the section
\ref{sec:MJscaling}, these observables at low multiplicity are highly
sensitive to $\alpha_s$ over a large range of energies and thus are
interesting to look at differentially. The perturbative corrections to
the $R_{3j/2j}$ ratio are known to be more stable for the average of
the leading and sub-leading jet $\tfrac{1}{2}(p_{T,j_1}+p_{T,j_2})$
\cite{Rubin:2010xp}.

\begin{table}
\begin{center}
  \begin{tabular}[h]{cccccccc}
    & \multicolumn{7}{c}{$j^n$} \\
    \hhline{~-------}
    & \multicolumn{3}{c}{$p_{T,j_1}>100$ GeV $p_T^{\rm min}> 50$ GeV} && \multicolumn{3}{c}{$p_{T,j_1}>1000$ GeV $p_T^{\rm min}> 500$ GeV} \\
    \hhline{~---~---}
    $n$ & LO & NLO & K && LO & NLO & K \\
    2 & $28.8^{+2.7}_{-2.6}$ $\mu$b & $54.1^{+5.9}_{-4.7}$ $\mu$b & $1.9^{+0.03}_{-0.00}$ && $3.4^{+0.7}_{-0.6}$ nb & $5.5^{+0.6}_{-0.5}$ nb & $1.6^{+0.13}_{-0.14}$ \\
    3 & $20.7^{+5.3}_{-4.0}$ $\mu$b & $7.5^{+3.2}_{-6.4}$ $\mu$b & $0.4^{+0.28}_{-0.32}$ && $1.1^{+0.4}_{-0.3}$ nb & $0.8^{+0.1}_{-0.2}$ nb & $0.7^{+0.28}_{-0.32}$ \\
    4 & $5.6^{+2.4}_{-1.6}$ $\mu$b & $1.7^{+1.1}_{-2.8}$ $\mu$b & $0.3^{+0.39}_{-0.45}$ && $106.8^{+47.6}_{-31.1}$ pb & $92.2^{+0.0}_{-20.0}$ pb & $0.9^{+0.34}_{-0.40}$
  \end{tabular}
\end{center}
\caption{Inclusive cross-sections for multijet production at NLO and LO using democratic cuts of $50$ and $500$ GeV together with an additional
  restriction on the leading jet of $100$ GeV or $1$ TeV. \label{tab:MJNLOXSpt1cut}}
\end{table}

In Table \ref{tab:MJNLOXSpt1cut} we look at representative NLO
$K$-factors with additional cuts on the highest $p_T$ jet, Even in
this hierarchical configuration there appears to be problems with the
scale choice at NLO. We examine this further in Figure
\ref{fig:MJ3jscalevar} looking the scale variation over a larger range
of values than the traditional factor of 2 from the central
value. This is done at 14 TeV for minimum $p_T^{\rm min} = 30,60,100$
GeV and at 100 TeV for $p_T^{\rm min} = 50,100,250$ GeV.  At NLO the
cross section will have a maximum value with the choice of scale. For
extremely low scales the cancellations between real and virtual
contributions become unstable which is clearly seen at both 14 TeV and
100 TeV. For the cases of $p_T>30$ GeV at 14 TeV and $p_T>50$ GeV at
100 TeV this caused the cross section to become negative. The stable
region of the cross-section occur near to the peak value, where it
also happens that the LO cross section agrees with the NLO. The
$K$-factors approach 1 for much higher values of the scale factor for
low $p_T$ cuts and the situation is exacerbated at 100 TeV.

We stress that having a $K$-factor of 1 is not the aim of this
analysis but that even at NLO scale variations can be large. For the
lower $p_T$ cuts in multi-jets at 100 TeV it appears a central scale
choice of $\hat{H}_T$ rather than $\hat{H}_T/2$ would give more
realistic predictions.

\begin{figure}[hbtp]
  \begin{center}
    \begin{tabular}[h]{cc}
      \includegraphics[width=7cm]{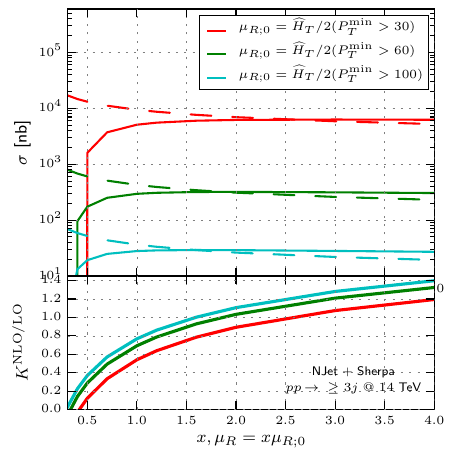} &
      \includegraphics[width=7cm]{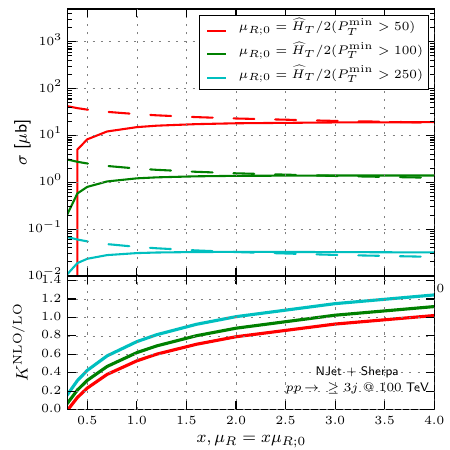} \\
      (a) & (b)
    \end{tabular}
  \end{center}
  \caption{Total cross sections for $pp\to\geq3j$ as a function of the
    scale choice for 3 different sets of minimum $p_T$ at both 100 and
    14 TeV. In the upper plot solid lines show NLO predictions while
    dashed lines show LO predictions. The lower plots show the NLO/LO
    K-factors.}
  \label{fig:MJ3jscalevar}
\end{figure}

Table \ref{tab:MJNLOXSRdep} shows the dependence on the K-factor with
respect to the anti-$k_T$ jet radius for $pp\rightarrow 3j$ using a
minimum $p_T>250$ GeV. Overall the dependence on the jet radius is
relatively mild. As expected even with a relatively high $p_T$ cut
perturbative stability is compromised for $R\lesssim0.3$.

\begin{table}
\begin{center}
  \begin{tabular}[h]{cccc}
    \multicolumn{4}{c}{$j^3 \qquad p_T>250$ GeV} \\
    \hhline{----}
    $R$ & LO & NLO & K \\
    0.2 & $45.6^{+14.1}_{-10.2}$ nb & $22.4^{+5.2}_{-12.9}$ nb & $0.5^{+0.29}_{-0.33}$ \\
    0.3 & $43.4^{+13.5}_{-9.7}$ nb & $27.6^{+3.2}_{-9.7}$ nb & $0.6^{+0.28}_{-0.32}$ \\
    0.4 & $41.8^{+13.0}_{-9.4}$ nb & $30.8^{+1.9}_{-7.5}$ nb & $0.7^{+0.27}_{-0.31}$ \\
    0.5 & $40.5^{+12.6}_{-9.1}$ nb & $33.5^{+0.8}_{-5.8}$ nb & $0.8^{+0.26}_{-0.31}$ \\
    0.6 & $39.3^{+12.3}_{-8.8}$ nb & $35.6^{+0.0}_{-4.3}$ nb & $0.9^{+0.26}_{-0.30}$ \\
    0.7 & $38.2^{+11.9}_{-8.6}$ nb & $36.8^{+0.0}_{-3.2}$ nb & $1.0^{+0.25}_{-0.29}$ \\
  \end{tabular}
\end{center}
\caption{Inclusive three jet cross-section as a function of jet radius at NLO and LO using democratic cuts on all jets of $250$ GeV.
  Renormalisation and factorisation scales are chosen equal with a central values $\mu_R = \mu_F = \hat{H}_T/2$
  with theoretical uncertainty estimated through variations over the range $[1/2,2]$. \label{tab:MJNLOXSRdep}}
\end{table}

\subsubsection{Photon and diphoton production in association with jets}

Representative NLO $K$-factors for photon plus jets final states are
presented in Table \ref{tab:AMJNLOXS} for two sets of minimum
$p_T^{\rm min}$, 50 and 500 GeV, applied to all photons and jets,
respectively.
\begin{table}[h!]
\begin{center}
  \begin{tabular}[h]{cccccccc}
    & \multicolumn{7}{c}{$\gamma+j^n$} \\
    \hhline{~-------}
    & \multicolumn{3}{c}{$p_T^{\rm min}> 50$ GeV} && \multicolumn{3}{c}{$p_T^{\rm min}> 500$ GeV} \\
    \hhline{~---~---}
    $n$ & LO & NLO & K && LO & NLO & K \\
    1 & $71.6^{+6.5}_{-8.2}$ nb & $115.5^{+5.0}_{-3.1}$ nb & $1.6^{+0.29}_{-0.17}$ && $39.4^{+2.3}_{-2.2}$ pb & $46.9^{+1.4}_{-1.2}$ pb & $1.2^{+0.04}_{-0.03}$ \\
    2 & $24.6^{+2.4}_{-2.2}$ nb & $32.7^{+1.7}_{-1.5}$ nb & $1.3^{+0.07}_{-0.06}$ && $3.9^{+0.8}_{-0.6}$ pb & $4.8^{+0.2}_{-0.3}$ pb & $1.2^{+0.15}_{-0.16}$ \\
    3 & $11.2^{+2.9}_{-2.2}$ nb & $10.1^{+0.0}_{-0.9}$ nb & $0.9^{+0.22}_{-0.25}$ && $654.1^{+206.0}_{-147.9}$ fb & $671.1^{+0.0}_{-32.1}$ fb & $1.0^{+0.24}_{-0.28}$
  \end{tabular}
\end{center}
\caption{Inclusive cross-sections photon plus multijet production as a function of jet
multiplicity at NLO and LO using democratic cuts on all jets of $50$ and $500$ GeV.
Renormalisation and factorisation scales are chosen equal with a central values $\mu_R = \mu_F = \hat{H}_T/2$
where the photon $p_T$ is included. The theoretical uncertainty estimated through variations over the range $[1/2,2]$.
\label{tab:AMJNLOXS}}
\end{table}
The appearance of additional production channels from LO at NLO gives
rise to large $K$-factors in the low multiplicity cases. For the high
energy cuts perturbative stability seems to be improved in all
cases. We note that similar effects are seen in W and Z plus jet
studies in section \ref{ref:Vnjets}.

Using the same set of cuts for di-photon production with up to two
additional jets shows a similar pattern.
\begin{table}[h!]
\begin{center}
  \begin{tabular}[h]{cccccccc}
    & \multicolumn{7}{c}{$\gamma\gamma+j^n$} \\
    \hhline{~-------}
    & \multicolumn{3}{c}{$p_T^{\rm min}> 50$ GeV} && \multicolumn{3}{c}{$p_T^{\rm min}> 500$ GeV} \\
    \hhline{~---~---}
    $n$ & LO & NLO & K && LO & NLO & K \\
    0 & $45.7^{+8.6}_{-8.5} $ pb & $100.1^{+3.8}_{-5.2} $ pb & $2.2^{+0.36}_{-0.28} $ && $49.0^{+1.4}_{-1.7} $ fb & $68.7^{+1.4}_{-1.1} $ fb & $1.4^{+0.08}_{-0.06} $ \\
    1 & $27.3^{+0.3}_{-0.6} $ pb & $61.7^{+3.7}_{-3.1} $ pb & $2.3^{+0.19}_{-0.13} $ && $7.3^{+0.7}_{-0.6} $ fb & $12.5^{+1.0}_{-0.9} $ fb & $1.7^{+0.04}_{-0.04} $ \\
    2 & $24.4^{+3.1}_{-2.7} $ pb & $31.6^{+1.3}_{-1.4} $ pb & $1.3^{+0.10}_{-0.10} $ && $2.5^{+0.5}_{-0.4} $ fb & $3.2^{+0.1}_{-0.2} $ fb & $1.3^{+0.15}_{-0.17} $
  \end{tabular}
\end{center}
\caption{Inclusive cross-sections diphoton plus multijet production as a function of jet
radius at NLO and LO using democratic cuts on all jets of $50$ and $500$ GeV.
Renormalisation and factorisation scales are chosen equal with a central values $\mu_R = \mu_F = \hat{H}_T'/2$
with theoretical uncertainty estimated through variations over the range $[1/2,2]$.
\label{tab:AAMJNLOXS}}
\end{table}
In this case the $K$-factors are very high as the number of additional
channels is more extreme than for the single photon case. Again the
low $p_T$ cut of 50 GeV appears to be disfavoured.

\subsection{Scaling behaviour in multi-jet production \label{sec:MJscaling}}

When considering hadron collisions at highest energies QCD jet
production processes are omnipresent. Even processes with very large
multiplicity of (associated) jets exhibit sizable rates.  Accurate
predictions for such final states pose a severe challenge for
Monte-Carlo event generators and one might have to resort to
approximate methods. One such approach is based on the scaling
behaviour of QCD jet rates with respect to jet multiplicity that this
section shall be focused on.

In Fig.~\ref{figDiffJetRatesLO14vs100} anti-$k_T$ jet rates at NLO QCD
differential in jet transverse momentum and additionally binned in jet
rapidity $y$ are presented. Results have been obtained with
\Blackhat\!\!+\Sherpa~\cite{Berger:2008sj}, renormalisation and
factorisation scale have been set to $\mu_R=\mu_F=\frac12 H_T$.
Comparing rates for $14$ and \SI{100}{TeV} centre-of-mass energy an
increase of about one order of magnitude for central jets with low and
moderate $p_T$ is observed. Considering larger $p_T$ values the
differences get more extreme, at $p_T=\SI{3.5}{TeV}$ the FCC rates are
more than three orders of magnitude larger than at the LHC.
\begin{figure}[hbtp]
  \centering
  \includegraphics[width=\linewidth]{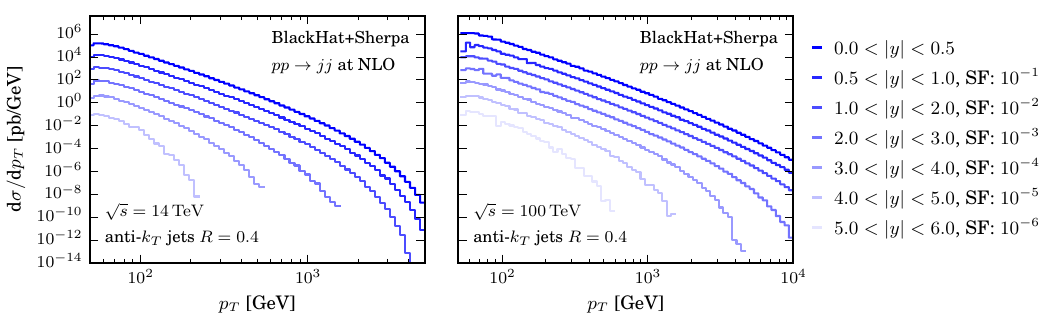}
  \caption{NLO QCD inclusive jet cross sections for LHC (left) and FCC
    (right) collision energies, differential in $p_T$ for different
    bins in jet rapidity $y$.  Note that for illustrative purpose
    results have been multiplied by variable scaling factors (SF), as
    indicated in the legend.}
  \label{figDiffJetRatesLO14vs100}
\end{figure}
In fact, the FCC provides substantial jet rates even for very large
rapidities: \SI{200}{GeV} jets with $5<|y|<6$ come with rates about
two orders of magnitude larger than those for \SI{200}{GeV} jets in
the $4<|y|<5$ bin at the LHC. From these rate estimates it can be
concluded that one can expect at least ten times more jets at the FCC
compared to the LHC, and this factor gets larger when looking into
high $p_T$ and/or high $|y|$ regions or demanding large jet
multiplicities. Accordingly, the rapidity coverage of general-purpose
detectors at the FCC should increase with respect to ATLAS or CMS.

The QCD jet production rates to be anticipated at the FCC demand
suitable theoretical methods even for very large jet
multiplicities. While a fixed-order prediction for a given jet process
is suitable to describe the corresponding jet multiplicity bin,
matrix-element parton-shower merging techniques provide inclusive
predictions, differential in the jet multiplicity, with high jet
multiplicities being modelled through the parton shower.
Alternatively, there has recently been progress in making
(semi-)analytical predictions for jet rates at hadron colliders that
account for small jet radii and high jet
counts~\cite{Gerwick:2012fw,Gerwick:2013haa,Dasgupta:2014yra}.

With the advent of such methods, the morphology of the entirety of the
jet-multiplicity distribution can be studied.  Guided by
phenomenological evidence, supported by both fixed-order calculations
and parton-shower simulations, certain jet-multiplicity scaling
patterns can be identified~\cite{Gerwick:2012hq} that find their
analogue in the analytical jet-rate
predictions~\cite{Gerwick:2012fw,Gerwick:2013haa}.

As already visible in Fig.~\ref{fig:JetRatesLO14vs100}, jet rates
differential in the number of jets exhibit a high degree of
regularity. To study this feature one considers the ratio
$R_{(n+1)/n}$ of the exclusive $n+1$ over the $n$-jet cross section,
i.e.
\begin{equation}
  R_{(n+1)/n} \equiv \frac{\sigma_{n+1}^\text{excl}}{\sigma_{n}^\text{excl}} \,.
\end{equation}
The approximately equal step size (on a logarithmic scale) between the
subsequent exclusive jet rates observed in
Fig.~\ref{fig:JetRatesLO14vs100} translates into a flat plateau for
$R_{(n+1)/n}$, i.e. $R_{(n+1)/n} \sim \text{constant}$, translating
into a simple exponential form of the jet-rate distribution. This
shape of the jet rates is called a Staircase Pattern. Another
regularity in jet rates found is named Poisson Pattern. Jet cross
sections following a simple Poisson statistics result in $R_{(n+1)/n}
\sim \bar{n}/(n+1)$, with the average number of jets given by
$\bar{n}$.

Both these patterns have been observed in LHC
data~\cite{Aad:2011tqa,Aad:2013ysa,Aad:2014qxa,Khachatryan:2014uva}
and in Monte-Carlo
studies~\cite{Englert:2011cg,Englert:2011pq,Bern:2014fea}.  They can
be understood as the limiting cases for the jet-emission probability:
for $\alpha_S/\pi \log^2 Q/Q_0 \ll 1$ a Staircase Pattern is expected
while for $\alpha_S/\pi \log^2 Q/Q_0 \gg 1$ a Poisson Scaling is
observed~\cite{Gerwick:2012fw,Gerwick:2012hq,Gerwick:2014koa}.  Here
$Q$ denotes the hard process scale and $Q_0$ is of the order of the
jet resolution scale, i.e. $Q_0\sim p_{T,\text{min}}$. The derivation
is based on the language of generating functionals for the jet rates.
The two distinct regimes correspond to additional parton emissions
being distributed either equally among all other partons or stemming
predominantly from a single hard parton line. The latter follows a
simple Sudakov decay-like model which results in a Poisson
distribution, as it is the case for photon emissions from a hard
electron line~\cite{Peskin:1995ev}.  The case of democratic emissions
(mainly gluons from gluons) on the other hand is exclusive to field
theories with a non-abelian group structure as QCD.

In realistic measurements jet patterns will be overlaid and cut off by
other effects, such as phase-space constraints. When the available
energy for further jet emission is being depleted or jets already
radiated cover a good fraction of the available solid
angle~\cite{Gerwick:2014koa}, then higher multiplicities will quickly
tend to zero. On the other hand, the first few emissions carry away
sizable parts of the total energy available, such that the increase in
the partonic momentum fractions at which any participating PDFs are
evaluated is comparably large. This leads to somewhat steeper decrease
of jet rates for the first few emissions and is known as the PDF
suppression effect~\cite{Gerwick:2012hq}.

In view of the enormous phase space available for producing additional
jets at the FCC collider, studies of the jet multiplicity distribution
based on scaling patterns provide a handle to estimate and probe the
tails of the distribution, where otherwise one has to largely rely on
parton-shower simulations alone. Based on these predictions background
subtractions for New Physics signatures resulting from decays of new
heavy coloured particles yielding a distinct imprint on the
multiplicity distribution might become
feasible~\cite{Englert:2011cg,Hedri:2013pvl}.

Of course jet patterns will be overlaid and cut off by other effects,
such as phase space effects: if the available energy is being depleted
or the existing jets already cover the available solid
angle~\cite{Gerwick:2014koa}, then higher multiplicities will quickly
tend to zero.

To study in how far simple jet scaling patterns describe the jet
multiplicity distributions at FCC energies fits of $R_{(n+1)/n}$ in
Monte-Carlo predictions are considered. For that purpose \Sherpa
Monte-Carlo samples for pure jet production are explored, triggering
scaling patterns using either democratic or hierarchical, i.e.
staggered, jet cuts. As mentioned before, here \emph{democratic}
reflects the fact that all jet $p_{T,\text{min}}$ are of the same
order, i.e. uniform, whereas \emph{hierarchical} refers to the
scenario where the cut on the leading jet,
$p_{T,\text{min}}^{\text{leading}}$, is significantly increased.

\begin{table}[hbtp]
\begin{center}
\begin{tabular}{@{}lrrlll@{}}
label & $p_{T,\text{min}}^\text{leading}\;\text{[GeV]}$ & $p_{T,\text{min}}\;\text{[GeV]}$
& fit function & fit region & fit parameters
\\
\hline \noalign{\smallskip}
S1 (democratic)    & 100   & 50  & $f_\text{Staircase}$ & $3 \leq n \leq 5$ & $c=0.342,\;m=0.006$       \\
S2 (democratic)   & 200   & 100 & $f_\text{Staircase}$ & $3 \leq n \leq 5$ & $c=0.274,\;m=0.003$       \\
P1 (hierarchical)  & 500   & 50  & $f_\text{Poisson}$ & $1 \leq n \leq 5$ & $\bar{n}=2.21,\;c=0.16$ \\
P2 (hierarchical)  & 2000  & 50  & $f_\text{Poisson}$ & $1 \leq n \leq 5$ & $\bar{n}=2.64,\;c=0.25$ \\
\end{tabular}
\caption{The jet-cut scenarios considered for pure jet production at FCC energies. Furthermore,
the fit hypothesis used, cf. Eqs.~(\ref{eq:fitfuncs1}) and (\ref{eq:fitfuncs2}), and the corresponding parameters are listed.}
\label{tabPureJetsScenarios}
\end{center}
\end{table}

The cut scenarios considered for pure jet production are listed in
Tab.~\ref{tabPureJetsScenarios}. In all case the $2\to 2$ core process
has been considered at \MCatNLO accuracy, furthermore LO matrix
elements for final-state multiplicities up to six partons are
included, all consistently merged with the parton shower. In
Fig.~\ref{figR} the resulting $R_{(n+1)/n}$ distributions are
presented for the four considered selections. Note, the index $n$
counts the number of jets radiated off the hard two-to-two core,
i.e. $n=1$ corresponds to the production of three final-state jets.

\begin{figure}[hbtp]
  \centering
  \includegraphics[width=0.49\textwidth]{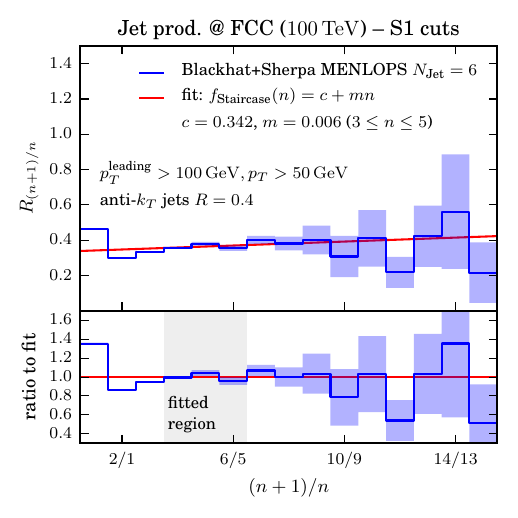}
  \includegraphics[width=0.49\textwidth]{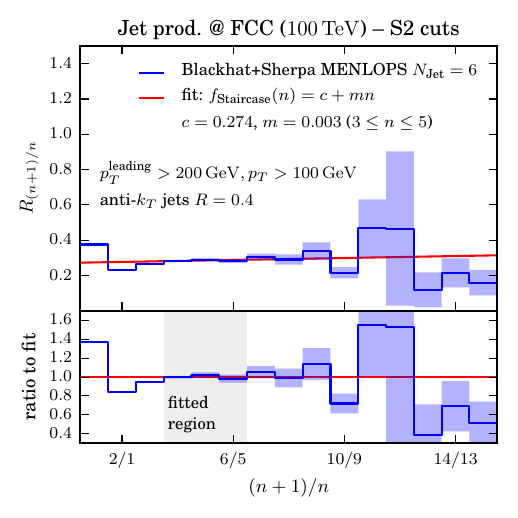}
  \\
  \includegraphics[width=0.49\textwidth]{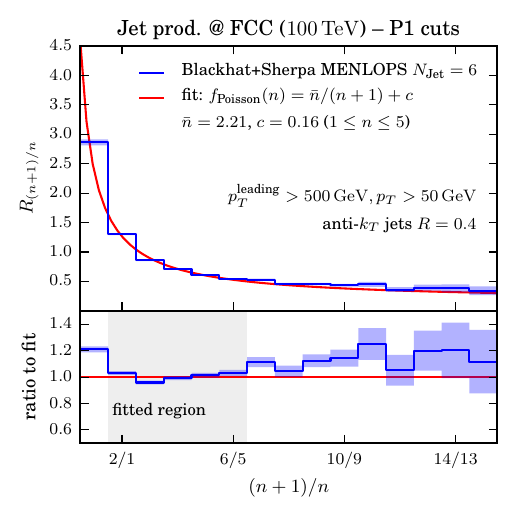}
  \includegraphics[width=0.49\textwidth]{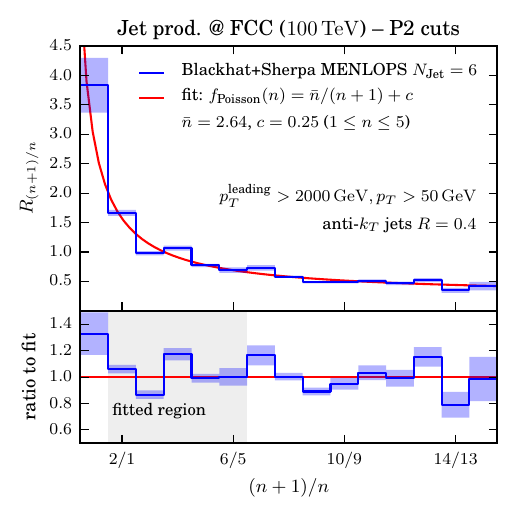}
  \caption{The exclusive jet multiplicity ratio $R_{(n+1)/n}$ in pure jet production at the FCC. 
  Results are presented for the four cut scenarios described in Tab.~\ref{tabPureJetsScenarios}
  with fits  for the Staircase and Poisson patterns,
  cf. Eqs.~(\ref{eq:fitfuncs1}), (\ref{eq:fitfuncs2}).}
  \label{figR}
\end{figure}

As discussed in~\cite{Englert:2011pq}, jets assigned to the core
process behave differently from jets emitted thereof, which is why
they have to be dismissed from pattern fits through the
data. Furthermore, PDF effects leave a non-universal imprint on the
first few bins. Therefore, for the Staircase-like patterns found for
the democratic cut scenarios, cf. the two upper panels of
Fig.~\ref{figR}, the fits are based on the values from $R_{4/3}$
through $R_{6/5}$. For the hierarchical cut scenarios PDF suppression
effect are less prominent, due to hard cut on the leading jet that
induces a much higher scale $Q$ for the core process. Accordingly, the
fits for the Poisson-like patterns, cf. the two lower panels in
Fig.~\ref{figR}, are based on $R_{2/1}$ up to $R_{6/5}$.  To quantify
the quality of the fits, term linear in $n$ for the Staircase pattern
and a constant term for the Poisson pattern have been added to the
ideal scaling hypotheses. The resulting fit functions for the two
scenarios read
\begin{align}
  f_\text{Staircase}(n) &= c + m \, n \,,\label{eq:fitfuncs1}\\
  f_\text{Poisson}(n)   &= \frac{\bar{n}}{n+1} + c\,.\label{eq:fitfuncs2}
\end{align}
All resulting fit parameters are listed in
Tab.~\ref{tabPureJetsScenarios}.  For all cut scenarios the fit
function and its extrapolation to higher jet bins describe the
simulated data very well. For the two democratic scenarios, the
constant $c$ decreases from 0.35 to 0.29 when we increase the jet
cuts, reflecting the fact that the {\em costs} for adding an
additional jet gets higher.

Poisson patterns are obtained when hierarchical cuts are applied.
Although the constant offset $c$ increases from 0.16 to 0.25 when
enlarging the gap between the leading jet cut and the overall jet cut
$p_{T,{\rm min}}$ one can see by eye that the fit quality is better
for the larger cut gap, i.e.  \SI{2000}{GeV} vs. \SI{50}{GeV}. For the
smaller cut gap, i.e.  \SI{500}{GeV} vs. \SI{50}{GeV} the fit
increasingly underestimates $R_{(n+1)/n}$ for growing $n$, which might
indicate a faster transition to a more Staircase-like behaviour.  As
expected the average jet multiplicity $\bar{n}$ found from the fit
increases with a larger leading jet cut (from $2.2$ to $2.6$). In
particular the S2 and P2 cut scenarios are very well modelled by the
simple scaling pattern hypotheses and allow for reliable
extrapolations where explicit calculations based on fixed order or
even parton shower simulations become computationally infeasible.

Both patterns can also be observed in $W$ production in association
with additional jets, as have been discussed
sec.~\ref{sec:Vjets-ratios}.

To further illustrate the universality of jet scaling patterns,
Fig.~\ref{figLOJetsMultiScaling} compiles the exclusive jet
multiplicity ratios for a variety of processes, including pure jets,
$\gamma+$jets, $t\bar{t}+$jets and $W/Z+$jets.  The predictions are
based on dedicated $n-$jet tree-level matrix element calculations,
without invoking parton showers. Democratic jet selection cuts are
applied, i.e. requiring $p_{T,j}>\SI{50}{GeV}$ in all processes. In
addition, the photon production processes are regulated by the
selection criteria $p_{T,\gamma}>\SI{50}{GeV}$ and $R_{j,\gamma} >
0.4$, with $R_{j,\gamma}$ the $\eta-\phi$ distance between all jets
and the photon.

There are a few remarkable aspects to note here. Apparently, for the
pure jets and $W+$jets processes these LO rate estimates nicely
reproduce the staircase scaling parameters found in the matrix-element
plus parton-shower samples for the analogous jet-selection cuts,
cf. Fig.~\ref{figR} (upper left panel).  This is supported by the fact
that for exact Staircase scaling the cross section ratios for
subsequent jet multiplicities are identical for exclusive and
inclusive cross sections~\cite{Gerwick:2012hq}, i.e. in this limit
\begin{equation}
\frac{\sigma_{n+1}^\text{excl}}{\sigma_{n}^\text{excl}}= \frac{\sigma_{n+1}^\text{incl}}{\sigma_{n}^\text{incl}}=R={\rm const.}
\end{equation}
Also note that the ratios of the three vector-boson production
processes, $W/Z/\gamma+$jets, are basically the same, illustrating the
fact that the actual gauge-boson mass does not yield a big imprint on
the jet-production probabilities. The production of a pair of
top-quarks, however, induces a large upper scale for subsequent jet
emission. Correspondingly, the jet rates for the first few emissions
are sizable, resulting in ratios $R_{(n+1)/n}>0.5$, indicating that a
pure leading-order approximation is inappropriate.

\begin{figure}[hbtp]
  \centering
  \includegraphics[width=0.6\linewidth]{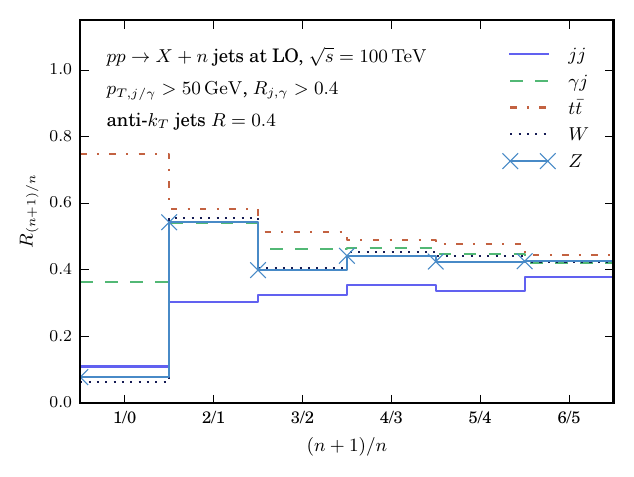}
  \caption{The jet multiplicity ratio $R_{(n+1)/n}$ for several processes
  calculated at LO for each final-state multiplicity. Note, the index 
  $n$ counts jets associated to the core process listed in the legend.
   }
  \label{figLOJetsMultiScaling}
\end{figure}

In conclusion to this section, it can be noted that it is possible to
fit jet multiplicities $n$ up to values of $n=15$ or even higher using
results for much lower $n$. The underlying fits are based on the
theoretical hypothesis of simple scaling pattern, namely Staircase and
Poisson scaling. These extrapolations allow trustworthy predictions to
be made for very high jet-multiplicity bins that will be populated by
a variety of production processes at FCC energies.  The methods
discussed enable the use of techniques that discriminate New Physics
signals and QCD backgrounds based on the shape of the jet multiplicity
distribution.

\clearpage
\section{Heavy flavour production\footnote{Editor: M.Cacciari}}
\label{sec:hvq}

\subsection{Inclusive bottom production}
\label{sec:hvq_b}
Inclusive production of $b$ hadrons in hadronic collisions offers
unlimited opportunities for flavour studies in the $b$ sector, as
shown very well by the Tevatron and LHC experiments.

The long-term interest in these studies will depend on what future
LHCb and Belle2 data will tell us, and on the flavour implications of
possible LHC discoveries in the high-$Q^2$ region. But it is likely
that heavy flavour studies will remain a pillar of the physics
programme at 100~TeV. The flavour-physics aspects of the 100~TeV
collider will be discussed in a future document.

The total $b\bar{b}$ production cross section at 100~TeV is about 3mb,
an increase by a factor of $\sim 5$ relative to the LHC, and it is
more than a 1\% fraction of the total $pp$ cross section. As discussed
above, a large fraction of the total rate comes from gluons at very
small $x$ values, where the knowledge of PDFs is today rather poor.
The upper plot of Fig.~\ref{fig:bot100x} shows that, for a detector
like LHCb, covering the rapidity region $2.5<y<5$, about 50\% of the
$b$ events produced at 100~TeV would originate from gluons with
momentum $x<10^{-5}$. This domain, at these rather large values of
$Q^2$, is almost unexplored, although the first
constraints~\cite{Zenaiev:2015rfa,Gauld:2015kvh,Cacciari:2015fta} are
emerging from forward charm and bottom production at the
LHC~\cite{Aaij:2015bpa,Aaij:2015rla} (see also Section~\ref{sec:pdf}
for a discussion of small-$x$ issues at 100 TeV).
\begin{figure}[h]
\centering
\includegraphics[width=0.65\textwidth]{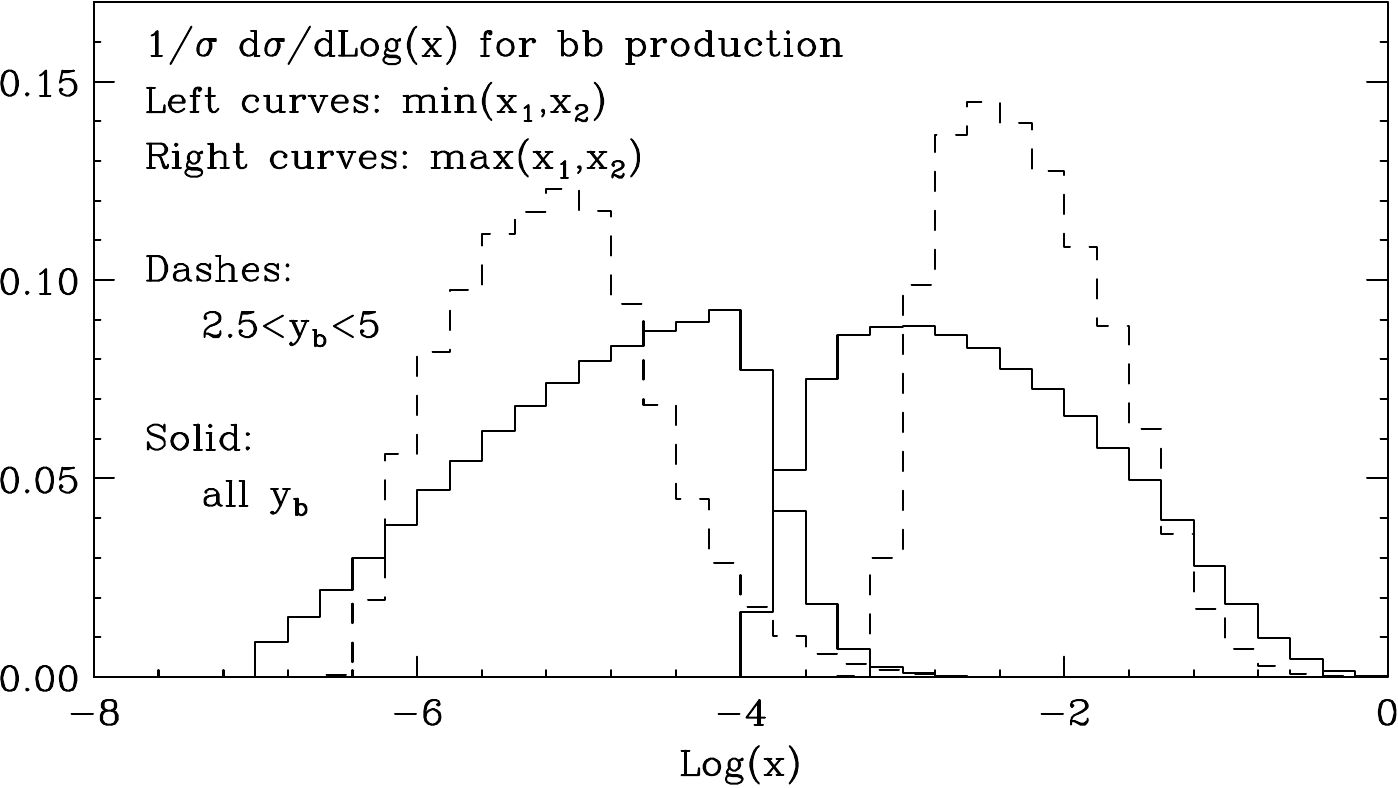}\\
\includegraphics[width=0.65\textwidth]{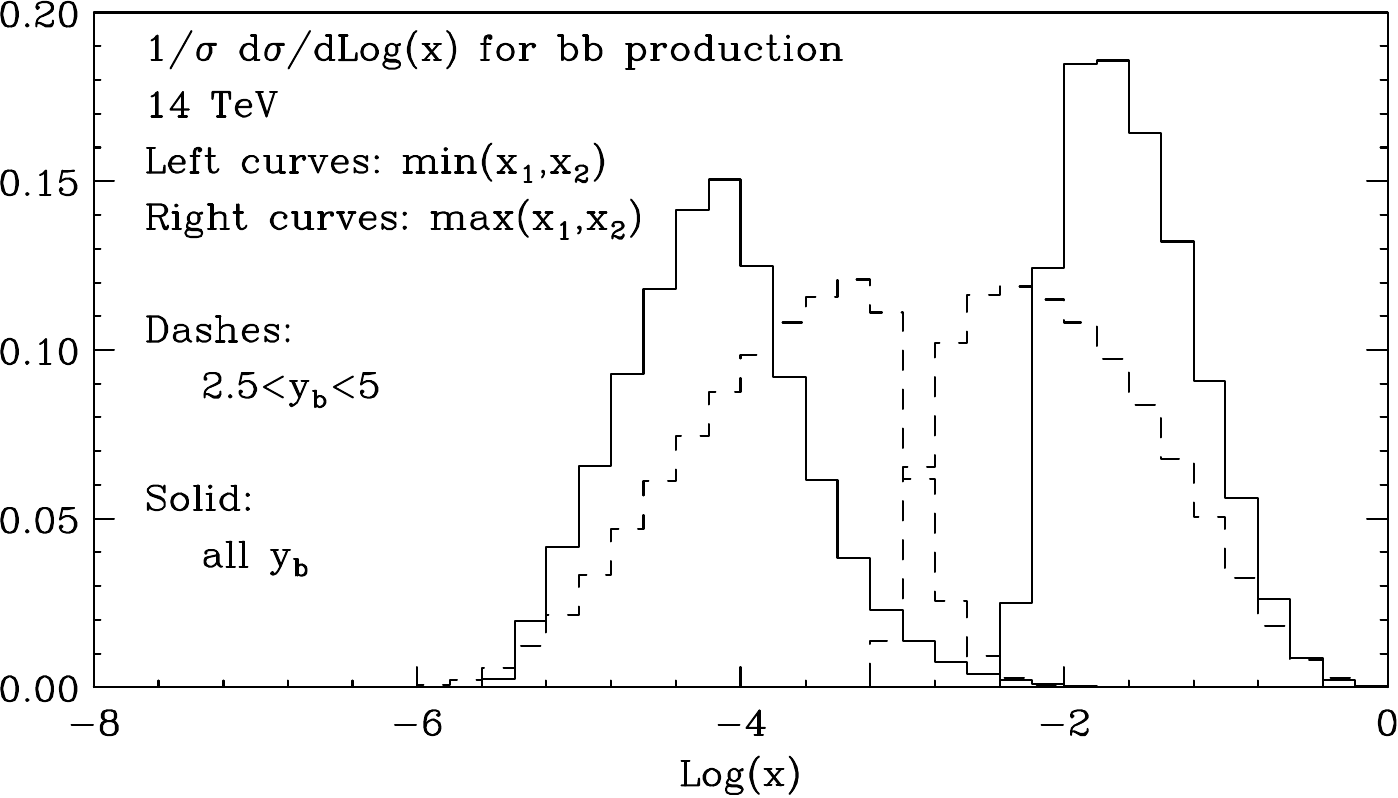}
\caption{Top (bottom) panel: distribution, at 100 (14)~TeV, of the
  smaller and larger values of the initial partons momentum fractions
  in inclusive $b\bar{b}$ events (solid) and in events with at least
  one $b$ in the rapidity range $2.5<\vert y \vert < 5$ (dashes).}
\label{fig:bot100x}
\end{figure}

\begin{figure}[h]
\centering
\includegraphics[width=0.45\textwidth]{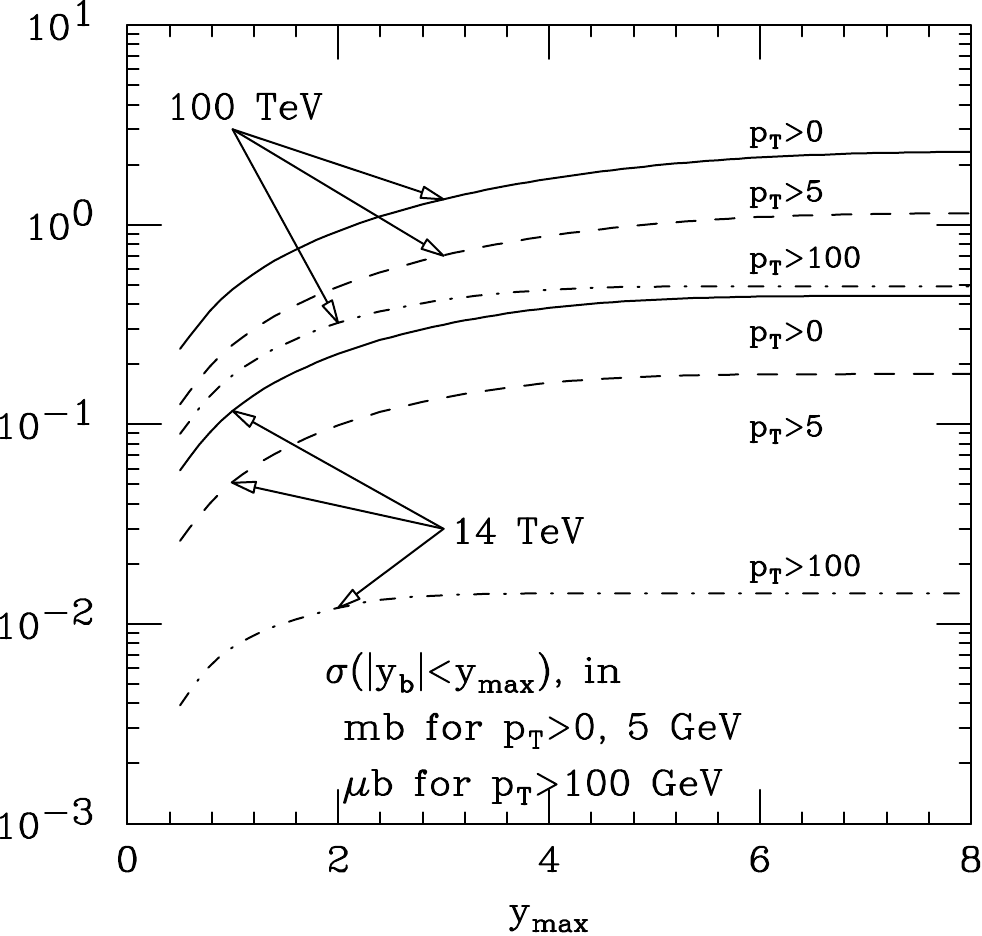}
\includegraphics[width=0.45\textwidth]{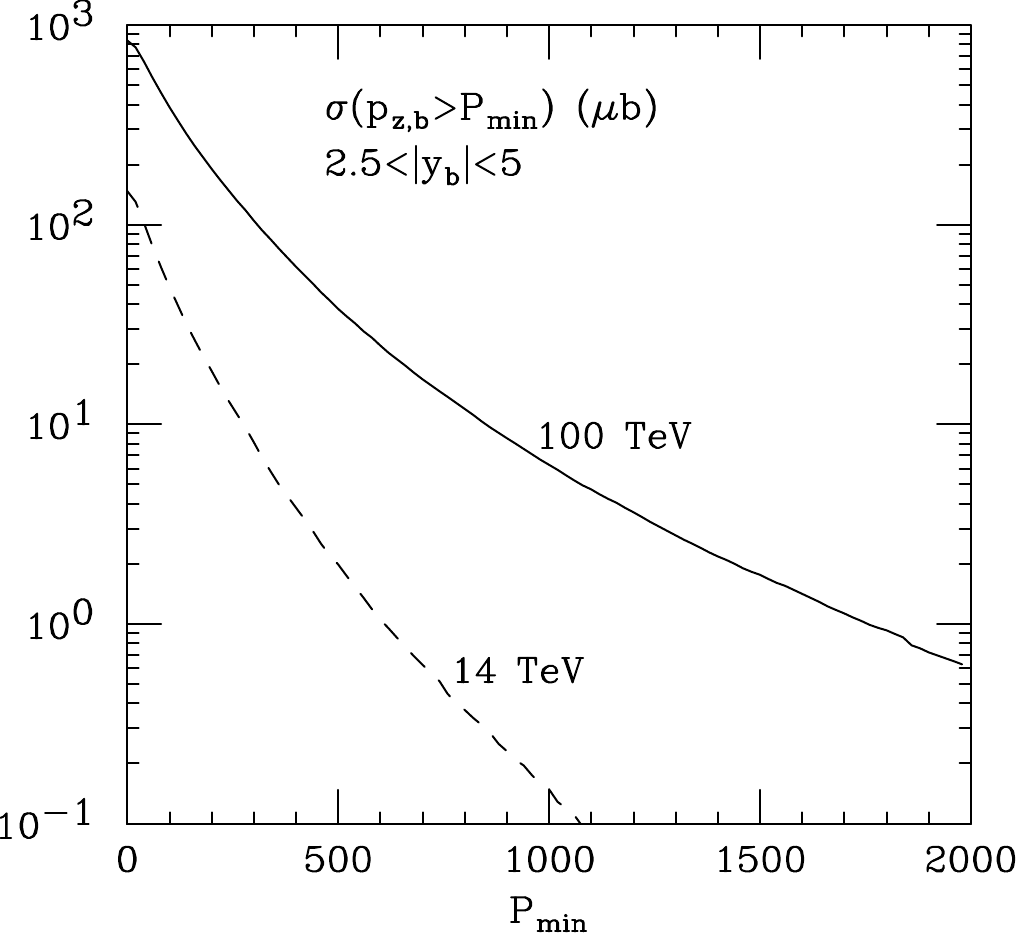}
\caption{Left: production rates for $b$
  quarks as a function of detection acceptance in $y$, for various
  $p_T$ thresholds (rates in $\mu$b for $p_T>100$~GeV, in mb
  otherwise). Right: forward $b$ production rates, as a function of
  the $b$ longitudinal momentum. }
\label{fig:bot100y}
\end{figure}

In Fig.~\ref{fig:bot100y} we show the rapidity distributions for $b$
quarks produced above some thresholds of $p_T$ and, for $b$ quarks
produced in the region $2.5<\vert y \vert < 5$, the integrated
spectrum in longitudinal momentum $p_z$, comparing results at 14 and
100 TeV. We note that, while the total production rate grows only by a
factor of $\sim 5$ from 14 to 100~TeV, the rate increase can be much
larger once kinematic cuts are imposed on the final state. For
example, at 100~TeV $b$ quarks are produced in the forward region
$2.5<\vert y \vert < 5$ with $p_z>1$~TeV at the astounding rate of
$10\mu$b, 100 times more than at the LHC. To what extent this opens
opportunities for new interesting measurements, to be exploited by the
future generation of detectors, remains to be studied.

\subsection{Inclusive top pair production}
\label{sec:hvq_t}
Table~\ref{tab:ttXS} shows the NNLO cross
section~\cite{Czakon:2013goa,Czakon:2011xx} for inclusive $t\bar{t}$
production at 100~TeV. For reference, the LO and NLO results obtained
with the appropriate PDF sets of the NNPDF3.0 group are 24.3~nb and
31.9~nb, respectively. This means $K$ factors of 1.3 (NLO/LO) and 1.1
(NNLO/NLO), indicating an excellent convergence and consistency of the
perturbative expansion. Together with the small size of PDF
uncertainties, this suggests that the predictions for top production
at 100 TeV are already today rather robust.
\begin{table}[h]
\begin{center}
\def\arraystretch{1.5}
\begin{tabular}{ l | c | c  c | c  c }
  PDF & $\sigma$(nb)  & $\delta_{scale}$(nb) & (\%) &
  $\delta_{PDF}$(nb) & (\%) \\ \hline
  CT14 & 34.692   
    & ${+ 1.000 \atop  - 1.649 }$
  & ${ (+ 2.9\%) \atop (-4.7 \%) }$ & ${+ 0.660 \atop  - 0.650 }$
  & ${ (+ 1.9\%) \atop (-1.9 \%) }$ 
  \\
  NNPDF3.0 & 34.810   
    & ${+ 1.002 \atop  - 1.653 }$
  & ${ (+ 2.9\%) \atop (-4.7 \%) }$ & ${+ 1.092 \atop  - 1.311 }$
  & ${ (+ 3.1\%) \atop (-3.8 \%) }$ 
  \\
  PDF4LHC15 & 34.733   
    & ${+ 1.001 \atop  - 1.650 }$
  & ${ (+ 2.9\%) \atop (-4.7 \%) }$ & $ \pm 0.590 $
  & $ (\pm 1.7\%) $ 
\end{tabular}
\end{center}
\caption{Total $t\bar{t}$ production cross sections, at NNLO, for
  various PDF choices. $m_{top}=173.3$~GeV. The scale uncertainty is
  derived from the 7 scale choices of $\mu_{R,F}=k m_{mtop}$, with
  $k=0.5, 1, 2$ and $1/2<\mu_R/\mu_F <2$. The
  PDF4LHC15~\cite{Butterworth:2015oua} recommendation combines the
  systematics from the following NLO PDF sets:
  NNPDF3.0~\cite{Ball:2014uwa}, MMHT2014~\cite{Harland-Lang:2014zoa}
  and CT14~\cite{Dulat:2015mca}.}
\label{tab:ttXS}
\end{table}
The $\sim 30$~nb inclusive rate is more than 30 times larger than at
14~TeV. For the planned total integrated luminosity of
$O(20)$\iab~\cite{Hinchliffe:2015qma}, two experiments would produce
of the order of $10^{12}$ (anti)top quarks.  The possible applications
emerging from this huge statistics have yet to be explored in
detail. It would be interesting to consider the potential of
experiments capable of recording all these events (only a small
fraction of top quarks produced at the LHC survives for the
analyses). Triggering on one of the tops, would allow for unbiased
studies of the properties of the other top and of its decay products:
studies of inclusive $W$ decays~\cite{Mangano:2014xta} (which are
impossible using the $W$'s produced via the Drell-Yan process), of
charm and $\tau$ leptons produced from those $W$ decays, of
flavour-tagged $b$'s from the top decay itself~\cite{Gedalia:2012sx}.

\begin{figure}[h!]
\centering
\includegraphics[width=0.47\textwidth]{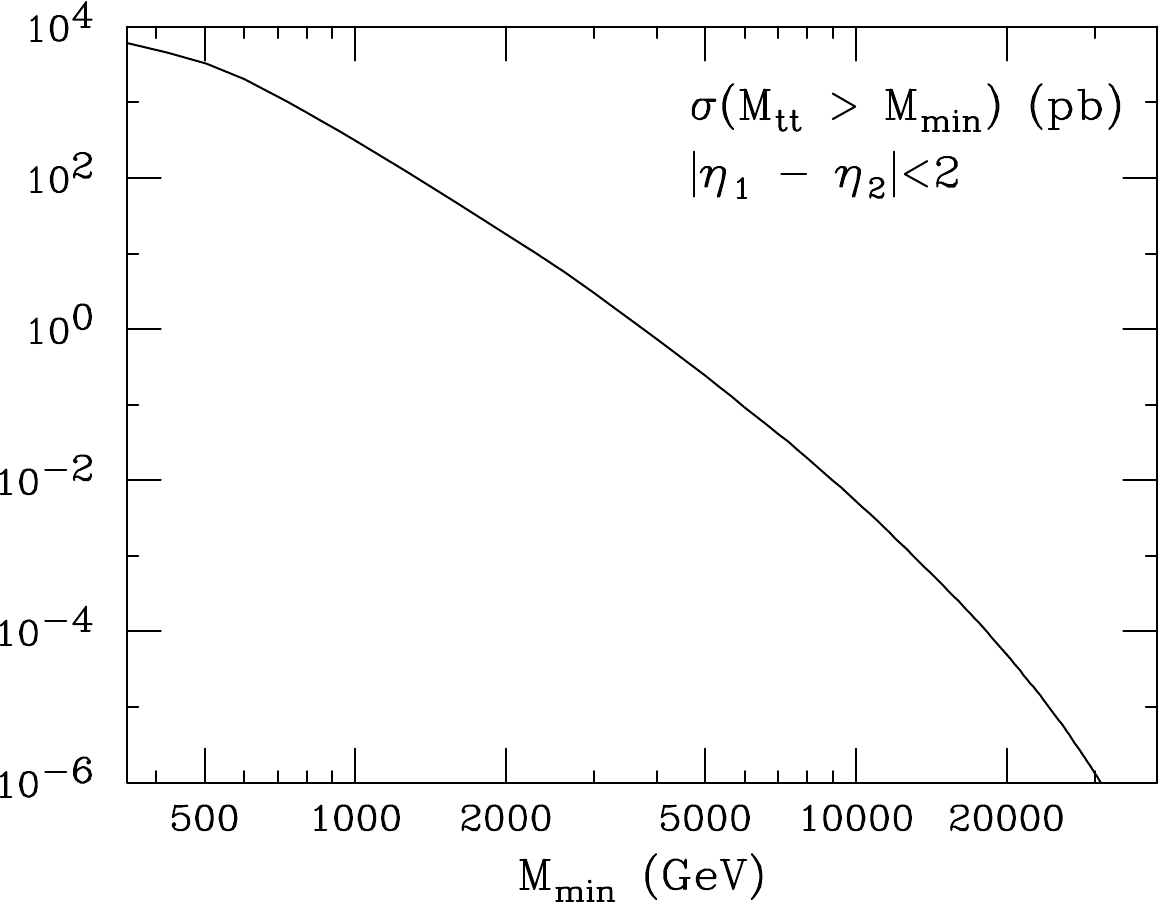}
\includegraphics[width=0.47\textwidth]{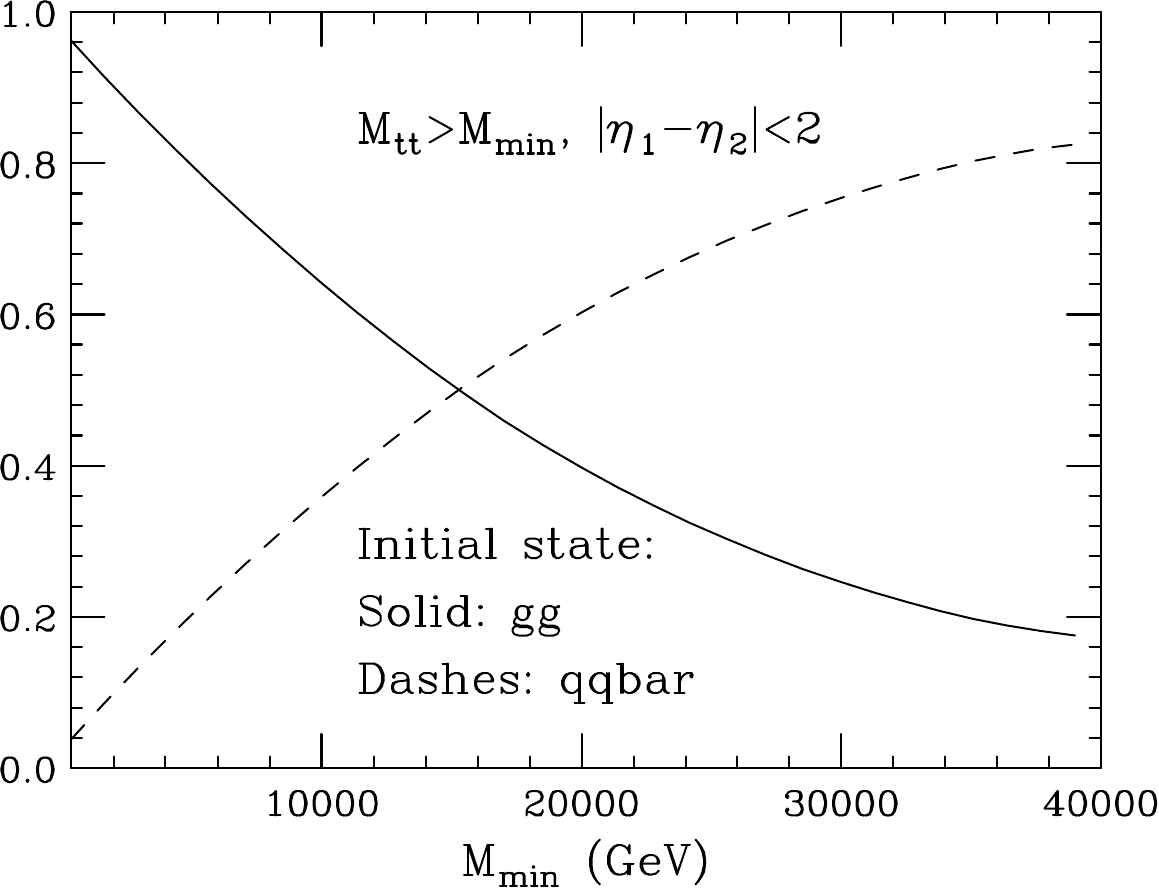}
\caption{Left: integrated invariant mass distribution for production
  of central $t\bar{t}$ quark pairs. Right: initial state composition
  as a function of the $t\bar{t}$ invariant mass.}
\label{fig:mtt}
\end{figure}

\subsection{Bottom and top production at large $Q^2$}
\label{sec:hvq_highpt}
Production of bottom and top quarks at large $Q^2$ is characterized by
two regimes. On one side we have final states where the heavy quark
and antiquark ($Q$ and $\bar{Q}$) give rise to separate jets, with a
very large dijet invariant mass $M_{QQ}$. These are the configurations
of relevance when, for example, we search for the $Q\bar{Q}$ decay of
massive resonances.  In the case of top quarks, the left-hand side of
Fig.~\ref{fig:mtt} shows the production rate for central $t\bar{t}$
pairs above a given invariant mass threshold.  At 100~TeV there will
be events well above $M_{tt}>30$~TeV. The right plot in
Fig.~\ref{fig:mtt} furthermore shows that, due to the absence at LO of
contributions from $qq$ or $qg$ initial states, $gg$ initial states
remain dominant up to very large mass, $M_{tt}\sim 15$~TeV. Well above
$M_{Q\bar Q}\sim$~TeV, the results for $b\bar{b}$ pair production are
similar to those of the top.

\begin{figure}
\centering
\includegraphics[width=0.48\textwidth]{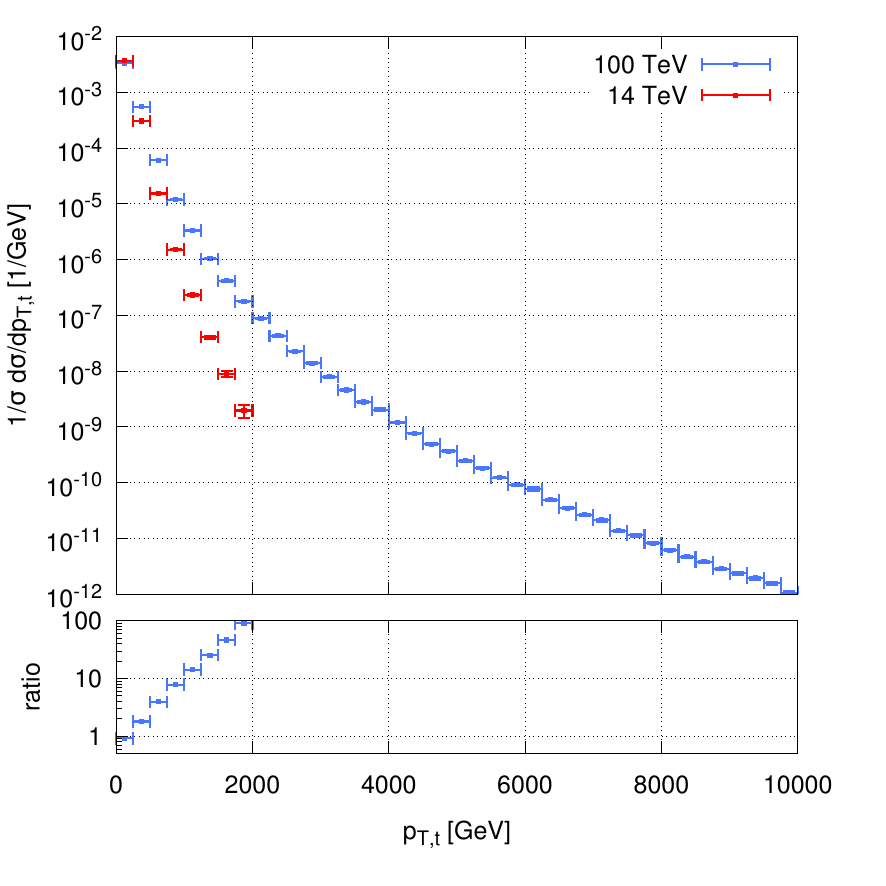}
\includegraphics[width=0.48\textwidth]{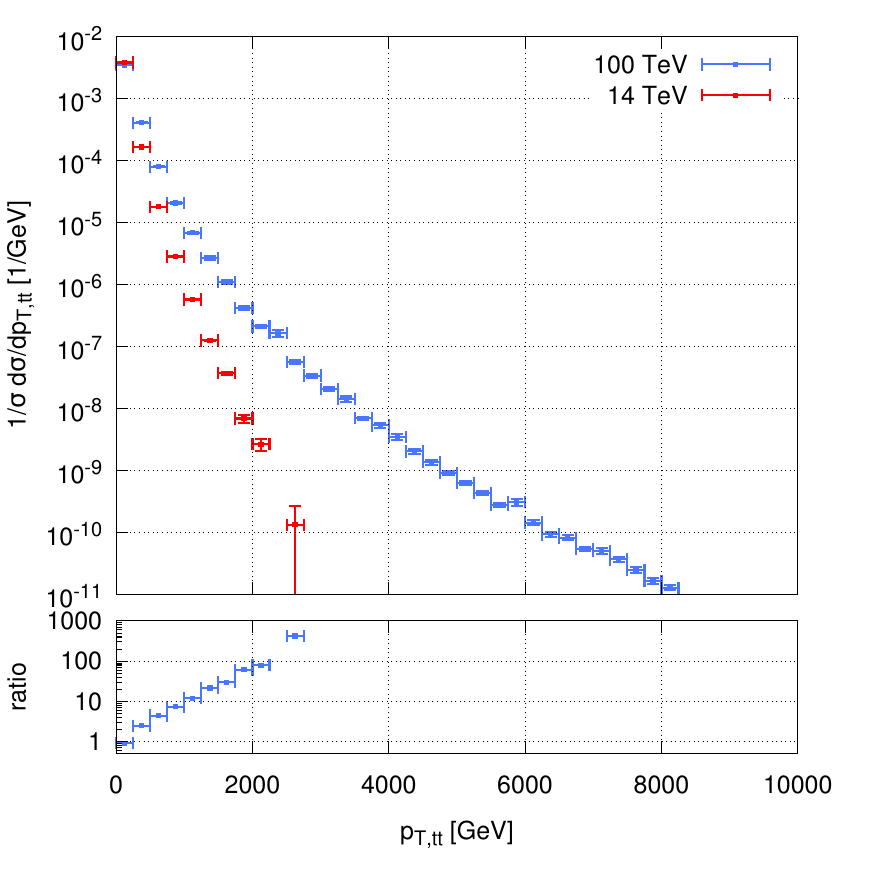}
\includegraphics[width=0.48\textwidth]{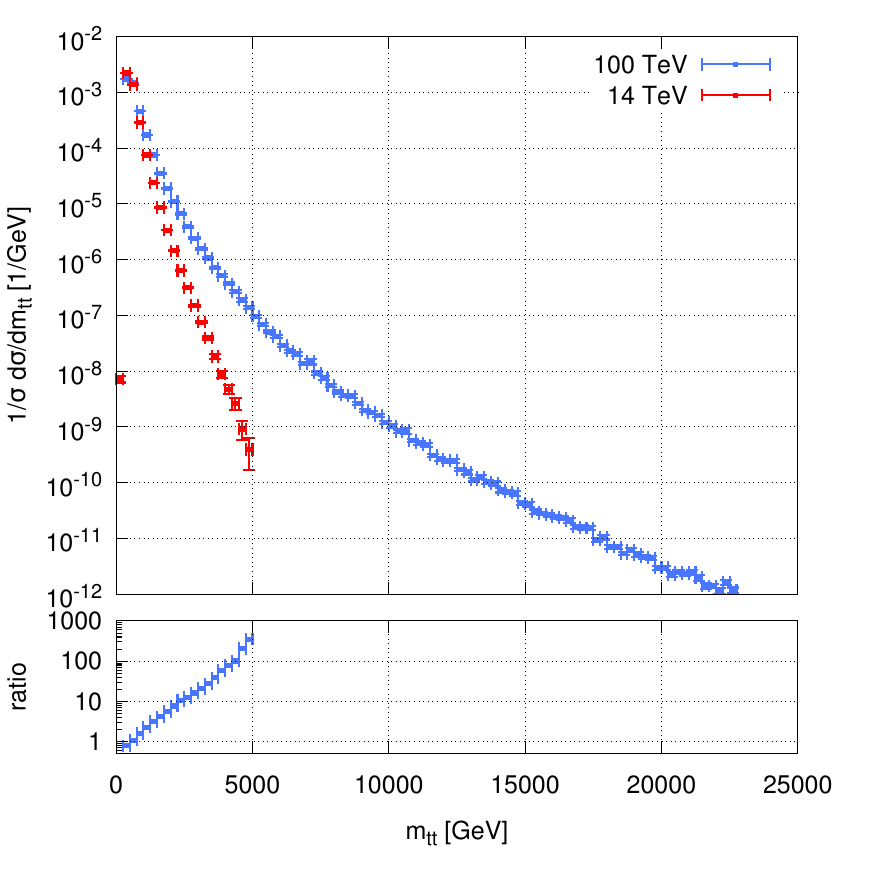}
\caption{\label{fig:top-distr}Normalised distributions for, from left
  to right, top quark transverse momentum, transverse momentum of the
  $t\bar t$ pair and its invariant mass, as evaluated by a NLO+PS
  calculation performed with the POWHEG-BOX implementation of heavy
  quark hadroproduction.}
\end{figure}

This first high-$Q^2$ regime can be further investigated by looking at
other differential distributions for the top quark beyond the
invariant mass. In the following we show results obtained using
POWHEG-BOX implementation of the NLO calculation for heavy quark
hadroproduction~\cite{Frixione:2007vw,Frixione:2007nw,Alioli:2010xd},
matched to the parton shower of PYTHIA~6~\cite{Sjostrand:2006za}
(without MPI).\footnote{In order to improve the generation at
  high-$p_T$ a POWHEG ``Born suppression factor'' $F(p_T) = [(p_T^2 +
  m^2)/(p_T^2 + m^2 + B^2)]^3$ with $B=10$~TeV has been used. $m$ is
  the top quark mass.} NNPDF30 PDFs are used throughout, and the
factorisation/renormalisation scales are set equal to the top
transverse mass. We first show, in Fig.~\ref{fig:top-distr}, the
distributions for the top transverse momentum, the transverse momentum
of the $t\bar t$ pair, and its invariant mass, both at 14 and at 100
TeV. In all three cases, as expected, the normalised distributions at
FCC100 are much harder than at LHC14: they are larger by a factor of
about 10 at a scale of 1 TeV, and of about 100 at 2 TeV.

\begin{figure}[t]
\centering
\includegraphics[width=0.48\textwidth,height=6cm]{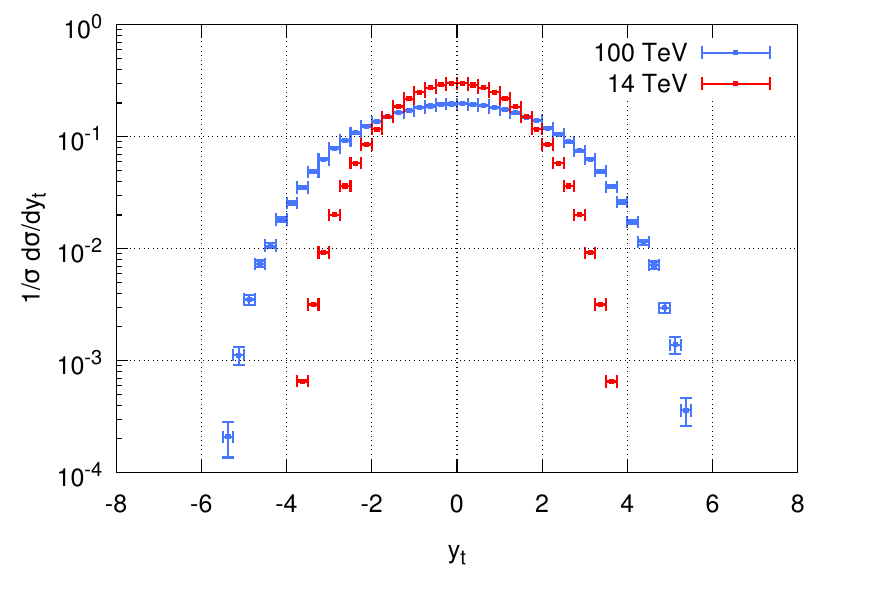}
\includegraphics[width=0.48\textwidth,height=6cm]{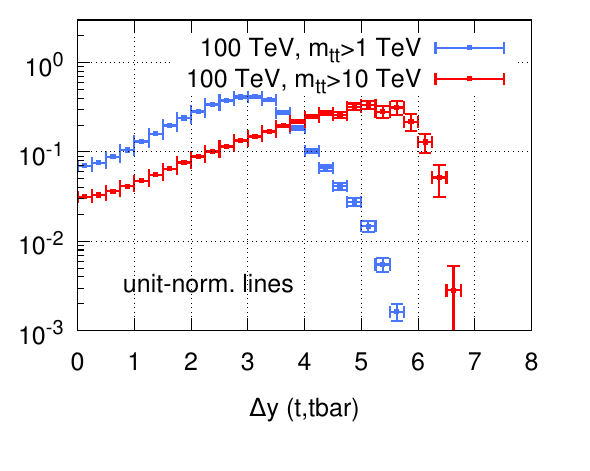}
\caption{\label{fig:top-rap} Left plot: normalised rapidity
  distribution of top quarks at FCC100 and LHC14. Right plot:
  distribution of the rapidity difference $\Delta y$ between the top
  and the anti-top at the FCC100, for two different invariant mass
  cuts. }
\end{figure}

Another characteristic of top distributions at high-$Q^2$ that one can
study is the rapidity dependence. The plots in Fig.~\ref{fig:top-rap}
show that at 100 TeV (and especially so at high invariant masses) top
quarks tend to be produced at larger rapidity than at 14 TeV, and with
a larger rapidity gap.
This suggests that the top quarks at 100 TeV will be a copious source
of large-rapidity lepton. Fig.~\ref{fig:top-decay} shows that this is
indeed the case: one can see that rapidity distributions for the $B$
hadrons and for the leptons produced by top decays are distributed
quite uniformly in rapidity until at least $y \simeq 3$, and only fall
off quite steeply beyond $y \simeq 4$.

\begin{figure}[t]
\centering
\includegraphics[width=0.49\textwidth]{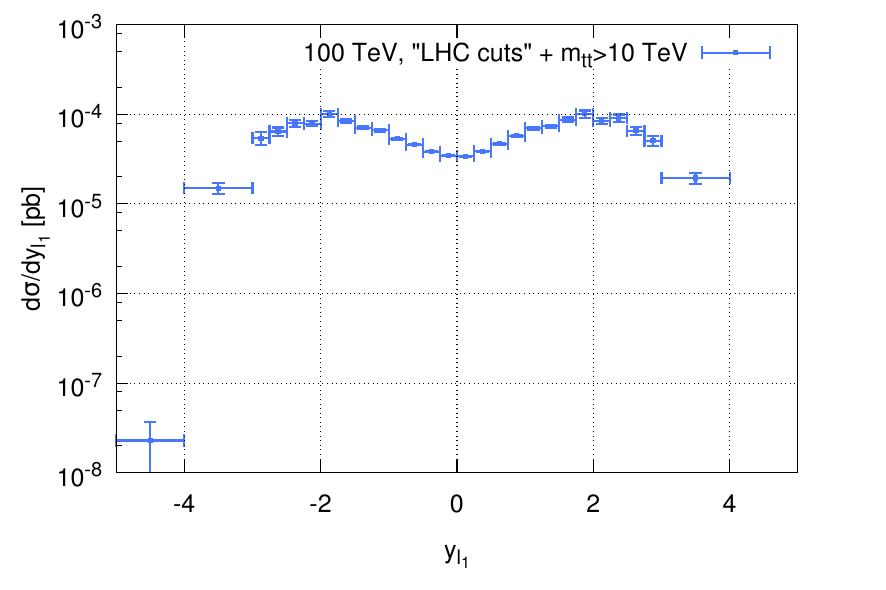}
\includegraphics[width=0.49\textwidth]{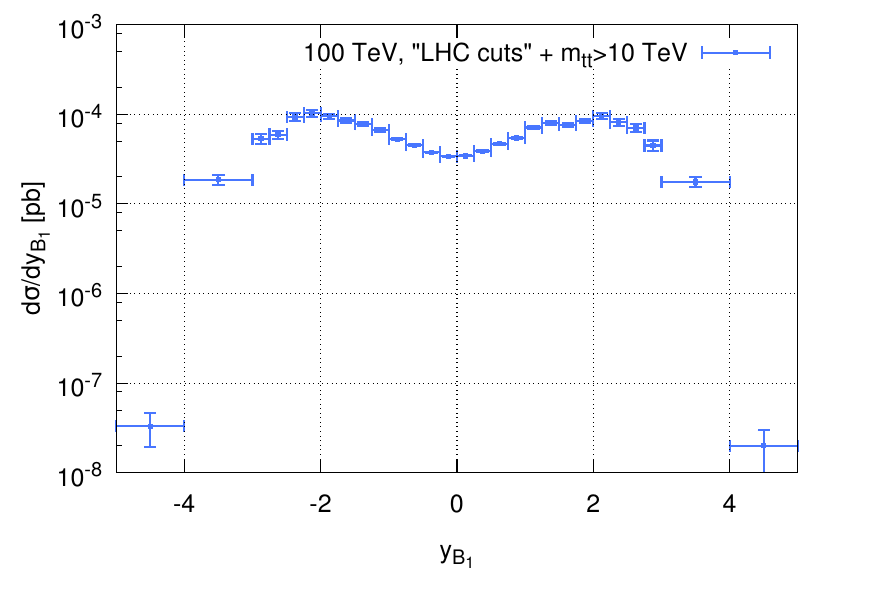}
\caption{\label{fig:top-decay} Rapidity distributions of leptons (left
  plot) and $B$ hadrons (right plot) from top decays.  LHC-like cuts
  for transverse momenta and missing energies ($p_{T,\ell} > 20$~GeV,
  $p_{T,B} > 20$~GeV, $E_{T,miss} > 20$~GeV) are used, but rapidity
  cuts were removed. An additional cut on the invariant mass of the
  $t\bar t$ pair, $m_{t\bar t} > 10$~TeV, is also included. }
\end{figure}

As a consequence of these wide rapidity distributions, ``LHC-like''
lepton cuts, where the leptons are only measured in a central
acceptance region $|y_\ell| < 2.5$, may turn out not to be ideal at
100 TeV. Moving this cut to at least $y_\ell=3$-3.5 would reduce the
cross-section loss by a non-negligible amount.

Since the top quark transverse momentum distribution is expected to
remain measurable at the FCC100 up to several TeV, it is worth
studying how the cross section at such large transverse momenta (much
larger than the top mass) is affected by multiple quasi-collinear
emissions of gluons off the top quarks. Techniques exist to resum
these emissions to all orders with next-to-leading logarithmic
accuracy~\cite{Mele:1990cw,Cacciari:1993mq}, leading to a more
reliable theoretical prediction. We show in Fig~\ref{fig:fonll-top}
predictions obtained using the FONLL approach~\cite{Cacciari:1998it},
compared to the next-to-leading order results. While the FONLL and the
NLO predictions are largely compatible within their respective
uncertainties (estimated varying the renormalisation and factorisation
scales), as expected the FONLL distribution is softer, and has a
smaller perturbative uncertainty.

\begin{figure}[t]
\centering
\includegraphics[width=0.48\textwidth]{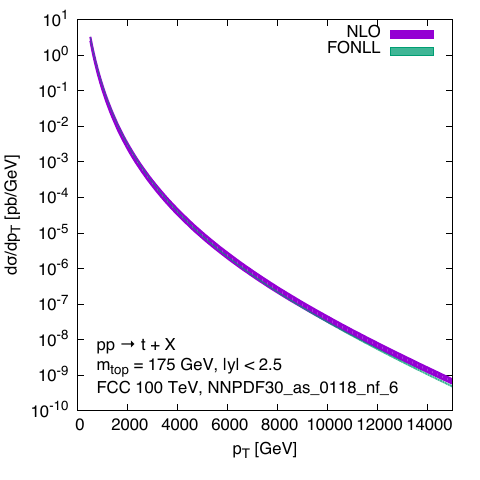}
\includegraphics[width=0.48\textwidth]{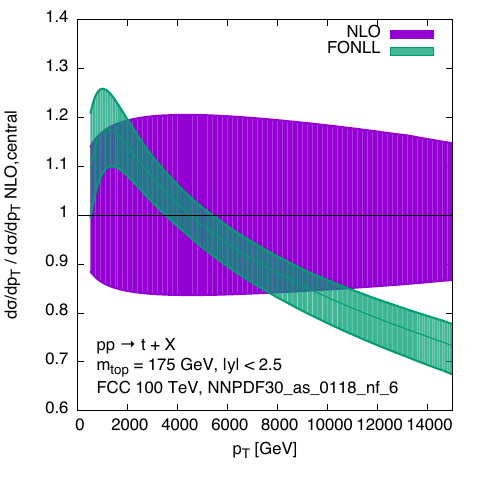}
\caption{\label{fig:fonll-top} Left plot: Transverse momentum
  distribution of top quarks in $t\bar t$ hadroproduction, calculated
  to NLO and also with the FONLL approach. Uncertainties are estimated
  by varying the renormalisation and factorisation scales within a
  factor of two around the top transverse mass, with the constraint
  $1/2 < \mu_R/\mu_F < 2$. Right plot: ratios to the NLO central
  prediction.}
\end{figure}

\begin{figure}[tp]
\centering
\includegraphics[width=0.47\textwidth]{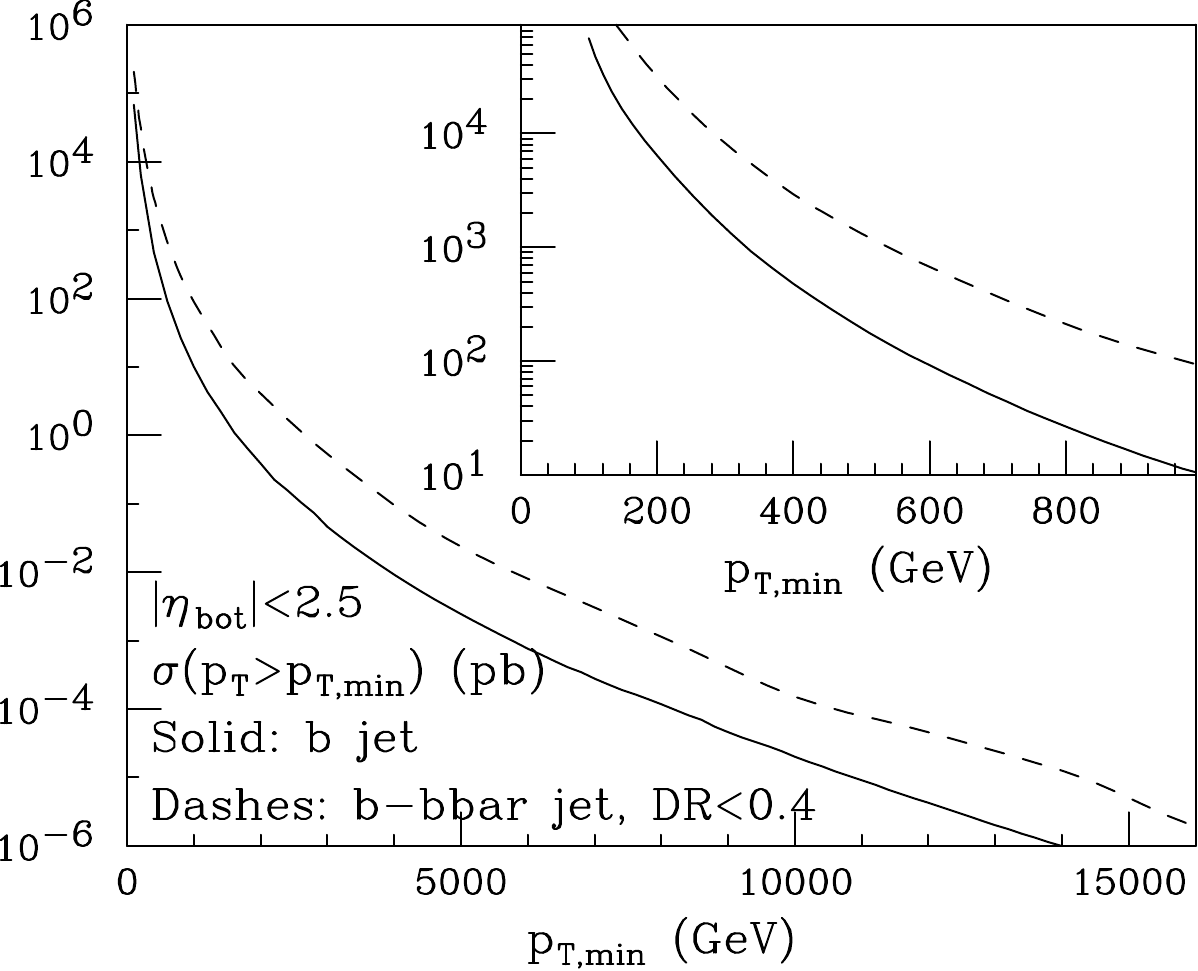}
\includegraphics[width=0.47\textwidth]{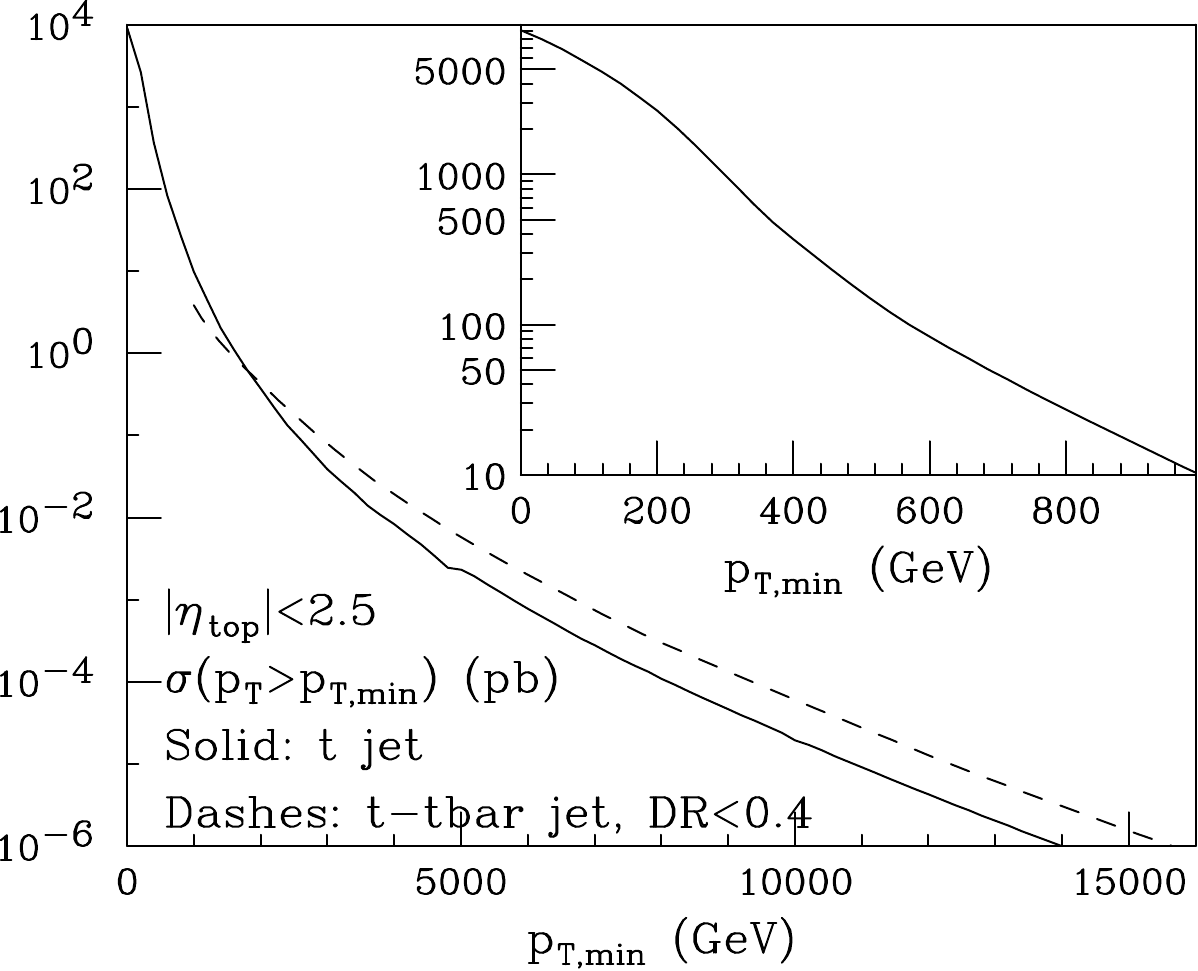}
\caption{Left: production rates for $b$ jets (solid), and for jets
  containing a $b\bar{b}$ pair within $\Delta R<0.4$ (dashes). Right:
  same, for top-quark jets (top treated as stable).  }
\label{fig:bbjet}
\end{figure}

The second regime of high-$Q^2$ production occurs when we request only
one jet to be tagged as containing a heavy quark. This could be of
interest, for example, in the context of high-$p_T$ studies of single
top production.  In this regime, configurations in which the heavy
quark pair arises from the splitting of a large-$p_T$ gluon are
enhanced.  The final state will then contain a jet formed by the
heavy-quark pair, recoiling against a gluon jet. An example of the
role of these processes is shown in Fig.~\ref{fig:bbjet}, where we
compare the $p_T$ spectrum of $b$ jets in events where the $b\bar{b}$
pair is produced back to back (as in the first case we discussed
above), and the spectrum of jets containing the $b$ pair (here jets
are defined by a cone size $R=0.4$). The latter is larger by
approximately one order of magnitude at the highest $p_T$ values,
leading to rates in excess of 1 event/\iab\ for $p_T>15$~TeV. Similar
considerations apply to the case of top quark production in this
multi-TeV regime, as shown in the right plot of
Fig.~\ref{fig:bbjet}. In this case the rate for $t\bar{t}$ jets is
only slightly larger than that for single-top jets, due to the much
larger mass of the top quark, which leads to a smaller probability of
$g\to t\bar{t}$ splitting.

\subsection{Single top production} 
\label{sec:hvq_singletop}

Like $t\bar t$ pairs, production of single top at 100 TeV is also
increased by large factors with respect to LHC. However, since single
top production is dominated by quark initiated $t-$channel production,
the total $t+\bar t$ production cross section grows by about a factor
25 with respect to the LHC13, compared to the growth of about 40 for
the $t\bar t$ cross-section (and of about 15 for its other major
background, $W+$jets).

\begin{figure}[t]
\centering
\includegraphics[width=0.6\textwidth]{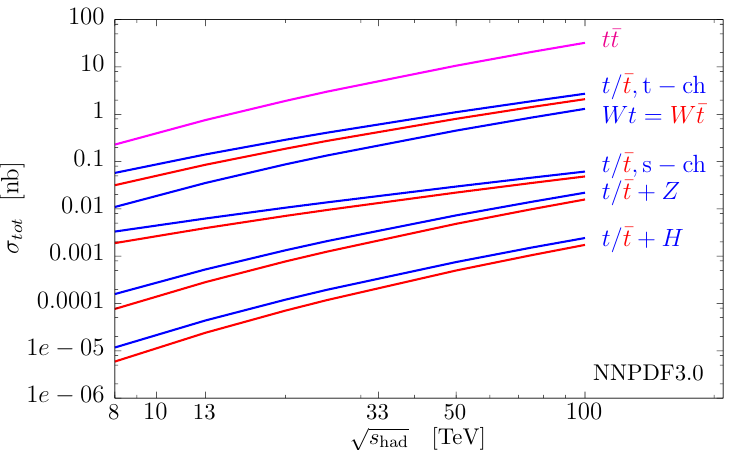}
\caption{\label{fig:singletop-xsect} {Cross sections for top processes
    as a function of proton-proton collider energy. See text for
    details.}}
\end{figure}

\begin{table}
\centering
\begin{tabular}{lcc}
& $pp$, 13 TeV & $pp$, 100 TeV \\
\hline
$\sigma^{t,~t-channel}_{\rm{NNLO}}$ [nb]  & 0.14 & 2.6 \\
$\sigma^{\bar t,~t-channel}_{\rm{NNLO}}$ [nb]  & 0.08 & 2.0 \\
\hline
$\sigma^{Wt}_{\rm{NLO}} = \sigma^{W\bar t}_{\rm{NLO}}$ [nb] & 0.035 & 1.3 \\
\hline
$\sigma^{t,~s-channel}_{\rm{NLO}}$ [pb]  & 6.3 & 61.5 \\
$\sigma^{\bar t,~s-channel}_{\rm{NLO}}$ [pb]  & 3.9 & 48.6 \\
\hline
$\sigma^{tZ}_{\rm{LO}}$ [pb] & 0.5 & 22.1 \\
$\sigma^{\bar tZ}_{\rm{LO}}$ [pb] & 0.3 & 15.8 \\
\hline
$\sigma^{tH}_{\rm{LO}}$ [pb] & 0.01 & 2.4 \\
$\sigma^{\bar tH}_{\rm{LO}}$ [pb] & 0.006 & 1.7 \\
\end{tabular}
\caption{Single top cross sections in $pp$ collisions at 13 TeV and 100 TeV. All
  values are for fully inclusive cross sections, with the exception of the $Wt$
  processes, see text for details.}
\label{table:singletop_xsec}
\end{table}

Fig.~\ref{fig:singletop-xsect} shows the total production cross
section for various channels as a function of the centre of mass
energy. $t\bar t$ and single top results are computed at NLO QCD,
while associated $tZ$ and $tH$ production are computed at LO
QCD\footnote{Predictions are obtained using HatHor~\cite{Kant:2014oha}
  and
  MCFM~~\cite{Campbell:1999ah,Campbell:2011bn,Campbell:2015qma}.}. For
(N)LO predictions (N)LO evolution of $\alpha_s$ and parton
distributions were employed. For all the results in this section we
used the NNPDF3.0 parton set~\cite{Ball:2014uwa}. Apart from
associated $Wt$ production, all results here are fully inclusive and
are computed with $\mu_r=\mu_f=m_t = 172.5$~GeV. For $Wt$ production,
a $b-$jet veto of $p_{b,t}=80$~GeV is applied on additional $b-$jet
radiation coming from $gg\to Wtb$ diagrams to separate this process
from the $t\bar t$ background, see~\cite{Campbell:2005bb} for
details. As suggested in~\cite{Campbell:2005bb}, we used in this case
a lower scale, $\mu=p_{b,t,{\rm veto}} = 80$~GeV. Results for 13 TeV
and 100 TeV are also summarized in
Tab.~\ref{table:singletop_xsec}\footnote{For the numbers in the table
  we computed $t-$channel production to NNLO
  QCD~\cite{Brucherseifer:2014ama}. The difference with respect of NLO
  is however irrelevant for the considerations here.}.

\begin{figure}[t]
\centering
\includegraphics[width=0.49\textwidth]{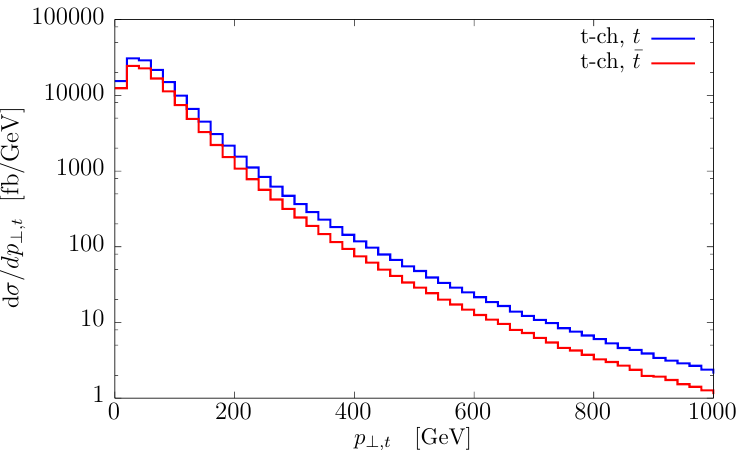}
\includegraphics[width=0.49\textwidth]{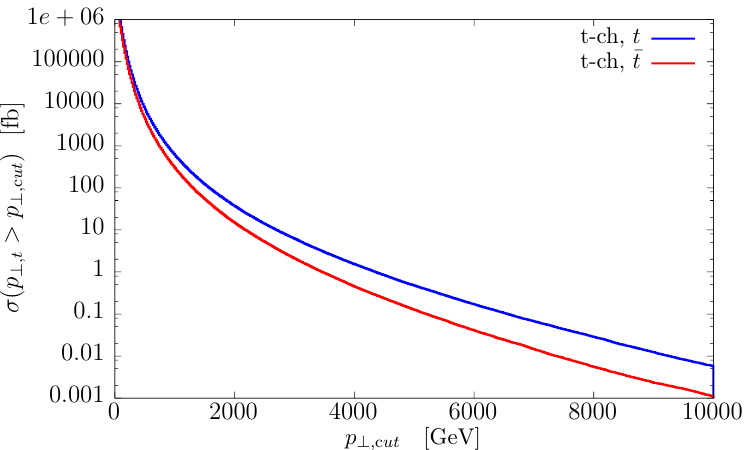}
\caption{\label{fig:singletoppt} {Left: LO $t$-channel single top
    transverse momentum distribution.  Right: LO cross section for
    $t$-channel production as a function of a cut on the top
    transverse momentum. See text for details.}  }
\end{figure}

\begin{table}[t]
\centering
\begin{tabular}{lccc}
 & $p_T^{min} = 0$ & $p_T^{min} = 1$~TeV & $p_T^{min} = 5$~TeV \\
\hline
$\sigma^{t,~t-channel}_{\rm{NLO}}(p_T > p_T^{min})$   & 2.7 nb& 1.0 pb & 0.5 fb\\
$\sigma^{\bar t,~t-channel}_{\rm{NLO}}(p_T > p_T^{min})$   & 2.0 nb& 0.57 pb & 0.2 fb\\
\hline
\end{tabular}
\caption{NLO cross section for $t-$channel single top production as a function of a cut on the
  top transverse momentum.
  See text for details.}
\label{table:singletoppt}
\end{table}

At 100 TeV, $t-$channel single top is about a factor of 20 larger than
at 13 TeV, while $s-$channel production is about a factor of 10
larger. Associated production (with Higgs, $Z$ or $W$) tends to
increase more, about a factor of 35 or so. A consequence of these
different behaviours as a function of the centre of mass energy is
that at 100 TeV the s-channel process becomes even less relevant,
decreasing from 3\% at LHC energy to 1\% of the total single top cross
section. This makes the (in principle unphysical) distinction between
$s-$ and $t-$ channels non problematic at the FCC. On the other hand,
the increased relative importance of $Wt$ associated production (from
20\% to 35\% of the total cross-section) calls for a proper treatment
of this process. This can be achieved by considering the physical
$WWb\bar b$ final state.

We also note that associated $Zt$ and $Ht$ production rates are
sizable at FCC100. The first process is an important background to
FCNC top decays. The second provides information on unitarity in the
Higgs/top sector. For example, modification of the top Yukawa coupling
can lead to unitarity violations in the few TeV regime, which can be
exposed using $Ht$ production. For more details on these processes and
their potential, we refer the reader
to~\cite{Maltoni:2001hu,Farina:2012xp,Campbell:2013yla}.

A study of differential cross sections in $t$-channel single top
production is shown in Fig.~\ref{fig:singletoppt} and in
Table~\ref{table:singletoppt}, where cross section values for $t$ and
$\bar t$ integrated over a given minimum transverse momentum are
given. Even above $p_T = 5$~TeV the integrated cross section remains
in the femtobarn range.

\clearpage
\section{Associated production of top quarks and gauge
  bosons\footnote{Editors: D.~Pagani, I.~Tsinikos}}
\label{ref:top}
\definecolor{rosso}{RGB}{210,0,0}
\def\perc{\%}
\def\sss{\scriptscriptstyle}

\def\mev{\,\textrm{MeV}}
\def\gev{\,\textrm{GeV}}
\def\TO{\rightarrow}

\def\pnote#1{\textcolor{rosso}{[Davide:} \textit{#1}\textcolor{rosso}{]}}
\def\hsnote#1{\textbf{[HS:}\textit{#1}\textbf{]}}
\def\mznote#1{\textbf{[MZ:}\textit{#1}\textbf{]}}
\def\tth{t\bar{t}H}
\def\tthc{\lambda_{\tth}}
\def\pptth{pp\TO\tth}
\def\ggtth{gg\TO\tth}
\def\qqtth{q\bar{q}\TO\tth}
\def\qqtthg{q\bar{q}\TO\tth g}
\def\qgtthq{qg\TO\tth q}
\def\asa#1#2{\alpha_s^{#1}\alpha^{#2}}
\def\tril{\lambda_{(3)}}
\def\trilsm{\tril^{\rm SM}}
\def\qual{\lambda_{(4)}}

\def\beq{\begin{equation}}
\def\beqn{\begin{eqnarray}}
\def\eeq{\end{equation}}
\def\eeqn{\end{eqnarray}}
\def\beal{\begin{align}}
\def\endal{\end{align}}
\def\abs#1{\left|#1\right|}

\def\mydot{\!\cdot\!}
\def\ep{\epsilon}
\def\half{\frac{1}{2}}
\def\quarter{\frac{1}{4}}
\def\ub{\bar{u}}
\def\db{\bar{d}}
\def\bb{\bar{b}}
\def\tb{\bar{t}}
\def\sqs{\sqrt{s}}
\def\epem{e^+e^-}
\def\mpmm{\mu^+\mu^-}
\def\irmv{\remove{i}{0.18}}
\def\jrmv{\remove{j}{0.21}}
\def\irmvb{\removeb{i}{0.18}{0.18}}
\def\jrmvb{\removeb{j}{0.21}{0.17}}
\def\isubrmv{\remove{i}{0.125}}
\def\jsubrmv{\remove{j}{0.145}}
\def\FKSpairs{{\cal P}_{\sss\rm FKS}}
\def\FKSpairsred{\overline{{\cal P}}_{\sss\rm FKS}}
\def\FKSelem{N_{\sss\rm FKS}}
\def\FKSelemred{\overline{N}_{\sss\rm FKS}}
\def\nchannels{N_{\rm ch}}
\def\proc{\qqtth}
\def\procB{r_{\sss B}}
\def\procR{r_{\sss R}}
\def\allproc{{\cal R}}
\def\allprocnpo{\allproc_{n+1}}
\def\allprocn{\allproc_{n}}
\def\BornME{{\cal B}}
\def\nini{n_{\sss I}}
\def\nlight{n_{\sss L}}
\def\nlightB{\nlight^{\sss (B)}}
\def\nlightR{\nlight^{\sss (R)}}
\def\nlightBorR{\nlight^{\sss (B/R)}}
\def\nheavy{n_{\sss H}}
\def\nzero{n_\emptyset}
\def\ident{{\cal I}}
\def\numofgr{N_d}
\def\amp{{\cal A}}
\def\ampmt{\amp^{(m,0)}}
\def\ampnt{\amp^{\asa{1}{1/2}}}
\def\ampntb{\amp^{\alpha^{3/2}}}
\def\ampnpot{\amp^{\asa{3/2}{1/2}}}
\def\ampnpotb{\amp^{\asa{1/2}{3/2}}}
\def\ampnl{\amp_{\proc,~\text {Loop}}^{\asa{2}{1/2}}}
\def\ampnlb{\amp_{\proc,~\text {Loop}}^{\asa{1}{3/2}}}
\def\ampsq{{\cal M}}
\def\ampsqmt{\ampsq^{(m,0)}}
\def\ampsqnt{\ampsq^{(n,0)}}
\def\ampsqnpot{\ampsq^{(n+1,0)}}
\def\ampsqnl{\ampsq^{(n,1)}}
\def\vampsqnl{{\cal V}^{(n,1)}}
\def\hvampsqnl{\hat{\cal V}^{(n,1)}}
\def\vampsqnlF{{\cal V}^{(n,1)}_{\sss FIN}}
\def\hvampsqnlF{\hat{\cal V}^{(n,1)}_{\sss FIN}}
\def\tampsq{\widetilde{\cal M}}
\def\tampsqnt{\tampsq^{(n,0)}}
\def\tampsqnpot{\tampsq^{(n+1,0)}}
\def\rone{r_{[1]}}
\def\rtwo{r_{[2]}}
\def\Ione{\ident_1}
\def\Itwo{\ident_2}
\def\xii{\xi_i}
\def\yij{y_{ij}}
\def\phii{\varphi_i}
\def\yi{y_i}
\def\xic{\left(\frac{1}{\xii}\right)_c}
\def\lxic{\left(\frac{\log\xii}{\xii}\right)_c}
\def\omyijd{\left(\frac{1}{1-\yij}\right)_\delta}
\def\omyid{\left(\frac{1}{1-\yi}\right)_\delta}
\def\opyid{\left(\frac{1}{1+\yi}\right)_\delta}
\def\Dfun{{\cal D}}
\def\Sfun{{\cal S}}
\def\Sfunij{\Sfun_{ij}}
\def\asfun{a_{\Sfun}}
\def\bsfun{b_{\Sfun}}
\def\stepf{\Theta}
\def\phsp{d\phi}
\def\phspn{\phsp_{n}}
\def\phspnpo{\phsp_{n+1}}
\def\tphsp{d\widetilde{\phi}}
\def\tphspn{\tphsp_{n}}
\def\tphspnij{\tphsp_{n}^{ij}}
\def\asotwopi{\frac{\as}{2\pi}}
\def\gs{g_{\sss S}}
\def\aW{\alpha_{\sss W}}
\def\gW{g_{\sss W}}
\def\aem{\alpha}
\def\xicut{\xi_{cut}}
\def\ximax{\xi_{\rm max}}
\def\deltaO{\delta_{\sss O}}
\def\deltaI{\delta_{\sss I}}
\def\NC{N_{\sss c}}
\def\CA{C_{\sss A}}
\def\CF{C_{\sss F}}
\def\TF{T_{\sss F}}
\def\DA{D_{\sss A}}
\def\eikint{{\cal E}}
\def\eikintD{\hat{\cal E}}
\def\APdamp{\overline{P}}
\def\Qdamp{\overline{Q}}
\def\Qop{\vec{Q}}
\def\JetsB{J^{\nlightB}}
\def\JetsR{J^{\nlightB+1}}
\def\kin{\left\{k_k\right\}}
\def\velkl{v_{kl}}
\def\alkl{\alpha_{kl}}
\def\avg{{\cal N}}
\def\symm{\varsigma}
\def\symmnij{\symm_{ij}^{(n)}}
\def\symmnpoij{\symm_{ij}^{(n+1)}}
\def\veck{\vec{k}}
\def\kbar{\bar{k}}
\def\kkdotkl{k_k\mydot k_l}
\def\polv{\varepsilon}
\def\polP{{\cal T}}
\def\polQ{{\cal W}}
\def\muF{\mu_{\sss F}}
\def\muR{\mu_{\sss R}}
\def\clH{{\mathbb H}}
\def\clS{{\mathbb S}}
\def\bt{\bar{t}}
\def\bq{\bar{q}}
\def\bqp{\bar{q}^\prime}
\def\aNLO{{\sc\small MadGraph5\_aMC@NLO}}
\def\UFO{{\sc\small UFO}}
\def\MLf{{\sc\small MadLoop5}}
\def\ML{{\sc\small MadLoop}}
\def\CutTools{{\sc\small CutTools}}
\def\OL{{\sc\small OpenLoops}}
\def\MadFKS{{\sc\small MadFKS}}
\def\Madspin{{\sc\small Madspin}}
\def\Pythiae{{\sc\small Pythia8}}
\def\FastJet{{\sc\small FastJet}}
\def\pt{p_{\sss T}}
\def\Ht{H_{\sss T}}
\def\ttbar{t\bar{t}}
\def\ttVV{\ttbar VV}
\def\ttWW{\ttbar W^{+}W^{-}}
\def\ttZZ{\ttbar ZZ}
\def\ttaa{\ttbar \gamma\gamma}
\def\ttWa{\ttbar W^{\pm}\gamma}
\def\ttWZ{\ttbar W^{\pm}Z}
\def\ttV{\ttbar V}
\def\ttZa{\ttbar Z\gamma}
\def\ttZ{\ttbar Z}
\def\ttW{\ttbar W^{\pm}}
\def\tta{\ttbar \gamma}
\def\tttt{\ttbar \ttbar}
\def\mua{\mu_{a}}
\def\mug{\mu_{g}}

At 100 TeV, heavy particles and high-multiplicity final states are
abundantly produced, giving the opportunity to scrutinise the dynamics
and the strength of the interactions among the heaviest known
particles: the gauge and Higgs bosons, and the top quark. The large
rates, and the very high energies at which these particles can be
produced, open new opportunities to test with greater precision and at
smaller distances the couplings of the top quark with the $W,Z$ and
Higgs bosons.

Final states involving the heaviest states of the SM are also an
important ingredient of physics at 100 TeV, since they naturally lead
to high-multiplicity final states (with or without missing transverse
momentum). These signatures are typical in BSM scenarios featuring new
heavy states decaying via long chains involving, e.g., dark matter
candidates.  Thus, whether as signal or as background processes,
predictions for this class of SM processes need to be known with the
best possible precision, to maximise the sensitivity to deviations
from the SM.

Table~\ref{tab:topXS} shows the NLO cross sections for the inclusive
production of two top quark pairs, and for production in association
with one and two gauge bosons.
\begin{table}[h]
\begin{center}
\def\arraystretch{1.5}
\begin{tabular}{ l  | c | c | c | c| c | c }
 &  $t\bar{t}t\bar{t}$ &$t\bar{t}\,W^\pm $ & $t\bar{t}\,Z^0$ & $t\bar{t}\,WW$ &
  $t\bar{t}\,W^\pm Z$ & $t\bar{t}\,ZZ$  
\\   \hline
$\sigma$(pb) & 4.93 & 20.5 & 64.2 & 1.34 & 0.21 & 0.20
\\
\end{tabular}
\end{center}
\caption{NLO cross sections for associated production of (multiple)
  top quark pairs and gauge bosons~\cite{Torrielli:2014rqa,Maltoni:2015ena}.}
\label{tab:topXS}
\end{table}
Comparing the rates for associated production, in
Table~\ref{tab:topXS}, with those in Table~\ref{tab:multiVB} for
multiple gauge boson production, and considering that each top quark
gives rise to a $W$ through its decay, we remark that top quark
processes at 100~TeV will provide the dominant source of final states
with multiple $W$ bosons, and thus with multiple leptons. This will
have important implications for the search of new physics signals
characterized by the presence of many gauge bosons or leptons from the
decay of the new heavy particles.

Notice also that $t\bar{t}Z^0$ production is more abundant than
$t\bar{t}W^\pm$, contrary to the usual rule that $W$ bosons are
produced more frequently than $Z^0$'s in hadronic collisions. This is
because the $t\bar{t}Z^0$ process is driven by the $gg$ initial state,
which for these values of $\hat{s}/S$ has a much larger luminosity
than the $q\bar{q}'$ initial state that produces $t\bar{t}W$. This
also implies that studies of top production via initial state light
quarks (e.g. in the context of $t$ vs $\bar{t}$ production
asymmetries) will benefit from a higher purity of the $q\bar{q}$
initial state w.r.t. $gg$ if one requires the presence of a $W$ boson
(see e.g. Ref.~\cite{Maltoni:2014zpa}).

In this section we discuss in some detail the associated production of
a top-quark pair with one boson ($\ttV$), covering a broad range of
kinematical regions.  Associated production with a Higgs boson is
discussed in more detail in the Higgs volume of this report.  We
review the impact of NLO QCD corrections, and the residual theoretical
uncertainties due to missing higher orders, by considering the
dependence of key observables on different definitions of central
renormalisation and factorisation scales and on their
variations. These results for 100~TeV mimic the detailed study
presented for 13~TeV in Ref.~\cite{Maltoni:2015ena}. We refer to this
paper for more details.

\subsection{$t\bar{t}V$ production}
\label{sec:ttv-davide}

\allowdisplaybreaks 

The NLO QCD corrections were calculated for $\ttbar H$ in
\cite{Beenakker:2001rj, Beenakker:2002nc, Dawson:2002tg,
  Dawson:2003zu}, for $\tta$ in \cite{Melnikov:2011ta,Hirschi:2011pa},
for $\ttZ$ in
\cite{Hirschi:2011pa,Lazopoulos:2008de,Garzelli:2011is,Kardos:2011na,Garzelli:2012bn},
for $\ttW$ in \cite{Hirschi:2011pa,Garzelli:2012bn,Campbell:2012dh,
  Maltoni:2014zpa} and for $\tttt$ in
\cite{Bevilacqua:2012em,Alwall:2014hca}. NLO electroweak and QCD
corrections have also already been calculated for $\ttbar H$ in
\cite{Frixione:2014qaa, Yu:2014cka, Frixione:2015zaa} and for $\ttW$
and $\ttZ$ in \cite{Frixione:2015zaa}. Moreover, in the case of the
$\ttbar H$ process, NLO QCD corrections have been matched to parton
showers \cite{Frederix:2011zi, Garzelli:2011vp} and calculated for
off-shell top (anti)quarks with leptonic decays in
\cite{Denner:2015yca}.

The results presented here have been obtained in the {\aNLO}
framework~\cite{Alwall:2014hca}.  We start by defining the approach
used to determine the theoretical systematic uncertainty, obtained
from the variation of renormalisation and factorisation scales.  Given
the broad kinematical range accessible at 100~TeV, in addition to
using a fixed scale we consider dynamical scales that depend on the
transverse masses $(m_{T,i})$ of the final-state particles. Following
Ref.~\cite{Maltoni:2015ena}, we consider the arithmetic mean of the
$m_{T,i}$ of the final-state particles ($\mua$) and the geometric mean
($\mug$), which are defined by:
\beqn
\mua=\frac{H_{T}}{N}:=\frac{1}{N}\sum_{
   i=1,N(+1)  }m_{T,i}\, ,\label{mua}\\
\mug:=\left(\prod_{\substack{i=1,N}} m_{T,i}\right)^{1/N}\, .\label{mug}
\eeqn
Here, $N$ is the number of final-state particles at LO and with
$N(+1)$ in eq. \eqref{mua} we understand that, for the real-emission
events contributing at NLO, the transverse mass of the emitted parton
is included.\footnote{This is not possible for $\mug$; soft real
  emissions would lead to $\mug\sim 0$. Conversely, $\mua$ can also be
  defined excluding partons from the real emission and, in the region
  where $m_{T,i}$'s are of the same order, is numerically equivalent
  to $\mug$. We remind that in {\aNLO} the renormalisation and
  factorisation scales are by default set equal to $H_T/2$.}

All the NLO and LO results have been produced with the {\sc\small
  MSTW2008} (68\% c.l.) PDFs \cite{Martin:2009iq} respectively at NLO
or LO accuracy, in the five-flavour-scheme (5FS) and with the
associated values of $\alpha_s$.  We use $m_t=173 \gev$, $m_H=125
\gev$ and the CKM matrix is considered as diagonal. NLO computations
assume the top quark and the vector bosons to be stable.  If not
stated otherwise photons are required to have a transverse momentum
larger than 50 GeV ($p_T(\gamma)> 50 \gev$) and Frixione isolation
\cite{Frixione:1998jh} is imposed for jets and additional photons,
with the technical cut $R_0=0.4$. The fine structure constant $\alpha$
is set equal to its corresponding value in the $G_{\mu}$-scheme for
all the processes.

As first step, we show for all the $\ttV$ processes the dependence of
the NLO total cross sections on the variation of the renormalisation
and factorisation scales $\mu_r$ and $\mu_f$. This dependence is shown
in Fig.\ref{fig:scales_ttVH_100} by keeping $\mu=\mu_r=\mu_f$ and
varying it by a factor eight around the central value $\mu=\mug$
(solid lines), $\mu=\mua$ (dashed lines) and $\mu=m_t$ (dotted
lines). The scales $\mua$ and $\mug$ are respectively defined in
eqs. \eqref{mua} and \eqref{mug}.
 
As $\mua$ is typically larger than $\mug$ and $m_t$, the bulk of the
cross sections originates from phase-space regions where
$\alpha_s(\mua)<\alpha_s(\mug) ,\, \alpha_s(m_t)$. Consequently, such
choice gives systematically smaller cross sections. On the other hand,
the dynamical-scale choice $\mug$ leads to results very close in shape
and normalisation to a fixed scale of order $m_t$. Note that the scale
dependence is monotone over this broad range for all scale
choices. This is due to the $qg$ initial states, which give a very
large contribution and appear only at NLO. Consequently, no
renormalisation and stabilisation of the $\mu_r$ is present for the
numerically dominant contribution.

As done in~\cite{Maltoni:2015ena}, in the following we use $\mug$ as
the reference scale, and vary $\mu_f$ and $\mu_r$ independently by a
factor of two around the central value $\mug$, $\mug/2<\mu_f ,
\mu_r<2\mug$, in order to estimate the uncertainty due to missing
higher orders.  This can be seen as a more conservative choice than
$\mua$ as central scale; as can be seen in
Fig.~\ref{fig:scales_ttVH_100}, the scale dependence in the range
$\mua/2<\mu_f , \mu_r<2\mua$ is smaller than in the $\mug/2<\mu_f ,
\mu_r<2\mug$ range.

Table~\ref{table:100tevttv} lists LO and NLO cross sections, with PDF
and scale uncertainties, and $K$-factors for the central values.  As
expected, the scale dependence is strongly reduced from LO to NLO
predictions.  $K$-factors are very similar and close to 1, with the
exception of $\ttW$ production. For this process, which at LO includes
only $q\bar{q}$ initial states, the opening of $gq$ channels in the
initial state has a huge effect.  Similar effects may be expected at
NNLO, i.e., the first perturbative order including the $gg$ initial
state.  However, as suggested by the detailed analysis presented in
this section for the case of the $p_T(t \bar{t})$ distributions, NNLO
corrections should not have such a large impact. For the $\tta$
process we also find that in general the dependence of the
cross-section scale variation is not strongly affected by the minimum
$p_T$ of the photon.

\begin{figure}[t]
\centering
\includegraphics[width=0.70\textwidth]{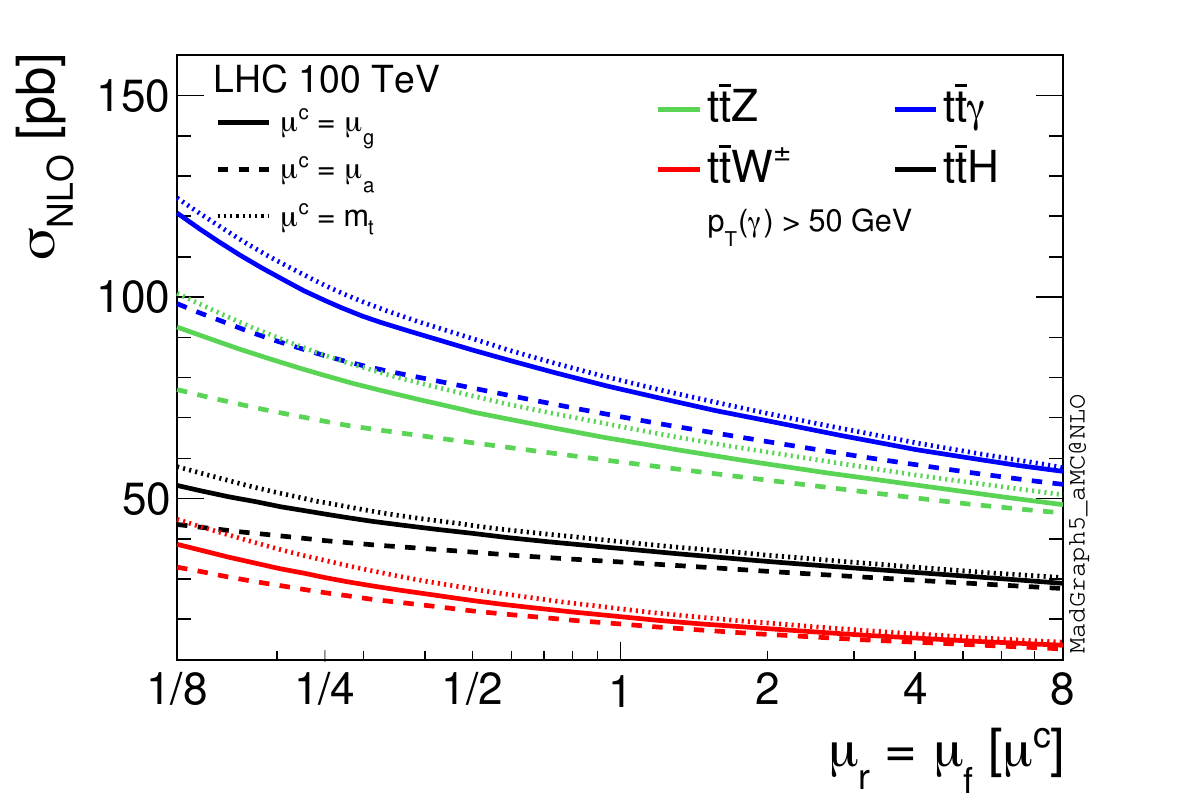}
\caption{Comparison of the NLO scale dependence in the interval
  $\mu^c/8<\mu<8\mu^c$ for the three different choices of the central
  value $\mu^c$: $\mu_g$, $\mu_a$, $m_t$.}
\label{fig:scales_ttVH_100}
\end{figure}
\noindent


\begin{table}[t]
\small
\renewcommand{\arraystretch}{1.5}
\begin{center}
\begin{tabular}{  c | c c c c c }
\hline\hline
100 TeV $ \sigma$[pb] & $t \bar t H$ &  $t \bar t Z$ & $t \bar tW^{\pm}$ & $t \bar t \gamma$ \\ [2pt]
\hline
NLO & $37.56^{+9.9 \%}_{-9.8 \%}~^{+1.0 \%}_{-1.3 \%}$ & $64.07^{+10.8 \%}_{-10.9 \%}~^{+0.9 \%}_{-1.2 \%}$ & $20.65^{+21.5 \%}_{-18.0 \%}~^{+1.1 \%}_{-0.8 \%}$ & $76.68^{+13.3 \%}_{-12.6 \%}~^{+0.9 \%}_{-1.2 \%}$ \\ [2pt]
\hline
LO & $34.26^{+25.6 \%}_{-19.6 \%}~^{+0.9 \%}_{-1.6 \%}$ & $54.57^{+25.3 \%}_{-19.3 \%}~^{+0.9 \%}_{-1.7 \%}$ & $9.39^{+34.1 \%}_{-25.1 \%}~^{+0.9 \%}_{-1.4 \%}$ & $61.51^{+26.8 \%}_{-20.3 \%}~^{+0.9 \%}_{-1.7 \%}$ \\ 
\hline
$K$-factor & 1.10 & 1.17 & 2.20 & 1.25 \\ [2pt]
\hline
\end{tabular}
\caption{NLO and LO cross sections for $\ttV$ processes and $\ttbar H$
  production for $\mu=\mu_g$. The first uncertainty is given by the
  scale variation within $\mug/2<\mu_f,\mu_r<2\mug$, the second one by
  PDFs (MSTW2008). The relative statistical integration error is equal
  or smaller than one permille.}
\label{table:100tevttv}
\end{center}
\end{table}

We now show the impact of NLO QCD corrections on important
distributions and we discuss their dependence on the scale variation
and on the definition of the scales.  For all the processes that we
analysed the distribution of the invariant mass of the top-quark pair
and the $\pt$ and the rapidity of the (anti)top quark, of the
top-quark pair and of the vector or scalar boson.  Here, we show only
representative results; all the distributions considered and
additional ones can be produced via the public code \aNLO.

For each figure, we display together the same type of distributions
for the four different processes considered: $\tta$, $\ttbar H$,
$\ttW$ and $\ttZ$. Most of the plots, for each individual process,
will be displayed in the format described in the following.

\begin{figure}[t]
\centering
\includegraphics[width=0.475\textwidth]{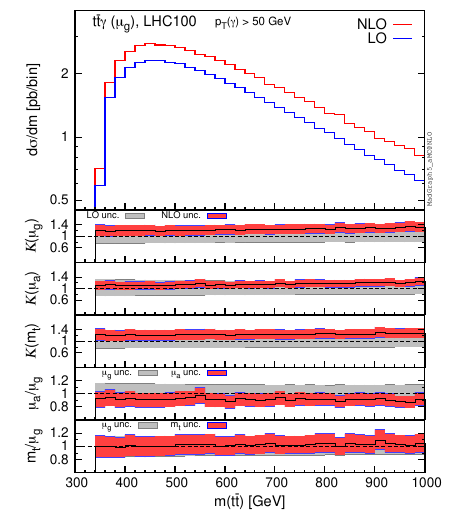}
\includegraphics[width=0.475\textwidth]{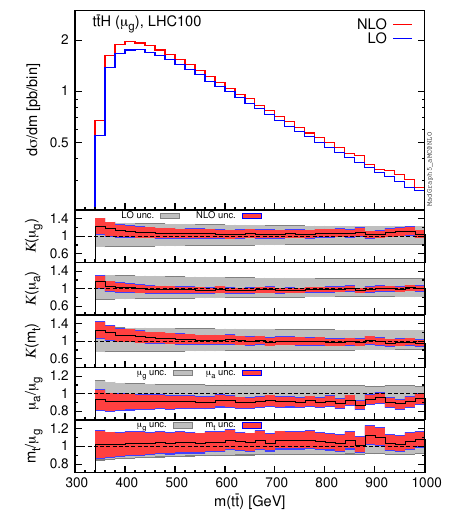}
\includegraphics[width=0.475\textwidth]{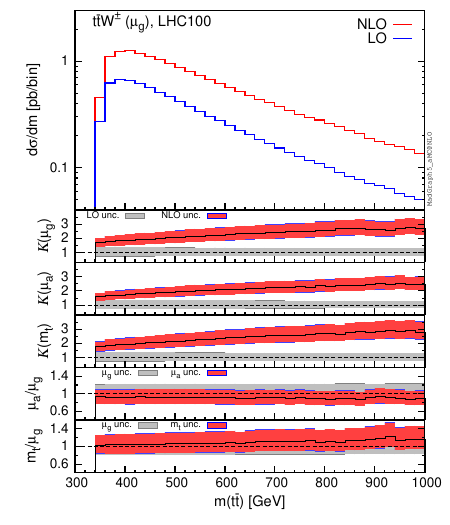}
\includegraphics[width=0.475\textwidth]{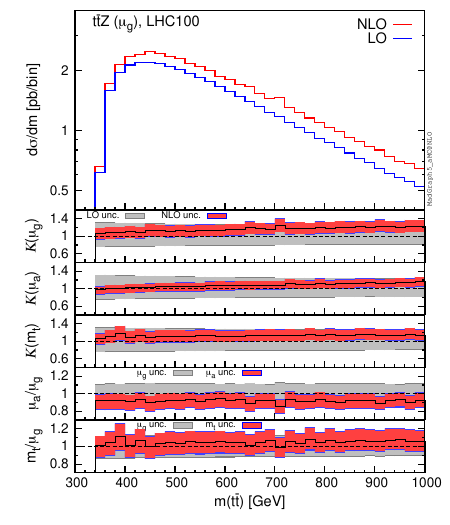}
\caption{Differential distributions for the invariant mass of
  top-quark pair, $m(\ttbar)$ at 100 TeV. The format of the plots is
  described in detail in the text.}
\label{fig:ttV_inv_100}
\end{figure}

\begin{figure}[t]
\centering
\includegraphics[width=0.475\textwidth]{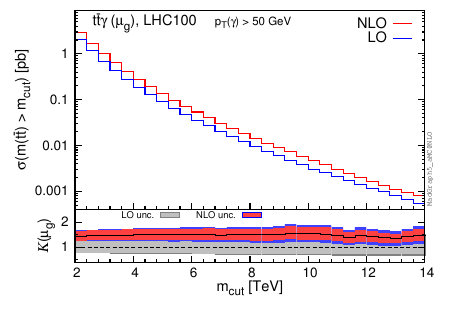}
\includegraphics[width=0.475\textwidth]{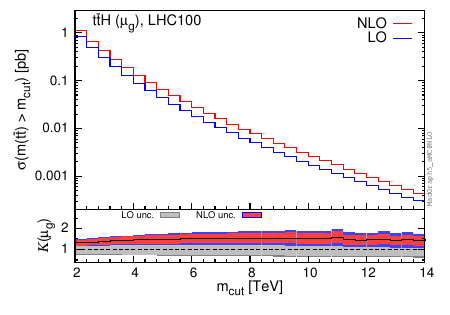}
\includegraphics[width=0.475\textwidth]{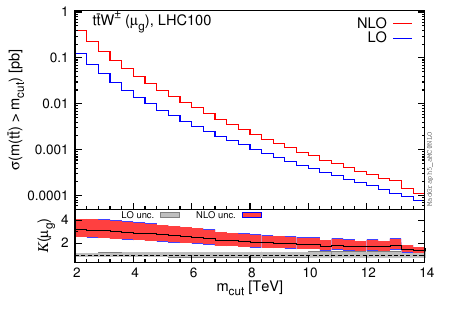}
\includegraphics[width=0.475\textwidth]{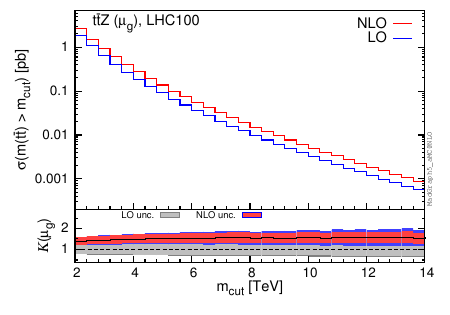}
\caption{Cumulative distributions for the invariant mass of top-quark
  pair, $m(\ttbar)$ at 100 TeV. The format of the plots is described
  in detail in the text.}
\label{fig:ttV_inv_100_cumul}
\end{figure}

In each plot, the main panel shows the distribution at LO (blue) and
NLO QCD (red) accuracy, with $\mu=\mu_f=\mu_r$ equal to the reference
scale $\mug$.  In the first inset we display the scale and PDF
uncertainties normalised to the blue curve, i.e., the LO with
$\mu=\mug$. The light-grey band indicates the scale variation at LO
for the standard range $\mug/2<\mu_f, \mu_r<2\mug$, while the
dark-grey band shows the PDF uncertainty. The black dashed line is the
central value of the light-grey band, thus it is by definition equal
to one.  The solid black line is the NLO QCD differential $K$-factor
for the scale $\mu=\mu_g$, the red band around it indicates the scale
variation in the standard range $\mug/2<\mu_f, \mu_r<2\mug$. The
additional blue borders show the PDF uncertainty.  We stress that in
the plots, as in the tables, scale uncertainties are always obtained
by the independent variation of the factorisation and renormalisation
scales, via the reweighting technique that has been introduced in
\cite{Frederix:2011ss}.  The second and third insets show the same
content of the first inset, but with different scale choices.  In the
second panel both LO and NLO have been evaluated with $\mu=\mua$,
while in the third panel with $\mu=m_t$.

The fourth and the fifth panels show a comparison of NLO QCD
predictions using the scale $\mug$ and, respectively, $\mu_a$ and
$m_t$. All curves are normalised to the red curve in the main panel,
i.e., the NLO with $\mu=\mug$. The light-grey band now indicates the
scale variation dependence of NLO QCD with $\mu=\mug$. Again the
dashed black line, the central value, is by definition equal to one
and the dark-grey borders include the PDF uncertainties. The black
solid line in the fourth panel is the ratio of the NLO QCD predictions
at the scales $\mua$ and $\mug$. The red band shows the scale
dependence of NLO QCD predictions at the scale $\mua$, normalised to
the central value of NLO QCD at the scale $\mug$. Blue bands indicate
the PDF uncertainties.  The fifth panel is completely analogous to the
fourth one, but it compares NLO QCD predictions with $\mug$ and
$m_t$ as central scales.\\

\begin{figure}[t]
\centering
\includegraphics[width=0.475\textwidth]{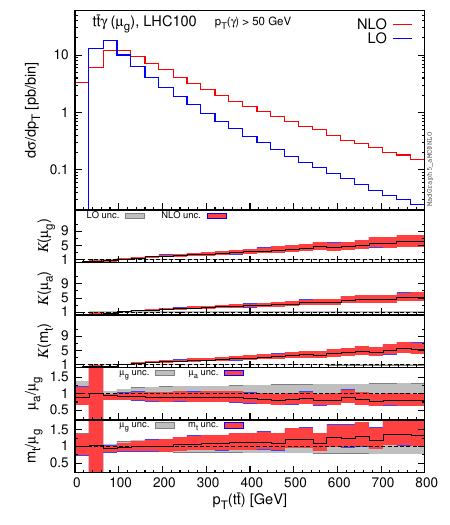}
\includegraphics[width=0.475\textwidth]{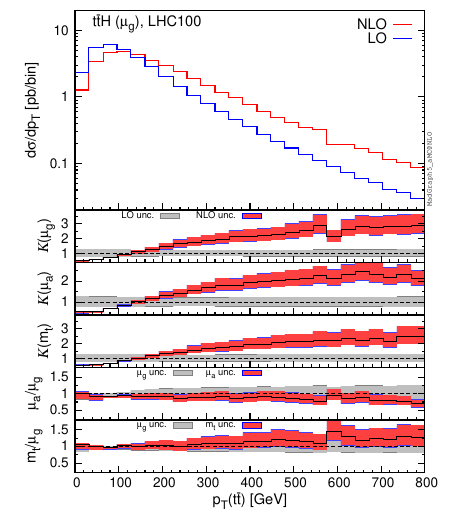}
\includegraphics[width=0.475\textwidth]{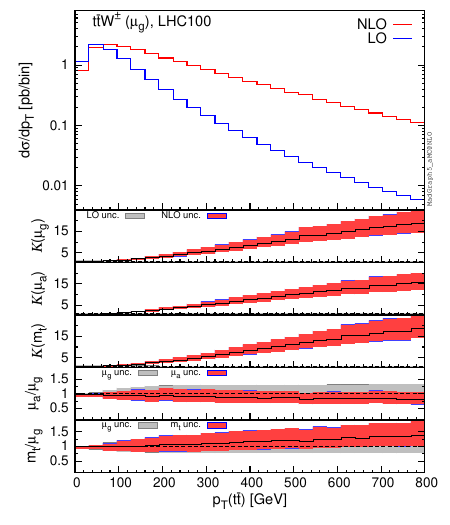}
\includegraphics[width=0.475\textwidth]{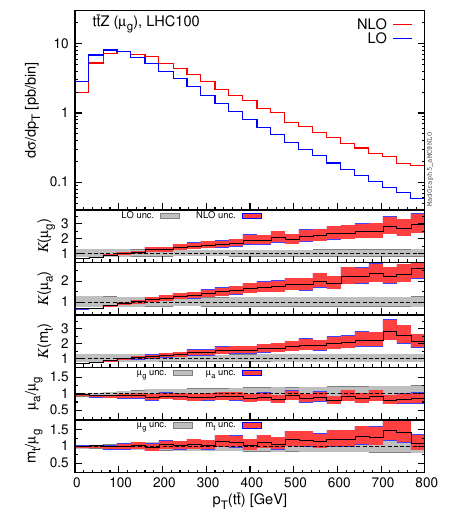}
\caption{Differential distributions for the $\pt$ of top-quark pair,
  $\pt(\ttbar)$ at 100 TeV. The format of the plots is described in
  detail in the text.}
\label{fig:ttV_ptttx_100}
\end{figure}

\begin{figure}[t]
\centering
\includegraphics[width=0.475\textwidth]{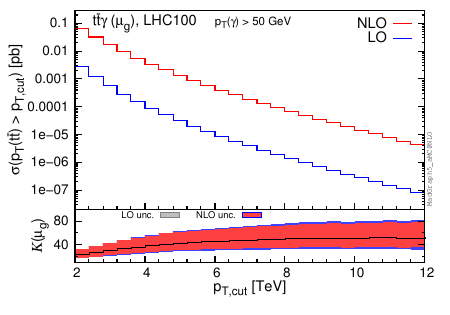}
\includegraphics[width=0.475\textwidth]{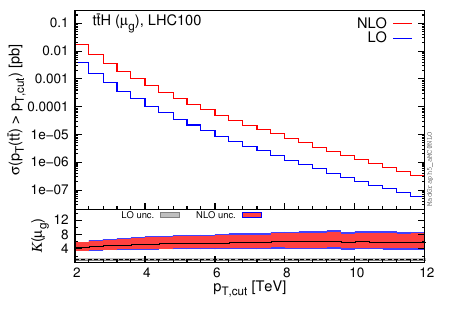}
\includegraphics[width=0.475\textwidth]{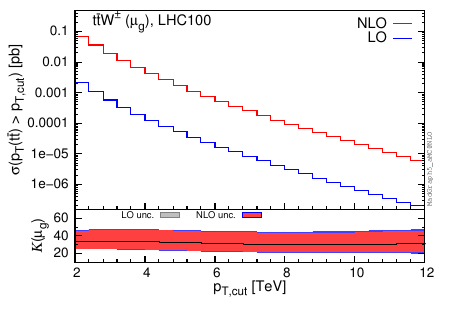}
\includegraphics[width=0.475\textwidth]{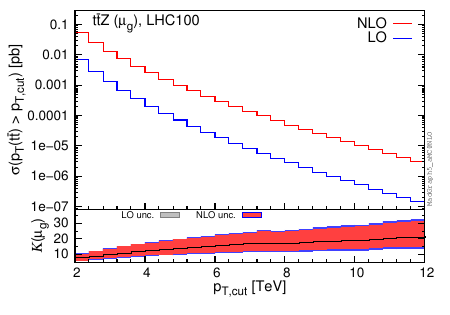}
\caption{Cumulative distributions for the $\pt$ of top-quark pair,
  $\pt(\ttbar)$ at 100 TeV. The format of the plots is described in
  detail in the text.}
\label{fig:ttV_ptttx_100_cumul}
\end{figure}

We start with fig.~\ref{fig:ttV_inv_100}, which shows the distribution
for the invariant mass of the top-quark pair ($m(\ttbar)$) for the
four production processes. From this distribution it is possible to
note some features that are typical for most of the distributions. As
can be seen in the fourth insets, the use of $\mu=\mua$ leads to NLO
values compatible with, but also systematically smaller than, those
obtained with $\mu=\mu_g$. Conversely, the use of $\mu=m_t$ leads to
scale uncertainties bands that overlap with those obtained with
$\mu=\mu_g$.  By comparing the first three insets for the four
different processes, it can be noted that the reduction of the scale
dependence from LO to NLO results is stronger in $\ttbar H$ production
than for the $\ttV$ processes. As said, all these features are not
peculiar for the $m(\ttbar)$ distribution, and they are consistent
with the total cross section analysis presented before, see
fig.~\ref{fig:scales_ttVH_100} and table \ref{table:100tevttv}. From
fig.~\ref{fig:ttV_inv_100} one can also see that the two dynamical
scales $\mug$ and $\mua$ yield slightly flatter $K$-factors than those
obtained with the fixed scale $m_t$, supporting a posteriori such a
reference scale.

However, at 100 TeV the $K-$factor for the ($m(\ttbar)$) distribution
in $\ttW$ production is not flat, independently of the scale
definition employed, as can be seen in
fig.~\ref{fig:ttV_inv_100}. This effect is induced by the
$qg(\bar{q}g)$ initial states, which have at 100 TeV a relative large
PDF luminosity also for high values of $m(\ttbar)$ and especially
$t$-channel-like diagrams for the top-quark pair, at variance with LO
$q\bar{q}'$ production.

In fig.~\ref{fig:ttV_inv_100_cumul} we display for the same observable
cumulative plots, i.e., we plot the dependence of the total cross
sections on the cut $m(\ttbar)>m_{\rm cut}$ by varying $m_{\rm
  cut}$. We can notice that at very high values of $m_{\rm cut}$ the
luminosities of the $qg(\bar{q}g)$ initial states are not the dominant
ones, for example the $K$-factor of $\ttW$ decreases accordingly. For
cumulative distributions we show in the plots only results obtained by
using $\mug$ as central scale.

For particular observables and processes, like the $\pt$ of the
top-quark pair ($\pt(\ttbar)$) in $\ttW$ and $\tta$ production, the
$K$ factors show a strong kinematic dependence. This is shown in
Figs.~\ref{fig:ttV_ptttx_100} and~\ref{fig:ttV_ptttx_100_cumul}.  The
origin of these effects is well
understood~\cite{Frixione:1992pj,Frixione:1993yp,Baglio:2013toa}.
Top-quark pairs with a large $\pt$ originate at LO from the recoil
against a hard vector or a hard scalar boson. Conversely, at NLO, in
this kinematical configuration the largest contribution emerges from
the recoil of the top-quark pair against a hard jet and a soft scalar
or vector boson. In particular, the cross section for a top-quark pair
with a large $\pt$ receives large corrections from the $qg$ initial
state, which appears for the first time only at NLO.

In the case of $\ttW$ production, for instance, the emission of a $W$
collinear to the final-state quark in $qg\rightarrow \ttW q'$ can be
approximated as the $qg\rightarrow \ttbar q$ process times the
$q\rightarrow q' W^\pm$ splitting. For the $W$ momentum, the splitting
involves a soft and collinear singularity that is regulated by the $W$
mass. Thus, once the $W$ momentum is integrated, the $qg\rightarrow
\ttW q'$ process yields a contribution to the $\pt(\ttbar)$
distributions that is proportional to
$\alpha_s\log^2\left[\pt(\ttbar)/m_W\right]$, leading to large
corrections. The same argument clearly applies also to $\ttZ$ for the
$q\rightarrow q Z$ splitting in $qg\rightarrow \ttZ q$. However, in
the case of $\ttW$, this effect is further enhanced also by a
different reason. Unlike the other production processes, $\ttW$
production does not originate at LO from the gluon--gluon initial
state, which has the largest partonic luminosity. Consequently, the
relative corrections induced by the quark--gluon initial states have a
larger impact.
 
The argument above clarifies the origin of the enhancement at high
$\pt$ of the $\ttbar$ pairs, yet it raises the question of the
reliability of NLO predictions for $\ttbar V$ in this region of the
phase space. In particular, the giant $K$-factors and the large scale
dependence call for better predictions. One could argue that only a
complete NNLO calculation for $\ttbar V$ would settle this
issue. However, since the dominant kinematic configurations involve a
hard jet, it is possible to start from the $\ttbar Vj$ final state and
reduce the problem to the analysis of NLO corrections to $\ttbar Vj$,
which can be automatically obtained with \aNLO.  We have therefore
computed results for different minimum $\pt$ for the additional jet
both at NLO and LO accuracy. In fig.~\ref{fig:ttVj_100}, we summarise
the most important features of the $\ttW(j)$ cross section as a
function of the $\pt(\ttbar)$ as obtained from different
calculations. Similar results, even though less extreme, hold for
$\ttZ$ and $\ttbar H$ final states and therefore we do not show them
for sake of brevity.  In fig.~\ref{fig:ttVj_100}, the solid blue and
red curves correspond to the predictions of $\pt(\ttbar)$ as obtained
from $\ttW$ calculation at LO and NLO accuracy, respectively. The
dashed light blue, purple and light-grey curves are obtained by
calculating $\ttW j$ at LO (with NLO PDFs and $\alpha_s$ and same
scale choice in order to consistently compare them with NLO $\ttW$
results) with a minimum $\pt$ cut for the jets of 50, 100, and 150
GeV, respectively. The three curves, while having a different
threshold behaviour, they all tend smoothly to the $\ttW$ prediction
at NLO at high $\pt(\ttbar)$, clearly illustrating that the dominant
contributions come from kinematic configurations with a hard jet.
Finally, the dashed green line is the $\pt(\ttbar)$ as obtained from
$\ttW j$ at NLO in QCD with the minimum $\pt$ cut of the jet of 100
GeV. This prediction for $\pt(\ttbar)$ at high $\pt$ is stable and
reliable, and in particular it does not feature any large $K$-factor,
as can be seen in the lower inset, which displays the differential
$K$-factor for $\ttW j$ production with the $\pt$ cut of the jet of
100 GeV. For large $\pt(\ttbar)$, NLO corrections to $\ttW j$ reduce
the scale dependence of the LO predictions, but do not increase their
central value. Consequently, since we do not expect large effects from
NNLO corrections in $\ttW$ production at large $\pt(\ttbar)$, a
simulation of NLO $\ttV$+jets merged sample \`a la
FxFx~\cite{Frederix:2012ps} should be sufficient in order to provide
reliable predictions over the entire phase space.

\begin{figure}[h]
\centering
\includegraphics[width=0.8\textwidth]{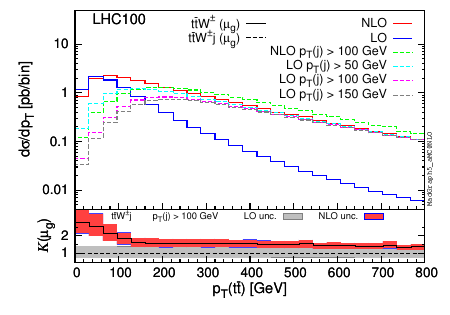}
\caption{Comparison at 100 TeV between differential distribution of
  the $t\bar t$ transverse momentum in $\ttW$ from calculations
  performed at different orders in QCD.  The blue and red solid
  histograms are obtained from the $\ttW$ calculation at LO and NLO,
  respectively. The dashed histograms are obtained from the $\ttW j$
  calculation at LO (light blue, purple, and light grey) and at NLO
  (green), for different minimum cuts (50, 100, 150 GeV) on the jet
  $\pt$. The lower inset shows the differential $K$-factor as well as
  the residual uncertainties given by the $\ttW j$ calculation.}
\label{fig:ttVj_100}
\end{figure}
\noindent

\begin{figure}[h]
\centering
\includegraphics[width=0.475\textwidth]{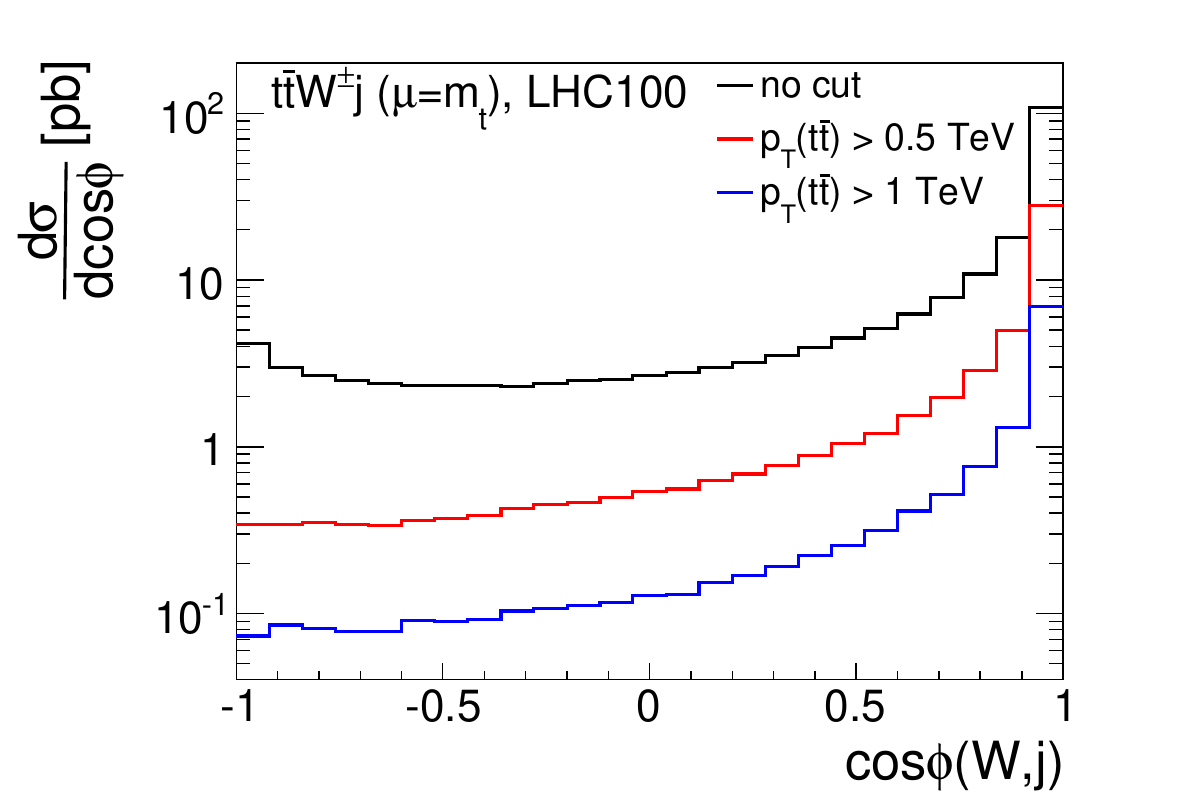}
\includegraphics[width=0.475\textwidth]{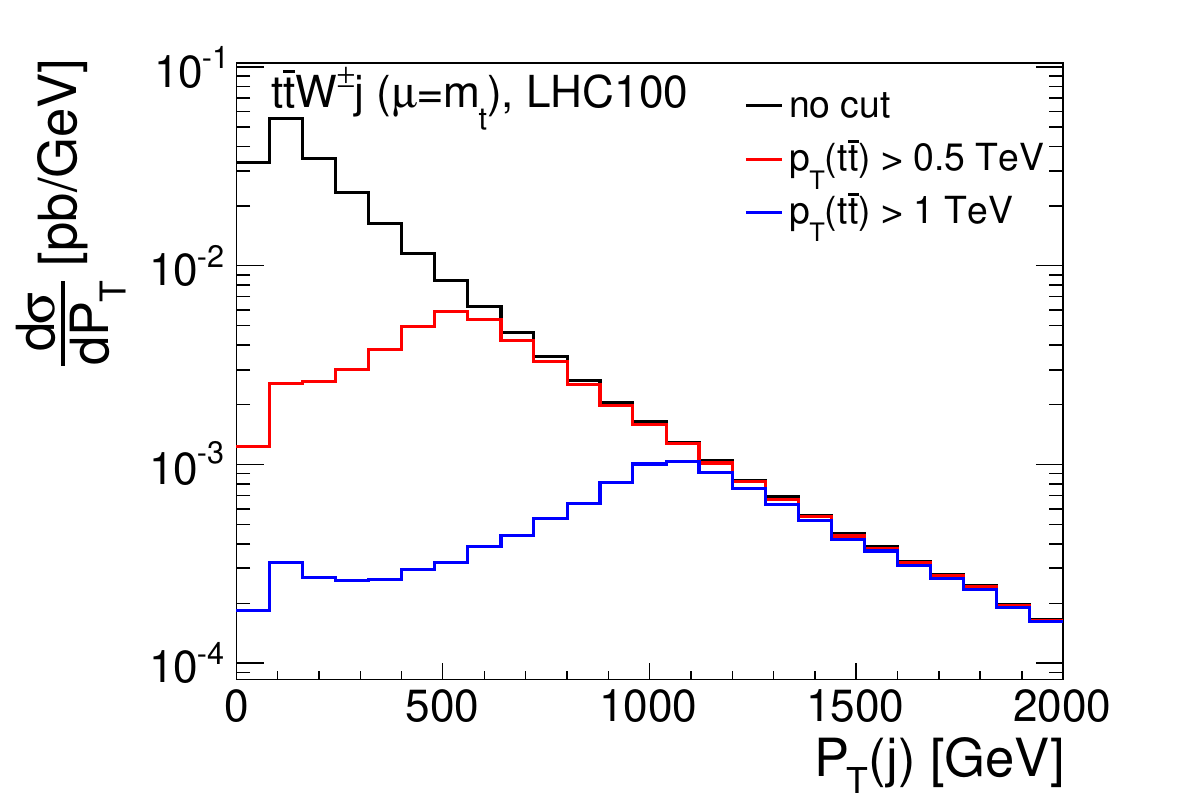}
\includegraphics[width=0.475\textwidth]{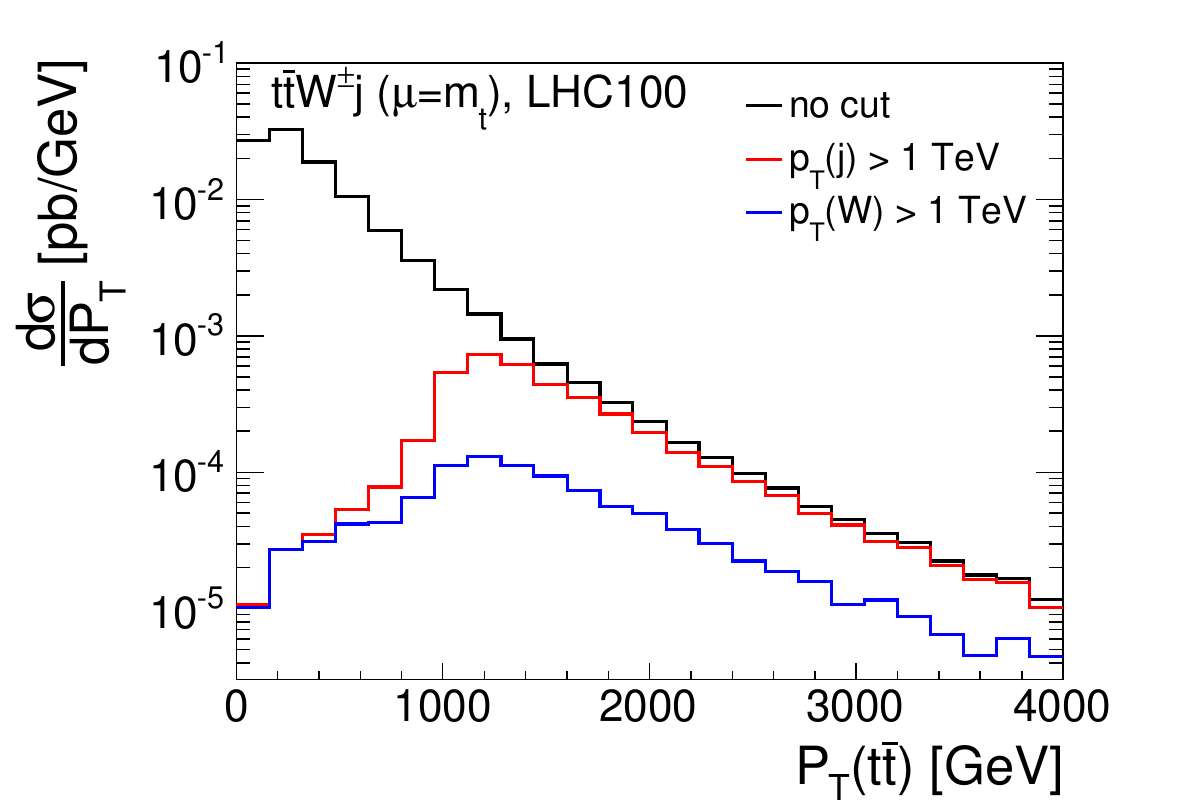}
\includegraphics[width=0.475\textwidth]{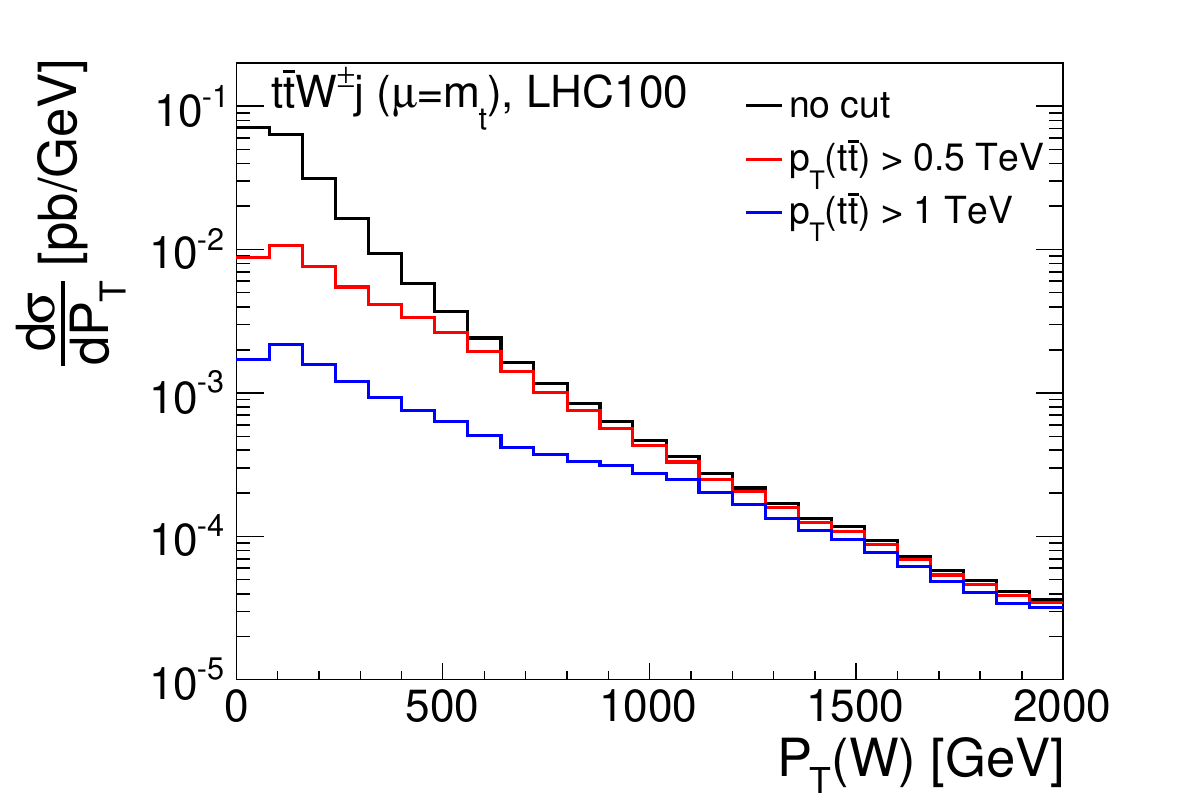}
\caption{Relevant distributions for $\ttW j$ production, where the
  fixed scale $\mu=m_t$ has been used. Black lines are without cuts,
  red and blue lines are with cuts. }
\label{fig:ttWj-various}
\end{figure}

\begin{table}[h]
\small
\renewcommand{\arraystretch}{1.5}
\begin{center}
\begin{tabular}{  c | c c c c }
\hline\hline
100 TeV $ \sigma$[pb] & $t \bar t Hj$ &  $t \bar t Zj$ & $t\bar t W^{\pm}j$ \\
\hline
NLO & $19.42^{+0.7 \%}_{-4.9 \%}~^{+1.0 \%}_{-1.2 \%}$  & $32.38^{+2.4 \%}_{-7.4 \%}~^{+0.9 \%}_{-1.1 \%}$ & $17.16^{+14.9 \%}_{-13.7 \%}~^{+0.7 \%}_{-0.6 \%}$ \\
\hline
LO & $27.02^{+39.3 \%}_{-26.4 \%}~^{+1.1 \%}_{-1.6 \%}$ & $39.81^{+39.8 \%}_{-26.7 \%}~^{+1.1 \%}_{-1.6 \%}$ & $15.67^{+37.7 \%}_{-25.5 \%}~^{+0.5 \%}_{-1.1 \%}$\\
\hline
$K$-factor & 0.72 & 0.81 & 1.10 \\
\hline
\end{tabular}
\caption{Cross sections with $p_T(j) > 100$ GeV. The renormalisation
  and factorisation scales are set equal to $\mu_g$ for the $t \bar t V$. The
  (N)LO cross sections are calculated with (N)LO PDFs, the relative
  statistical integration error is equal or smaller than one permille.
  \label{tab:ttVj_100} }
\end{center}
\end{table}

For completeness, we provide in table \ref{tab:ttVj_100} the total
cross sections at LO and NLO accuracy for $\ttW j$, $\ttZ j$ and
$\ttbar H j$ production, with a cut $p_T(j) > 100$ GeV.  At variance
with what has been done in Fig.\ref{fig:ttVj_100} LO cross sections
are calculated with LO PDFs and the corresponding $\alpha_s$.

In fig. \ref{fig:ttWj-various} we show additional proofs for the
argument discussed so far. We plot relevant distributions for the
$\ttW j$ production. One can see that the $W$ and the jet tends to be
collinear, especially for large $\pt(\ttbar)$, and that the $W$ is
typically soft.

The mechanism discussed in detail in previous paragraphs is also the
source of the giant $K$-factors for large $\pt(\ttbar)$ in $\tta$
production, see fig.~\ref{fig:ttV_ptttx_100}. This process can
originate from the $gg$ initial state at LO. However, the emission of
a photon involves soft and collinear singularities that are not
regulated by physical masses. When the photon is collinear to the
final-state quark, the $qg\rightarrow \tta q$ process can be
approximated as the $qg\rightarrow \ttbar q$ process times a
$q\rightarrow q \gamma$ splitting. In this case, soft and collinear
divergences are regulated by both the cut on the $\pt$ of the photon
($\pt^{\rm cut}$) and the Frixione-isolation parameter $R_0$. We have
checked that, increasing the values of $\pt^{\rm cut}$ and/or $R_0$,
the size of the $K$-factors is reduced.  It is interesting to note
that also corrections in the tail are much larger for $\mu=\mug$ than
$\mu=\mua$. This is due to the fact that the softest photons, which
give the largest contributions, sizeably reduce the value of the scale
$\mug$, whereas $\mua$ is by construction larger than
$2\pt(\ttbar)$. This also suggests that $\mug$ might be an appropriate
scale choice for this process only when the minimum $\pt$ cut and the
isolation parameters on the photon are harder.

\begin{figure}[t]
\centering
\includegraphics[width=0.475\textwidth]{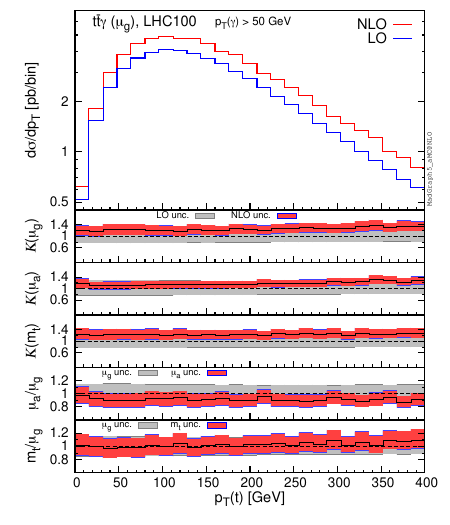}
\includegraphics[width=0.475\textwidth]{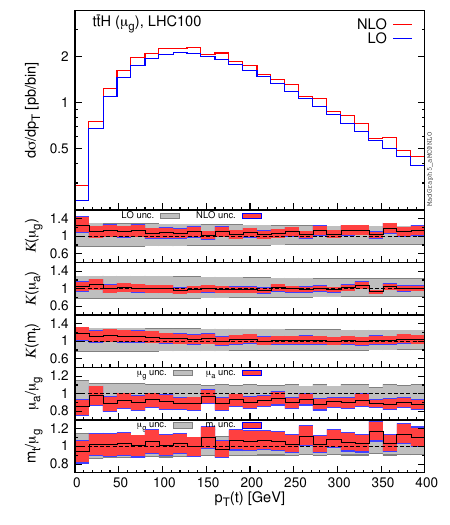}
\includegraphics[width=0.475\textwidth]{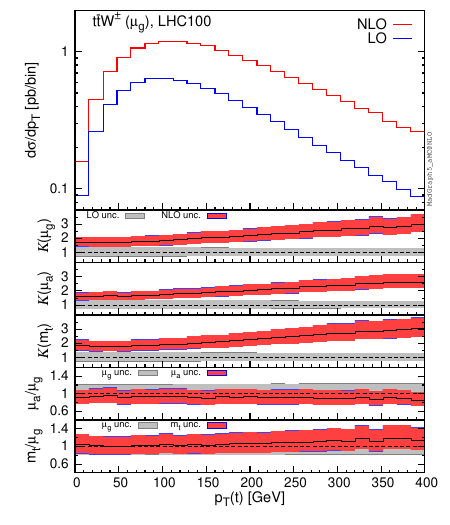}
\includegraphics[width=0.475\textwidth]{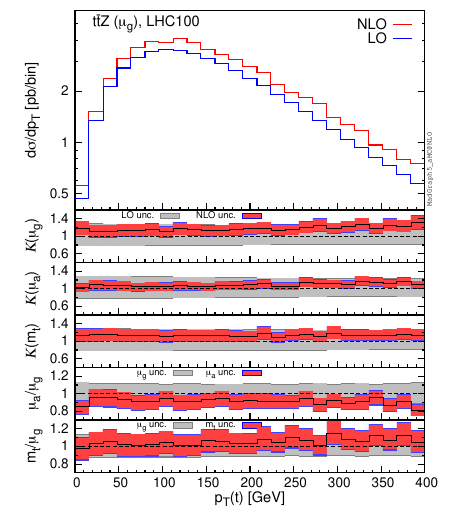}
\caption{Differential distributions for the $\pt$ of top-quark, $\pt(t)$ at 100 TeV. The format of the plots is described in detail in the text.}
\label{fig:ttV_ptt_100}
\end{figure}


\begin{figure}[t]
\centering
\includegraphics[width=0.475\textwidth]{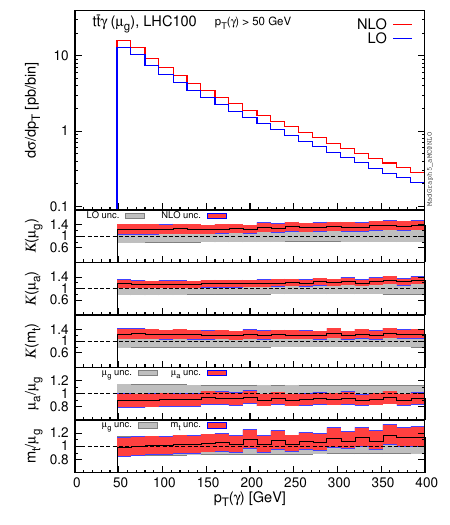}
\includegraphics[width=0.475\textwidth]{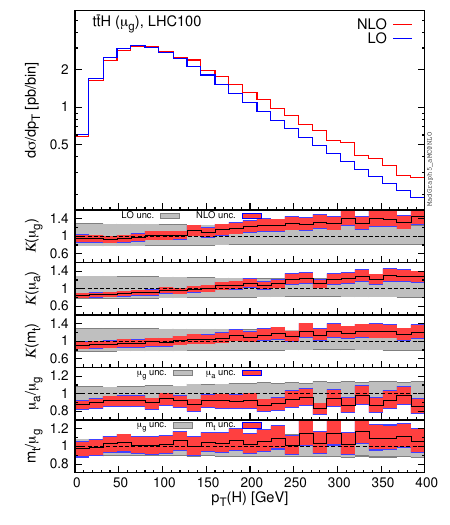}
\includegraphics[width=0.475\textwidth]{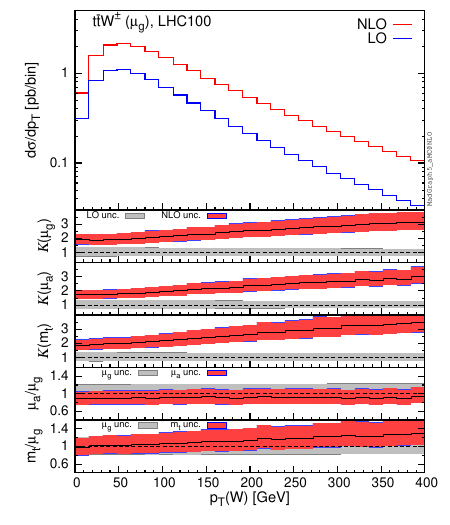}
\includegraphics[width=0.475\textwidth]{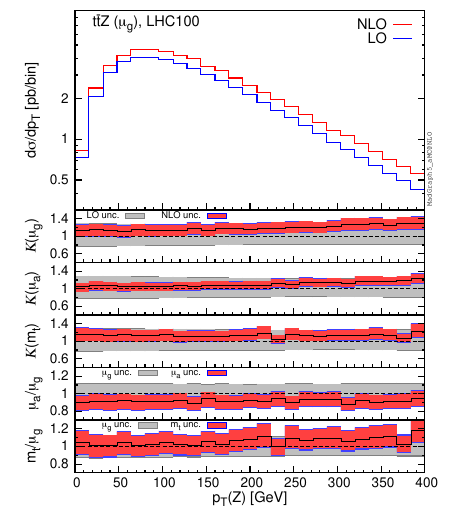}
\caption{Differential distributions for the $\pt$ of the vector or scalar boson, $\pt(V)$ at 100 TeV. The format of the plots is described in detail in the text.}
\label{fig:ttV_ptV_100}
\end{figure}

\begin{figure}[t]
\centering
\includegraphics[width=0.475\textwidth]{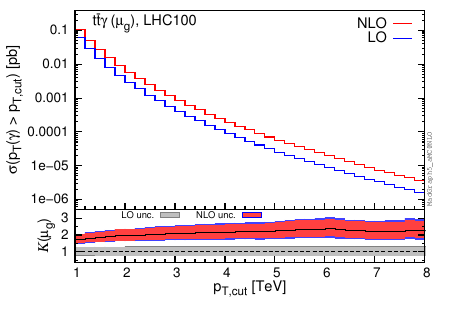}
\includegraphics[width=0.475\textwidth]{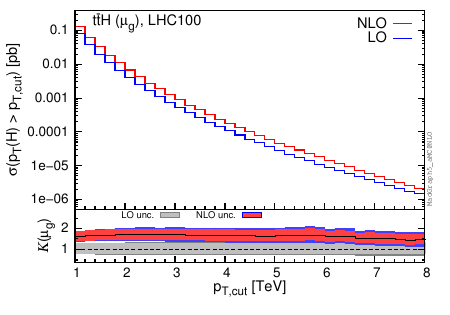}
\includegraphics[width=0.475\textwidth]{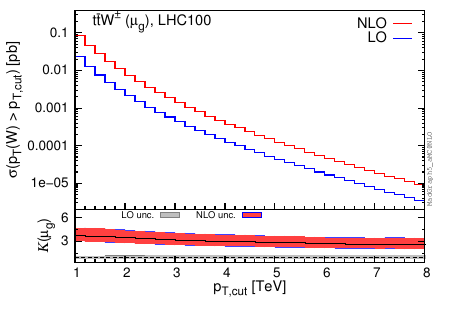}
\includegraphics[width=0.475\textwidth]{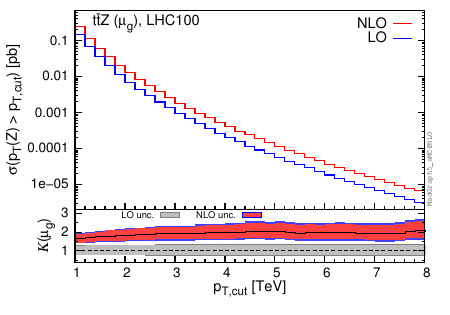}
\caption{Cumulative distributions for the $\pt$ of the vector or scalar boson, $\pt(V)$ at 100 TeV. The format of the plots is described in detail in the text.}
\label{fig:ttV_ptV_100_cumul}
\end{figure}


\begin{figure}[t]
\centering
\includegraphics[width=0.475\textwidth]{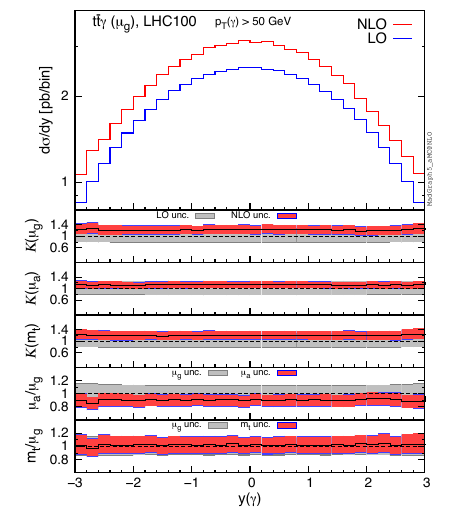}
\includegraphics[width=0.475\textwidth]{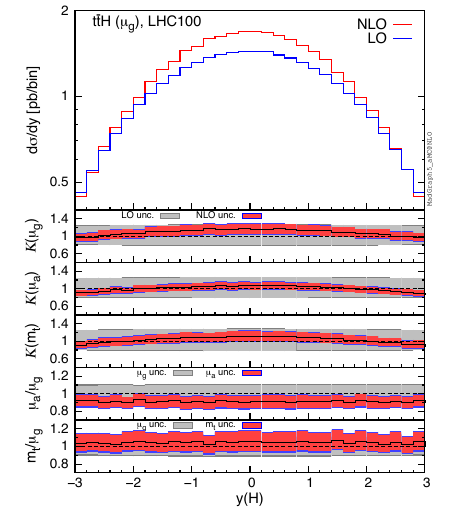}
\includegraphics[width=0.475\textwidth]{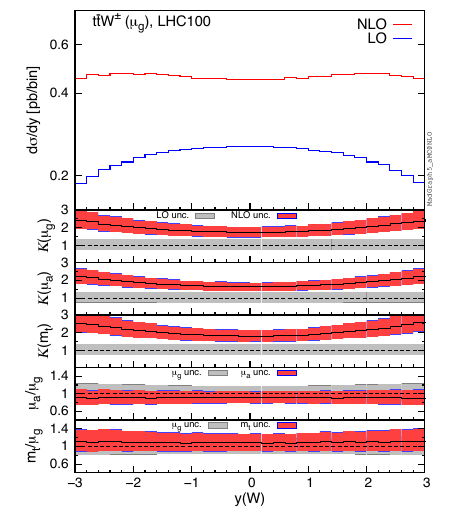}
\includegraphics[width=0.475\textwidth]{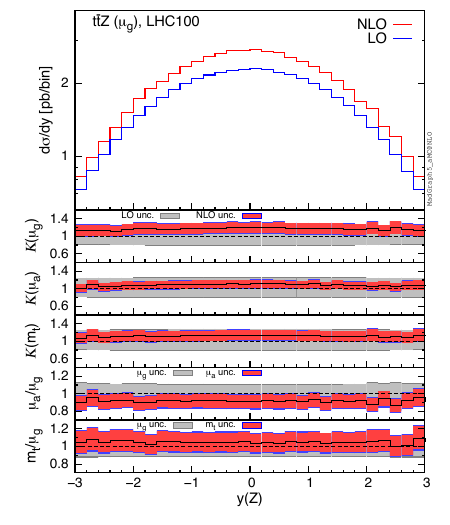}
\caption{Differential distributions for the rapidity of the vector or scalar boson, $y(V)$ at 100 TeV. The format of the plots is described in detail in the text.}
\label{fig:ttV_rapV_100}
\end{figure}

\begin{figure}[t]
\centering
\includegraphics[width=0.475\textwidth]{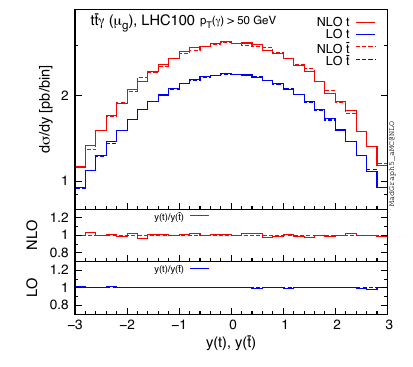}
\includegraphics[width=0.475\textwidth]{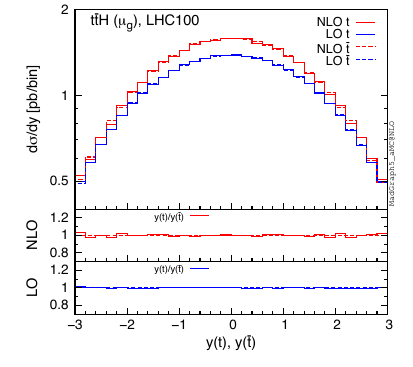}
\includegraphics[width=0.475\textwidth]{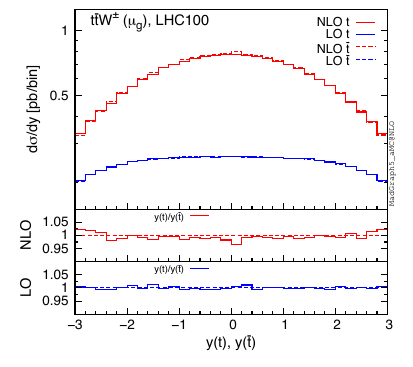}
\includegraphics[width=0.475\textwidth]{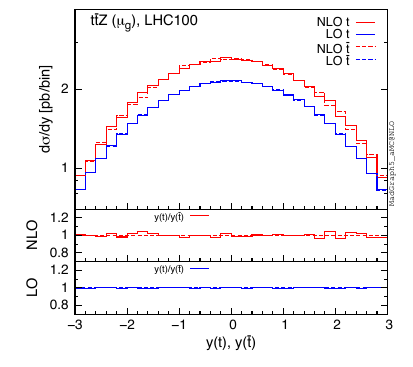}
\caption{Differential distributions for the rapidity of the top quark and antiquark, $y(t)$ and $y(\bar t)$ at 100 TeV.}
\label{fig:ttV_ptt_and_pttx_100}
\end{figure}

In figs.~\ref{fig:ttV_ptt_100} and \ref{fig:ttV_ptV_100} we
respectively show the $\pt$ distributions for the top quark and the
vector or scalar boson, $\pt(t)$ and $\pt(V)$. For these two
observables, we find the general features that have already been
addressed for the $m(\ttbar)$ distributions in
fig.~\ref{fig:ttV_inv_100}.  We display in
fig. \ref{fig:ttV_ptV_100_cumul} cumulative distributions for
$\pt(V)$.

In fig.~\ref{fig:ttV_rapV_100} we display the distributions for the
rapidity of the vector or scalar boson, $y(V)$. For the four processes
considered here, the vector or scalar boson is radiated in different
ways at LO. In $\ttbar H$ production, the Higgs boson is not radiated
from the initial state. In $\ttZ$ and $\tta$ production, in the
quark--antiquark channels the vector boson can be emitted from the
initial and final states, but in the gluon--gluon channel it can be
radiated only from the final state. In $\ttW$ production, the $W$ is
always emitted from the initial-state quarks.  The initial-state
radiation of a vector boson is enhanced in the forward and backward
directions, i.e., when it is collinear to the beam-pipe
axis. Consequently, the vector boson is more peripherally distributed
in $\ttW$ production, which involves only initial state radiation,
with respect to $\tta$ and especially $\ttZ$ production. In $\ttbar H$
production, large values of $|y(V)|$ are not related to any
matrix-element enhancement and indeed the $y(V)$ distribution is much
more central than in $\ttV$ processes.  With NLO QCD corrections, in
$\ttW$ production the vector boson is even more peripherally
distributed. On the contrary, NLO QCD corrections make the
distribution of the rapidity of the Higgs boson even more central.
In fig.~\ref{fig:ttV_rapV_100} one can also notice how the reduction
of the scale dependence from LO to NLO results is much higher in
$\ttbar H$ production than in $\ttV$ type processes.  Furthermore, for
this observable, $K$-factors are in general not flat also with the use
of dynamical scales in the case of $\ttW$ and $t \bar{t} H$. From a
phenomenological point of view, this is particularly important for
$\ttW$, since the cross section originating from the peripheral region
is not suppressed.
 
In fig.~\ref{fig:ttV_ptt_and_pttx_100} we show distributions for the
rapidities of the top quark and antiquark, $y(t)$ and $y(\bar t)$. In
this case we use a different format for the plots. In the main panel,
as in the previous plots, we show LO results in blue and NLO results
in red. Solid lines correspond to $y(t)$, while dashed lines refer to
$y(\bar t)$. In the first and second insets we plot the ratio of the
$y(t)$ and $y(\bar t)$ distributions at NLO and LO accuracy,
respectively. These ratios are in principle useful to identify which
distribution is more central(peripheral) and if there is a central
asymmetry for the top-quark pair.

In the case of $\ttbar$ production the charge asymmetry $A_c$, which
in proton--proton collisions corresponds to a central asymmetry
defined as
\begin{equation}\label{eq:asymmetry_AC}
A_c=\frac{\sigma(|y_t|>|y_{\bar{t}}|)-\sigma(|y_t|<|y_{\bar{t}}|)}{\sigma(|y_t|>|y_{\bar{t}}|)+\sigma(|y_t|<|y_{\bar{t}}|)}\, ,
\end{equation}
or to a forward-backward asymmetry in proton--antiproton collisions,
originates from QCD and EW corrections. At NLO, the asymmetry receives
contributions from the interference of initial- and final-state
radiation of neutral vector bosons (gluon in QCD corrections, and
photons or $Z$ bosons in EW corrections)
\cite{Kuhn:1998jr,Kuhn:1998kw,Bernreuther:2010ny,Hollik:2011ps,Kuhn:2011ri,Bernreuther:2012sx}. Thus,
the real-radiation contributions involve, at LO, the processes
$pp\rightarrow \ttZ$ and $pp\rightarrow \tta$, which are analysed here
both at LO and at NLO accuracy.  The $\tta$ production yields an
asymmetry already at LO, and this feature has been studied in
\cite{Aguilar-Saavedra:2014vta}. The $\ttZ$ production central
asymmetry is also expected to be non vanishing at LO.
The asymmetry is instead analytically zero at LO for $\ttW$ ($\ttbar
H$) production, where the interference of initial- and final-state
$W$(Higgs) bosons is not possible.\footnote{In principle, when the
  couplings of light-flavour quarks are considered as non-vanishing,
  the initial-state radiation of a Higgs boson is possible and also a
  very small asymmetry is generated. However, this possibility is
  ignored here.}

Conversely, at NLO all the $\ttV$ processes and the $\ttbar H$
production have an asymmetry. However, both at LO and NLO asymmetric
effects on $y(t)$ and $y(\bar t)$ distributions are small at 100 TeV
and difficult to be seen in fig.~\ref{fig:ttV_ptt_and_pttx_100}. These
effects can be better quantified by looking directly to the asymmetry
$ A_c$ defined in eq.~\eqref{eq:asymmetry_AC}.  NLO and LO results for
$ A_c$ are listed in table \ref{table:100tevttv_asymm}, which clearly
demonstrates, once again, that NLO QCD effects cannot be neglected in
the predictions of the asymmetries. For $\ttW$ and $\ttbar H$
production, an asymmetry is actually generated only at NLO. The case
of $\ttW$ production has been studied in detail in
\cite{Maltoni:2014zpa}, also for 100 TeV collisions. Furthermore, NLO
QCD corrections largely increase the asymmetry in $\ttZ$ production,
and decrease it by $\sim 40\%$ in $\tta$ production.

\begin{table}[t]
\small
\renewcommand{\arraystretch}{1.5}
\begin{center}
\begin{tabular}{  c | c c c c c }
\hline\hline
100 TeV $ A_c$ [\%]  & $\ttW$ & $t \bar t \gamma$ \\
\hline
LO & - & $ (-0.70\pm 0.05 )^{ +0.04 }_{ -0.04 }~^{ +0.03 }_{ -0.02 } $ \\
\hline
NLO   & $ (1.3\pm 0.1) ^{ +0.23 }_{ -0.16 }~^{ +0.05 }_{ -0.03 }$  & $
(-0.45\pm 0.04)^{ +0.05 }_{ -0.04 }~^{ +0.01 }_{ -0.02 } $ \\
\hline\hline
100 TeV $ A_c$ [\%] & $t \bar t H$ &  $t \bar t Z$  \\
\hline
LO & - & $ (0.03\pm 0.05)^{ +0.001 }_{ -0.004 }~^{ +0.003}_{ -0.01 } $  \\
\hline
NLO & $ (0.17\pm 0.01) ^{ +0.06 }_{ -0.04 }~^{ +0.01 }_{ -0.01 }$ &
$(0.22 \pm 0.04)^{ +0.06 }_{ -0.04 }~^{ +0.01 }_{ -0.01 } $  \\
\hline
\end{tabular}
\caption{NLO and LO central asymmetries for $\ttV$-type processes and
  $\ttbar H$ production at 100 TeV for $\mu=\mu_g$. The first
  uncertainty is due to the limited integration statistics. The second and
  third uncertainties reflect the scale variation within
  $\mug/2<\mu_f,\mu_r<2\mug$, and the PDFs. These were obtained by
  reweighting the distributions, during integration, on an event-by-event basis.}
\label{table:100tevttv_asymm}
\end{center}
\end{table}


\subsection{Photon emission off the top quark decay products}

It is interesting to note that in $t\bar{t}+\gamma$ final states the
photon is not only radiated in the production stage ({\it i.e.} before
the top quarks go on-shell), it is also emitted off the top quark
decay products (after one of the top quarks has gone on-shell).  The
branching $t \to bW + \gamma$ has a kinematically large phase space
with allowed photon energies $p_{\perp,\mathrm{cut}}^\gamma \le
E_\gamma \le m_t-M_W \approx 92$~GeV in the top quark rest frame.  The
small masses of the $b$-quark and $W$ decay products lead to
additional collinear enhancements.  As a result, radiative top quark
decays yield a large contribution to $W^+ W^- b \bar{b} + \gamma$
final states (with intermediate on-shell $t\bar{t}$ pairs).
\begin{figure}[h]
\centering
\includegraphics[scale=0.65]{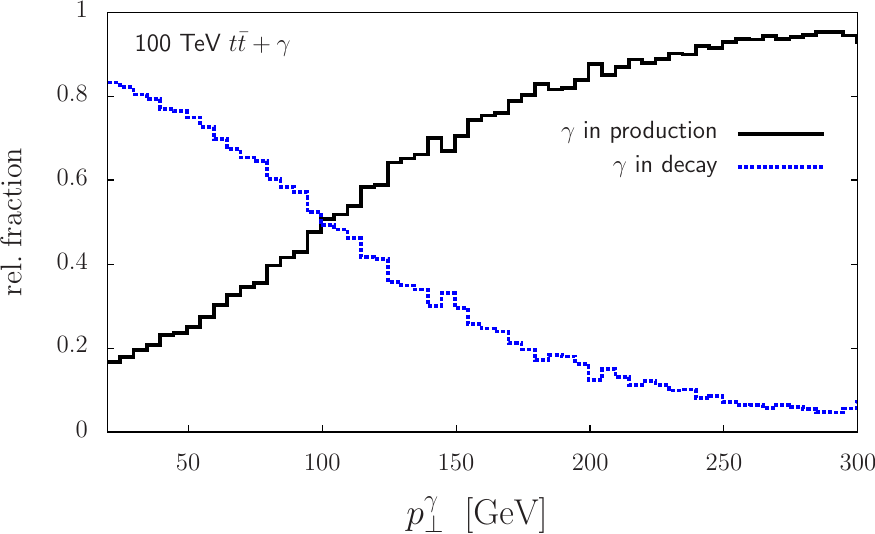}
\caption{Relative contribution of photons from the top quark
  production (black) and decay (blue) stage in $W^+W^-
  b\bar{b}+\gamma$ final states at 100 TeV.  Photons are required to
  have $p_\perp \ge 20$~GeV and be separated from jets and leptons by
  $\Delta R_{\gamma\ell}=\Delta R_{\gamma j}=0.2$.}
\label{fig:ttbgammadecay}
\end{figure}
In Fig.~\ref{fig:ttbgammadecay} we show their relative contribution to
the total cross section and compare them to photons radiated in the
production process.  In this study, we assume photons with
$p_\perp^\gamma \ge 20$~GeV and require a separation of $\Delta R=0.2$
between photons and leptons or jets.  At moderate photon transverse
momenta (20-60~GeV), the contribution from radiative top quark decays
dominates the total cross section with more than 70\%.  Beyond
$p_\perp^\gamma \approx 100$~GeV the contribution from the
$t\bar{t}+\gamma$ process with subsequent top quark decays takes over,
but radiative top quark decays still matter at the 10\% level up to
transverse photon momenta of 300~GeV.  Dedicated selection cuts to
remove the radiative top quark decay process have been presented in
Refs.~\cite{Baur:2001si,Aguilar-Saavedra:2014vta}.  However, at NLO
QCD the fraction of $t \to bW + \gamma$ events that still pass these
cuts can be as large as 10\% \cite{Melnikov:2011ta}, the same order of
magnitude as the NLO corrections themselves.  \\ Because of these
features specific to top quark pair production in association with a
photon, certain care has to be taken when describing a realistic final
state of $W^+ W^- b \bar{b} + \gamma$.  Recent experimental analyses
at the LHC \cite{Aad:2015uwa,CMS:2014wma} apply typical selection cuts
on leptons, jets, missing energy and the photon, but do not explicitly
suppress radiative top quark decays.  Hence, neglecting this
contribution in the theoretical description can lead to an
underestimation of the event rate by a factor of up to 3.

\clearpage
\section{Top properties\footnote{Editors: M~Schulze, J.A.~Aguilar Saavedra}}
\label{ref:tprop}
In the SM, the top quark is possibly the particle whose production and
decay properties are simpler.  It lacks the rich phenomenology of
hadronic spectroscopy characteristic of all other quarks; its decay is
dominated by the $Wb$ final state, with a tiny contamination of $Ws$
and $Wd$, and all other SM-allowed decays (FCNC, $t\to WZb$, etc)
being so small as to be beyond the experimental reach.  On the other
hand, its large mass implies a particular sensitivity to the mechanism
of electroweak symmetry breaking.  Thus, precision studies of the
top-Higgs couplings, as well as the couplings of the top to the
electroweak gauge bosons, are of great importance in understanding
electroweak symmetry breaking and possibly challenging its SM
realization.  Furthermore, new physics unrelated to the mechanism of
electroweak symmetry breaking might be revealed through modifications
of SM interactions rather than through a direct discovery of new
particles. For a general introduction to the study of top quark
properties in hadronic collisions, we refer to the old report on SM
physics at the LHC, Ref.~\cite{Beneke:2000hk}.

We avoid here a discussion of the determination of the top mass at
100~TeV: any progress relative to what will be known at the end of the
LHC will depend on theoretical progress that is hard to anticipate
now, and on a very precise definition of the future experimental
environment and detector performance.  We focus in this section on the
prospects to measure precisely the top couplings to EW bosons and to
gluons, and to constrain possible deviations from the SM expectations.

The anomalous chromomagnetic and chromoelectric dipole moments $d_V$
and $d_A$ in
\beq
 \mathcal{L}= \mathcal{L}_\mathrm{QCD} + \frac{g_s}{m_t} \, \bar{t} \sigma^{\mu\nu} ( d_V + \mathrm{i} \, d_A \gamma_5 ) \frac{\lambda_a}{2} \, t \, G^a_{\mu\nu}
\eeq
modify the couplings of top quarks to gluons and hence they affect any
observable involving final state tops.  Since top quark pairs are
copiously produced in hadronic $pp$ collisions, and since the
production and decay dynamics of this process are very well
understood, $pp \to t\bar{t}$ is ideally suited to an investigation of
the top-gluon interactions.  In particular, the chromodipole moments
are expected to have an important impact on the high energy behavior
of this process.  Numerous studies have investigated these effects in
the LHC environment and a large number of sensitive observables have
been
described~\cite{Martinez:2001qs,Atwood:1994vm,Martinez:2007qf,Hioki:2009hm,Hioki:2010zu,Hioki:2013hva,Kamenik:2011dk,Bernreuther:2013aga,Franzosi:2015osa}.
High energy production rates will be even more accessible at the 100
TeV FCC.  A cross section analysis suggests that using
$m_{t\bar{t}}\gtrsim 10$ TeV at the FCC offers the best balance
between the sensitivity of the high energy behavior and the statistics
in this regime~\cite{Aguilar-Saavedra:2014iga}.  This leads to an
improvement of the chromodipole moment constraints by an order of
magnitude, as compared with a similar analysis for the high energy LHC
run, see Fig.~\ref{fig:chromo}.
\begin{figure}[t]
\centering
\includegraphics[scale=0.3]{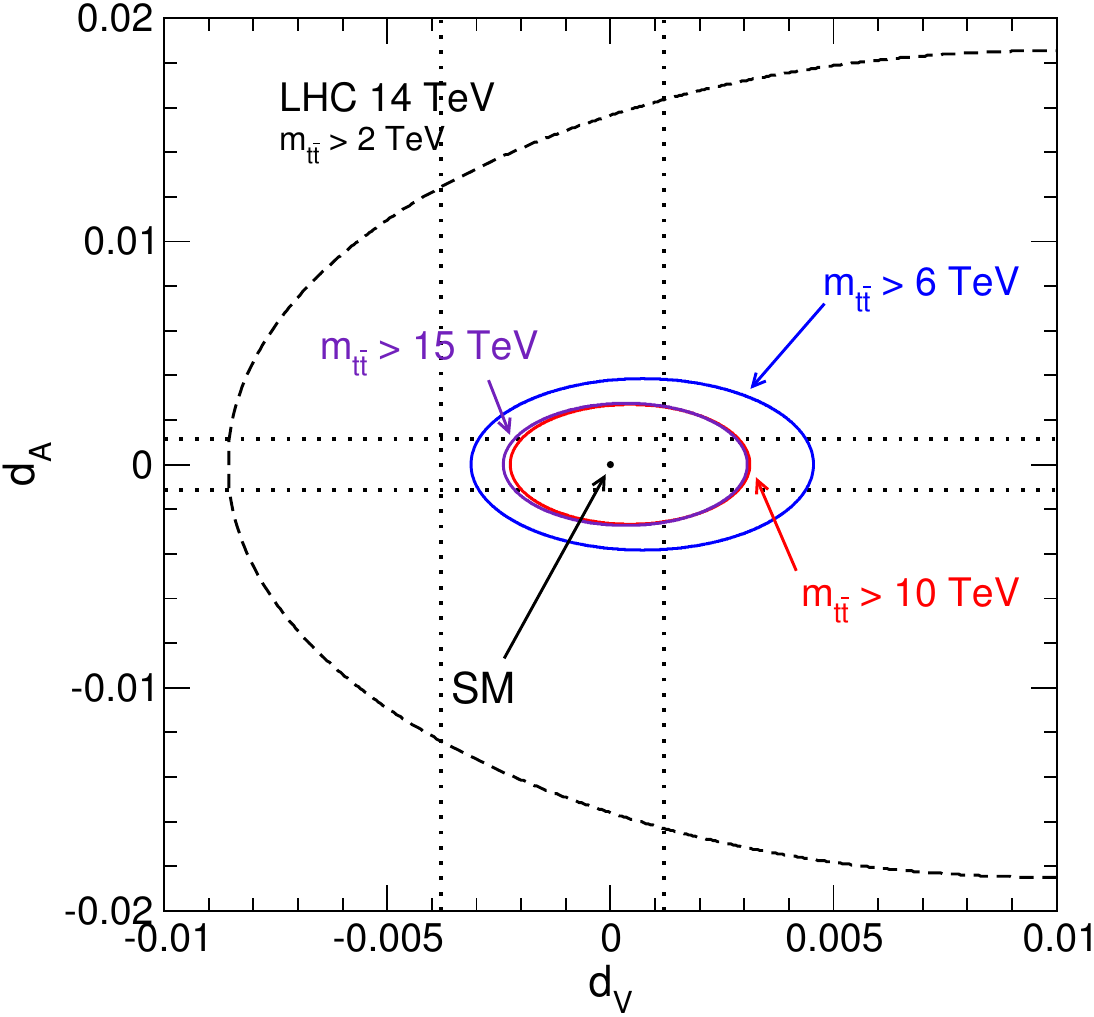}
\includegraphics[scale=0.3]{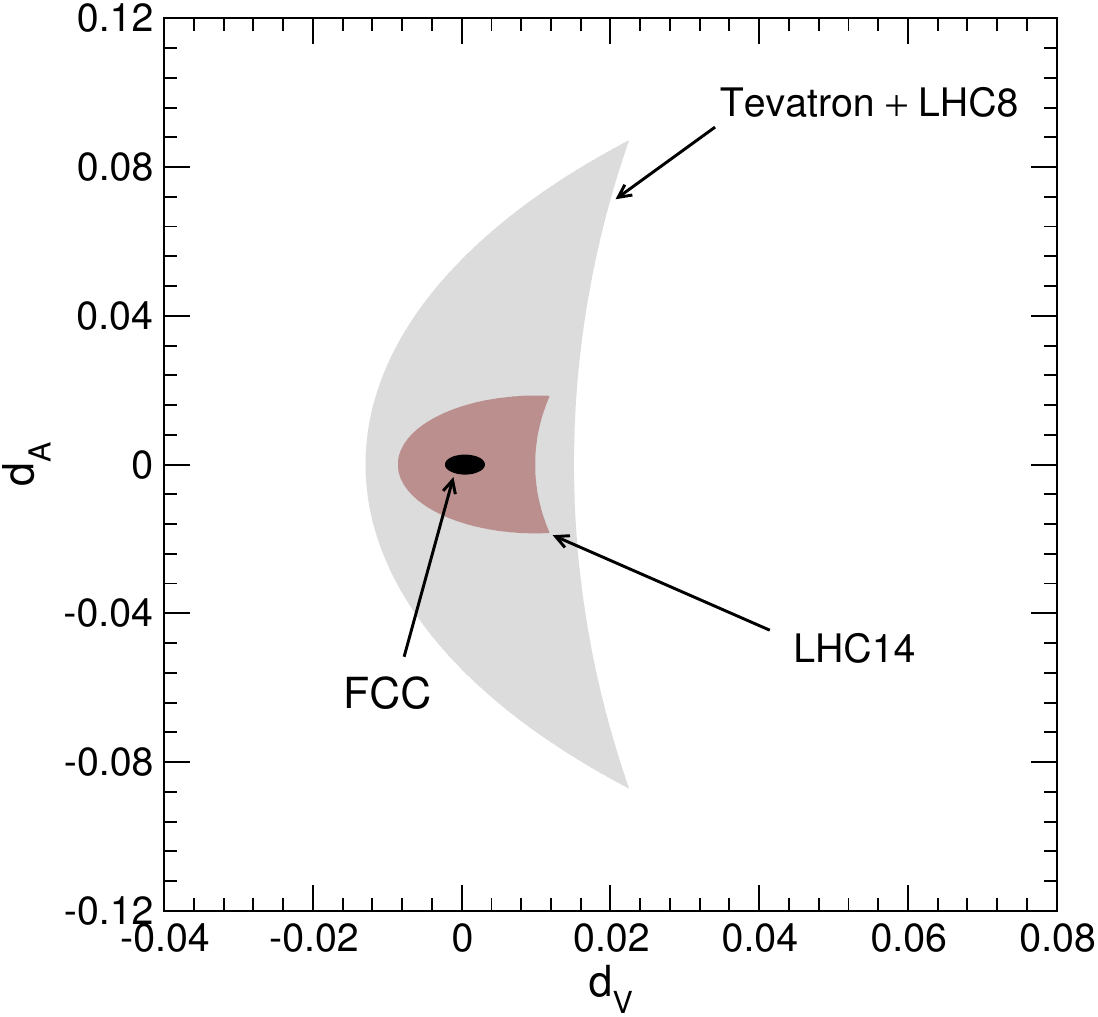}
\caption{(Left) Sensitivity of the $\sqrt{s}=14$ TeV LHC, and the $\sqrt{s}=100$ TeV FCC to the chromomagnetic and chromoelectric dipole moments 
$d_V$ and $d_A$ from  $t\bar{t}$ production. Three different definitions for the boosted regime at the FCC are shown. 
(Right) A comparison of constraints on $d_V$ and $d_A$ from past, present, and future hadron colliders. For more details, see Ref.~\cite{Aguilar-Saavedra:2014iga} \label{fig:chromo}}
\end{figure}
\\

The abundant production of top quark pairs at the FCC will also
improve the limits on top rare decays, for example those mediated by
top flavour-changing neutral couplings to the gauge bosons,
\begin{eqnarray}
 \mathcal{L} & = &  \frac{g}{2 c_W} \, \bar{q} \left[ \gamma^\mu (X_{qt}^L P_L + X_{qt}^R P_R ) 
 + \frac{\mathrm{i} \sigma^{\mu \nu} q_{\nu}}{M_Z} ( \kappa_{qt}^L P_L + \kappa_{qt}^R P_R)
  \right] t Z_\mu \notag \\
& &  + e \, \bar{q} \frac{\mathrm{i} \sigma^{\mu \nu} q_{\nu}}{m_t} ( \lambda_{qt}^L P_L + \lambda_{qt}^R P_R) t A_\mu + \mathrm{h.c.} \,,
\end{eqnarray}
with $q=u,c$. There are not yet dedicated studies of the FCC sensitivity
to such processes. Performing a naive rescaling of the LHC
expectations for $\sqrt s = 14$ TeV and 100
fb$^{-1}$~\cite{Han:1995pk,AguilarSaavedra:2004wm} and assuming a
luminosity of 10 ab$^{-1}$ for the FCC, one would expect an
improvement of almost two orders of magnitude, reaching a sensitivity
of $\text{Br}(t \to qZ,q\gamma) \simeq 10^{-7}$. However, at such a
level of precision the systematic uncertainties in the background
predictions will likely be dominant, and a more reliable estimation of
the sensitivity requires a detailed analysis.

Let us now turn to the discussion of final states with top quarks in
association with electroweak bosons.  These processes yield direct
sensitivity to the top quark electroweak couplings and are copiously
produced in 100~TeV collisions.  We postpone to the Higgs volume of
this Report the more detailed discussion of top production with a
Higgs boson and the determination of the top Yukawa coupling.  Studies
of the couplings of the top quark to the electroweak gauge bosons are
complementary to studies of the top-Higgs interactions.  The couplings
of the neutral gauge bosons $Z$ and $\gamma$ to the top quark are
fixed by the SM quantum numbers and gauge symmetries.  Weak and
electromagnetic dipole moments of the top quark arise effectively
through loop corrections but are very
small~\cite{Bernabeu:1995gs,Czarnecki:1996rx,Hollik:1998vz} in the SM.
Possible anomalous contributions from physics beyond the SM can modify
any of these couplings and are best studied in associated production
with a top pair or single top.  The sensitivity of $t\bar{t}Z$ and
$t\bar{t} \gamma$ at the LHC to the top-electroweak couplings was
first explored at LO in Ref.~\cite{Baur:2004uw,Baur:2005wi}, and more
recently $t\bar{t}Z$ studies at NLO QCD have been presented in
Refs.~\cite{Rontsch:2014cca,Rontsch:2015una,Bylund:2016phk}, and for
$t\bar{t}\gamma$ with photon from the production process in
Ref.~\cite{Bylund:2016phk}.  The transverse momentum of the vector
boson, and, in the case of $t\bar{t}Z$ production, the azimuthal angle
between the leptons arising from the decay of the $Z$ boson, are
particularly sensitive to the top-electroweak couplings.  These
couplings may also be probed through the charge asymmetry in
$t\bar{t}\gamma$ production, which appears at LO due to the $q\bar{q}$
initial state~\cite{Aguilar-Saavedra:2014vta}.  Similar to $t\bar{t}H$
production, the cross section for $t\bar{t}Z$ production increases by
a factor of about 50 at the FCC as compared to the $\sqrt{s}=13$ TeV
LHC.  Using the coupling parametrization \beq \mathcal{L}_{t\bar{t}Z}
= e \bar{\psi}_t\biggl[ \gamma^{\mu} \bigl(C_{1,V} + \gamma_5 C_{1,A}
\bigr) + \frac{\mathrm{i} \sigma^{\mu \nu} q_{\nu}}{M_Z} \bigl(C_{2,V}
+ \mathrm{i} \gamma_5 C_{2,A} \bigr) \biggr] \psi_t Z_{\mu}, \eeq
possible constraints on the couplings $C_{1/2,V/A}$ at the
$\sqrt{s}=13$ TeV LHC with $3~\mathrm{ab}^{-1}$ of data has been
presented in Refs.~\cite{Rontsch:2014cca,Rontsch:2015una} and are
shown in Fig.~\ref{fig:ttZcoupl} and Table~\ref{tab:constraints}
together with constraints achievable at the 100 TeV FCC with
$10~\mathrm{ab}^{-1}$.  These analyses take account of the theoretical
uncertainty, currently at 15\% but projected to decrease to 5\% by the
time the FCC is operational.  Driven by the larger statistics and
reduction of the theoretical uncertainties, the sensitivity of the FCC
to the top-$Z$ couplings is anticipated to exceed that of the LHC by
factors of 3-10.  Moreover, the construction of cross section ratios
to cancel various uncertainties has been proposed in
Ref.~\cite{Schulze:2016qas} and can further boost sensitivity by
factors of 2-4.
\begin{figure}[t]
\centering
\includegraphics[scale=0.4]{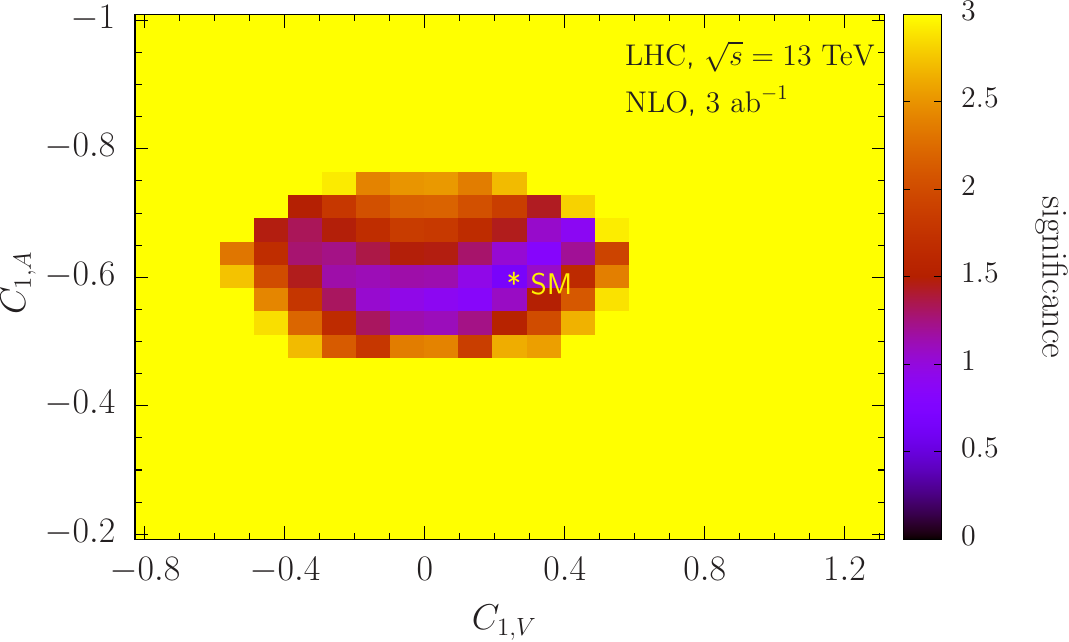}\hspace{10mm}
\includegraphics[scale=0.4]{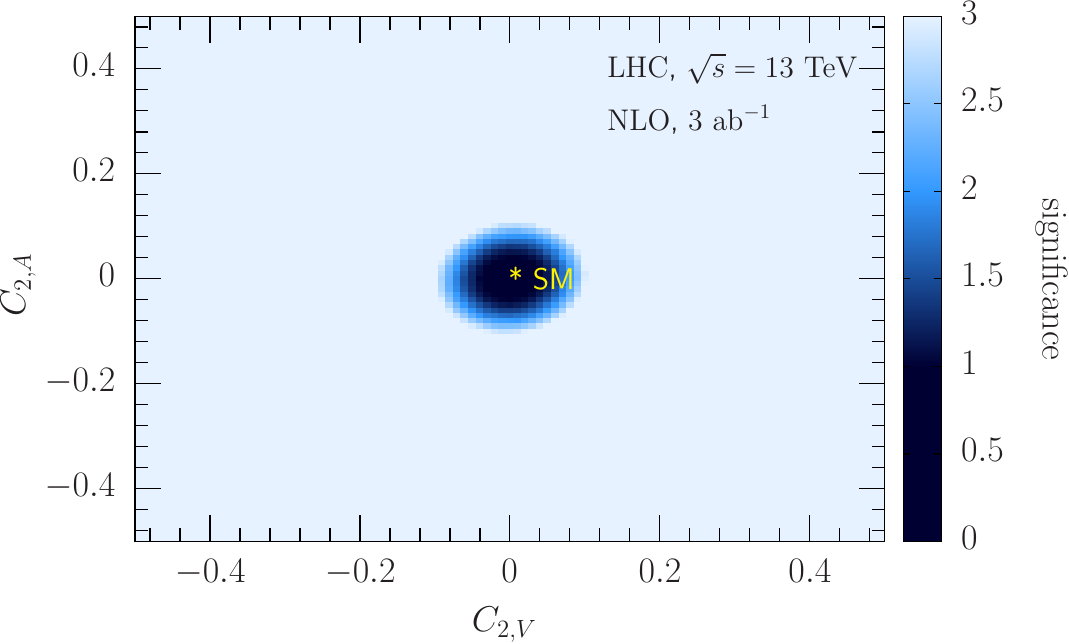}\\[5mm]
\includegraphics[scale=0.4]{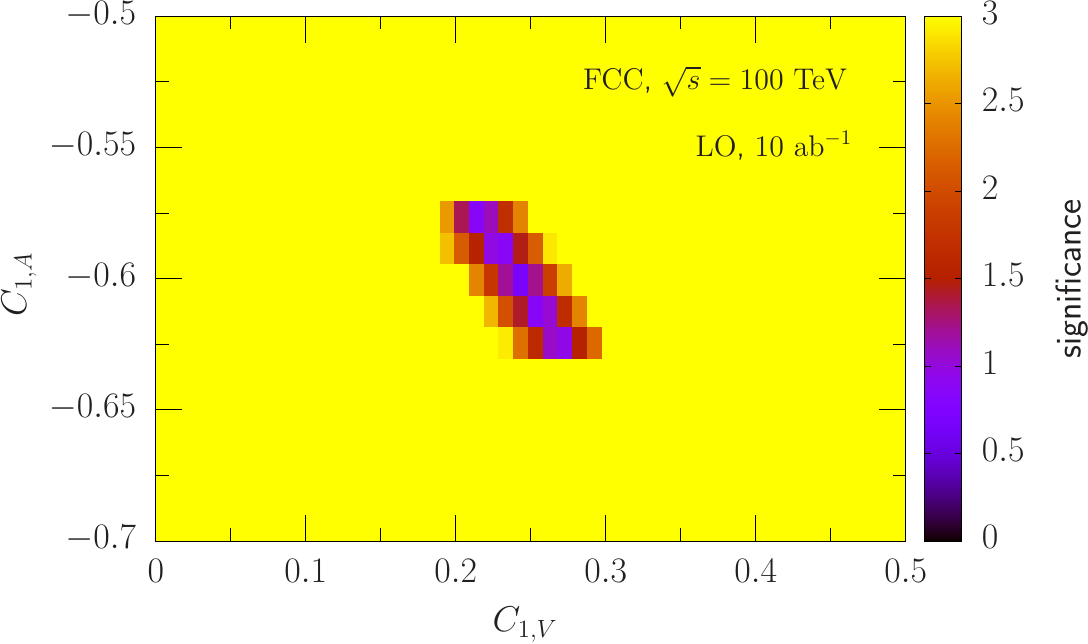}\hspace{8mm}
\includegraphics[scale=0.4]{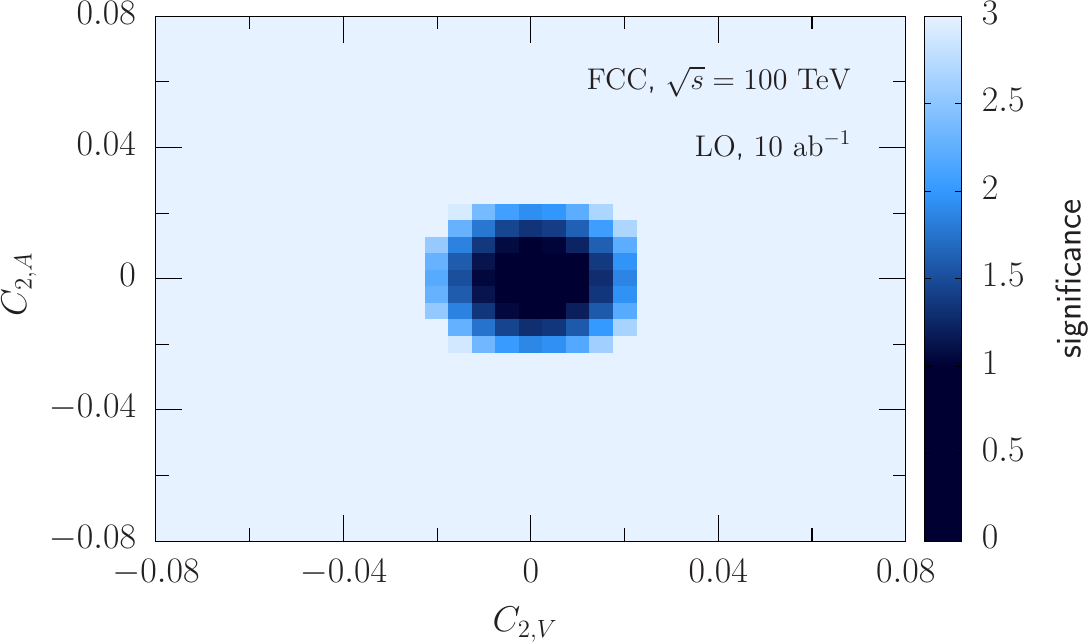}
\caption{Comparison of potential constraints on couplings $C_{1/2,V/A}$ achievable at the LHC and FCC. For further details, see Refs.~\cite{Rontsch:2014cca,Rontsch:2015una}. \label{fig:ttZcoupl} }
\end{figure}
\\

The process $t\bar{t}+W$ is peculiar in this context as it does not
yield an enhanced sensitivity to the $Wtb$ coupling.  The reason is
the simple fact that the $W$ boson can only be radiated off the $q
\bar{q}$ initial state.  This also prohibits a $gg$-initiated process
and, therefore, the production cross section is small with 587~fb at
the 13~TeV LHC and 19~pb at the FCC~\cite{Maltoni:2014zpa}, before
branching of the top quarks and the $W$ boson.  Nevertheless, the
authors of Ref.~\cite{Maltoni:2014zpa} pointed out that these
particular features allow for the study of a charge asymmetry as the
top quarks largely inherit the polarization of the initial state.  At
a 100 TeV collider, a SM asymmetry of about +2\% is expected and can
be used to discriminate against new physics scenarios of axigluons
\cite{Ferrario:2009bz,Frampton:2009rk} which induce asymmetries of
$\mathcal{O}(10\%)$ for axigluon masses in the few TeV range
\cite{Maltoni:2014zpa}.  It was shown in
Ref.~\cite{Aguilar-Saavedra:2014vta} that similar axigluon models can
also be probed through asymmetries in $t\bar{t}+\gamma$
production. (See Ref.~\cite{Aguilar-Saavedra:2014kpa} for a review.)

\renewcommand{\arraystretch}{1.5}  
\begin{table}[h]
\centering
\begin{tabular}{ l | c | c | c | c |}
                         &   $C_{1,V}$   &   $C_{1,A}$   &   $C_{2,V}$    &   $C_{2,A}$  \\ \hline
     SM value            &   $0.24$        &     $-0.60$     &      $ < 0.001$     &    $\ll 0.001$    \\ \hline   
13\,TeV, 3\,ab$^{-1}$    & $[-0.4,+0.5]$ & $[-0.5,-0.7]$ & $[-0.08,+0.08]$ & $[-0.08,+0.08]$ \\ 
100\,TeV, 10\,ab$^{-1}$  & $[+0.2,+0.28]$ & $[-0.63,-0.57]$ & $[-0.02,+0.02]$ & $[-0.02,+0.02]$  \\ 
\end{tabular}
\caption{ \label{tab:constraints}
Possible constraints on anomalous vector and axial couplings ($C_{1,V/A}$) and weak dipole moment couplings ($C_{2,V/A}$)
in $pp \to t\bar{t}+Z$ production at the LHC and FCC. The bounds correspond to the 95\,\% C.L. exclusion for one coupling when all others are marginalized over.
For further details, see Ref.~\cite{Rontsch:2015una}. }
\end{table}
\renewcommand{\arraystretch}{1.0}  

As yet, no studies of the sensitivity of the single top + $Z/\gamma$
processes to the flavor-conserving top couplings exist, despite the
fact that associated production with a single top is known to have a
comparable rate to production with a top
pair~\cite{Campbell:2013yla}. Single top production plus a $Z$ boson
or a photon can also be mediated by top flavour-changing neutral
couplings~\cite{delAguila:1999kfp}, in the processes $gq \to Zt/
\gamma t$, with $q=u,c$.  At the LHC, the potential of these processes
to probe $u-t$ couplings is similar to $t\bar t$ production followed
by a flavour-changing decay $t \to
uZ/u\gamma$~\cite{AguilarSaavedra:2004wm} but the sensitivity to $c-t$
couplings is much worse, due to the lower parton luminosity for charm
quarks. At the FCC, the $gu\to Zt/\gamma t$ cross sections increase by
a factor of $15$ with respect to the LHC, and by a factor of 50 (40)
for $gc \to Zt (\gamma t)$. (We assume here that $Ztu$ and $Ztc$
couplings have tensor structure.) The larger enhancement for
charm-initiated processes leads to a comparable sensitivity to $u-t$
and $c-t$ couplings. But, more interestingly, the production cross
section for highly-energetic $Zt / \gamma t$ pairs does not decrease
as fast as for the SM backgrounds, due to the momentum dependence of
the $\sigma^{\mu \nu}$ vertex, as it is shown in
fig.~\ref{fig:tZ}. (The differential distributions for SM backgrounds
are expected to be similar to the ones for $gu \to Zt$ mediated by
$\gamma^\mu$ couplings, shown in fig.~\ref{fig:tZ}.) With the large
cross sections and luminosities expected for the FCC, it will be
possible to explore the highly-boosted $Zt /\gamma t$ regime, where SM
backgrounds are small. It is then expected that the sensitivity to top
flavour-changing neutral couplings will be excellent, though a
quantitative statement and a comparison with $t \bar t$ decays
requires a detailed analysis.

\begin{figure}[t]
\begin{center}
\begin{tabular}{cc}
\includegraphics[height=5cm]{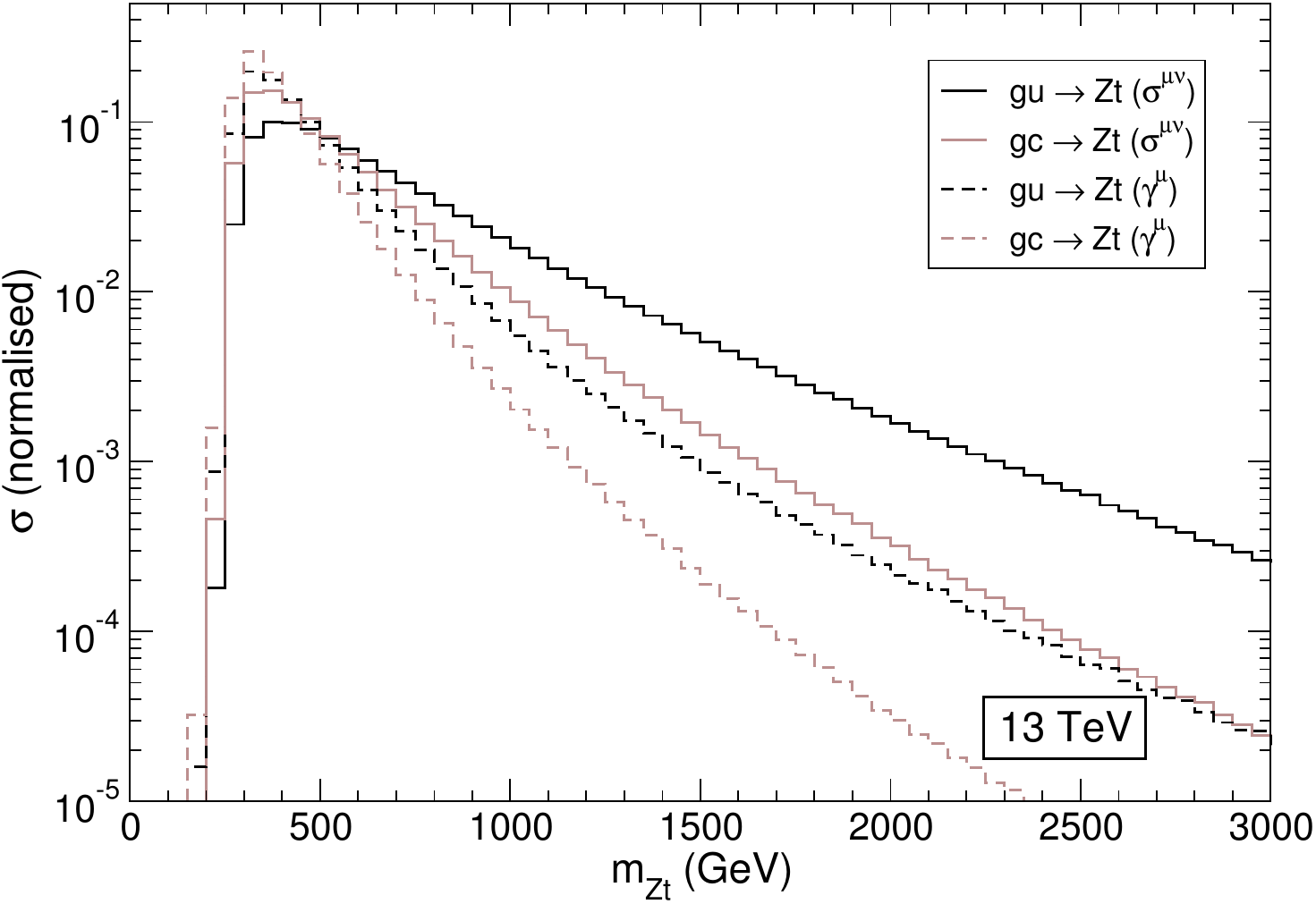} &
\includegraphics[height=5cm]{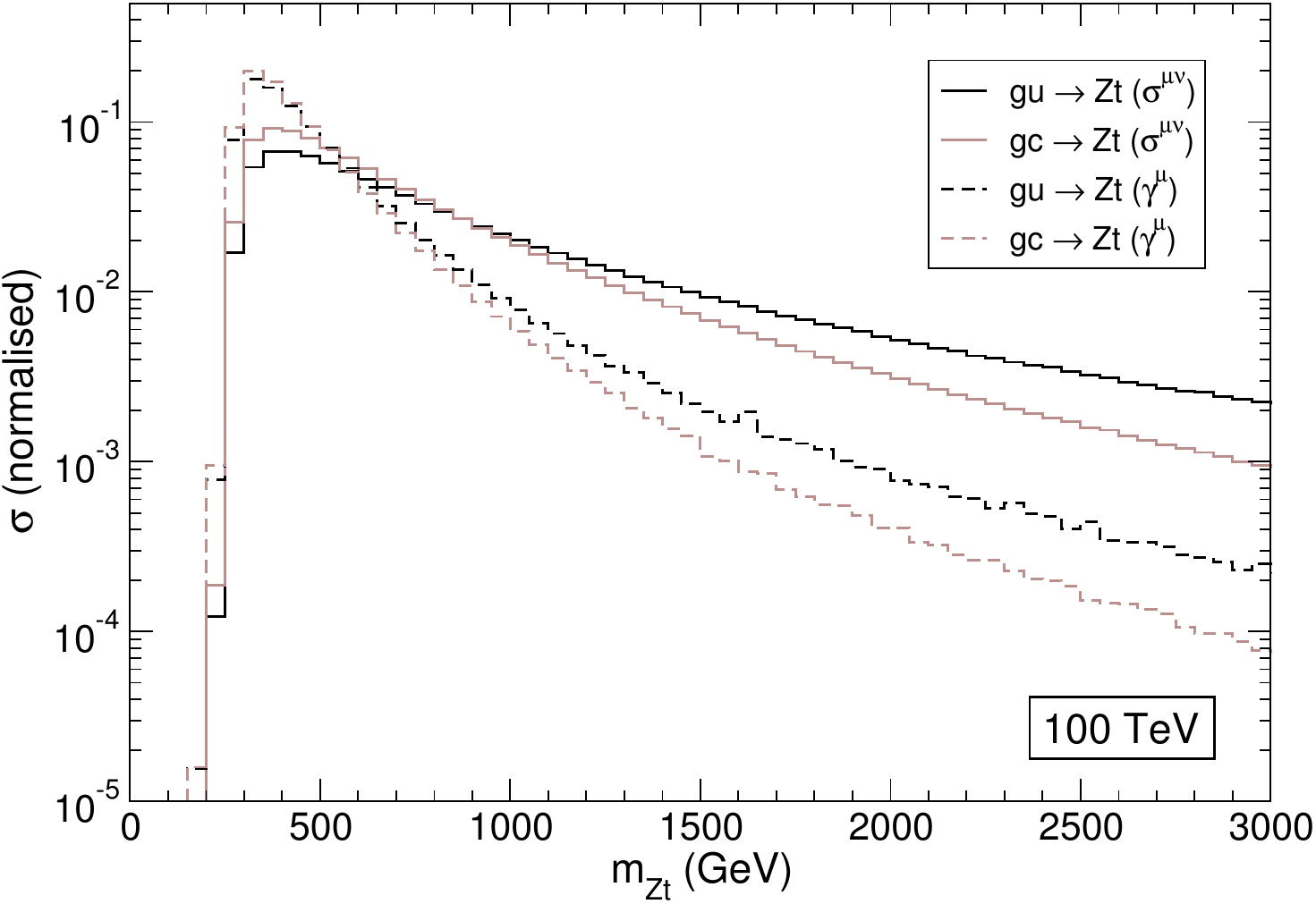} \\
\includegraphics[height=5cm]{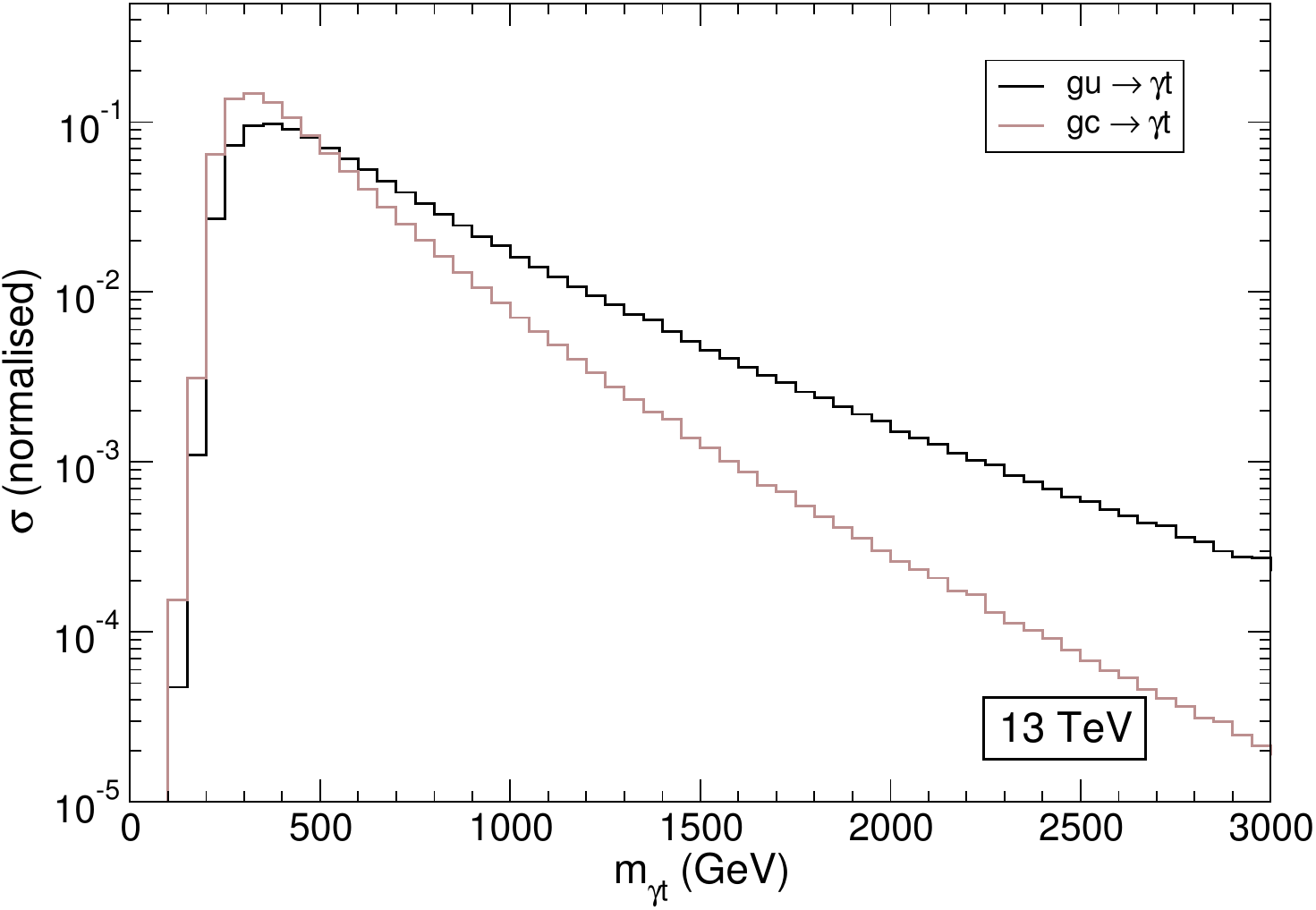} &
\includegraphics[height=5cm]{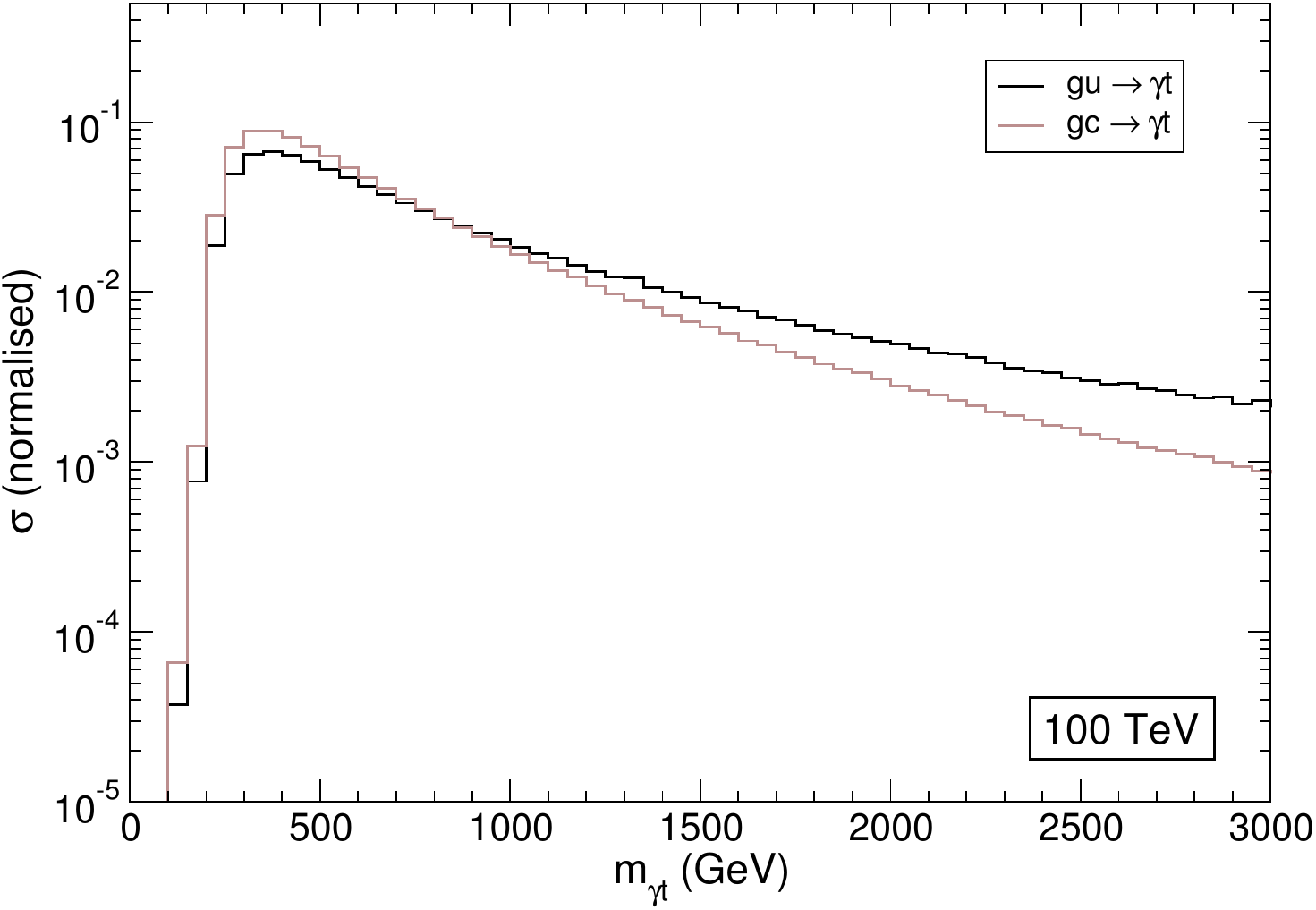}
\end{tabular}
\end{center}
\caption{Normalised invariant mass distributions for $Zt$ and $\gamma
  t$ production mediated by top flavour-changing couplings to the $Z$
  boson, at the LHC (left) and FCC (right). The pseudo-rapidities of
  the top quark and the $Z/\gamma$ boson are required to be in the
  range $|\eta| \leq 2.5$.} \label{fig:tZ}
\end{figure}

\clearpage
\section{Production of multiple heavy objects\footnote{Editor: P.~Torrielli}}
\label{sec:multi}
Standard Model processes featuring many heavy particles in the final
state are challenging at colliders. On one side, the presence of many
particles is indicative of the dynamic complexity of these processes,
that entail several powers of the strong and/or of the electroweak
coupling constant; on the other hand, the production of such heavy
states requires considerable energy owing to the high mass
thresholds. These effects are responsible for their small rates,
which, together with the experiential difficulty in reconstructing
such complicated topologies, makes their measurement a formidable
task.

Nevertheless, from this very complexity stem the main reasons of
interest in these processes. Their dynamic and kinematic structure is
so rich that the measurement of one of them may probe several
properties of the underlying theory at the same time; these reactions
are typically sensitive to couplings of different nature which make
them ideal tools for understanding in detail the interplay among
different particle sectors. Moreover, their complex kinematics may
lend them unique features, which allow cleaner signal extraction
through the definition of elaborated event-selection strategies.
Finally, they very often appear as important backgrounds to many BSM
signals, for example those featuring heavy intermediate new-physics
states with long decay chains to SM light particles.

While some of these reactions are out of reach at present colliders, a
substantial increase in centre-of-mass energy and in luminosity may
render them accessible at future accelerators, with a consequent step
up in the level of detail to which fundamental interactions can be
probed. A future 100 TeV hadronic collider may thus unleash the
potential of some of these channels to measure SM parameters with
unprecedented accuracy, to possibly discover new physics through rare
production mechanisms, and to constrain BSM parameter spaces in new,
more and more elaborated manners.

In the following, some of the processes that today are considered as
`rare' are presented, categorised according to their matter content,
together with some physics opportunities they may give once their
yield will be statistically significant at a 100-TeV collider. The
rates shown in the tables and figures of this section are at the NLO
in QCD, and have been obtained in \cite{Torrielli:2014rqa} with the
automatic code {\tt MadGraph5\_aMC@NLO} \cite{Alwall:2014hca}. The
setup employed is summarised below.
\begin{itemize}
\item Non-zero particle masses are $m_t=173$ GeV, $m_H=125$ GeV,
  $m_Z=91.188$ GeV, $m_W=80.419$ GeV. The bottom-quark mass is set to
  $m_b=4.7$ GeV in the four-flavour-scheme (4FS) simulations, and to
  $m_b=0$ in the five-flavour-scheme (5FS) ones. The CKM matrix is
  $V_{\tiny{\mbox{CKM}}}=1$, and the fine-structure constant is
  $\alpha=1/132.507$.
\item Renormalisation and factorisation scales are chosen as
  $\mu_R=\mu_F=\frac12\sum_k m_T^{(k)}$, $m_T^{(k)}$ being the
  transverse mass of the $k$-th final-state particle. Independent
  variation of $\mu_R$ and $\mu_F$ in the range $[1/2,2]$ is obtained
  in an exact way without rerunning the code, through the reweighting
  technique described in \cite{Frederix:2011ss}. The uncertainty
  associated with this variation is shown as a dark band in the plots
  of the section.
\item As PDFs, the MSTW 2008 NLO (68\%~c.l.) sets \cite{Martin:2009iq}
  are used, relevant to four or five active flavours, depending on the
  flavour scheme employed in the simulation. PDF uncertainties are
  estimated according to the asymmetric-hessian prescription provided
  by the PDF set, and obtained automatically as in explained in
  \cite{Frederix:2011ss}. They are shown as a light band in the plots
  of the section. The value and the running of the strong coupling
  constant $\alpha_{\tiny{\mbox{S}}}$ are as well set according to the
  PDF set.
\item Whenever relevant, photons are isolated by means of the Frixione
  smooth-cone criterion \cite{Frixione:1998jh}, with parameters
  $R_0=0.4$, $p_T(\gamma)>20$ GeV, $\epsilon_\gamma=n=1$.
\end{itemize}

\subsection{Production of multiple gauge bosons}
Production processes featuring many gauge bosons in the final state
are important for diverse reasons. On one hand they are backgrounds in
many searches for BSM signals, characterised by multi-lepton
signatures, with or without missing transverse energy (like for
example SUSY \cite{Baer:1995va} and extra dimensions
\cite{Cheng:2002ab}), or in searches for SM signals like $VH$, see for
example \cite{Hoeche:2014rya}. On the other hand, and even more
importantly, viewed themselves as signals they provide key tests of
the SM, in that they are particularly sensitive to the gauge structure
of its interactions.

In the SM, the couplings for triple and quadruple gauge-boson vertices
are fixed as a consequence of its non-abelian gauge symmetry. Possible
new physics in the gauge sector can be parametrised in a
model-independent way through a set of higher-dimension operators
involving gauge vectors, see for example
\cite{Belanger:1992qh,Eboli:2003nq,Eboli:2006wa}
\begin{eqnarray}
\mathcal L=\mathcal L_{\mbox{\tiny{SM}}}+\sum_i \frac{c_i}{\Lambda^2}\mathcal O_{3V,i}+\sum_j \frac{f_j}{\Lambda^4}\mathcal O_{4V,j}+\cdots,
\end{eqnarray}
giving rise to anomalous triple gauge couplings, (a)TGC's, anomalous
quartic gauge couplings, (a)QGC's, and so on. The presence of
anomalous couplings results in modified rates and spectra for
multi-boson production processes, which are thus an ideal ground to
set constraints on the gauge interactions of BSM models and on the
scale $\Lambda$ of possible new physics.

\begin{table}[h!]
\begin{center}
\begin{small}
\begin{tabular}{rl | l | l | c}
\hline\hline
 &Process& $\sigma_{\tiny{\mbox{NLO}}}$(8 TeV) [fb]& $\sigma_{\tiny{\mbox{NLO}}}$(100 TeV) [fb]\vspace{1mm}&$\rho$\\\hline
$pp~~\to$ & $W^+W^-W^\pm~(\mbox{4FS})$ & $8.73\cdot 10^1~{}^{+6\%}_{-4\%}~{}^{+2\%}_{-2\%}$ & $4.25\cdot 10^3~{}^{+9\%}_{-9\%}~{}^{+1\%}_{-1\%}$ & 49\vspace{1mm}\\\hline\vspace{1mm}
$pp~~\to$ & $W^+W^-Z~(\mbox{4FS})$ & $6.41\cdot 10^1~{}^{+7\%}_{-5\%}~{}^{+2\%}_{-2\%}$ & $4.01\cdot 10^3~{}^{+9\%}_{-9\%}~{}^{+1\%}_{-1\%}$ & 63\vspace{1mm}\\\hline\vspace{1mm}
$pp~~\to$ & $W^\pm ZZ$ & $2.16\cdot 10^1~{}^{+7\%}_{-6\%}~{}^{+2\%}_{-2\%}$ & $1.36\cdot 10^3~{}^{+10\%}_{-10\%}~{}^{+1\%}_{-1\%}$ & 63\vspace{1mm}\\\hline\vspace{1mm}
$pp~~\to$ & $ZZZ$ & $5.97\cdot 10^0~{}^{+3\%}_{-3\%}~{}^{+2\%}_{-2\%}$ & $2.55\cdot 10^2~{}^{+5\%}_{-7\%}~{}^{+2\%}_{-1\%}$ & 43\vspace{1mm}\\\hline\hline\vspace{1mm}
$pp~~\to$ & $W^+W^-W^\pm Z~(\mbox{4FS})$ & $3.48\cdot 10^{-1}~{}^{+8\%}_{-7\%}~{}^{+2\%}_{-2\%}$ & $5.95\cdot 10^1~{}^{+7\%}_{-7\%}~{}^{+1\%}_{-1\%}$ & 171\vspace{1mm}\\\hline\vspace{1mm}
$pp~~\to$ & $W^+W^-W^+W^-~(\mbox{4FS})$ & $3.01\cdot 10^{-1}~{}^{+7\%}_{-6\%}~{}^{+2\%}_{-2\%}$ & $4.11\cdot 10^1~{}^{+7\%}_{-6\%}~{}^{+1\%}_{-1\%}$ & 137\vspace{1mm}\\\hline\vspace{1mm}
$pp~~\to$ & $W^+W^-ZZ~(\mbox{4FS})$ & $2.01\cdot 10^{-1}~{}^{+7\%}_{-6\%}~{}^{+2\%}_{-2\%}$ & $3.34\cdot 10^1~{}^{+6\%}_{-6\%}~{}^{+1\%}_{-1\%}$ & 166\vspace{1mm}\\\hline\vspace{1mm}
$pp~~\to$ & $W^\pm ZZZ$ & $3.40\cdot 10^{-2}~{}^{+10\%}_{-8\%}~{}^{+2\%}_{-2\%}$ & $7.06\cdot 10^0~{}^{+8\%}_{-7\%}~{}^{+1\%}_{-1\%}$ & 208\vspace{1mm}\\\hline\vspace{1mm}
$pp~~\to$ & $ZZZZ$ & $8.72\cdot 10^{-3}~{}^{+4\%}_{-4\%}~{}^{+3\%}_{-2\%}$ & $8.05\cdot 10^{-1}~{}^{+4\%}_{-4\%}~{}^{+2\%}_{-1\%}$ & 92\vspace{1mm}\\\hline\hline\vspace{1mm}
$pp~~\to$ & $ZZZZZ$ & $1.07\cdot 10^{-5}~{}^{+5\%}_{-4\%}~{}^{+3\%}_{-2\%}$ & $2.04\cdot 10^{-3}~{}^{+3\%}_{-3\%}~{}^{+2\%}_{-1\%}$ & 191\vspace{1mm}\\\hline\hline
\end{tabular}
\end{small}
\end{center}
\caption{\label{tab:multiVB} Production of multiple vector bosons at NLO in QCD at 8 and 100 TeV from ref.~\cite{Torrielli:2014rqa}. The rightmost column reports the ratio $\rho$ of 100-TeV to 8-TeV cross sections. Theoretical uncertainties are due to scale and PDF variations, respectively.}
\end{table}

\begin{figure}[h!]
\begin{minipage}{0.49\textwidth}
\centering
\includegraphics[width=1\textwidth]{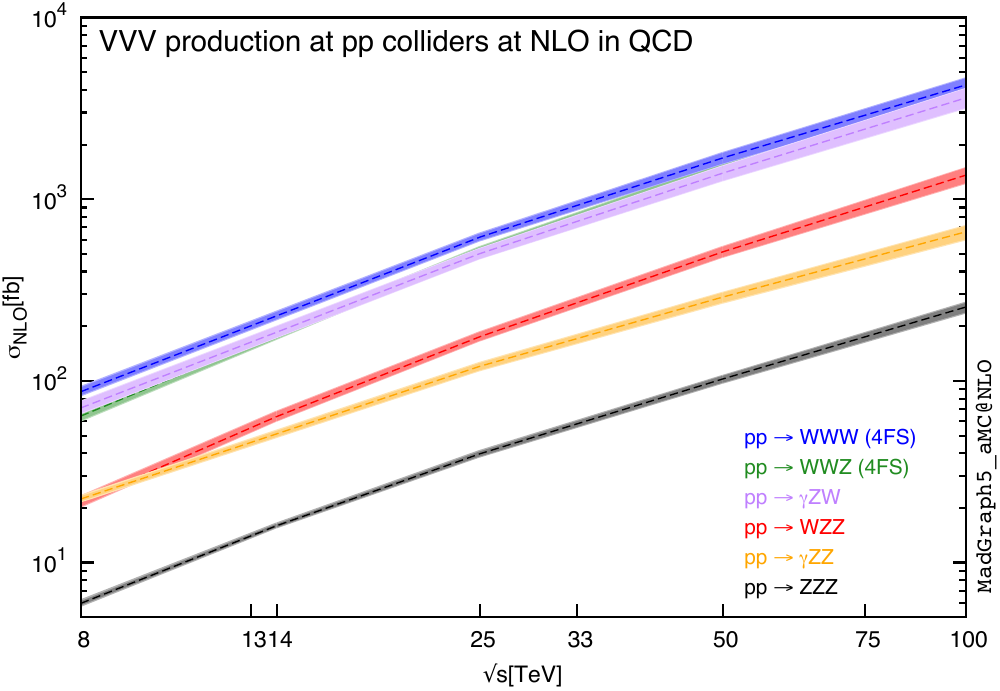}
\end{minipage}
\begin{minipage}{0.49\textwidth}
\centering
\includegraphics[width=1\textwidth]{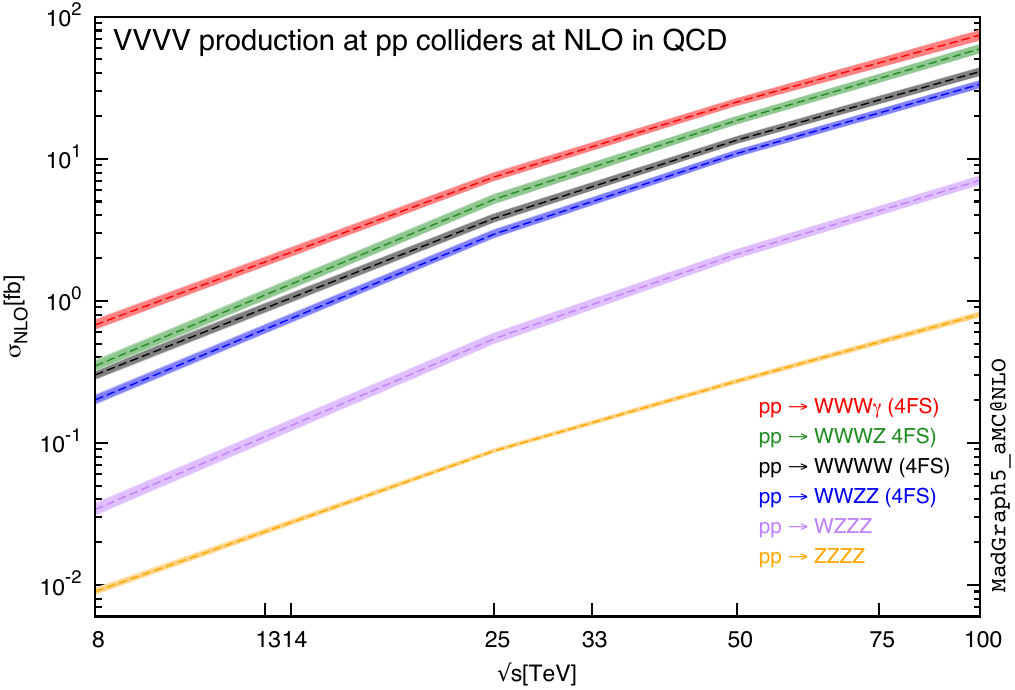}
\end{minipage}
\caption{\label{fig:V}NLO total cross section for production of three (left panel) and four (right panel) electroweak bosons, as a function of the hadronic-collider centre-of-mass energy.}
\end{figure}

In the first two sections of table \ref{tab:multiVB} and in figure
\ref{fig:V} \cite{Torrielli:2014rqa}, sample cross sections are
reported for the production of up to four undecayed electroweak vector
bosons. Three-boson final states are abundantly produced at 100 TeV,
and final states with four bosons are in principle observable even
upon including branching ratios for the best leptonic decays of each
boson.

The addition of a gauge boson brings production rates down typically
by a factor of the order of, or slightly larger than $\alpha$,
compatibly with the perturbative counting~\cite{Arkani-Hamed:2015vfh},
and with the fact that an extra massive particle in the final state
constrains the scattering to a region of larger Bjorken-$x$,
suppressing the cross section. The rate increase $\rho$ from 8 to 100
TeV ranges from few tens to few hundreds, with larger values for
larger multiplicities. It is relatively mild, owing to the fact that
all of these channels proceed through $q\bar q$
scattering. Theoretical uncertainties on the total cross sections,
stemming from renormalisation/factorisation-scale variations and from
PDFs, range between 5\% and 10\%.

Three-boson production is crucial to probe aQGC's. Although these
couplings involve complicated topologies, featuring more bosons in the
final state with respect to aTGC's, the information they carry is not
a mere replica of the one contained in the latter. In some cases
\cite{Ye:2013psa,Eboli:2003nq}, the exchange of heavy bosons can
contribute at tree level to four-boson couplings while giving only a
suppressed one-loop contribution to triple-boson vertices. In such
scenarios, only QCG's would significantly deviate from the SM
expectation, and could result mandatory to probe new
physics. Moreover, in case aTCG's are observed at a 100-TeV machine,
the measurement of aQCG's will acquire an even more relevant role, as
capable of providing complementary insight about the strength,
structure, and scale of new-physics forces.

A particularly interesting channel in this respect is $W^\pm W^\pm
W^\mp$, which has the largest cross section among the triple-boson
reactions, as displayed in table \ref{tab:multiVB} and figure
\ref{fig:V}.  At 100 TeV the sensitivity to the dimension-8 operator
$f_{T0}/\Lambda^4\mbox{Tr}[\hat W_{\mu\nu}\hat
W^{\mu\nu}]\mbox{Tr}[\hat W_{\alpha\beta}\hat W^{\alpha\beta}]$
increases by a factor of 300 with respect to LHC-8, and of 25 with
respect to LHC-14, assuming a common luminosity of 3000 fb$^{-1}$
\cite{Degrande:2013yda,Wen:2014mha}. The enhancement in sensitivity at
100 TeV is affected by the application of a unitarity-violation bound
\cite{Degrande:2013yda}, which indicates that this channel is
sensitive to the direct production of the heavy states integrated out
in the effective field theory. This is expected to hold generically
for three-boson production induced by dimension-8 operators, where the
growth of the rate with energy is more rapid than with dimension-6
operators.

Four-boson production can in principle constrain yet higher-order
(quintic, in this case) anomalous couplings, on top of carrying
further complementary information on aTCG's and aQCG's. Production
rates at 100 TeV range from few units to few tens of femtobarns. The
sensitivity of the various channels has to be carefully assessed after
inclusion of branching ratios for the bosons. In this respect,
reactions with one or more photons in the final state could be useful
if they have sufficient rate after selection cuts, as they are less
affected by BR's.

Five-boson final states, of which an example is reported in the third
section of table \ref{tab:multiVB}, will be inaccessible at 100 TeV
under the assumption of SM couplings, even with $\mathcal O$(10)
ab$^{-1}$ luminosity, as they feature sub-femtobarn cross section.

\subsection{Multi-top and top-vector-boson associated production}
Processes with many top pairs, and associated top-pair vector-boson
productions offer another remarkable set of tests of the structure of
SM interactions, and of the mechanism of electroweak-symmetry
breaking. The top quark plays a special role in this programme, as its
large mass and its quantum numbers allow it to couple significantly
with all of the bosons in the theory, hence to connect the
interactions of different sectors. The accurate measurement and
understanding of its properties is moreover believed to be an
important mean to indirectly probe possible BSM physics, in case new
states elude direct detection \cite{Aguilar-Saavedra:2014iga}, owing
to the closeness of its mass to the electroweak scale.

The cross sections for the production of two top-antitop pairs at 100
TeV is detailed in the first section of table
\ref{tab:multitopVB}. Its very sizable growth $\rho$ with the
collider energy is due to the fact that this reaction predominantly
proceeds through $gg$ scattering \cite{Maltoni:2015ena}, with a gluon
PDF growing much faster than the quark ones at small $x$. Theoretical
uncertainties are quite large, of the order of $\pm25\%$ at 100 TeV,
mainly due to the presence of four powers of
$\alpha_{\tiny{\mbox{S}}}$ at the LO. The PDF uncertainty is reduced
at 100 TeV, again due to the gluon PDF being probed at much smaller
$x$ than at 8 TeV. The study of this final state is interesting at
hadron colliders as a probe of the nature of EWSB, see
e.g.~\cite{Spira:1997ce}, and of many BSM models with modified
symmetry-breaking sectors \cite{Cacciapaglia:2011kz}.

The final state with three top-antitop pairs has a cross section of
the order of 1 fb at 100 TeV \cite{Deandrea:2014raa}, hence, taking
branching ratios into account, it cannot be seen directly with the
luminosities usually assumed. The absence of the observation of this
signal, which is also enhanced in many BSM scenarios with top
partners, see e.g.~\cite{Cacciapaglia:2010vn}, can be used to
constrain the parameter space of these models, as nowadays is done
with two top-antitop pairs at the LHC \cite{CMS:2013xma}.

The second part of table \ref{tab:multitopVB} reports cross sections
for $t\bar tV$ production, with $V=W^\pm,Z$. Viewed as signals, these
channels are interesting in their own right as excellent tests for the
SM, probing top couplings to the gauge sector, and thus giving direct
insights on the mechanism of symmetry breaking. On the other hand,
they are prominent backgrounds for many BSM signals, on top of playing
an important role in $t\bar tH$ searches \cite{Maltoni:2015ena} in
case of multi-lepton signatures.

The rates for these processes make them well visible at 100 TeV. A
comparison between these cross sections and those in table
\ref{tab:multiVB} for multiple gauge-boson production shows
\cite{Arkani-Hamed:2015vfh} that top-quark processes at 100 TeV will
provide the dominant source of multi-$W$ and thus multi-lepton final
states, since each top gives rise to a $W$ through its decay. This
will have important implications for the search of new-physics signals
characterised by the presence of many gauge bosons or leptons from the
decay of the new heavy particles.

The larger growth $\rho$ for the neutral channel $t\bar tZ$ with
respect to $t\bar tW^\pm$ is again driven by the fact that the former
proceeds through $gg$ as opposed to $q\bar q$ (see
\cite{Maltoni:2015ena} for details). The absence of the $gg$
contribution, although disadvantageous in terms of total number of
expected events, makes $t\bar tW^{\pm}$ particularly interesting as a
handle to constrain new physics through asymmetry and polarisation
effects \cite{Maltoni:2014zpa}: charge asymmetry between $t$ and $\bar
t$ is significantly enhanced in $t\bar tW^\pm$ with respect to
inclusive $t\bar t$ production, and the final-state products display
very asymmetric rapidity distributions, induced by the $W$ acting as a
polariser of the initial state. In this respect, a 100 TeV energy will
be highly beneficial, allowing to reach few-percent statistical
precision for these asymmetries (down to 3\% for a luminosity of 3000
fb$^{-1}$, compared to 14\% at the LHC-14 \cite{Maltoni:2014zpa}),
that could thus become precision measurements of the properties of QCD
and powerful discriminators of BSM models.

The $t\bar tZ$ channel is also interesting for various reasons. The
weak electric and magnetic dipole moments of $tZ$ interactions are an
excellent probe of new physics given their small SM values
\cite{Rontsch:2015una}. For this purpose, the large rate at 100 TeV
will improve the constraints on these moments by a factor of 3 to 10
compared to the LHC, at 3000 fb$^{-1}$. Moreover the $t\bar tZ$
channel can be exploited to measure the top Yukawa coupling $y_t$ down
to 1\% accuracy at 100 TeV, through the ratio $\sigma(t\bar
tH)/\sigma(t\bar tZ)$ \cite{Plehn:2015cta}.

The third part of table \ref{tab:multitopVB} details the rates for
$t\bar tVV$ production at NLO in QCD
\cite{Torrielli:2014rqa,Maltoni:2015ena,Frixione:2015zaa}. The rate
growth with collider energy follows the expected pattern, with the
neutral channels, $gg$-dominated, displaying larger $\rho$ with
respect to $ttW^\pm Z$. Theoretical uncertainties for these channels
(as well as for $t\bar t V$) are under better control with respect to
$t\bar tt\bar t$, due to the presence of only two powers of the strong
coupling at the LO. These processes, elusive at the LHC, will be
accessible at 100 TeV, having cross sections in the 10$^2$ to 10$^3$
fb range. Exploiting asymmetry and polarisation effects to probe new
physics is possible for this category as well \cite{Maltoni:2015ena},
but the potential of this kind of observables for a 100-TeV collider
still needs to be studied in detail.
\begin{table}[h!]
\begin{center}
\begin{small}
\begin{tabular}{rl | l | l | c}
\hline\hline
 &Process& $\sigma_{\tiny{\mbox{NLO}}}$(8 TeV) [fb]& $\sigma_{\tiny{\mbox{NLO}}}$(100 TeV) [fb]\vspace{1mm}&$\rho$\\\hline
$pp~~\to$ & $t\bar tt \bar t$ & $1.71\cdot 10^0~{}^{+25\%}_{-26\%}~{}^{+8\%}_{-8\%}$ & $4.93\cdot 10^3~{}^{+25\%}_{-21\%}~{}^{+2\%}_{-2\%}$ & 2883\vspace{1mm}\\\hline\hline\vspace{1mm}
$pp~~\to$ & $t\bar tZ$ & $1.99\cdot 10^2~{}^{+10\%}_{-12\%}~{}^{+3\%}_{-3\%}$ & $5.63\cdot 10^4~{}^{+9\%}_{-10\%}~{}^{+1\%}_{-1\%}$ & 282\vspace{1mm}\\\hline\vspace{1mm}
$pp~~\to$ & $t\bar tW^\pm$ & $2.05\cdot 10^2~{}^{+9\%}_{-10\%}~{}^{+2\%}_{-2\%}$ & $1.68\cdot 10^4~{}^{+18\%}_{-16\%}~{}^{+1\%}_{-1\%}$ & 82\vspace{1mm}\\\hline\hline\vspace{1mm}
$pp~~\to$ & $t\bar tW^+W^-~(\mbox{4FS})$ & $2.27\cdot 10^0~{}^{+11\%}_{-13\%}~{}^{+3\%}_{-3\%}$ & $1.10\cdot 10^3~{}^{+9\%}_{-9\%}~{}^{+1\%}_{-1\%}$ & 486\vspace{1mm}\\\hline\vspace{1mm}
$pp~~\to$ & $t\bar tW^\pm Z$ & $9.71\cdot 10^{-1}~{}^{+10\%}_{-11\%}~{}^{+3\%}_{-2\%}$ & $1.68\cdot 10^2
~{}^{+16\%}_{-13\%}~{}^{+1\%}_{-1\%}$ & 173\vspace{1mm}\\\hline\vspace{1mm}
$pp~~\to$ & $t\bar tZZ$ & $4.47\cdot 10^{-1}~{}^{+8\%}_{-10\%}~{}^{+3\%}_{-2\%}$ & $1.58\cdot 10^2~{}^{+15\%}_{-12\%}~{}^{+1\%}_{-1\%}$ & 353\vspace{1mm}\\\hline\hline
\end{tabular}
\end{small}
\end{center}
\caption{\label{tab:multitopVB}
  Production of two top-antitop pairs, and of a top-antitop pair in association with up to two electroweak vector bosons at 8 and 100 TeV \cite{Maltoni:2015ena,Torrielli:2014rqa}. The rightmost column reports the ratio $\rho$ of the 100-TeV to the 8-TeV cross sections. Theoretical uncertainties are due to scale and PDF variations, respectively. Production of $t\bar tt\bar t$ is with the setup of ref.~\cite{Maltoni:2015ena}.}
\end{table}

\begin{figure}[h!]
\centering
\includegraphics[width=1\textwidth]{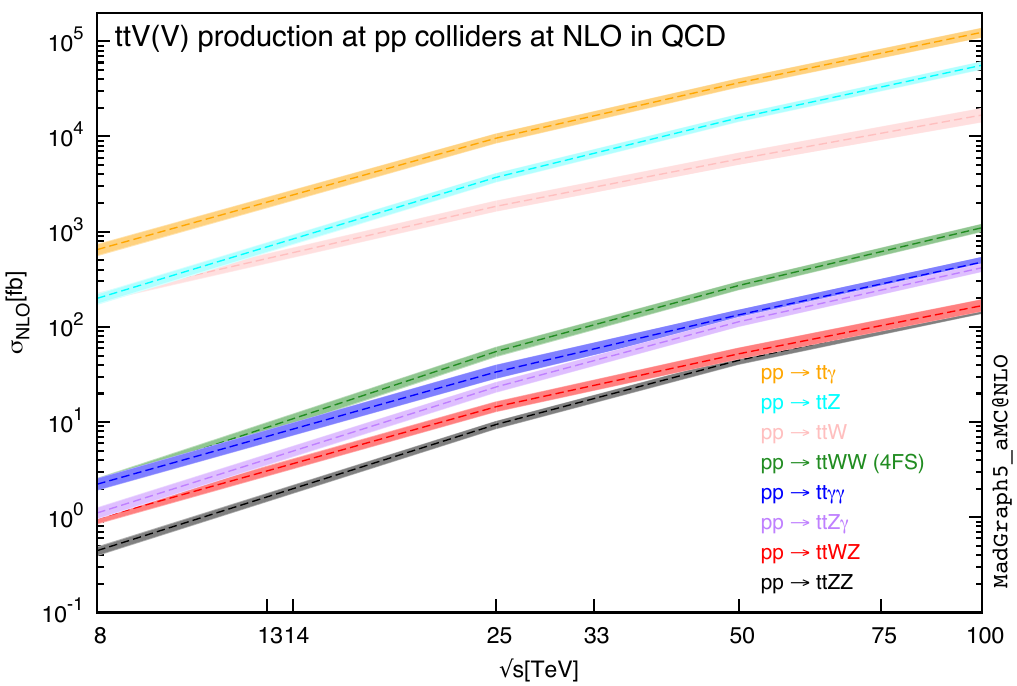}
\caption{\label{fig:t}NLO total cross section for production of a
  top-antitop pair in association with up to two electroweak bosons
  \cite{Torrielli:2014rqa}.}
\end{figure}

\subsection{Multi Higgs boson production by gluon fusion and VBF}
Processes featuring many Higgs bosons in the final state are of the
utmost importance at colliders, as they offer direct information about
Higgs self-interactions, which at present have not been observed at
the LHC. These processes offer a unique handle on the nature of the
Higgs potential, with crucial implications not only for SM and BSM
phenomenology, but also for more fundamental questions like the origin
of electroweak-symmetry breaking and the stability of the vacuum
\cite{Degrassi:2012ry}.

In the SM the Higgs potential is
\begin{eqnarray*}
V(H)=\frac12m_H^2H^2+\lambda_{3H}vH^3+\frac14\lambda_{4H}H^4,
\end{eqnarray*}
with triple and quadruple Higgs couplings equal to each other and
predicted in terms of the Higgs mass and VEV,
$\lambda_{3H}=\lambda_{4H}\equiv\lambda_{\tiny{\mbox{SM}}}=m_H^2/2v^2$;
measurement of multi-Higgs final states is thus the most direct way to
confirm or disprove this prediction, and for example to provide
information about the possible existence of a richer scalar sector,
featuring additional scalar fields.

The dominant production mechanisms of a Higgs pair in the SM are
displayed in table \ref{tab:HH} and in figure \ref{fig:HH}
\cite{Frederix:2014hta}, where the total rate at the NLO in QCD is
shown as a function of the hadron-collider energy. The dominant
channel is gluon fusion, as it is for single Higgs, followed by VBF,
with a cross section smaller by more than an order of magnitude.

\begin{figure}[h!]
\centering
\includegraphics[width=1\textwidth]{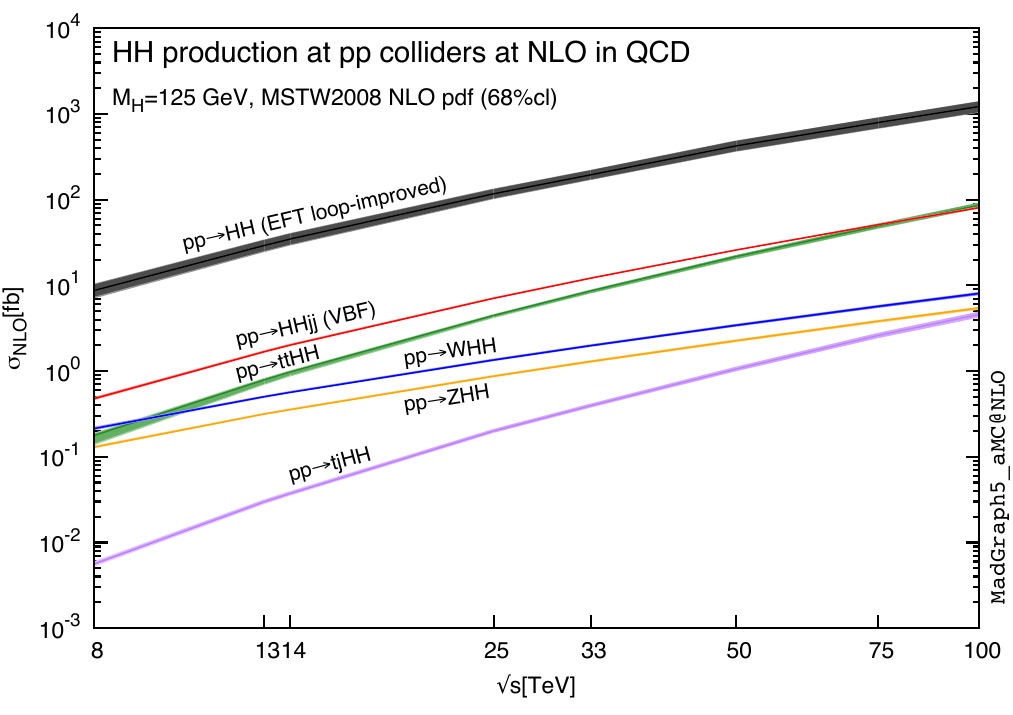}
\caption{\label{fig:HH}NLO total cross section for the dominant
  production channels of a Higgs pair \cite{Frederix:2014hta}.}
\end{figure}

\begin{table}[h!]
\begin{center}
\begin{small}
\begin{tabular}{rl | l c}
\hline\hline
 &Process&  $\sigma_{\tiny{\mbox{NLO}}}$(100 TeV) [fb]\\\hline
$pp~~\to$ & $HH$ & $1.23\cdot 10^3~{}^{+14\%}_{-14\%}~{}^{+1\%}_{-2\%}$\vspace{1mm}\\\hline\vspace{1mm}
$pp~~\to$ & $t\bar tHH$ & $8.62\cdot 10^1~{}^{+7\%}_{-7\%}~{}^{+1\%}_{-1\%}$\vspace{1mm}\\\hline\vspace{1mm}
$pp~~\to$ & $jjHH$ (VBF) & $8.09\cdot 10^1~{}^{+1\%}_{-1\%}~{}^{+2\%}_{-2\%}$\vspace{1mm}\\\hline\vspace{1mm}
$pp~~\to$ & $W^\pm HH$ & $8.09\cdot 10^0~{}^{+2\%}_{-3\%}~{}^{+2\%}_{-1\%}$\vspace{1mm}\\\hline\vspace{1mm}
$pp~~\to$ & $ZHH$ & $5.46\cdot 10^0~{}^{+2\%}_{-4\%}~{}^{+2\%}_{-1\%}$\vspace{1mm}\\\hline\vspace{1mm}
$pp~~\to$ & $tjHH$ & $4.58\cdot 10^0~{}^{+8\%}_{-8\%}~{}^{+0\%}_{-1\%}$\vspace{1mm}\\\hline\vspace{1mm}
\end{tabular}
\end{small}
\end{center}
\caption{\label{tab:HH}
NLO total cross section for the dominant production channels of a Higgs pair at 100 TeV \cite{Frederix:2014hta}.}
\end{table}

The cross section for gluon fusion is in excess of 1.5 pb at 100 TeV,
see for example
\cite{Maltoni:2014eza,deFlorian:2013jea,Grigo:2014jma}. This rate is
expected to provide a clear signal in the $HH\to (b\bar
b)(\gamma\gamma)$ channel and to allow determination of $\lambda_{3H}$
with an accuracy of $30-40\%$ with a luminosity of 3 ab$^{-1}$, and of
$5-10\%$ with a luminosity of 30 ab$^{-1}$
\cite{Barr:2014sga,Azatov:2015oxa,He:2015spf}. A rare decay channel
which is potentially interesting is $HH\to (b\bar b)(ZZ)\to(b\bar
b)(4l)$, with a few expected signal events against $\mathcal O(10)$
background events at 3 ab$^{-1}$ \cite{Papaefstathiou:2015iba}.

Given the similarity of single- and double-Higgs production
mechanisms, the cross-section ratio $\sigma(gg\to HH)/\sigma(gg\to H)$
has been advocated \cite{Goertz:2013kp} as a good observable to
constrain $\lambda_{3H}$ at the LHC, being more theoretically stable
than the cross section itself. The similarity of these two processes
renders double-Higgs production also a good tool to lift the
degeneracy in the parameter space of Higgs anomalous couplings that
currently affects the precise measurement of $gg\to H$
\cite{Azatov:2015oxa,Cao:2015oaa}. The considerations at the basis of
these statements are expected to be largely independent of collider
energy, making $gg\to HH$ a golden channel for precision Higgs physics
at 100 TeV.
 
Vector-boson fusion is the second production mechanism for Higgs
pairs, as well as for single Higgs. The relevance of this channel is
twofold: on one side, it provides an independent way to constrain
$\lambda_{3H}$; on the other hand, it is the main channel that is
sensitive to the Higgs-gauge couplings $W^+W^-HH$ and $ZZHH$. The
cross section for this process, computed up to NNLO in QCD, is 80 fb
at 100 TeV \cite{Liu-Sheng:2014gxa}. Despite the smaller cross section
(by a factor of roughly 20) with respect to gluon fusion, VBF has a
clear experimental signature, with the Higgs pair produced at central
rapidity and two hard jets in the forward/backward region, hence it
makes background reduction feasible. Moreover, its sensitivity to
$\lambda_{3H}$ is quite high, so that a deviation of this coupling
from its SM value can significantly enhance the VBF cross section (see
for example figure 5 of \cite{Liu-Sheng:2014gxa}).

The cross sections for triple-Higgs production processes are obviously
much smaller than those for double-Higgs production, both due to the
presence of an extra weak coupling, and to the fact that an extra
massive particle implies larger $x$. The gluon-fusion channel is again
the dominant one, but compatibly with what just outlined, its cross
section at 100 TeV is of the order of 5 fb \cite{Maltoni:2014eza},
i.e.~more than 300 times smaller than double-Higgs production, which
makes it a challenging process. This channel is in principle sensitive
to both triple and quadruple Higgs self interactions, but the
contribution from the triangle diagrams, the ones featuring
$\lambda_{4H}$, is particularly small \cite{Maltoni:2014eza}: the
production rate indeed depends very mildly on the quartic coupling,
with a variation of only $\pm10\%$ upon varying the quartic in the
range $[0,m_H^2/v^2]$, and assuming
$\lambda_{3H}=\lambda_{\tiny{\mbox{SM}}}$
\cite{Papaefstathiou:2015paa}. The extraction of $\lambda_{4H}$ from
triple-Higgs production is thus unlikely at 100 TeV. The $HHH\to(b\bar
b)(b\bar b)(\gamma\gamma)$ decay channel could in principle be
exploited to constrain a dimension-6 operator
$c_6\lambda_{\tiny{\mbox{SM}}}H^6/\Lambda^2$, but it turns out to be
effective only in a possible high-luminosity phase (of the order of 30
ab$^{-1}$) of the 100-TeV collider \cite{Papaefstathiou:2015paa}.

\subsection{Multi Higgs boson production in association with top quarks or gauge bosons} 
Associated production of a Higgs pair with a top-antitop pair or with
a vector boson are the main subdominant double-Higgs production
channels. Inspection of figure \ref{fig:HH} \cite{Frederix:2014hta}
shows that while at the LHC the cross sections for these three
channels are of the same order (within a factor of two), at 100 TeV
$t\bar tHH$ production grows roughly ten times more than $VHH$, since
it proceeds through $gg$. This fact causes its cross section to be
very close to (or even slightly larger than) that for VBF, roughly 85
fb. Detailed analyses \cite{Englert:2014uqa,Liu:2014rva} show that
this channel can provide significant statistical power to increase the
sensitivity to $\lambda_{3H}$, and that the presence of the top pair
is crucial for a substantial reduction of the backgrounds with respect
to gluon fusion.

$VHH$ processes are also relevant for the determination of
$\lambda_{3H}$. Studies of these channels show a good sensitivity to
$\lambda_{3H}$ already at the HL-LHC \cite{Cao:2015oxx}, and the
cross-section increase, which is modest with respect to $t\bar tHH$
but still of a factor of roughly 40 from 8 to 100 TeV, should further
extend their potential, especially in a high-luminosity phase.
  
Production of a $tjHH$ final state, namely a single top in association
with a Higgs pair, is also potentially interesting at 100 TeV, and
completes the programme for the determination of the trilinear. While
at 8 TeV its cross section is below 10$^{-2}$ fb, which makes it
phenomenologically irrelevant for the present, at 100 TeV its rate
grows by roughly a factor of $10^3$ and becomes comparable to that for
$VHH$, see figure \ref{fig:HH}. This process is of interest because it
has the largest sensitivity to $\lambda_{3H}$ among the double-Higgs
channels, see fig.~3 of \cite{Frederix:2014hta}, and it may become
clearly visible at 100 TeV in case the trilinear significantly
deviates from the SM expectation. In addition to that, it is sensitive
to couplings to both vector bosons and top quarks, and to their
relative phases \cite{Frederix:2014hta}.

\clearpage
\section{Loop-induced processes\footnote{Editors: V.~Hirschi, O.~Mattelaer}}
\label{sec:loop}

\subsection{Cross-sections at 100 TeV}
In this section, we present the cross sections for various
loop-induced SM processes involving associated production of Higgs and
gauge bosons.  The calculations are performed in the four-flavour
scheme with the SM parameters described in the
Table~\ref{table:LIParams}.  Whenever relevant, photons are isolated
by means of the Frixione smooth-cone criterion \cite{Frixione:1998jh},
with parameters $R_0=0.4$, $p_T(\gamma)>20$ GeV,
$\epsilon_\gamma=n=1$.  In the case of the pair production of heavy
boson, a technical cut of 1 GeV on the transverse momenta of the final
state bosons is applied in order to avoid the integrable singularity
at $p_{t}^V\rightarrow 0$.

\begin{table}[ht]
\begin{center}
\begin{tabular}{ll|ll}
Parameter & value & Parameter & value
\\
$\alpha_{S}(m_Z^2)$ & $ 0.13355  $ & $n_{lf}$ & \tt{4}
\\
$\mu_R=\mu_F$ & $\tt{\hat{\mu}} = \frac{H_T}{2}$& $m_{b}=y_{b}$ & \tt{4.7}
\\
$m_{t}=y_{t}$ & \tt{173.0} & $\Gamma_{t}$ & \tt{0}
\\
$G_F$ & \tt{1.16639e-05}  & $\alpha^{-1}$ & \tt{132.507}
\\
$m_Z$ & \tt{91.188} & $\Gamma_{Z}$ & \tt{2.4414}
\\
$m_{W}$ & $\frac{M_Z}{\sqrt{2}}\sqrt{1 + \sqrt{1 - \frac{4\pi}{\sqrt{2}}\frac{\alpha}{G_F M_Z^2}} }$ & $\Gamma_{W}$ & \tt{2.0476}
\\
$m_H$ & \tt{125.0} & $\Gamma_{H}$ & \tt{0.00638}
\\
$V^{CKM}_{ij}$ & $\delta_{ij}$  & $m_{e^\pm}=m_{\mu^\pm}$ & \tt{0.0}
\\
$m_{\tau^\pm}=y_{\tau^\pm}$ & $1.777$  & $\Gamma_{\tau^\pm}$ & \tt{0.0}\\
\end{tabular}
\end{center}
\caption{\label{table:LIParams} SM parameters used for
obtaining the results presented in
table~\ref{LI:table:cross1}. Dimensionful parameters are
given in powers of GeV.} 
\end{table}

\begin{figure}[h!]
        \centering
        \includegraphics[trim=10 0 40 20,clip,scale=0.40]{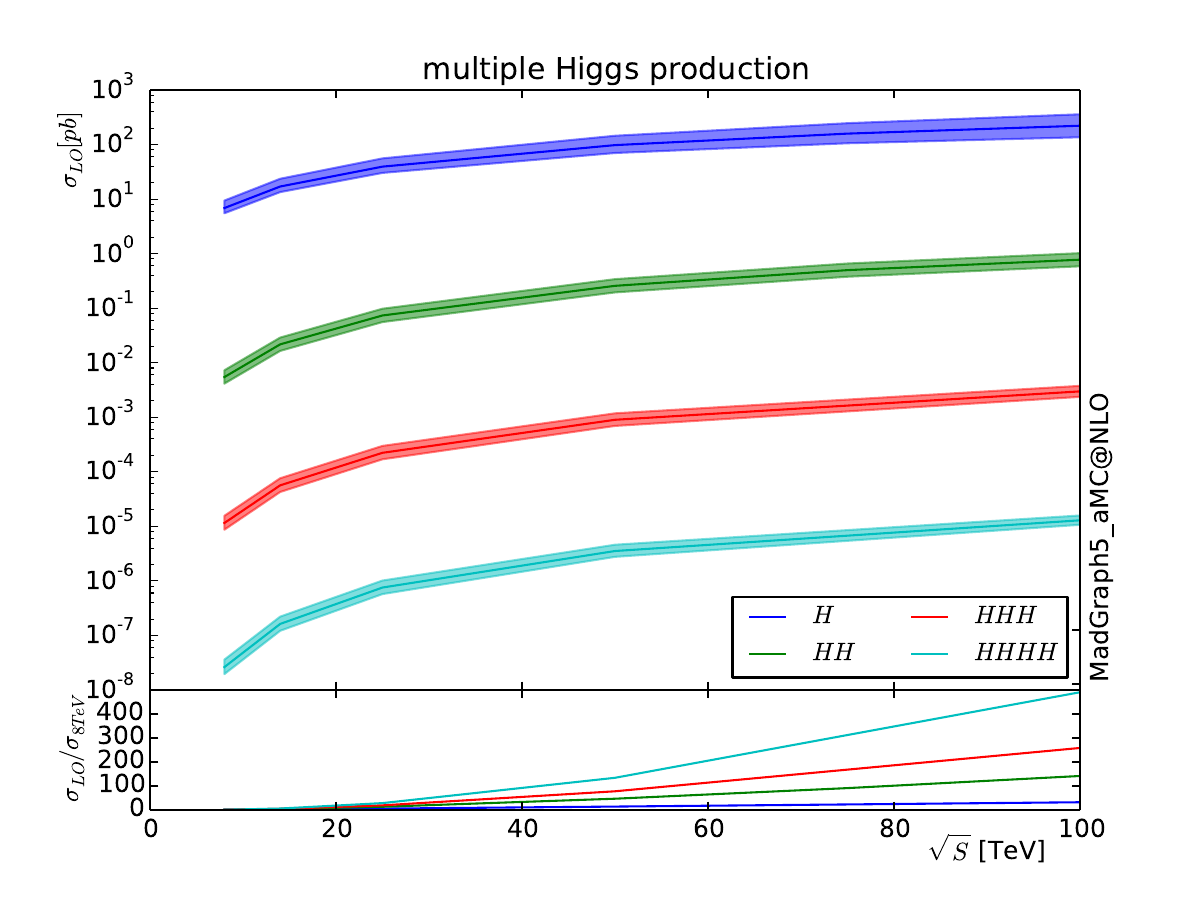}
        \includegraphics[trim=10 0 40 20,clip,scale=0.40]{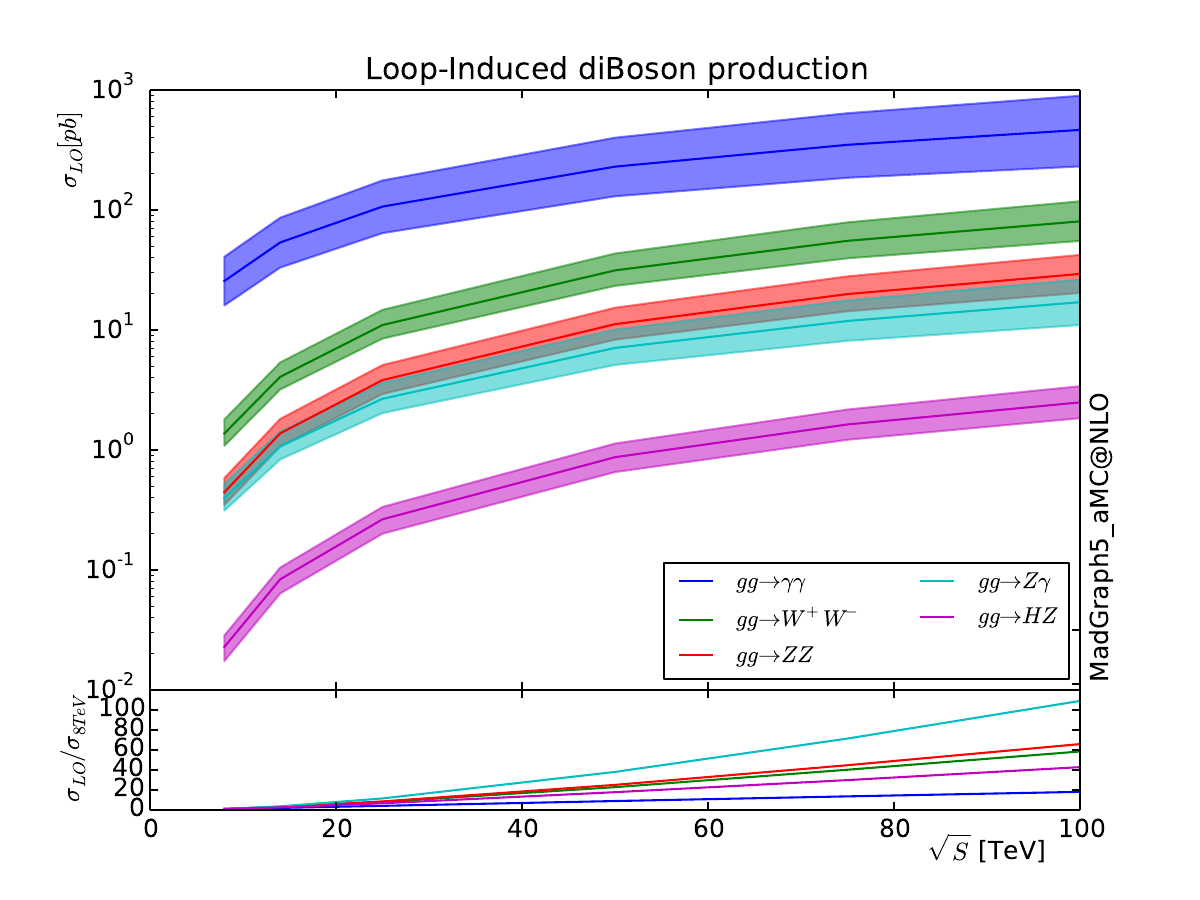}
        \caption{\label{fig:LI_8to100} Increase of the LO cross
        section with the collider energy,  for various loop-induced
        processes with initial-state gluons.} 
\end{figure}

The evolution of the cross-sections with the collider energy is shown
in Fig~\ref{fig:LI_8to100} for the production of multiple Higgs
(left)\cite{Degrande:2016oan} and various di-boson production
processes (right). In order to be able to easily compare the
cross-sections, they are all computed at exact LO, even for the ones
of lower multiplicity which are available in the literature at higher
QCD orders; all those cross-sections are expected to have a large NLO
QCD K-factor of around two. As expected, the cross-section increases
with the energy of the collider, and it does so at about the same rate
for all Higgs multiplicities. As a rule of thumb (rather accurate at
higher energies), producing an additional Higgs in the final state
costs three orders of magnitude in the production
cross-section. Increasing the energy is therefore required in order to
be able to observe multiple Higgs production processes.  The case of
double vector boson production processes is different because the
opening of the phase-space at larger energy is less relevant and the
corresponding factor
$\rho=\frac{\sigma_{100\text{TeV}}}{\sigma_{8\text{TeV}}}$ is
therefore smaller. The shape of the cross-section increase with the
collider energy is quite different for the processes $g g \rightarrow
\gamma \gamma$ and $g g \rightarrow Z \gamma$ because they do not
receive contributions from three-point loop diagrams.

A collection of results for the 100~TeV energy is given in
Table~\ref{LI:table:cross1}. 
\begin{table}[h!]
\begin{center}
\begin{small}
\begin{tabular}{rl | l || rl | l }
\hline\hline
\multicolumn{2}{c}{Loop Induced Process} &
  $\sigma_{\tiny{\mbox{LO}}}$(100 TeV) [fb] &
\multicolumn{2}{c}{Loop Induced Process} &
  $\sigma_{\tiny{\mbox{LO}}}$(100 TeV) [fb]
\\
\hline
 \phantom{{\Big |}}  $gg~~\to$ & $H$
 &
 $2.21\cdot 10^{+5}~{}^{+58 \% }_{-39 \% } \,\, {}^{+1 \% }_{-1 \% }$
 &
  \phantom{{\Big |}} $gg~~\to$ & $HZ$    & 
 $2.50\cdot 10^{+3}~{}^{+35 \% }_{-26 \% } \,\, {}^{+1 \% }_{-1 \% }$
\\
\phantom{{\Big |}} $gg~~\to$ & $Hj$  
&
$2.77\cdot 10^{+5}~{}^{+67 \% }_{-39 \% } \,\, {}^{+24 \% }_{+22 \%}$
&
 \phantom{{\Big |}} $gg~~\to$ & $Hjj$    &
 $2.02\cdot 10^{+5}~{}^{+66 \% }_{-38 \% } \,\, {}^{+0 \% }_{-1 \% }$
 \\
  \phantom{{\Big |}} $gg~~\to$ & $HW^+W^-$    &
 $16.8\phantom{\, \cdot 10^{+0}d}~{}^{+31 \% }_{-23 \% } \,\,
 {}^{+8 \% }_{+6 \% }$
 &  
  \phantom{{\Big |}} $gg~~\to$ & $HZZ$   &
 $7.29\phantom{\, \cdot 10^{+0}d}~{}^{+28 \% }_{-22 \% } \,\,
 {}^{+0 \% }_{-1 \% }$
 \\
\phantom{{\Big |}} $gg~~\to$ & $HZ\gamma$     &
 $0.279\phantom{0^{+0}d}~{}^{+33 \% }_{-25 \% } \,\, {}^{+0 \%
 }_{-1 \% }$
 &
\phantom{{\Big |}} $gg~~\to$ & $H\gamma\gamma$    &
 $0.374\phantom{0^{+0}d}~{}^{+33 \% }_{-25 \% } \,\, {}^{+10 \%
 }_{+9 \% }$
\\
\phantom{{\Big |}} $gg~~\to$ & $HH$  &
 $7.74\cdot 10^{+2}~{}^{+32 \% }_{-24 \% } \,\, {}^{+0 \% }_{-1 \% }$
 & 
 \phantom{{\Big |}} $gg~~\to$ & $HHZ$  & $3.35\phantom{\, \cdot
 10^{+0}d}~{}^{+29 \% }_{-22 \% } \,\, {}^{+0 \% }_{-1 \% }$ 
\\
\phantom{{\Big |}} $gg~~\to$ & $HHH$  &
  $2.99\phantom{\, \cdot 10^{+0}d}~{}^{+29 \% }_{-22 \% } \,\,
  {}^{+5 \% }_{+4 \% }$
&  
 \phantom{{\Big |}} $gg~~\to$ & $HHHH$  &
 $1.30\cdot 10^{-2}~{}^{+23 \% }_{-18 \% } \,\, {}^{+1 \% }_{-1 \% }$
\\
\phantom{{\Big |}} $gg~~\to$ & $W^+W^-$    &
 $8.06\cdot 10^{+4}~{}^{+48 \% }_{-33 \% } \,\, {}^{+31 \% }_{+29 \%}$
&
\phantom{{\Big |}} $gg~~\to$ & $ZZ$   & $2.92\cdot
 10^{+4}~{}^{+42 \% }_{-30 \% } \,\, {}^{+1 \% }_{-1 \% }$
 \\
\phantom{{\Big |}} $gg~~\to$ & $Z\gamma$   &
 $1.70\cdot 10^{+4}~{}^{+52 \% }_{-35 \% } \,\, {}^{+1 \% }_{-1 \% }$
 & 
\phantom{{\Big |}} $gg~~\to$ & $\gamma\gamma$  &
 $4.59\cdot 10^{+5}~{}^{+89 \% }_{-50 \% } \,\, {}^{+3 \% }_{-3 \% }$
 \\
 \phantom{{\Big |}} $gg~~\to$ & $W^+W^-Z$   &
 $4.71\cdot 10^{+2}~{}^{-100 \% }_{-100 \% } \,\, {}^{+0 \% }_{+0 \%}$
 &
 \phantom{{\Big |}} $gg~~\to$ & $ZZZ$  &
 $4.00\phantom{\, \cdot 10^{+0}d}~{}^{+30 \% }_{-23 \% } \,\,
 {}^{+0 \% }_{-1 \% }$
\\ 
\phantom{{\Big |}} $gg~~\to$ & $\gamma ZZ$   &
 $0.13\phantom{\, \cdot 10^{+0}d}~{}^{+34 \% }_{-25 \% } \,\,
 {}^{+1 \% }_{-1 \% }$ 
&
 \phantom{{\Big |}} $gg~~\to$ & $Z\gamma\gamma$ &
 $3.42\phantom{\, \cdot 10^{+0}d}~{}^{+44 \% }_{-31 \% } \,\,
 {}^{+1 \% }_{-1 \% }$
 \\
\end{tabular}
\end{small}
\end{center}
\caption{\label{LI:table:cross1} Cross sections for loop-induced
 associated production of gauge and Higgs bosons, at 100~TeV.
 Theoretical uncertainties describe scale and PDF variations,
 respectively. The numerical integration error is always smaller than
 theoretical uncertainties, and is not shown. Jets are within
 $\vert \eta \vert < 5$ and have $p_T>20$~GeV. For $pp\to HV jj$,
 furthermore, $m(jj)> 100$ GeV.} 
\end{table}

\clearpage
\section{Electroweak corrections\footnote{Editor: F.~Piccinini}}

\label{sec:EW}
\label{sec:EWintro}
The electroweak coupling constant $\alpha$ is more than a factor of 10
smaller than the QCD coupling constant $\alpha_s$, and therefore
perturbative corrections from QCD are typically much larger than those
from electroweak effects.  From the size of the coupling constants,
one can expect that NLO electroweak corrections are roughly comparable
to NNLO QCD corrections. For colliders at relatively low energies
(such that the partonic center of mass energy does not exceed the
electroweak scale significantly), this scaling in general holds, but
of course depends on the process under consideration.

At partonic energies which far exceed the electroweak scale, however,
electroweak corrections receive a logarithmic enhancement. For each
power in the electroweak coupling constant, one finds two powers of
the logarithm
\begin{equation}
  L_V(s) = \ln \frac{m_V^2}{s}
  \,,\end{equation}
where $\sqrt{s}$ is the partonic center of mass of the hard collision. This implies 
that each order in perturbation theory gives a factor 
\begin{equation}
  \frac{\alpha}{4\pi} L_V^2(s)
  \,.\end{equation}
For a concrete example, consider the Drell-Yan process $pp \to \ell^+ \ell^-$, 
where $\ell$ denotes either an electron or a muon. At fixed order, the electroweak 
corrections due to the exchanges of $W$ and $Z$ bosons are given by
\begin{equation}
\frac{\sigma^{\rm NLO}(s)}{\sigma^{\rm LO}(s)} = 1 - \frac{\alpha}{4\pi} \left[ 1.56 \, 
L_W^2(s) + 1.78 \, L_Z^2(s) + \ldots \right]
\end{equation}
where we have only kept the terms enhanced by two powers of $L_V(s)$.
For $\sqrt{s} \gtrsim 1$~TeV electroweak corrections are at the 10\%
level and above, and for $\sqrt{s} \gtrsim$ 10 TeV the corrections
become larger than the Born results, such that fixed order electroweak
perturbation theory is expected to break down completely.

Note that the virtual corrections in the above results are infrared
(IR) finite by themselves. This is contrary to virtual corrections
involving massless gauge bosons, which are IR divergent, and only
yield finite answers when they are combined with the soft and
collinear radiation of real massless gauge bosons. The reason is that
the soft and collinear divergences that are present for massless gauge
theories are regulated by the masses of the vector bosons, such that
both the virtual and the real radiation are separately finite, albeit
with logarithmic sensitivity to the gauge boson masses. This makes of
course physical sense; the real radiation of massive gauge bosons
(even those with infinitely soft or collinear momentum) can always be
observed experimentally, such that both the virtual and real
contributions lead to experimentally observable cross sections and
therefore they have to be finite by themselves. The logarithmic
sensitivity on the gauge boson masses is a consequence of the fact
that in the massless limit we have to recover the usual result where
both virtual and real corrections are separately divergent.

From the above argument it of course follows that not only the virtual
corrections are logarithmically sensitive to the masses of the gauge
bosons, but the real corrections have to be as well. This logarithmic
sensitivity should cancel for completely inclusive quantities. To see
this, let us consider the process
\begin{equation}
  \sigma_{q_1 q_2} \equiv \sum_{l_1, l_2} \sigma_{q_1 q_2 \to l_1 l_2} 
  + \sum_{l_1, l_2, V} \sigma_{q_1 q_2 \to l_1 l_2 V}  
  \,.\end{equation}
To double logarithmic accuracy for the NLO correction $\delta
\sigma_{q_1 q_2}$ one finds
\begin{eqnarray}
  \delta \sigma_{u \bar u}(s) = \delta \sigma_{d \bar d}(s) 
  = -\delta \sigma_{u \bar d}(s) = -\delta \sigma_{d \bar u}(s)
  \,.\end{eqnarray}
Thus, if we sum over the flavors of the initial state on top of the
flavor of the final state (thus calculating a completely inclusive
quantity), all double logarithms cancel
\begin{eqnarray}
\label{KLNCancellation}
\delta \sigma &\equiv&  \sum_{q_1, q_2} \delta \sigma_{q_1 q_2} =  0  
+ {\cal O}\left(\frac{\alpha}{4\pi} \right)\, .
\end{eqnarray}
The result of Eq.~(\ref{KLNCancellation}) is of course the result of
the KLN theorem, which states that all IR sensitivity will cancel in
completely inclusive observables.  However, the sum over initial
states as performed in Eq.~(\ref{KLNCancellation}) is of course not
possible for a hadron collider, since each cross-section is weighted
by their parton luminosities which are not equal to one another. This
gives the important result that at a hadron collider the logarithmic
sensitivity on the gauge boson masses do not cancel, even for
completely inclusive observables. This can be understood easily by
noting that the protons in the initial state are not singlets under
the SU(2) gauge group, such that the initial state breaks the
inclusivity of the observable.

The fixed order results can be calculated using standard techniques
for NLO calculations, however the presence of several mass scales
means that the required calculations are typically more difficult than
the corresponding calculation for massless gauge theories such as QCD.
Much effort has been put into understanding the electroweak logarithms
arising from virtual
corrections~\cite{Beccaria:1998qe,Ciafaloni:1998xg,Fadin:1999bq,Melles:2000gw,Melles:2000ed,Melles:2000ia,
  Melles:2001mr,Melles:2001ye,Ciafaloni:1999ub,Ciafaloni:2000df,Ciafaloni:2000rp,Ciafaloni:2000gm,Denner:2000jv,
  Melles:2001dh,Ciafaloni:2001vt,Ciafaloni:2001vu,Denner:2001gw,Denner:2003wi,Denner:2004iz,Denner:2006jr,
  Chiu:2007yn,Chiu:2007dg,Chiu:2008vv,Denner:2008yn,Chiu:2009mg} and
the structure of the logarithmic terms at one and two loops was
derived for a general process in
Refs.~\cite{Denner:2000jv,Denner:2001gw}
and~\cite{Denner:2003wi,Denner:2004iz,Denner:2006jr,Denner:2008yn},
respectively. The issue of real weak boson emission has been addressed
in Refs.~\cite{Ciafaloni:2001vt,
  Ciafaloni:2000rp,Ciafaloni:2000df,Ciafaloni:2001vu,Ciafaloni:2001mu,Ciafaloni:2003xf,Ciafaloni:2005fm,Ciafaloni:2006qu,Ciafaloni:2008cr,Bell:2010gi,Ciafaloni:2010ti,Stirling:2012ak,Bauer:2016kkv}
and, on a more phenomenological ground, in
Refs.~\cite{Baur:2006sn,Bell:2010gi,Bern:2012vx,Chiesa:2013yma,Christiansen:2014kba,Krauss:2014yaa,Frixione:2014qaa,Frixione:2015zaa}.
Since as discussed the logarithmic terms dominate over the terms not
logarithmically enhanced, this general result provides a good
approximation to the exact NLO corrections at sufficiently high
partonic center of mass energies. This approximation is often called
the Sudakov log approximation.

The resummation of the logarithmic terms that arise in the virtual
exchanges of $W$ and $Z$ has been the subject of a considerable amount
of work over the past
decade~\cite{Kuhn:1999nn,Fadin:1999bq,Ciafaloni:1999ub,Beccaria:2000jz,Hori:2000tm,Ciafaloni:2000df,Denner:2000jv,Denner:2001gw,Melles:2001ye,Beenakker:2001kf,Denner:2003wi,Pozzorini:2004rm,Feucht:2004rp,Jantzen:2005xi,Jantzen:2005az,Jantzen:2006jv,Chiu:2007yn,Chiu:2008vv,Manohar:2012rs}.
In~\cite{Chiu:2007yn,Chiu:2008vv} a completely general method to resum
these logarithms for an arbitrary process was developed, using
soft-collinear effective
theory~\cite{Bauer:2000ew,Bauer:2000yr,Bauer:2001ct,Bauer:2001yt}. The
resummation of the real radiation has so far not received much
attention yet, even though the large logarithms originating from the
real radiation are by the KLN theorem as large as those resulting from
virtual exchanges. In a recent paper~\cite{Bauer:2016kkv}, it was
shown how to resum the double logarithmic corrections from the real
radiation of $W$ and $Z$ bosons.

From the above discussion it is clear that a real paradigm shift is
happening with regards to electroweak corrections when partonic center
of mass energies exceed a few TeV, which can easily happen at a future
100 TeV machine. Thus, at such a machine, electroweak corrections are
{\it much more} important than at current colliders.  While at past
and current colliders electroweak corrections were usually computed to
obtain high precision for a few observables, electroweak corrections
at a 100 TeV collider are required even to get rough estimates of the
expected cross-sections at the highest available phase space corners.
Furthermore, at high enough center of mass energies, not only are the
electroweak corrections required at fixed order accuracy, but the
leading logarithms need to be resummed.

In this section we give a brief account of the available tools and
algorithms for the calculation of electroweak corrections at hadronic
colliders and discuss the phenomenological impact of electroweak
corrections (at $\sqrt{s} = 100$~TeV) to the following benchmark
Standard Model processes: Drell-Yan, weak boson pairs ($WW$, $WZ$,
$WH$ and $ZH$), $V + $~jets, dijet production, $t \bar t$, $t \bar t
H$ and $t \bar t+$~jets.  A last section will be devoted to the issue
of the inclusion of real radiation.

\subsection{Tools}\label{sec:EW-tools}

In the past exact NLO corrections have been calculated and implemented
in simulation tools for a limited class of hadronic collision
processes.  In particular for charged and neutral Drell-Yan, the most
important processes for the precision physics program of Tevatron and
LHC, a number of dedicated codes have been developed, such as
{\sc{HORACE}}~\cite{CarloniCalame:2006zq,CarloniCalame:2007cd},
{\sc{RADY}}~\cite{Dittmaier:2001ay,Brensing:2007qm,Dittmaier:2009cr},
{\sc{SANC}}~\cite{Arbuzov:2005dd,Arbuzov:2007db},
{\sc{WGRAD}}~\cite{Baur:2004ig}, {\sc{WINHAC}}
\cite{Placzek:2003zg,Bardin:2008fn} and
{\sc{ZGRAD}}~\cite{Baur:2001ze}.  In particular {\sc{SANC}} also
includes NLO QCD corrections, while {\sc{HORACE}} includes the effect
of all order photon radiation properly matched to the $\mathcal{O}(
\alpha )$ corrections. NLO EW corrections are added to the
$\mathcal{O}( \alpha_S^2 )$ ones in the {\sc{FEWZ}}
code~\cite{Li:2012wna}, while factorized NLO EW and NLO QCD
corrections to the single $W$ and $Z$ production matched with QED and
QCD parton shower have been implemented in the {\sc{POWHEG-BOX}} Monte
Carlo event
generator~\cite{Barze:2012tt,Bernaciak:2012hj,Barze':2013yca}.

The large center of mass energy and the high luminosity of LHC run II
and beyond will require the inclusion of at least $\mathcal{O}( \alpha
)$ corrections in theoretical predictions for several processes, as
documented, for example, in the Les Houches wish-list in Tables~1-3 of
Ref.~\cite{Butterworth:2014efa}.

Besides Drell Yan processes, full one loop electroweak (EW)
corrections have been calculated also for $V+1$~jet ($V=Z$, $W$,
$\gamma$)~\cite{Maina:2004rb,Kuhn:2005gv,Kuhn:2005az,Kuhn:2007cv},
dilepton+jets~\cite{Denner:2009gj,Denner:2011vu,Denner:2012ts}, single
top~\cite{Beccaria:2006ir,Beccaria:2008av,Bardin:2010mz}, $t
\overline{t}$~\cite{Beenakker:1993yr,Kuhn:2005it,Moretti:2006nf,Bernreuther:2006vg,Kuhn:2006vh,Hollik:2007sw,Bernreuther:2008md},
dijet~\cite{Moretti:2005ut,Moretti:2006ea,Dittmaier:2012kx},
$Z/W+H$~\cite{Denner:2011id,Ciccolini:2003jy} (including the $Z/W$
decay products), $H$ production in vector boson
fusion~\cite{Ciccolini:2007ec,Ciccolini:2007jr}, $V V'$ (with on-shell
vector bosons or in pole approximation)~
\cite{Accomando:2005ra,Bierweiler:2012kw,Billoni:2013aba,Bierweiler:2013dja,Gieseke:2014gka,Baglio:2013toa},
$W W+1$~jet~\cite{Wei-Hua:2015gaa}, $WZZ$~\cite{Nhung:2013jta},
$WWZ$~\cite{Yong-Bai:2015xna}, $W \gamma$
production~\cite{Denner:2014bna}, $Z \gamma$
production~\cite{Denner:2015fca} $t \overline{t}
+H$~\cite{Frixione:2014qaa,Yu:2014cka}.  All these calculations have
been carried out on a process-by-process basis.  In the QCD sector,
during last ten years we have witnessed the so called ``NLO
revolution'': several groups succeeded in building new codes able to
calculate NLO QCD corrections in a completely automatic way, such as
{\sc{BlackHat}}~\cite{Berger:2008sj},
{\sc{GoSam}}~\cite{Cullen:2011ac,Cullen:2014yla},
{\sc{HELAC-NLO}}~\cite{Bevilacqua:2011xh},
{\sc{MadLoop}}~\cite{Hirschi:2011pa}/
{\sc{MadGraph}}/{\sc{MadGraph5\_aMC@NLO}}~\cite{Alwall:2011uj,Alwall:2014hca},
{\sc{Njet}}~\cite{Badger:2012pg},
{\sc{OpenLoops}}~\cite{Cascioli:2011va} and
{\sc{RECOLA}}~\cite{Actis:2012qn}. In various cases the automation of
hadron collider simulations is realised in combination with the Sherpa
Monte Carlo~\cite{Gleisberg:2008ta}.  These developments towards the
automatic computation of NLO QCD corrections allowed recent progress
in the calculation of NLO EW corrections, despite the difficulties of
virtual EW corrections, mainly due to the presence of several mass
scales and of unstable particles in the loops, as well as of the
chiral structure of electroweak interactions. For a recent review on
these items see Ref.~\cite{Chiesa:2015mya}.  With these automatic
algorithms, exact $\mathcal{O}( \alpha )$ corrections to $Z(\to l
\overline{l})+2$~jets~\cite{Denner:2014ina}, $W+n$~jets ($n \le
3$)~\cite{Kallweit:2014xda}, $W+n / Z+n$~jets ($n \le 2$ including
off-shell vector boson decays and matching with Parton
Shower)~\cite{Kallweit:2015dum}, $t \overline{t}
+H/Z/W$~\cite{Frixione:2014qaa,Frixione:2015zaa} and $\mu^+ \mu^- e^+
e^-$~\cite{Biedermann:2016yvs} have been computed for the first time.

As far as only the Sudakov regime is concerned, the universality of
the infrared limit of weak corrections can be exploited to develop
general algorithms for the calculation of the EW corrections in the
logarithmic approximation
~\cite{Denner:2000jv,Denner:2001gw,Fadin:1999bq,Chiu:2008vv}.
Following this approach, the Sudakov corrections to
diboson~\cite{Accomando:2004de,Accomando:2005ra,Accomando:2005xp,Fuhrer:2010eu},
vector boson plus one~\cite{Becher:2013zua} or several
jets~\cite{Chiesa:2013yma}, $t
\overline{t}$+jets~\cite{Manohar:2012rs},
$H$~\cite{Fuhrer:2010vi,Siringo:2012mi} and
$H+$jet~\cite{Fuhrer:2010eu} production have been computed pointing
out further the phenomenological impact of the EW corrections at high
energies. Order $\alpha$ corrections to dijet, Drell-Yan and
$t\overline{t}$ production have been recently included in the
{\sc{MCFM}} Monte Carlo program~\cite{Campbell:2010ff}: both the
Sudakov approximation and the full one loop corrections have been
implemented in order to provide a tool for the fast evaluation of the
approximated $\mathcal{O}(\alpha)$ corrections that also allows to
asses the validity of the approximated
results~\cite{Campbell:2015vua}.  Following the work presented in
Ref.~\cite{Chiesa:2013yma}, the algorithm for double and single logs
has been implemented in the {\sc{ALPGEN}}~\cite{Mangano:2002ea} LO
matrix element event generator, for the processes $V + $~multijets,
QCD multijet and heavy flavour plus jets.

\subsection{Drell-Yan}\label{sec:dyew}
We consider the processes $p p \to W^{+,*} \to \mu^+ \nu_{\mu} + (X)$
and $p p \to \gamma^*/Z^* \to \mu^+ \mu^- + (X)$, at the c.m. energy
$\sqrt{s} = 100$~TeV and using the NNPDF 2.3QED PDF
set~\cite{Ball:2013hta} with factorization/renormalization scale $\mu
= M({\bar \ell} \ell^{(\prime)} \gamma)$.  We applied the following
acceptance cuts:
\begin{eqnarray}
p_\perp^\mu, \, p_\perp^\nu  \geq 25 \, {\rm GeV} \, , \qquad \qquad |\eta_\mu| \leq 2.5 \, .
\end{eqnarray} 
Muons are considered ``bare" ({\it i.e.} without photon
recombination). In order to focus on the high energy dynamics, we
further impose the additional cut on the transverse mass $M_T \geq
5~$TeV, where $M_T$ is defined as $M_T = \sqrt{2 p_\perp^\ell
  p_\perp^\nu (1 - \cos\phi)}$ (with $\phi$ the angle between lepton
and neutrino in the transverse plane), for the charged Drell-Yan
process and $M(\ell^+ \ell^-) \geq 5~$TeV for the neutral current
process.

The results, with NLO accuracy in the electroweak coupling, have been
obtained with the code
{\sc{HORACE}}~\cite{CarloniCalame:2006zq,CarloniCalame:2007cd} using
the $G_\mu$ scheme and the following input parameters
\begin{center}
\begin{tabular}{lll}
$G_{\mu} = 1.1663787~10^{-5}$ GeV$^{-2}$ & 
$M_W = 80.385$~GeV&
$M_Z = 91.1876$~GeV \\
$\Gamma_W = 2.4952$~GeV & 
$\sin^2\theta_W = 1 - M_W^2/M_Z^2$&
$M_{\rm Higgs} = 125$~GeV\\
$m_e=510.998928$~KeV &
$m_{\mu}=105.6583715$~MeV &
$m_{\tau}=1.77682$~GeV \\
$m_u = 69.83$~MeV &
$m_c = 1.2$~GeV &
$m_t = 173$~GeV \\
$m_d = 69.83$~MeV &
$m_s = 150$~MeV &
$m_b = 4.6$~GeV \\
$V_{ud}=0.975$ &
$V_{us}=0.222$ &
$V_{ub}=0$ \\
$V_{cd}=0.222$ &
$V_{cs}= 0.975$ &
$V_{cb}=0$ \\
$V_{td}=0$ &
$V_{ts}=0$ &
$V_{tb}=1$ \\
\end{tabular}
\end{center}
For the coupling of external photons to charged particles needed for
the evaluation of photonic corrections we use $\alpha = \alpha (0)
=1/137.03599911$ and $\alpha_s (\mu)$ from the PDF set.

In Fig.~\ref{fig:dyWpmt} we present the integrated transverse mass $M_T$ and charged lepton transverse 
momentum $p_\perp^\ell$ (integrated) distributions in the window $[5 - 25]$~TeV. 
The effects of the NLO EW corrections with respect to the LO predictions are huge 
and negative, exceeding 60\% in absolute value for $M_\perp^\ell \geq 10$~TeV 
(red line in the lower left panel). The shaded bands 
around the lines give the estimate of the PDF uncertainty according to the NNPDF prescription, which 
is contained within 10\% level.  
In the same window of $[5 - 25]$~TeV the corrections to $p_\perp^\ell$ are even larger, 
because a given bin of the $p_\perp^\ell$ 
distribution corresponds roughly to a bin twice larger in the transverse mass distribution. 

A general issue regarding EW corrections is the relevance of the
inverse bremsstrahlung (also called ``gamma''-induced) processes which
are a contribution to the real radiation.  In fact the elementary
scattering process is $\gamma q \to \mu^+ \nu_{\mu} q^{\prime}$, whose
amplitude can be obtained by crossing symmetry from the standard real
radiation amplitude $q \bar {q^\prime} \to \mu^+ \nu_{\mu}
\gamma$. These contributions have been discussed in the
literature~\cite{Dittmaier:2009cr,Arbuzov:2007kp,CarloniCalame:2007cd,Buttar:2006zd},
with particular reference to LHC.  An essential ingredient is the
photon PDF, which is discussed in Section~\ref{sec:pdf}.  At present
it is affected by very large uncertainties, which blow up at large
energy scales.  However, in the future, these uncertainties will be
constrained by LHC data.  For the transverse mass,
Fig.~\ref{fig:dyWpmt}, left panel, blue line, the central value of the
inverse bremsstrahlung contribution is positive, at the \% level up to
10~TeV and increases up to values of around 10\% at 25~TeV. These
effects should be considered with caution because of the above
mentioned large uncertainties. In fact for $p_\perp^\ell$ not only the
uncertainty but also the central value blows up. For comparison, we
have included also the effect of higher order photonic corrections
(violet line) ~\footnote{In {\sc{HORACE}} the higher order effects are
  included by means of a proper matching between fixed order
  calculation and all orders Parton Shower. Other approaches can be
  adopted for the simulation of higher order photonic corrections,
  such as, for instance, the YFS formalism used in
  Refs.~\cite{Jadach:2013aha,Yost:2013boa,Davidson:2010ew,Golonka:2005pn}.},
which are positive and become of the order of 10\% at scales of the
order of 20~TeV.  In Fig.~\ref{fig:dyWpetal}, left panel, we show the
effects of the same higher order contributions discussed above for the
integrated lepton pseudorapidity distribution, where no particular
shape is present except for the overall normalization effects.
\begin{figure}[t]
\begin{center}
\includegraphics[width=0.48\textwidth]{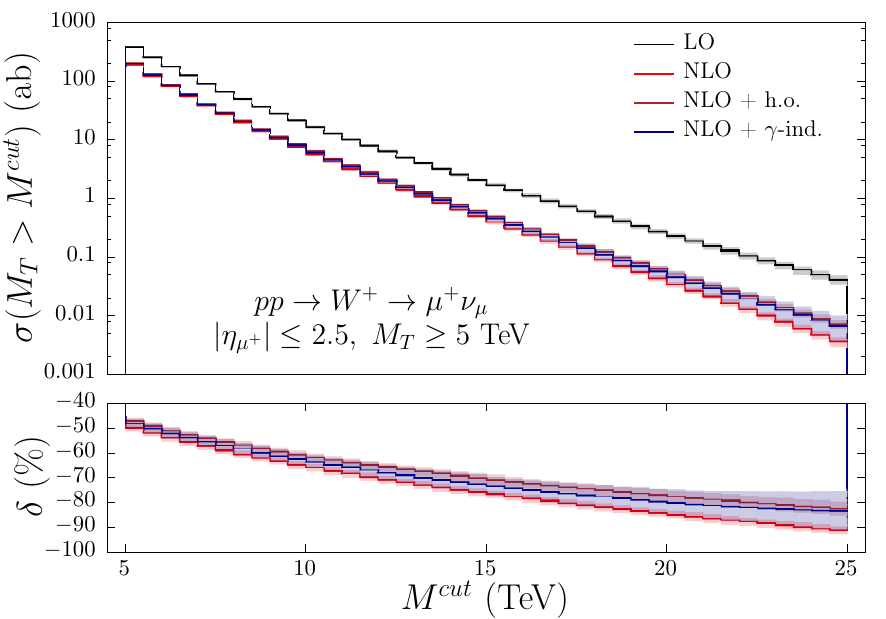}
\includegraphics[width=0.48\textwidth]{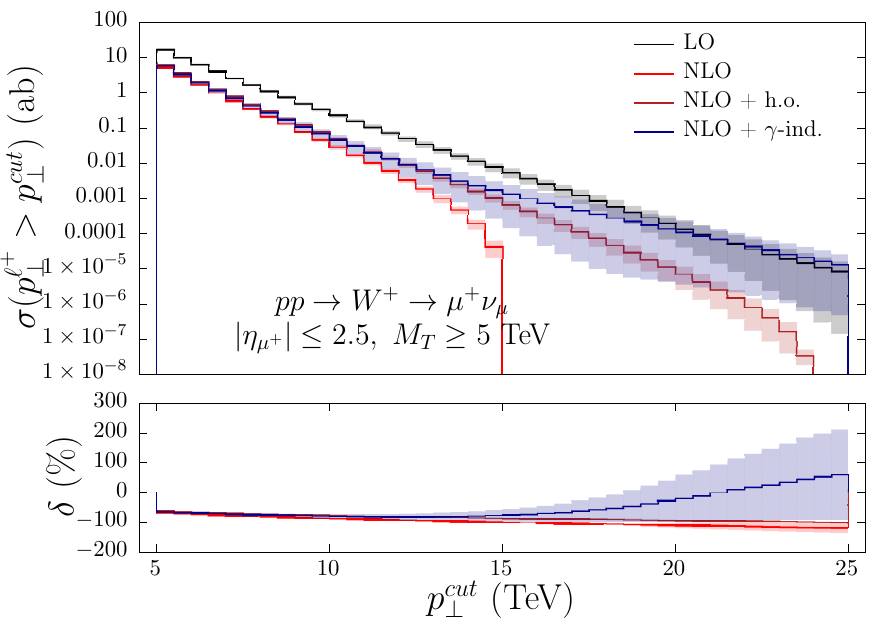}
\end{center}
\caption{Left: the distribution of the transverse mass for $W^+$
  production.  Right: the distribution of the charged lepton
  transverse momentum.  The black lines represent the LO predictions,
  the red lines give the NLO EW predictions, the violet lines include
  also higher order photonic corrections and the blue lines include
  the contribution of the $\gamma$-induced processes, in addition to
  the NLO EW corrections. The lower panels contain the relative
  deviations of the various levels of approximation with respect to
  the tree-level predictions.  }
\label{fig:dyWpmt}
\end{figure}

In Fig.~\ref{fig:dyWpetal}, right panel, and Figs.~\ref{fig:dyZptl}
and \ref{fig:dyZetal} we plot the predictions for the neutral Drell
Yan process $p p \to \gamma^*/Z^* \to \mu^+ \mu^- + (X)$. In
particular, Fig.~\ref{fig:dyWpetal}, right panel, displays the
invariant mass $M_{\mu^+ \mu^-}$ integrated distribution, while
Fig.~\ref{fig:dyZptl} contains the leading (left panel) and softest
(right panel) lepton transverse momentum integrated distribution.  For
the invariant mass the NLO EW corrections are slightly smaller than
for the charged Drell Yan case: they reach the size of $60$\% at
scales above 20~TeV. The corrections are of the same order for the
leading lepton transverse momentum, while they are larger for the
softest one, as can be expected with phase space arguments.

For the neutral current Drell Yan process there is a contribution from
$\gamma$-induced processes already at tree level, i.e. from $\gamma
\gamma \to \mu^+ \mu^-$.  For the considered observables, we plot
separately the LO prediction including/excluding (green/black lines)
the tree-level $\gamma \gamma \to \mu^+ \mu^-$ contribution and the
NLO prediction with and without (blue and red lines, respectively)
$\gamma$-induced contributions. The blue lines include both the
tree-level and the radiative $\gamma$-induced processes. As can be
seen in Figs.~\ref{fig:dyWpetal},~\ref{fig:dyZptl},~\ref{fig:dyZetal}
the largest effects come from the tree-level $\gamma \gamma \to \mu^+
\mu^-$ process, ranging from few \% at $M_{\ell^+ \ell^-} = 5$~TeV to
a factor of two at $M_{\ell^+ \ell^-} = 20$~TeV.  The effects of the
radiative $\gamma$-induced processes can be inferred by looking at the
difference between the blue and red lines of the lower panels.  They
are positive and moderate in size, reaching about 50\% at scales above
20~TeV.  However, given the very large uncertainties of photon PDF's,
all the predictions involving $\gamma$-induced processes should be
taken with caution.
\begin{figure}[t]
\begin{center}
\includegraphics[width=0.48\textwidth]{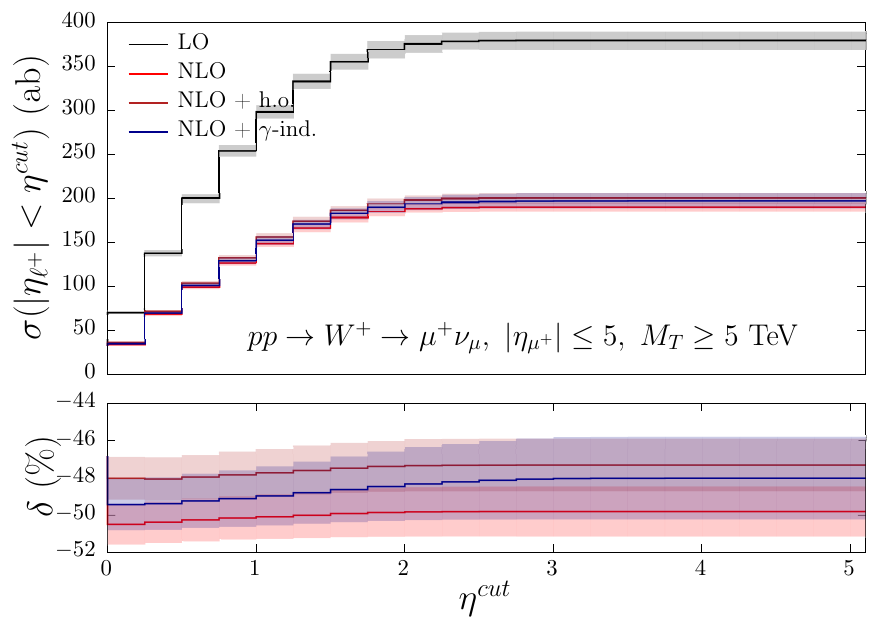}
\includegraphics[width=0.48\textwidth]{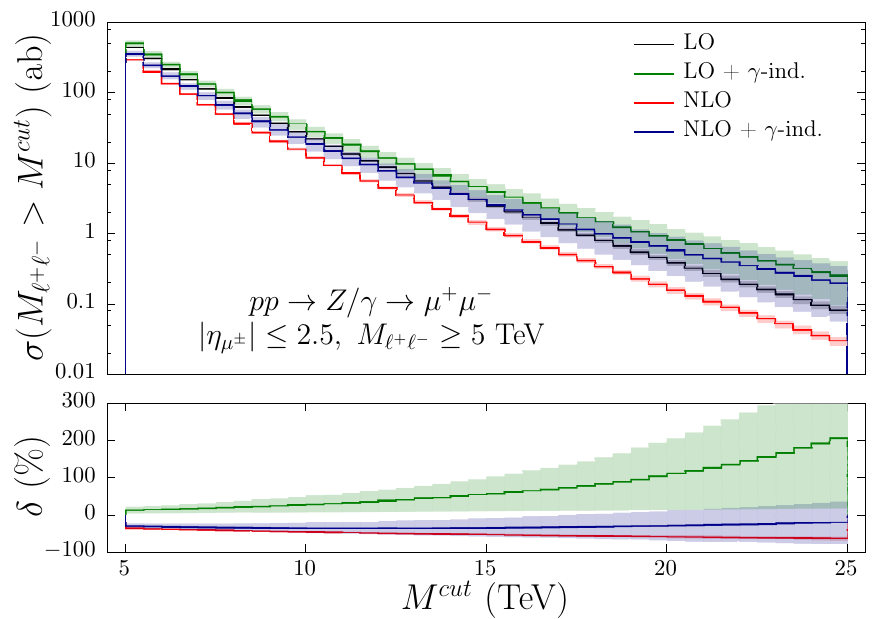}
\end{center}
\caption{Left: the distribution of the lepton pseudorapidity for $W^+$
  production. The meaning of the lines is the same as for
  Fig.~\ref{fig:dyWpmt}.  Right: the distribution of the invariant
  mass for $\ell^+ \ell^-$ production.  In the right panel the black
  line represents the LO predictions while the green line includes the
  LO ${\cal O}(\alpha^3) \gamma$-induced process. The red line shows
  the EW NLO predictions while the blue line includes the NLO ${\cal
    O}(\alpha^3) \gamma$-induced processes (both tree-level and
  radiative diagrams).  In the lower panel the green line is the
  relative deviation of the LO prediction including the tree-level
  $\gamma \gamma \to \mu^+ \mu^-$ process with respect to the pure LO.
  The red line gives the size of the EW NLO corrections excluding
  $\gamma$-induced processes; the blue line quantifies the deviation
  of the complete EW NLO corrections (including all $\gamma$-induced
  processes) with respect to the the LO order predictions which
  include the tree-level $\gamma \gamma \to \mu^+ \mu^-$ process.}
\label{fig:dyWpetal}
\end{figure}

\begin{figure}[t]
\begin{center}
\includegraphics[width=0.48\textwidth]{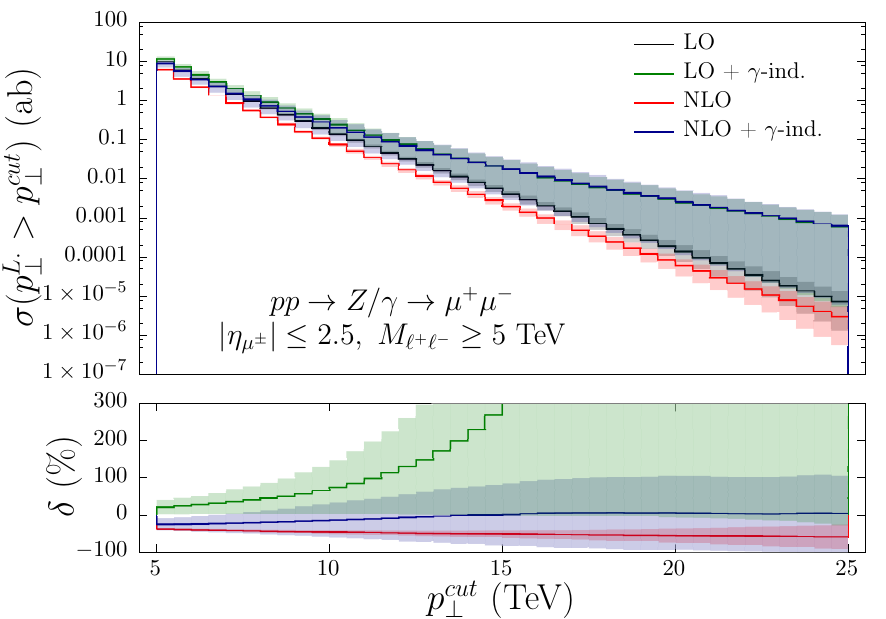}
\includegraphics[width=0.48\textwidth]{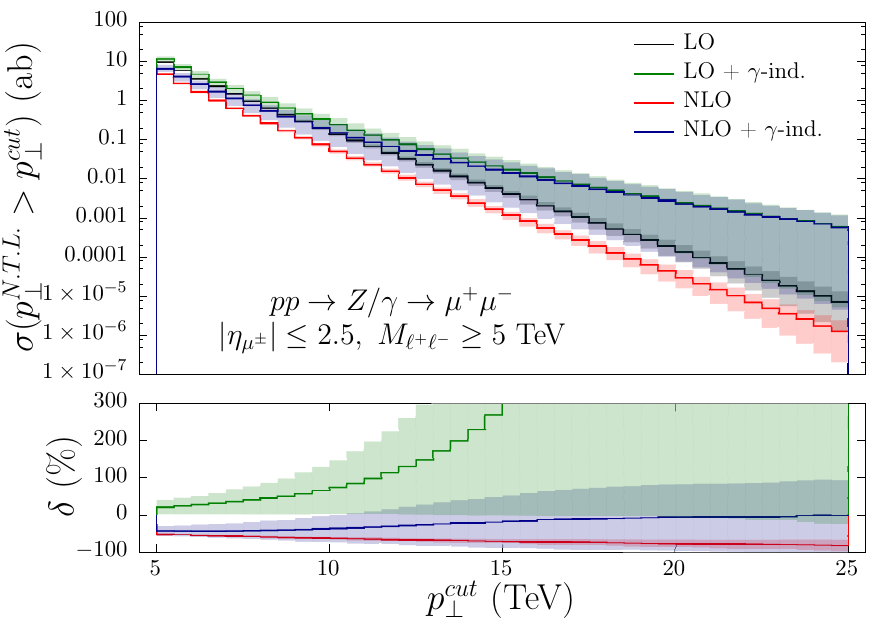}
\end{center}
\caption{The distribution of the leading (left) and softest (right)
  lepton pt for $\ell^+ \ell^-$ production.  The meaning of the lines
  is the same as in Fig.~\ref{fig:dyWpetal}, right panel.}
\label{fig:dyZptl}
\end{figure}

\begin{figure}[t]
\begin{center}
\includegraphics[width=0.48\textwidth]{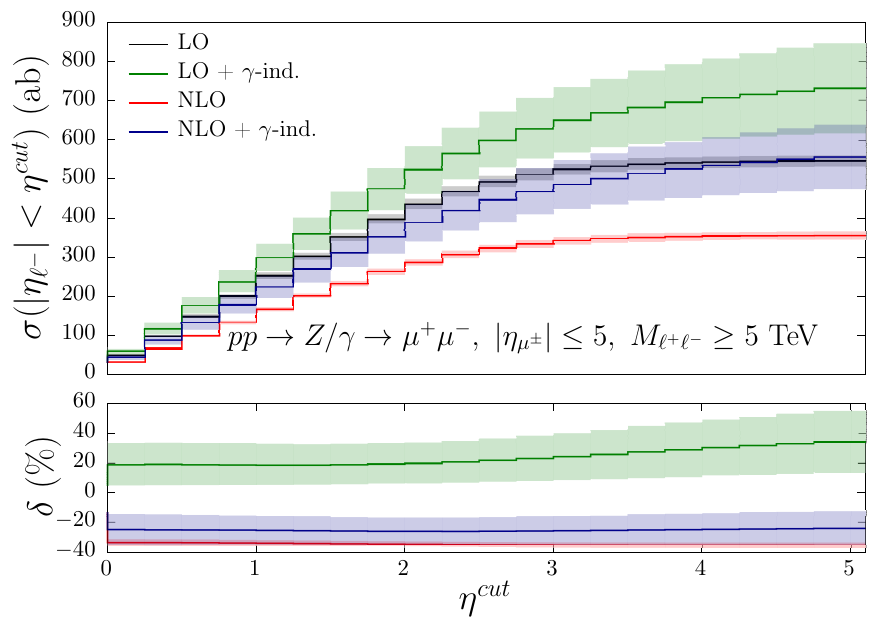}
\includegraphics[width=0.48\textwidth]{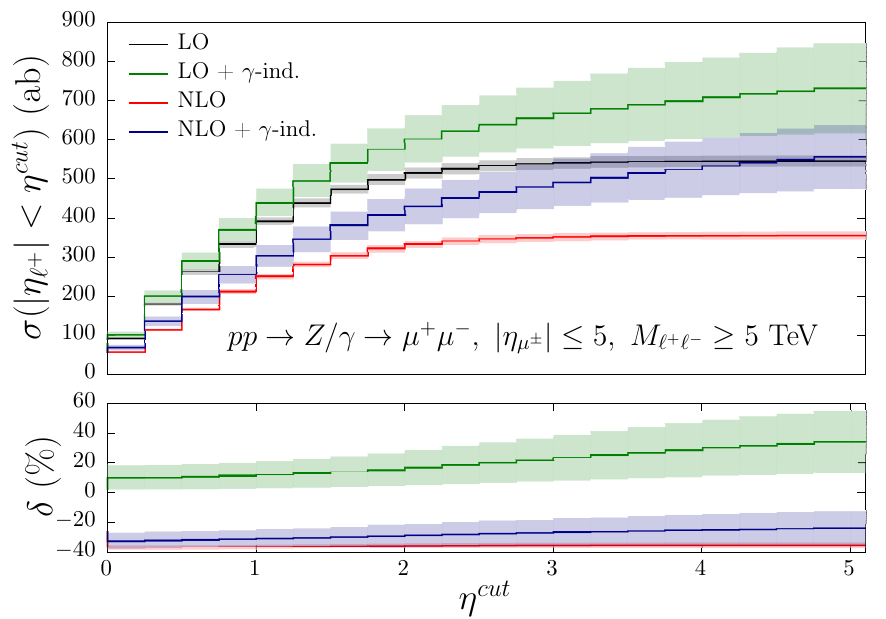}
\end{center}
\caption{The distribution of $\mu^-$ (left) and $\mu^+$ (right)
  pseudorapidity for $\ell^+ \ell^-$ production.  The meaning of the
  lines is the same as in Fig.~\ref{fig:dyWpetal}, right panel.}
\label{fig:dyZetal}
\end{figure}
In summary, the effects of the EW NLO corrections on Drell-Yan
processes at a future hadron collider at 100~TeV are very large,
spoiling the stability of fixed order perturbative calculations and
calling for resummed approaches, in order to obtain reliable
predictions.  The inverse bremsstrahlung processes could have a
relevant impact, even if at present it is difficult to put on a
quantitative ground. To this aim, reliable photon PDF's would be
necessary, which will be available after the scheduled LHC runs.

\subsection{Gauge boson pairs and Higgsstrahlung}
\label{sec:dibson}
In the present section we focus on the EW and QCD corrections to the
diboson production at the FCC at 100~TeV. In particular, we discuss
the impact of NLO QCD and EW corrections to the processes $V V'$
(\emph{i.e.} $W^+ W^-$, $ZW^\pm$ and $ZZ$) and $HV$ ($V=W^\pm$, $Z$)
computed by means of the automated tool
{\sc{MadGraph5\_aMC@NLO}}~\cite{Alwall:2014hca} and a currently
private extension that allows to calculate NLO QCD and EW corrections
\cite{Frixione:2014qaa, Frixione:2015zaa}.  We work in the $G_{\mu}$
scheme with:
\begin{equation}
\label{eq:vvinput}
G_{\mu}= 1.16639\, 10^{-5} \; {\rm GeV}^{-2}, \qquad M_W =80.385 \; {\rm GeV}, \qquad M_Z = 91.188 \;  {\rm GeV} ,
\end{equation}    
the top quark and Higgs boson masses being set to $173.3$~GeV and
$125$~GeV, respectively.  We use the NNPDF 2.3QED PDF
set~\cite{Ball:2013hta} with the following factorization and
renormalization scales:
\begin{equation}
\label{eq:vvinput1}
\mu_F =\mu_R = \mu = \frac{H_T}{2}, \qquad H_T = \sum_{i} \sqrt{m_i^2 + p_{T,\,i}^2},
\end{equation}
where the index $i$ runs over all the final state particles. Scale
uncertainties are estimated by varying independently the scales
$\mu_F$ and $\mu_R$ in the range $[ \mu / 2, 2 \mu ]$. Massive
external particles are treated as stable and no cuts are applied at
the analysis level.

One loop EW corrections to $VV'$ production at hadron colliders have
been computed in the Sudakov approximation in
Refs.~\cite{Accomando:2004de,Accomando:2005xp,Fuhrer:2010eu}, while
the full $\mathcal{O}(\alpha)$ results can be found in
Refs.~\cite{Bierweiler:2013dja,Baglio:2013toa} for on-shell $V$ and
$V'$ and in
Refs.~\cite{Billoni:2013aba,Gieseke:2014gka,Biedermann:2016yvs}
including vector boson decays. Here, we show predictions at NLO QCD
and EW accuracy, taking into account the contribution from
initial-state photons and evaluating the corresponding PDF
uncertainties, which are expected to be large.

In figure \ref{fig:Zwmew} we show predictions at NLO QCD and NLO EW
accuracy for cumulative distributions in $ZW^-$ production (results
for $ZW^+$ are qualitatively identical).  In the upper row we show the
dependence of the cross section on a cut on the $Z$ transverse
momentum ($p_T(Z)$).  In the lower row we show the dependence on a cut
on the $ZW^-$ invariant mass ($m(ZW^-)$). The plots on the left do not
include any contribution from the photon in the initial state, {\it
  i.e.}, the photon PDF has been artificially set to zero. On the
contrary, the plots on the right do include these contributions. The
left plots allow to better identify the negative contributions due to
the Sudakov logarithms and disentangle them from those related to
photon-initiated processes (quark radiation from $\gamma q$
initial-states). Instead, the plots on the right include this kind of
processes and thus the comparison with those on the right is useful
for estimate the photon-induced contributions, which typically have
huge PDF uncertainties and are very large with opposite sign w.r.t the
Sudakov logarithms. It is important to note that the plots on the
right strongly depend on the PDF set used and their large uncertainty
are due to the currently poor determinations of the photon PDF.

\begin{figure}[t]
\begin{center}
\includegraphics[width=0.48\textwidth]{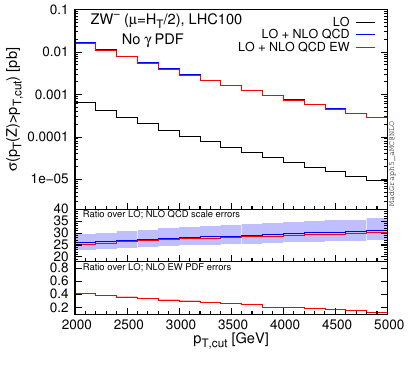}\hspace{10pt}
\includegraphics[width=0.48\textwidth]{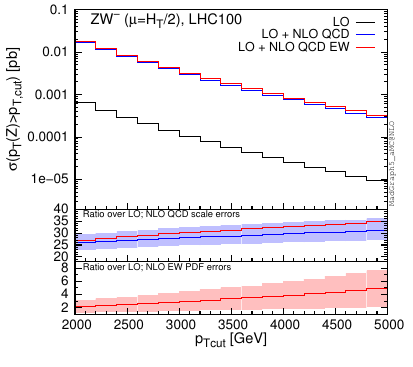}
\includegraphics[width=0.48\textwidth]{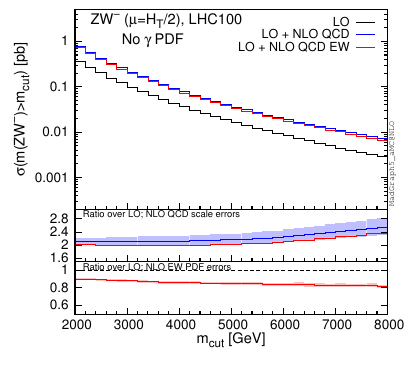}\hspace{10pt}
\includegraphics[width=0.48\textwidth]{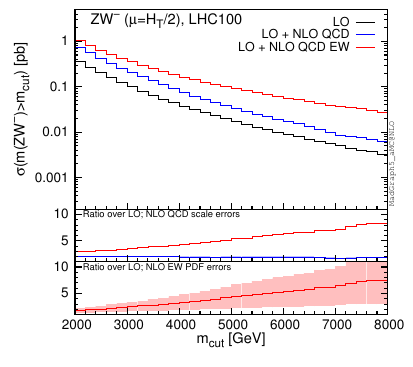}
\end{center}
\caption{Cumulative distributions for $ZW^-$ production at NLO QCD and
  NLO EW accuracy. The upper plots show the dependence of the total
  cross sections on a cut on the $p_T(Z)$, the lower ones on a cut on
  the $m(ZW^-)$.  The plots on the left do not include contribution
  from photons in the initial state; they are included in the right
  plots.  See text for further details.}
\label{fig:Zwmew}
\end{figure}
In each plot we display in the main panel LO (black), LO + NLO QCD
(blue) and LO + NLO QCD + NLO EW (red) distributions.  In the first
inset we show the (LO + NLO QCD)/LO ratio with scale uncertainties
(blue band), {\it i.e.} the QCD $K$-factor, and the (LO + NLO QCD +
NLO EW)/ LO ratio (red line). In the second inset we show only the (LO
+ NLO EW)/LO ratio without NLO QCD contributions, but including the
PDF uncertainties for the numerator in order to enlighten the
qualitative difference for the EW corrections in the cases with
(right) and without (left) photons in the initial state.  As can be
seen in the left plots, the effect of Sudakov logarithms is very
large; for $p_T(Z)>5~$TeV the NLO EW corrections are $ \sim - 80\%$ of
the LO, while for $m(ZW^-)>8~$TeV they are $ \sim - 20\%$. The origin
of the huge $K$-factor in the QCD corrections has already been studied
in the literature ~\cite{Frixione:1992pj,Frixione:1993yp}. At LO a
hard $Z$ has to recoil against a hard $W^-$, while at NLO QCD the
dominant kinematic configuration is given by a hard $Z$ recoiling
against a hard jet and a soft $W$. In the case of quark radiation,
this kinematical configurations involve corrections $\sim \alpha_s
\log^2(p_T(Z)/m_W)$ that are further enhanced by the large gluon PDF
luminosity at the 100 TeV collider.~\footnote{Similar arguments are
  present for the $p_T(t\bar t)$ in NLO QCD corrections in $t \bar t
  V$ production and are discussed in some details in section
  \ref{sec:ttv-davide} of this report.}  A similar dynamic is present
also in the photon-initiated corrections (left plots), where a
correction $\sim \alpha \log^2(p_T(Z)/m_W)$ is present for the same
reason ~\cite{Baglio:2013toa}. However, on top of that, the photon in
the initial state can also directly couple to the $W$ boson
originating new $t$-channel configurations, which on the contrary are
not present in NLO QCD. This effect compensates the suppression due to
the $\alpha$ coupling and explains also why photon-induced
contributions, at variance to NLO QCD corrections, are large and
strongly depend also on the $m(ZW^-)$ cut; if no rapidity cuts are
applied $t$-channel configurations are much less suppressed at high
invariant masses w.r.t $s$-channel ones. These photon-induced
contributions strongly depend on the PDF set employed and, in the case
of the NNPDF 2.3QED used here, they have large uncertainties. Moreover
it is clear that a possible jet-veto, as in LHC analyses, would not
only decrease the NLO QCD $K$-factor and its dependence on $p_T(Z)$,
but it would also strongly suppress the large photon-initiated
contribution.
\begin{figure}[t]
\begin{center}
\includegraphics[width=0.48\textwidth]{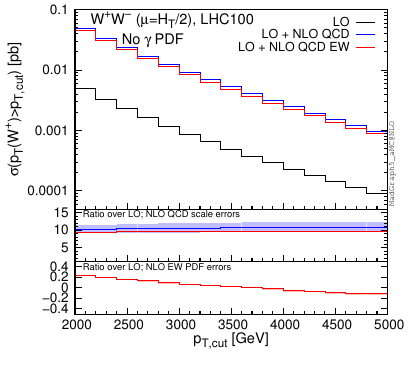}\hspace{10pt}
\includegraphics[width=0.48\textwidth]{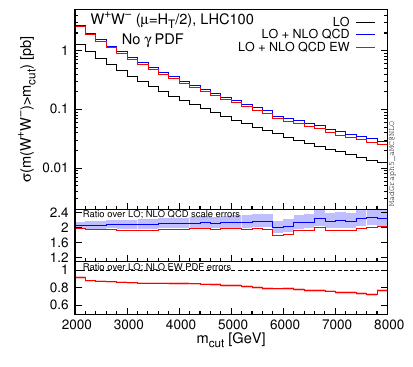}
\end{center}
\caption{Cumulative distributions for $W^+W^-$ production at NLO QCD
  and NLO EW accuracy. Both plots do not include contribution from
  photons in the initial state.  The left plot shows the dependence of
  the total cross sections on a cut on the $p_T(W^+)$, the right one
  on a cut on the $m(W+W^-)$.  See text for further details.}
\label{fig:wwew}
\end{figure}

In figure \ref{fig:wwew} we show similar cumulative distributions for
$W^+W^-$ production. In this case we show only results with the photon
PDF set equal to zero.  However, $W^+W^-$ production receives
contribution from initial-state photons already at LO via the $\gamma
\gamma$ initial states and their impact on $m(W^-W^-)$ distributions
is discussed in section \ref{sec:pdf_photon} and shown in figure
\ref{fig:ww-integrated}.  As can be seen in figure \ref{fig:wwew} NLO
QCD corrections shows the typical behaviour of $V V '$ production,
with a large dependence on the $p_T$ of the vector boson. The NLO EW
corrections involve very large Sudakov logarithms that are $ \sim -
120\%$ of the LO at $p_T(W^+)>5~$TeV and thus they have to be
resummed.

In the case of $ZZ$ production, which we do not show here, NLO QCD and
NLO EW corrections are qualitatively similar to the $ZW^\pm$ and $W^+
W^-$ production. However, since the photon cannot directly couple to
the $Z$ boson, no new $t$-channel is created for $\gamma q$ initial
state and, as consequence, their contribution is relatively much
smaller w.r.t the case of $ZW^\pm$ and $W^+ W^-$ production.
 
\begin{figure}[t]
\begin{center}
\includegraphics[width=0.48\textwidth]{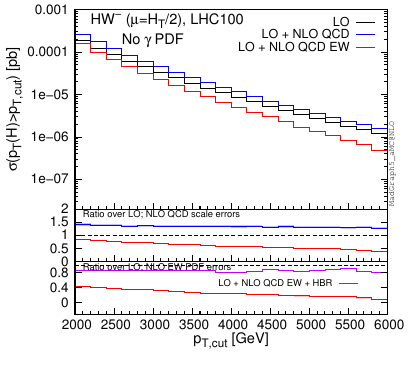}\hspace{10pt}
\includegraphics[width=0.48\textwidth]{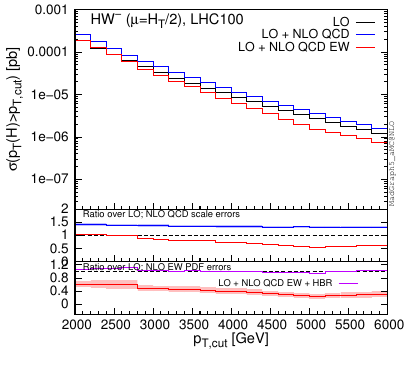}
\includegraphics[width=0.48\textwidth]{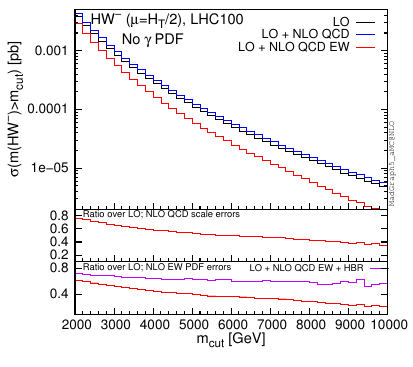}\hspace{10pt}
\includegraphics[width=0.48\textwidth]{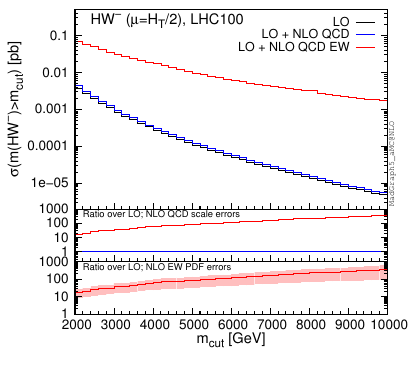}
\end{center}
\caption{Cumulative distributions for $HW^-$ production at NLO QCD and
  NLO EW accuracy. The upper plots show the dependence of the total
  cross sections on a cut on the $p_T(H)$, the lower ones on a cut on
  the $m(HW^-)$.  The plots on the left do not include contribution
  from photons in the initial state; they are included in the right
  plots. See text for further details.}

\label{fig:hwmew}
\end{figure}

We now turn to the case of $HV$ production.  One loop EW corrections
to $HV$ production have been computed in Ref.~\cite{Ciccolini:2003jy}
for on-shell $V$ and in Ref.~\cite{Denner:2011id} including the
off-shell decay of the vector boson.  Here, we show predictions at NLO
QCD and EW accuracy, taking into account the contribution of
initial-state photons and evaluating their large PDF uncertainties.

Figure \ref{fig:hwmew} is analogous to figure \ref{fig:Zwmew} and
displays the corresponding quantities for $HW^-$ production (the
$HW^+$ case is qualitatively identical).  In the second insets the
purple line, which is not present only in the bottom-right plot, is
the ratio (NLO EW + HBR)/LO where with HBR (Heavy-Boson-Radiation) we
denote the emission of an extra Heavy-Boson. From the comparison with
the red lines in the same insets we can notice a partial cancellation
of the effects due to the Sudakov logarithms, which also in this case
are very large: $ \sim - 100\%$ of the LO for $p_T(H)>6~$TeV and $
\sim - 80\%$ for $m(HW^-)>10~$TeV.

At variance with the $ZW^{-}$ case, neither NLO QCD and
photon-initiated contributions in NLO EW corrections contain terms
proportional to $\log^2(p_T(H)/m_W)$.  \footnote{The reason is that
  $Hj$ production is not possible at the tree-level. Thus, in the real
  quark radiation, the limit of a $W$ collinear to a final-state jet
  cannot be decomposed into $Hj$ production times an integrated
  $q\rightarrow q' W$ splitting that leads to a $\log^2(p_T(H)/m_W)$
  enhancement. } However, initial-state photons can couple directly to
the $W$ and open $t$-channel configuration for the $HW^-$ pair.  Since
at LO no $t$-channel diagrams are present at all, the LO contribution
is much more suppressed at high $m(HW^-)$ w.r.t. the NLO EW, to the
point that NLO EW corrections are $\sim$ 400 times larger than the LO
for $m(HW^-)>10~$TeV.  It is worth to notice that this estimate
strongly depends on the PDF set used and on possible additional cuts.
For instance, we explicitly verified that by simply requiring
$|\eta(H)|,|\eta(W^-)|< 4$ the NLO EW $K$-factor for $m(HW^-)>10~$TeV
is reduced from $\sim$ 400 to $\sim$ 10. A possible additional
jet-veto would further suppress the photon-induced contribution.

The case of $HZ$ production is similar to $HW^-$ production, but
photons cannot couple directly to the $Z$ and consequently, without
new $t$-channel production channels, no large enhancement from photon
induced processes is present for large invariant masses.

\subsection{V $+$ jets}\label{sec:vjets}
The production of a vector boson $V$ ($V=Z,W,\gamma$) in association
with jets is a process of great interest at hadron colliders and
precise theoretical predictions for $V+$multijets are mandatory. In
the literature, the one loop weak corrections to $V+1$~jet ($V=Z$,
$\gamma$) have been computed in
Refs~\cite{Maina:2004rb,Kuhn:2005gv,Kuhn:2005az}, while the full NLO
EW corrections have been computed for the processes
$W+1$~jet~\cite{Kuhn:2007cv}, monojet~\cite{Denner:2012ts} and
dilepton+jets production~\cite{Denner:2009gj,Denner:2011vu}. Besides
the exact calculations, the $\mathcal{O}(\alpha)$ corrections to
$V+1$~jet in the Sudakov approximation have been computed in
Refs.~\cite{Kuhn:2004em,Kuhn:2007qc,Kuhn:2005gv} by means of the
algorithm of Refs.~\cite{Denner:2000jv,Denner:2001gw} and in
Ref.~\cite{Becher:2013zua} in the SCET framework~\cite{Chiu:2008vv},
while in Refs.~\cite{Chiesa:2013yma,Mishra:2013una,Campbell:2013qaa}
the phenomenological impact of the $\mathcal{O}(\alpha)$ corrections
to $Z+2/3$~jets in the Sudakov limit has been investigated in the
context of the direct searches for New Physics at the LHC and at
higher energy future colliders.  More recently, the exact NLO EW and
QCD corrections to the processes $Z(\to
l^+l^-)+2$~jets~\cite{Denner:2014ina} and $W+n$~jets
($n=1,2,3$)~\cite{Kallweit:2014xda} have been computed by means of the
automated tools {\sc{ RECOLA}}~\cite{Actis:2012qn} and {\sc{
    Munich/Sherpa+OpenLoops}}~\cite{Cascioli:2011va}, respectively.
The full $\mathcal{O}(\alpha)$ to $W^++2$~jets have also been computed
in Ref.~\cite{Chiesa:2015mya} in the {\sc{ GoSam+MadDipole}}
framework. In Ref.~\cite{Kallweit:2015dum} the NLO QCD and EW
corrections to $Z/W+0,1,2$~jets including the effect of off-shell
vector boson decays and multijet merging have been computed.

In this section we study the phenomenological impact of the
$\mathcal{O}(\alpha)$ corrections on some distributions of interest
for the production of a vector boson in association with up to three
jets. We work in the $G_{\mu}$ scheme with input parameters:
\begin{equation}
\label{eq:vjinput}
G_{\mu}= 1.16637\, 10^{-5} \; {\rm GeV}^{-2}, \qquad M_W =80.385 \; {\rm GeV}, \qquad M_Z = 91.1876 \;  {\rm GeV} .
\end{equation}                            
We use the PDF set NNPDF2.3QED with factorization and renormalization
scale set to:
\begin{equation}
\label{eq:vjinput1}
\mu = \frac{H'_T}{2}, \qquad H'_T = \sum_{j=1}^{N {\rm jets}} p_{T,\,j} +p_{T,\,\gamma} +\sqrt{M_V^2 + p_{T,\,V}^2},
\end{equation}                            
where $p_{T,\,\gamma}$ stands for the transverse momentum of the
photon in the real radiation contribution (for the exact
$\mathcal{O}(\alpha)$ predictions). In order to evaluate the NLO EW
corrections we set the remaining as follows:
\begin{equation}
\label{eq:vjinput2}
M_H = 126 \; {\rm GeV} , \qquad M_{\rm top} = 173.2 \; {\rm GeV},
\end{equation}                            
while all the light fermions are massless. We consider the following
set of cuts:
\begin{equation}
\label{eq:vjcutj}
p_{T,\,j} \ge 300 \; {\rm GeV}, \qquad | \eta_j| \le 4.5, 
\end{equation}                            
where the jets are selected according to the anti-$k_T$
algorithm~\cite{Cacciari:2008gp} with $R$ separation $0.4$ for the
exact $\mathcal{O}(\alpha)$ predictions, while for the calculation in
the Sudakov approximation we simply require $\Delta R_{\rm min}(jj)
\ge 0.4$, as the number of partons is fixed and no real corrections
are included. The results for $\gamma+1,2,3$~jets have been obtained
by imposing the additional cuts on the photon:
\begin{equation}
\label{eq:vjcutph}
p_{T,\,\gamma} \ge 300 \; {\rm GeV}, \qquad | \eta_{\gamma}| \le 4.5 , \qquad 
\Delta R (j-\gamma) > 0.4 .
\end{equation}                            
No cuts are applied on the massive vector bosons that are treated as
stable particles.

\begin{figure}[t]
\begin{center}
\includegraphics[width=0.48\textwidth]{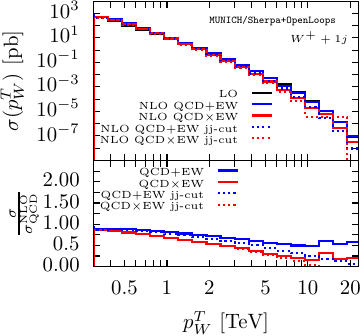}\hspace{10pt}
\includegraphics[width=0.48\textwidth]{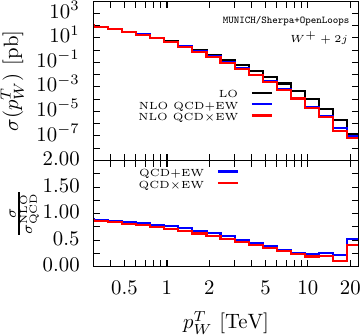}
\end{center}
\caption{Integrated $p_{T\;W}$ distribution for $W^++1$~jet (left
  panel) and $W^++2$~jet (right panel) at the FCC at $100$~TeV. LO
  predictions (solid black lines) correspond to the leading ${\cal
    O}(\alpha_S \alpha)$ and ${\cal O}(\alpha_S^2 \alpha)$ tree level
  contributions to $W^++1$~jet and $W^++2$~jets, respectively. The
  predictions including the full NLO EW and QCD corrections ({\sc{
      Munich/Sherpa+OpenLoops}}) according to the additive and to the
  multiplicative prescriptions correspond to the solid red and blue
  lines, respectively.  The results for $W^++1$~jet after imposing the
  veto on the dijet-like configuration correspond to dashed lines in
  the left plot. Lower panels: effect of the full NLO EW and QCD
  corrections in both the additive and the multiplicative
  prescriptions normalized to the full one loop QCD corrections.}
\label{fig:vjetabsptw}
\end{figure}
\begin{figure}[t]
\begin{center}
\includegraphics[width=0.48\textwidth]{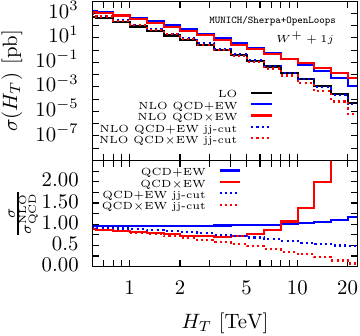}\hspace{10pt}
\includegraphics[width=0.48\textwidth]{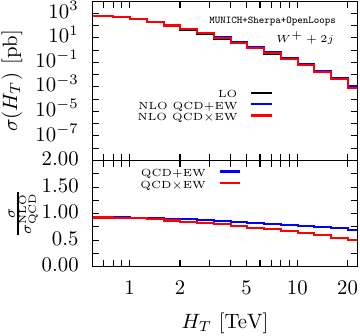}
\end{center}
\caption{Integrated $H_{T}$ distribution for $W+1$~jet (left panel)
  and $W+2$~jet (right panel) at the FCC at $100$~TeV. Same notations
  and conventions as in Fig.~[\ref{fig:vjetabsptw}].}
\label{fig:vjetabsht}
\end{figure}
\begin{figure}[t]
\begin{center}
\includegraphics[width=0.48\textwidth]{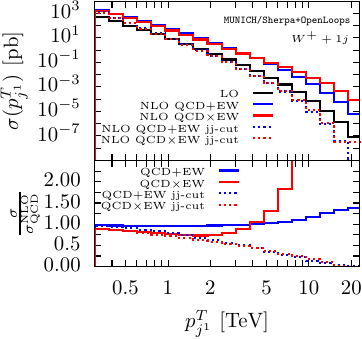}\hspace{10pt}
\includegraphics[width=0.48\textwidth]{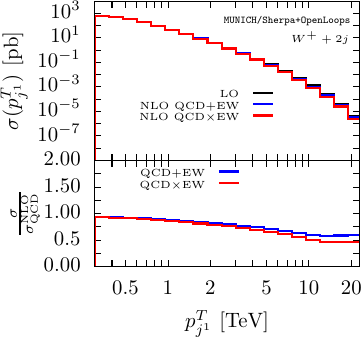}
\end{center}
\caption{Integrated $p_{T\,j}$ distribution for the leading jet for
  the processes $W+1$~jet (left panel) and $W+2$~jet (right panel) at
  the FCC at $100$~TeV. Same notations and conventions as in
  Fig.~[\ref{fig:vjetabsptw}].}
\label{fig:vjetabspj}
\end{figure}
\noindent
We collect in Figs.~\ref{fig:vjetabsptw}-\ref{fig:vjetabspj} the
theoretical predictions for the production of a $W^+$ boson in
association with one and two jets at the FCC at 100~TeV both at LO
accuracy and including the effect of the full one NLO EW and QCD
corrections computed by means of the program {\sc{
    OpenLoops}}~\cite{hepforge} interfaced with {\sc{
    Sherpa}}~\cite{Gleisberg:2007md,Gleisberg:2008ta} and {\sc{
    MUNICH}}~\cite{munich}. In particular, we focus on the following
distributions: $W$ boson $p_T$, leading jet $p_T$ and the total
transverse activity defined as $H_T=\sum_{\rm jets} p_{T\;j} +
p_{T\;V}$. NLO results are obtained by combining QCD and EW results
according to the additive prescription:
\begin{equation}
\label{eq:vjopenloop1}
\sigma^{ \rm NLO}_{ \rm QCD + EW}= \sigma^{ \rm LO } + \delta \sigma_{ \rm QCD } +\delta \sigma_{ \rm EW},
\end{equation}                            
as well as in the factorized prescription:
\begin{equation}
\label{eq:vjopenloop2}
\sigma^{\rm NLO}_{ \rm QCD \times EW}=  
\sigma^{\rm NLO}_{ \rm QCD} \Big( 1 + \frac{\sigma^{\rm NLO}_{ \rm EW}}{\sigma^{\rm LO}} \Big) =
\sigma^{\rm NLO}_{ \rm EW } \Big( 1 + \frac{\sigma^{\rm NLO}_{ \rm QCD}}{\sigma^{\rm LO}} \Big),
\end{equation}                            
taking the difference of the two results as an estimate of the
uncertainty related to the missing higher order terms.  The
corrections are shown normalized to the QCD NLO results: this
corresponds to the usual definition of $\delta_{\rm EW}$ for the
factorized corrections.

In figures~\ref{fig:vjetabsptw}-\ref{fig:vjetabspj} we consider only
the leading $\mathcal{O}(\alpha_S \alpha)$ and $\mathcal{O}(\alpha_S^2
\alpha)$ terms contributing to $W+1$~jet and $W+2$~jets, respectively,
while the NLO QCD and the NLO EW corrections to the process of
$\mathcal{O}(\alpha_S^m \alpha^n)$ are defined as the sum of the one
loop virtual and real contribution of $\mathcal{O}(\alpha_S^{m+1}
\alpha^n)$ and $\mathcal{O}(\alpha_S^{m} \alpha^{n+1})$,
respectively. In particular, in the case of $W+1$~jets, this implies
that the real corrections receive contributions from the interference
of amplitudes of $\mathcal{O}(\alpha^{3/2} )$ with ones of
$\mathcal{O}(\alpha_S \alpha^{1/2})$: at this order these mixed
interference terms are finite because of color flow, but in general do
not vanish in the presence of identical quarks.

As can be seen from Figs.~\ref{fig:vjetabsptw}-\ref{fig:vjetabspj} ,
the one loop corrections to $W+2$~jets are negative and large,
reaching the order of $-50$~\% in the tails of the distributions under
consideration. The same behaviour can be observed in the NLO
corrections to $W+1$~jet for the $p_{T\;W}$ distribution, where the
corrections are however larger than in the case of $W+2$~jets and can
become of order $-100$~\% for $p_{T\;W} \simeq 20$~TeV. The picture is
different if we consider the NLO predictions for the leading
$p_{T\;j}$ and the $H_T$ distributions for $W+1$~jets: in fact, the
corrections become positive for $p_{T\;j} \simeq 5$~TeV and $H_{T}
\simeq 9$~TeV, respectively.  The increase in the cross section
results from a new kinematical configuration which is available for
$W+2$~jets and has no LO counterpart: namely, the one where the
leading jet $p_T$ is balanced by a second hard jet and the vector
bosons tend to be soft. This part of the cross section can be
separated by applying a veto on the events with an angular separation
between the two jets larger then $3 \pi /4$ (jj-cut in the plots).
Once the veto on the dijet-like configurations is imposed on the
corrections to $W+1$~jet, the effects on the leading jet $p_T$ and on
the $H_T$ distributions become similar to the ones on the $p_{T\;W}$
distribution.

\begin{figure}[t]
\begin{center}
\includegraphics[width=0.48\textwidth]{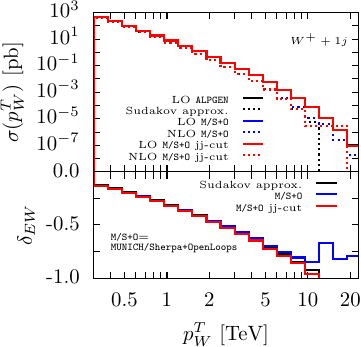}\hspace{10pt}
\includegraphics[width=0.48\textwidth]{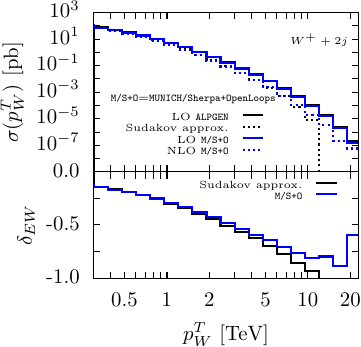}
\end{center}
\caption{Integrated $p_{T\; W}$ distribution for $W+1$~jet (left
  panel) and $W+2$~jet (right panel) at the FCC at
  $100$~TeV. Comparison between the the exact $\mathcal{O}(\alpha)$
  predictions ({\sc{ Munich/Sherpa+OpenLoops}}) and the ones computed
  in the Sudakov approximation ({\sc{ALPGEN}}). For $W+1$~jets, the
  {\sc{ Munich/Sherpa+OpenLoops}} predictions are shown both with and
  without the veto on the dijet-like configurations.}
\label{fig:vjetcompptw}
\end{figure}

\begin{figure}[t]
\begin{center}
\includegraphics[width=0.48\textwidth]{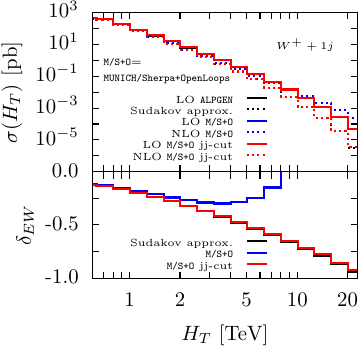}\hspace{10pt}
\includegraphics[width=0.48\textwidth]{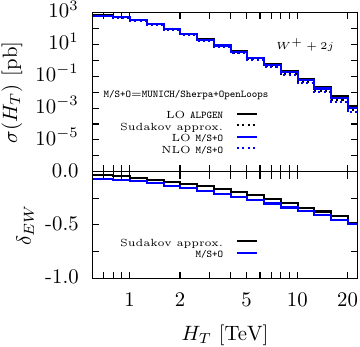} \\
\includegraphics[width=0.48\textwidth]{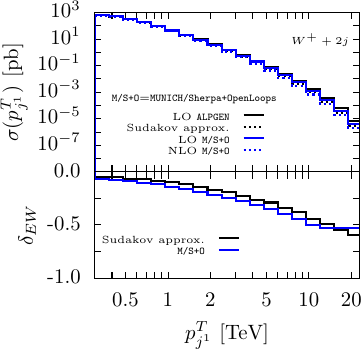}
\end{center}
\caption{Integrated $H_{T}$ distribution for $W+1$~jet (left panel)
  and $W+2$~jet (right panel) at the FCC at $100$~TeV. Comparison
  between the the exact $\mathcal{O}(\alpha)$ predictions ({\sc{
      Munich/Sherpa+OpenLoops}}) and the ones computed in the Sudakov
  approximation ({\sc{ALPGEN}}). For $W+1$~jets, the {\sc{
      Munich/Sherpa+OpenLoops}} predictions are shown both with and
  without the veto on the dijet-like configurations. Lower plot:
  Comparison between the the exact $\mathcal{O}(\alpha)$ predictions
  ({\sc{ Munich/Sherpa+OpenLoops}}) and the ones computed in the
  Sudakov approximation ({\sc{ALPGEN}}) for the leading jet $p_T$
  distribution.}
\label{fig:vjetcompht}
\end{figure}

In Refs.~\cite{Denner:2000jv,Denner:2001gw} a process independent
algorithm for the computation of NLO EW corrections in the Sudakov
approximation has been developed.  According to the algorithm, the
$\mathcal{O}(\alpha)$ corrections to a generic process involving $N$
external particles of flavour $i_1, \cdots, i_N$ in the high energy
limit factorize as follows:
\begin{equation}
\label{eq:dlsl}
\delta \mathcal{M}_{i_{1}\cdots i_{n}}^{NLL} \Bigg|_{\rm Sudakov}=
\sum_{k=1}^{N}\sum_{l>k}\delta_{kl}^{DL}\mathcal{M}_{i_{1}\cdots j_{k}\cdots j_{l}\cdots i_{n}}^{LO} +
\sum_{k=1}^{N}\delta_{k}^{SL}\mathcal{M}_{i_{1}\cdots j_{k}\cdots i_{n}}^{LO}+
\delta^{PR}\mathcal{M}_{i_{1}\cdots i_{n}}^{NLL}. 
\end{equation}
In Eq.~(\ref{eq:dlsl}), the radiator functions $\delta_{kl}^{DL}$ and
$\delta_{k}^{SL}$ contain the Sudakov double and single logarithmic
contributions, respectively.  They depend only on the flavour and on
the kinematics of the external particles. These terms multiply leading
order matrix elements that are obtained from the one of the process
$i_1, \cdots, i_N$ under $SU(2)$ transformations of pair or single
external legs, $j_k$ being in Eq.~(\ref{eq:dlsl}) the $SU(2)$
transformed of the particle $i_k$.  The last term in
Eq.~(\ref{eq:dlsl}) comes from parameter renormalization:
\begin{equation}
\label{eq:slpr}
\delta^{PR}\mathcal{M}_{i_{1}\cdots i_{n}}^{NLL}=
\delta e\frac{\delta\mathcal{M}_{i_{1}\cdots i_{n}}^{LO}}{\delta e}+
\delta c_{W}\frac{\delta\mathcal{M}_{i_{1}\cdots i_{n}}^{LO}}{\delta c_{W}}+
\delta h_{t}\frac{\delta\mathcal{M}_{i_{1}\cdots i_{n}}^{LO}}{\delta h_{t}}+
\delta h_{H}\frac{\delta\mathcal{M}_{i_{1}\cdots i_{n}}^{LO}}{\delta h_{H}},
\end{equation}
where $h_t=m_t/M_W$, $h_H=M_H^2/M_W^2$ and $c_W=M_W/M_Z$.  In
Ref.~\cite{Chiesa:2013yma}, the algorithm of
Refs.~\cite{Denner:2000jv,Denner:2001gw} has been implemented in the
{\sc{ALPGEN}}~\cite{Mangano:2002ea} event generator: the analytic
expressions of the process-independent radiator functions have been
coded and all the required leading order matrix elements are computed
numerically by means of the ALPHA
algorithm~\cite{Caravaglios:1995cd}. According to
Refs.~\cite{Denner:2000jv,Denner:2001gw} the purely weak part of the
corrections can be isolated by setting to $M_W$ the value of the
photon mass in the virtual corrections.  In the following we consider
only this part of the correction neglecting the QED part: this in
particular means that no real radiation contribution is included in
the approximated results. The results for the weak corrections in the
Sudakov limit computed by means of the modified version of
{\sc{ALPGEN}} described above are compared to the exact
$\mathcal{O}(\alpha)$ predictions by {\sc{ Munich/Sherpa+OpenLoops}}
in Figs.~\ref{fig:vjetcompptw}-\ref{fig:vjetcompht}: as can be seen,
the approximated results are in good agreement with the exact ones for
$W+2$~jets and $W+1$~jet once the veto on the dijet-like
configurations is imposed. The differences between the predictions by
{\sc{ALPGEN}} and the ones by {\sc{ Munich/Sherpa+OpenLoops}} for
$W+1$~jet when the veto is not imposed come from the fact that the
approximated results do not include real corrections and in particular
no mixed interference terms: these terms could however be included as
separate tree level-like contributions regardless of the Sudakov
approximation. We point out, however, that in order to obtain more
reliable predictions, especially at high jet $p_T$, it is important to
include EW$\times$QCD interference effects (which are neglected
throughout in this section) and to merge NLO QCD$+$EW predictions with
different jet multiplicities~\cite{Kallweit:2015dum}.

\begin{figure}[t]
\begin{center}
\includegraphics[width=0.48\textwidth]{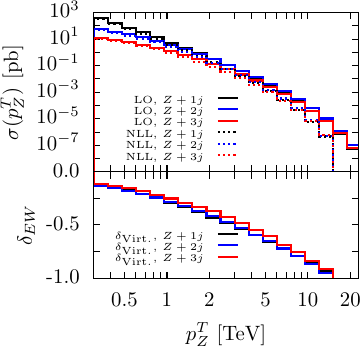}\hspace{10pt}
\includegraphics[width=0.48\textwidth]{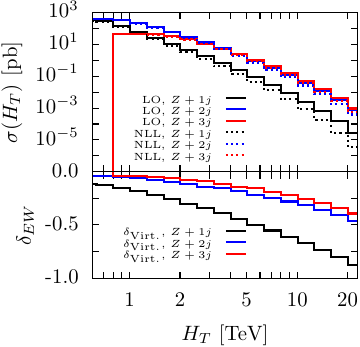} \\
\includegraphics[width=0.48\textwidth]{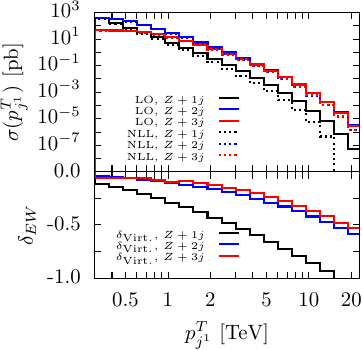}
\end{center}
\caption{Upper panels: integrated $p_T^Z$ (upper left plot), $H_T$
  (upper right plot) and leading jet $p_T$ (lower plot) distributions
  for the processes $Z+1$, 2 and 3~jets at LO and approximated NLO
  accuracy (solid and dashed lines, respectively) at the FCC at
  100~TeV. Lower panels: relative corrections $\delta_{\rm EW}$.}
\label{fig:vjetreszz}
\end{figure}

\begin{figure}[t]
\begin{center}
\includegraphics[width=0.48\textwidth]{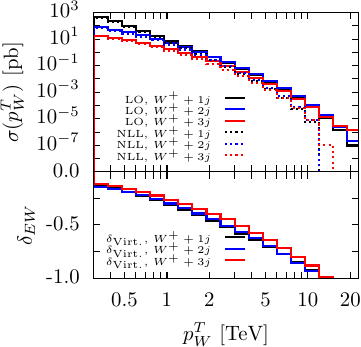}\hspace{10pt}
\includegraphics[width=0.48\textwidth]{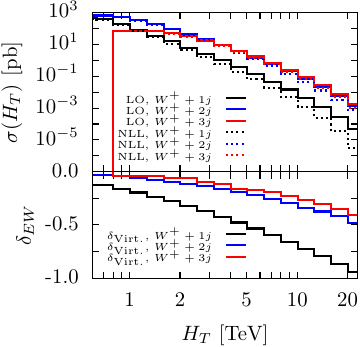} \\
\includegraphics[width=0.48\textwidth]{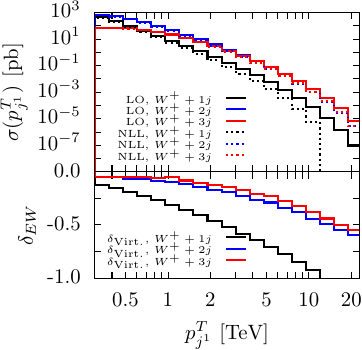}
\end{center}
\caption{Upper panels: integrated $p_T^W$ (upper left plot), $H_T$
  (upper right plot) and leading jet $p_T$ (lower plot) distributions
  for the processes $W^++1$, 2 and 3~jets at LO and approximated NLO
  accuracy (solid and dashed lines, respectively) at the FCC at
  100~TeV. Lower panels: relative corrections $\delta_{\rm EW}$. }
\label{fig:vjetresww}
\end{figure}

\begin{figure}[t]
\begin{center}
\includegraphics[width=0.48\textwidth]{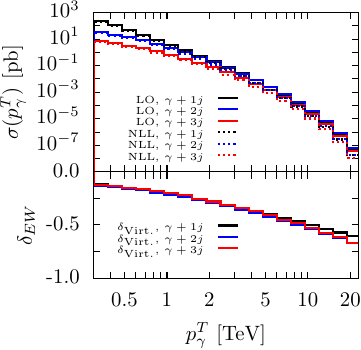}\hspace{10pt}
\includegraphics[width=0.48\textwidth]{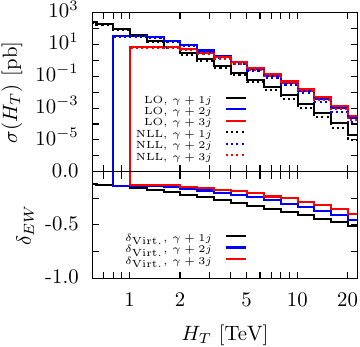} \\
\includegraphics[width=0.48\textwidth]{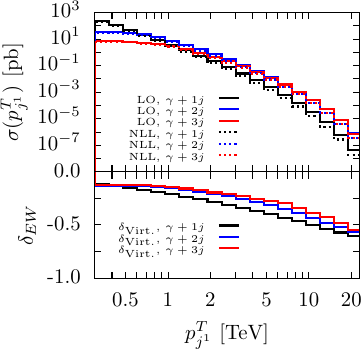}
\end{center}
\caption{Upper panels: integrated $p_T^{\gamma}$ (upper left plot),
  $H_T$ (upper right plot) and leading jet $p_T$ (lower plot)
  distributions for the processes $\gamma+1$, 2 and 3~jets at LO and
  approximated NLO accuracy (solid and dashed lines, respectively) at
  the FCC at 100~TeV. Lower panels: relative corrections $\delta_{\rm
    EW}$. }
\label{fig:vjetresaa}
\end{figure}

The Sudakov approximation and the exact $\mathcal{O}(\alpha)$
calculation basically agree for $W+$~jets. Having assessed the
validity of the logarithmic approximation, in
Figs.~\ref{fig:vjetreszz}-\ref{fig:vjetresaa} we show the predictions
for the NLO EW corrections in the Sudakov limit to the production of a
vector boson $V$ ($Z,$ $W^+$, $\gamma$) in association with 1, 2 and
3~jets. Looking at the $p_{T\;V}$ distributions, we can notice that
the corrections in the high energy limit are negative, large and
independent of the jet multiplicity. Conversely, if we consider the
$p_{T\;j}$ and $H_T$ distributions for $Z/W+n$~jets, the corrections
for $n=2$ and $n=3$ are similar, while the ones for $n=1$ turn out to
be larger. This is a consequence of the event selection in
eq.~(\ref{eq:vjcutj}) where no cuts are imposed on the massive vector
bosons: as a result, while for $Z/W+1$~jet the high $p_{T\;j}$ or
$H_T$ region corresponds to the kinematical configurations where the
vector boson is hard, this is no longer the case for high jet
multiplicities. On the contrary, when the same cuts are imposed on
both the vector boson and on the jets, as in the case of
$\gamma+$~jets production, the EW corrections are in general weakly
dependent on the jet multiplicity for all the observables under
consideration. At the FCC the size of the NLO EW corrections to
$V+$~multijet production turns out to be large, reaching the order of
$-100$~\% for $p_{T\;V}$ around 10~TeV (with the exception of
$p_{T\;V}$ which receives smaller corrections): this is an indication
that the NLO approximation is not reliable anymore in these regions of
phase space and higher order effects should be included in theoretical
predictions.
\begin{figure}[t]
\begin{center}
\includegraphics[width=0.48\textwidth]{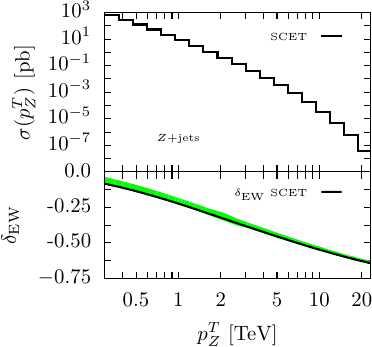}\hspace{10pt}
\includegraphics[width=0.48\textwidth]{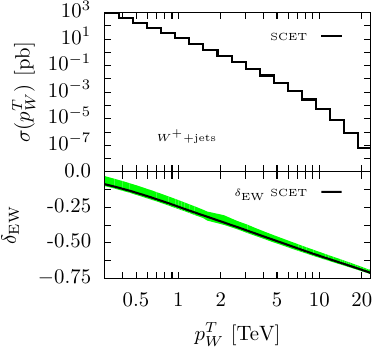} \\
\includegraphics[width=0.48\textwidth]{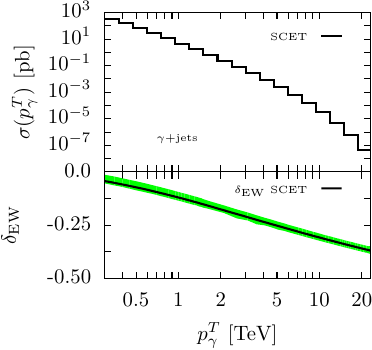}
\end{center}
\caption{Predictions within the SCET framework of resummed leading EW
  corrections to the integrated inclusive $Z$ boson $p_T$ (upper left
  plot), $W$ boson $p_T$ (upper right plot) and $p_T^{\gamma}$ (lower
  plot) at the FCC at 100~TeV. The lower panels display the relative
  corrections $\delta_{\rm EW}$. The bands have been obtained by
  varying the EW matching scale.}
\label{fig:vjetptscet}
\end{figure}
The EW corrections to the vector boson $p_T$ distribution are shown 
in Fig.~\ref{fig:vjetptscet}, once the leading logarithmic terms have been 
resummed in the framework of SCET~\cite{Becher:2013zua,Chiu:2008vv}: 
the resummed corrections are smaller than the ones 
computed at fixed order, even though they remain large, becoming of the order 
of $-50$~\% for $p_{T\;V} \simeq 20$~TeV.

\begin{figure}[t]
\begin{center}
\includegraphics[width=0.48\textwidth]{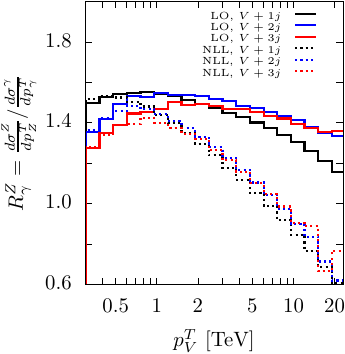}\hspace{10pt}
\includegraphics[width=0.48\textwidth]{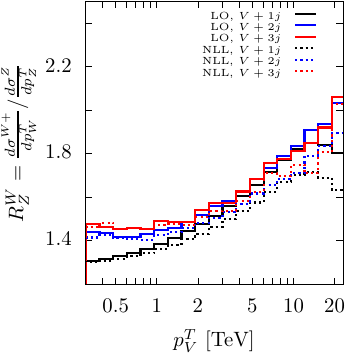}
\end{center}
\caption{ Differential distributions of the ratios $R^Z_{\gamma}$
  (left plot) and $R^W_Z$ (right plot) for $V+1$, 2 and 3~jets at the
  FCC at 100~TeV. Solid lines and dotted lines correspond to the LO
  and the approximated NLO predictions, respectively. The results for
  the $R^Z_{\gamma}$ ratio have been obtained imposing the same cuts
  of eq.~(\ref{eq:vjcutph}) on both the photon and the $Z$ boson.}
\label{fig:vjetratio}
\end{figure}

We conclude this section studying the impact of the NLO EW corrections
on the ratio of differential distributions for $Z$, $W$ and $\gamma$
plus jets. More precisely, in Fig.~\ref{fig:vjetratio} we consider the
ratios:
\begin{equation}
\label{eq:vjratio}
R^Z_{\gamma} = \frac{d \sigma^Z }{ d p^T_{Z} } / \frac{d \sigma^{\gamma} }{ d p^T_{\gamma} } 
\qquad {\rm and } \qquad 
R^W_Z=\frac{d \sigma^{W+} }{ d p^T_{W} } / \frac{d \sigma^{Z} }{ d p^T_{Z} } \;. 
\end{equation}
for the three different jet multiplicities.  In
Figs.~\ref{fig:vjetreszz}-\ref{fig:vjetresaa} it is shown how the EW
corrections for $Z+$~multijets and $W+$~multijets are similar, while
the ones for $\gamma+$~multijets are smaller.  As a result, the
$R^W_Z$ ratio is basically unaffected by the EW corrections, while
this is not the case for the $R^Z_{\gamma}$ ratio, where the Sudakov
corrections change significantly the shape of the $R^Z_{\gamma}$
distribution. It is worth mentioning that a reliable prediction for
$R^Z_{\gamma}$ should also include NLO QCD corrections, as for small
transverse momenta mass effects are not negligible.

\subsection{Di-jets}\label{sec:dijet}
The electroweak contributions to di-jet production can be safely
classified according to the coupling constant power of ${\cal
  O}(\alpha_s^2 \alpha)$.  While the tree-level processes involving
two or more gluons are of ${\cal O}(\alpha_s^2)$, the processes with
four quarks can proceed through the exchange of electroweak gauge
bosons.  This implies that there are two classes of contributions at
${\cal O}(\alpha_s^2 \alpha)$: the one-loop virtual EW NLO corrections
on QCD tree-level processes of ${\cal O}(\alpha_s^2)$ (involving
tree-level diagrams with two as well as four quark legs) and the QCD
NLO corrections to the interferences between ${\cal O}(\alpha_s)$
tree-level diagrams with ${\cal O}(\alpha)$ tree-level diagrams (the
colour structure allows a non-zero contribution from this interference
only between u- and t-channel diagrams).  The former can be calculated
in a gauge invariant way separating the genuine weak corrections
(which involve $W$ and $Z$ exchanges in the loops and are the
interesting contributions at high energies) from the photonic
corrections. The complete calculation of the ${\cal O}(\alpha_s^2
\alpha)$ contributions, neglecting photonic corrections, have been
presented in Ref.~\cite{Dittmaier:2012kx}. Previous results have been
presented in
Refs.~\cite{Moretti:2005ut,Moretti:2006ea,Scharf:2009sp,Kuhn:2009nf}.
The investigation of the phenomenological impact of ${\cal
  O}(\alpha_s^2 \alpha)$ terms at the LHC and higher energy future
colliders have been presented in
Refs.~\cite{Dittmaier:2012kx,Dittmaier:2013hha,Mishra:2013una,Campbell:2013qaa},
for jet-jet invariant mass and jet transverse momentum distributions.
 
The ${\cal O}(\alpha_s^2 \alpha)$ interference terms are positive and
tend to partially cancel the negative effect of the virtual
corrections to the ${\cal O}(\alpha_s^2)$ LO contribution.  However,
as can be seen from Figs. [1-3] of Ref.~\cite{Mishra:2013una}, the
relevance of the tree-level interferences tends to decrease with the
increase of the collider energy, being of the order of 1\% for jet
transverse momenta of 3~TeV.  At variance with the interference terms,
the virtual corrections to the ${\cal O}(\alpha_s^2)$ LO
contributions, instead, are negative and grow up in size at the level
of 10\%.  It is worth remarking that the corrections on the di-jet
invariant mass are smaller with respect to the jet transverse momenta,
due to the fact that the requirement of large jet transverse momenta
guarantees that all invariants are much larger than the weak vector
boson masses ({\em i.e.} Sudakov logs are dominating), while the high
invariant mass region is dominated by the forward region, where the
$t$ and $u$ invariants remain small. This feature is not present for
event selections with tight cuts on the jet transverse momenta. In
this subsection we present results for di-jet invariant mass
$(m_{jj})$ and leading $p_T$ jet distributions $(p_T^{j1})$,
integrated from the lower edge to the kinematical limit, as obtained
with the modified version of {\sc{ALPGEN v2.14}} to include NLO EW
corrections to pure QCD LO contributions with logarithmic
accuracy~\cite{Chiesa:2013yma}, up to single logs.  While the
approximation is not expected to be fully reliable for the di-jet
invariant mass distribution, as discussed above, it should give
reliable predictions for $p_T^{j1}$, where the condition for the
validity of the Sudakov approximation is fully satisfied.  The
numerical results have been obtained at parton level, with the
following set of cuts:
\begin{eqnarray}
p_T^j \geq 25 \, {\rm GeV} \, , \qquad \qquad |\eta_j| \leq 2.5 \, \qquad \qquad 
\Delta R_{jj} \geq 0.6.
\end{eqnarray} 
The running parameters have been kept the default ones of {\sc{ALPGEN
    v2.14}}, in particular the renormalization and factorization
scales are $\mu_F = \mu_R = \sqrt{\left({p_T^{j_1}}\right)^2 +
  \left({p_T^{j_2}}\right)^2}$.
\begin{figure}[t]
\begin{center}
\includegraphics[width=7cm]{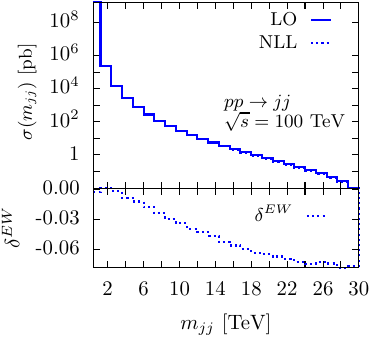}
\hspace{1cm}
\includegraphics[width=7cm]{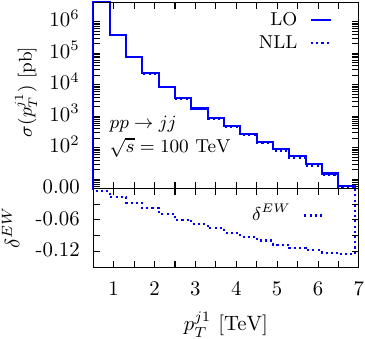}
\end{center}
\caption{Integrated invariant mass distribution (left panel) and
  Integrated leading jet $p_T$ distribution for the process $p p \to j
  j$ at $\sqrt{s} = 100$~TeV. The lower panels give the relative
  effects of the EW corrections (calculated with leading logarithmic
  accuracy, as described in the text) with respect to the LO order
  predictions.}
\label{fig:dijet}
\end{figure}
As can be seen, the estimated corrections are moderate (at the level
of about 10\%), even at the multi-TeV scales accessible at the
$\sqrt{s} = 100$~TeV collider.


\subsection{$t \bar t$, $t \bar t\, + $~jets and $t \bar t H$}\label{sec:tt-ttjets}
In this subsection we present exact NLO results for the EW corrections
to $t \bar t$ pair production and NLO Weak corrections to $t \bar t H$
production. Also, approximate results, using the logarithmic
approximation, for $t \bar t + n$~jets, with $n=0,1,2,3$ are given. As
for the case of dijet production, we have both ${\cal O}(\alpha_s^2)$
as well as ${\cal O}(\alpha_s \alpha)$ tree-level diagrams. The EW
corrections (or some subset of them) have been calculated in the
literature~\cite{Beenakker:1993yr,Kuhn:2005it,Moretti:2006nf,Bernreuther:2006vg,
  Kuhn:2006vh,Hollik:2007sw,Bernreuther:2008md} and have been included
recently in the Monte Carlo code~\cite{Campbell:2010ff}.

The numerical results for $t \bar t$ production at 100~TeV have been
obtained by means of the automatic code
{\sc{Madgraph5\_aMC@NLO}}~\cite{Alwall:2014hca} and a currently
private extension that allows to calculate NLO QCD and EW corrections
\cite{Frixione:2014qaa, Frixione:2015zaa}.  In
Fig.~\ref{fig:ttbar-MG5aMC} we present the integrated distributions of
the $t \bar t$ invariant mass in the range [8-30]~TeV (left panel) and
of the top quark transverse momentum in the range [4-17]~TeV (right
panel). The format of the plots is the same of those in
Fig.~\ref{fig:Zwmew}, which is described in the text. As can be seen
in the lower panels, the effect of NLO EW radiative corrections is
negative and moderate for the $t \bar t$ invariant mass, ranging from
$\sim - 10$\% for $M(t \bar t) > 10$~TeV to $\sim - 20$\% for $M(t
\bar t) > 30$~TeV. QCD corrections are also presented and display a
positive effect of the order of 60\%, almost flat over the entire
invariant mass range. The uncertainty estimate given by the scale
variation is at the 10\% level, dominating with respect to the PDF
uncertainty.

A more pronounced effects of the EW corrections is present on the transverse 
momentum distribution, where it ranges from $-30$\% for $p_T^{\rm min} = 4$~TeV to 
$-50$\% for $p_T^{\rm min} = 17$~TeV. In the same interval the QCD corrections range from 
a factor of almost 2.5 over the LO predictions to about $+50$\%, giving rise to 
large cancellation between the two kinds of corrections in the very large tail of the 
distribution. 

In these plots we did not include the effect from photon-initiated
processes. However, at ${\cal O}(\alpha_s \alpha)$ also the
contribution from the $\gamma g $ initial state is present
\cite{Hollik:2007sw}. We explicitly verified that, with NNPDF2.3QED,
this contribution would compensate the NLO EW corrections for $M(t
\bar t) > 20$~TeV and would relatively grow for even larger $M(t \bar
t)$ values.  However, the $\gamma g $ initial-state has huge PDF
uncertainties in this region $(\sim 100\%)$ and the growth is totally
given by the PDF luminosity and not by matrix-element enhancements. In
the case of cumulative top quark transverse momentum we observe a
milder effect from the $\gamma g $ initial state; it compensates the
NLO EW corrections only at the end of the explored range. Additional
plots, including effects due the photon PDF, can be found in
ref.~\cite{Pagani:2016caq}, where a detailed discussion on the
compensation of EW Sudakov logarithms and photon-induced contributions
is presented.

\begin{figure}[t]
\begin{center}
\includegraphics[width=0.48\textwidth]{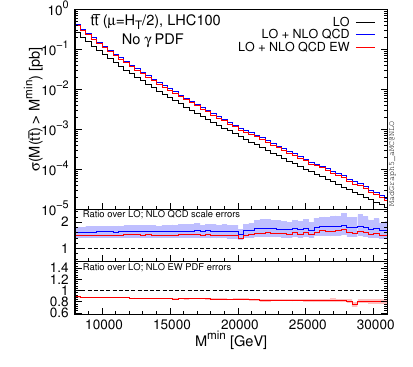}
\includegraphics[width=0.48\textwidth]{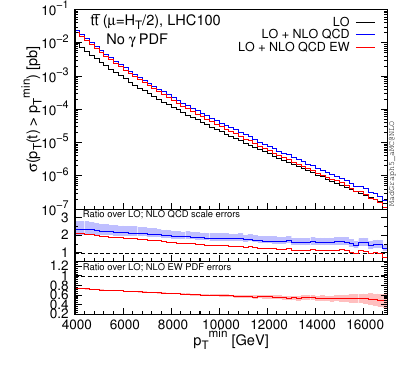}
\end{center}
\caption{Left: the cumulative distribution of the $t \bar t $
  invariant mass.  Right: the cumulative distribution of the $p_T(t )
  $.  as obtained with {\sc{Madgraph5\_aMC@NLO}}. The blue lines
  display the effects of NLO QCD corrections, while the red lines
  correspond to the predictions with NLO EW corrections on top of NLO
  QCD accurate distributions. In the lower panels, which display the
  relative effects of the NLO QCD and EW corrections, also the effects
  of scale variations and PDF uncertainties are considered.}
\label{fig:ttbar-MG5aMC}
\end{figure}

For the signature $t \bar t + n$~jets (with $n$ up to 3), the results
have been obtained with the upgraded version of {\sc{ALPGEN}}
mentioned in Subsection~\ref{sec:EW-tools} and briefly described in
Subsection~\ref{sec:vjets}. Since the logarithmic approximation has
been shown to largely overestimate the EW corrections for the $t \bar
t$ invariant mass~\cite{Campbell:2015vua}, when no cut on the
transverse momentum of the top quarks is imposed, we present only
results for the inclusive transverse momentum of the top quarks
(Fig.~\ref{fig:pt-top-alp}, left panel) and for the transverse
momentum of the $t \bar t$ pair (Fig.~\ref{fig:pt-top-alp}, right
panel), in the range [0-6]~TeV.  The numerical results, obtained with
{\sc{ALPGEN}}, are based on the default parameters of the version
{\sc{v2.14}} of the code, with a minimum transverse momentum threshold
of $50$~GeV both for top quarks and light partons. In addition, for
the light partons a cut of 5 units in the maximum pseudorapidity is
required, and a separation in $\Delta R$ of 0.7. The
factorization/renormalization scale is taken as $\mu_F = \mu_R =
\sqrt{{E_1}_{T}^2 + {E_2}_{T}^2 + \sum_i {p_i}_{T}^2}$, where ${E_i}_{T}^2$
is the transverse energy of the $i-th$ top quark.  By comparison of
the black dotted line of the left-lower panel of
Fig.~\ref{fig:pt-top-alp} with the red line of the right-lower panel
of Fig.~\ref{fig:ttbar-MG5aMC}, we can see a nice agreement between
the logarithmic approximation and the exact NLO calculation for
$n_{jets} = 0$. The effect of the EW corrections for higher parton
multiplicities are slightly smaller. For the $t \bar t$ pair
transverse momentum the effect of the corrections is almost the same
for all the studied multiplicities.
\begin{figure}[t]
\begin{center}
\includegraphics[width=0.48\textwidth]{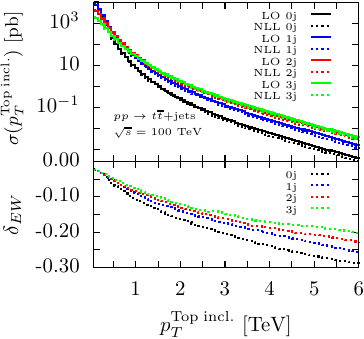}
\includegraphics[width=0.48\textwidth]{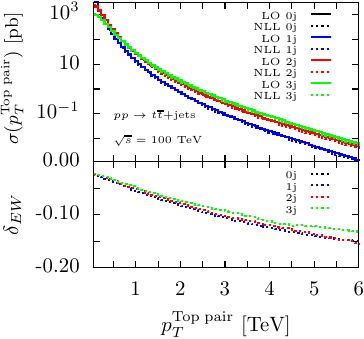}
\end{center}
\caption{Left: the distribution of the inclusive top quark transverse
  momentum for the final states $t \bar t + n$~jets, with $n=0,1,2,3$.
  Right: the distribution of the $t \bar t$ pair transverse momentum
  for the final states $t \bar t + n$~jets, with $n=0,1,2,3$.  The
  lower panels give the relative effects of the EW corrections
  (calculated with leading logarithmic accuracy, as described in the
  text) with respect to the LO order predictions.}
\label{fig:pt-top-alp}
\end{figure}

As a last comment, we observe that the EW corrections to $t \bar t$
and $t \bar t + $~jets are moderate even if larger than the ones for
dijet production.  This can be qualitatively understood because of the
presence of processes involving only gluons and due to the average
over flavours in dijet production with respect to $t \bar t$.

We also show NLO Weak differential corrections for the production of a
top-quark pair in association with the Higgs boson ($t \bar t H$). For
this process, NLO Weak and QED corrections are separately gauge
invariant and the former, which contain all the Sudakov logarithms,
have been calculated in \cite{Frixione:2014qaa}, from where plots in
Fig.~\ref{fig:ttbarHweak} have been directly taken. Results for NLO EW
(Weak+QED) corrections at 100 TeV for $t \bar{t}H$, $t \bar{t}Z$ and
$t \bar{t} W^\pm$ total cross sections con be found in
\cite{Frixione:2015zaa}. In the main panel of plots in
Fig.~\ref{fig:ttbarHweak} we show the NLO Weak contributions from each
partonic subprocess and their sum (in black) and also the contribution
from Heavy-Boson-Radiation (HBR), {\it i.e.}, $t \bar t HV$ with
$V=H,W^\pm,Z$. In the lower inset we display the ratios of the
quantities in the main panel with the LO prediction, using the
corresponding colors. The left and right plots show the $p_T(H)$ and
the $m(t \bar t H)$ distributions, respectively. Weak corrections
reach $\sim - 10 \%$ level in the range explored and are almost
completely given by the $gg$ initial state, which is dominant due to
the larger value of the gluon PDF. It is worth to note also that HBR
contributions lead to a partial cancellation of the Sudakov logarithms
from NLO Weak corrections.

\begin{figure}[t]
\begin{center}
\includegraphics[width=0.48\textwidth]{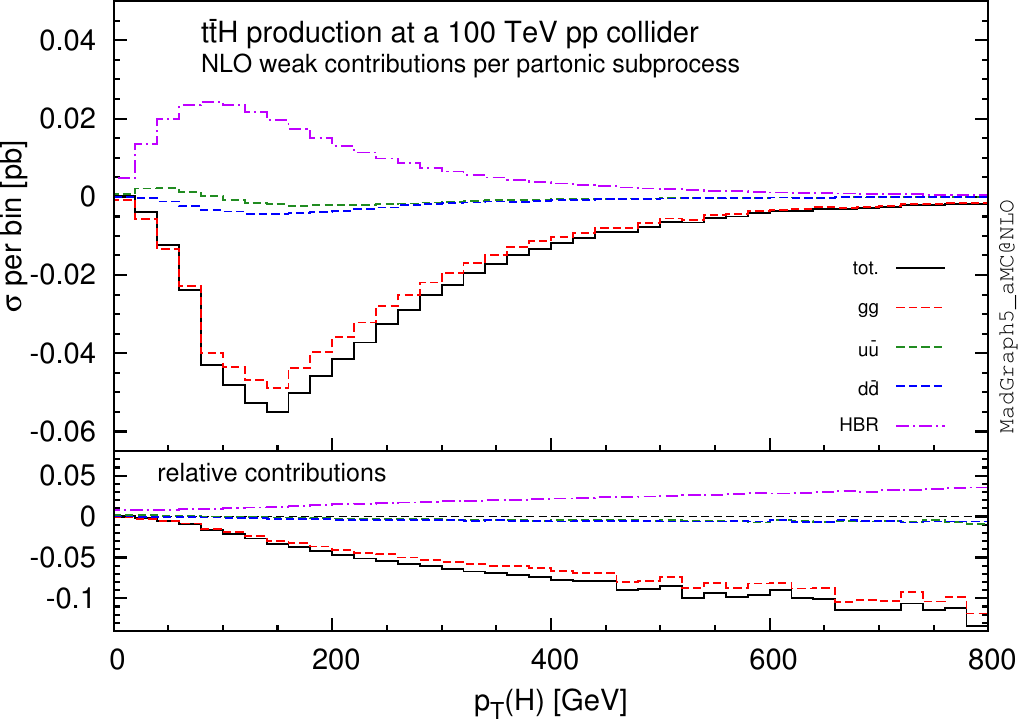}
\includegraphics[width=0.48\textwidth]{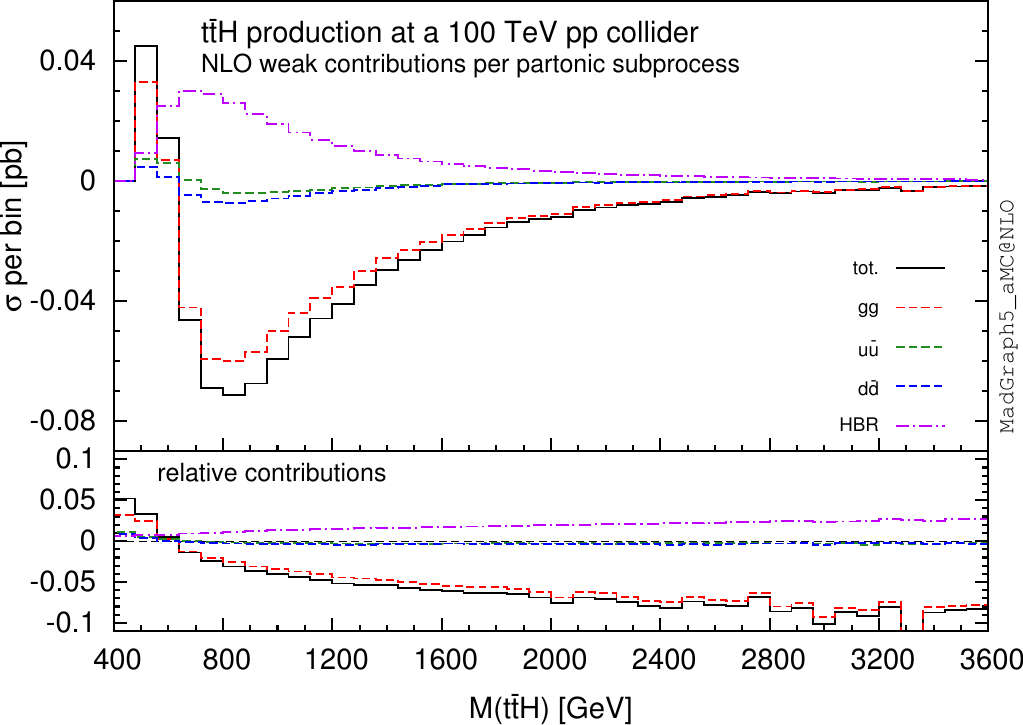}
\end{center}
\caption{NLO Weak corrections for the distribution of the $p_T(H)$
  (left) and of the $m(t \bar t H)$ (right). Plots are taken from
  \cite{Frixione:2014qaa} and explained in the text. }
\label{fig:ttbarHweak}
\end{figure}


\subsection{Real radiation}\label{sec:realradiation}
As discussed in Section~\ref{sec:EWintro}, the electroweak corrections
grow with the center of mass energy $Q = \sqrt{s}$ of the partonic
collision, due to the appearance of two powers of $L_V$ for each order
in perturbation theory. Thus, as the center of mass energy grows, the
convergence of electroweak perturbative theory gets worse, until it
breaks down completely for
\begin{align}
  \alpha_{\rm ew} L_V^2 \sim 1 \,.\end{align} As one can see from the
results in this chapter, perturbative electroweak corrections at the
100 TeV FCC become very large at high center of mass energies. A
consistent resummation of these Sudakov logarithms improves the
convergences of perturbation theory significantly, and becomes
crucially important for measurements at the highest energies
available.

As already mentioned in the introduction, Sudakov logarithms in
exclusive cross-sections (without extra radiation of additional
electroweak gauge bosons) have been resummed for many processes. A
general formalism based on soft-collinear effective
theory~\cite{Bauer:2001ct,Bauer:2000ew,Bauer:2000yr,Bauer:2001yt} was
developed to perform this resummation at NLL (and in principle to
higher accuracy as well).  However, for the energies that can be
reached at the FCC, it is only the leading logarithms that need to be
resummed; the subleading logarithms can still be treated in fixed
order perturbation theory until the partonic center of mass energy
becomes large enough such that $\alpha_{\rm ew} L_V \sim 1$.

To LL accuracy, the resummation of the double logarithms can be
obtained using the coherent branching
formalism~\cite{Marchesini:1983bm,Sjostrand:2003wg,Bahr:2008pv} that
underlies parton shower algorithms, used extensively to describe the
emissions of extra particles in the strong interaction. In this
approach~\cite{Fadin:1999bq,Bauer:2016kkv} one uses the fact that, to LL
accuracy, the cross-sections factorize into products of emission
probabilities. These emission probabilities are given by the
Altarelli-Parisi splitting functions, supplemented by a so-called
no-branching probability, given by an integral over the
Altarelli-Parisi splitting functions.

This approach reproduces the known resummation of the exclusive
results, but can also be used to obtain resummed results for the real
radiation of $W$ and $Z$ bosons. In the remainder of this section, we
will provide the results for the production of lepton pairs at the
FCC, but the method can be applied to any other process as well. The
results given below are taken from~\cite{Bauer:2016kkv}, where details
on their derivation can be found.

The partonic cross-sections for the exclusive cross section $q_1 \bar
q_2 \to \ell_1 \ell_2$ at partonic center of mass energy $s = Q^2 =
x_a x_b S$, where $\ell_i$ denotes either a charged lepton or a
neutrino, are given by
\begin{align}
\label{eq:virtualAsShower}
\hat \sigma^{\rm LL}_{q_1^H q_2^H \to \ell_1^H \ell_2^H}(s) &= \hat
\sigma^{B}_{q_1^H q_2^H \to \ell_1^H \ell_2^H}
\Delta_{q_1^Hq_2^H\ell_1^H\ell_2^H}(m_V^2, s; s) \, \Delta^{\rm
  em}_{q_1^Hq_2^H\ell_1^H\ell_2^H}(\Lambda^2, m_V^2; s) \,.\end{align}
where the superscript $H$ denotes the helicity of the fermions.  The
Born cross-sections are given by
\begin{align}
\hat \sigma^{B}_{q^{H_q} q^{H_q} \to \ell^{H_\ell} \ell^{H_\ell}} &= \frac{\pi}{8 \,N_C \,s} \,  \frac{8\left(\alpha_2 \, T^3_{q^{H_q}} T^3_{\ell^{H_\ell}} + \alpha_1 \, Y_{q^{H_q}} Y_{\ell^{H_\ell}}\right)^2}{3}
\nonumber\\
\hat \sigma^{B}_{q_1^L q_2^L \to \ell_1^L \ell_2^L} &= \frac{\pi}{8 \,N_C \, s} \,  \frac{2\alpha_2^2 }{3}
\,,\end{align}
where $T^3_{f^H}$ denotes the weak isospin of the fermion $f = q /
\ell$ with helicity $H$, $Y_{f^H}$ its weak hypercharge with
normalization $Y_i = Q_i - T^3_i$, and $\alpha_2$ and $\alpha_1$ are
the couplings of the SU(2) and U(1)$_Y$ gauge group, respectively.

The Sudakov factor $\Delta_{q_1^Hq_2^H\ell_1^H\ell_2^H}(m_V^2, s; s)$
describes the evolution from $s$ to $m_V^2$ and factors into two
pieces, one for the SU(2) and one for the U(1) symmetry
\begin{align}
\Delta_{q_1^Hq_2^H\ell_1^H\ell_2^H}(m_V^2, s; s) &=\Delta_{q_1^Hq_2^H\ell_1^H\ell_2^H}^{SU(2)}(m_V^2, s; s) \, \Delta_{q_1^Hq_2^H\ell_1^H\ell_2^H}^{U(1)}(m_V^2, s; s)
\,.\end{align}
The SU(2) and U(1) contributions are given by 
\begin{align}
\Delta^{\rm SU(2)}_{q_1^Hq_2^H\ell_1^H\ell_2^H}(m_V^2, s; s) &= \exp\left[ - \frac{A^{\rm SU(2)}_{q_1^H q_1^H\ell_1^H\ell_2^H}}{2}\ln^2 \frac{m_V^2}{s}\right]
\nonumber\\
 \Delta_{q_1^Hq_2^H\ell_1^H\ell_2^H}^{U(1)}(m_V^2, s; s)) &=  \exp\left[ - \frac{A_{q_1^Hq_2^H\ell_1^H\ell_2^H}^{U(1)}}{2}\ln^2 \frac{m_V^2}{s} \right]
\,,\end{align}
with
\begin{align}
\label{eq:AfHdefU1}
A^{\rm SU(2)}_{q_1^H q_1^H\ell_1^H\ell_2^H} = \frac{\alpha_2}{2\pi} \sum_iT_{i}^2
\,,\qquad
A^{\rm U(1)}_{q_1^H q_1^H\ell_1^H\ell_2^H} = \frac{\alpha_1}{2\pi} \sum_iY_{i}^2
\,,\end{align}
and the sum is running over all fermions $i \in \{q_1^H, q_1^H,
\ell_1^H, \ell_2^H \}$. For scales below $m_V$ only photons can be
exchanged, and the resulting Sudakov factor
\begin{align}
\Delta^{\rm em}_{q_1^Hq_2^H\ell_1^H\ell_2^H}(\Lambda^2, m_V^2; s) =  \exp\left[ - \frac{\alpha(Q) Q_{\rm tot}^2}{4\pi}\left(\ln^2\frac{\Lambda^2}{s} - \ln^2\frac{m_V^2}{s} \right)\right]
\,,\end{align}
only depends on the electromagnetic charges of the fermions 
\begin{align}
Q_{\rm tot}^2 = \sum_i Q_{i}^2
\,.\end{align}
This agrees with the results of~\cite{Fadin:1999bq}, and summing over
all possible helicity structures, one reproduce the resummed results
of \cite{Chiu:2007yn,Chiu:2007dg,Chiu:2008vv}.

Using the same coherent branching formalism, one can calculate the
partonic real radiation cross section $q_i q_j \to \ell \ell V$, where, 
as before, $\ell$ denotes either a charged lepton or a neutrino, $V$
denotes either a $Z$ or a $W$ boson. Following the results
of~\cite{Bauer:2016kkv} one obtains for the exclusive emission of a
$W^\pm$ boson (exclusive here means that emissions of additional gauge
bosons are vetoed)
\begin{align}
&\hat \sigma^{\rm LL}_{q_1^H q_2^H \to \ell_1^H \ell_2^H + W^\pm} 
\nonumber\\
&\quad
=  \left[ \Delta_{q_1^Hq_2^H\ell_1^H\ell_2^H}(m_V^2, s; s) \, \Delta^{\rm em}_{q_1^Hq_2^H\ell_1^H\ell_2^H W^\pm}(\Lambda^2, m_V^2; s) \int_{m_V^2}^{s} \!  \frac{{\rm d} k_T^2}{k_T^2} \ln \frac{s}{k_T^2} \Delta_{V}(m_V^2, k_T^2; k_T^2) \right]
\nonumber\\
&\quad\quad \times\left( \hat \sigma^{B}_{{q'}_1^H q_2^H \to \ell_1^H \ell_2^H} A^{W^\pm}_{q_1^H} + \hat \sigma^{B}_{q_1^H {q'}_2^H \to \ell_1^H \ell_2^H} A^{W^\pm}_{q_2^H} + \hat \sigma^{B}_{q_1^H q_2^H \to {\ell'}_1^H \ell_2^H} A^{W^\pm}_{\ell_1^H}+ \hat \sigma^{B}_{q_1^H q_2^H \to \ell_1^H {\ell'}_2^H} A^{W^\pm}_{\ell_2^H} \right) 
 \nonumber\\
&\quad
=  \left[ \Delta_{q_1^Hq_2^H\ell_1^H\ell_2^H}(m_V^2, s; s) \, \Delta^{\rm em}_{q_1^Hq_2^H\ell_1^H\ell_2^H W^\pm}(\Lambda^2, m_V^2; s) \, I_1 (m_V^2, s) \right]
\nonumber\\
&\quad\quad \times\left( \hat \sigma^{B}_{{q'}_1^H q_2^H \to \ell_1^H \ell_2^H} A^{W^\pm}_{q_1^H} + \hat \sigma^{B}_{q_1^H {q'}_2^H \to \ell_1^H \ell_2^H} A^{W^\pm}_{q_2^H} + \hat \sigma^{B}_{q_1^H q_2^H \to {\ell'}_1^H \ell_2^H} A^{W^\pm}_{\ell_1^H}+ \hat \sigma^{B}_{q_1^H q_2^H \to \ell_1^H {\ell'}_2^H} A^{W^\pm}_{\ell_2^H} \right) 
\,,
 \nonumber\\
 \label{eq:RealWFinal}
 \end{align}
 where $f'$ is the fermion $f$ becomes after having radiated a $W^\pm$
 that is $u'=d$, $d'=u$, $l'=\nu$ and $\nu'=l$ and for any flavor set
 which allows a $W^\pm$ emission there is one of the Born cross
 sections which is zero because its electromagnetic charge is not
 conserved. The Sudakov factor
\begin{align}
\Delta_V(m_V^2, k_T^2; k_T^2) = \exp\left[ - \frac{\alpha_2 \, C_A }{4\pi}\ln^2 \frac{m_V^2}{k_T^2} \right] 
\end{align}
is written in term of the Casimir $C_A = 2$ for the SU(2) gauge group
and describes the no-branching probability of the extra $W$ boson
radiated.  We have also defined $\Lambda$ to be the scale below which
a photon becomes unresolved, as well as
\begin{align}
A^{W^{\pm}}_{f^L} = \frac{\alpha_2(Q)}{4\pi} 
\,,\qquad
A^{W^{\pm}}_{f^R} = 0 
\,.
\end{align}
The integral over $k_T^2$ has been performed using the general result
\begin{align}
\label{eq:Idef}
I_\beta(m_V^2, s) &\equiv \int_{m_V^2}^{s} \! \frac{{\rm d} k_T^2}{k_T^2} \ln \frac{s}{k_T^2} \,\left[\Delta_{V}(m_V^2, k_T^2; k_T^2) \right]^\beta
\\
&= \frac{2\pi}{\alpha_2 \, \beta \, C_A} \left[ \frac{\sqrt{\alpha_2 \, \beta \, C_A}}{2} \ln\frac{m_V^2}{s} \, {\rm Erf}\left( \sqrt{\frac{\alpha_2 \, \beta \, C_A}{4\pi}} \, \ln\frac{m_V^2}{s}\right)   + \Delta_V^\beta(m_V^2, s; s) - 1 
\right] 
\,.
\nonumber
\end{align}

For the emissions of a $Z$ bosons and photon, one needs to take into
account the mixing between the third component of SU(2) gauge symmetry
and the U(1) gauge symmetry. After a few lines of algebra
(see~\cite{Bauer:2016kkv}) one finds for the emission of a $Z$ boson
\begin{align}
\hat \sigma^{\rm LL}_{q_1^H q_2^H \to \ell_1^H \ell_2^H + Z} 
&=  \hat \sigma^{B}_{q_1^H q_2^H \to \ell_1^H \ell_2^H} \, \Delta_{q_1^Hq_2^H\ell_1^H\ell_2^H}(m_V^2, s; s)  \, \Delta^{\rm em}_{q_1^Hq_2^H\ell_1^H\ell_2^H}(\Lambda^2, m_V^2; s)
\nonumber\\
& \qquad
\times \int_{m_V^2}^{s} \! \frac{{\rm d} k_T^2}{k_T^2} \ln \frac{s}{k_T^2}
\Bigg{(}
s_W^2 A^{\rm U(1)}_{q_1^Hq_2^H\ell_1^H\ell_2^H} - A^{\text{mixing}}_{q_1^Hq_2^H\ell_1^H\ell_2^H} \sqrt{\Delta_{W}(m_V^2, k_T^2; k_T^2)}
 \nonumber\\
& \qquad\qquad\qquad
 +c_W^2 A^{\rm SU(2)}_{q_1^Hq_2^H\ell_1^H\ell_2^H} \Delta_{W}(m_V^2, k_T^2; k_T^2) \Bigg{)}  \nonumber\\
&= \hat \sigma^{B}_{q_1^H q_2^H \to \ell_1^H \ell_2^H} \, \Delta_{q_1^Hq_2^H\ell_1^H\ell_2^H}(m_V^2, s; s)  \, \Delta^{\rm em}_{q_1^Hq_2^H\ell_1^H\ell_2^H}(\Lambda^2, m_V^2; s)\, \nonumber\\ & \qquad\qquad
\Bigg{(} s_W^2 A^{\rm U(1)}_{q_1^Hq_2^H\ell_1^H\ell_2^H} \, \frac{1}{2} \ln^2 \frac{m_V^2}{s}
  - A^{\text{mixing}}_{q_1^Hq_2^H\ell_1^H\ell_2^H}
\, I_{\frac{1}{2}}(m_V^2,s)
\nonumber\\
& \qquad \qquad \qquad 
+
c_W^2 \, A^{\rm SU(2)}_{q_1^Hq_2^H\ell_1^H\ell_2^H} \, I_{1}(m_V^2,s)
\Bigg{)} \,.
 \label{eq:RealZFinal}
\end{align}
The factors $A^{\rm SU(2)}$ and $A^{\rm U(1)}$ are given in
Eq.~(\ref{eq:AfHdefU1}), and the term arising from the mixing of the
$W^3$ and $B$ is given by
\begin{align}
A^{\text{mixing}}_{q_1^Hq_2^H\ell_1^H\ell_2^H} = \frac{\alpha_{\rm em}}{\pi} \sum_{i}  T^3_{i} Y_{i}
\,.
\end{align}

The emission of a photon is obtained in a similar manner, but one has
to include the extra emissions that can happen for scales $\Lambda <
\mu < m_V$. This gives
\begin{align}
\hat \sigma^{\rm LL}_{q_1^H q_2^H \to \ell_1^H \ell_2^H + \gamma} 
&= \hat \sigma^{B}_{q_1^H q_2^H \to \ell_1^H \ell_2^H} \, \Delta_{q_1^Hq_2^H\ell_1^H\ell_2^H}(m_V^2, s; s)
\nonumber\\
& \qquad \times \Bigg{[}
\Delta^{\rm em}_{q_1^Hq_2^H\ell_1^H\ell_2^H}(\Lambda^2, m_V^2; s)
\bigg{(}
c_W^2 A^{\rm U(1)}_{q_1^Hq_2^H\ell_1^H\ell_2^H} \frac{1}{2}\log^2(\frac{m_V^2}{s})
\nonumber\\
& \qquad  \qquad + A^{\text{mixing}}_{q_1^Hq_2^H\ell_1^H\ell_2^H}
I_{\frac{1}{2}}(m_V^2,s) +
s_W^2 A^{\rm SU(2)}_{q_1^Hq_2^H\ell_1^H\ell_2^H} 
I_{1}(m_V^2,s)
\bigg{)} +
\nonumber\\
& \qquad \qquad \int_{\Lambda^2}^{m_V^2} \! {\rm d} k_T^2 \! \frac{{\rm d}}{{\rm d} k_T^2} \left[\Delta^{\rm em}_{q_1^Hq_2^H\ell_1^H\ell_2^H}(k_T^2, m_V^2; s) \right] \Delta^{\rm em}_{q_1^Hq_2^H\ell_1^H\ell_2^H}(\Lambda^2, k_T^2; s)
\Bigg{]} 
\nonumber\\
&= \hat \sigma^{B}_{q_1^H q_2^H \to \ell_1^H \ell_2^H} \, \Delta_{q_1^Hq_2^H\ell_1^H\ell_2^H}(m_V^2, s; s) \Delta^{\rm em}_{q_1^Hq_2^H\ell_1^H\ell_2^H}(\Lambda^2, m_V^2; s)
\nonumber\\
& \qquad \times \Bigg{[}
c_W^2 A^{\rm U(1)}_{q_1^Hq_2^H\ell_1^H\ell_2^H} \frac{1}{2}\log^2(\frac{m_V^2}{s})+ A^{\text{mixing}}_{q_1^Hq_2^H\ell_1^H\ell_2^H}
I_{\frac{1}{2}}(m_V^2,s) +
\nonumber\\
& \qquad  \qquad  s_W^2 A^{\rm SU(2)}_{q_1^Hq_2^H\ell_1^H\ell_2^H} 
I_{1}(m_V^2,s) + \frac{\alpha Q^2_{\rm tot}}{4 \pi} \left( \ln^2 \frac{\Lambda^2}{s} - \ln^2 \frac{m_V^2}{s} \right)
\Bigg{]} 
\,.
\label{eq:RealPhotonFinal}
\end{align}

In Fig.~\ref{fig:Inclusive100} we show the resulting cross-sections as
function of the center of mass energy of the system [$s = (p_{\ell_1}
+ p_{\ell_2} + p_V)^2$]. In the top plot we show the size of the
perturbative corrections relative to the Born cross-section, where the
virtual corrections are in black, while the real radiation of a $Z$,
$\gamma$, $W^+$ and $W^-$ are shown in green, orange, red and blue,
respectively. The fixed order results (only including the double
logarithmic term) are shown by the dashed lines, while the resummed
results by the solid lines. One can clearly see that the size of the
perturbative corrections grows as the center of mass energy is raised,
and that the resummation decreases the overall size.  To illustrate
the importance of resummation, we show in the lower plot the
difference between the fixed order and resummed result, normalized to
the resummed. One can clearly see that the effect of resummation is
very important for large center of mass energies, and that the
resummation of the real corrections is even more important than the
resummation of the virtual corrections.

The results of this section have shown that resummation of electroweak
Sudakov logarithms becomes crucial for center of mass energies in the
multi-TeV range, which are easily reached at the FCC. This is true not
only for exclusive cross-sections where extra radiation is vetoed, but
also for the radiation of additional massive gauge bosons. Including
these effects in theoretical calculations will be crucial, not only
for precision studies.

\begin{figure}[h!]
\centering
\includegraphics[width=0.9\textwidth]{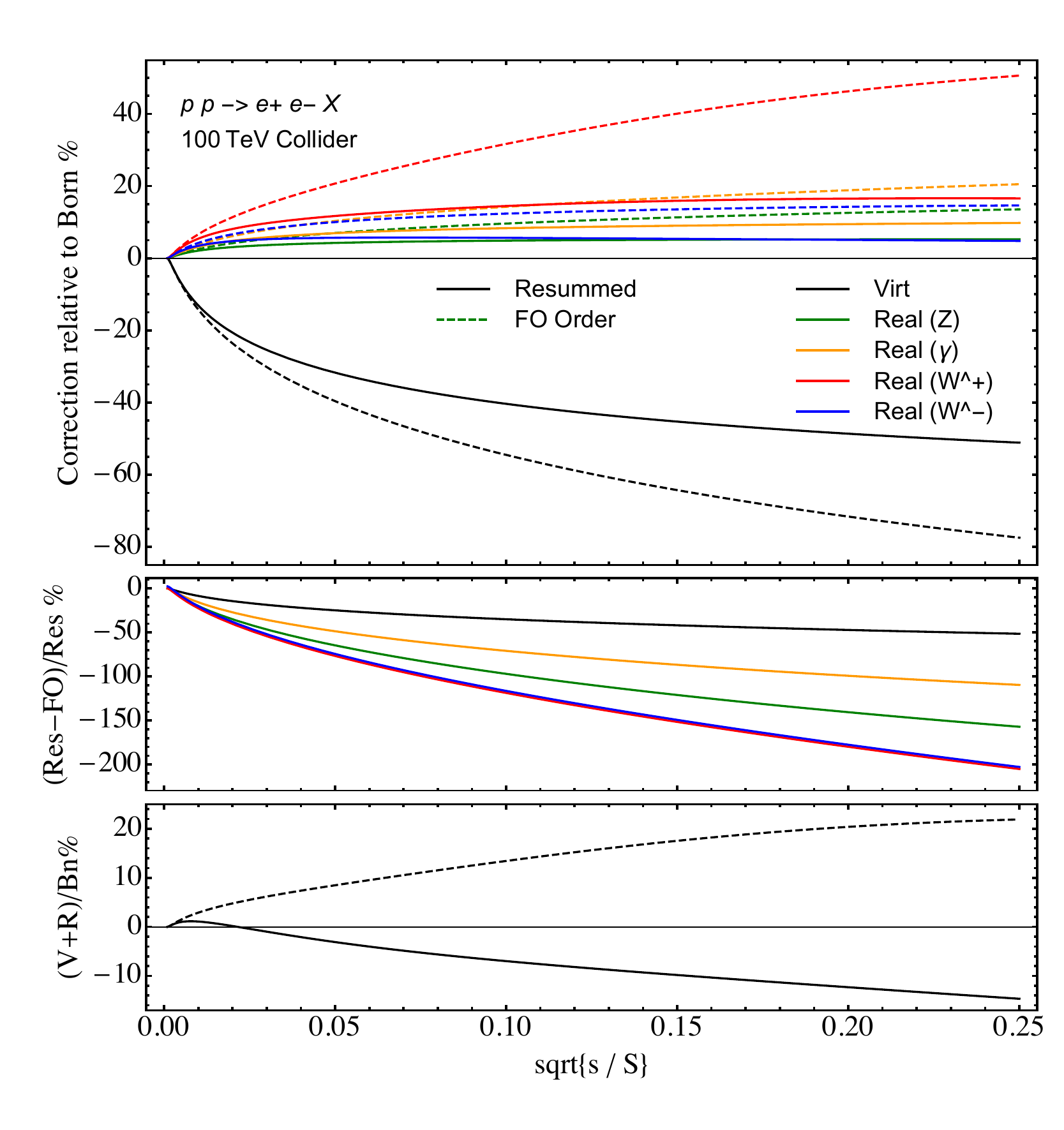}
\caption{The cross-section for $pp \to e^+ e^-+X$. Virtual corrections
  are shown in black, while real corrections are shown in green,
  orange, red and blue for $Z$, $\gamma$, $W^+$ and $W^-$ emissions,
  . Resummed corrections are shown in solid lines, while fixed order
  results are dashed.}
\label{fig:Inclusive100}
\end{figure}

\clearpage
\section{Sources of missing transverse energy}
\label{sec:met}
\begin{figure}[h!]
\centering
\includegraphics[width=0.49\textwidth]{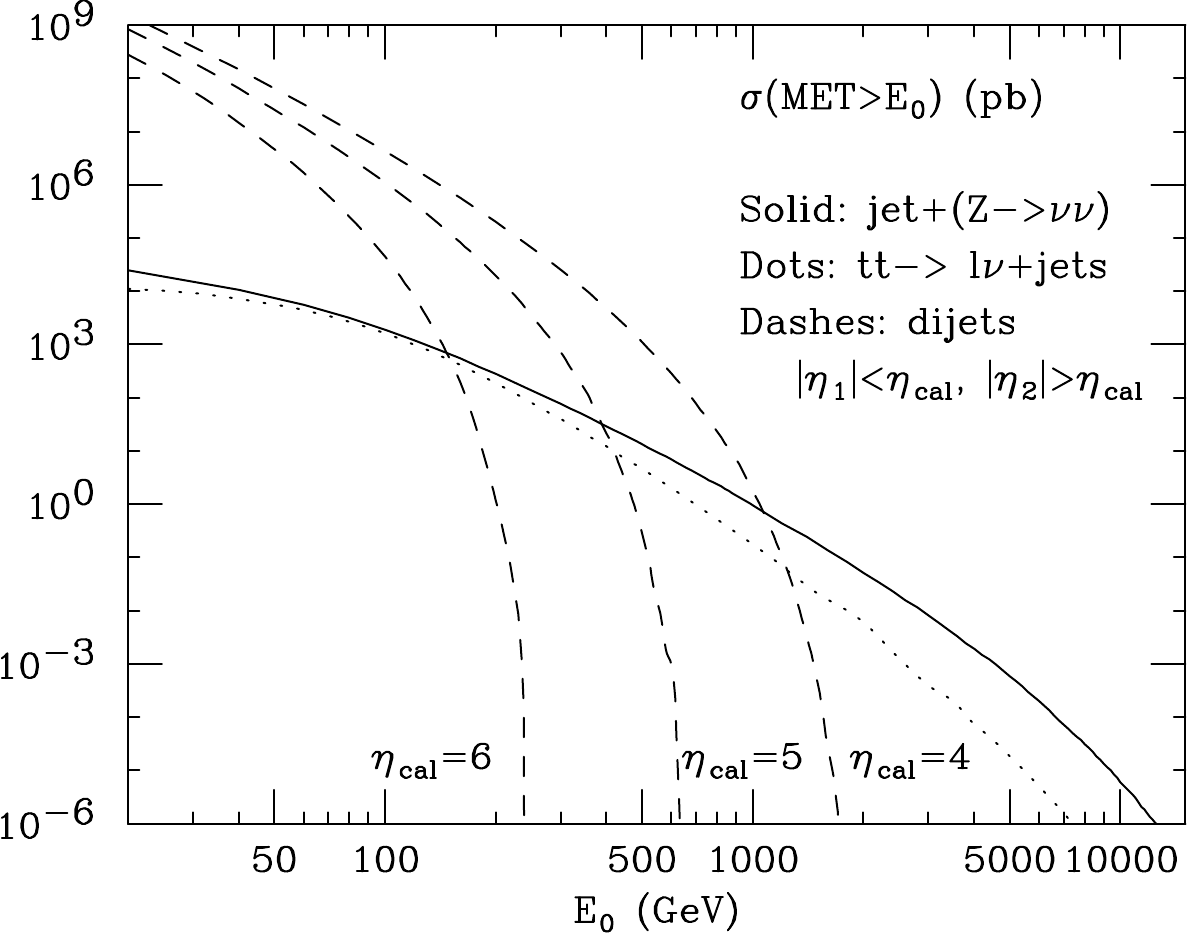}
\includegraphics[width=0.49\textwidth]{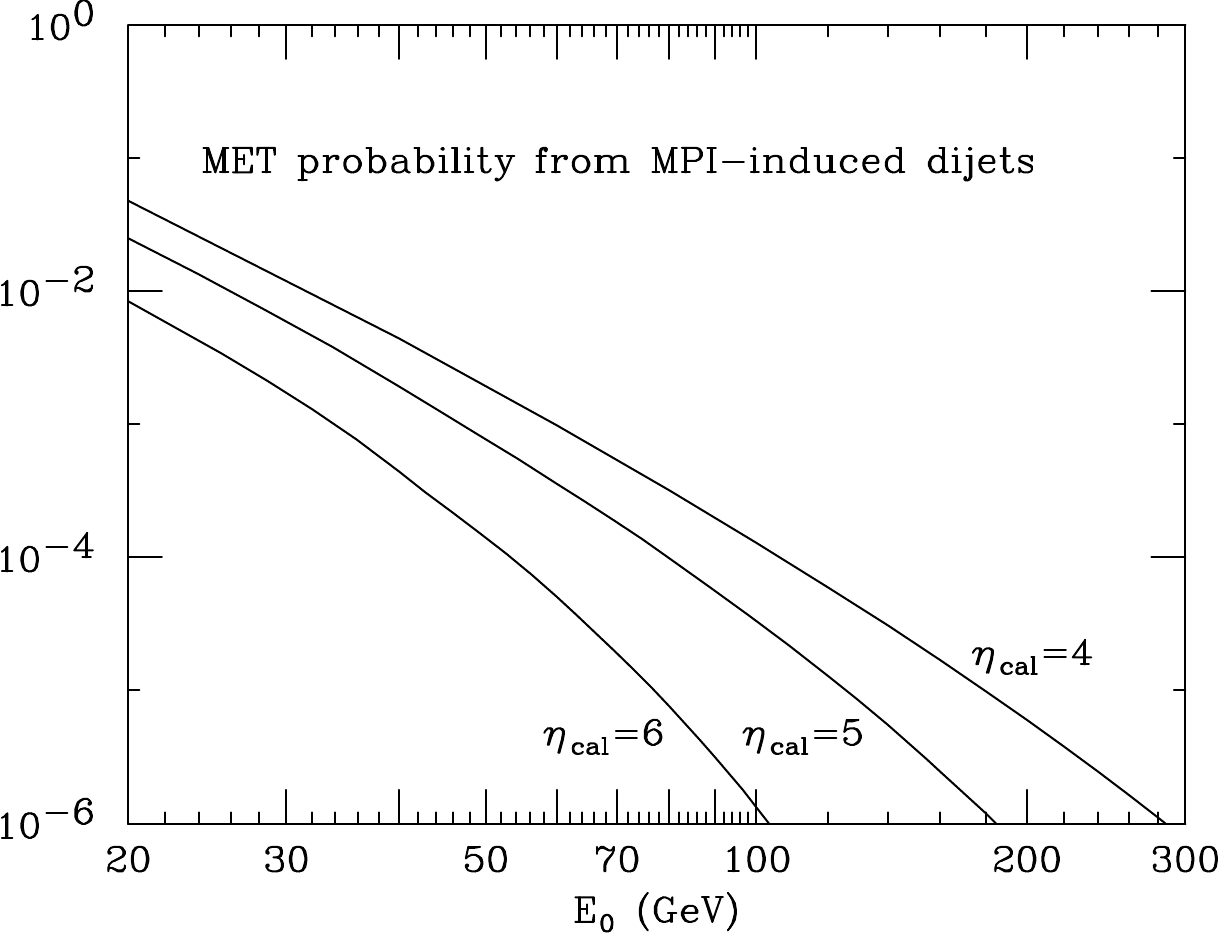}
\caption{Left: Missing transverse energy rates, from jet+$(Z\to
  \nu\bar{\nu})$ events and from dijets, with a jet escaping
  undetected at large rapidity. Right: Missing transverse energy
  probability induced by multiple-parton interactions, for different
  values of the jet rapidity acceptance. }
\label{fig:met}
\end{figure}
Missing transverse energy ($\met$) is an important signature for many
BSM processes. At 100~TeV, SM sources of $\met$ can contribute with
very large rates of irreducible backgrounds. We consider here, for
illustration, the effect of three of the leading sources of
irreducible $\met$: the associated production of jets and a $Z^0$
boson decaying to neutrinos, the semileptonic decay of top quarks, and
the production of jets outside the calorimeter acceptance. The latter
channel is important, since the high energy available in the CM allows
for the production of large $p_T$ jets at very forward
rapidities. This is shown in Fig.~\ref{fig:met}, where the dashed
lines correspond to the rate of dijet events in which one jet is
within the calorimeter acceptance (defined by the $\eta_{cal}$ label),
and the other is outside. With the standard LHC calorimeter coverage,
$\eta_{cal}=5$, dijets would give a $\met$ signal larger than $Z$+jets
for $\met$ up to $\sim 400$~GeV. This is reduced to $\sim 150$~GeV
with a calorimeter extending out to $\eta_{cal}=6$.

It must be noticed that the limited calorimeter acceptance can induce
a $\met$ signal in any hard process, due to the finite probability of
the coincidence of a multiparton interaction. Multiparton interactions
are hard scatterings taking place among the partons not engaged in the
primary hard process, and cannot be separated experimentally since the
resulting particles emerge from exactly the same vertex as the primary
scattering. The probability that a multiparton interaction leads to a
secondary hard process $X$ in addition to the primary one is
parametrized as $\sigma(X)/\sigma_0$, where $\sigma_0$ is a
process-independent parameter. The right plot of Fig.~\ref{fig:met}
shows the probability of multiparton interactions leading to dijet
final states, with one jet inside the calorimeter and the other
outside. For this example we chose $\sigma_0=30$~mb, a number
consistent with the direct experimental determinations from Tevatron
and LHC data. $\met$ signals in the range of 30-70~GeV are induced
with probability of about $10^{-3}$ if $\eta_{cal}$ is in the range 4
to 6, stressing once again the need to instrument the detectors with a
calorimetric coverage more extended than at the LHC.

\clearpage
\subsubsection*{Acknowledgements}

C.W.~B.~is supported by the Office of High Energy Physics 
of the U.S. Department of Energy under contract DE-AC02-05CH11231.
The work of M.B., V.B. and J.R. is
supported by an European Research Council
Starting Grant {\it ``PDF4BSM''}.
J.R. is also supported by  STFC Rutherford Fellowship
and Grant ST/K005227/1 and ST/M003787/1.
The work of D.P. and M.L.M. is supported by the European Research
Council advanced grant 291377 {\it ``LHCtheory'': Theoretical predictions and
analyses of LHC physics: advancing the precision frontier}. 
The work of S.~W. has been partially supported
by the U.S. National Science Foundation under Grant PHY-1212635.
The research of WA, WB, and GZ is supported by the ERC consolidator
grant 614577 ``HICCUP -- High Impact Cross Section Calculations for
Ultimate Precision''
The work of M.~Z. is supported by the European Union's Horizon 2020
research and innovation programme under the Marie Sklodovska-Curie
grant agreement N. 660171.
M.~C. and M.~Z. are supported in part by the ILP LABEX
(ANR-10-LABX-63), in turn supported by French state funds managed by
the ANR within the ``Investissements d'Avenir'' programme under
reference ANR-11-IDEX-0004-02.
H.I.'s work is supported by a Marie Sklodowska-Curie Action
Career-Integration Grant PCIG12-GA-2012-334228 of the European Union.  
The work of F.F.C. is supported by the Alexander von Humboldt Foundation, in the
framework of the Sofja Kovalevskaja Award 2014, endowed by the German Federal
Ministry of Education and Research.
The work of F.Piccinini has been partially supported by CERN TH-Unit. 
F.Petriello is supported by the U.S. DOE grants DE-FG02-91ER40684 and
DE-AC02-06CH11357.
R.B. is supported by the DOE contract DE-AC02-06CH11357
The work of R.B. and F.Petriello used resources of the Argonne Leadership
Computing Facility, which is a DOE Office of Science User Facility
supported under Contract DE-AC02-06CH11357.
P.~S.~is the recipient of an Australian Research Council Future
Fellowship, FT130100744. This work was also supported in part by the
ARC Centre of Excellence for Particle Physics at the Terascale.
The work of M.S. is supported by the National Fund for Scientific Research
(F.R.S.-FNRS Belgium), by the IISN ``MadGraph'' convention 4.4511.10,
by the IISN ``Fundamental interactions'' convention 4.4517.08, and in
part by the Belgian Federal Science Policy Office through the
Interuniversity Attraction Pole P7/37.
This research used computing resources from the Rechenzentrum Garching.

\bibliographystyle{report}
\bibliography{report}

\end{document}